\documentclass[twocolumn]{aastex63}
\usepackage{amsmath}
\usepackage{graphicx}
\usepackage{tabularx}
\usepackage{booktabs}

\usepackage{placeins}

\shorttitle{Constraining Circum-burst Environments of GRBs}
\shortauthors{Xin et al.}

\begin{document}

\title{Constraining Circum-burst Environments of GRBs with Jet Break Features in X-ray Afterglows}

\author{Si-Ji Xin}
\affiliation{School of Physics and Physical Engineering, Qufu Normal University, Qufu 273165, China; yisx2015@qfnu.edu.cn}
\author{Yu-Qi Zhou}
\affiliation{School of Physics and Physical Engineering, Qufu Normal University, Qufu 273165, China; yisx2015@qfnu.edu.cn}
\author{Sheng-Jin Sun}
\affiliation{School of Physics and Physical Engineering, Qufu Normal University, Qufu 273165, China; yisx2015@qfnu.edu.cn}
\author{Cheng-Jie Sun}
\affiliation{School of Physics and Physical Engineering, Qufu Normal University, Qufu 273165, China; yisx2015@qfnu.edu.cn}
\author{Shuang-Xi Yi}
\affiliation{School of Physics and Physical Engineering, Qufu Normal University, Qufu 273165, China; yisx2015@qfnu.edu.cn}
\author{Yuan-Chuan Zou}
\affiliation{Department of Astronomy, School of Physics, Huazhong University of Science and Technology, Wuhan 430074, China; zouyc@hust.edu.cn}
\author{Yu-Peng Yang}
\affiliation{School of Physics and Physical Engineering, Qufu Normal University, Qufu 273165, China; yisx2015@qfnu.edu.cn}
\author{Yan-Kun Qu}
\affiliation{School of Physics and Physical Engineering, Qufu Normal University, Qufu 273165, China; yisx2015@qfnu.edu.cn}
\author{Fa-Yin Wang}
\affiliation{School of Astronomy and Space Science, Nanjing University, Nanjing 210023, China}

\begin{abstract}
The nature of the circum-burst medium serves as a key diagnostic for probing the progenitor systems and the physics of relativistic jet propagation in gamma-ray bursts (GRBs). In this work, we systematically infer the density profile index $k$ (where $n \propto r^{-k}$) from the change in the temporal decay index at the jet break ($\Delta\alpha$). Within the framework of the uniform jet model, the two quantities are linked by the relation $\Delta\alpha = (3 - k)/(4 - k)$. We apply this diagnostic to a substantial and uniformly selected sample of 170 GRBs with clear jet breaks, identified from over 1,400 Swift/XRT X-ray afterglows observed from 2004 to 2024. By fitting the light curves with a broken power-law model, we obtain $\Delta\alpha$ for each burst and subsequently derive the corresponding $k$ value. We then use the derived $k$ values to classify the circum-burst environment of each GRB. Our results reveal a near-even split: 82 bursts ($\sim48\%$) are consistent with a constant-density interstellar medium (ISM, $k \approx 0$), while 88 bursts ($\sim52\%$) favor a wind environment ($k \approx 2$). For the 35 bursts with optical data, our X-ray-based classifications are generally consistent with independent multi-band analyses. Additionally, we derive jet opening angles and true beaming-corrected energies for bursts with known redshifts.
\end{abstract}
\section{Introduction}
Gamma-ray bursts (GRBs) are brief and intense flashes of gamma-rays originating from cosmological distances, marking some of the most energetic explosive events in the Universe. A GRB event typically comprises two distinct phases: an initial prompt emission phase followed by a long-lasting afterglow phase \citep{1997Natur.387..783C, 1997ApJ...476..232M, 1997ApJ...489L..37S}. The widely adopted theoretical framework for explaining these phenomena is the fireball model \citep{2004RvMP...76.1143P, 2006RPPh...69.2259M, 2007ChJAA...7....1Z, 2015PhR...561....1K}. Within this model, the prompt emission is widely attributed to radiation from internal shocks, which are produced when shells of material with different velocities within the relativistic jet collide \citep{1994ApJ...430L..93R, 1996AIPC..384..772M}, while the afterglow emission arises from external shocks, generated by the interaction of the jet with the circum-burst medium \citep{1993ApJ...405..278M, 1997ApJ...476..232M, 1998ApJ...497L..17S, 2005MNRAS.363...93Z, 2013NewAR..57..141G}.

The production of GRB afterglows is intimately connected to the dynamical evolution of the fireball. As an ultra-relativistic outflow ($\Gamma\gtrsim100$) propagates into the circum-burst medium, its deceleration drives a forward shock. This shock accelerates electrons, which then produce broad-band afterglow emission via synchrotron radiation \citep{1999PhR...314..575P}. In the early theoretical studies of GRB afterglows, the shock propagation was often simplified using a spherical symmetry assumption. However, multi-wavelength observations have consistently revealed a key feature in many GRBs: a distinct break in the late-time afterglow light curve (typically at $10^4$ to $10^6$ s), after which the temporal decay index steepens to $\alpha\sim2$ or even larger. This steepening cannot be explained by a spherical shock model but provides strong evidence for jet collimation, indicating that the GRB energy is channeled into a narrow conical jet \citep{1999ApJ...525..737R,1999ApJ...523L.121H,2001ApJ...562L..55F,1999ApJ...519L..17S,2002bjgr.conf..146L}. Understanding the jet structure is therefore crucial for GRB physics. It is essential not only for estimating the true energy budget of GRBs and thus constraining the central engine, but also for probing the properties of the progenitor, the shock physics, and the jet-environment interaction. Various jet models have been proposed to explain the observations, including the uniform jet model \citep{1998ApJ...499..301M,2004MNRAS.354...86R,2013ApJ...776L...9D,2018ApJ...856L..18M}; the Gaussian jet model \citep{2002ApJ...571..876Z,2002MNRAS.332..945R,2004ApJ...601L.119Z,2020MNRAS.499.3158C,2020ApJ...891..124H}; and structured jet models \citep{2021MNRAS.500.3511G,2021AAS...23832704T}.

Among various jet models, the uniform jet model has become a foundational framework for interpreting jet breaks, owing to its simplicity and clear physical predictions. Also known as the `top-hat' model, it assumes a conical jet with a sharp edge defined by a half-opening angle $\theta_{j}$. Within this cone, the energy per solid angle and the initial Lorentz factor are uniformly distributed and drop sharply beyond $\theta_{j}$. The jet break in the afterglow light curve within this model arises from the combination of two primary effects:

\begin{itemize}
\item[1)]
\textbf{Edge visibility effect.} Relativistic beaming initially confines the observed emission to a narrow cone along the line of sight. When the bulk Lorentz factor decreases to $\Gamma \sim \theta_{j}^{-1}$, the jet's edge enters the observer's field of view. If lateral spreading is negligible, the visible solid angle exceeds the angular area occupied by the jet material, leading to a deficit in received flux and a consequent steepening of the light curve \citep{1998ApJ...503..314P,1999MNRAS.306L..39M,1999ApJ...519L..17S,2007MNRAS.380..374P,2011MNRAS.410.2016V}.

\item[2)]
\textbf{Fast lateral spreading effect.} Once $\Gamma$ falls below $\theta_{j}^{-1}$, the jet center establishes causal contact with its edge. The jet can then, in principle, begin to expand laterally at a significant rate, leading to a sharp decline in the observed flux as the jet spreads out of the observer's line of sight, which produces a distinct break in the light curve \citep{1999ApJ...517L.105H,2000ApJ...543...90H,2001A&A...368..464G,2001ChJAA...1..433M,2004ChJAA...4..455W,2008ApJ...680..531K,2009ApJ...698...43R,2016ApJ...818...18G}.
\end{itemize}

However, for the uniform jet model, numerical studies indicate that lateral spreading remains very modest as long as the jet is highly relativistic \citep{2007RMxAC..27..140G}. Consequently, the jet break is predominantly caused by the edge visibility effect. Under this assumption of negligible lateral spreading, the change in the temporal decay index across the jet break is given by $\Delta\alpha = (3 - k)/(4 - k)$ \citep{2007RMxAC..27..140G, 2022A&A...658A..11M}. Here, the parameter $k$ characterizes the density profile of the circum-burst medium, assumed to follow $n \propto r^{-k}$. A constant-density interstellar medium (ISM) corresponds to $k = 0$ \citep{1998ApJ...497L..17S,2000ApJ...542..819K,2004MNRAS.353..511P}, yielding $\Delta\alpha = 0.75$, while a stellar wind environment corresponds to $k = 2$ \citep{1998MNRAS.298...87D,1998ApJ...499..301M,2000ApJ...543...66P,2004MNRAS.353..511P,2000ApJ...536..195C,2003MNRAS.342.1131W,2004ChJAA...4..455W,2003ApJ...597..455K,2005MNRAS.363...93Z}, yielding $\Delta\alpha = 0.5$. Thus, within this framework, the observed $\Delta\alpha$ provides a direct diagnostic for the density profile index $k$ and, consequently, offers constraints on the type of circum-burst environment.

While the $\Delta\alpha$-based diagnostic has been applied to interpret individual events \citep{2001ApJ...546..127J,2006ApJ...641L..13G,2010arXiv1012.5101G,2011MNRAS.410.2016V,2022A&A...658A..11M,2025ApJ...994L..17O}, a systematic, statistical application to a substantial and uniformly selected sample of GRBs exhibiting jet breaks has not yet been undertaken. This gap presents a practical challenge: without a prior constraint on the environment, the derivation of key physical parameters, such as the circum-burst density and the total energy budget, as well as the choice of applicable closure relations, depends on the assumed density profile (i.e., the value of $k$). Consequently, analyses of individual bursts must often consider multiple scenarios, complicating their interpretation. A systematic classification of circum-burst environments for a substantial sample would help to mitigate this ambiguity, offering clearer constraints for studies of progenitor systems, energy dissipation mechanisms, and related physics.

In this work, we utilize the rich sample of GRB X-ray afterglow observations provided by the Swift satellite up to the end of 2024. From this dataset, we systematically identify bursts that exhibit clear jet break signatures in their light curves. For each selected GRB, we model the jet break segment using a broken power-law function, with the model parameters fitted via a Markov Chain Monte Carlo (MCMC) method implemented in the \texttt{emcee} package \citep{2013PASP..125..306F}. This allows us to robustly measure the temporal decay indices before and after the break and thus derive $\Delta\alpha$ for each event. Based on the $\Delta\alpha$ values measured for individual bursts, we employ, for the first time, the relation $\Delta\alpha = (3 - k)/(4 - k)$ to statistically constrain the density profile index $k$ of the circum-burst medium for a large population of GRBs. We further present constraints on other derived physical parameters.

This paper is organized as follows. Section~\ref{sec:Data and Method} describes the sample selection and analysis methodology. In Section~\ref{sec:Distribution}, we present the distributions of key derived parameters, which include the jet break time $t_b$, the change in temporal decay index across the break ($\Delta\alpha$), and the density profile index $k$. For bursts with known redshifts, we also present the distributions of the derived jet half-opening angle $\theta_j$ and the true beaming-corrected energy $E_{\gamma}$. In Section~\ref{sec:discussion}, we discuss the origin of the light curve break and the methods for determining the circum-burst environment. A combined analysis of bursts with multi-band observations is also provided in this section. Finally, Section~\ref{sec:summary} provides a summary of our findings and their implications.
\section{SAMPLE SELECTION AND LIGHT CURVE FITTING}\label{sec:Data and Method}
\subsection{Sample Selection}
We used all X-ray afterglow data from the Swift X-ray Telescope (XRT) repository. Our sample spans from GRB 041223 to GRB 241228A, encompassing more than 1,400 GRBs. We first performed an initial screening of all GRB light curves. To fulfill the aim of using late time jet breaks to constrain circum-burst environments, we excluded bursts whose light curves lacked a clear break feature or had insufficient data points for reliable analysis.

For GRBs exhibiting potential break features, we applied the following selection criteria:
\begin{itemize}
    \item[1)] The X-ray afterglow light curve must show a distinct jet break feature, typically occurring at late times. This expectation arises because the jet break, attributed primarily to an edge effect, becomes pronounced only after the bulk Lorentz factor $\Gamma$ has decreased sufficiently, such that $\Gamma \sim \theta_{j}^{-1}$ \citep{1999ApJ...519L..17S}.
    \item[2)] The temporal decay index should be approximately $-1$ before the break and steepen to approximately $-2$ afterwards. This pattern corresponds to the transition from the standard afterglow decay phase to the post jet break phase \citep{1999ApJ...525..737R,1999ApJ...519L..17S,2001ApJ...562L..55F,2004ChJAA...4..455W,2008ApJ...675..528L,2009ApJ...698...43R,2018ApJ...859..160W, 2019pgrb.book.....Z}.
    \item[3)] If one or more flares are present in the light curve, we identified their temporal range using the XRT data and excluded these intervals from the fitting process to avoid contamination from flare activity.
\end{itemize}

During the sample selection, we first consulted the Swift/XRT light curve catalogue, which provides preliminary fits indicating the likely number of breaks and approximate decay slopes for the majority of GRBs. These reference fits guided our initial screening for bursts with clear jet break signatures. Furthermore, we also examined GRBs that exhibited a break-like feature in their late-time light curve, even if such a break was not formally identified in the catalogue. A light curve segment showing two distinct power-law behaviors in its final phase was considered a potential signature of a jet break. For these bursts, identification likewise began with an initial screening. Once a candidate was selected, we fitted the broadest possible temporal interval that encompassed the apparent break, ensuring alignment with the distinct power-law trends observed before and after it. This approach helped to average out short-timescale fluctuations, such as those from late-time re-brightening features.
\subsection{Light Curve Fitting}
For the GRBs selected through the above criteria, we performed detailed temporal fitting as described below. We employed a broken power-law (BPL) model to characterize the jet break segment in each X-ray afterglow light curve:
\begin{equation}\label{eq:BPL}
F(t) =
\begin{cases}
    F_b \times  \left( \dfrac{t}{t_b} \right)^{-\alpha_1}, & t < t_b \\
    F_b \times  \left( \dfrac{t}{t_b} \right)^{-\alpha_2}, & t \geq t_b \\
\end{cases}
\end{equation}
\noindent where $t_b$ is the break time, $F_b$ is the flux at the break time, $\alpha_1$ and $\alpha_2$ are the temporal decay indices before and after the break. For bursts with available optical data, we modeled the segments that exhibit a power-law decay. If an optical light curve (or a distinct segment of it) exhibits a single power-law (SPL) decay, we fit it using the model:
\begin{equation}\label{eq:SPL}
F = F_0 t^{-\alpha},
\end{equation}
\noindent where $\alpha$ is the temporal decay index. For the less common cases where the optical afterglow shows a clear break with sufficient data coverage, we applied the BPL model (Equation~\ref{eq:BPL}) in an analogous manner to the X-ray analysis.

For these multi-wavelength cases, our analysis primarily targets the final, common break feature. Two fitting approaches are possible for joint X-ray and optical data: (i) independent fitting of each band to compare the derived break times, and (ii) a constrained approach that assumes an achromatic jet break. Given our goal of combining multi-band information to constrain the circum-burst environment type, we adopted the second method, which explicitly enforces achromaticity. Specifically, we first determined the break time $t_b$ by fitting the more densely sampled X-ray data, then held $t_b$ fixed when fitting the optical data to obtain the corresponding decay indices for the same epoch.

For all fits, we evaluated the goodness of fit using the reduced $\chi^{2}$ statistic. The best-fitting parameters, including the break time $t_b$, the temporal decay indices $\alpha_1$ and $\alpha_2$, and the flux at the break $F_b$, were derived using the MCMC method implemented via the \texttt{emcee} package. All fitted light curves, including those for bursts with multi-band (X-ray and optical) data, are presented in Figures~\ref{fig:figure1} and \ref{fig:figure2}. The multi-band light curves are shown at the end of each figure. The complete set of fitting results, together with the duration ($T_{90}$) for each burst, the derived density profile index $k$, and our resulting classification of the circum-burst environment, are compiled in Table \ref{tab:Fitting Results}. We note that the calculation yields $k$ values less than zero for a small number of bursts. In our analysis, these cases are assigned to the ISM category. The rationale and implications of this choice are discussed in Section~\ref{sec:method}.
\section{Sample Overview and Distribution of Parameters}\label{sec:Distribution}
\subsection{Sample Overview}
The application of our selection and fitting methodology (Section~\ref{sec:Data and Method}) yields the following sample for analysis. Our analysis of GRBs detected by Swift from 2004 to 2024 identified a total of 170 events that display clear jet break features and satisfy our selection criteria. Within this sample, we find that the circum-burst environment for 82 bursts (48\%) is consistent with an ISM environment, while 88 bursts (52\%) favor a wind medium. Optical data are available for 35 of these events. Among these, 12 bursts exhibit an achromatic break across both X-ray and optical bands. For 8 of these 12 multi-wavelength cases, the classifications inferred separately from the two bands converge on the same environment type. For the remaining 4, differences in data quality or more complex physical processes (e.g., residual energy injection) may account for the discrepancy, a point we revisit in Section~\ref{sec:multiband}. This sample provides the basis for the physical consistency checks and parameter derivations that follow.
\subsection{Distribution of Parameters}
Based on our constrained classification of circum-burst environments, we further derived the geometric and energetic properties of the jets. The distribution of jet break times for our sample of 170 GRBs is presented in Figure~\ref{fig:figure3} and is fitted with a Gaussian function. The break times span from $5\times10^{2}$ s to $10^{6}$ s, with a mean value of $20.4$ ks, and the distribution is approximately Gaussian.

In Figure~\ref{fig:figure4}, we present the distributions of both the fitted $\Delta\alpha$ and the derived density profile index $k$ for our entire sample. The $\Delta\alpha$ distribution shows a bimodal structure with peaks around 0.5 and 0.75, corresponding to the canonical wind and ISM environments, respectively. The distribution of $k$ exhibits three notable peaks near $k \approx 0$, $k \approx 1$, and $k \approx 2$. The peaks at $k \approx 0$ and $k \approx 2$ correspond to bursts residing in the standard ISM and wind environments, as expected from the theoretical relation. The intermediate peak at $k \approx 1$ is consistent with findings from earlier studies \citep{2013ApJ...776..120Y,2020ApJ...895...94Y}; we discuss this feature in detail in Section~\ref{sec:interpretation}.

For GRBs with known redshifts, we obtained their isotropic gamma-ray energy ($E_{\gamma,\mathrm{iso}}$). Based on the fitted break time and the classified circum-burst environment type, we calculated the jet half-opening angle $\theta_j$. For bursts in an ISM environment, we applied the relation \citep{1999ApJ...525..737R,1999ApJ...519L..17S,2001ApJ...562L..55F,2015ApJ...807...92Y}:
\begin{equation}\label{eq:Theta_jet-ISM}
\begin{aligned}
\theta_{j}^{\text{ISM}} =&\:0.076\,\text{rad}\left ( \frac{t_b}{\rm{1\,day}} \right ) ^{3/8}\left ( \frac{1+z}{2} \right ) ^{-3/8} \\
& \times E_{\gamma ,\mathrm{iso},53}^{-1/8}\left ( \frac{\eta}{0.2} \right ) ^{1/8}\left ( \frac{n}{\rm{1\,cm^{-3}}} \right ) ^{1/8},
\end{aligned}
\end{equation}
and for those in a wind environment, we used \citep{2000ApJ...536..195C,2003ApJ...594..674B,2015ApJ...807...92Y}:
\begin{equation}\label{eq:Theta_jet-wind}
\begin{aligned}
\theta_j^{\text{wind}} =&\:0.12\,\text{rad}\left ( \frac{t_b}{\rm{1\,day}} \right ) ^{1/4}\left ( \frac{1+z}{2} \right ) ^{-1/4} \\
& \times E_{\gamma ,\mathrm{iso},52}^{-1/4}\left ( \frac{\eta}{0.2} \right ) ^{1/4}A_{*}^{1/4}.
\end{aligned}
\end{equation}
We adopted standard values for the radiative efficiency $\eta = 0.2$, the ambient density $n = 1\,\rm{cm^{-3}}$ (for ISM), and the wind parameter $A_* = 1$ (for wind). The wind parameter $A_*$ is defined via the wind density profile $\rho(r) = A r^{-2}$, where the normalization constant $A = \dot{M}_w / 4\pi V_w$. This constant can be written as $A = 5 \times 10^{11} A_* \, {\rm g~cm}^{-1}$, and the fiducial value $A_* = 1$ corresponds to a mass-loss rate $\dot{M}_w = 1 \times 10^{-5} \, M_\odot \, {\rm yr}^{-1}$ and a wind velocity $V_w = 10^3 \, {\rm km~s}^{-1}$. The true beaming-corrected gamma-ray energy was then estimated as $E_{\gamma} = E_{\gamma,\mathrm{iso}}(1-\cos\theta_j)$. The distributions of the derived jet half-opening angle $\theta_j$ and the true beaming-corrected gamma-ray energy $E_{\gamma}$ are presented together in Figure~\ref{fig:figure5}. The isotropic gamma-ray energies span from $10^{49}$ erg to $10^{55}$ erg, the jet half-opening angles are predominantly smaller than $0.2$ rad, and the distribution of $E_{\gamma}$ peaks around $10^{50}$ erg. These statistical properties are consistent with previous studies of GRBs with jet breaks \citep{2020ApJ...900..112Z}. As $\eta$, $n$, and $A_*$ are fixed with certain values for all the bursts, the derived values shown in Figure~\ref{fig:figure5} and Table~\ref{tab:Parameters} are illustrative and should be regarded as order-of-magnitude estimates only; they should therefore be used with caution.
\section{Discussion}\label{sec:discussion}
\subsection{The Origin of the Break}\label{sec:origin}
A typical GRB X-ray afterglow light curve often shows a complex, multi-segment structure, in which breaks are key features for revealing its physical evolution \citep{2006ApJ...642..354Z}. The late time break analyzed in this work is most likely attributed to a jet break, marking the jet's dynamical transition from relativistic, quasi-spherical expansion to non-relativistic lateral spreading. To substantiate this interpretation, we have systematically evaluated other physical mechanisms that could produce similar breaks in the light curve.

Regarding the energy injection break: the shallow-decay segment preceding a break is generally attributed to continuous energy injection from the central engine into the forward shock, typically producing a temporal slope of approximately $-0.5$ or shallower \citep{2006ApJ...642..354Z}. Our sample selection required the pre-break slope to be around $-1$, thereby favoring bursts in the normal decay phase and significantly reducing the inclusion of typical energy injection plateaus with such flat slopes. Nevertheless, we cannot rule out the possibility that some breaks with slightly shallower pre-break slopes may still originate from energy injection. This kind of break could still be produced by an energy injection break masquerading as a jet break (see \citealt{2015ApJ...805...13L}).

Another possible origin is a break following a flare. Flares are characterized by a rapid rise and decay, with the underlying afterglow light curve typically remaining unchanged before and after the flare event \citep{2006ApJ...642..389N,2006ApJ...647.1213O,2007ChJAA...7....1Z,2010MNRAS.406.2113C,2016ApJS..224...20Y}. This can produce a break between the steep decay phase of the flare and the subsequent normal afterglow decay. However, the decline slope of X-ray flares is typically $\lesssim -2$ \citep{2005Sci...309.1833B,2006ApJ...641.1010F,2007ApJ...671.1921F}, which is inconsistent with the typical post jet break slope. Furthermore, the Swift/XRT catalogue explicitly flags the time intervals of identified flares. We excluded these intervals from our analysis and fitting. For these reasons, we consider it unlikely that the breaks in our sample are produced by the transition from a flare decay phase to the normal afterglow phase.

Another mechanism to consider is a spectral break, typically caused by the passage of the cooling frequency $\nu_c$ through the X-ray band. However, when this occurs, the predicted change in the temporal decay index is $\Delta\alpha = 0.25$ and the spectral index change is $\Delta\beta = 0.5$, regardless of the circum-burst environment (ISM or wind) \citep{1999ApJ...519L..17S,1999ApJ...520L..29C}. This prediction is inconsistent with our selection criteria. Furthermore, we find no significant spectral evolution across the breaks: the distributions of the spectral index $\beta$ before and after the break, shown in Figure~\ref{fig:figure6}, are both well-described by Gaussian functions with central values of 0.87 and 0.91, respectively, and are statistically indistinguishable for most bursts. The lack of a significant $\Delta\beta$ further argues against a spectral break origin for our sample \citep{2006MNRAS.365.1031D,2006ApJ...642..389N}. We note that a similar analysis of spectral evolution across the break could not be performed for the optical band due to the lack of optical data with sufficient temporal sampling to resolve the pre- and post-break phases for the relevant bursts.

Off-axis observation is another possible origin for breaks in afterglow light curves. This scenario applies when the viewing angle $\theta_v > 0$, and its impact depends on the jet structure and the relation between $\theta_v$ and the jet half-opening angle $\theta_j$. For a slightly off-axis observer ($\theta_v < \theta_j$), the light curve is largely similar to the on-axis case ($\theta_v = 0$). In contrast, a significantly off-axis geometry ($\theta_v > \theta_j$) typically produces a distinct early rising light curve due to relativistic beaming effects \citep{2002ApJ...570L..61G,2025ApJ...983..111C}. Since the vast majority of bursts in our sample exhibit a normal decay from the earliest observed times without an early rising phase, the significantly off-axis scenario is unlikely to be the dominant cause of the breaks in our sample.

In addition, breaks could potentially arise from changes in the density distribution of the circum-burst medium (e.g., a density jump), possibly caused by the activity of the progenitor star. However, this scenario is disfavored for our sample because such density variations are typically accompanied by significant spectral evolution across the break \citep{2025ApJ...989...19D}. In contrast, most of our bursts show negligible change in the spectral index ($\Delta\beta$ is very small). Moreover, density jumps usually induce only minor changes in the temporal slope, further inconsistent with the prominent breaks we observe. Taken together, these considerations render a density-variation origin unlikely for the majority of our sample.

Finally, we consider the possibility that the observed steepening could be caused by a temporal evolution of the shock microphysical parameters. Recent studies have shown that the fractions of shock energy given to accelerated electrons ($\epsilon_e$) and to the magnetic field ($\epsilon_B$), if evolving with time (e.g., $\epsilon \propto t^\lambda$), can indeed modify the temporal decay indices of afterglows \citep{2021MNRAS.504.5685M, 2024MNRAS.527.1884F}. Nevertheless, we argue that this mechanism is unlikely to be the dominant cause of the breaks in our sample for the following reasons. First, our sample selection explicitly targets events exhibiting the canonical signatures of a geometric jet break in the standard external shock model, namely pre-break decay slopes of $\sim -1$ ($\alpha_1 \sim 1$) and post-break slopes of $\sim -2$ ($\alpha_2 \sim 2$). Within the evolving-parameter framework, the closure relations derived by \citet{2024MNRAS.527.1884F} show that the temporal decay indices are functions of the evolutionary parameters (i.e., the power-law indices of $\epsilon_e$ and $\epsilon_B$). This implies that, if the steepening were dominated by microphysical evolution, the resulting decay indices would depend on the specific values of these parameters rather than being fixed to the canonical values expected for a geometric jet break. Consequently, bursts significantly affected by microphysical evolution would be expected to exhibit decay indices that deviate from the standard-model predictions at various afterglow phases, and are thus unlikely to satisfy our selection criteria. Second, the theoretical and observational evidence for significant microphysical evolution has been predominantly established in the GeV--TeV band, where synchrotron self-Compton radiation is highly sensitive to variations in $\epsilon_e$ and $\epsilon_B$ \citep{2023MNRAS.525.1630F, 2024MNRAS.527.1884F}. In fact, \citet{2024MNRAS.527.1884F} showed that, within their framework, the typical observational signatures produced by evolving microphysical parameters are plateaus and peaks in the high-energy light curves, rather than sharp jet-break-like transitions. In contrast, our study focuses on X-ray synchrotron emission, which is the standard component widely used to identify jet breaks in the external shock model. While \citet{2021MNRAS.504.5685M} demonstrated that time-varying microphysics is required to simultaneously explain the multi-band (including X-ray) data of GRB 190114C, their analysis did not suggest that such evolution alone produces sharp, jet-break-like features in the X-ray band. Third, within our subsample of bursts with multi-band data, the majority of well-sampled events exhibit achromatic breaks across the X-ray and optical bands (Section~\ref{sec:multiband}). This achromaticity is a hallmark of a geometric jet break and is not a generic prediction of microphysical evolution models, which may produce band-dependent effects. We therefore consider that, although microphysical parameter evolution may play a role in shaping the detailed features of some individual afterglows, particularly at high energies, it is unlikely to be the primary mechanism producing the prominent, well-defined breaks analyzed in this work, which are most plausibly attributed to a geometric jet break.
\subsection{Analysis of X-ray and Optical Data}\label{sec:multiband}
Our sample includes 35 GRBs with both X-ray and optical afterglow light curves (presented in Figures~\ref{fig:figure1} and \ref{fig:figure2}). For the final classification of the circum-burst environment, we relied primarily on the X-ray fitting results. This choice is motivated by the fact that optical data are more susceptible to complex influences such as interstellar extinction, atmospheric conditions, host galaxy contamination, and redshift effects \citep{2000A&A...364L..54A, 2002ApJ...565..174R}, whereas the X-ray band provides a cleaner probe of the shock dynamics.

Among these 35 bursts, 12 exhibit an achromatic break across both X-ray and optical bands. This subset offers an ideal opportunity to validate our environmental diagnostic method using two independent bands for the same event. A detailed analysis of these 12 bursts yields the following results:

Two cases are particularly noteworthy and provide strong support for our methodology. GRB 160131A is an archetypal example in which a clear achromatic jet break at $\sim 1.2$ d has been robustly identified from its well-sampled X-ray and optical light curves \citep{2022A&A...658A..11M}. Detailed broadband modeling of this event, which includes a rich data set spanning from radio to X-rays, conclusively favors an ISM-like circum-burst environment. This conclusion is drawn primarily from the observed spectral index and its consistency with the closure relations of the standard afterglow model, effectively ruling out a wind-like density profile. In analyzing the jet break, the study also explicitly employed the change in the temporal decay index, $\Delta\alpha$, as a diagnostic tool. Starting from the observed pre-break index ($\alpha_\text{pre}\sim -1.25$) and using the relation $\Delta\alpha = (3-k)/(4-k)$ within the uniform jet model, the authors inferred that the expected post-break index in an ISM environment should be $\alpha_\text{post}\sim -2.0$, which agrees with the observed value ($\alpha_\text{obs}\sim -1.8$) within the uncertainties, thereby providing additional evidence for the ISM classification. The physical reasoning and diagnostic logic underlying this $\Delta\alpha$ based inference are fully aligned with the $\Delta\alpha$ environmental diagnostic method adopted in our study. GRB 060614, a confirmed kilonova-associated merger event that is also classified as occurring in an ISM environment \citep{2015NatCo...6.7323Y, 2016EPJWC.10908002J}, further validates the physical relevance of our diagnostic method.

Concordant Cases (8 bursts): For eight bursts, the environment type inferred from the two bands is fully consistent. Specifically, GRB 060614, GRB 080413B, GRB 160131A, and GRB 210610B are classified as ISM, while GRB 070411, GRB 200829A, GRB 140423A, and GRB 210619B are classified as wind, with all classifications consistent across the X-ray and optical data \citep{2006A&A...456..509A,2018ApJ...857..140Z,2019A&A...622A.138K,2011A&A...526A.113F,2009A&A...508..593K,2016A&A...589A..37V,2022A&A...658A..11M}.

Discordant Cases (4 bursts): The remaining four bursts (ISM-classified GRB 050922C and GRB 060927; wind-classified GRB 071112C and GRB 091127) show a systematic trend: the temporal decay is steeper in the X-ray band than in the optical band, either after the break (for ISM types) or before it (for wind types). This ``faster X-ray decay'' phenomenon is consistent with findings from some previous multi-band studies and may indicate more complex emission zone structures or energy injection histories \citep{2011MNRAS.412..561O,2015ApJ...805...13L}.

For the remaining bursts whose optical data do not fully cover the break, we cannot rule out the presence of a jet break. We therefore performed a SPL (Equation~\ref{eq:SPL}) fit to the portion of the optical light curve that was available (either pre- or post-break) and included these results in the corresponding figures.

To statistically assess the prevalence of the external forward shock model, we plot in Figure~\ref{fig:figure7} the temporal decay indices from the two bands for all bursts with simultaneous measurements. The theoretical expectation zones for the standard forward shock model (offset range $\pm\;0.25$) and the energy injection model (offset range $\pm\;0.5$) are indicated (see Tables 3 and 4 in \citealt{2015ApJS..219....9W}). Within the uncertainties, the vast majority of data points (28 out of 29) are consistent with the standard external shock model. The sole outlier is the pre-break point of GRB 091127, which lies near the boundary of the expected region. This might suggest the need for an additional emission component at this specific phase. Such multi-component scenarios have been invoked in detailed studies of similar complex afterglows (e.g., \citealt{2015ApJS..219....9W}).

\subsection{Physical Interpretation of Intermediate Density Profiles}\label{sec:interpretation}

A notable feature of the $k$ distribution derived from our $\Delta\alpha$ diagnostic (Figure~\ref{fig:figure4}) is not only the bimodal peaks at $k \approx 0$ and $k \approx 2$, but also a distinct concentration of bursts around intermediate values, particularly $k \approx 1$. Such intermediate density profiles are not anomalous; they have been robustly identified in several independent studies. For instance, by modeling the forward and reverse shock emission in a stratified medium, \citet{2013ApJ...776..120Y} found density profile indices in the range $0.4 \le k \le 1.4$ for 19 GRBs, with a typical value of $k \sim 1$. \citet{2013ApJ...774...13L} reported consistent results from a larger sample. More recently, \citet{2023MNRAS.525.1630F} utilized synchrotron self-Compton closure relations in a stratified environment to analyze the Second \textit{Fermi}-LAT Gamma-Ray Burst Catalog \citep[2FLGC;][]{2019ApJ...878...52A}, finding that intermediate values such as $k \approx 0.5$ and $k \approx 1$ are often favored. Additionally, \citet{2013ApJ...779L...1K} demonstrated through multi-wavelength modeling of GRB 130427A that an intermediate density profile between the ISM ($k=0$) and stellar wind ($k=2$) provides a better description of the data. The prevalence of intermediate $k$ values in our sample is thus consistent with these earlier findings and warrants a physical discussion.

We consider the following three scenarios, which are not mutually exclusive, that could give rise to an observationally inferred density profile index with $0 < k < 2$:

\begin{itemize}
\item \textbf{Non-ideal stellar wind profiles.}
We propose that the peak at $k \sim 1$ likely reflects a mass-loss history of the progenitors of these GRBs that deviates from the standard steady stellar wind. The canonical wind model assumes a constant mass-loss rate and a constant wind velocity, leading to $n \propto r^{-2}$. However, the evolution of massive stars may involve episodic mass ejections, time variations in wind velocity, or a pre-collapse eruptive phase driven by opacity (as discussed for GRB~130427A by \citealt{2013ApJ...779L...1K}). These processes can shape a circum-burst medium with an effectively flatter density profile (i.e., $k<2$). Consequently, the concentration of bursts with $k \approx 1$ in our sample likely represents a population of progenitors with non-standard, variable mass-loss histories, whose circum-burst environments resemble a ``weak wind'' or a ``modified wind''. It is worth noting that the continuous distribution of $k$ should be interpreted as an \textit{effective} description of the complexity of the circum-burst medium, rather than a strict identification of a single ideal power-law density profile for each burst.

\item \textbf{Relaxation of the idealized assumptions in the uniform jet model.}
The relation $\Delta\alpha = (3-k)/(4-k)$ is derived under several idealized assumptions, including a uniform jet, negligible lateral spreading, and time-independent microphysical parameters. Relaxing these assumptions (e.g., by considering a structured jet, late-time energy injection, evolution of microphysical parameters, or non-isotropic lateral spreading) could systematically shift the observed $\Delta\alpha$ and hence the inferred $k$ value.

\item \textbf{Transition from a stellar wind to a uniform ISM.}
One might argue that an observed effective $k$ between the two canonical values arises from a jet that initially propagates in a stellar wind region ($k=2$) and later moves into an outer uniform ISM ($k=0$). Although such a transition is theoretically possible and has been invoked in broadband modeling of individual GRBs (e.g., \citealt{2003ApJ...591L..21D,2007ApJ...664L...5K,2009MNRAS.400.1829J,2020ApJ...900..176L,2017ApJ...848...15F,2019ApJ...879L..26F}), we consider it unlikely to be the dominant explanation for the large number of bursts with intermediate $k$ values in our sample for the following reasons. First, a smooth wind-to-ISM transition requires that the transition occurs within a specific radius range ($R \sim 10^{18}$--$10^{20}$ cm), where the ambient density can jump by a factor of several or more \citep{2006ApJ...643.1036P}, and that the jet break happens to take place exactly within this transition zone. However, even if such a coincidence occurs, the break feature predicted by the medium transition model is typically a steep-to-shallow transition, which is completely opposite to the shallow-to-steep transition characteristic of jet breaks required by our sample selection. Second, and more importantly, our $\Delta\alpha$ diagnostic essentially probes the local density gradient of the circum-burst medium at the precise moment of the jet break (i.e., when the fireball decelerates to $\Gamma \sim \theta_j^{-1}$). If the jet break occurs while the jet is still in the wind-dominated region, one would obtain $k \approx 2$; if it occurs after the jet has entered the ISM-dominated region, one would obtain $k \approx 0$. Therefore, this method does not naturally yield a continuous range of intermediate $k$ values.

It is worth noting that, for GRB 140423A in our sample, \citet{2020ApJ...900..176L} previously suggested, through multi-band modeling, that a wind-to-ISM transition occurred at $\sim 5\times10^3$ s, accompanied by energy injection over the interval of $10^2$--$10^6$ s. In contrast, our analysis identifies a jet break at $\sim 2.6\times10^4$ s that meets our selection criteria, and our $\Delta\alpha$ method constrains its circum-burst environment to be wind-like ($k \approx 2$). Moreover, a simultaneous break is also detected in the optical band, with a slope change of $\Delta\alpha \approx 0.47$, consistent with the theoretical expectation for a jet break in a wind environment. We therefore conclude that the circum-burst environment of GRB 140423A is indeed wind-like, and that the change in the light-curve behavior at $\sim 5\times10^3$ s is more likely attributable to energy injection rather than a medium transition.
\end{itemize}

For these reasons, we conclude that non-ideal stellar wind profiles provide a more plausible and dominant mechanism for explaining the concentration of bursts around $k \approx 1$ in our sample.
\subsection{Analysis of Methods for Limiting the Circum-burst Environment}\label{sec:method}
Building upon the physical interpretation of intermediate density profiles presented in Section~\ref{sec:interpretation}, we now turn to a critical examination of the $\Delta\alpha$-based diagnostic method itself.

The evolution of a GRB afterglow is critically governed by its circum-burst environment, which is shaped by the mass-loss history of the progenitor star. Therefore, determining both the type (i.e., whether it is a constant-density interstellar medium, ISM, or a stellar wind) and the detailed density profile of this environment is fundamental for constraining progenitor properties, understanding jet dynamics, and identifying the dominant radiation mechanisms.

Several approaches exist to diagnose the environment from afterglow data. The classic method uses spectral indices and closure relations \citep{2006ApJ...642..354Z,2006MNRAS.367L..42P,2009ApJ...698...43R,2011A&A...526A..23S}. While conceptually straightforward, its accuracy is highly sensitive to spectral quality and can be compromised by energy injection or jet structure. The most robust, though demanding, approach is the joint multi-band fitting of the complete afterglow evolution, which provides a self-consistent set of physical parameters \citep{2022A&A...658A..11M}. Other methods analyze specific features, such as the smooth onset of early X-ray afterglows \citep{2022ApJ...925...54T}, the peak evolution in radio bands \citep{2022ApJ...927...84Z}, or early synchrotron flashes from jet-wind bubble interactions \citep{2024ApJ...976...55P}. However, these methods face practical limitations when applied to substantial statistical samples, often due to stringent data requirements or sensitivity to model complexities. This motivates the need for a diagnostic that balances reliability with broad applicability.

We employ a rapid diagnostic method based on the change in the temporal decay index across the jet break, $\Delta\alpha$. While this relation has been previously applied to explain individual events, this work presents its first application as the primary tool for a systematic, statistical classification. Specifically, by applying the relation $\Delta\alpha = (3 - k)/(4 - k)$, we infer the density profile index $k$ for each burst, thereby enabling a statistical classification of circum-burst environments for a substantial X-ray afterglow sample. The key advantage of the method is its reliance solely on the widely available and homogeneous X-ray light curves. By fitting the power-law segments before and after the jet break, a preliminary constraint on the GRB's circum-burst environment can be directly obtained.

We applied this method to all samples that satisfied our jet break selection criteria. As a consistency check, for those bursts in our sample with existing detailed multi-band studies, our $\Delta\alpha$-based classifications show agreement with the environment types derived from more complex modeling (e.g., GRBs 050721, 060614, 071112C, 080413B, 080710, 121024A, and 160131A). In classifying the environment broadly, we did not require $\Delta\alpha$ to match exactly 0.5 or 0.75. This pragmatic approach accounts for the expected deviations between an idealized theoretical model and complex observational data. The observed scatter in $\Delta\alpha$ can be attributed to several factors:

\begin{enumerate}
\item \textbf{Observational and fitting uncertainties.} This includes fluctuations in the light curve itself, uncertainties in background subtraction, and the large photometric errors typical of late-time afterglow observations, all of which can bias the measurement of the decay indices $\alpha_1$ and $\alpha_2$.

\item \textbf{Relaxation of the theoretical model assumptions.} The theoretical relation $\Delta\alpha = (3 - k)/(4 - k)$ is derived within a uniform jet model that neglects both energy injection and significant lateral spreading. In reality, different jet structures (e.g., structured jets), the presence of late-time energy injection, or non-negligible lateral spreading can all lead to a steeper post-break decay, thereby increasing the observed $\Delta\alpha$ value \citep{2007RMxAC..27..140G}. A notable example is the extremely bright GRB 221009A. Despite its final break showing a $\Delta\alpha \approx 1$, which is larger than the canonical value of 0.75, this feature is still interpreted as a jet break within a more complex physical scenario \citep{2023Sci...380.1390L}.

\item \textbf{The intrinsic complexity of the circum-burst medium.} The theoretical values (0.75 for ISM, 0.50 for wind) correspond to idealized, single-power-law density profiles ($k=0$ and $k=2$). The continuous distribution of $\Delta\alpha$ (and thus of derived $k$ values) observed in our sample suggests that a significant fraction of GRB progenitors reside in media that cannot be adequately described by either canonical density profile. A detailed discussion of the physical origin of these intermediate $k$ values, and a comparison with previous studies, is presented in Section~\ref{sec:interpretation}.
\end{enumerate}

It is also for these reasons that a portion of bursts in our sample have $\Delta\alpha > 0.75$. We find that the relation $\Delta\alpha = (3-k)/(4-k)$ cannot be used to constrain $k$ for these bursts, as it would yield a negative value. Therefore, in our table, we have marked the $k$ values for these bursts as ``uncertain''. However, considering the fitting error of $\Delta\alpha$ in conjunction with the reasons listed above, we still classify the circum-burst environment of these bursts as ISM and regard cases with $\Delta\alpha > 0.75$ as jet breaks.

An additional point warrants discussion. Among our entire sample, there are 7 short-duration GRBs (SGRBs). Five of these are classified into the wind environment based on our $k$-value diagnostics. This appears inconsistent with the common view that SGRBs originate from compact binary mergers, which are expected to occur in constant-density ISM environments. This discrepancy may stem from prior assumptions embedded in our analysis. In the absence of additional data, some of these light curves could be explained either by early reverse--forward shock emission or by a late external shock with a jet break, which could bias the inferred $k$ values toward the wind solution. Therefore, although our $\Delta\alpha$-$k$ diagnostic provides valuable preliminary constraints on the circum-burst environment for the majority of bursts in our sample, achieving a more accurate classification of both the density profile and the type of the circum-burst medium will require the joint analysis of multi-wavelength data and the consideration of more complex physical models.

\section{Summary and Conclusion}\label{sec:summary}
We analyzed X-ray afterglow light curves observed by Swift/XRT from 2004 to 2024 and constructed a sample of 170 GRBs with clear jet break features. By performing broken power-law fitting to the pre- and post-break segments for each burst, we obtained robust measurements of the temporal decay index change $\Delta\alpha$. The distribution of the fitted $\Delta\alpha$ values shows a bimodal structure, with peaks around 0.5 and 0.75, corresponding to the canonical wind and ISM environments, respectively. Using the theoretical relation $\Delta\alpha = (3 - k)/(4 - k)$, we then derived the density profile index $k$ for each burst. The derived $k$ values exhibit a distribution with three notable peaks near $k \approx 0$, $k \approx 1$, and $k \approx 2$. Based on the derived $k$ values, we find that among the 170 GRBs, the circum-burst environment of 82 bursts is consistent with an interstellar medium (ISM) profile, while 88 bursts favor a wind environment.
By systematically evaluating alternative origins for the observed breaks, such as energy injection, spectral evolution, off-axis viewing, and density jumps, we conclude that a jet break remains the most plausible explanation for the majority of our sample.

We present the distribution of jet break times $t_b$ for our sample. The break times span from $5\times10^{2}$ s to $10^{6}$ s, with a mean value of $20.4$ ks, and the distribution is approximately Gaussian. For bursts with known redshifts, we derived the jet half-opening angles $\theta_j$ and the true beaming-corrected energies $E_{\gamma}$. The resulting distributions show that $\theta_j$ is predominantly smaller than $0.2$ rad and $E_{\gamma}$ peaks around $10^{50}$ erg, consistent with previous studies of jetted GRBs.

Among the 35 bursts in our sample with multi-band data, 12 exhibit an achromatic break across both X-ray and optical bands. For 8 of these, the multi-band fitting results yield consistent environmental classifications, while the remaining 4 show discrepancies between the X-ray and optical inferences. We investigated the possible causes of these inconsistencies, such as differences in data quality, residual energy injection, or more complex emission zone structures. Furthermore, we statistically assessed the prevalence of the external forward-shock model. Within the uncertainties, the temporal decay slopes (before and after the break) for the vast majority of bursts (28 out of 29) are consistent with the standard external shock model. Only the pre-break slope of GRB 091127 lies near the boundary of the expected region, possibly indicating the need for an additional emission component at this specific phase.

The $\Delta\alpha$-$k$ diagnostic method employed in this work provides an efficient and observationally accessible approach for constraining GRB circum-burst environments directly from X-ray afterglow light curves. While founded on the uniform jet model, this diagnostic reliably distinguishes between ISM and wind environments for the majority of well-sampled jet breaks, particularly for bursts with $\Delta\alpha$ near the canonical values of 0.75 and 0.5. The continuous distributions of both $\Delta\alpha$ and the derived $k$ values further indicate that while most GRBs likely reside in either standard ISM or wind environments, a subset inhabits media with intermediate density profiles ($0 < k < 2$). These bursts represent promising targets for future multi-wavelength follow-up observations. The catalog of environmental classifications and derived physical parameters presented here offers a valuable statistical resource for advancing our understanding of GRB progenitors, jet physics, and their interaction with diverse stellar environments.
\section*{Acknowledgments}
We acknowledge the use of public data from the \textit{Swift} data archives. This work is supported by the Natural Science Foundation of Jiangxi Province of China (grant No. 20242BAB26012) and the Shandong Provincial Natural Science Foundation (grant ZR2025MS47).

\FloatBarrier
\bibliographystyle{aasjournal}
\bibliography{paper8.bib}

\clearpage
\suppressfloats[t]

\begin{figure*}
\centering
\resizebox{45mm}{!}{\includegraphics[]{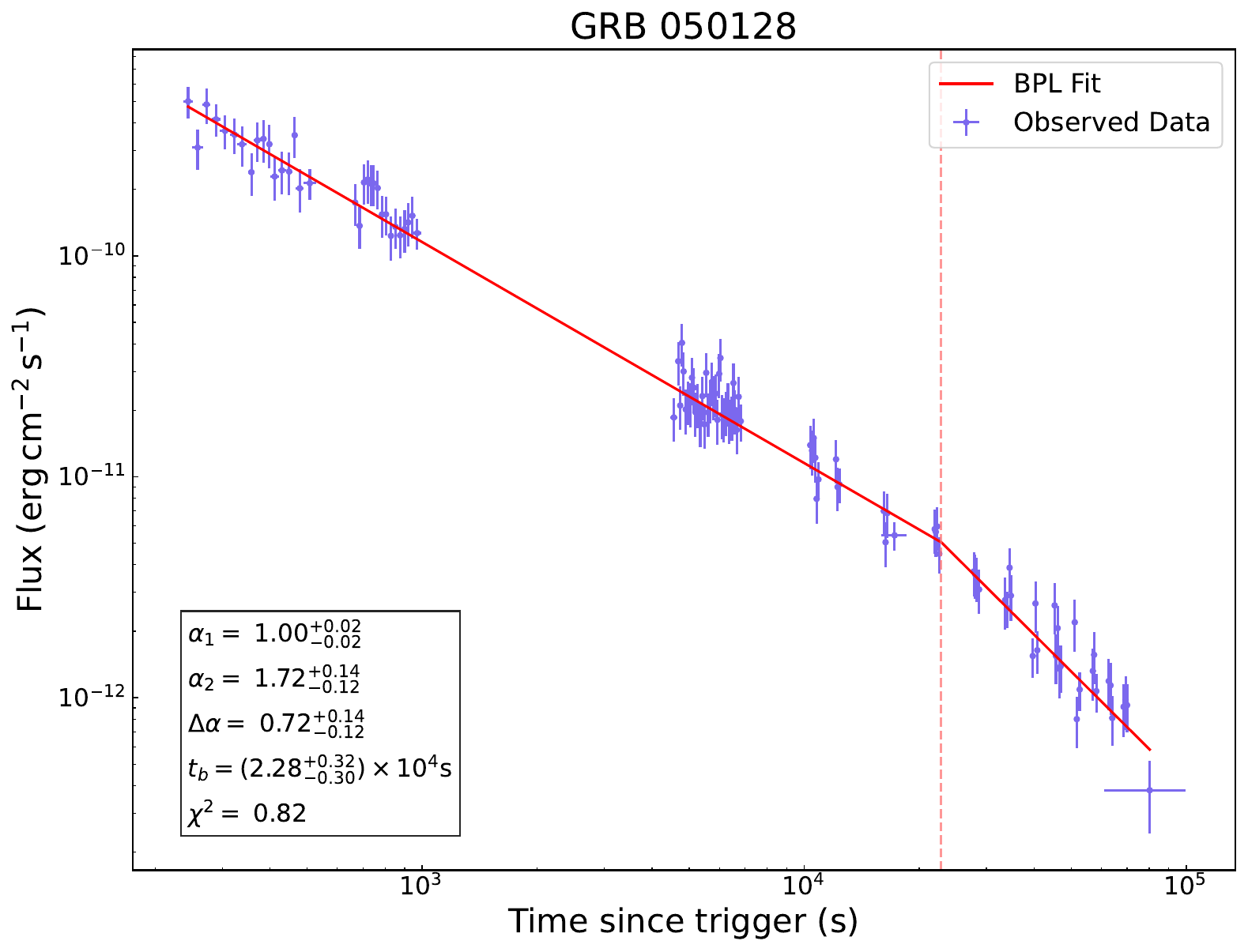}}%
\resizebox{45mm}{!}{\includegraphics[]{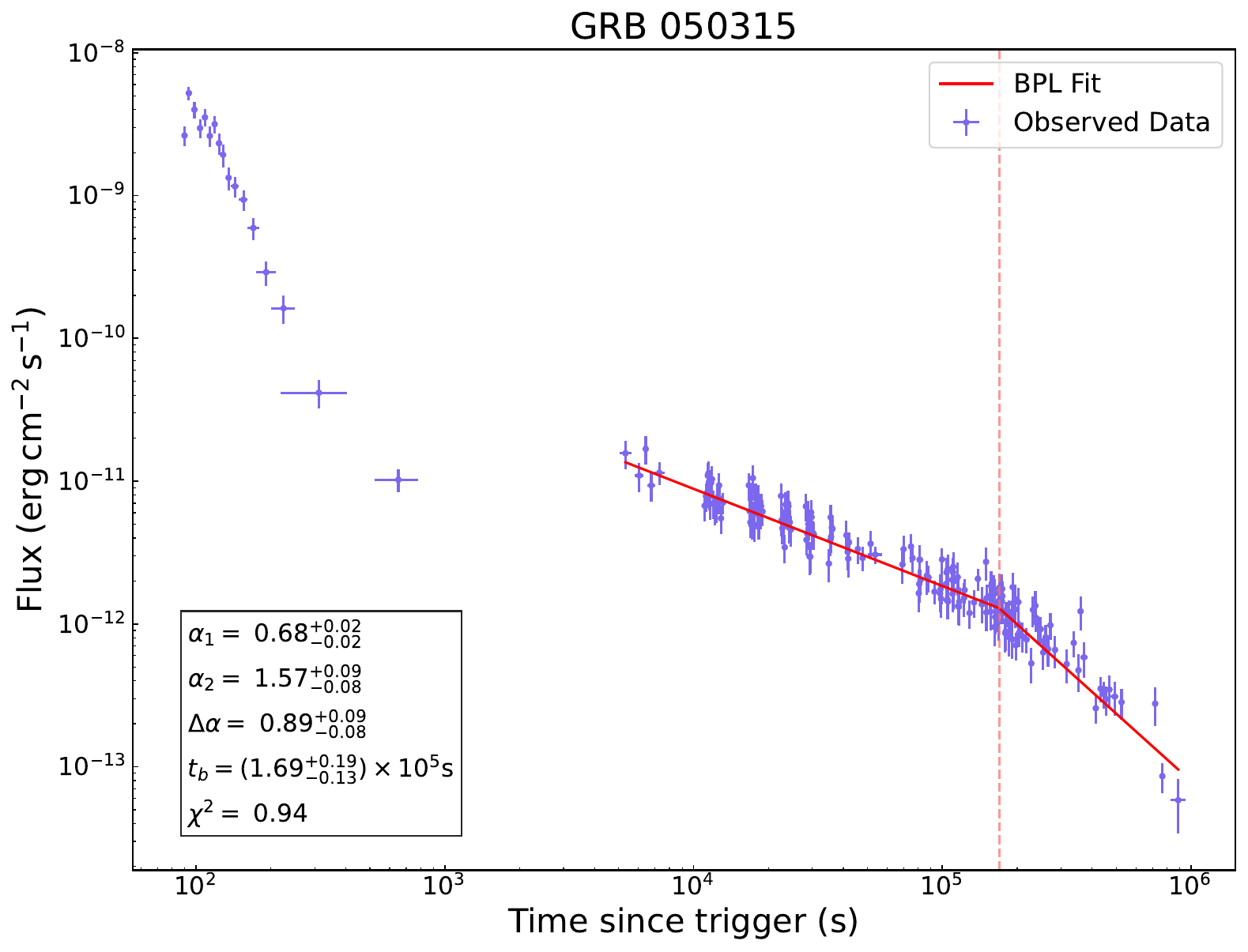}}%
\resizebox{45mm}{!}{\includegraphics[]{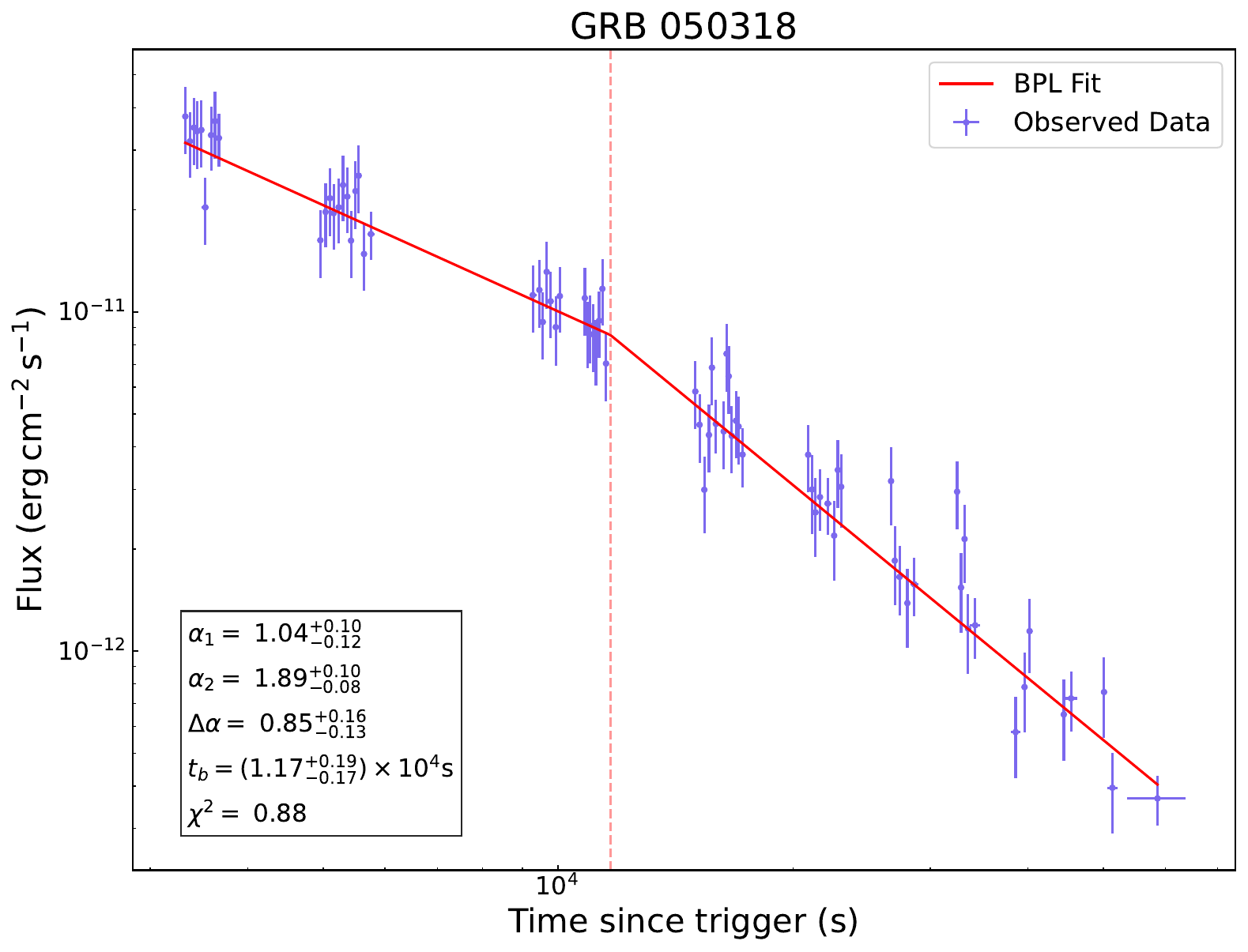}}%
\resizebox{45mm}{!}{\includegraphics[]{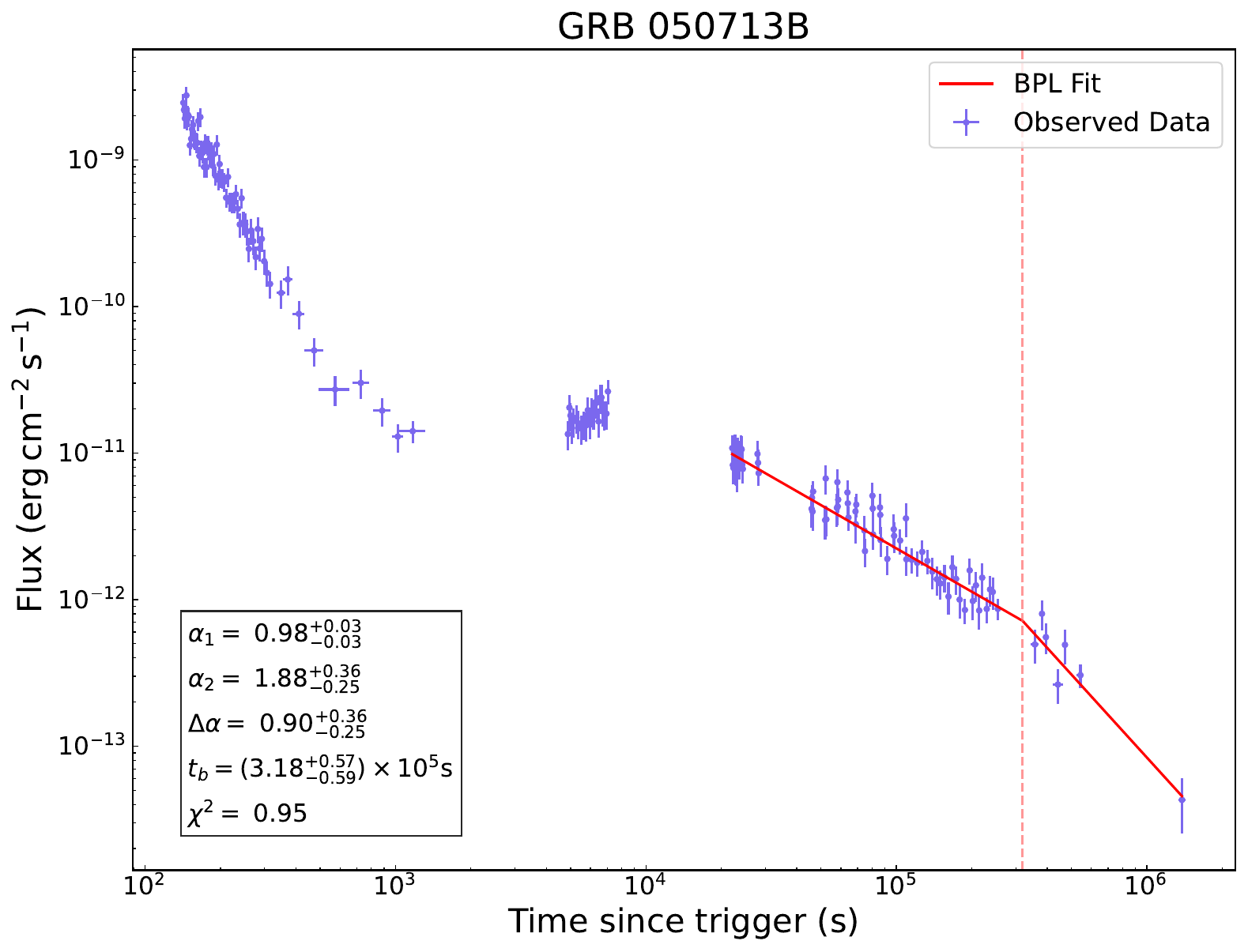}}\\
\resizebox{45mm}{!}{\includegraphics[]{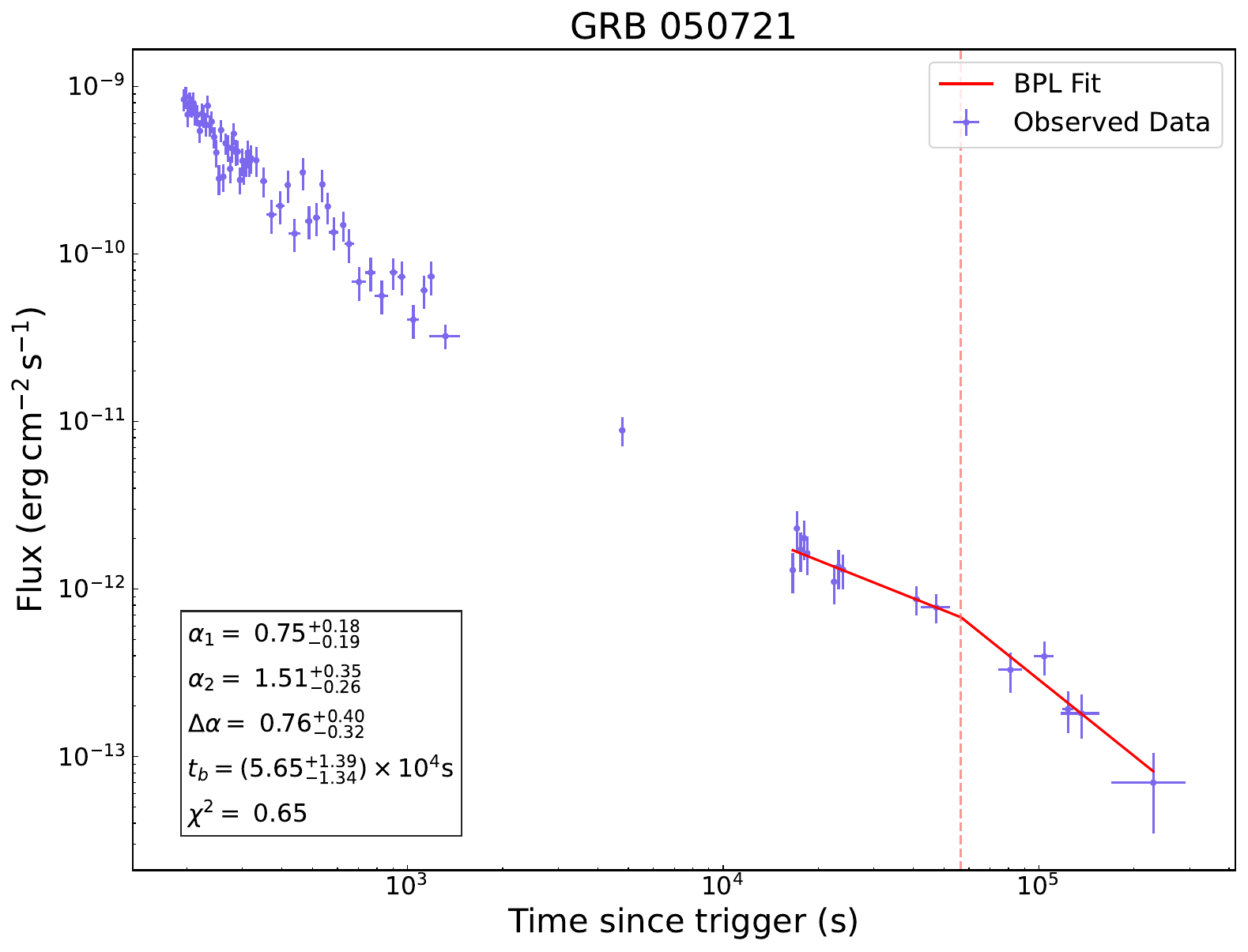}}%
\resizebox{45mm}{!}{\includegraphics[]{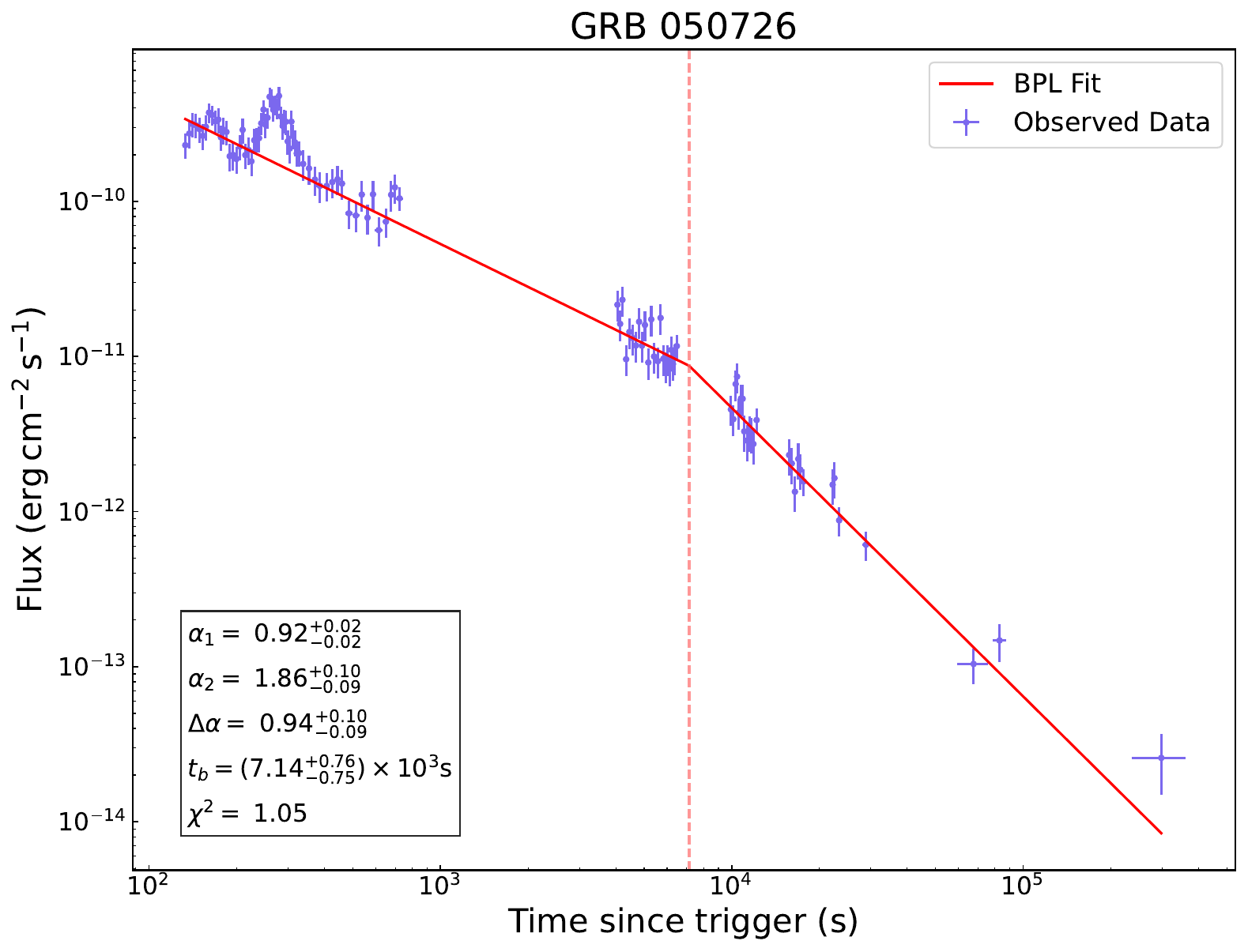}}%
\resizebox{45mm}{!}{\includegraphics[]{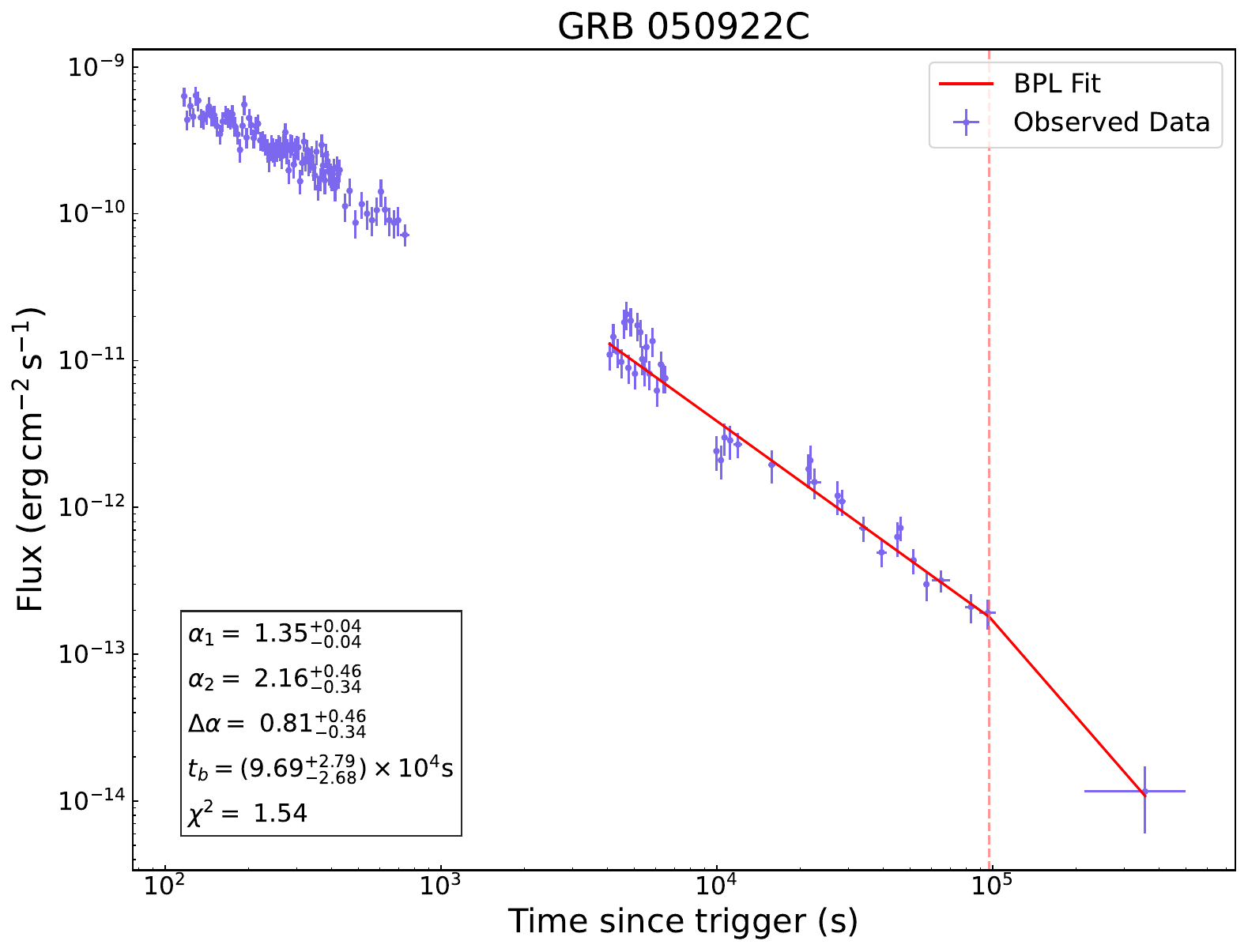}}%
\resizebox{45mm}{!}{\includegraphics[]{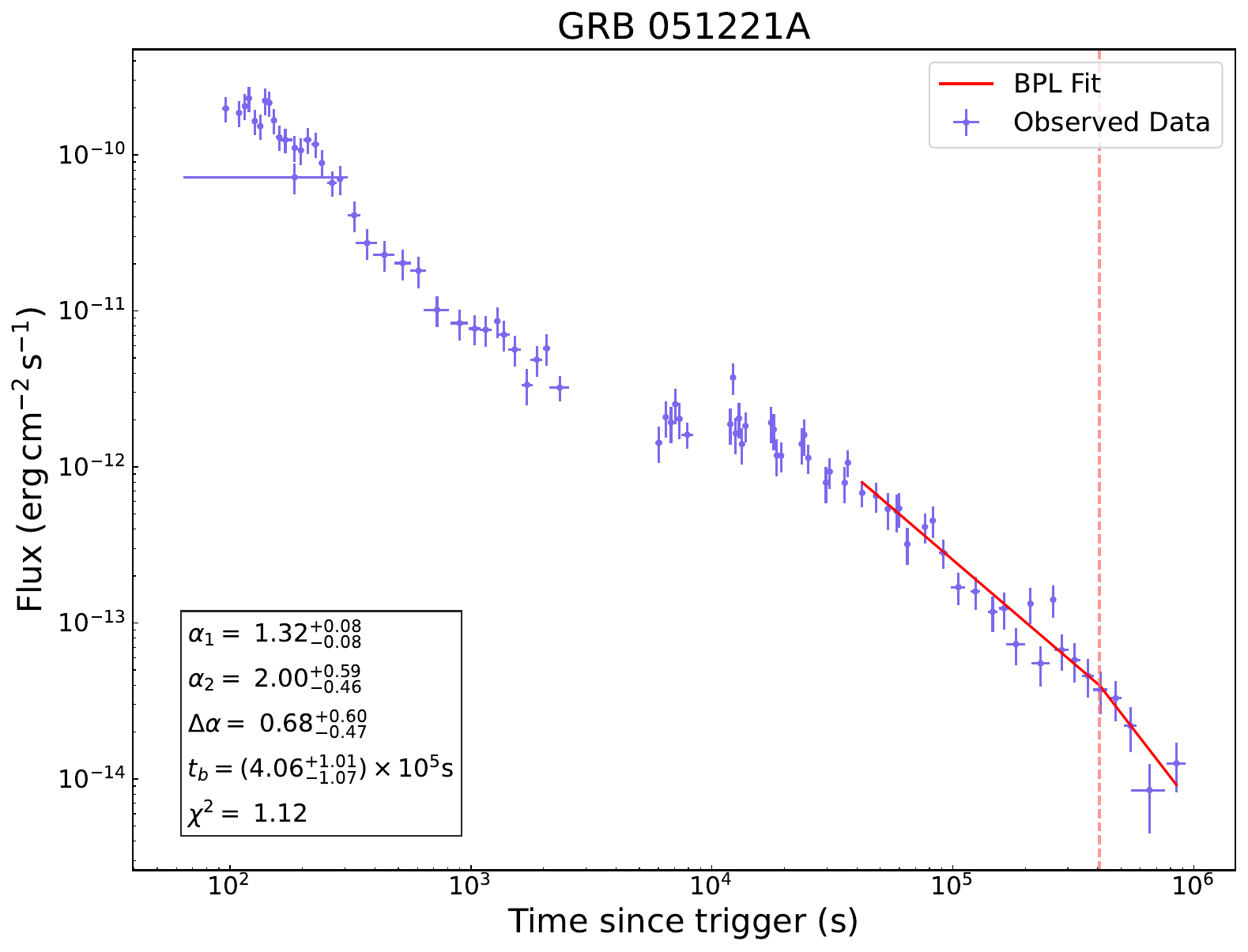}}\\
\resizebox{45mm}{!}{\includegraphics[]{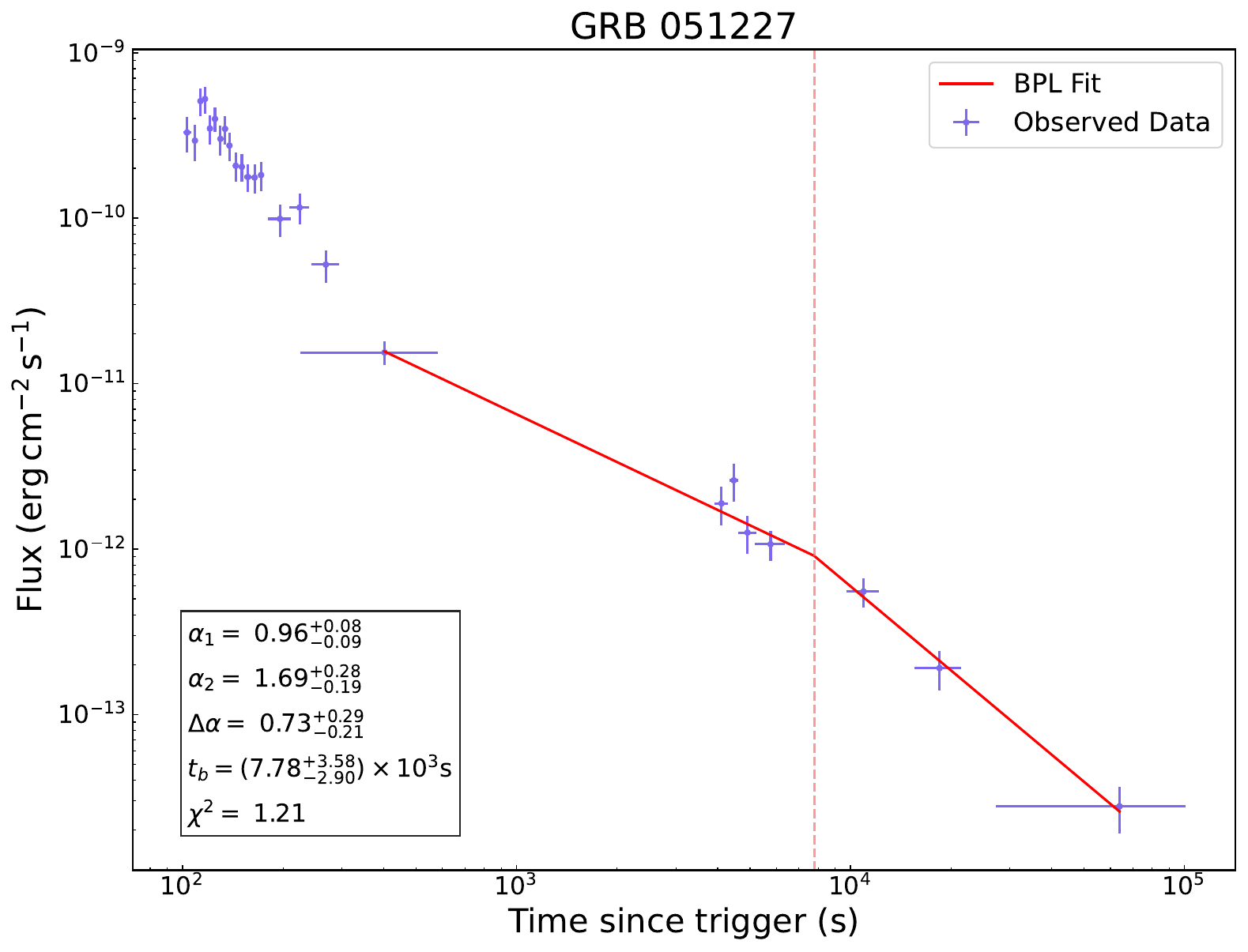}}%
\resizebox{45mm}{!}{\includegraphics[]{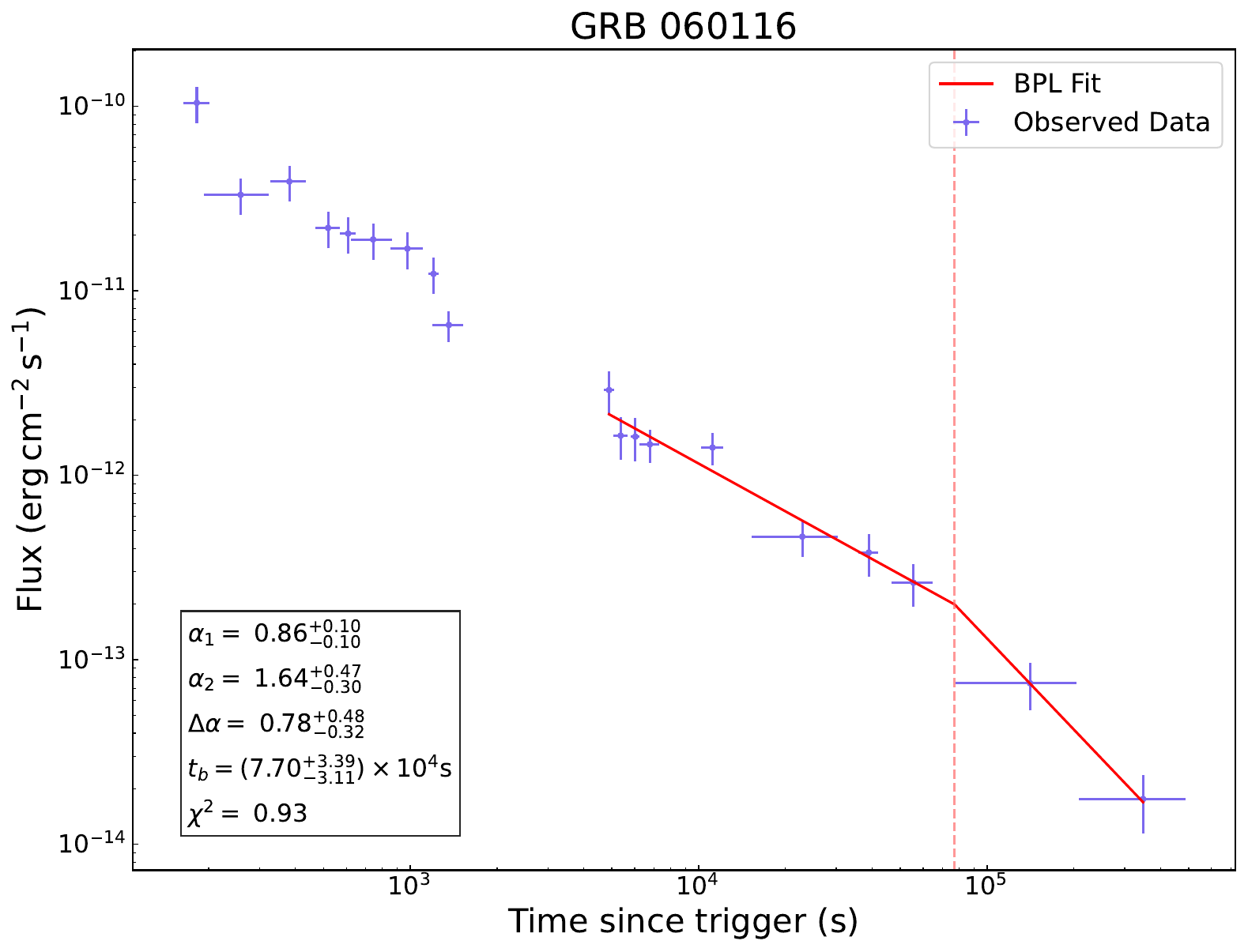}}%
\resizebox{45mm}{!}{\includegraphics[]{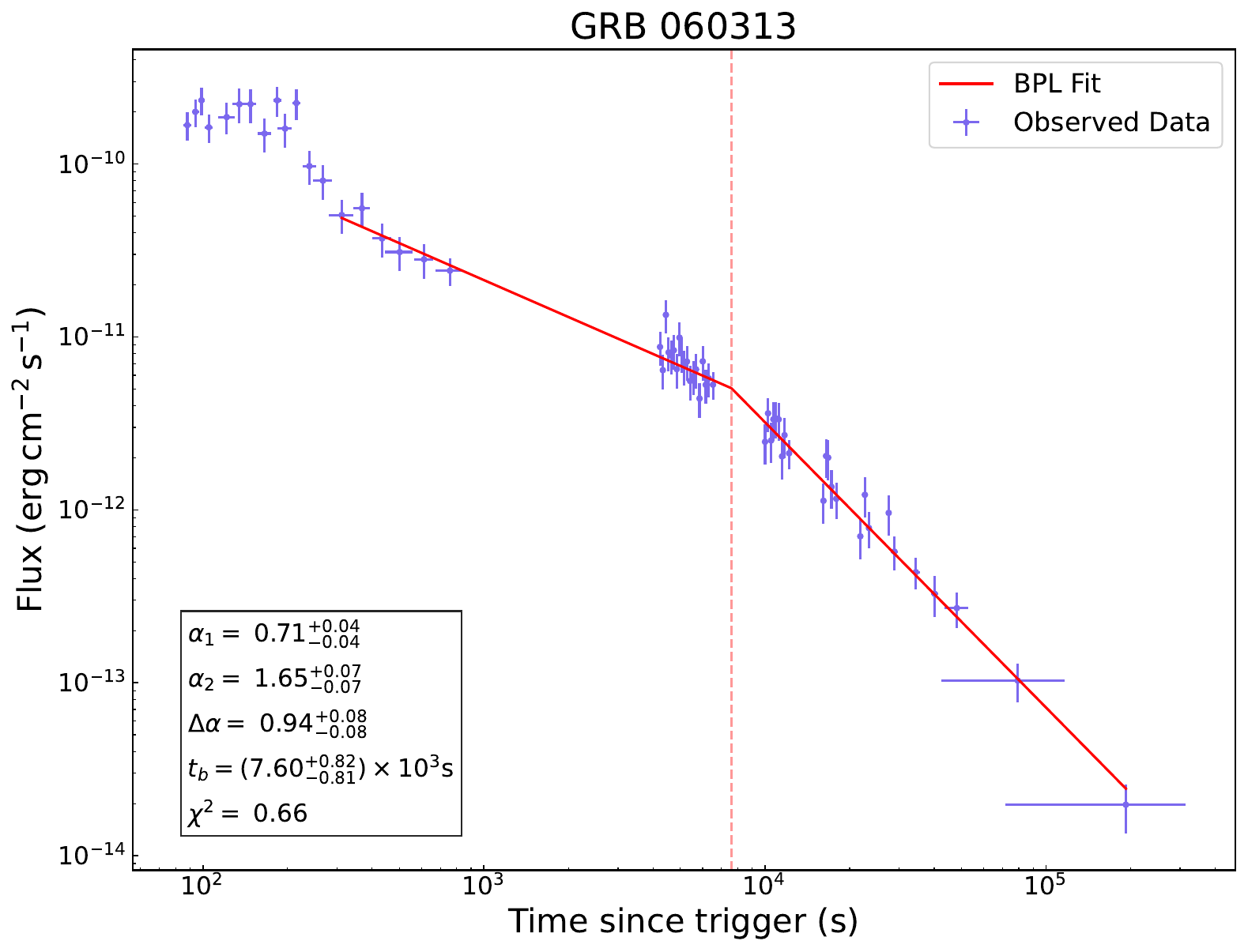}}%
\resizebox{45mm}{!}{\includegraphics[]{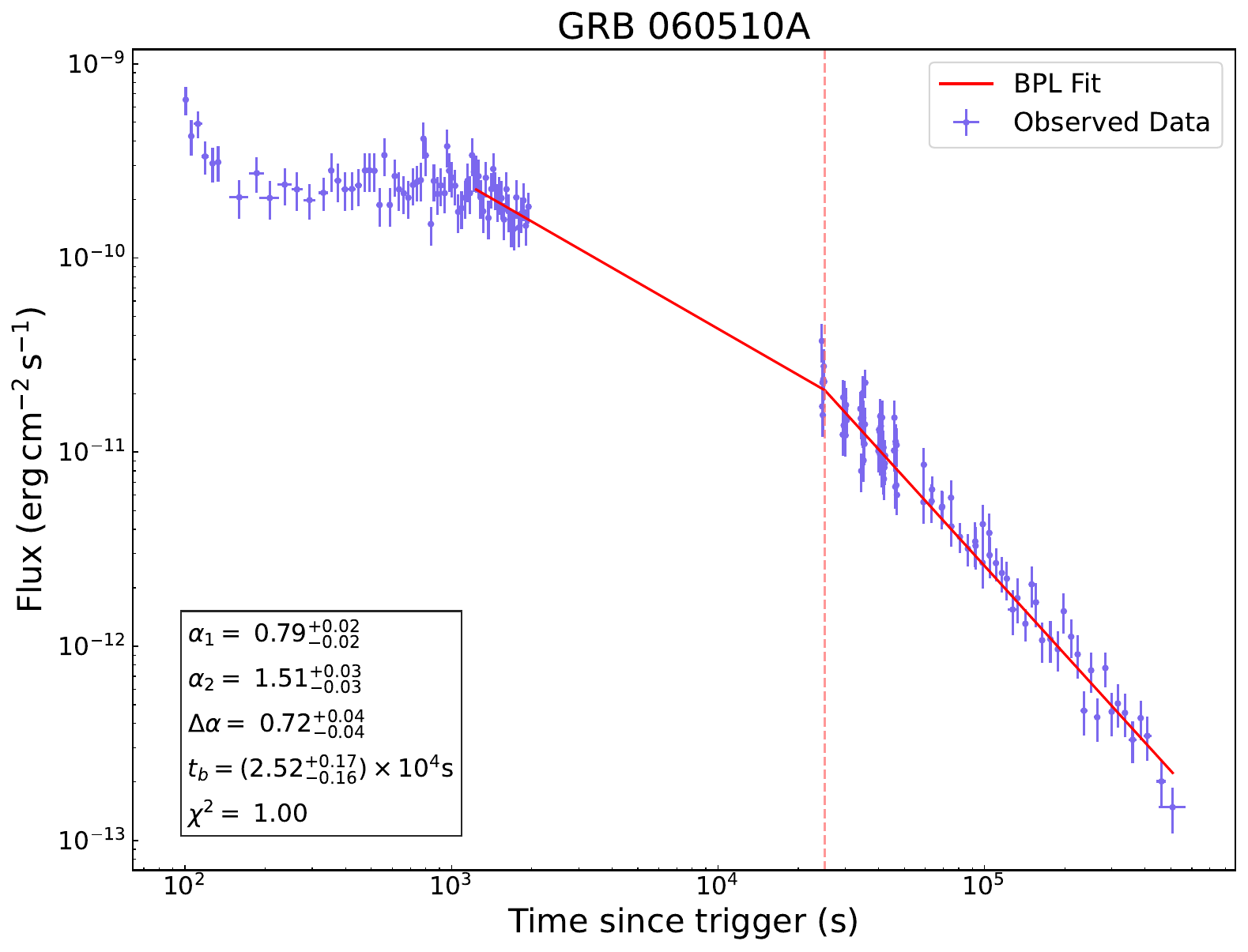}}\\
\resizebox{45mm}{!}{\includegraphics[]{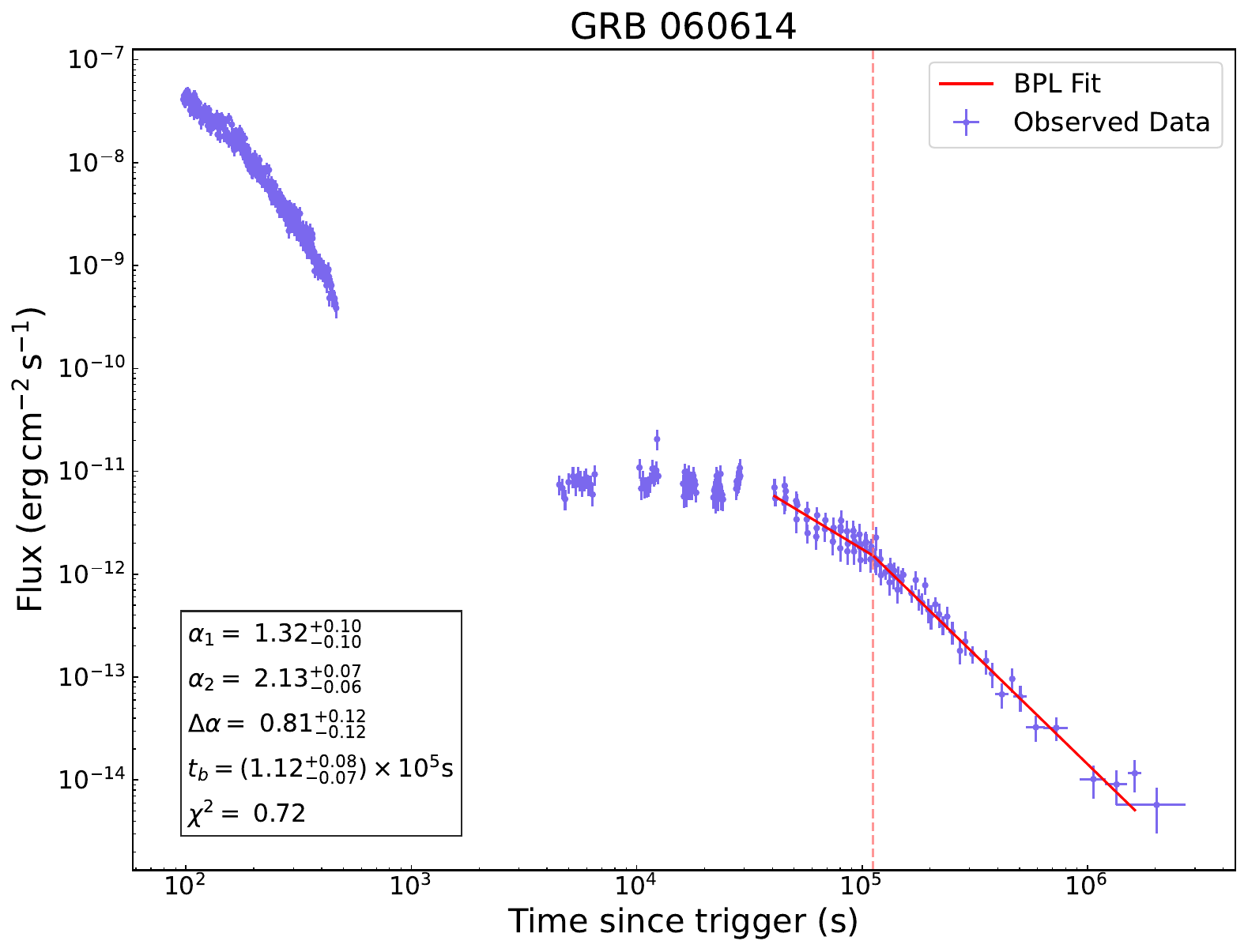}}%
\resizebox{45mm}{!}{\includegraphics[]{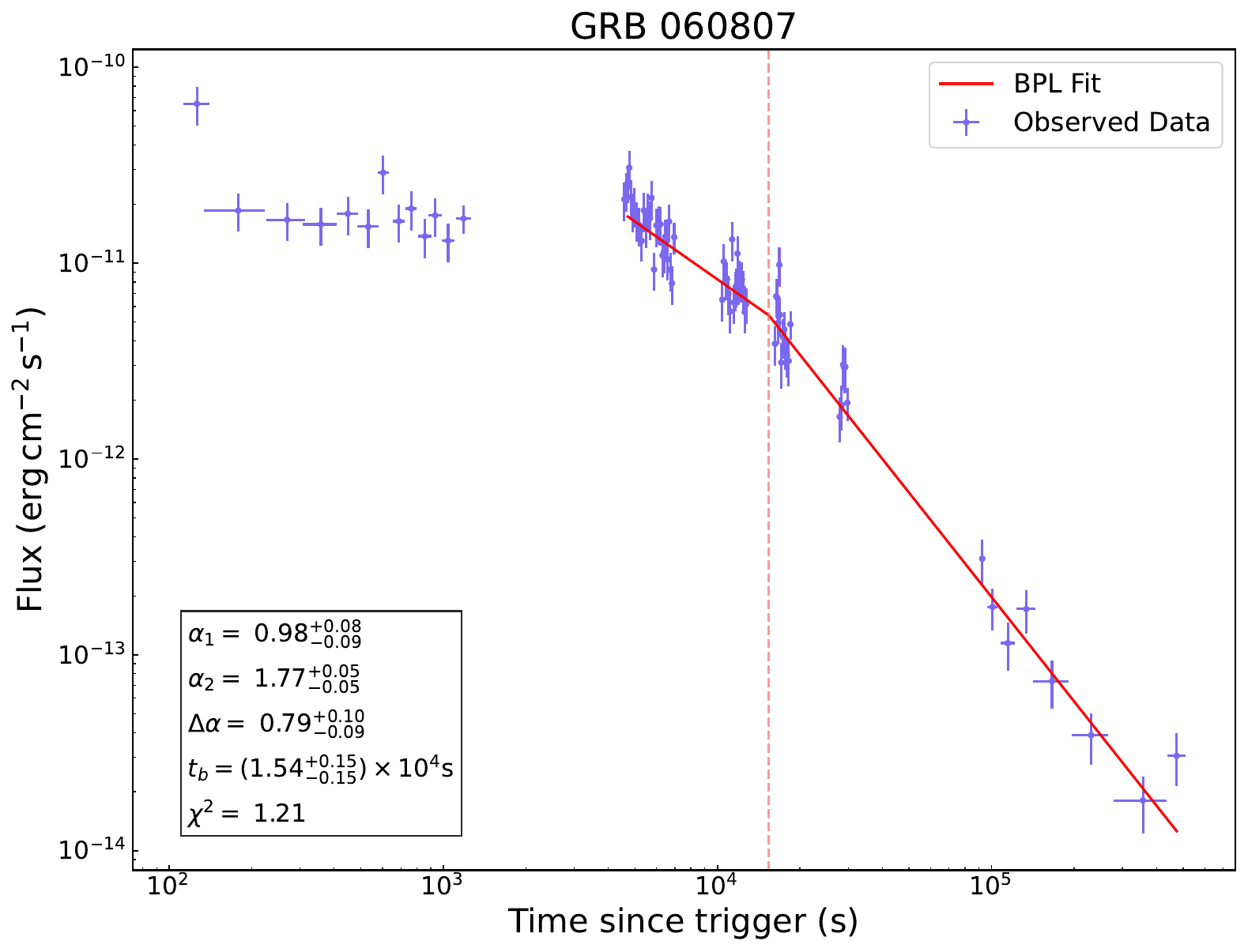}}%
\resizebox{45mm}{!}{\includegraphics[]{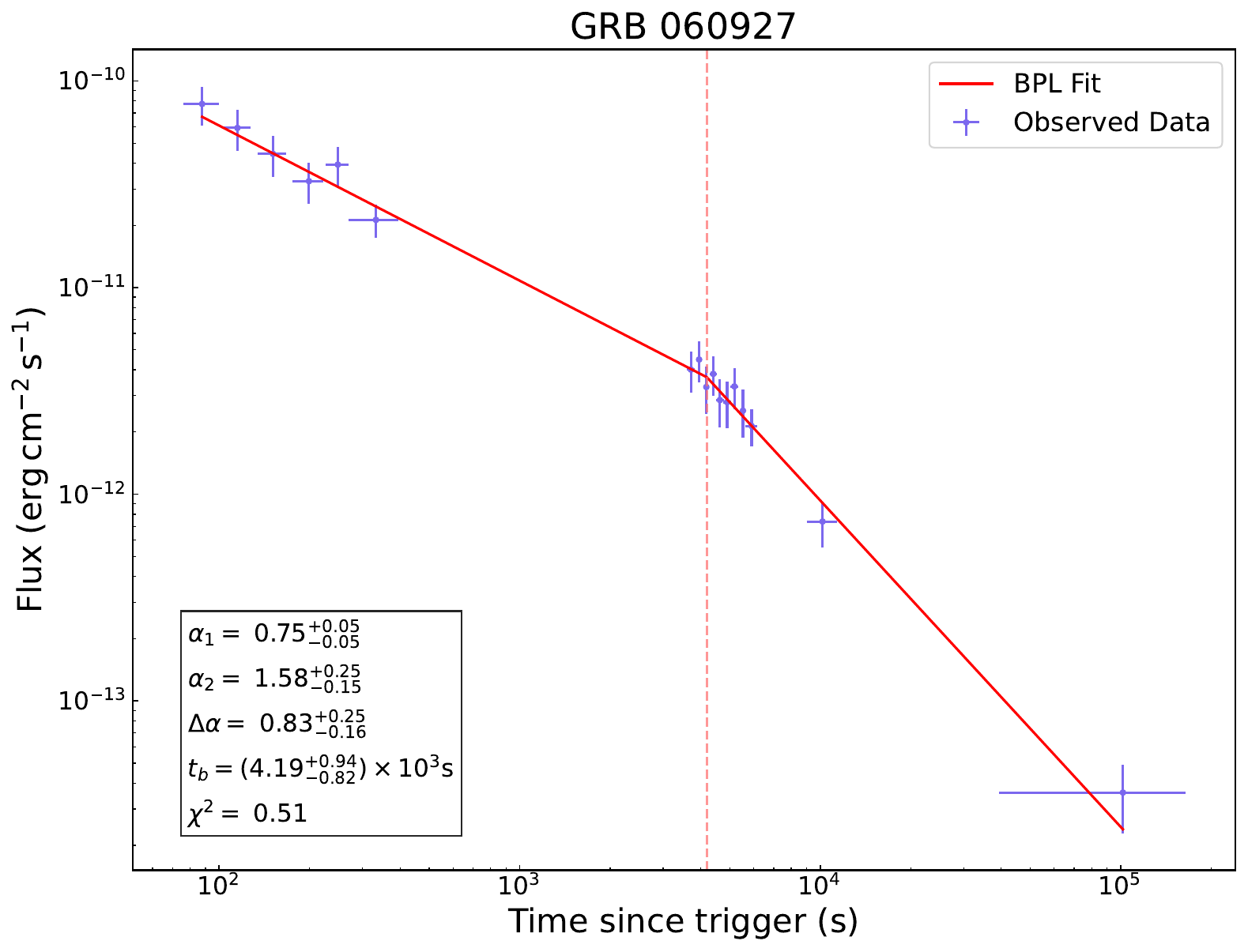}}%
\resizebox{45mm}{!}{\includegraphics[]{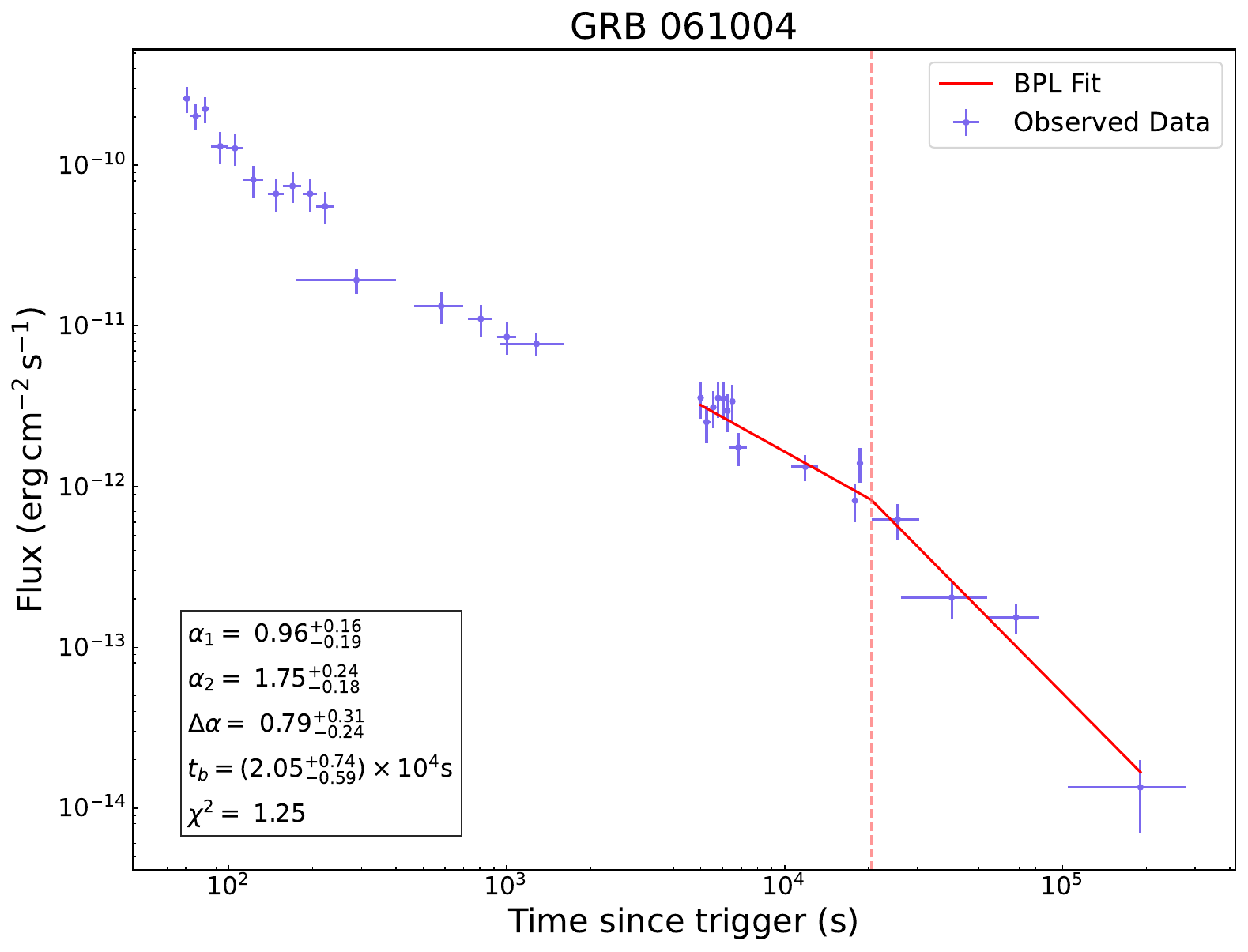}}\\
\resizebox{45mm}{!}{\includegraphics[]{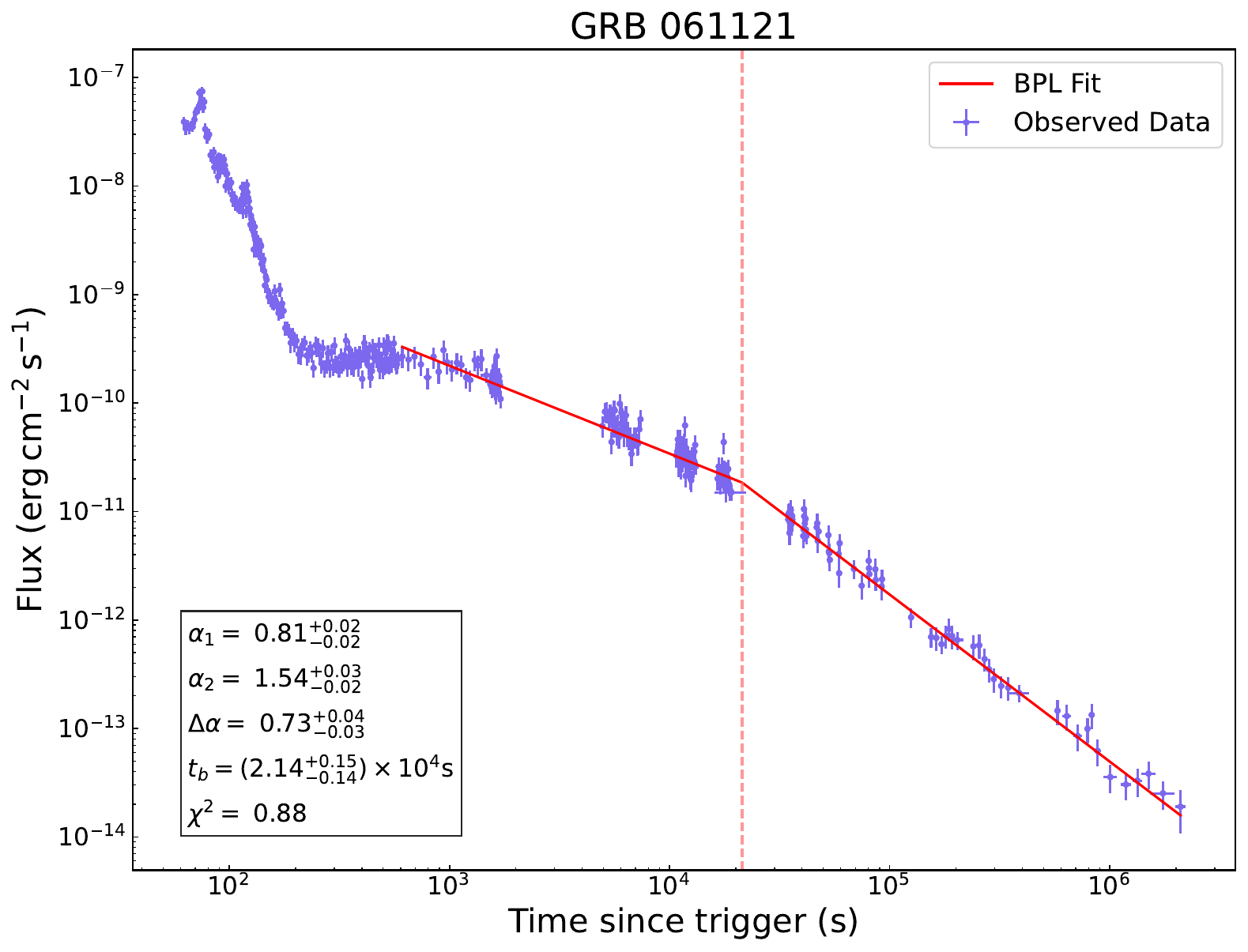}}%
\resizebox{45mm}{!}{\includegraphics[]{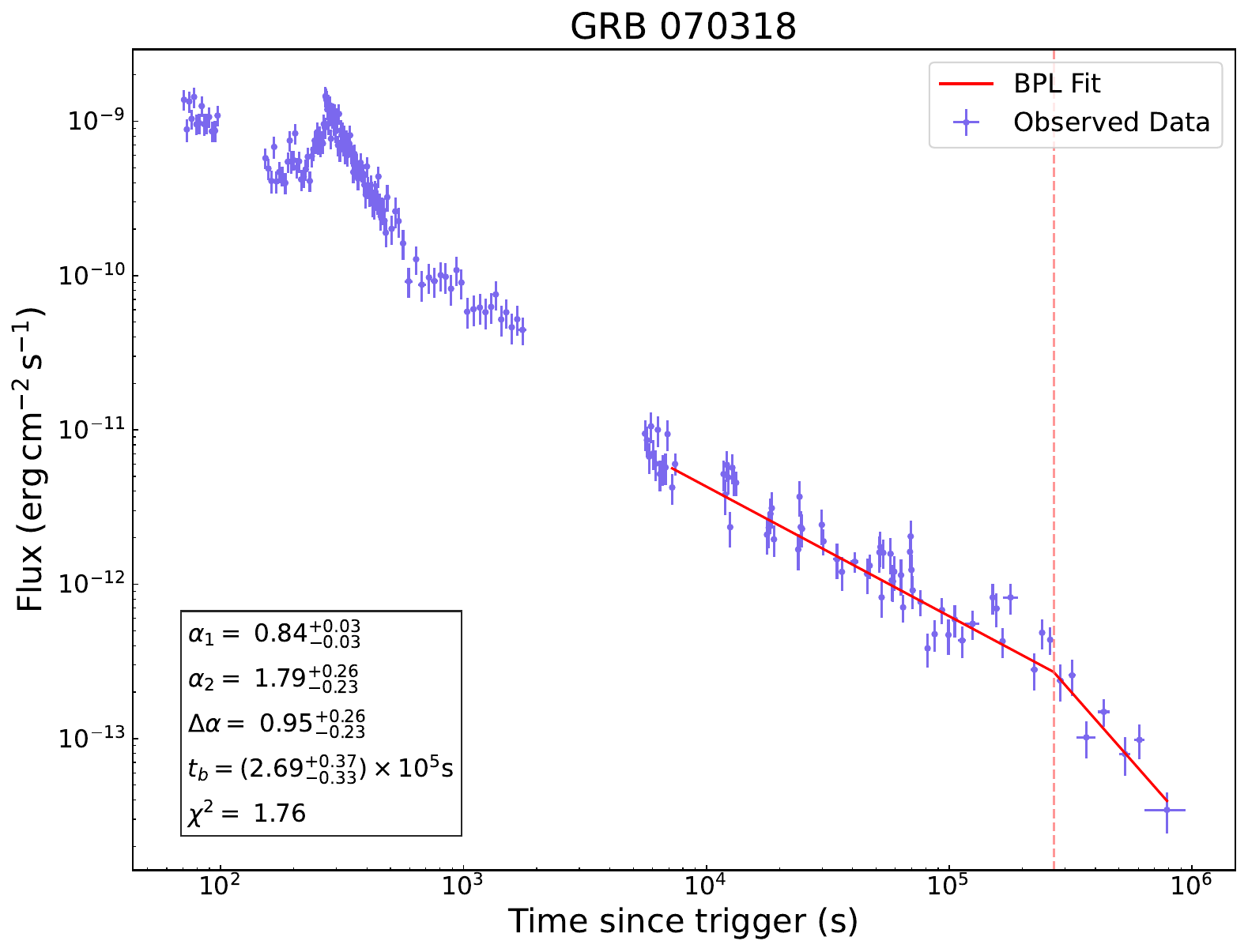}}%
\resizebox{45mm}{!}{\includegraphics[]{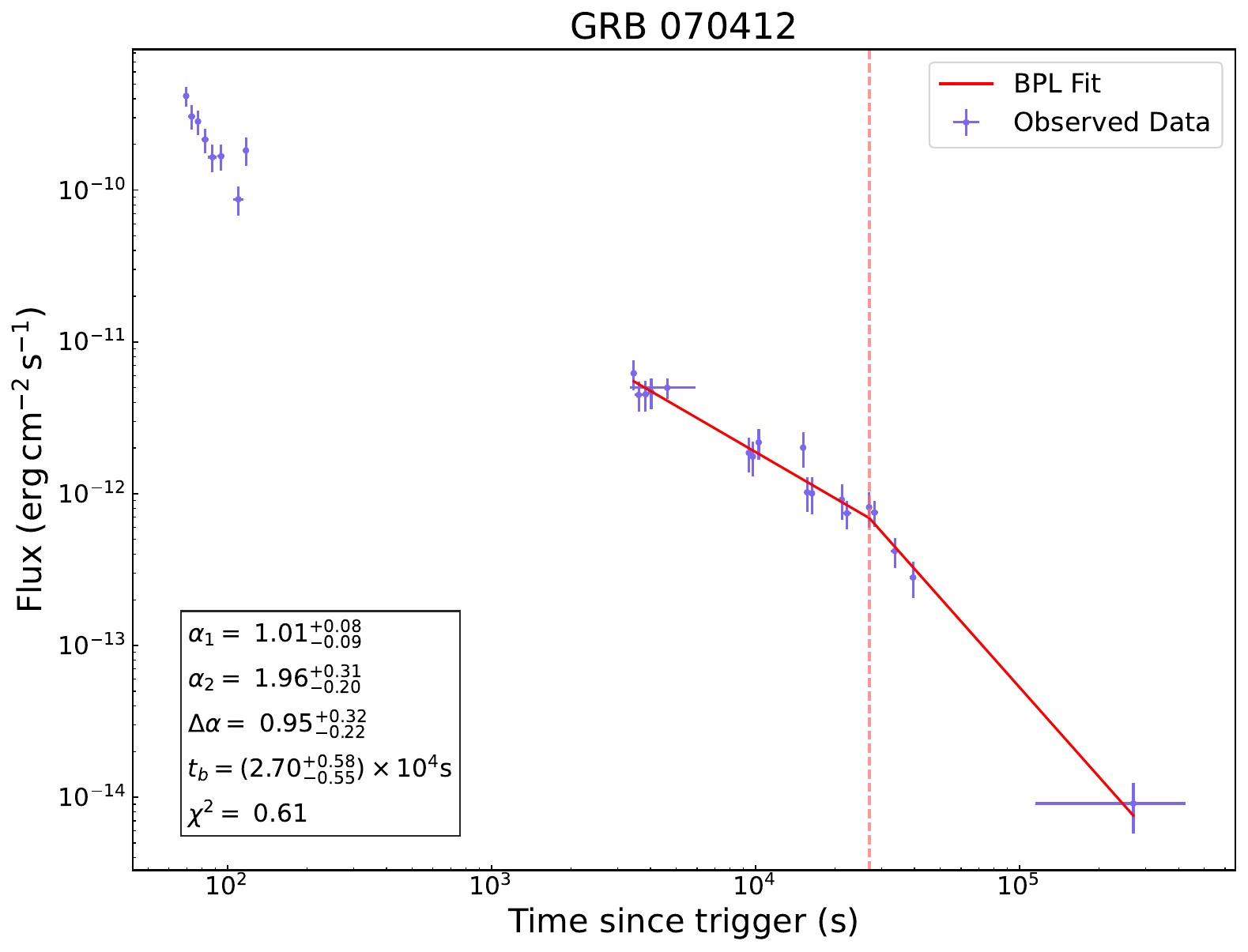}}%
\resizebox{45mm}{!}{\includegraphics[]{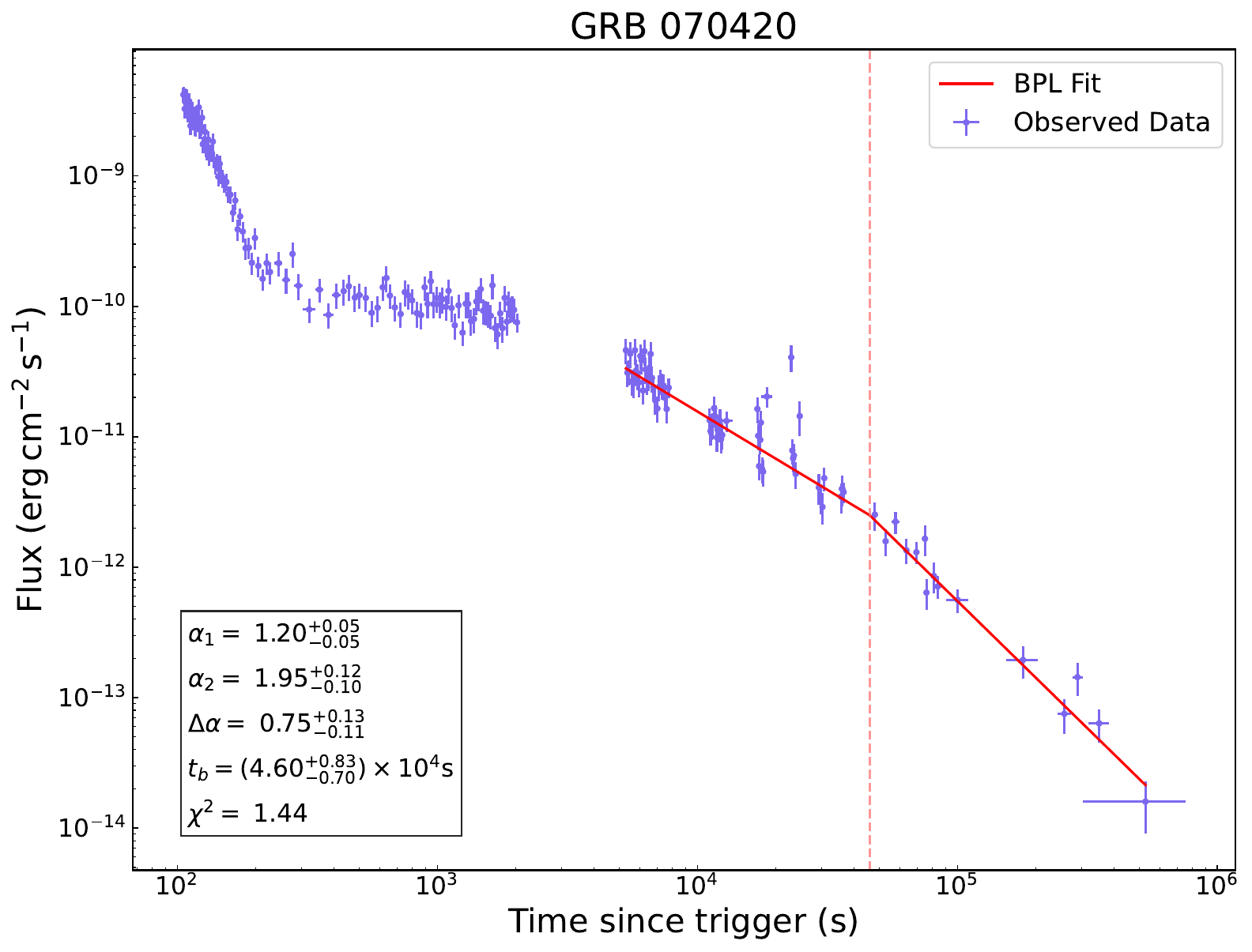}}\\
\resizebox{45mm}{!}{\includegraphics[]{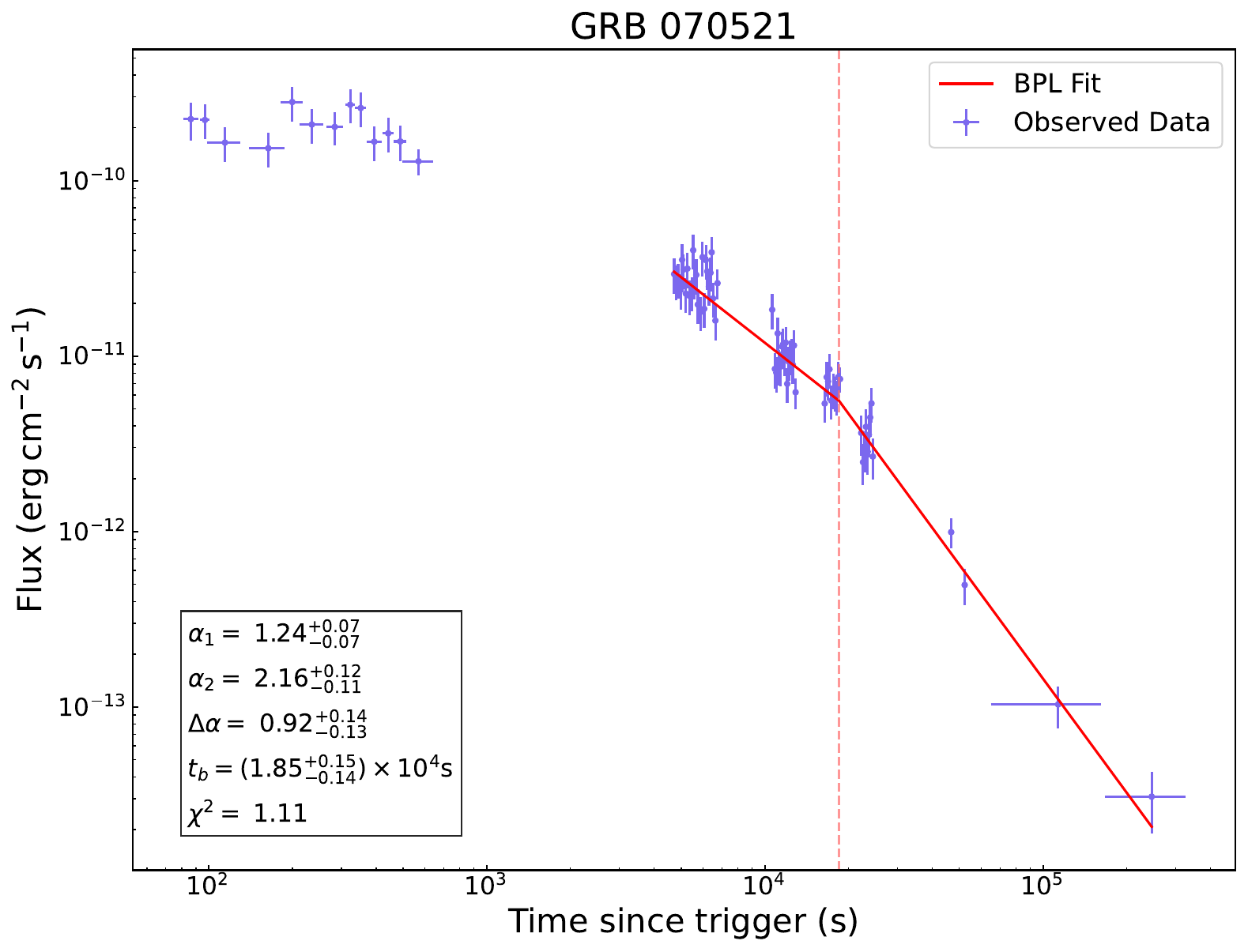}}%
\resizebox{45mm}{!}{\includegraphics[]{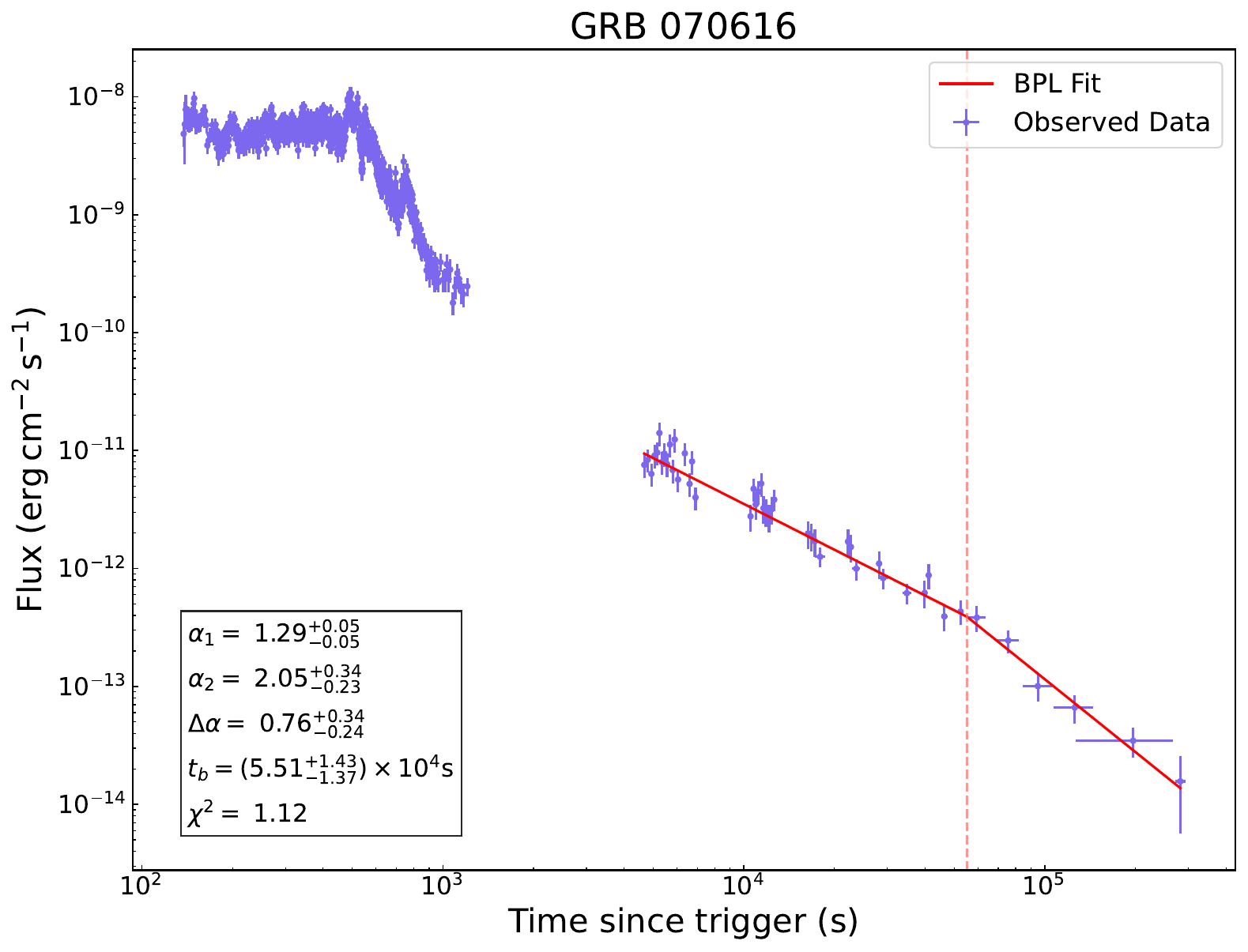}}%
\resizebox{45mm}{!}{\includegraphics[]{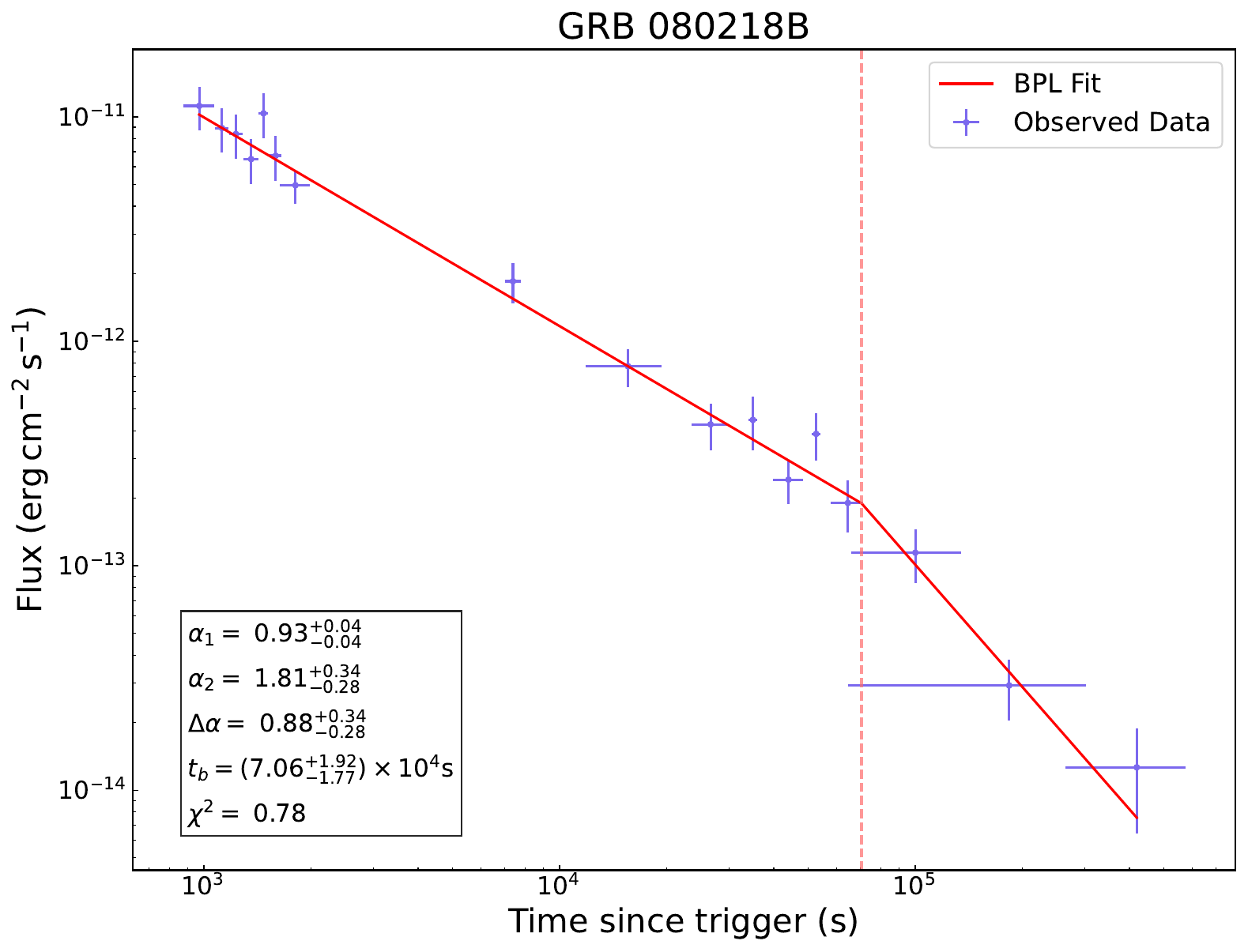}}%
\resizebox{45mm}{!}{\includegraphics[]{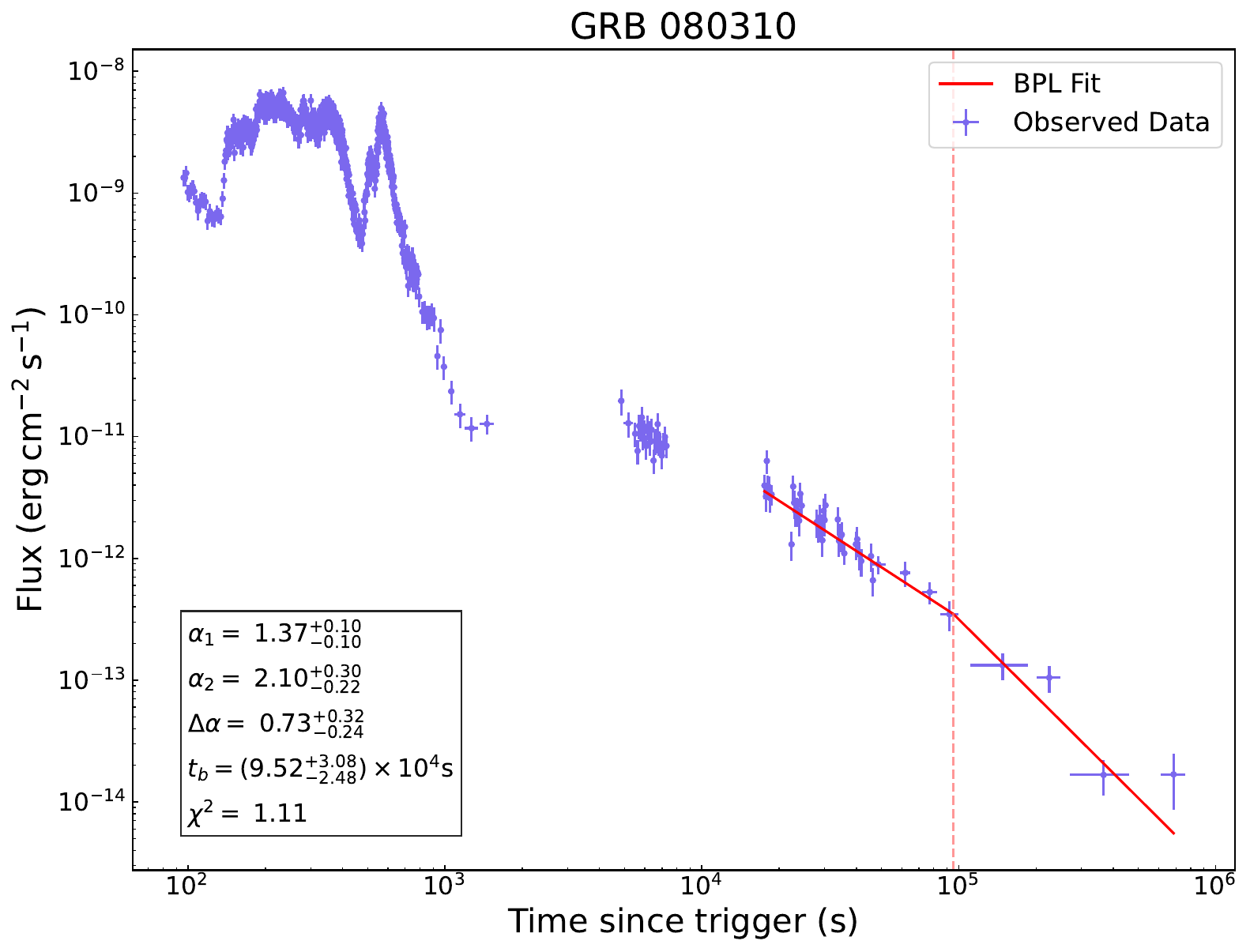}}\\
\resizebox{45mm}{!}{\includegraphics[]{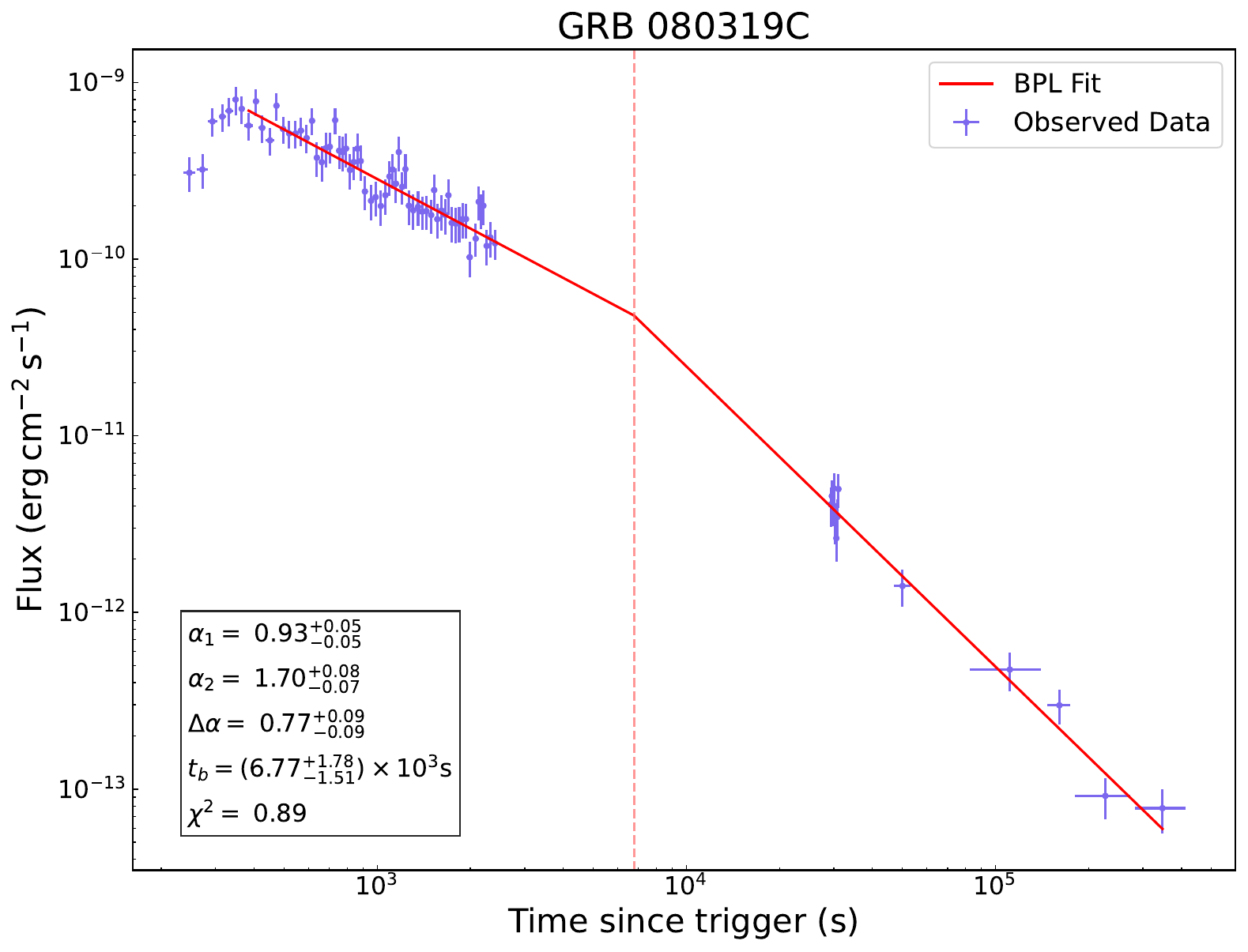}}%
\resizebox{45mm}{!}{\includegraphics[]{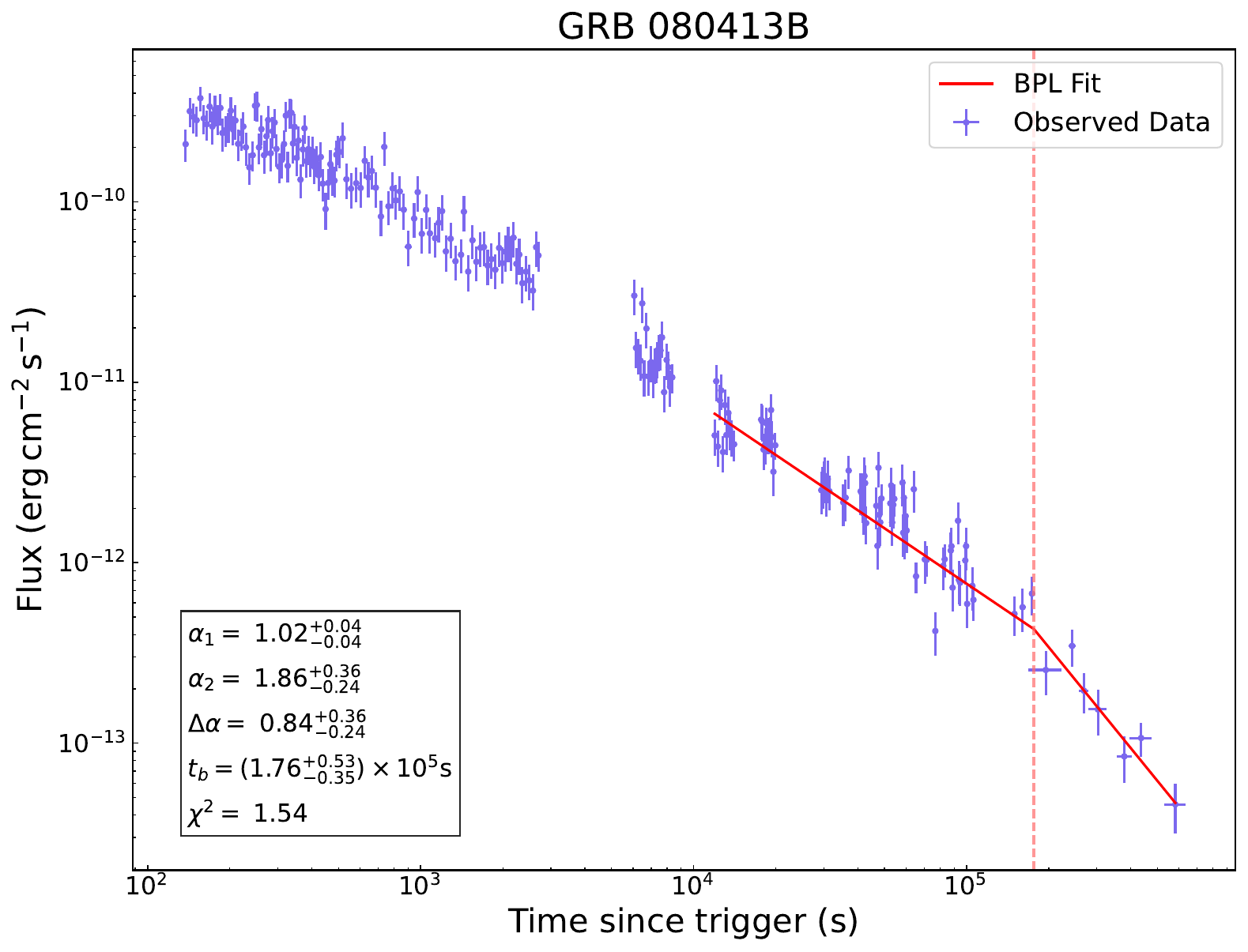}}%
\resizebox{45mm}{!}{\includegraphics[]{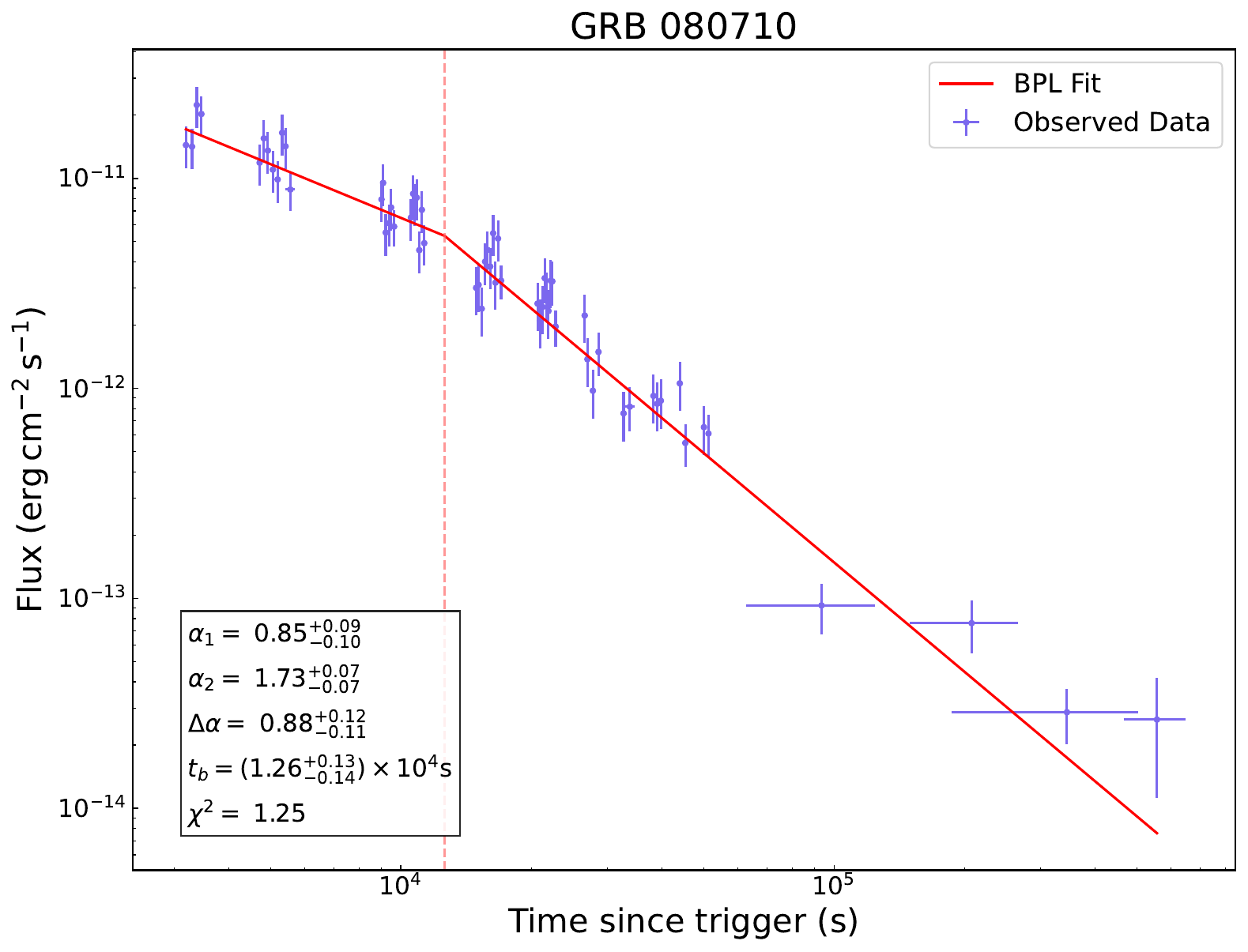}}%
\resizebox{45mm}{!}{\includegraphics[]{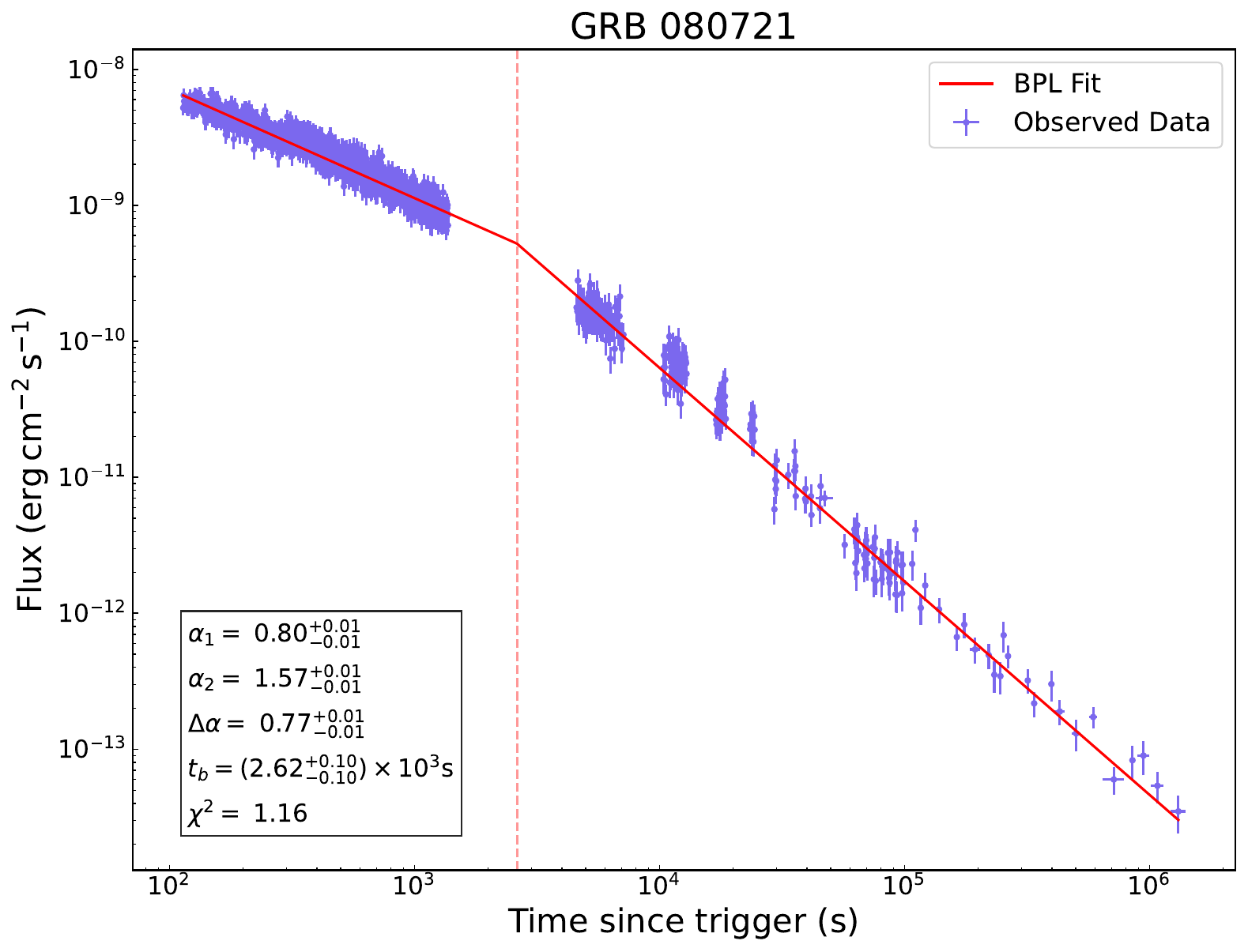}}\\
\caption{X-ray afterglow light curves for the 82 GRBs in our sample classified as having an ISM circum-burst environment. Blue points show the Swift/XRT observational data, and the solid red curves represent the best-fit broken power-law models. Multi-band (X-ray and optical) light curves are appended at the end.}
\label{fig:figure1}
\end{figure*}

\clearpage
\addtocounter{figure}{-1}
\begin{figure}
\centering
\resizebox{45mm}{!}{\includegraphics[]{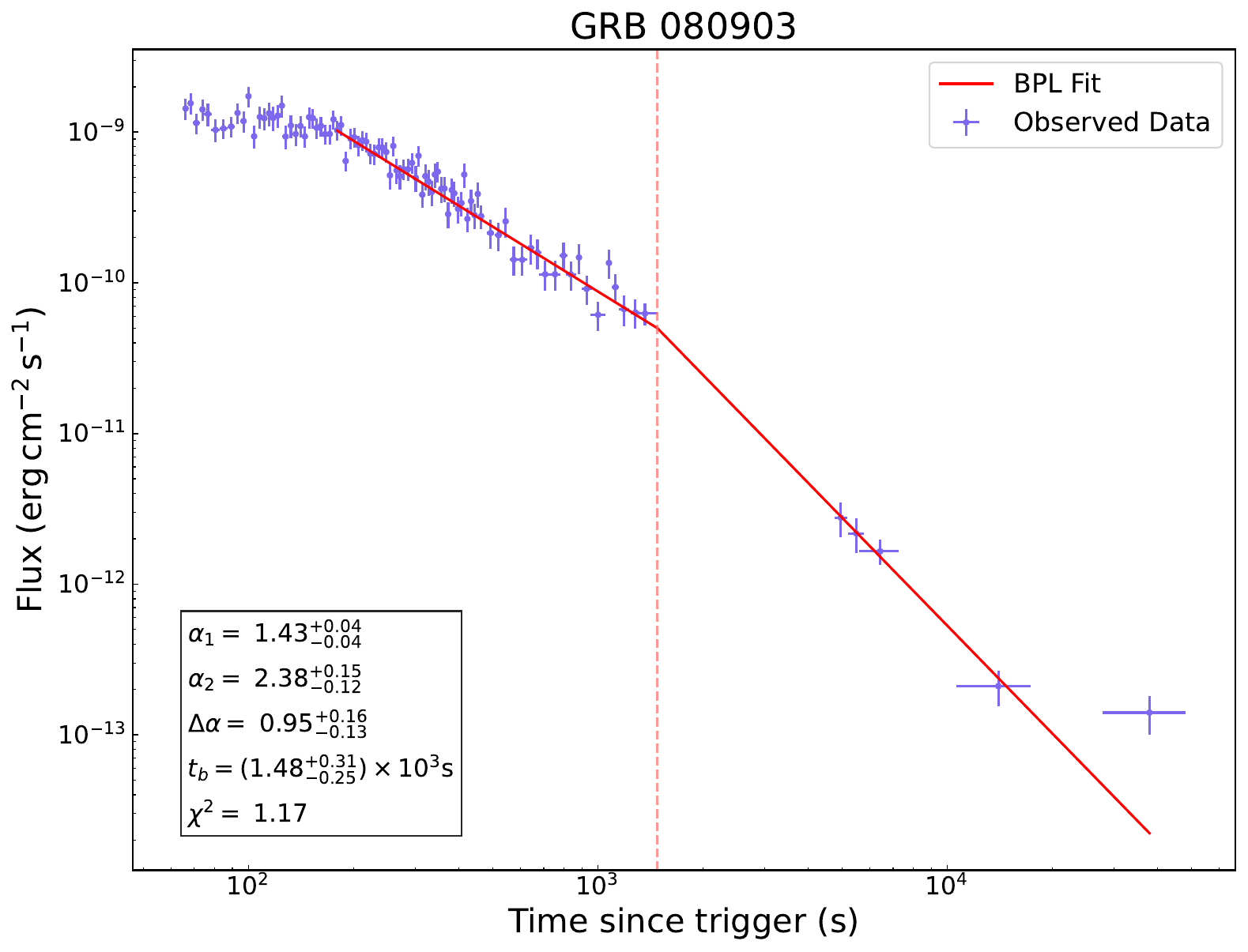}}%
\resizebox{45mm}{!}{\includegraphics[]{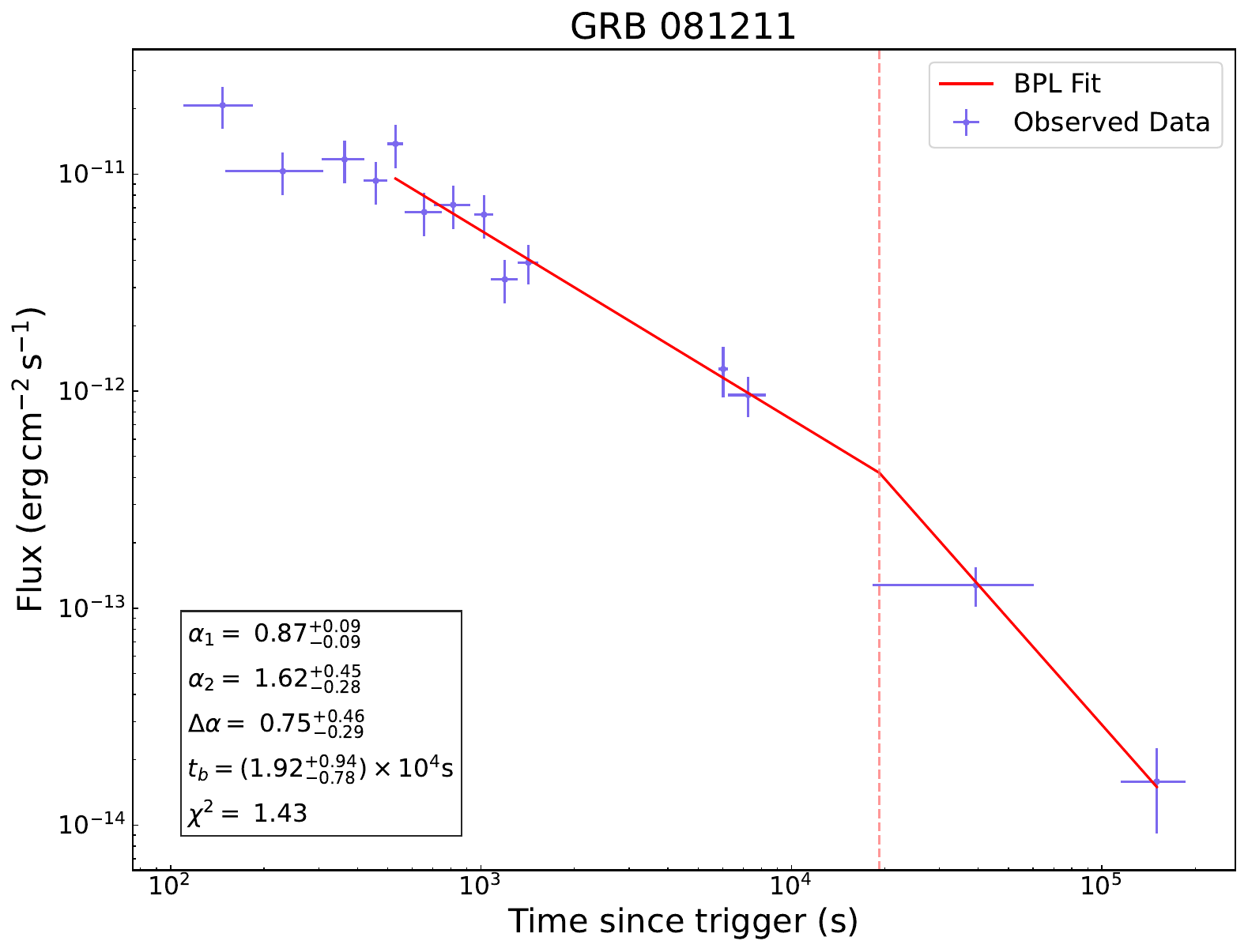}}%
\resizebox{45mm}{!}{\includegraphics[]{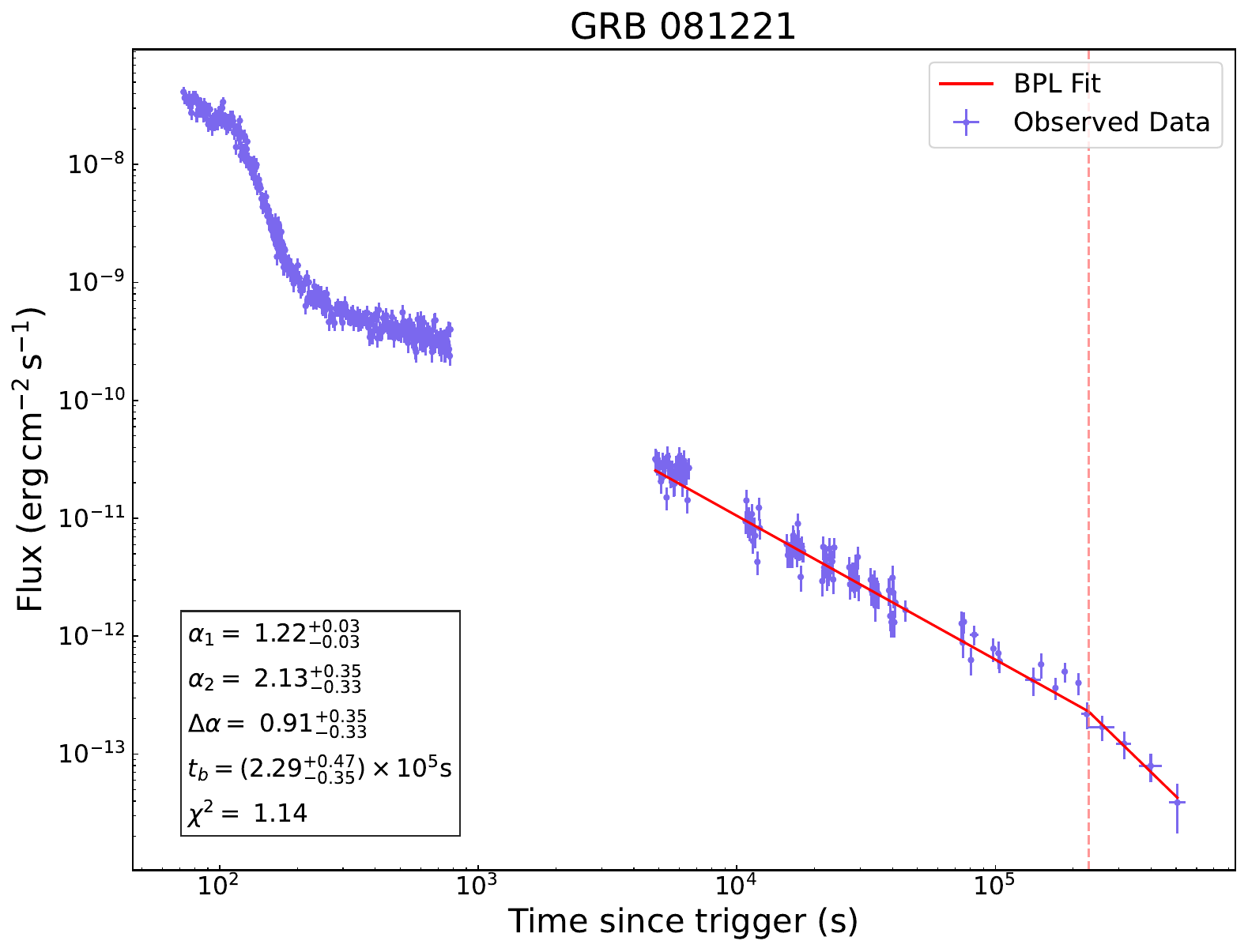}}%
\resizebox{45mm}{!}{\includegraphics[]{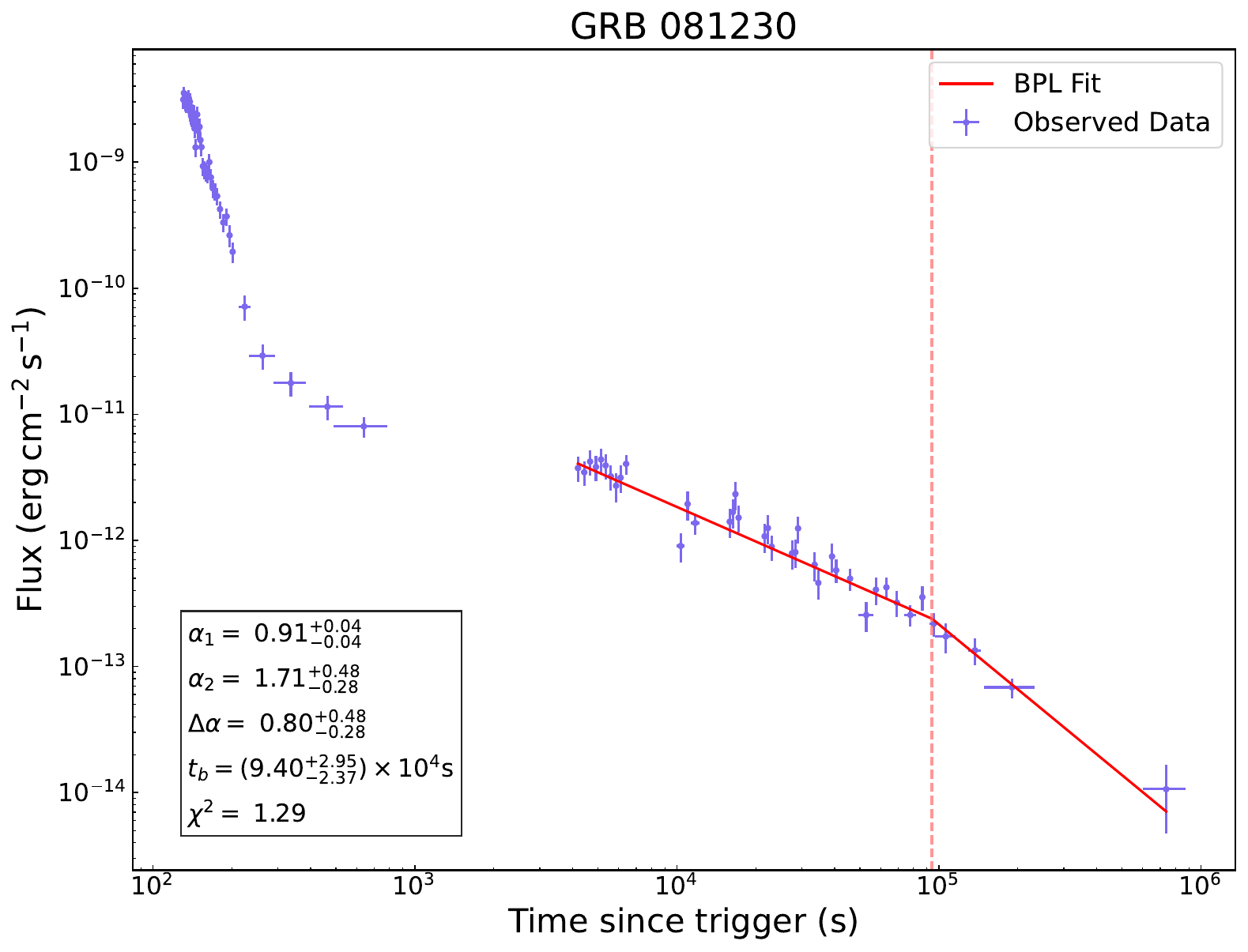}}\\
\resizebox{45mm}{!}{\includegraphics[]{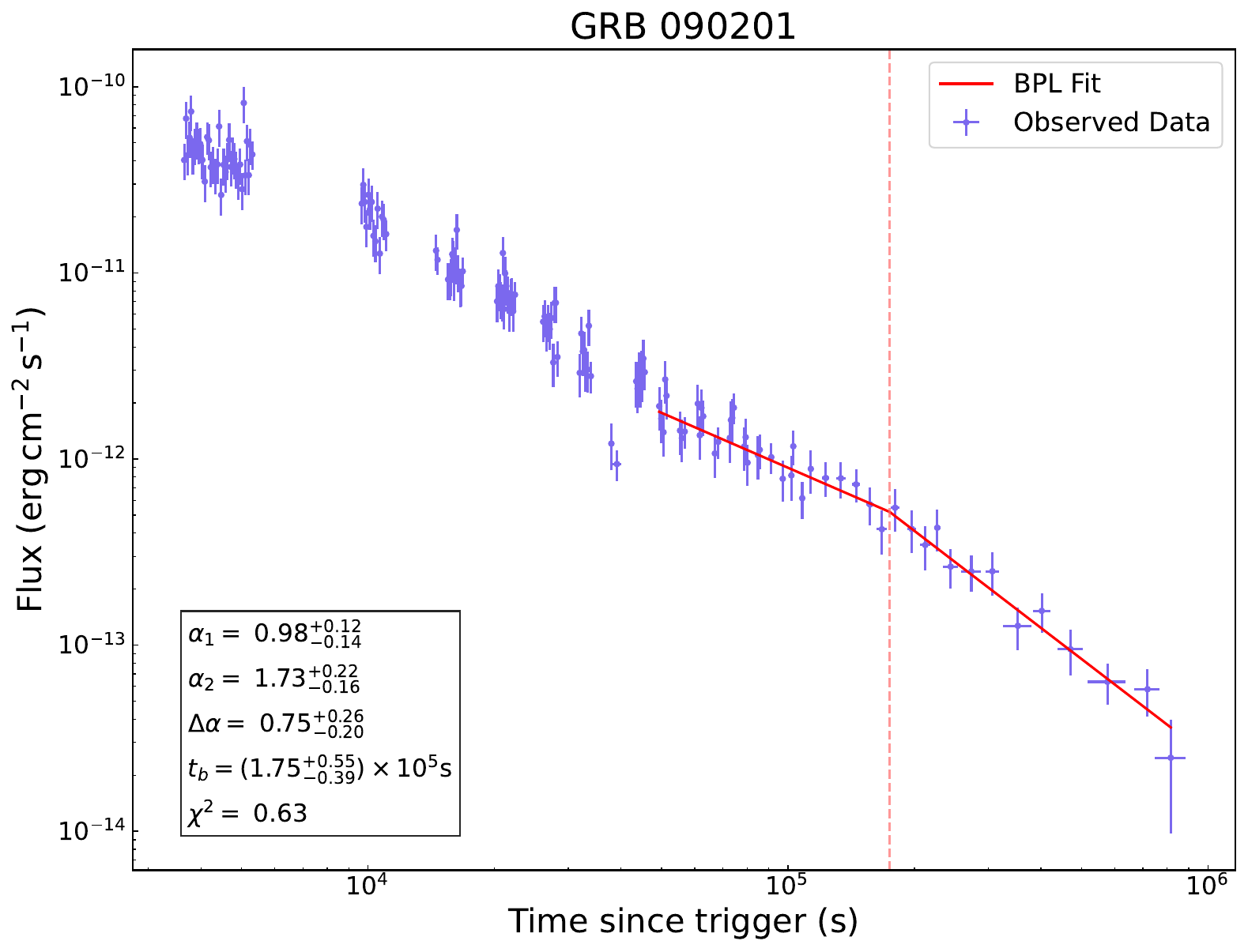}}%
\resizebox{45mm}{!}{\includegraphics[]{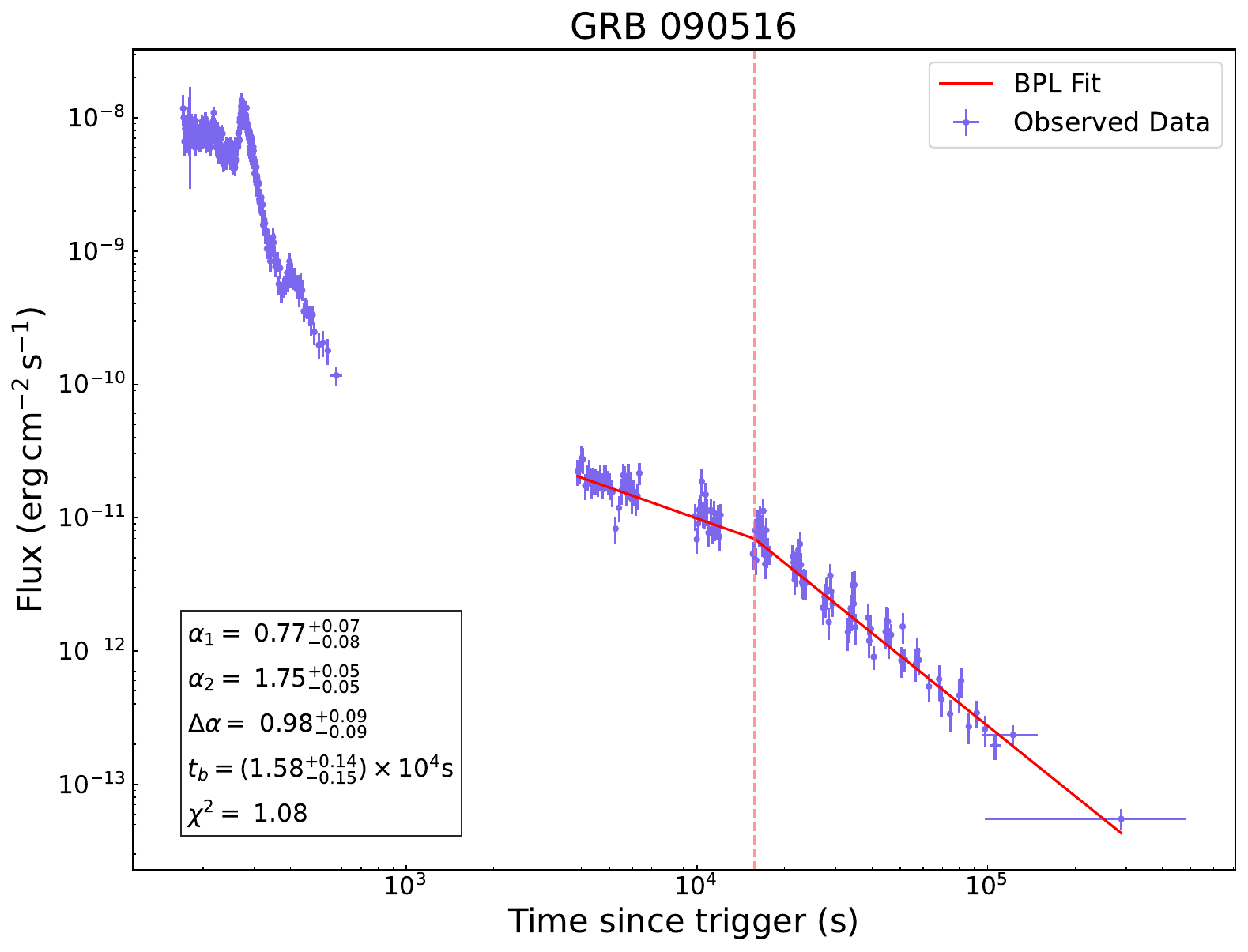}}%
\resizebox{45mm}{!}{\includegraphics[]{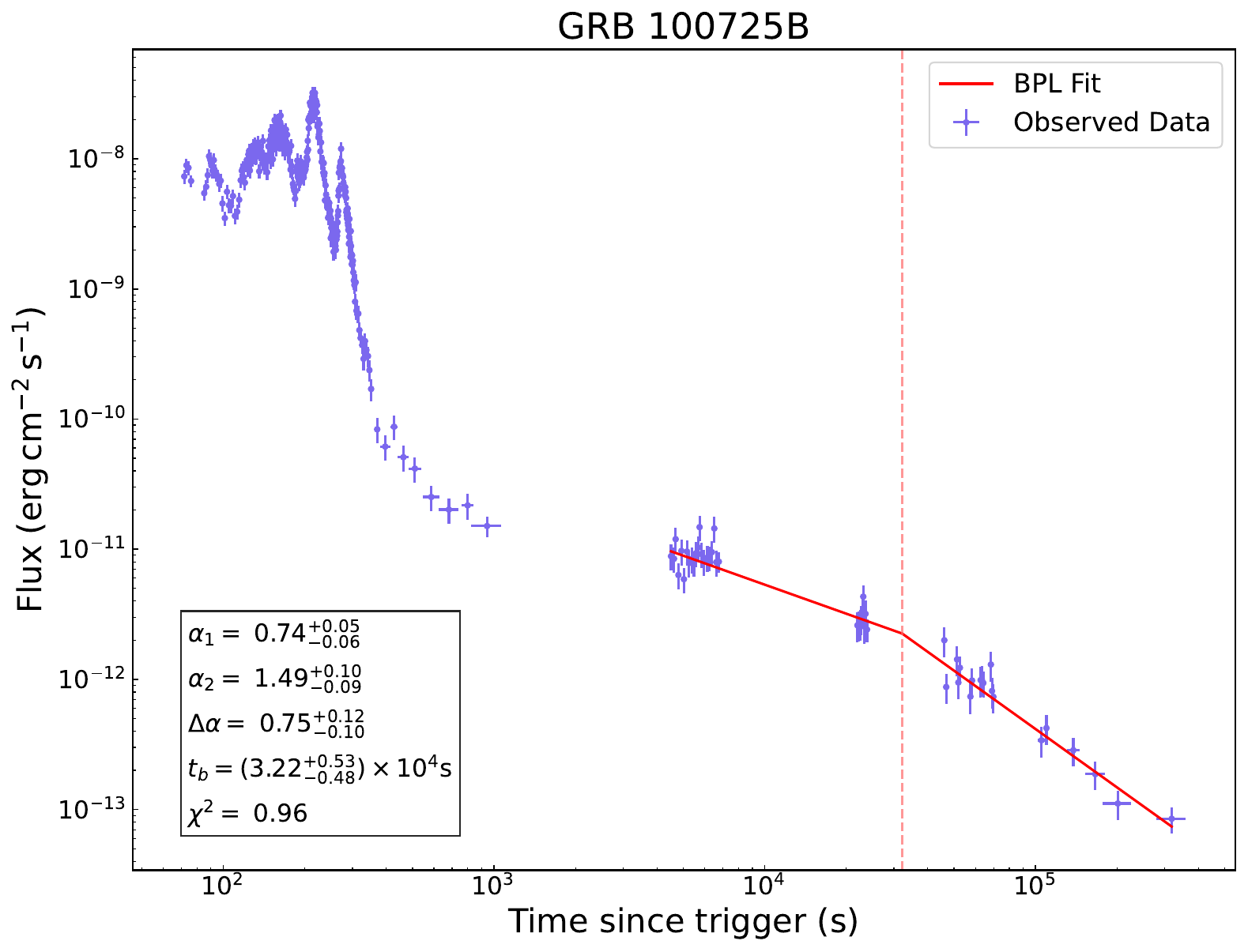}}%
\resizebox{45mm}{!}{\includegraphics[]{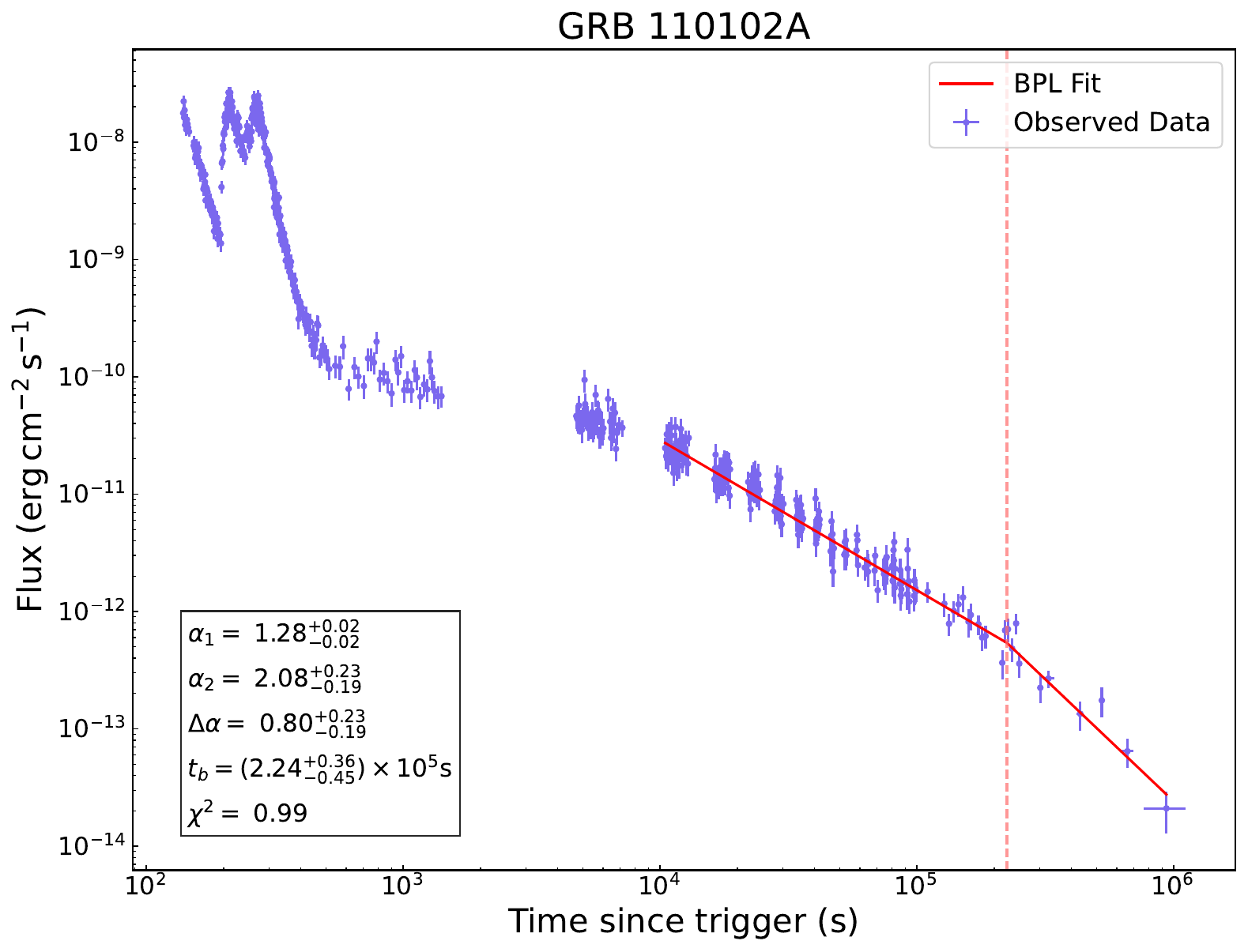}}\\
\resizebox{45mm}{!}{\includegraphics[]{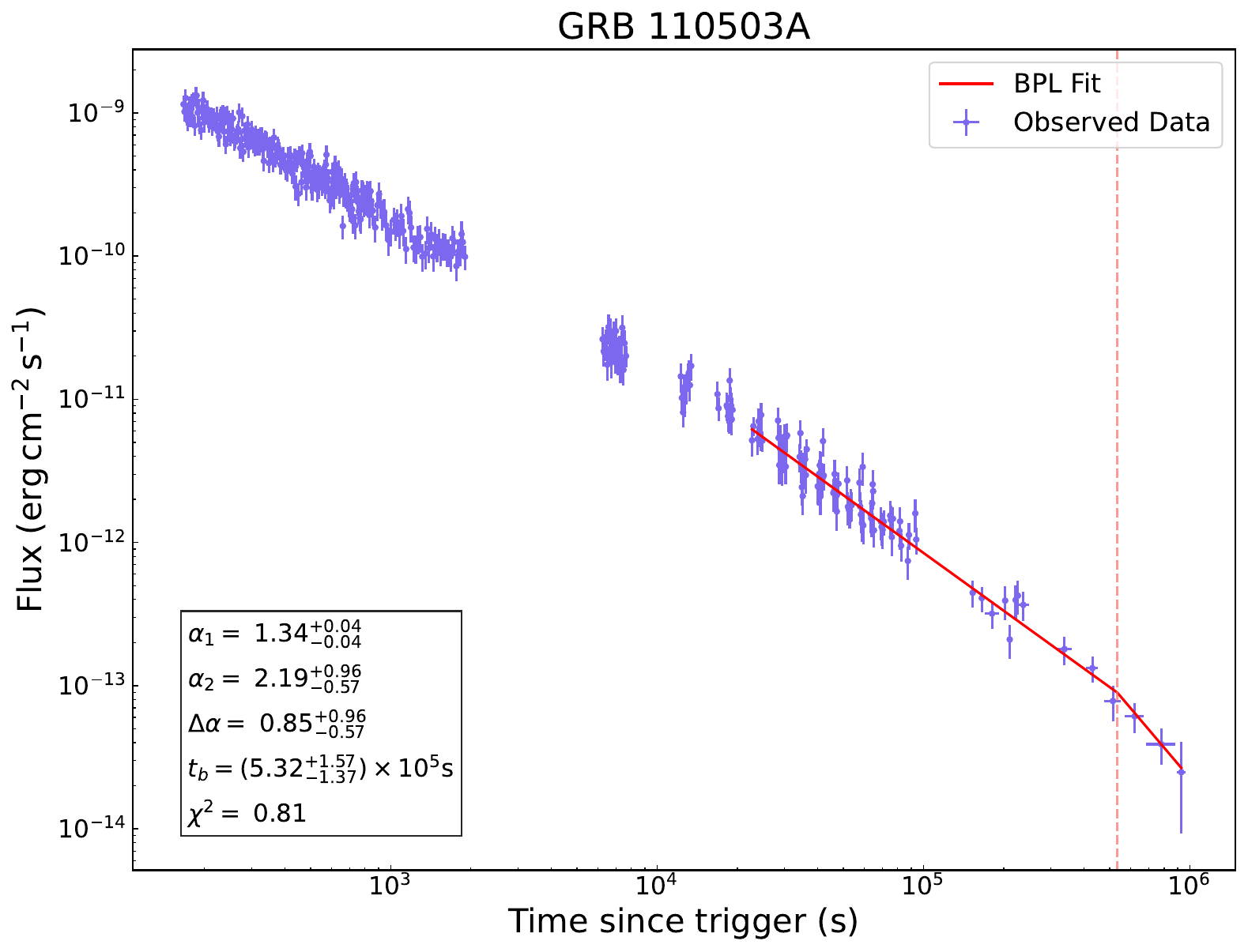}}%
\resizebox{45mm}{!}{\includegraphics[]{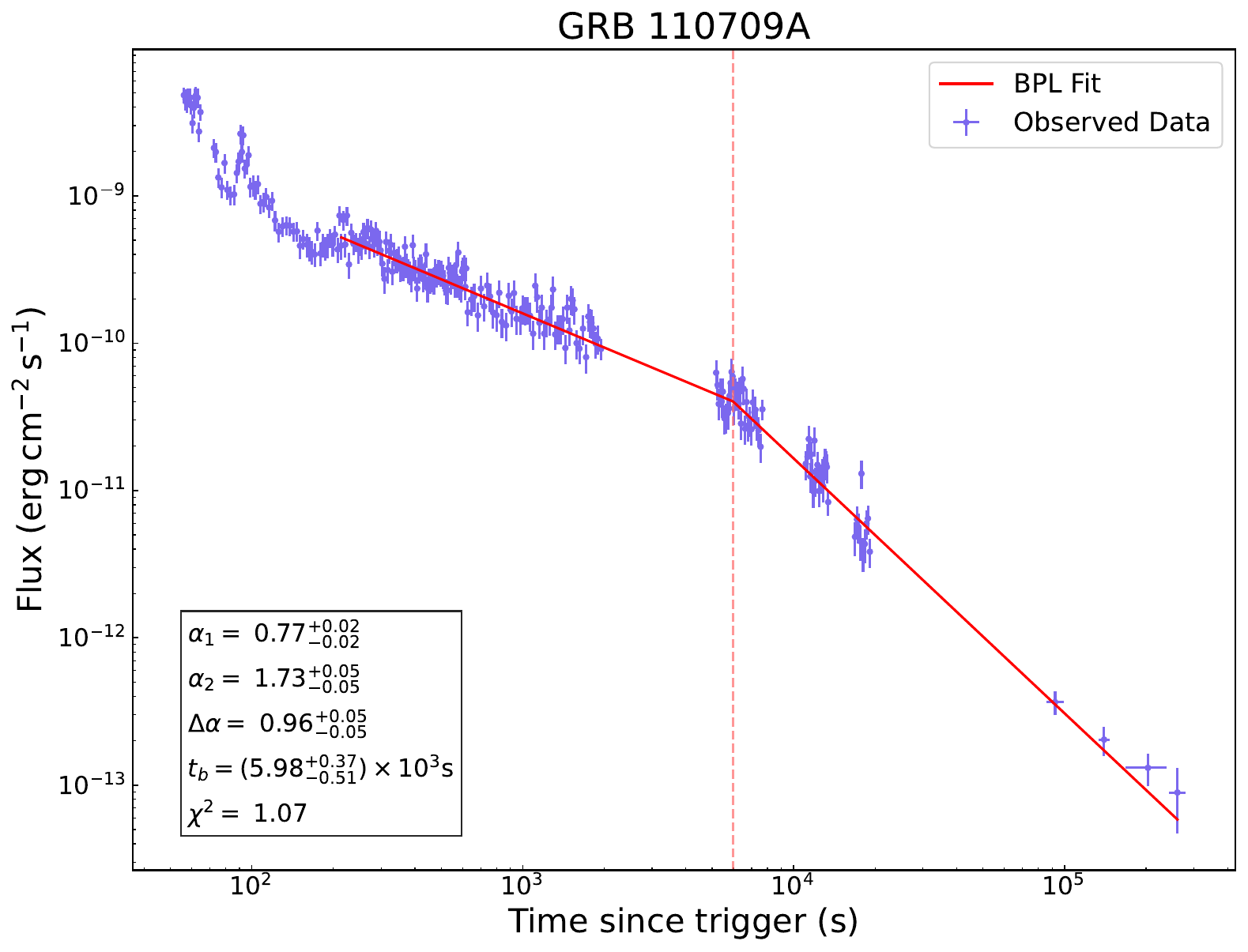}}%
\resizebox{45mm}{!}{\includegraphics[]{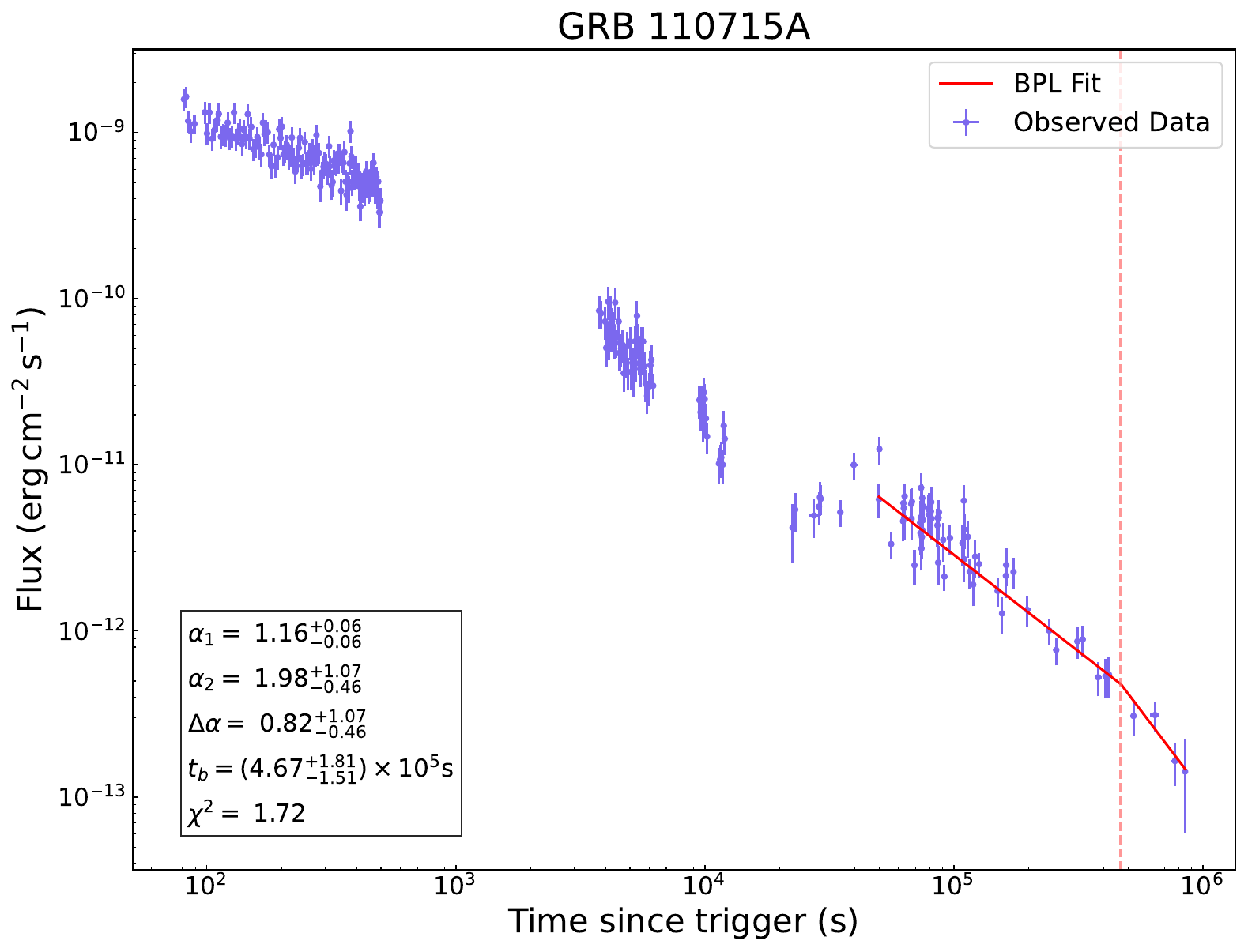}}%
\resizebox{45mm}{!}{\includegraphics[]{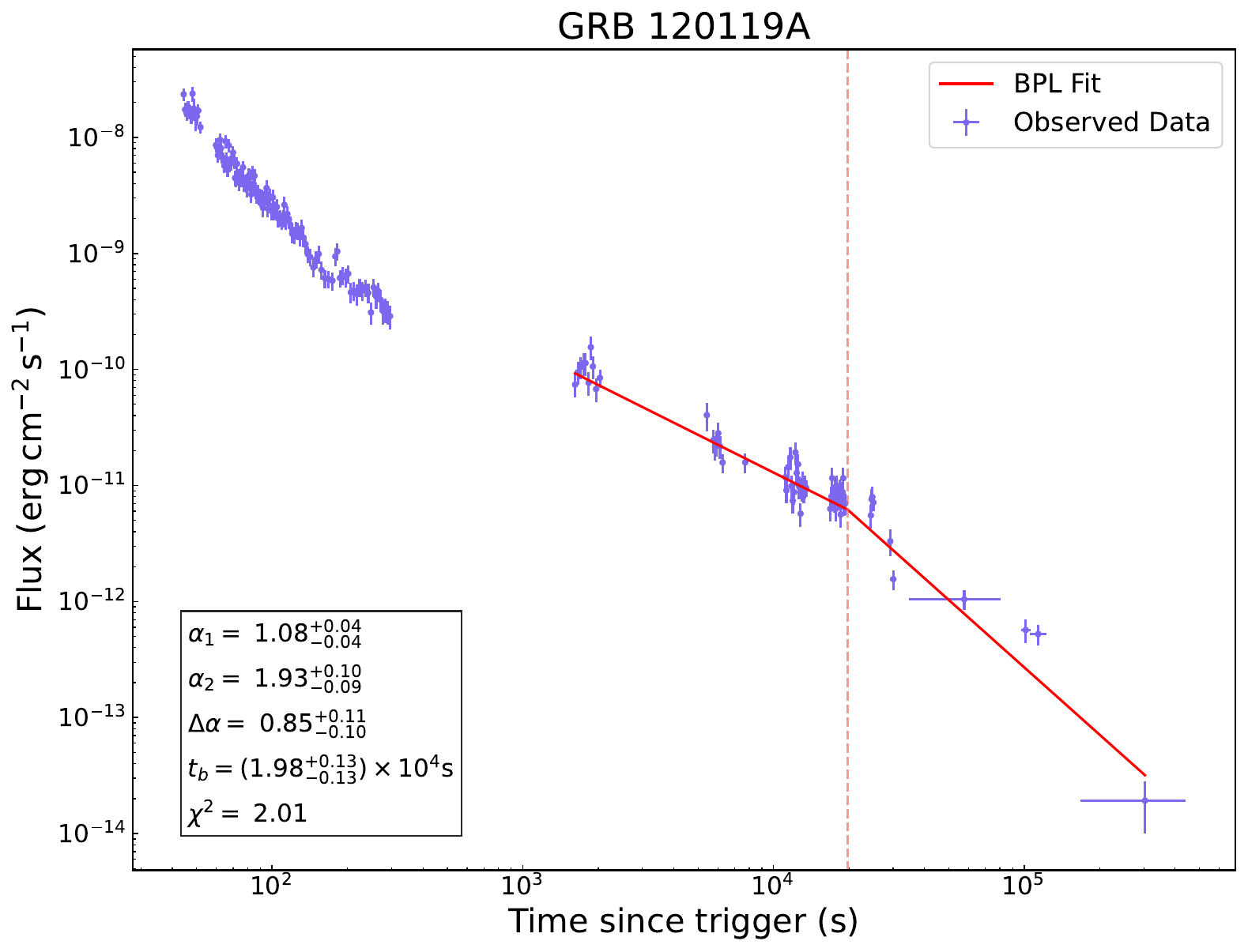}}\\
\resizebox{45mm}{!}{\includegraphics[]{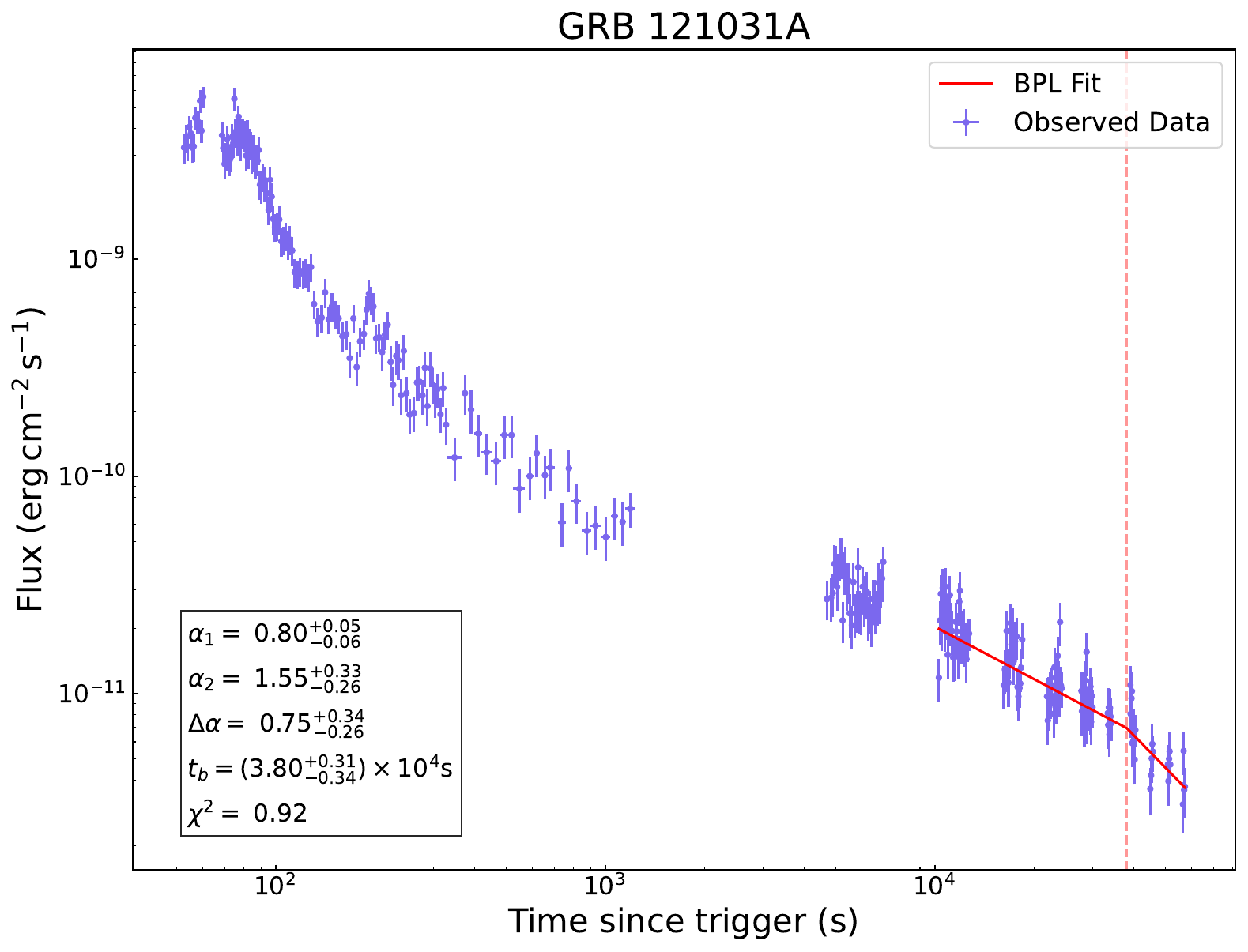}}%
\resizebox{45mm}{!}{\includegraphics[]{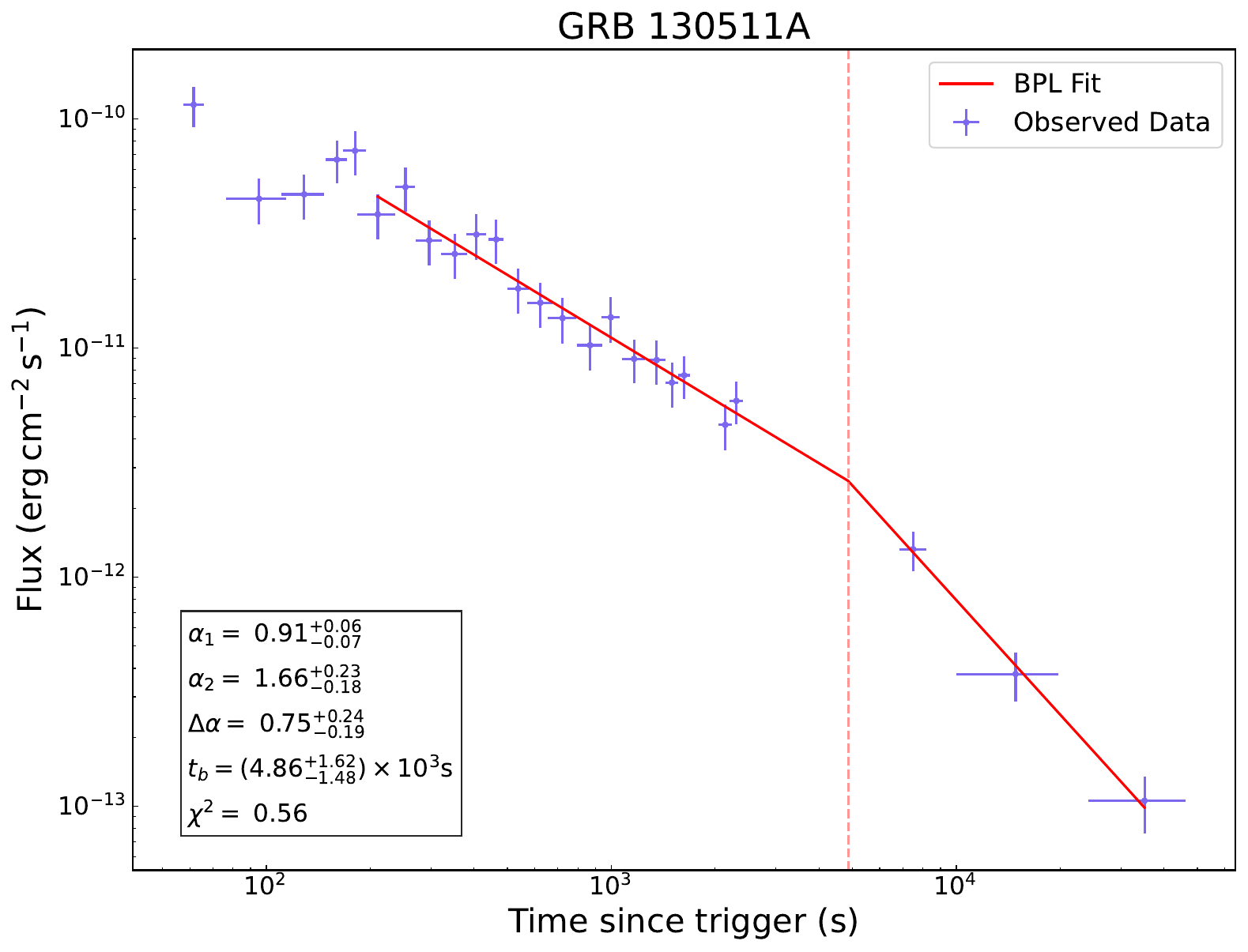}}%
\resizebox{45mm}{!}{\includegraphics[]{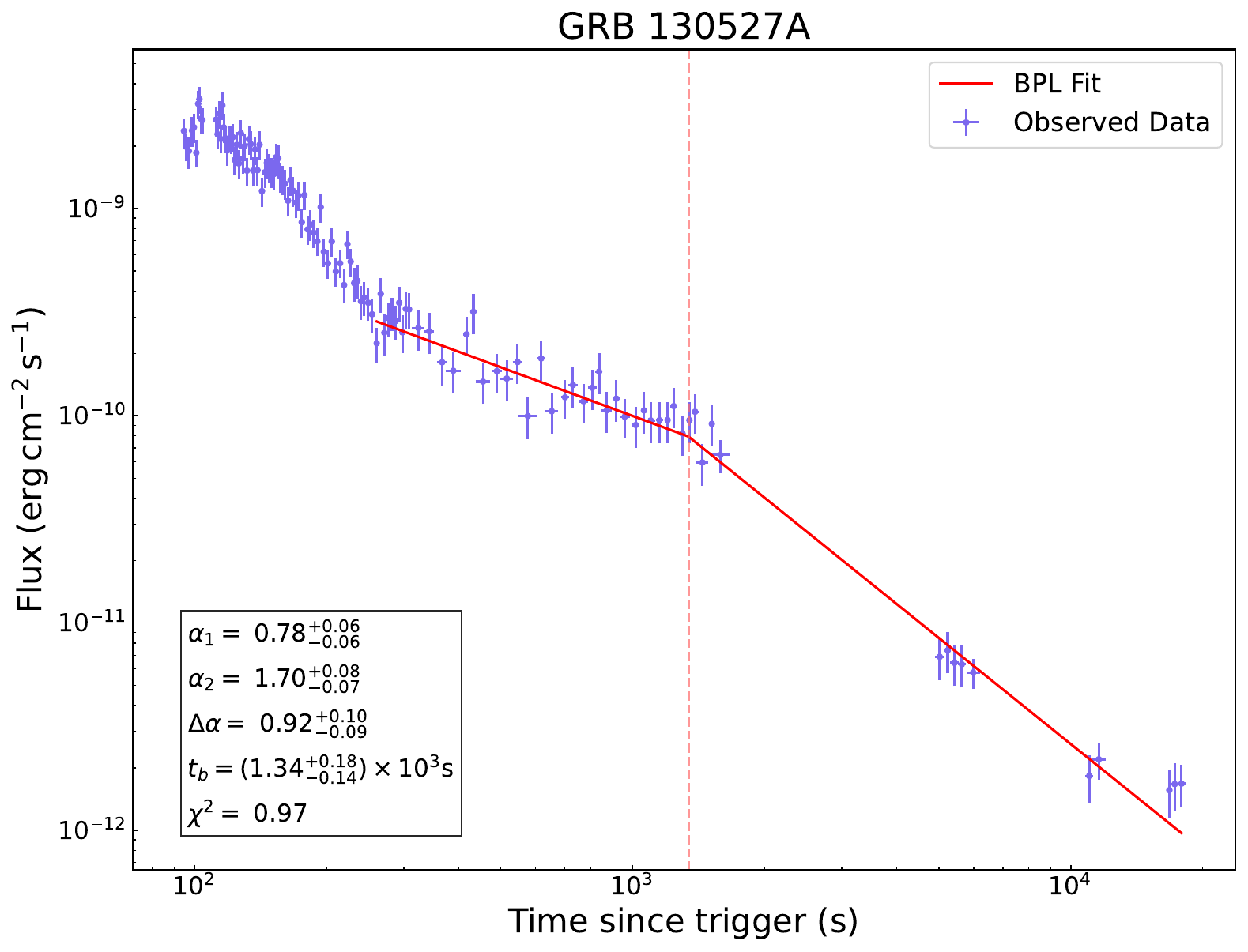}}%
\resizebox{45mm}{!}{\includegraphics[]{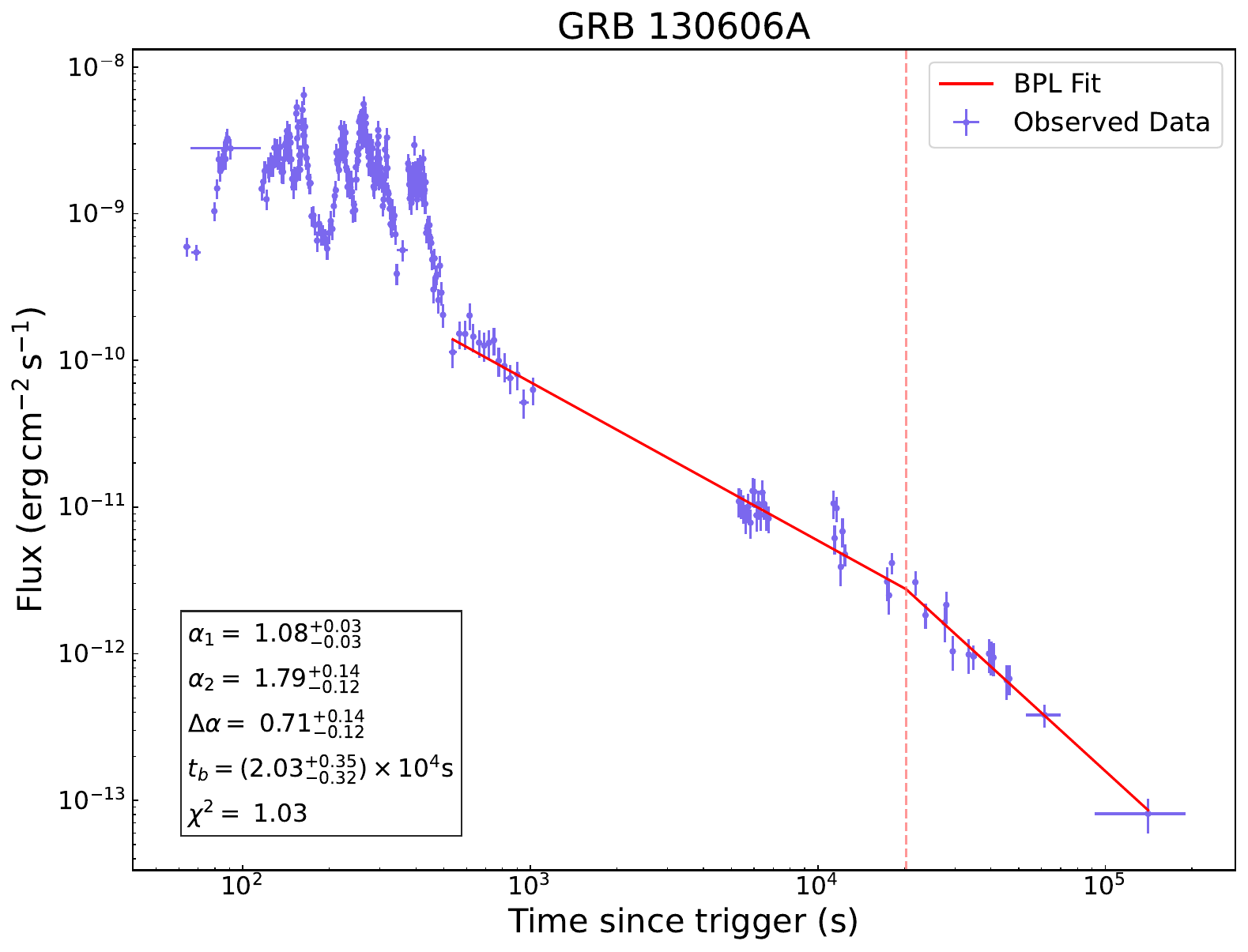}}\\
\resizebox{45mm}{!}{\includegraphics[]{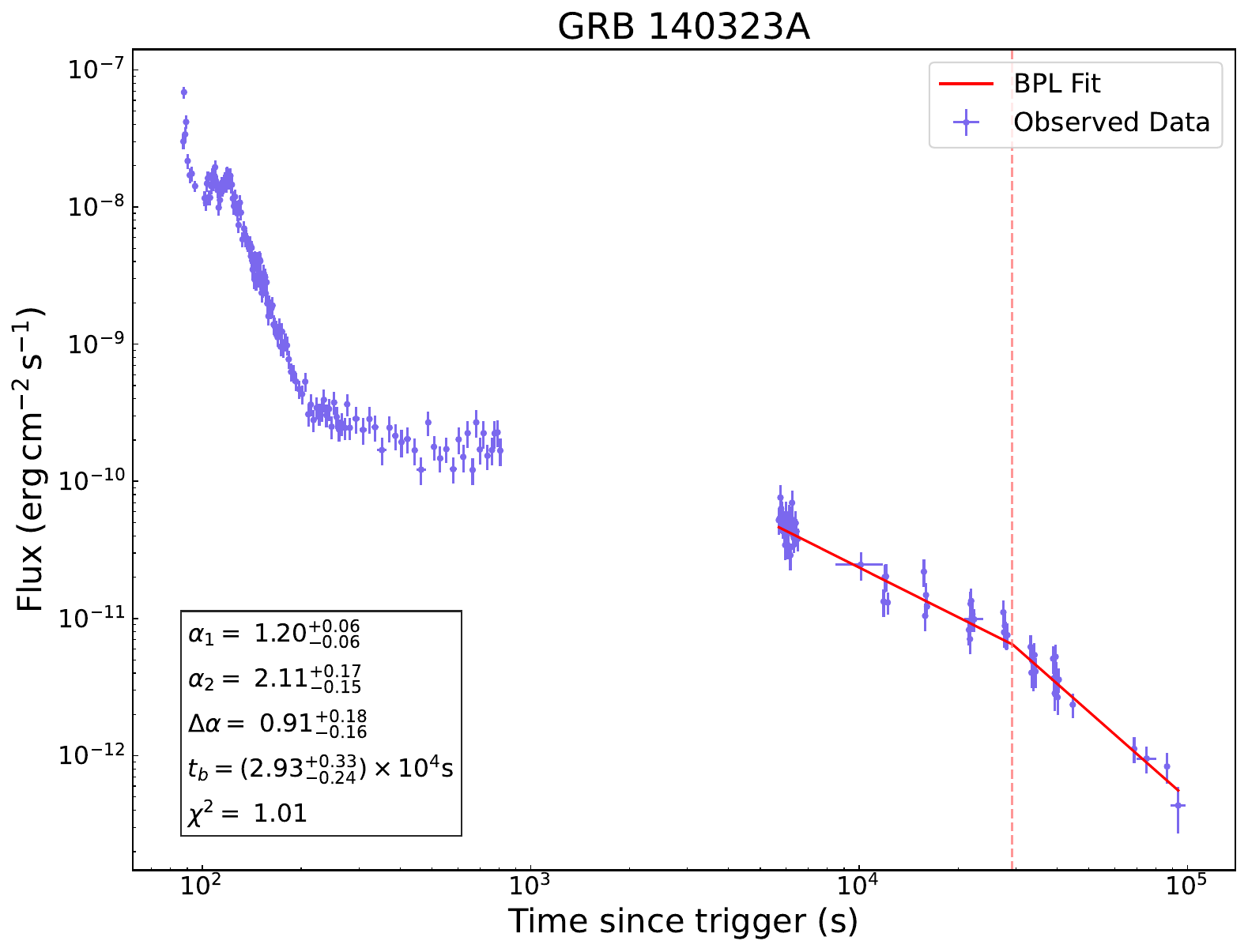}}%
\resizebox{45mm}{!}{\includegraphics[]{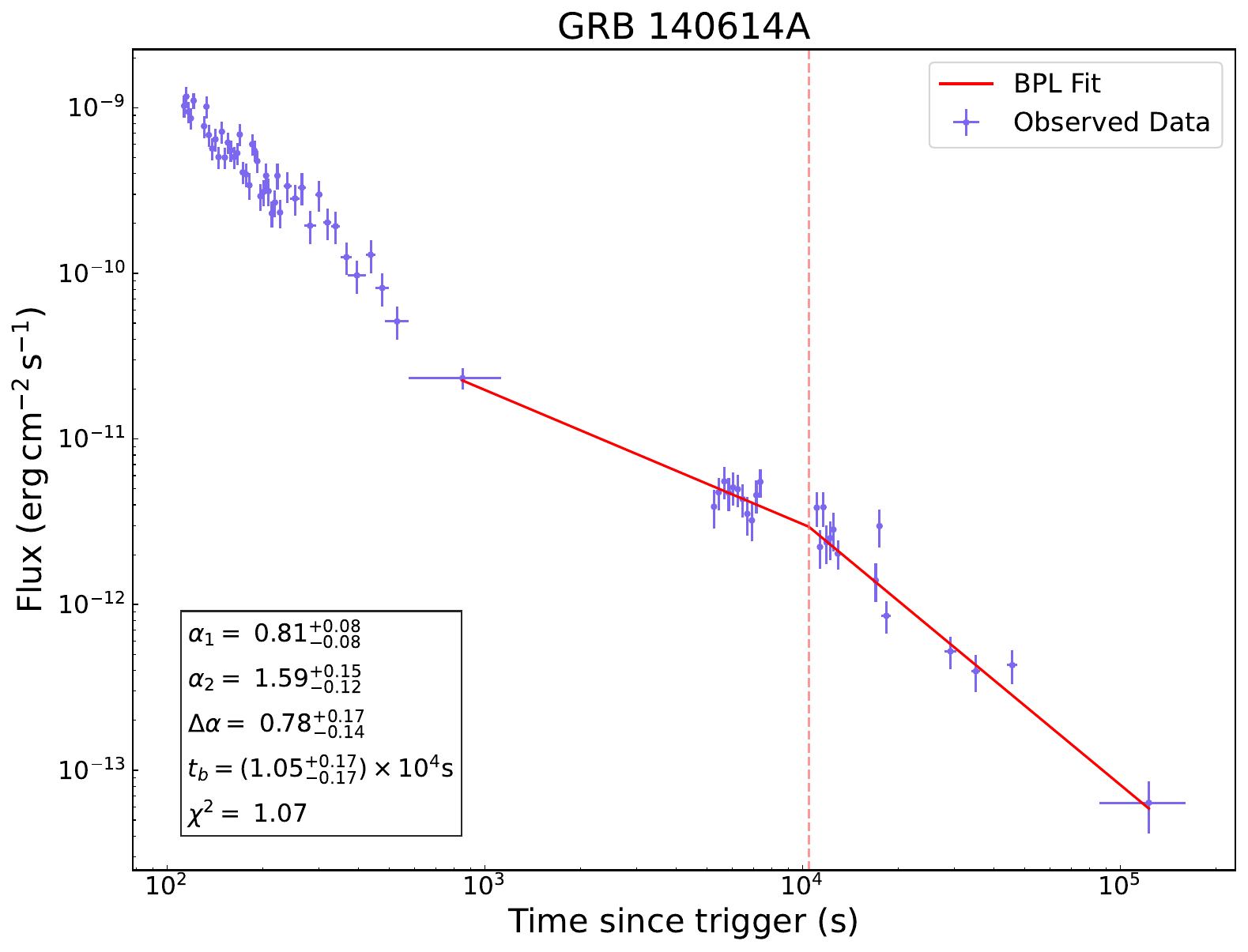}}%
\resizebox{45mm}{!}{\includegraphics[]{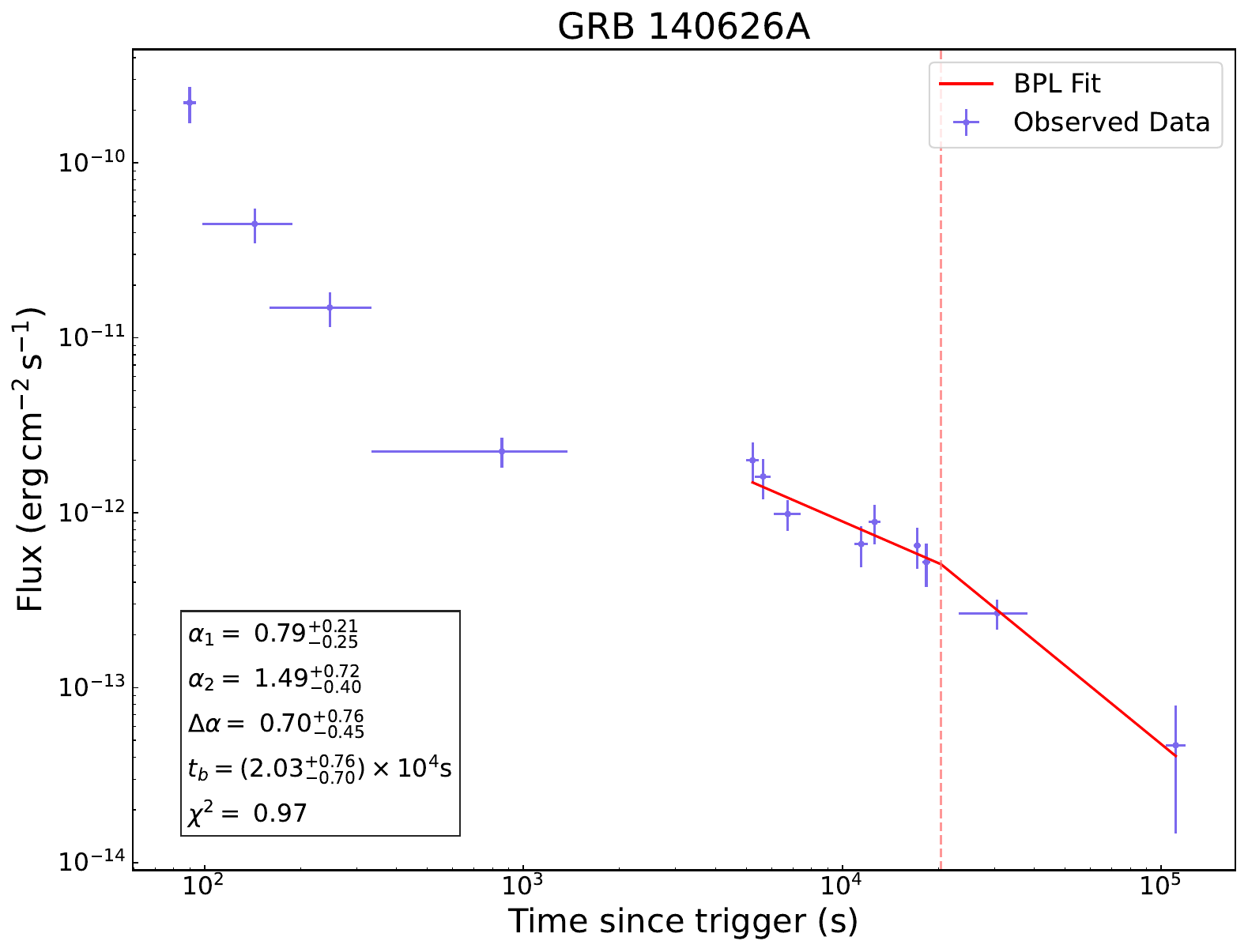}}%
\resizebox{45mm}{!}{\includegraphics[]{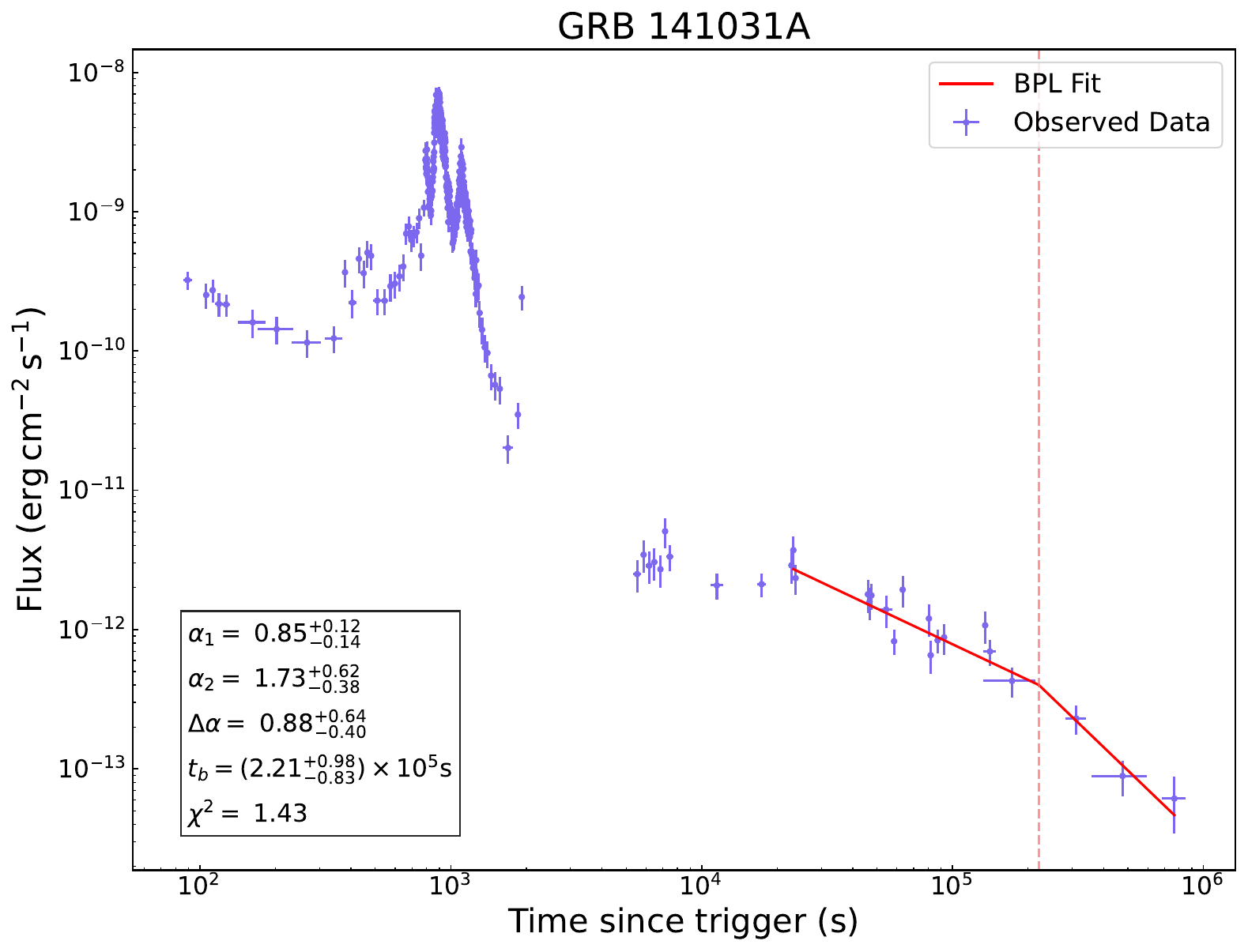}}\\
\resizebox{45mm}{!}{\includegraphics[]{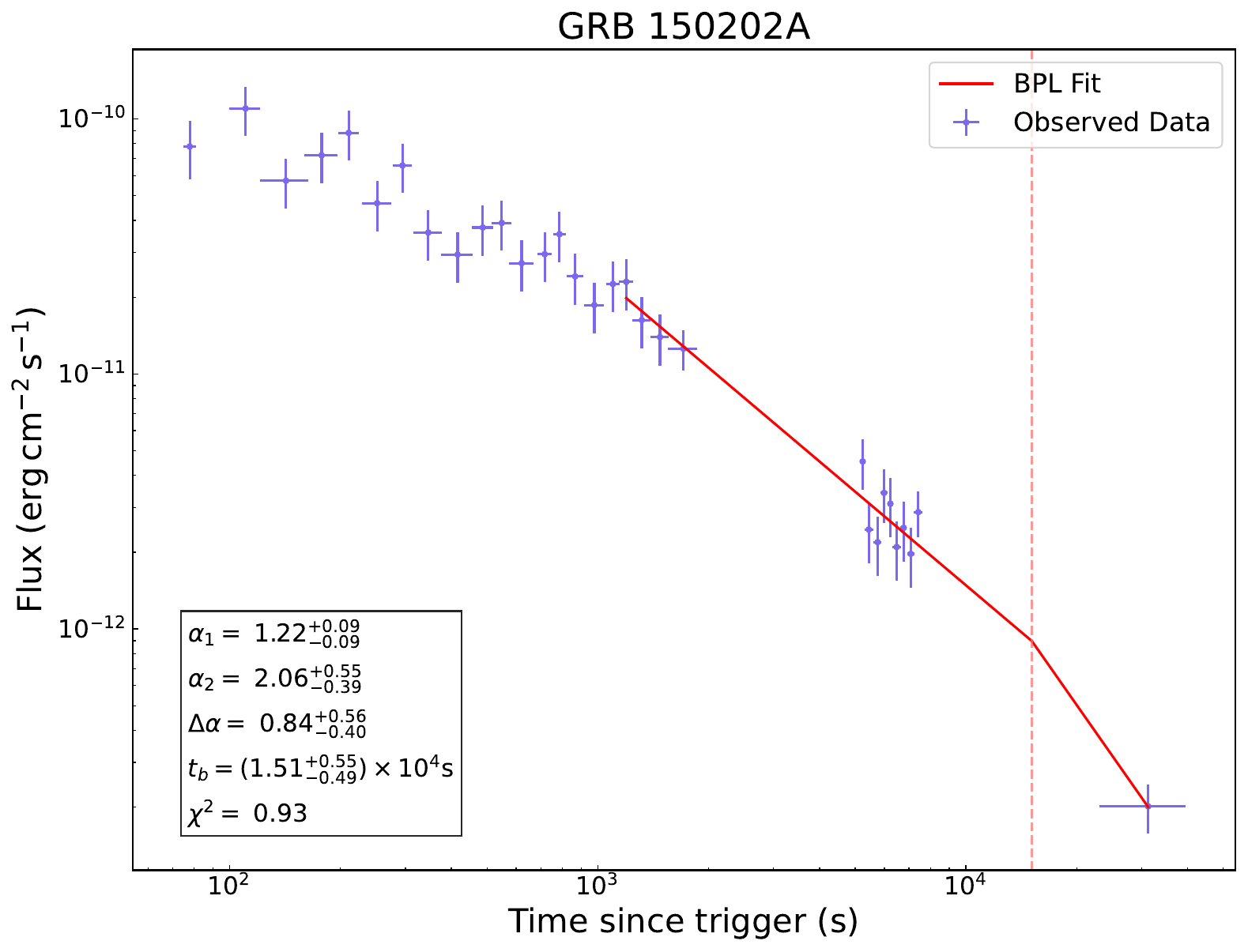}}%
\resizebox{45mm}{!}{\includegraphics[]{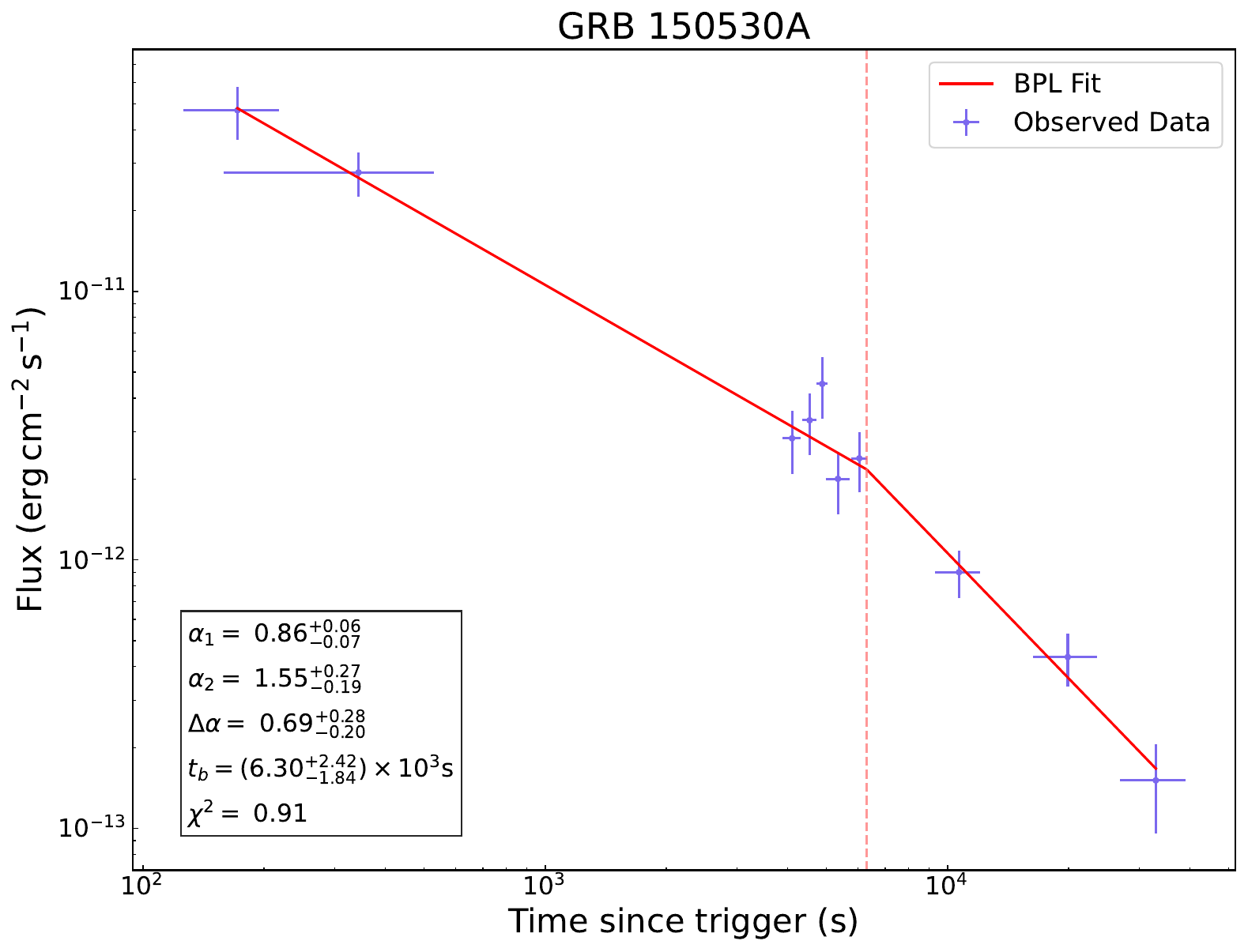}}%
\resizebox{45mm}{!}{\includegraphics[]{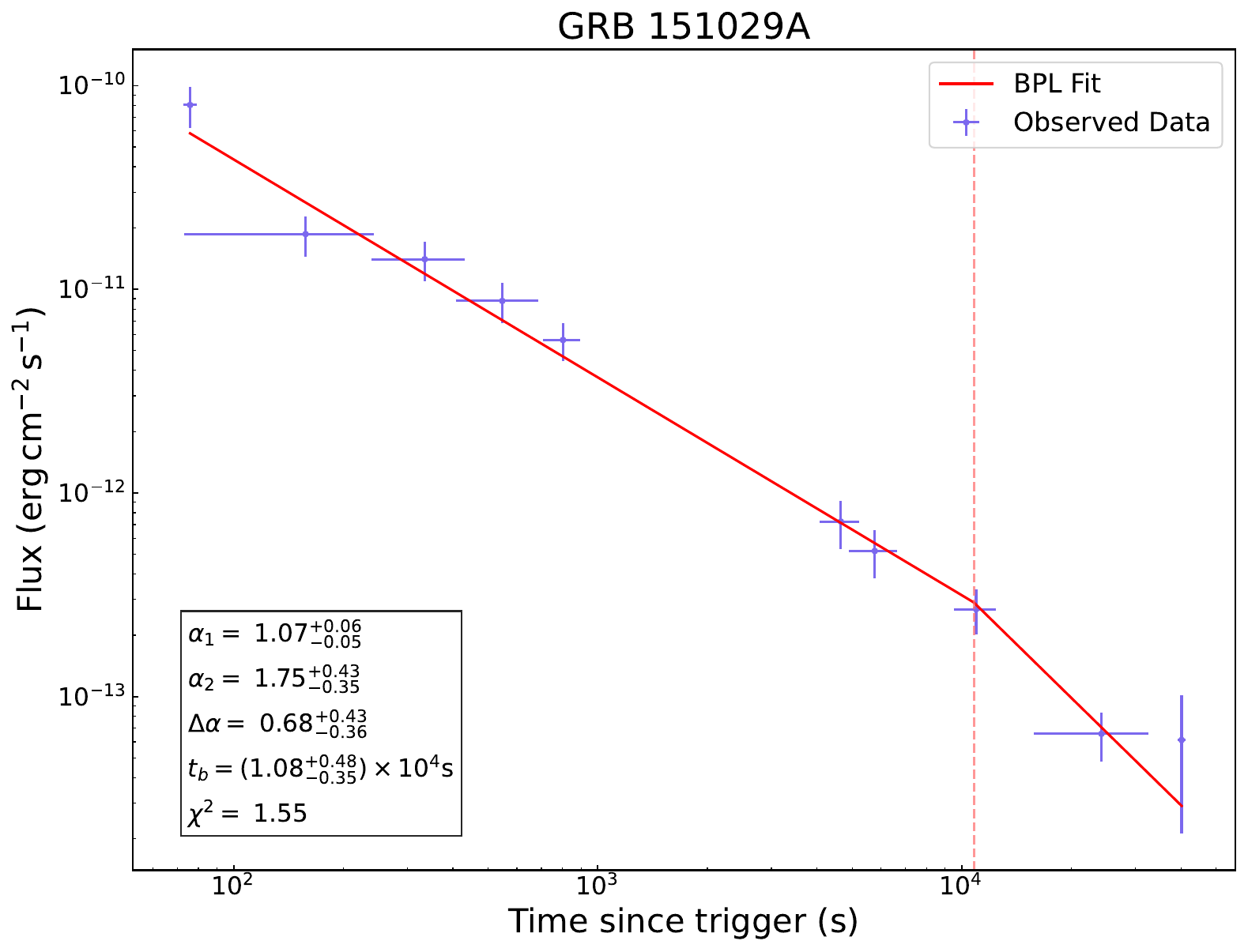}}%
\resizebox{45mm}{!}{\includegraphics[]{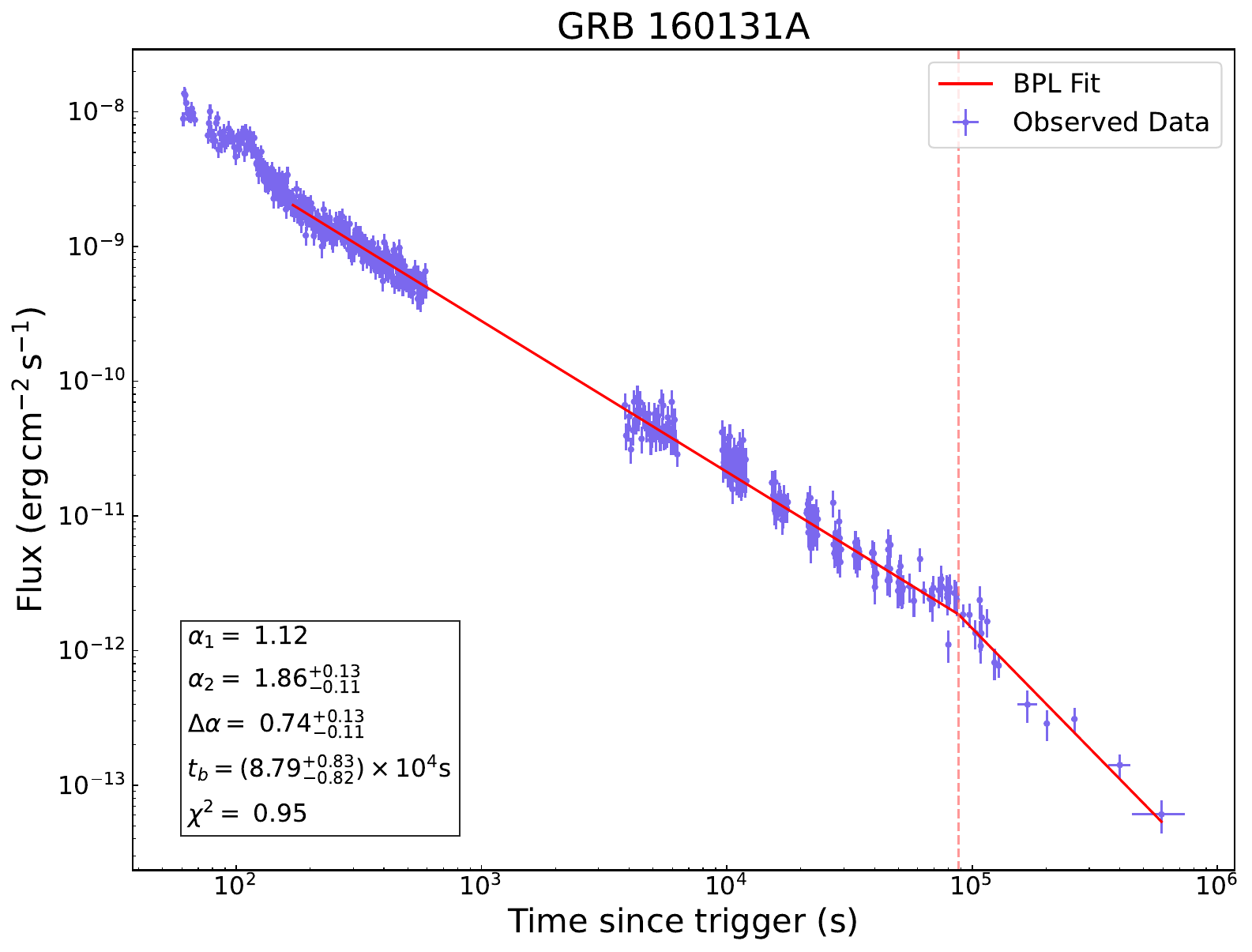}}\\
\resizebox{45mm}{!}{\includegraphics[]{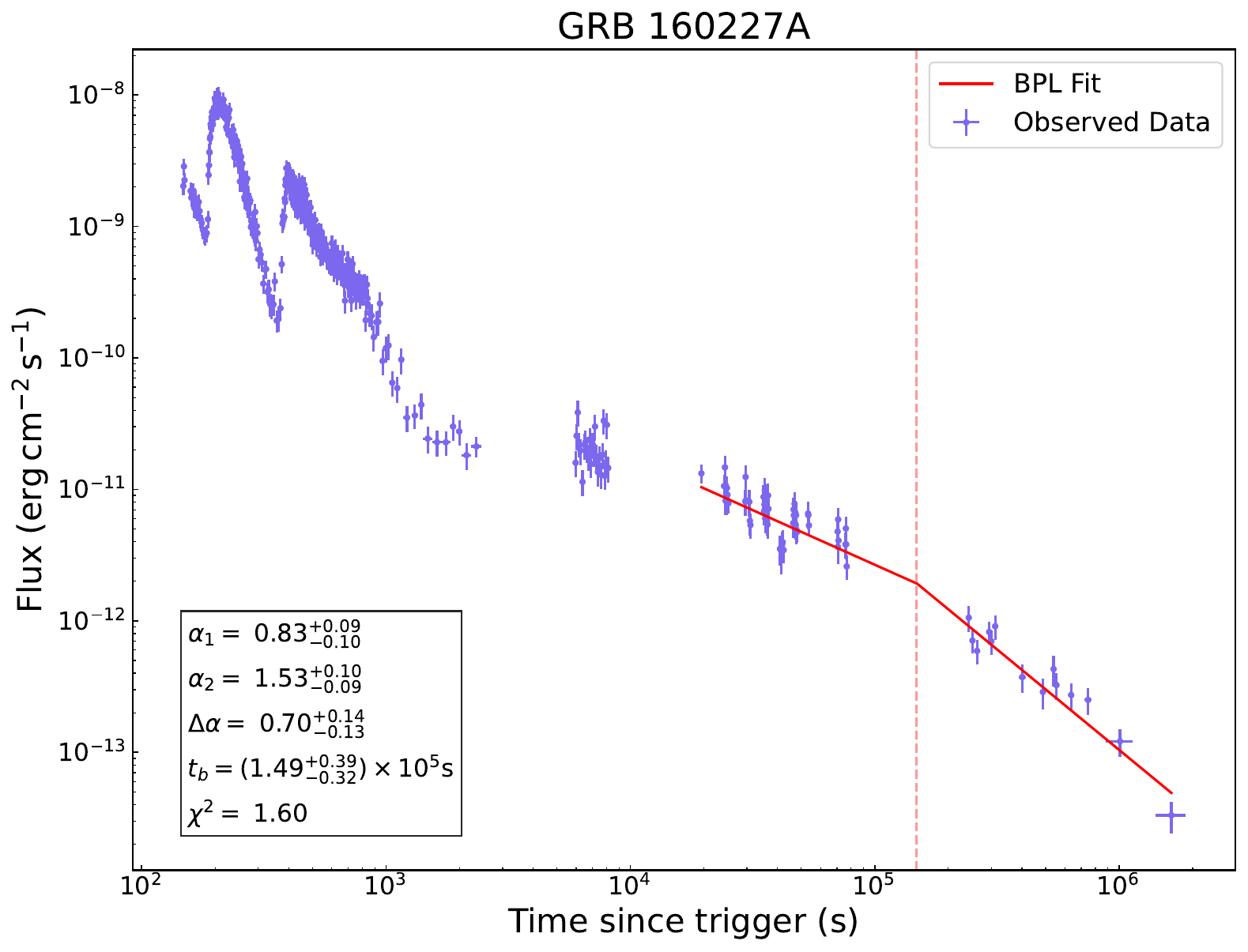}}%
\resizebox{45mm}{!}{\includegraphics[]{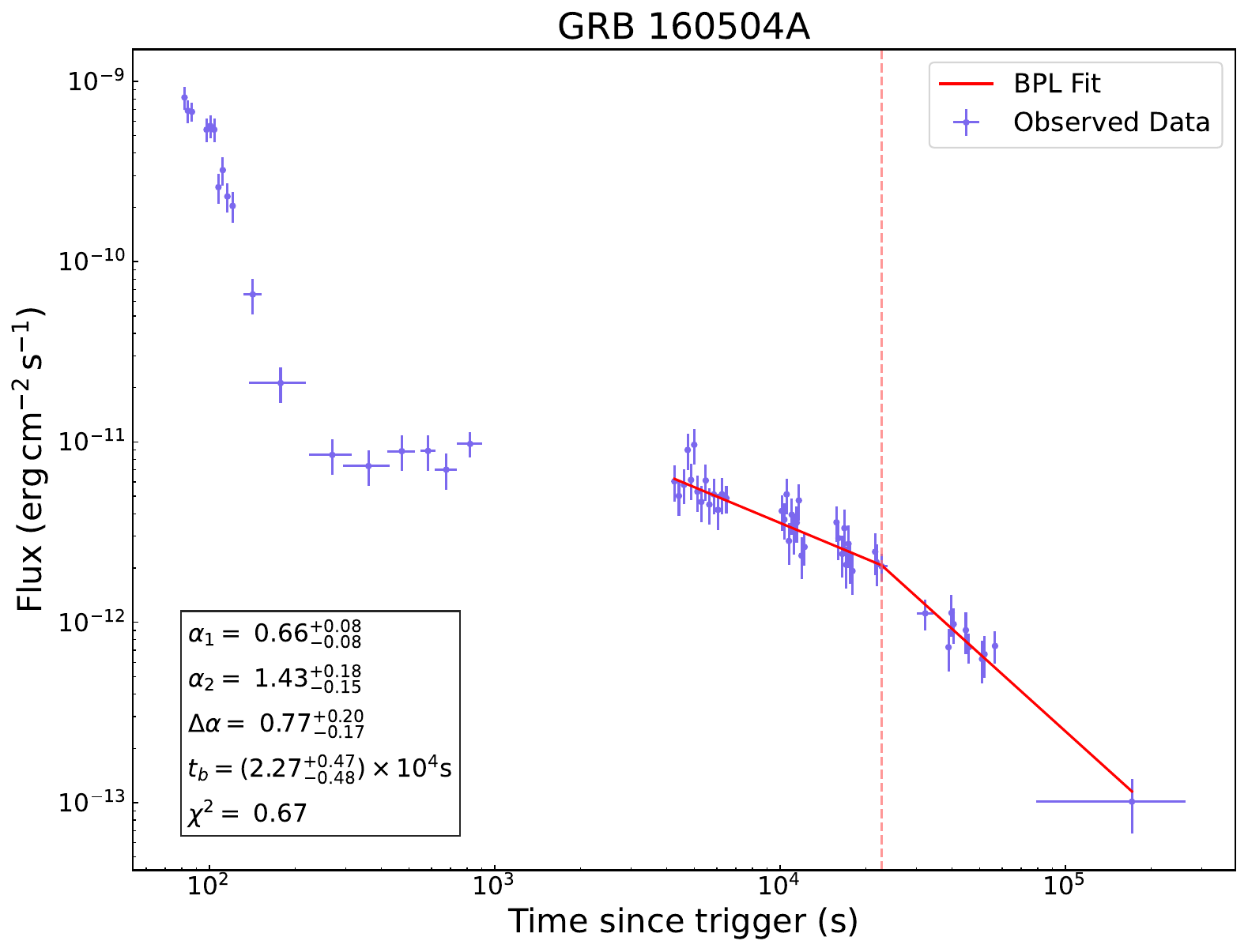}}%
\resizebox{45mm}{!}{\includegraphics[]{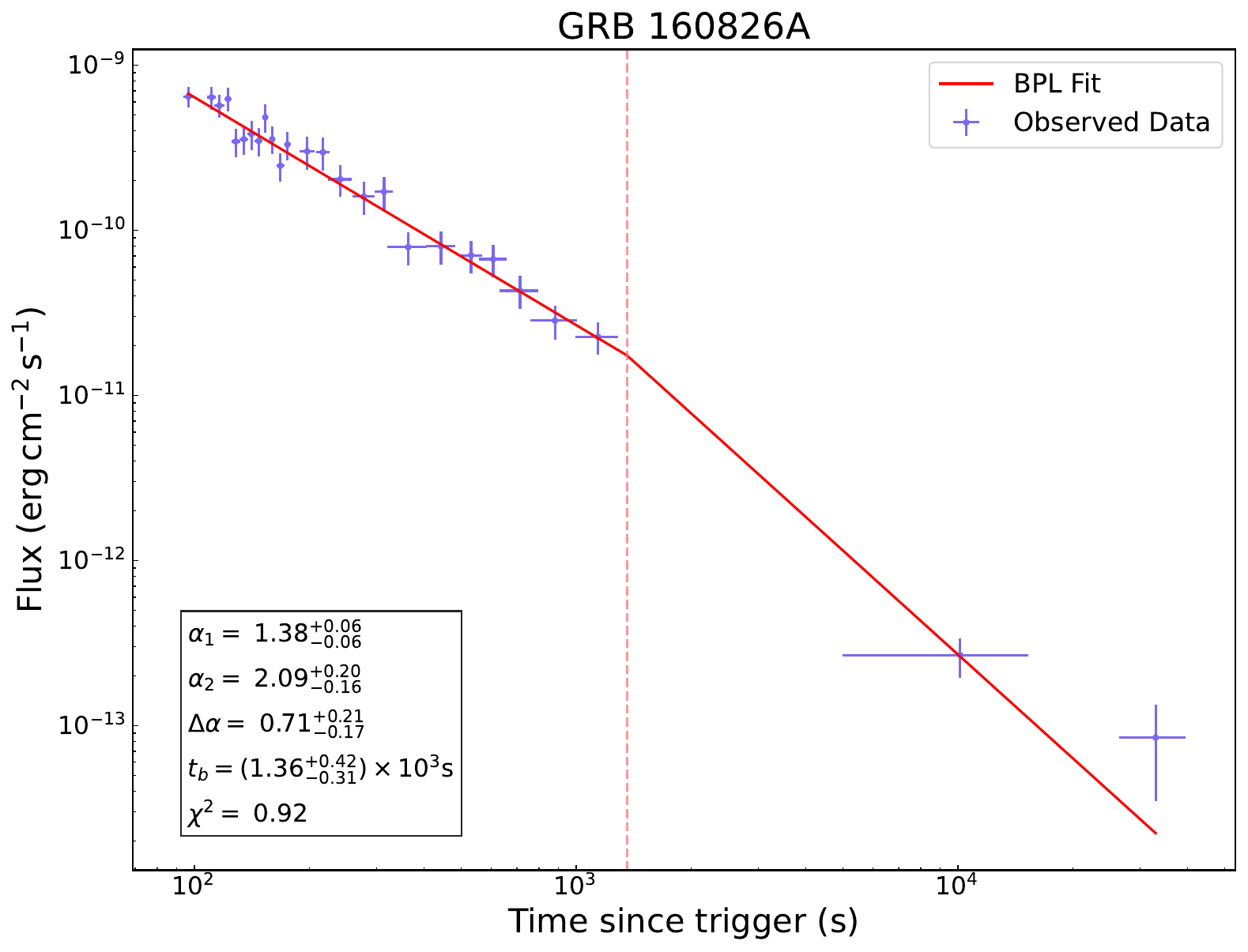}}%
\resizebox{45mm}{!}{\includegraphics[]{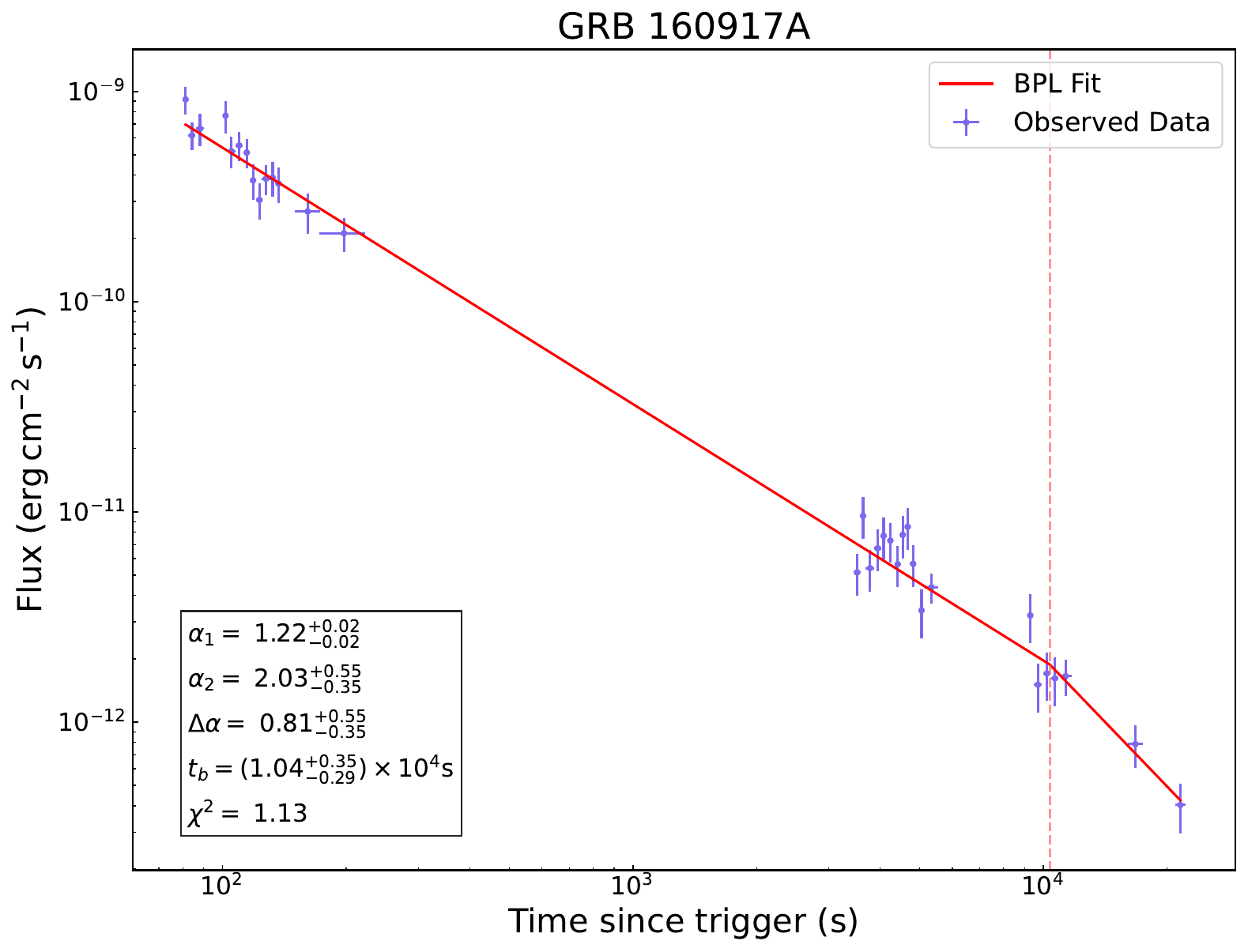}}\\
\caption{(Continued)}
\end{figure}

\clearpage
\addtocounter{figure}{-1}
\begin{figure}
\centering
\resizebox{45mm}{!}{\includegraphics[]{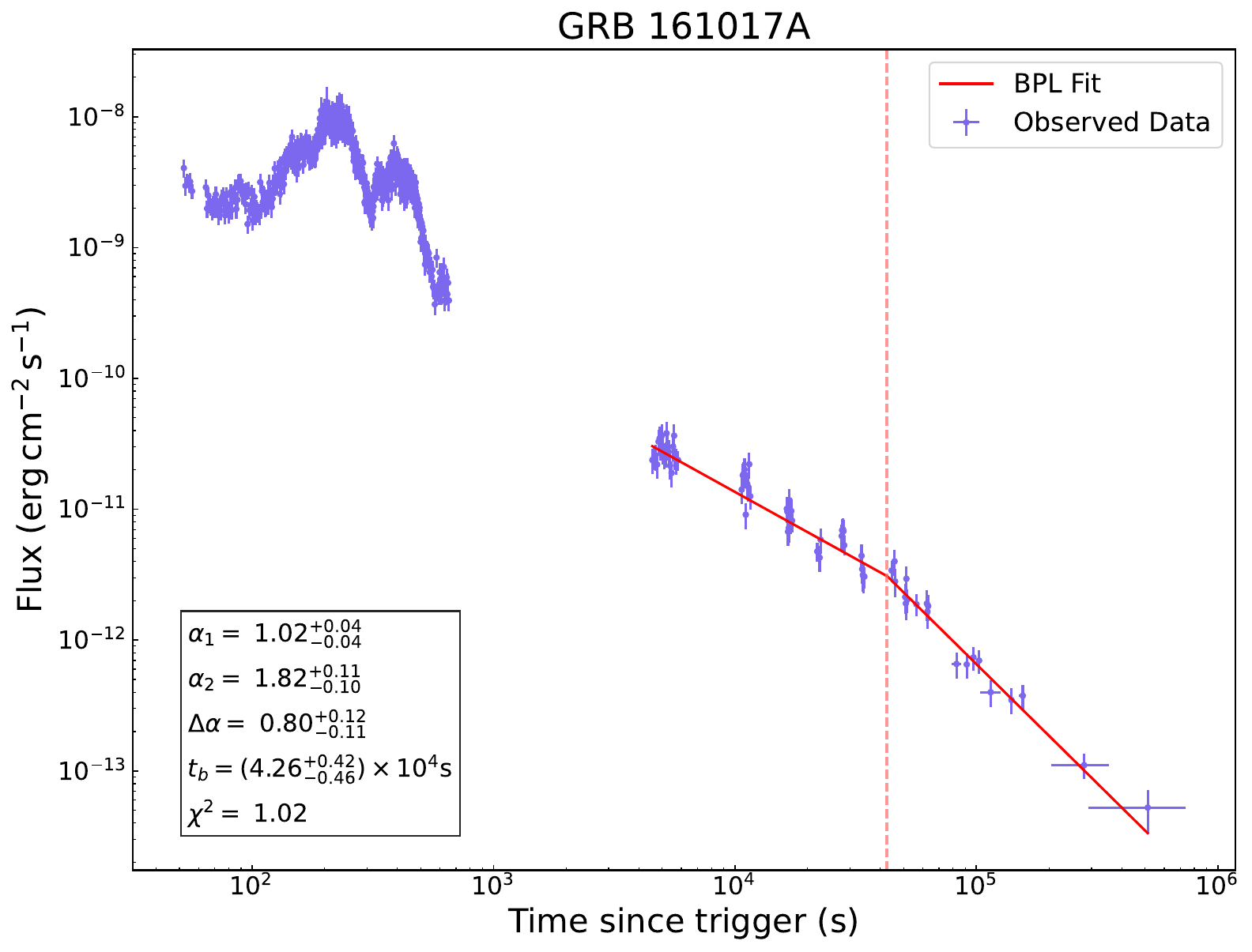}}%
\resizebox{45mm}{!}{\includegraphics[]{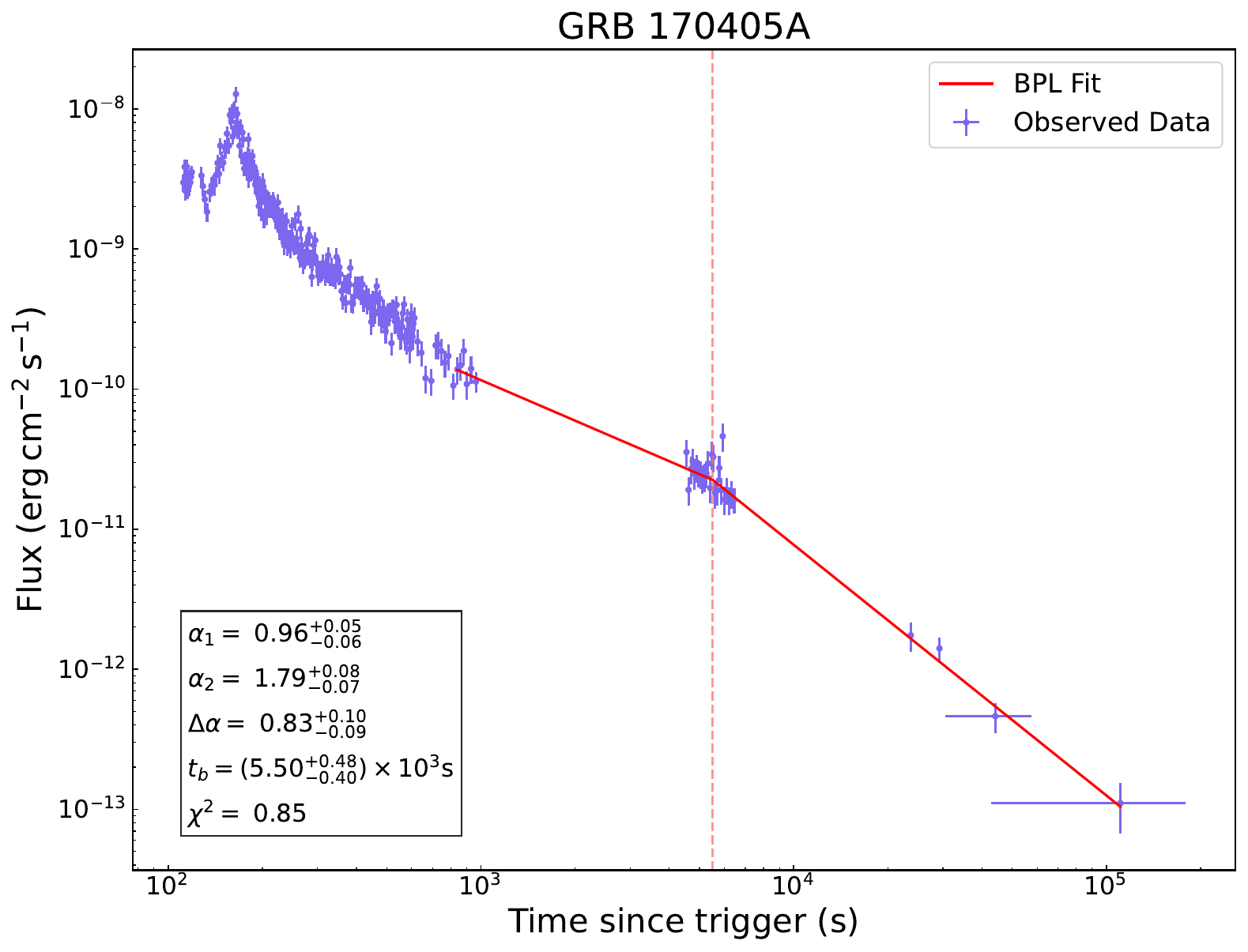}}%
\resizebox{45mm}{!}{\includegraphics[]{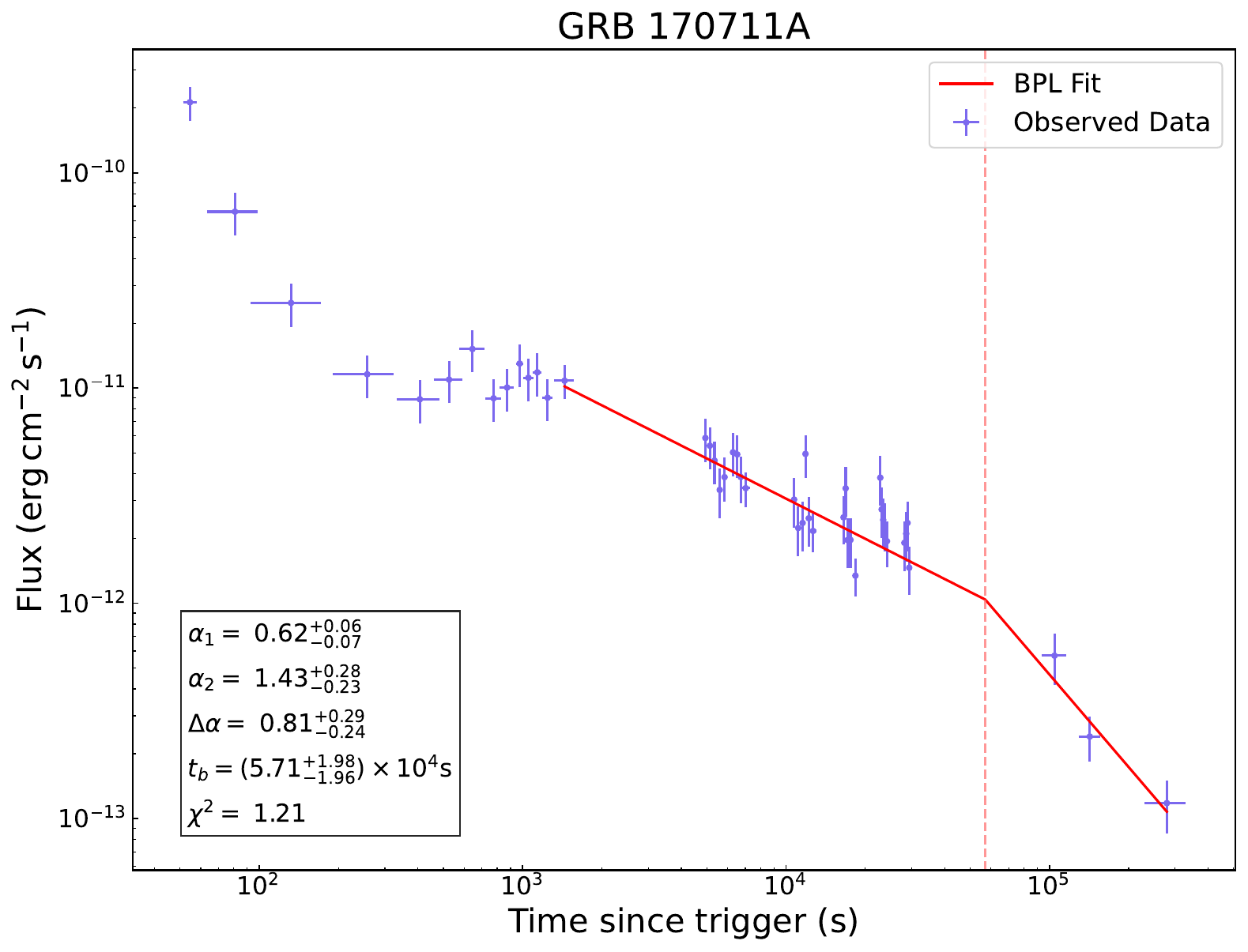}}%
\resizebox{45mm}{!}{\includegraphics[]{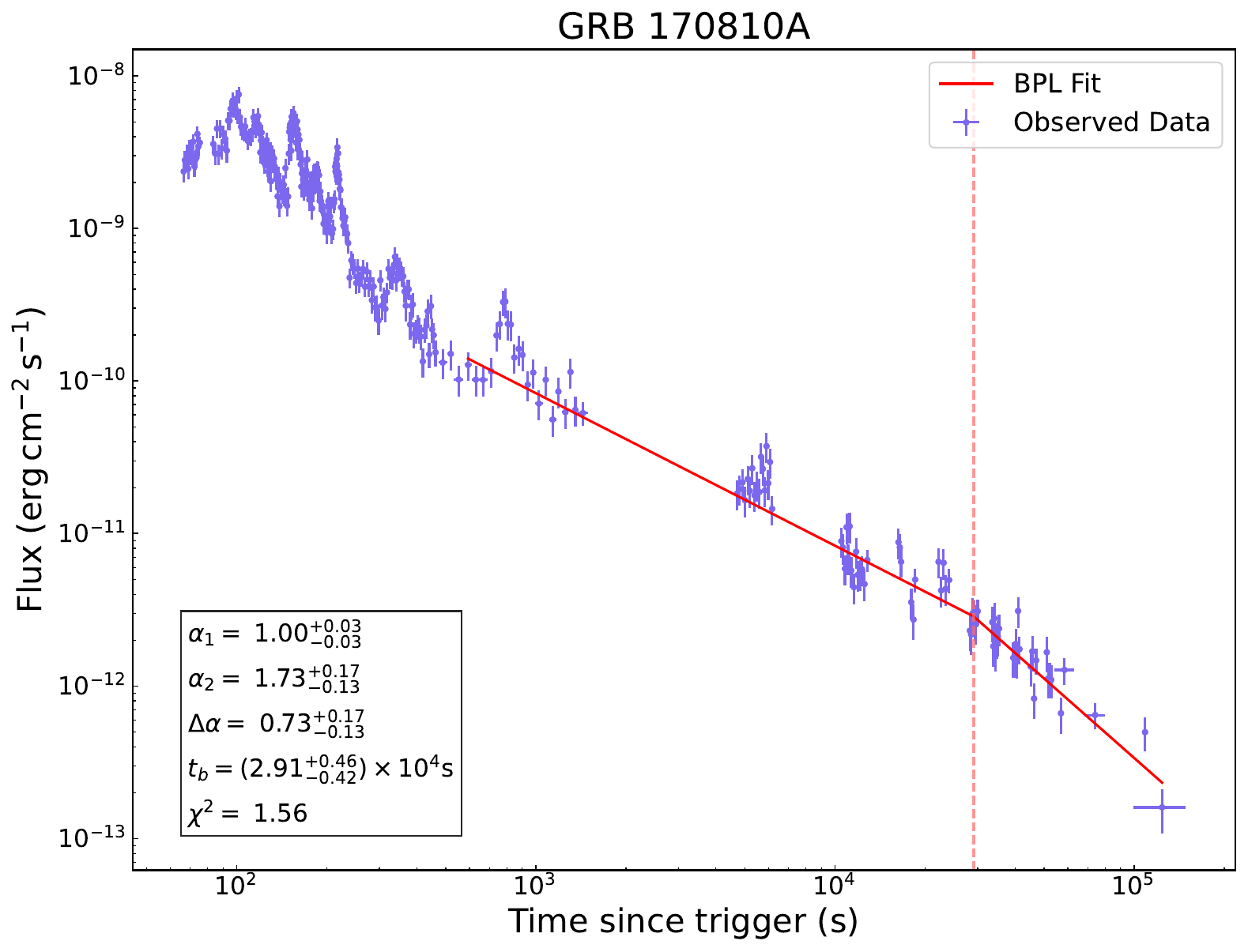}}\\
\resizebox{45mm}{!}{\includegraphics[]{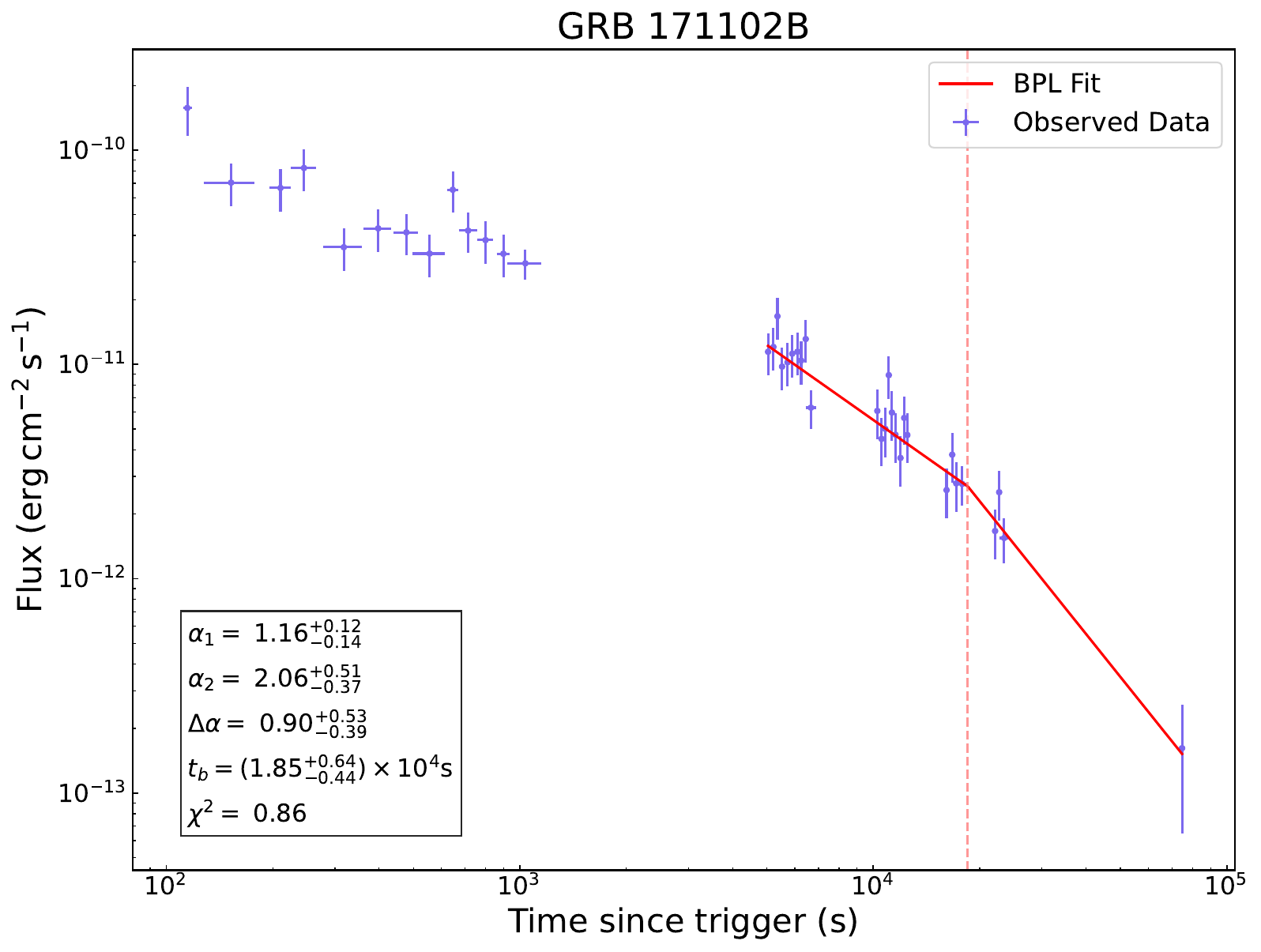}}%
\resizebox{45mm}{!}{\includegraphics[]{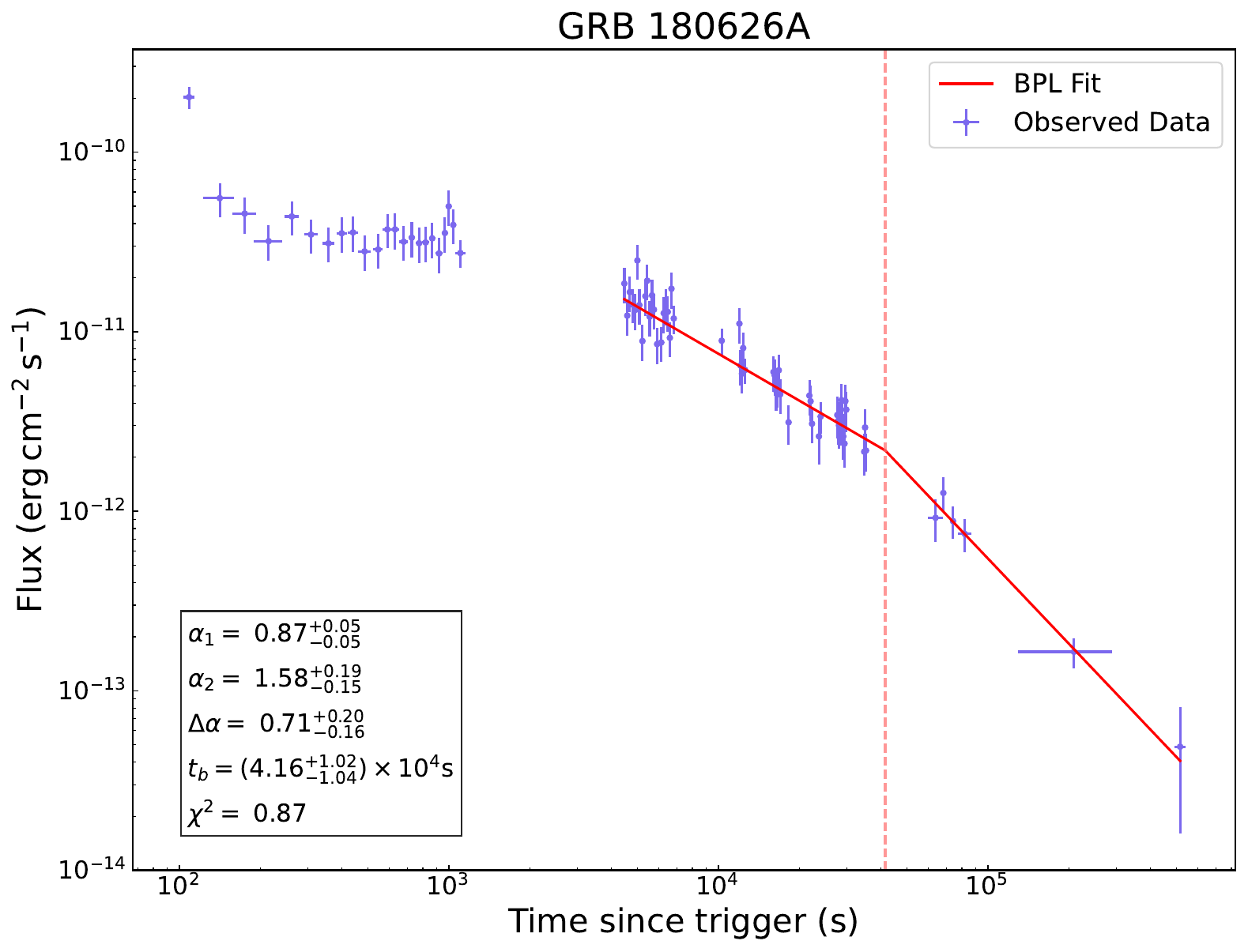}}%
\resizebox{45mm}{!}{\includegraphics[]{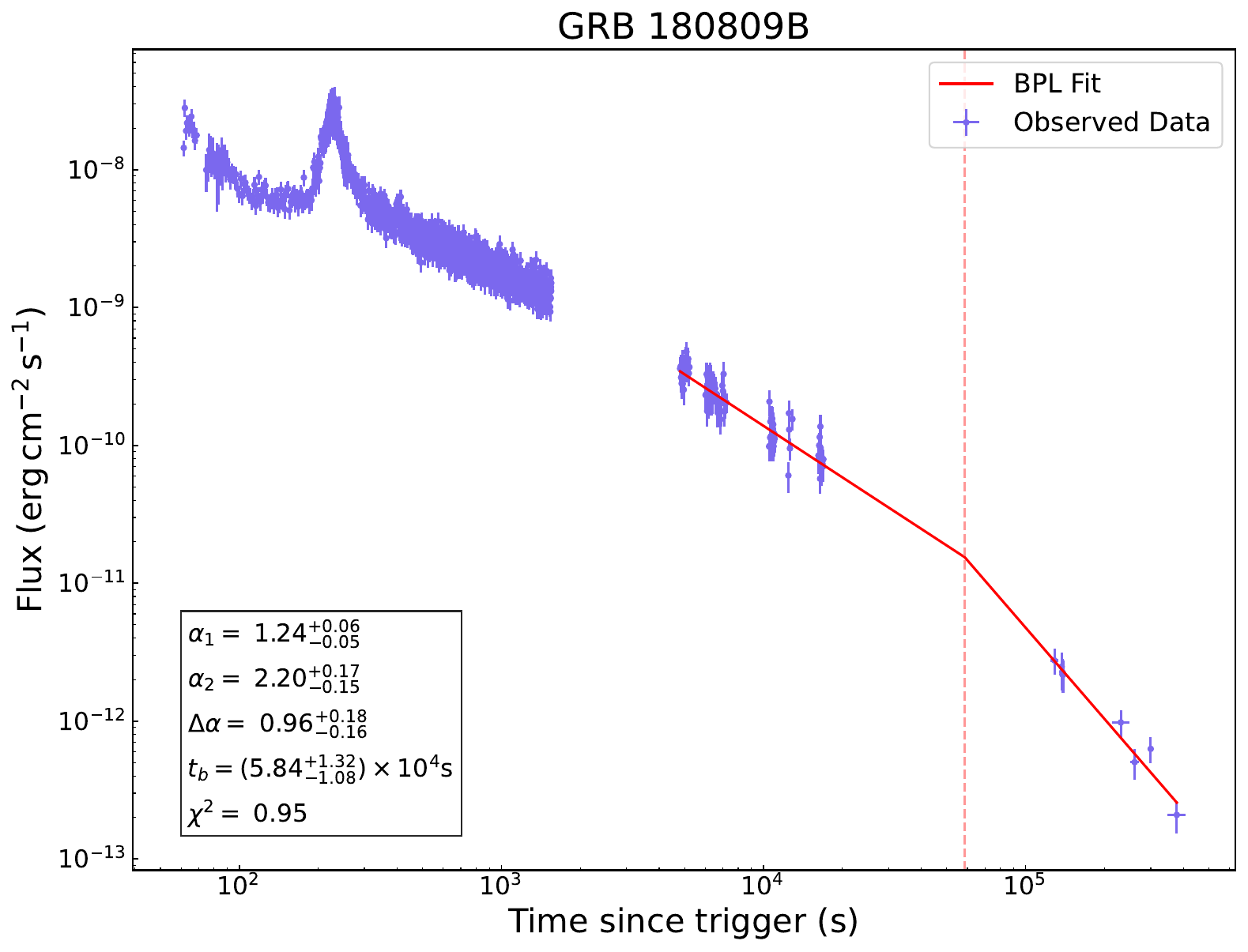}}%
\resizebox{45mm}{!}{\includegraphics[]{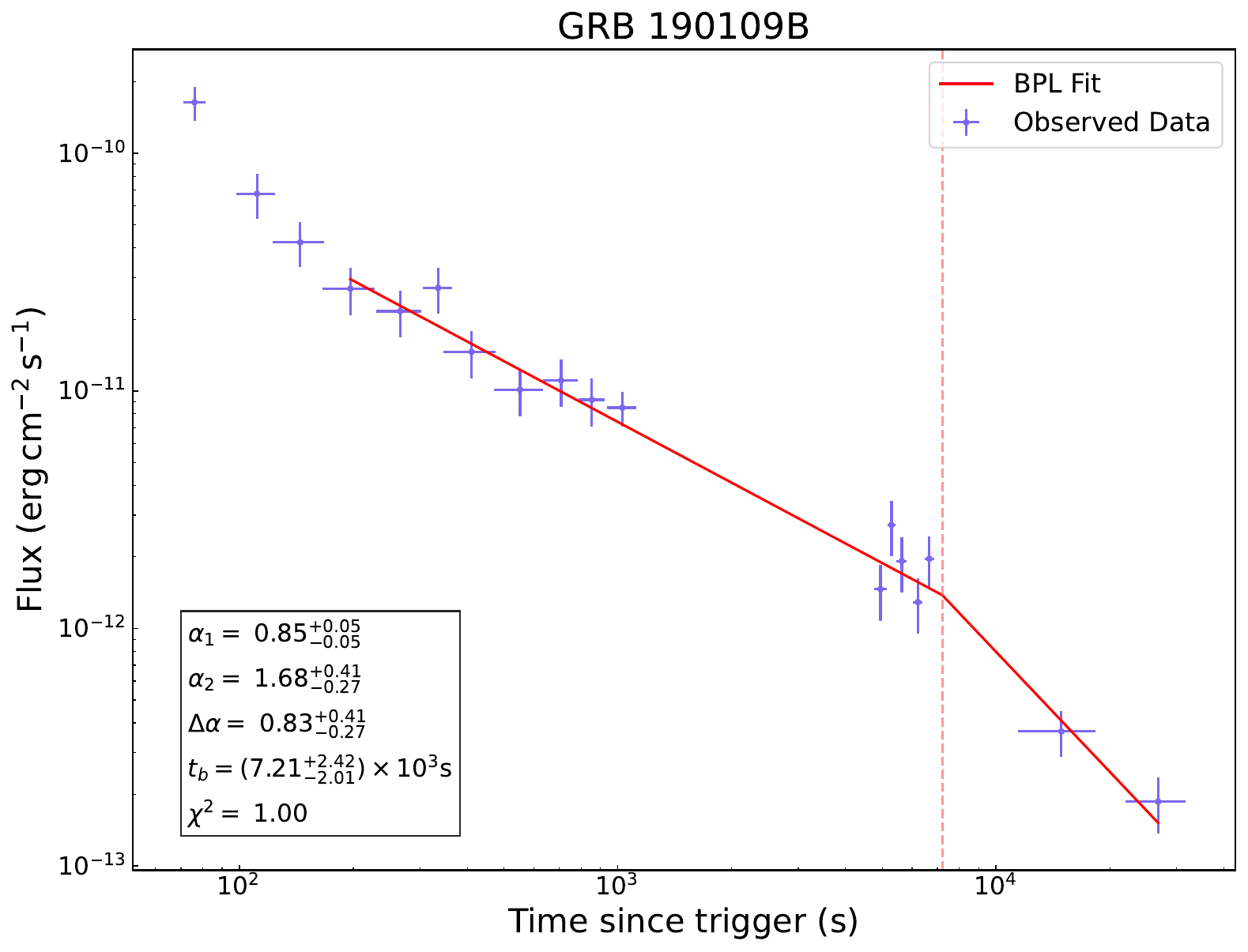}}\\
\resizebox{45mm}{!}{\includegraphics[]{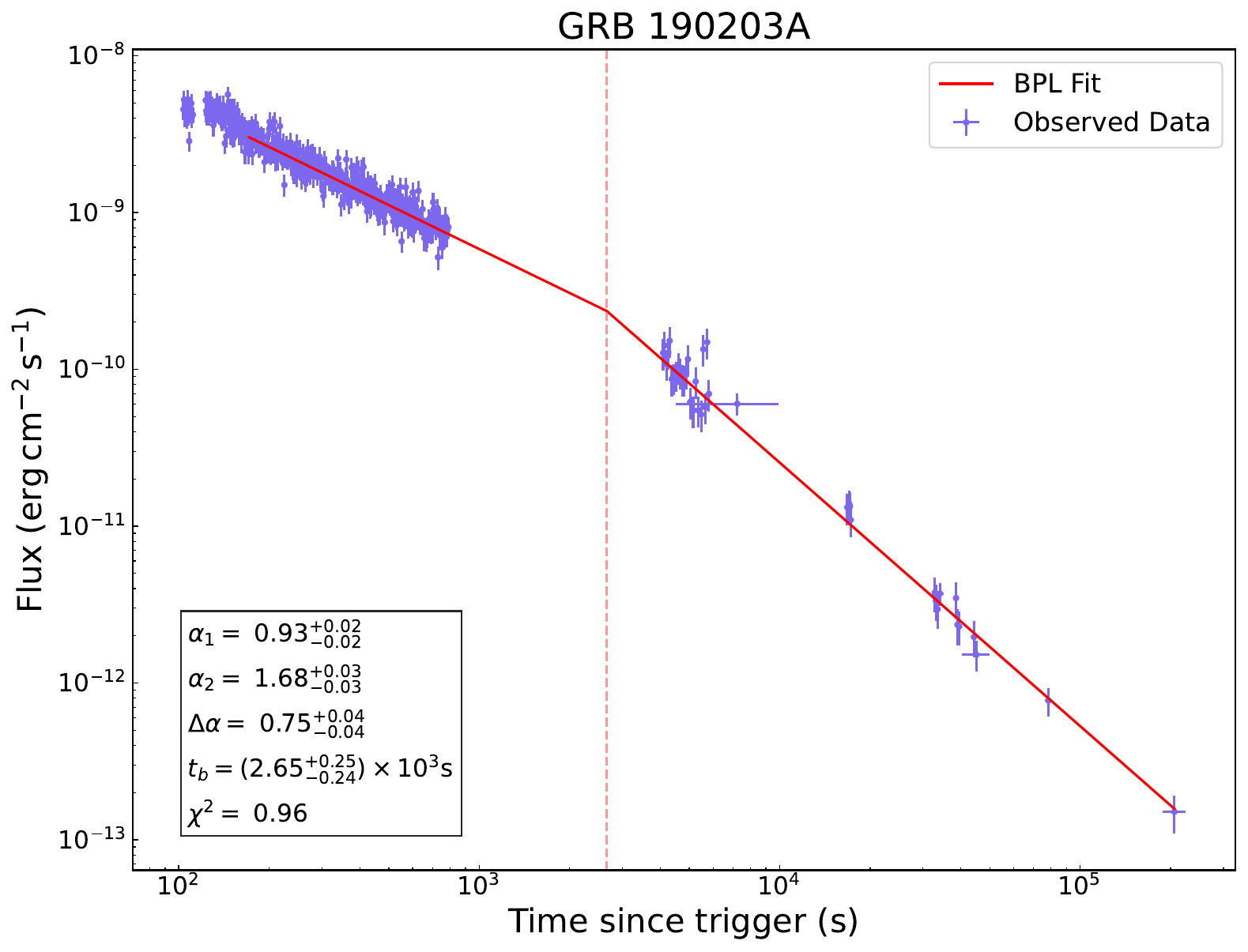}}%
\resizebox{45mm}{!}{\includegraphics[]{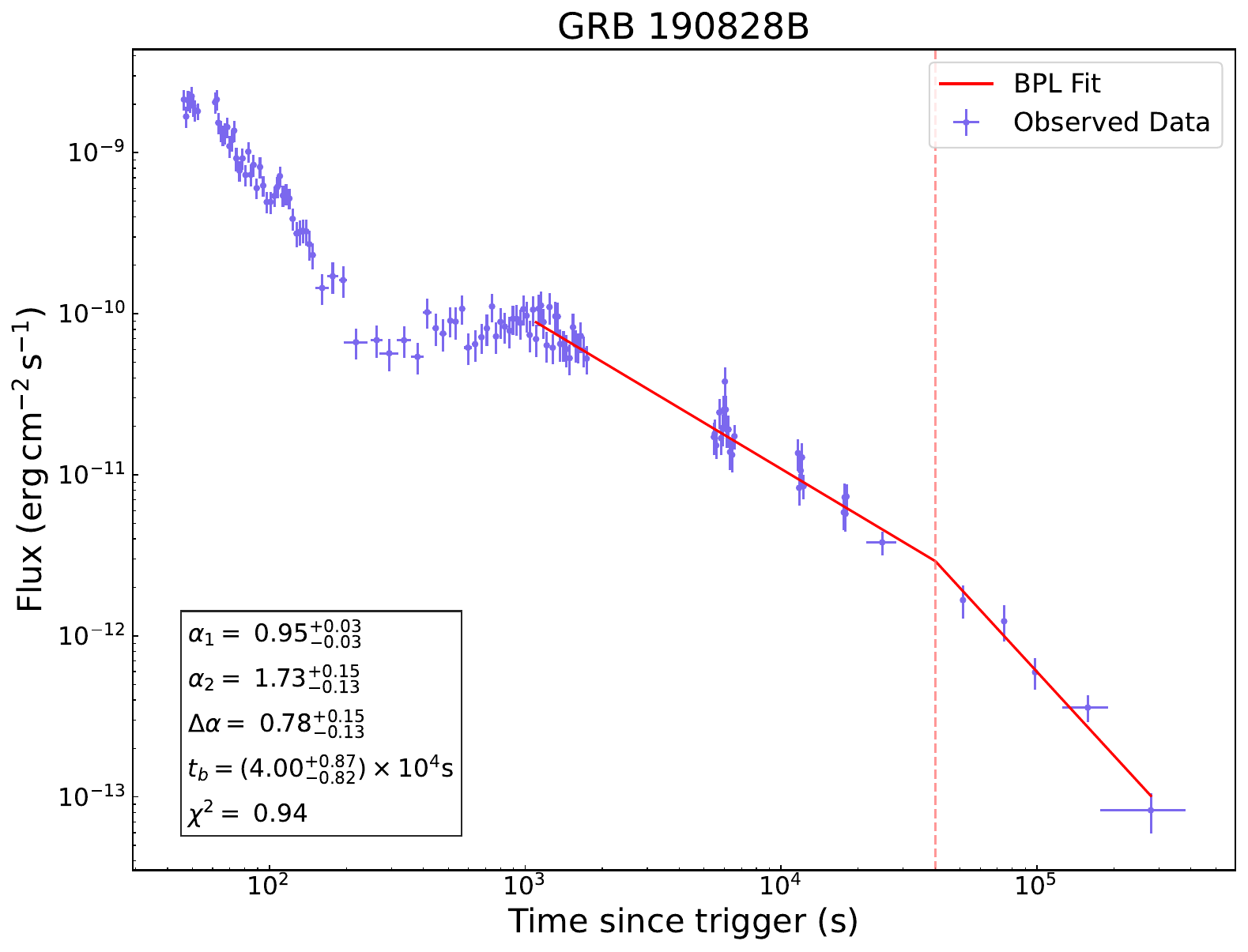}}%
\resizebox{45mm}{!}{\includegraphics[]{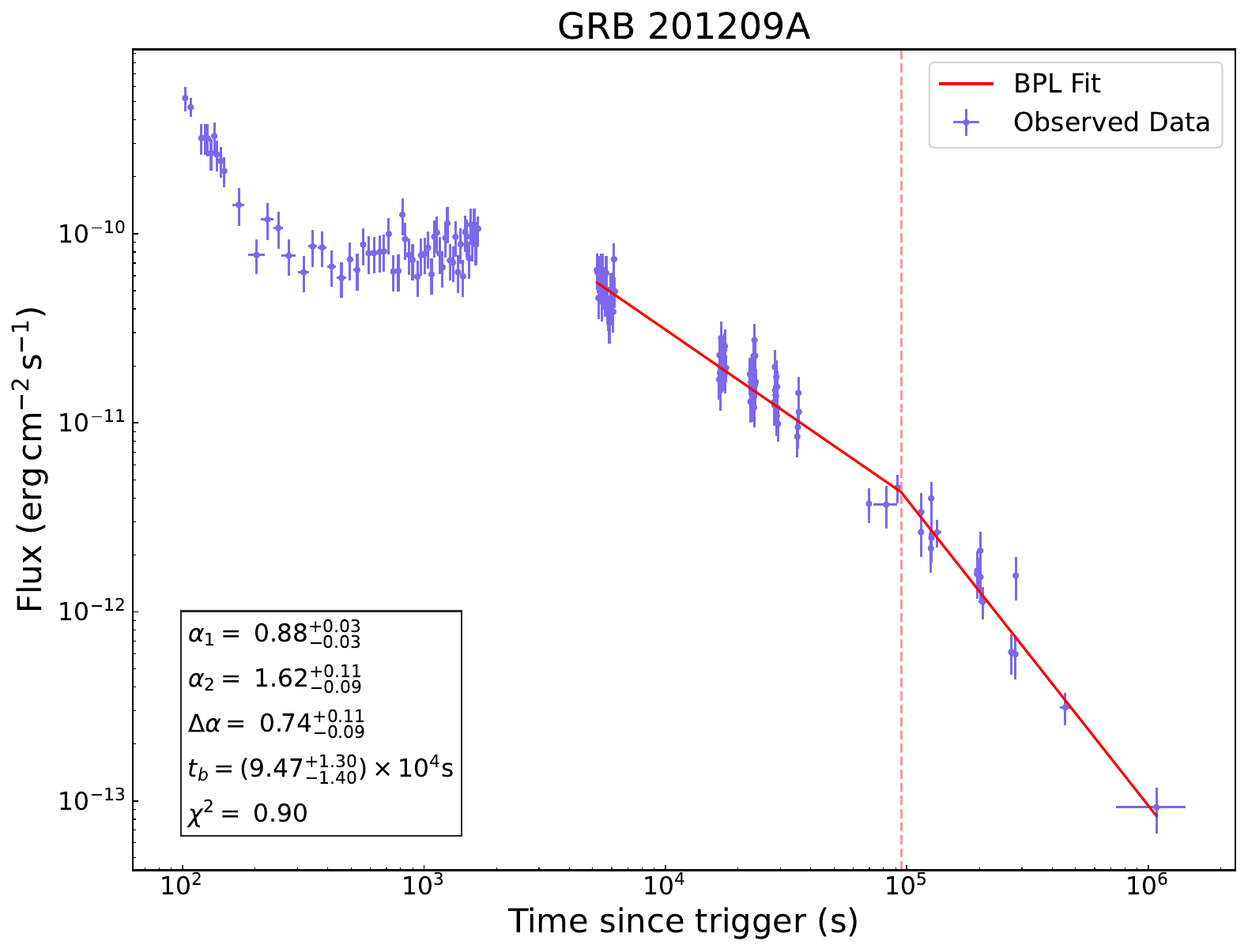}}%
\resizebox{45mm}{!}{\includegraphics[]{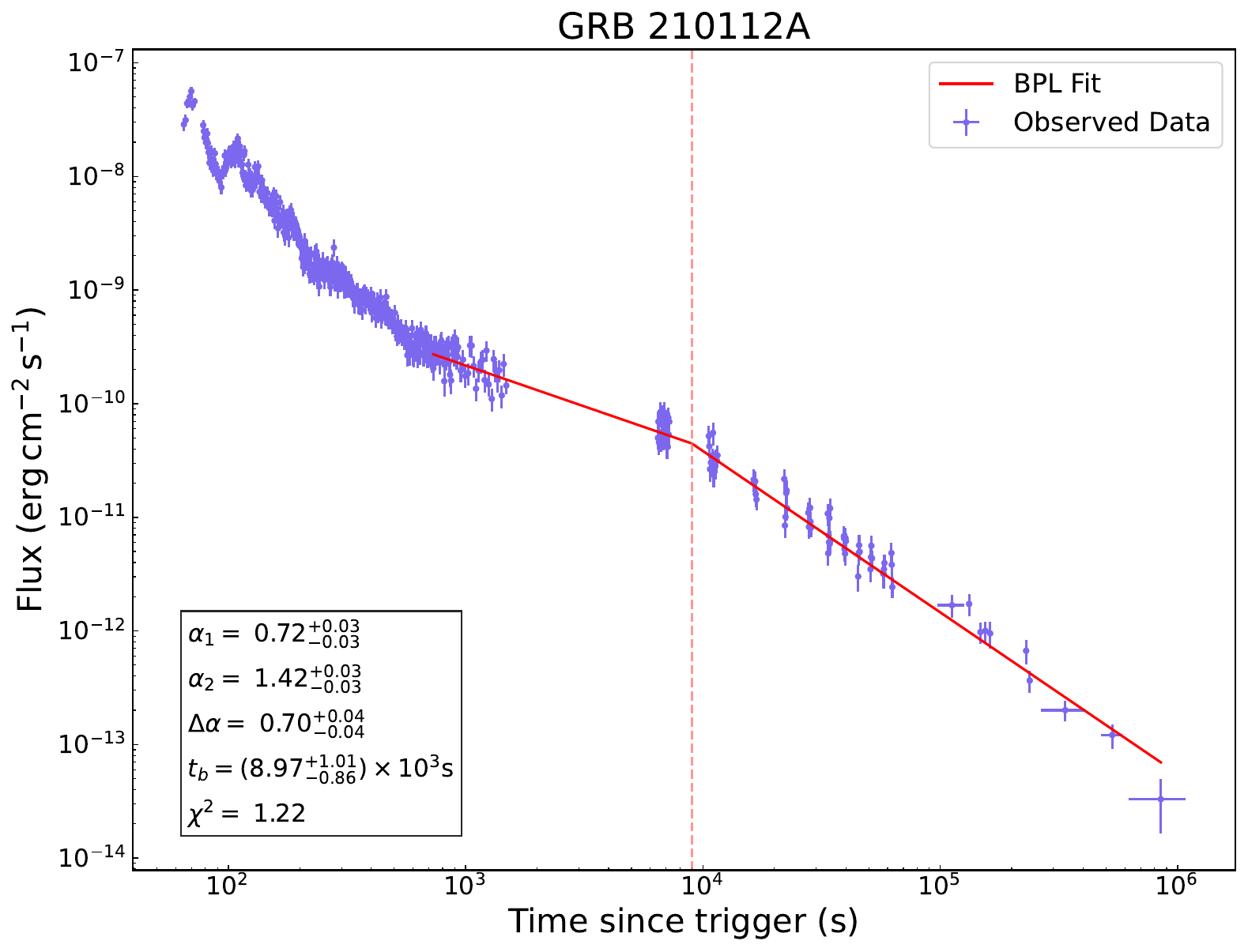}}\\
\resizebox{45mm}{!}{\includegraphics[]{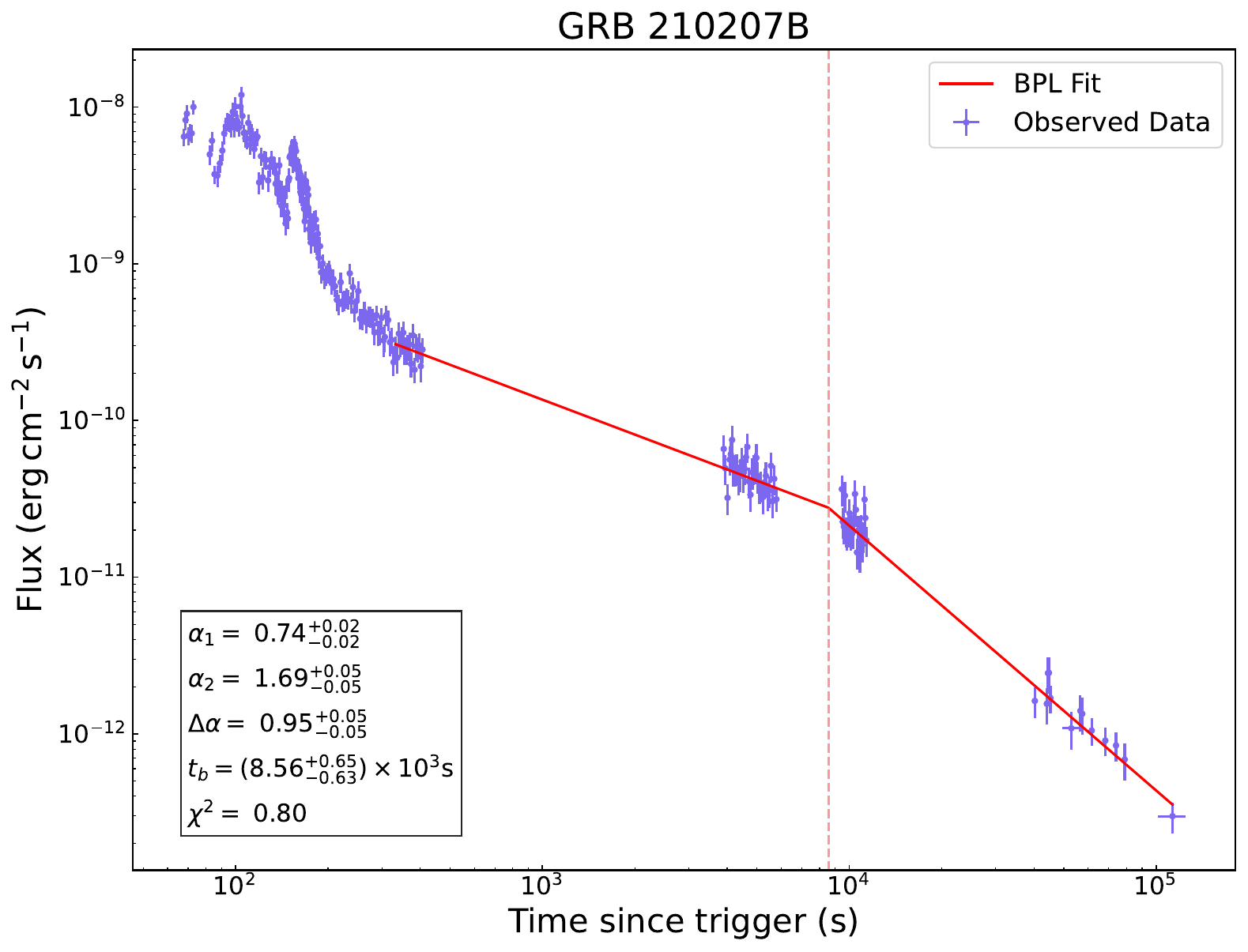}}%
\resizebox{45mm}{!}{\includegraphics[]{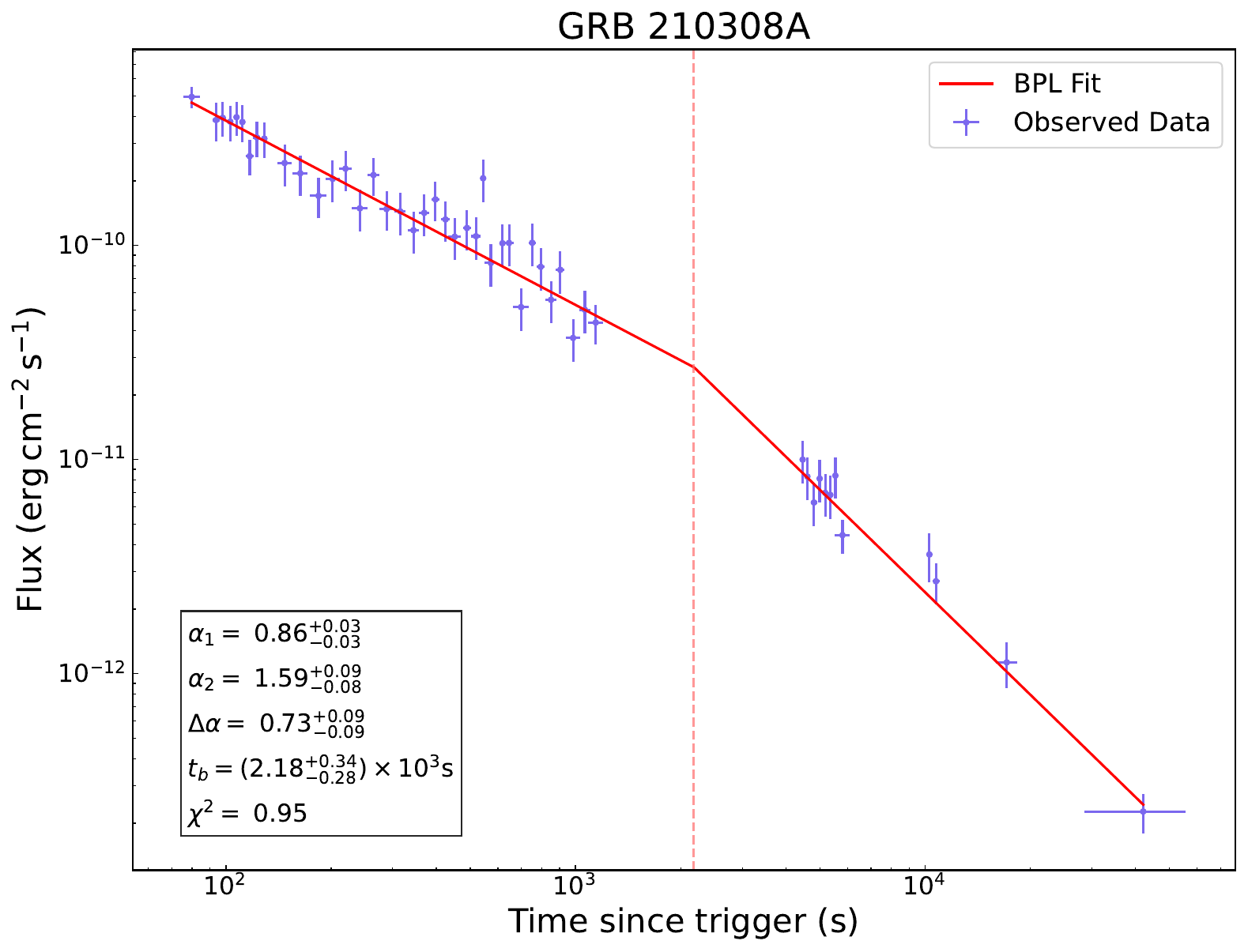}}%
\resizebox{45mm}{!}{\includegraphics[]{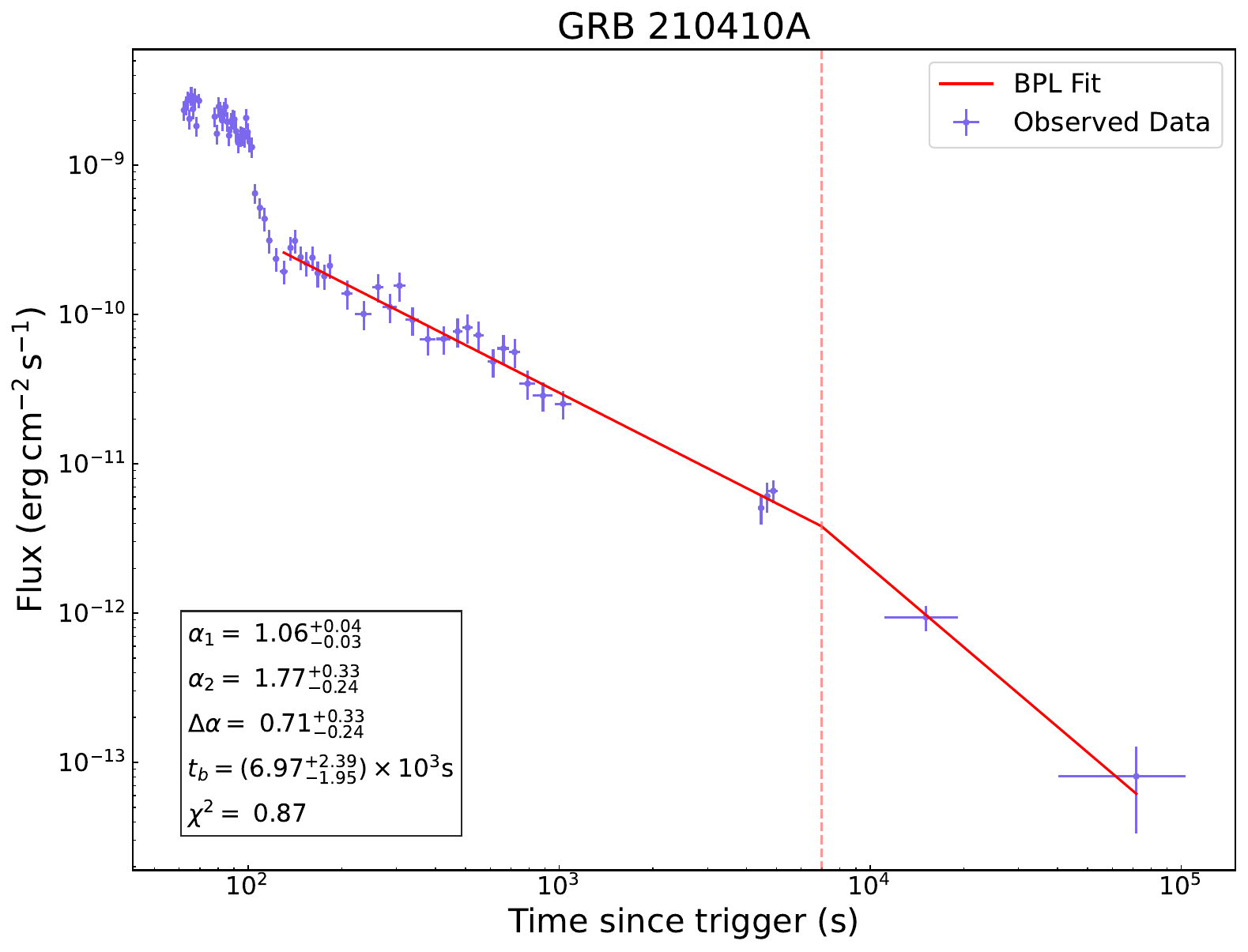}}%
\resizebox{45mm}{!}{\includegraphics[]{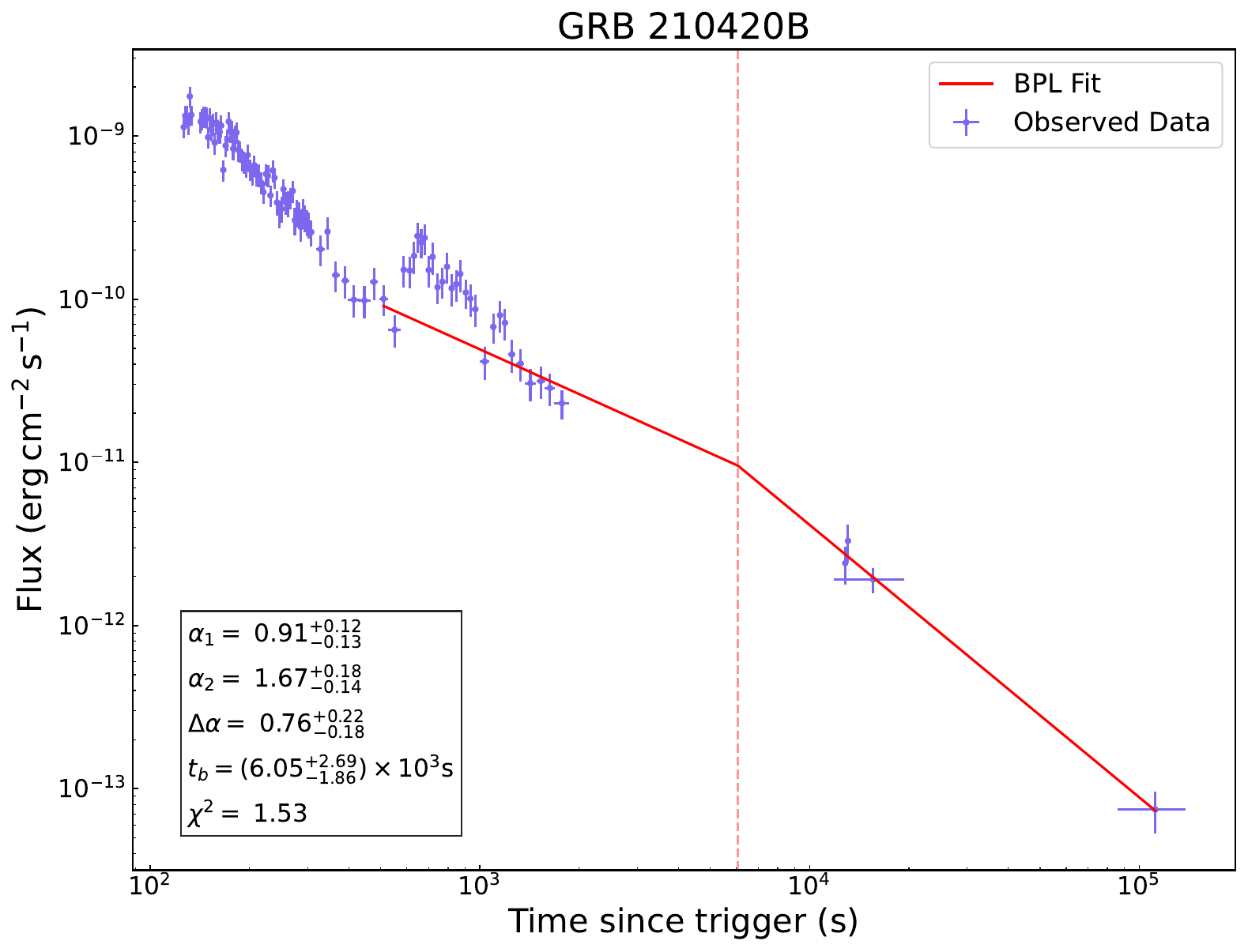}}\\
\resizebox{45mm}{!}{\includegraphics[]{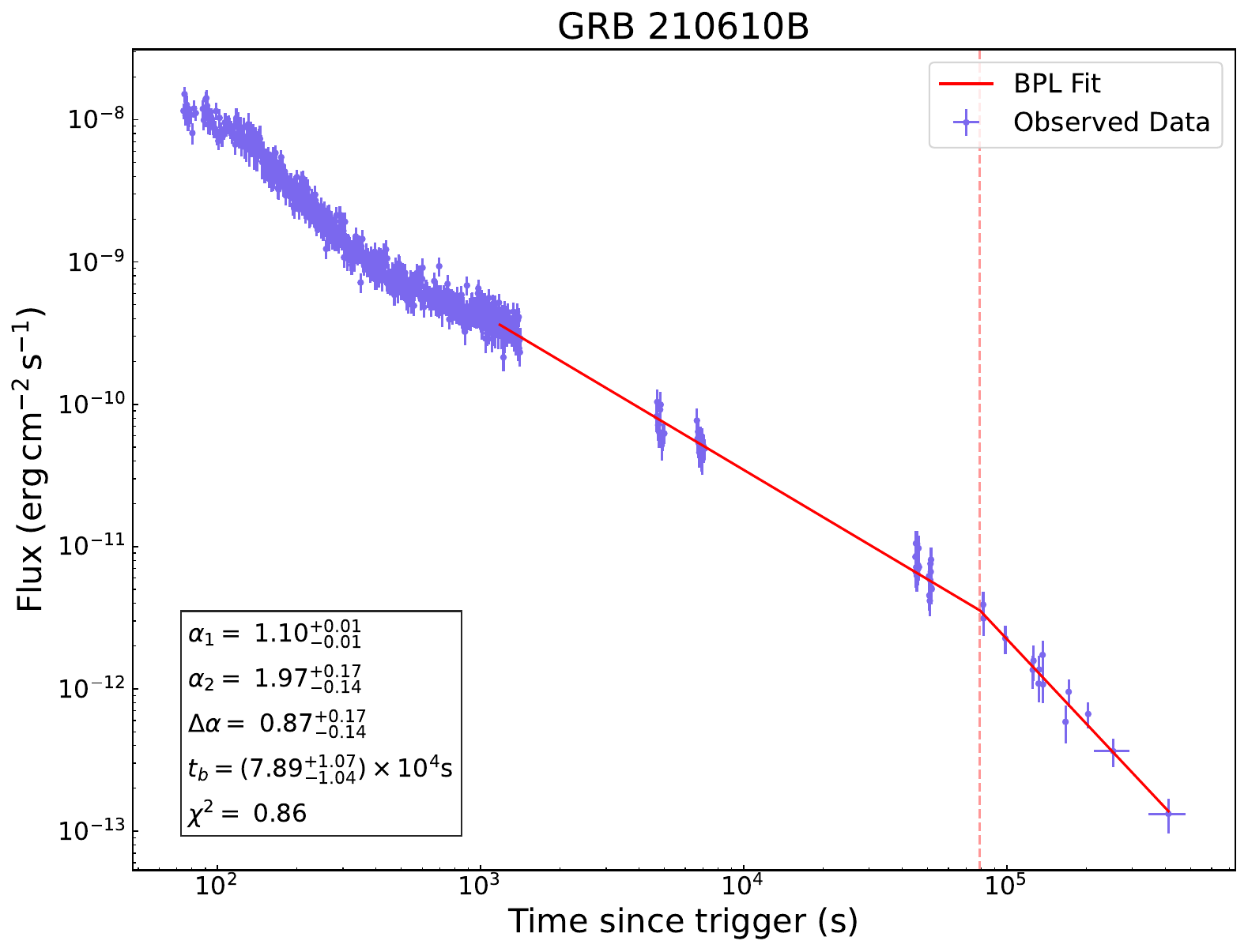}}%
\resizebox{45mm}{!}{\includegraphics[]{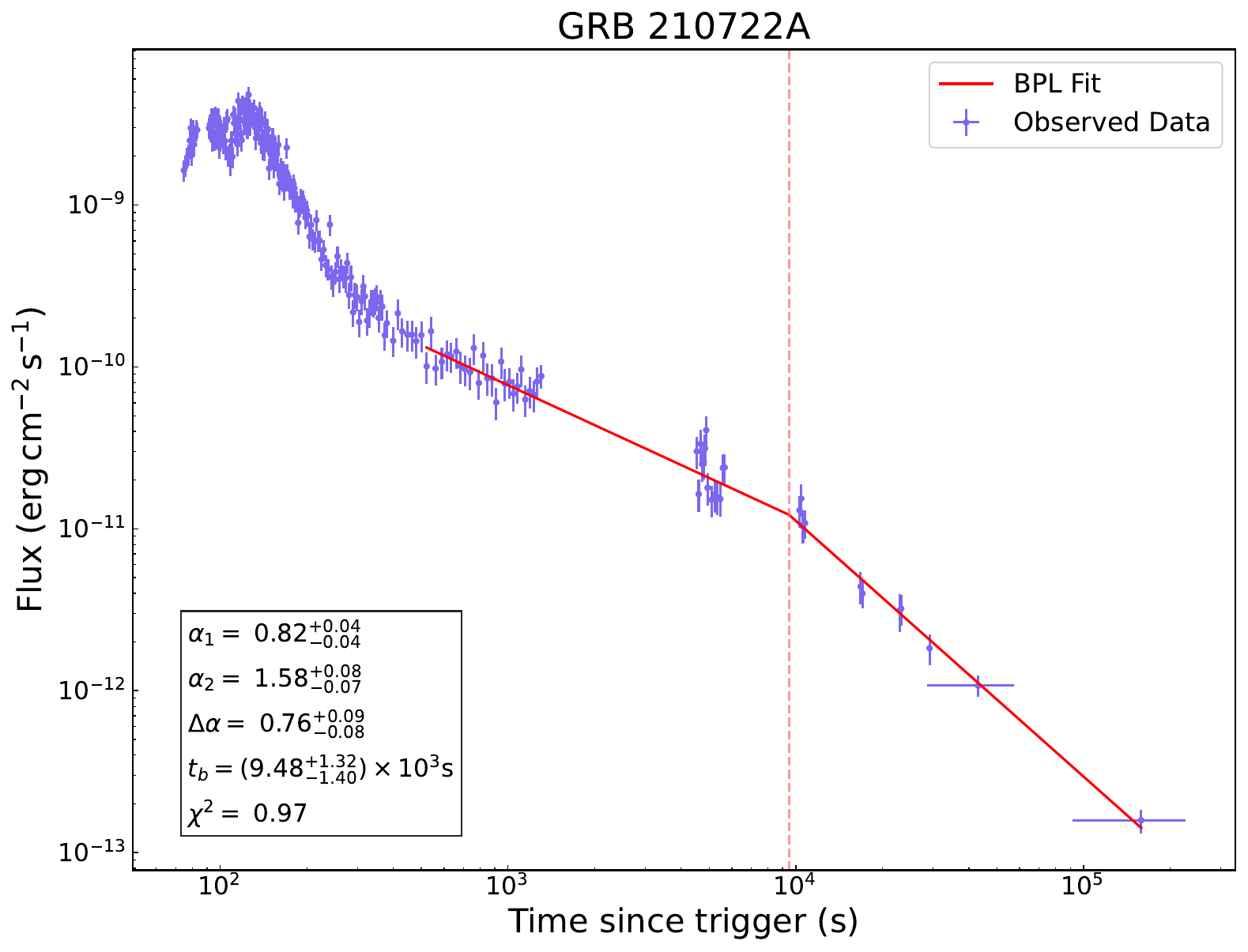}}%
\resizebox{45mm}{!}{\includegraphics[]{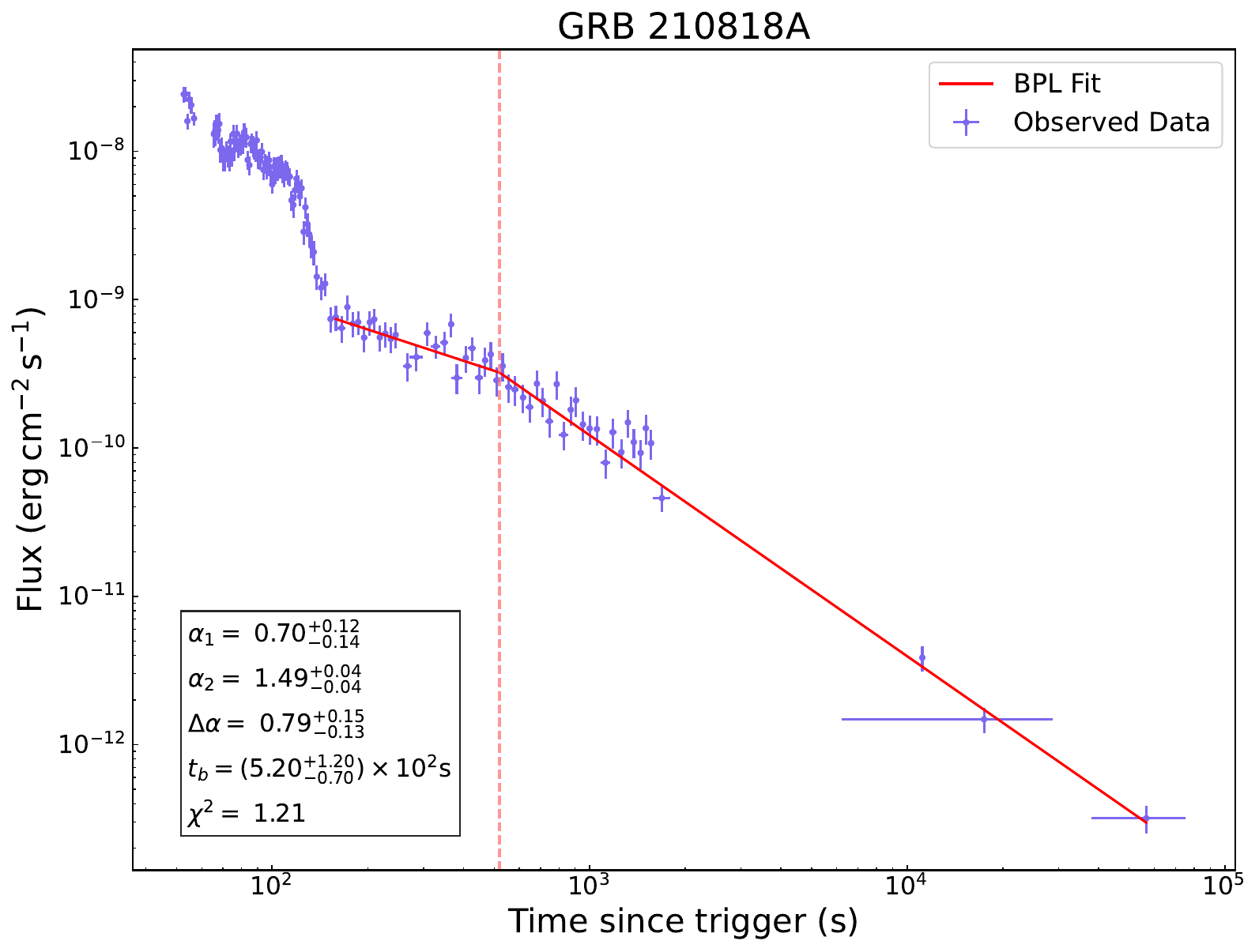}}%
\resizebox{45mm}{!}{\includegraphics[]{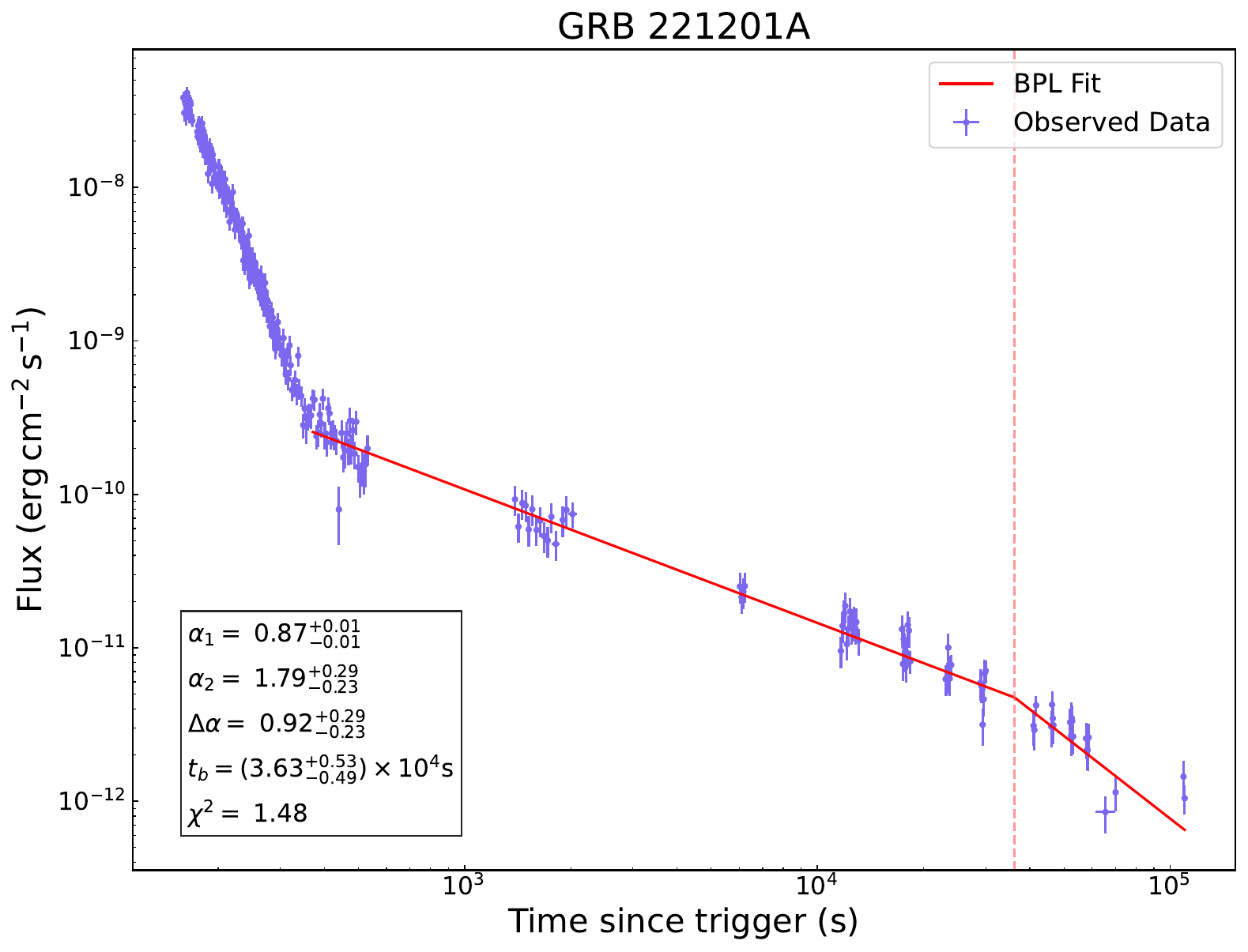}}\\
\resizebox{45mm}{!}{\includegraphics[]{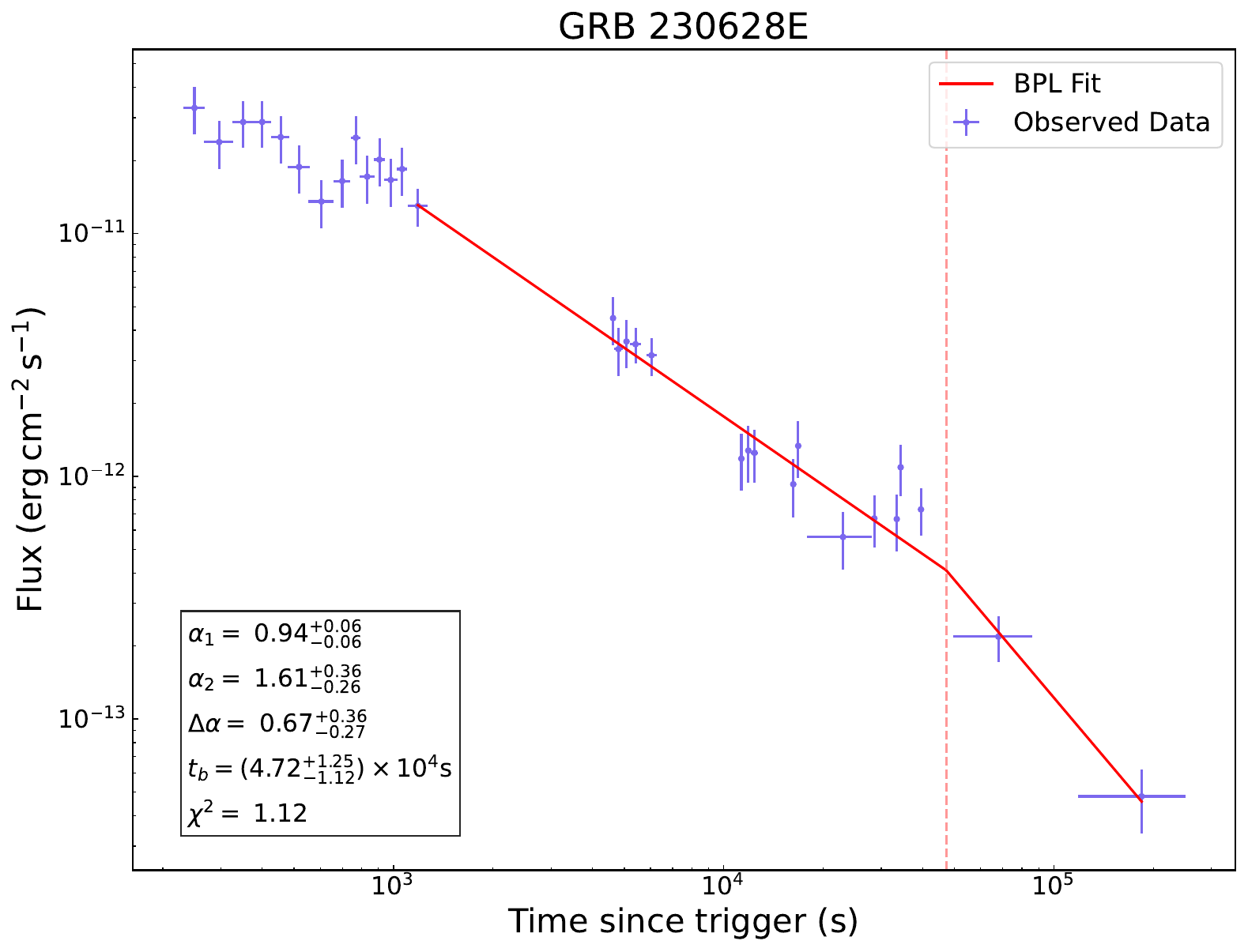}}%
\resizebox{45mm}{!}{\includegraphics[]{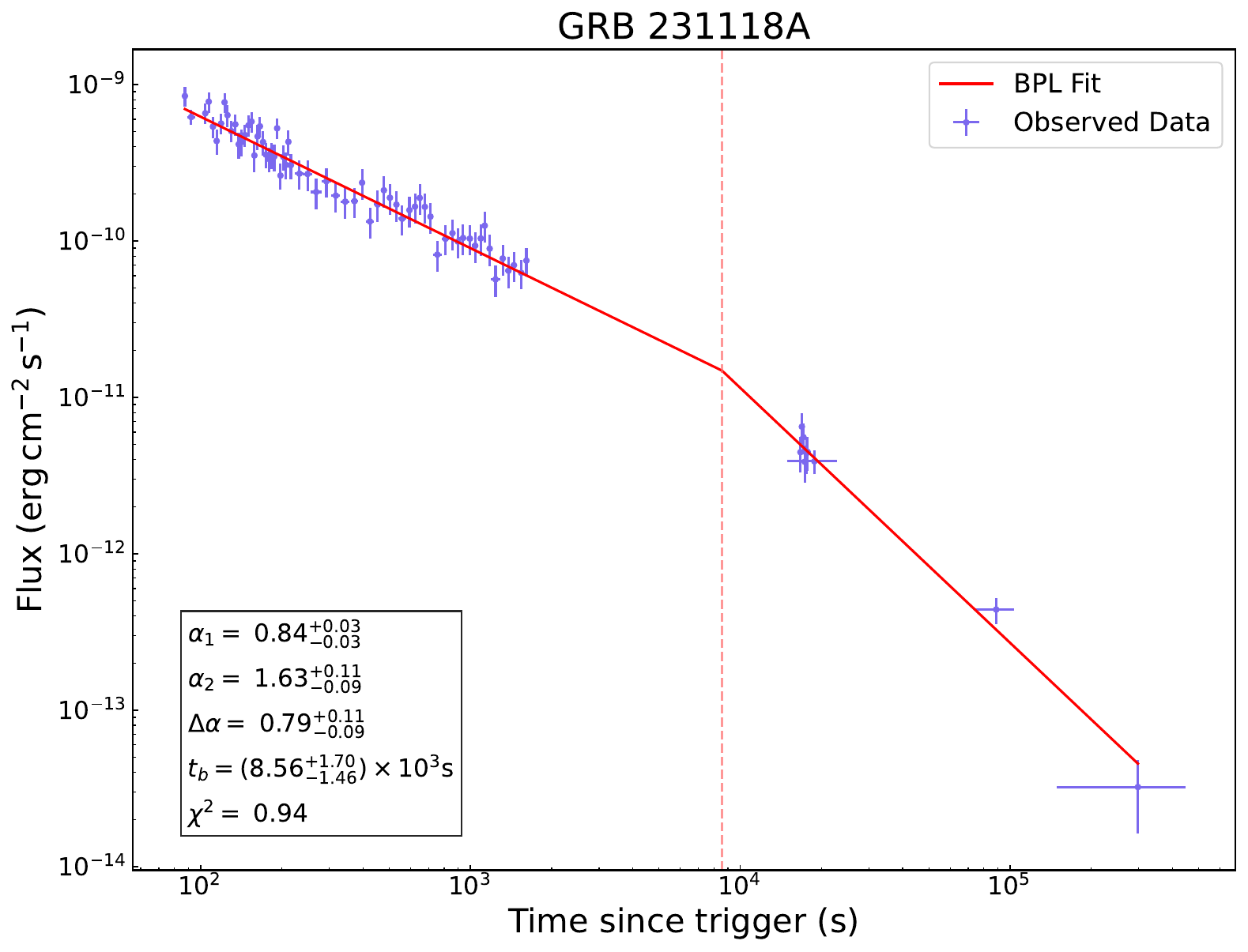}}%
\resizebox{45mm}{!}{\includegraphics[]{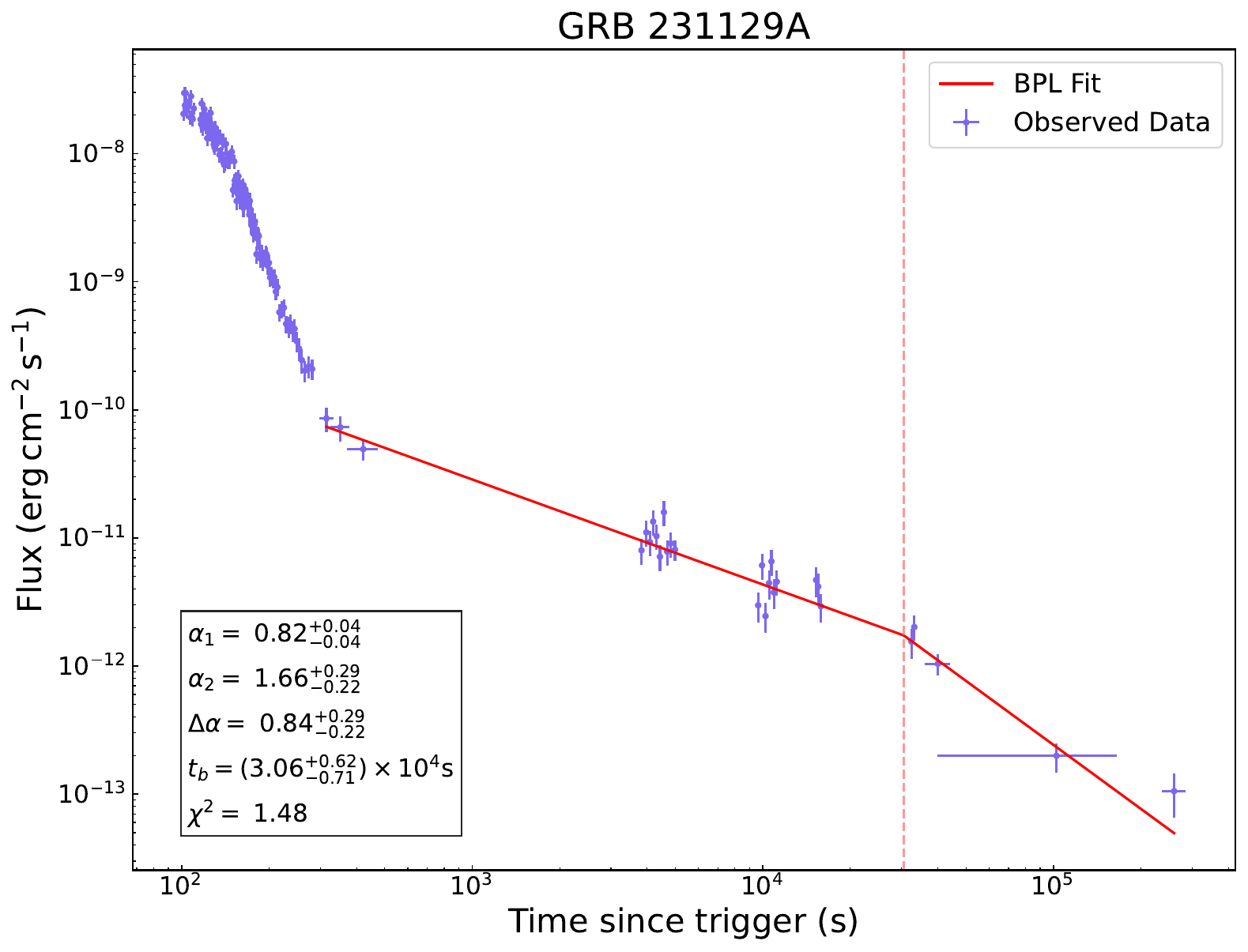}}%
\resizebox{45mm}{!}{\includegraphics[]{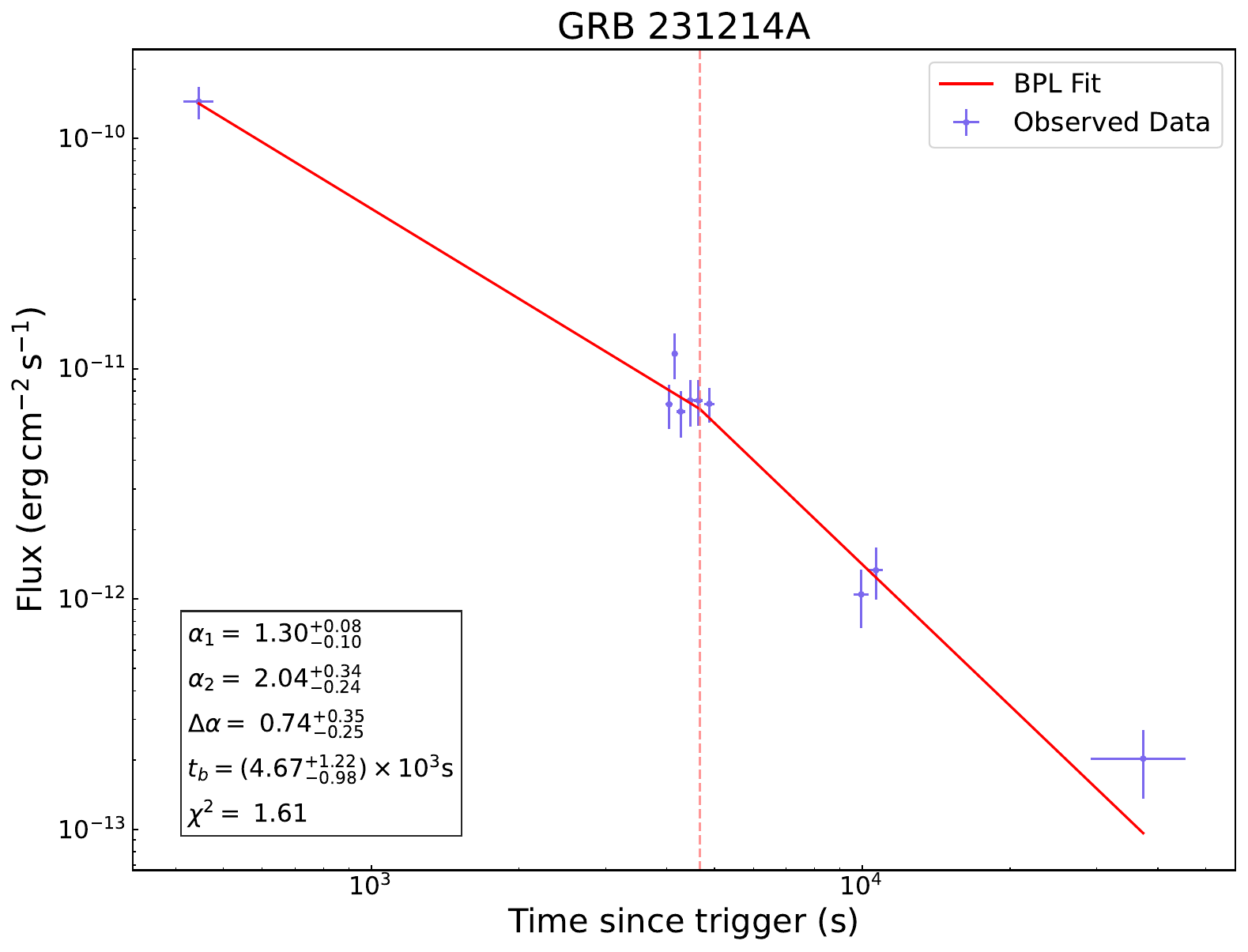}}\\
\resizebox{45mm}{!}{\includegraphics[]{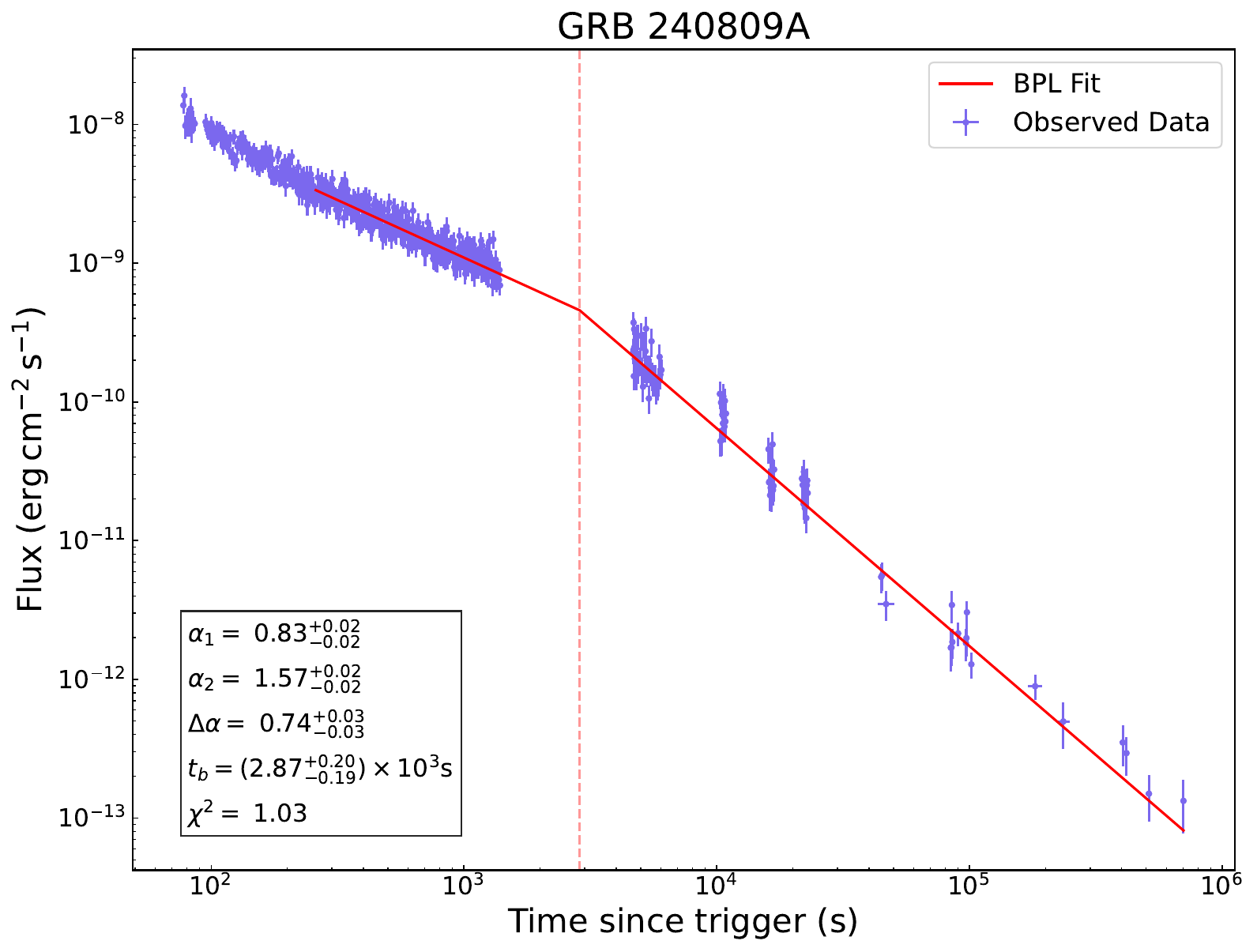}}%
\resizebox{45mm}{!}{\includegraphics[]{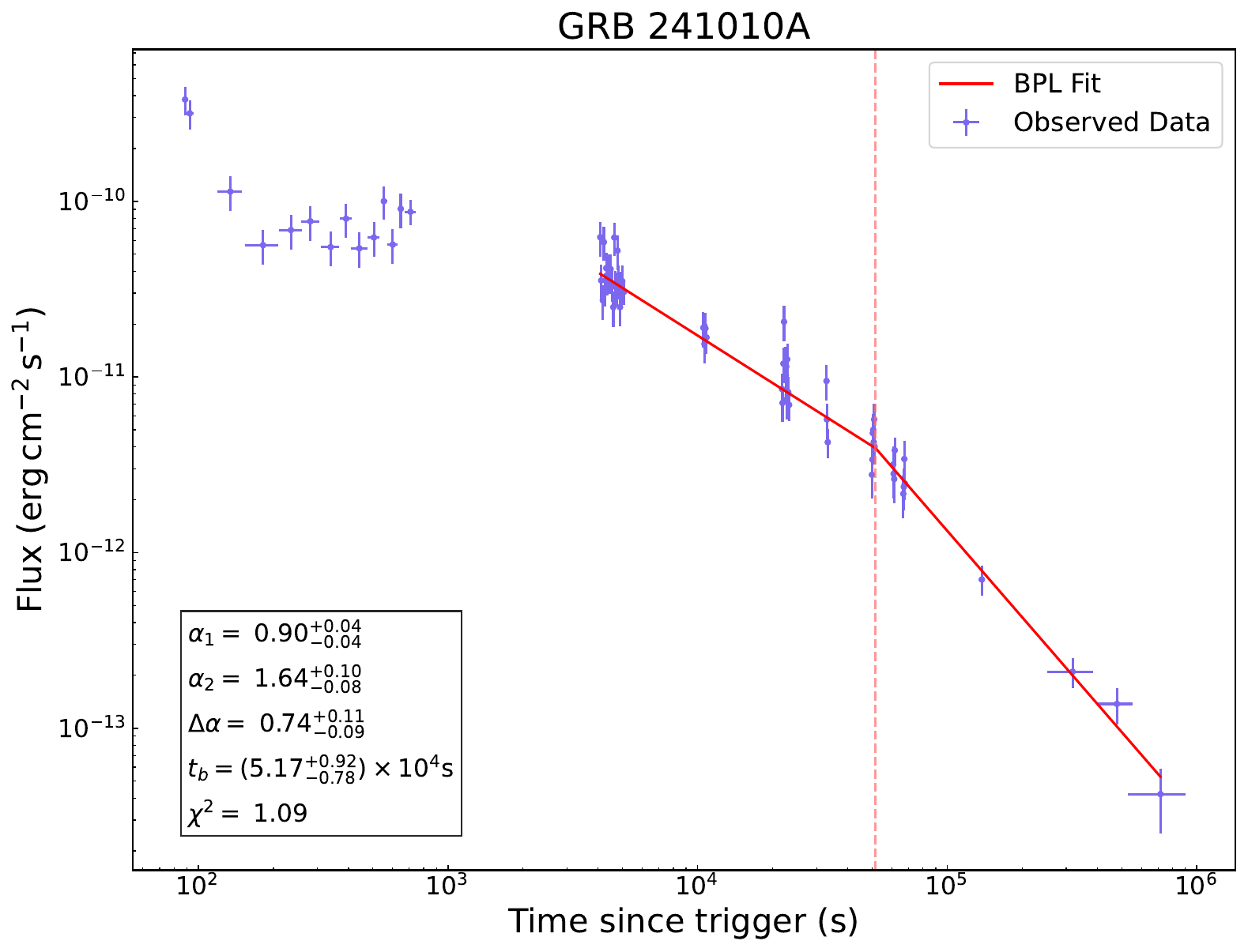}}%
\resizebox{45mm}{!}{\includegraphics[]{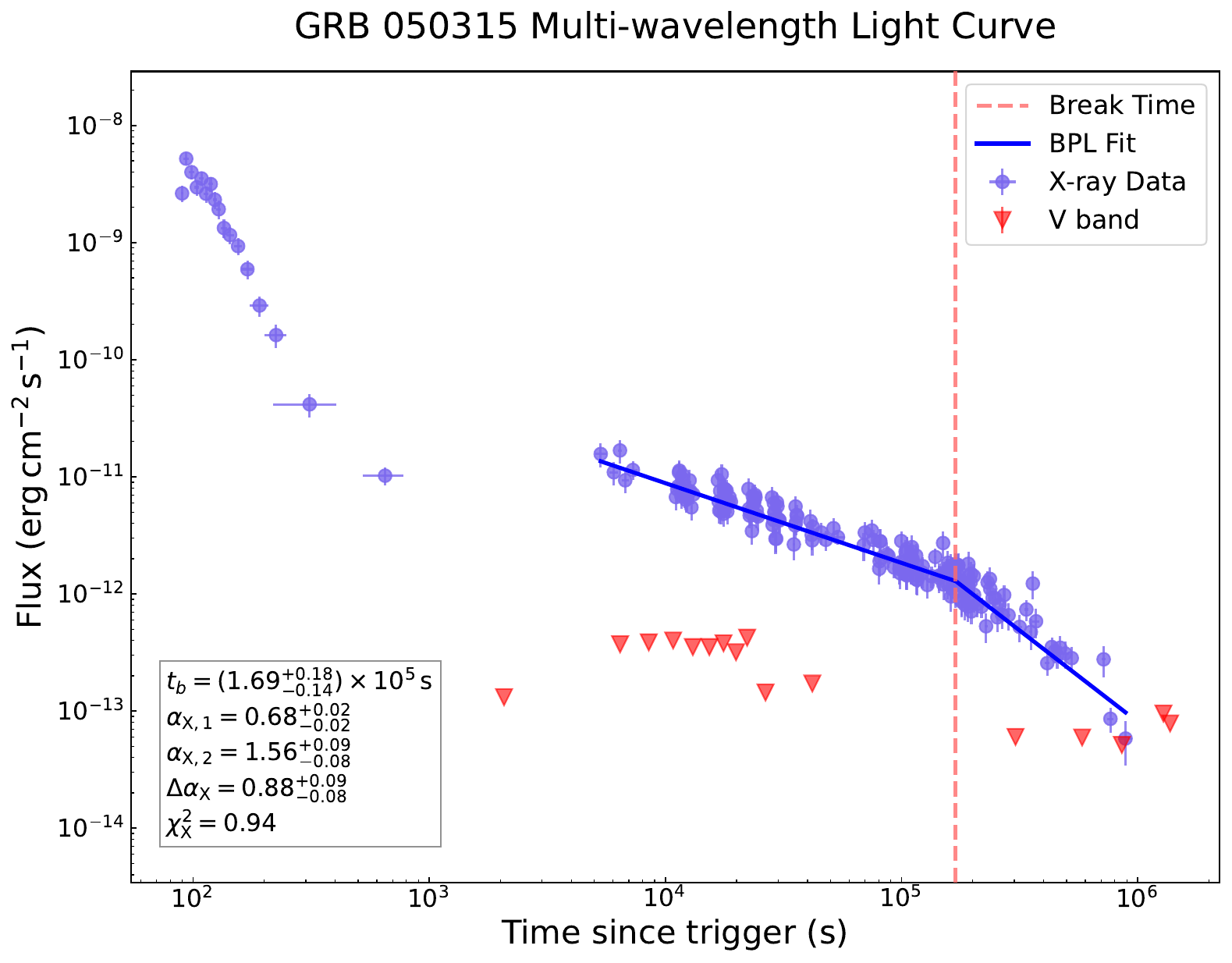}}%
\resizebox{45mm}{!}{\includegraphics[]{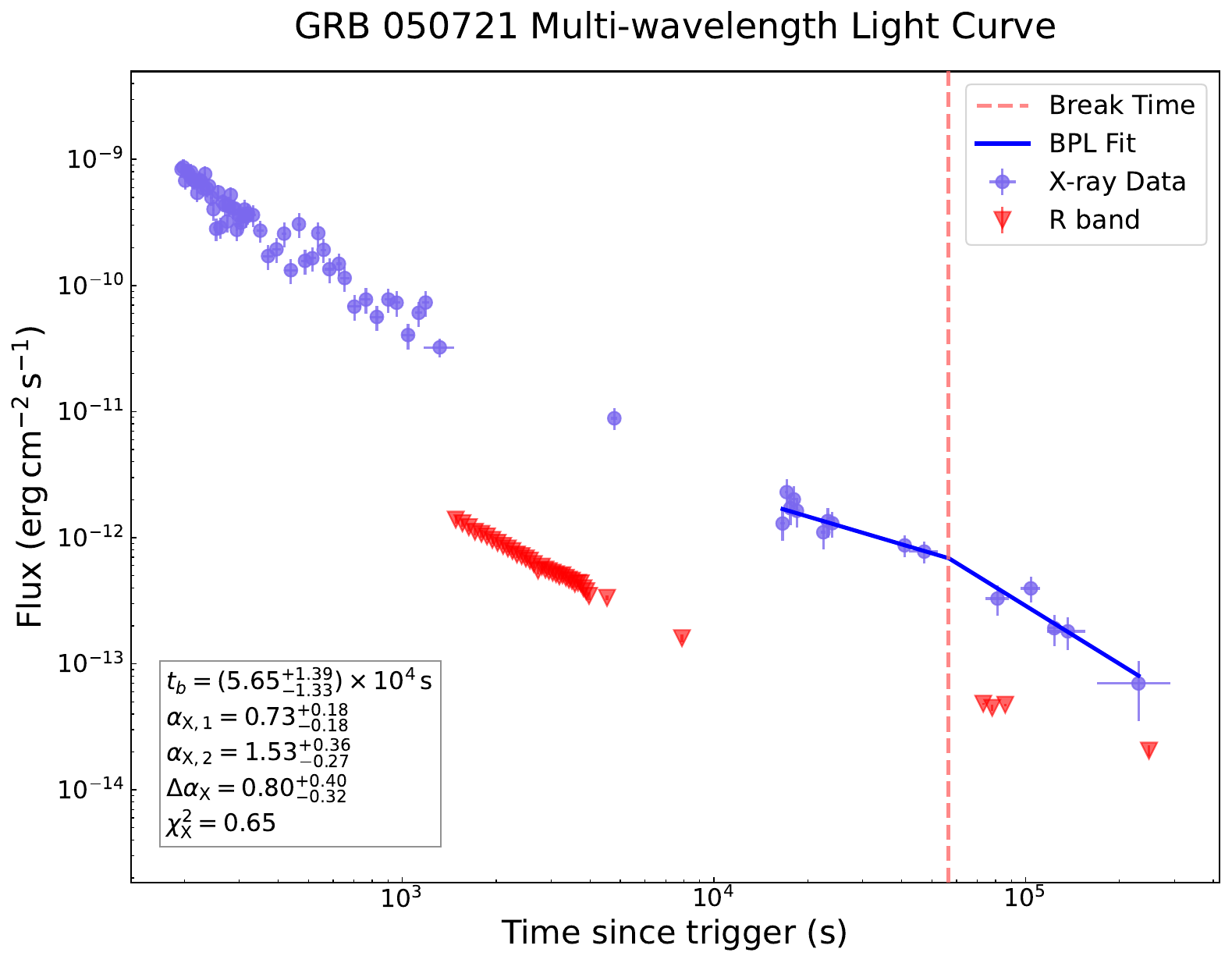}}\\
\caption{(Continued)}
\end{figure}

\clearpage
\addtocounter{figure}{-1}
\begin{figure}
\centering
\resizebox{45mm}{!}{\includegraphics[]{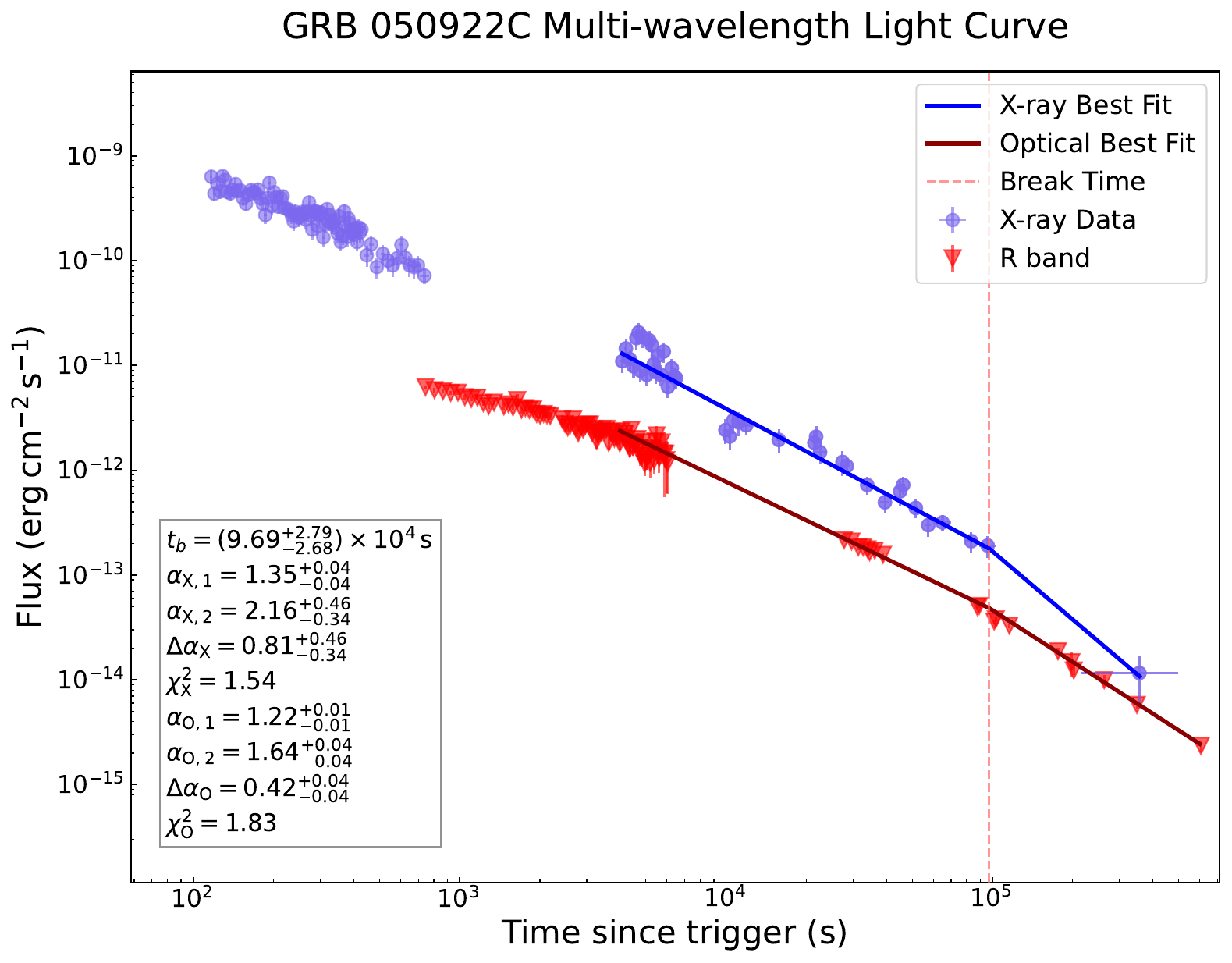}}%
\resizebox{45mm}{!}{\includegraphics[]{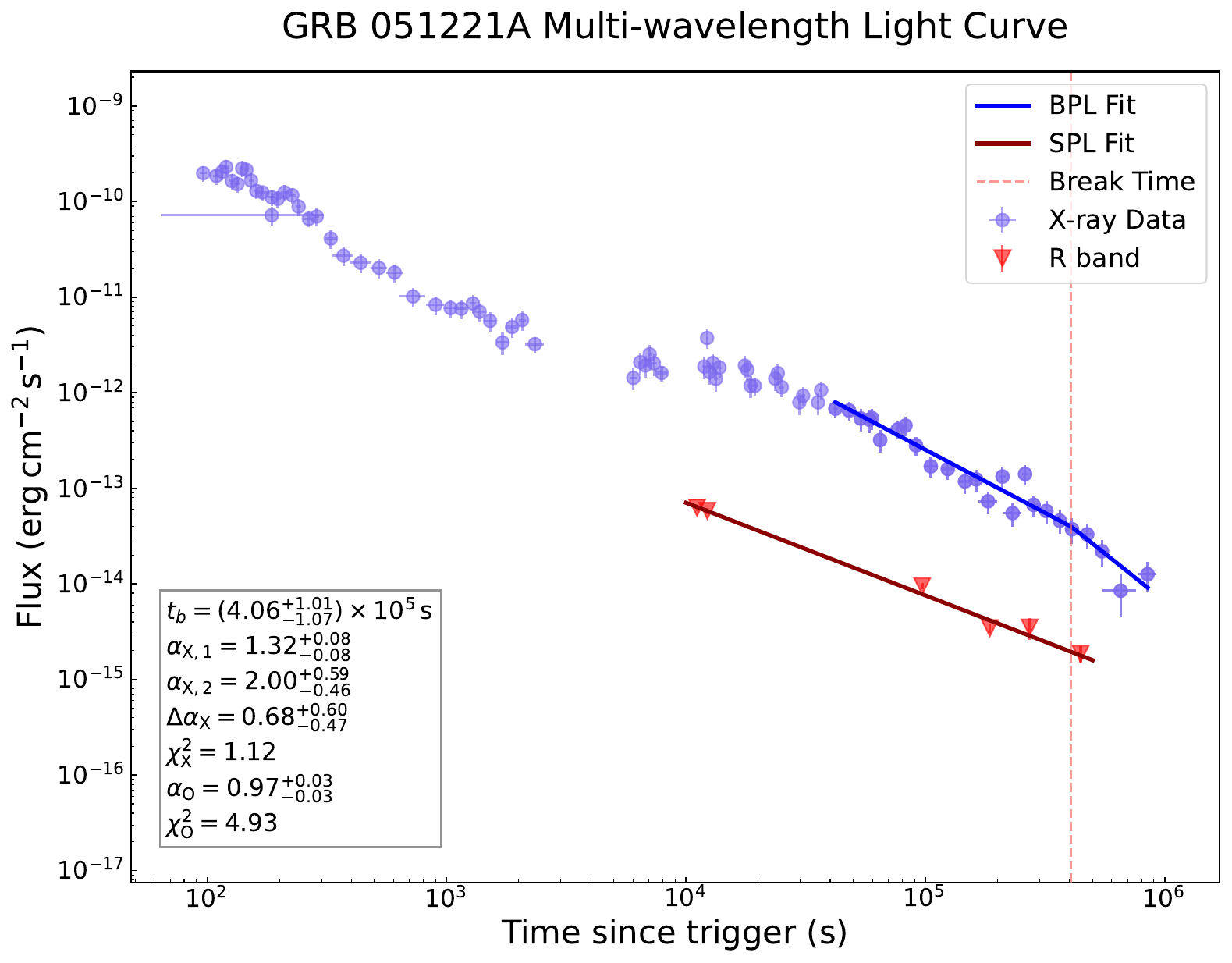}}%
\resizebox{45mm}{!}{\includegraphics[]{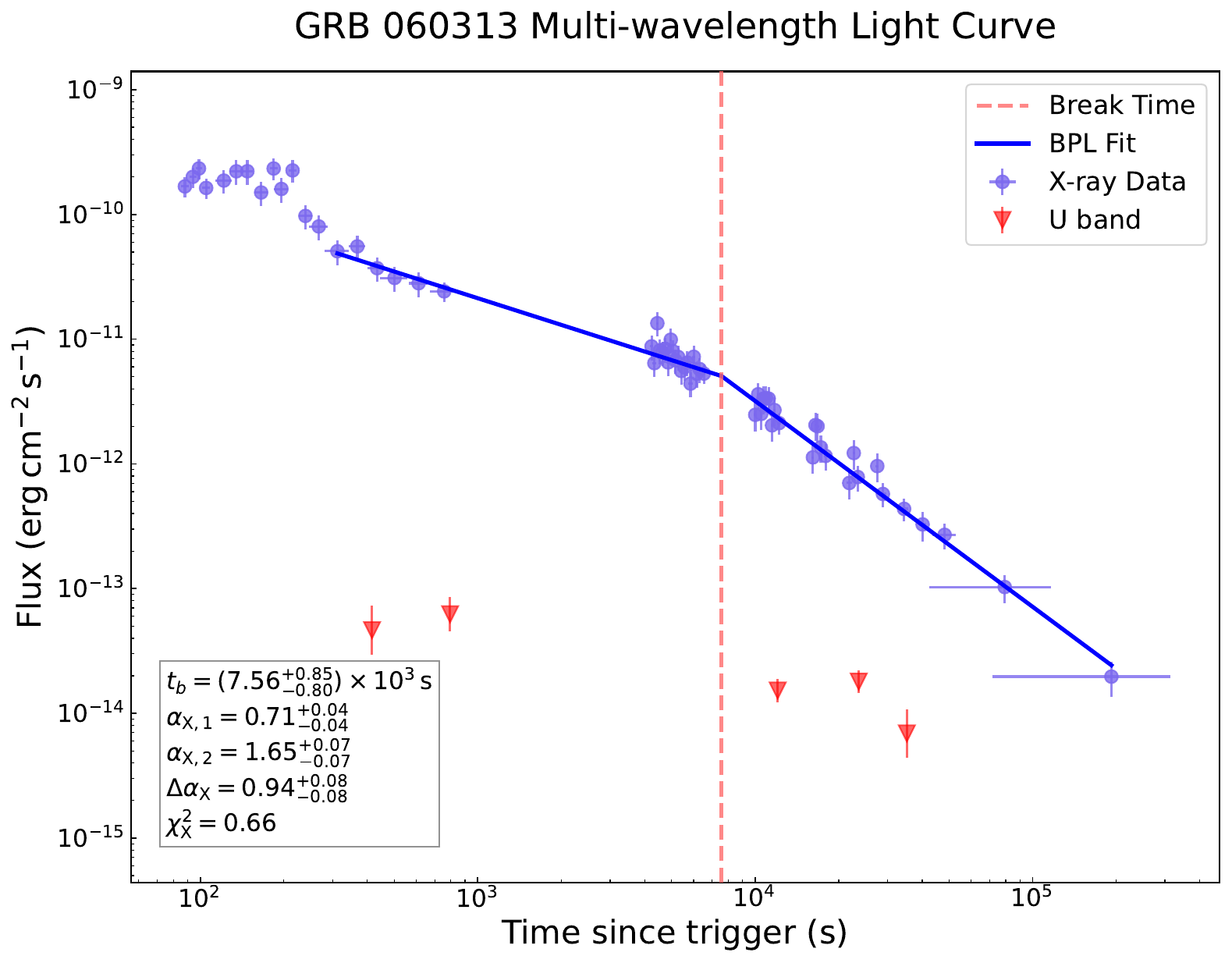}}%
\resizebox{45mm}{!}{\includegraphics[]{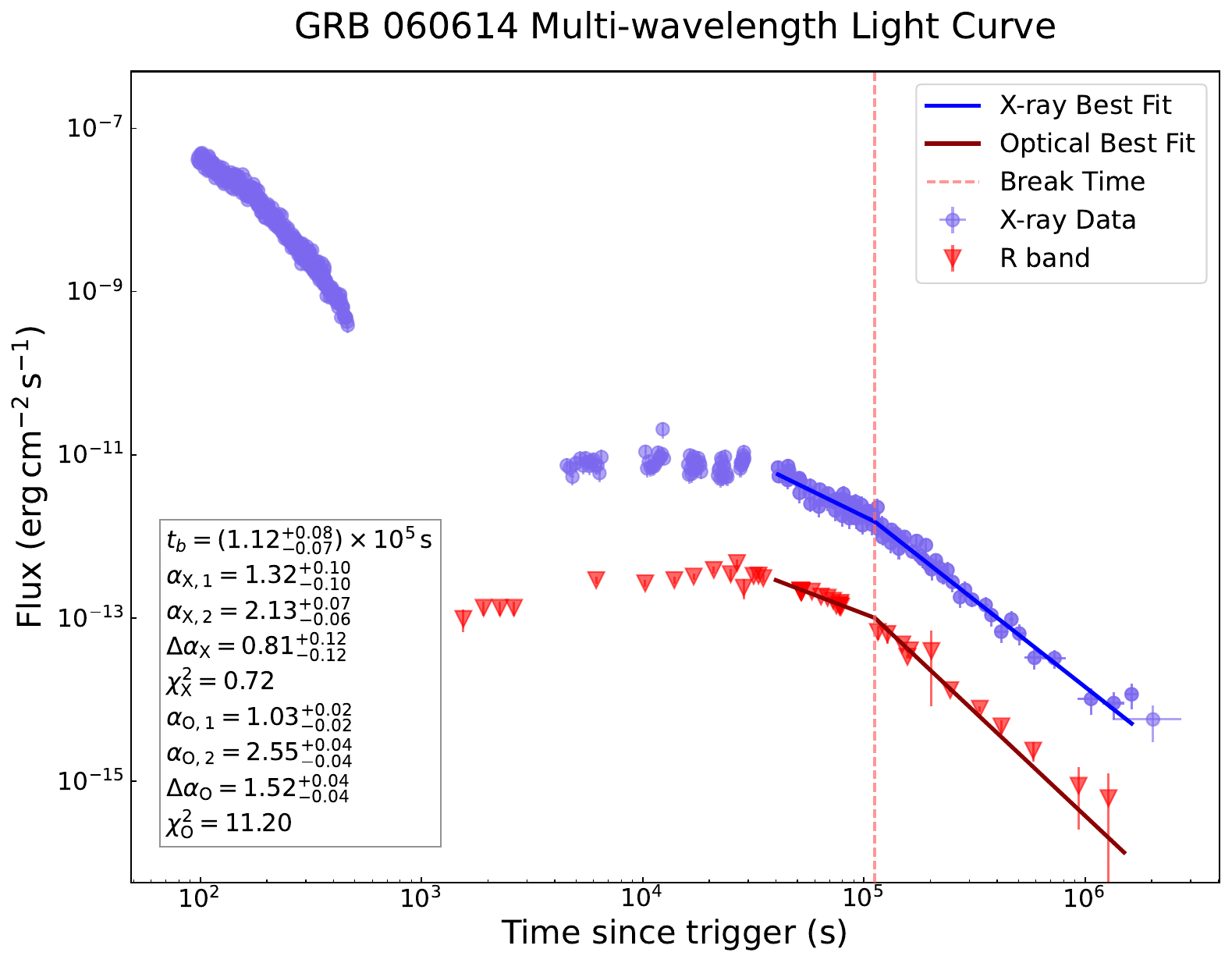}}\\
\resizebox{45mm}{!}{\includegraphics[]{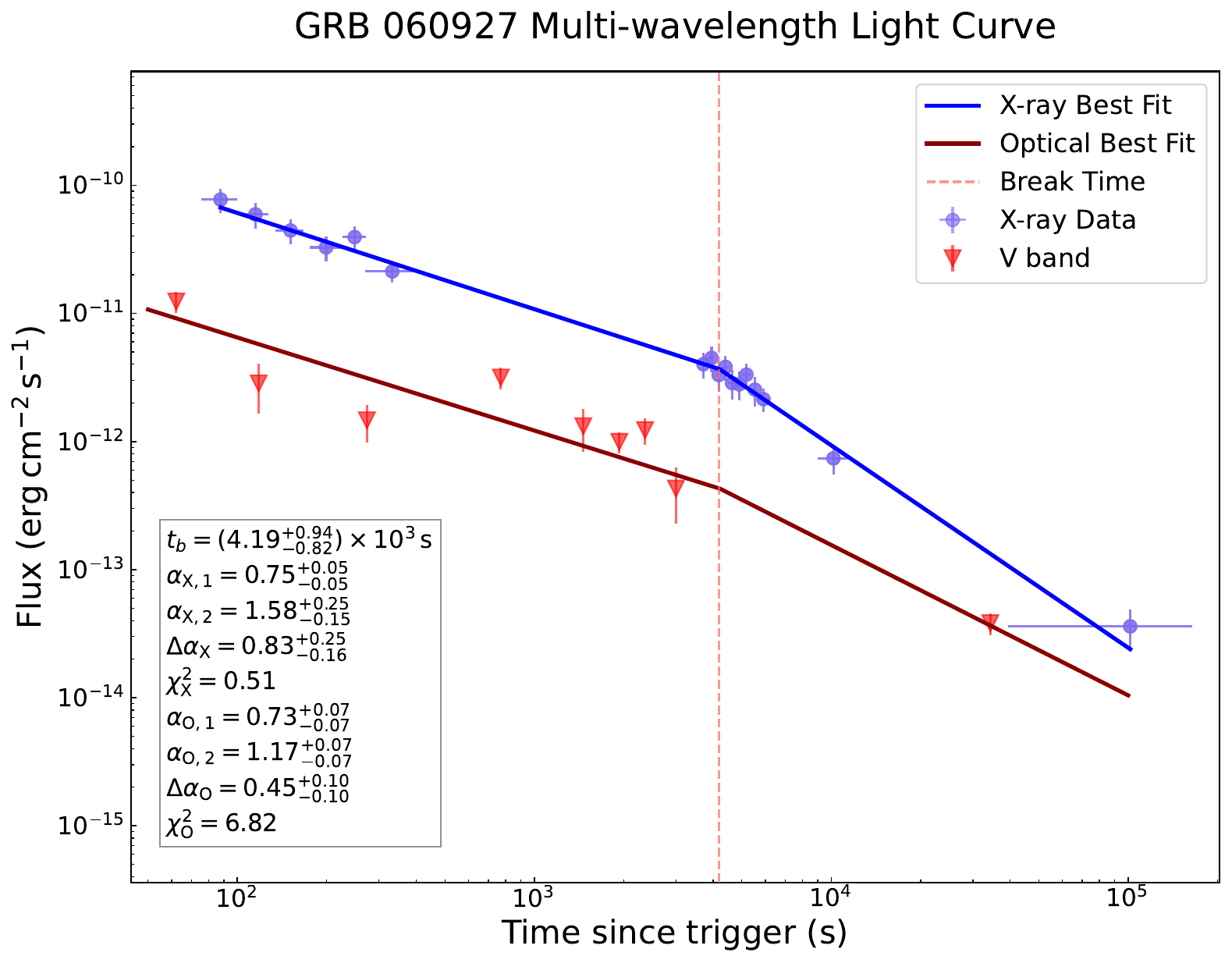}}%
\resizebox{45mm}{!}{\includegraphics[]{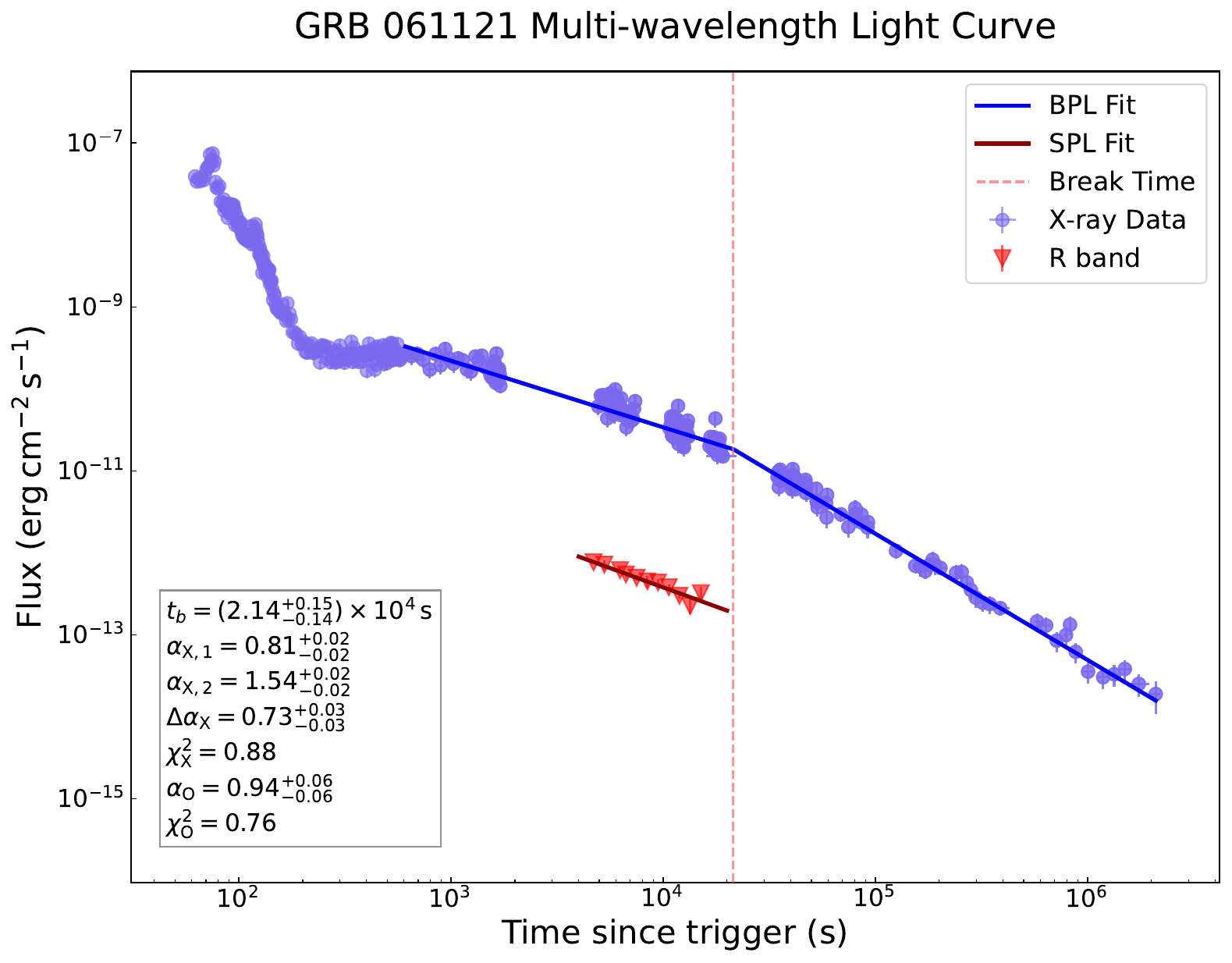}}%
\resizebox{45mm}{!}{\includegraphics[]{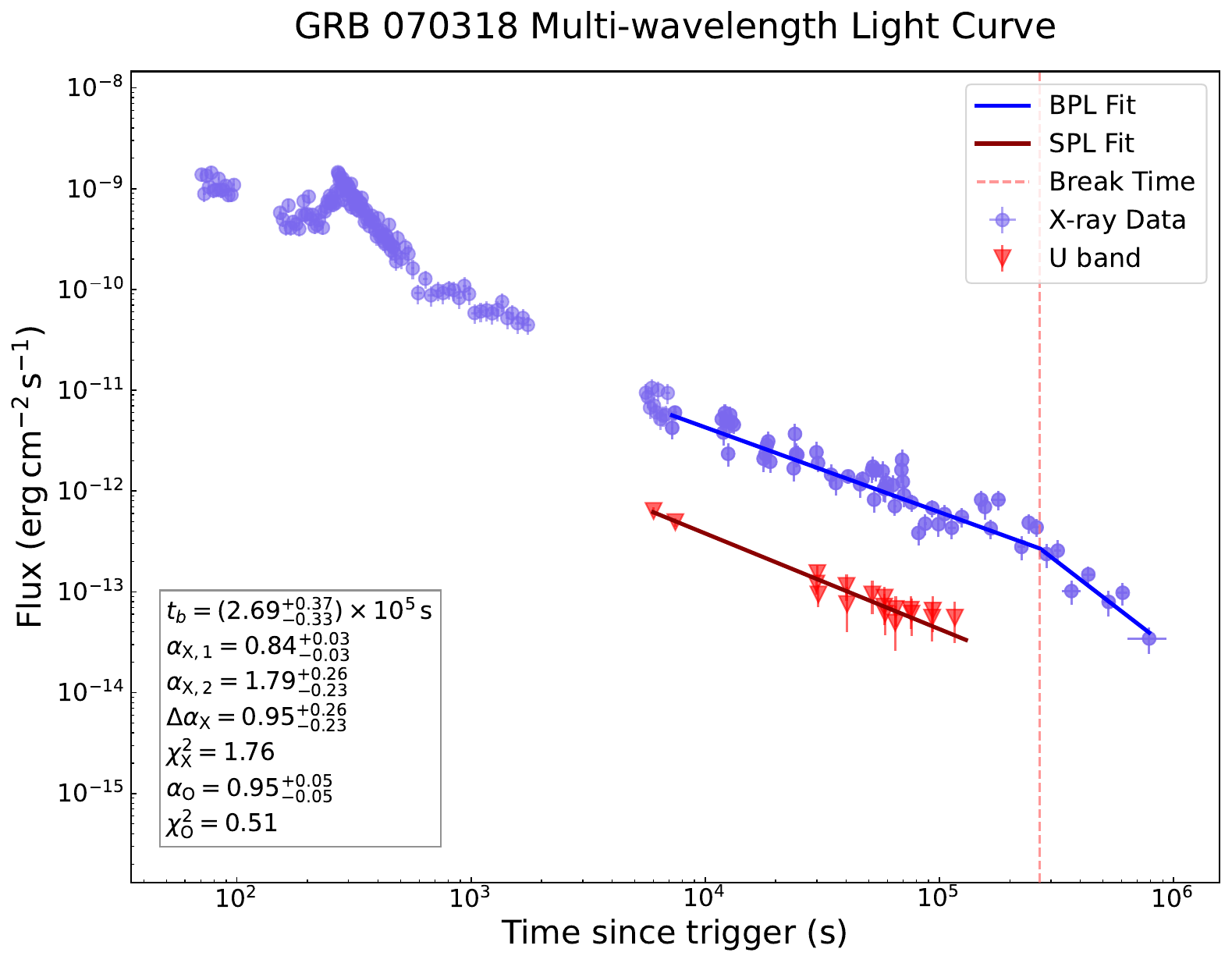}}%
\resizebox{45mm}{!}{\includegraphics[]{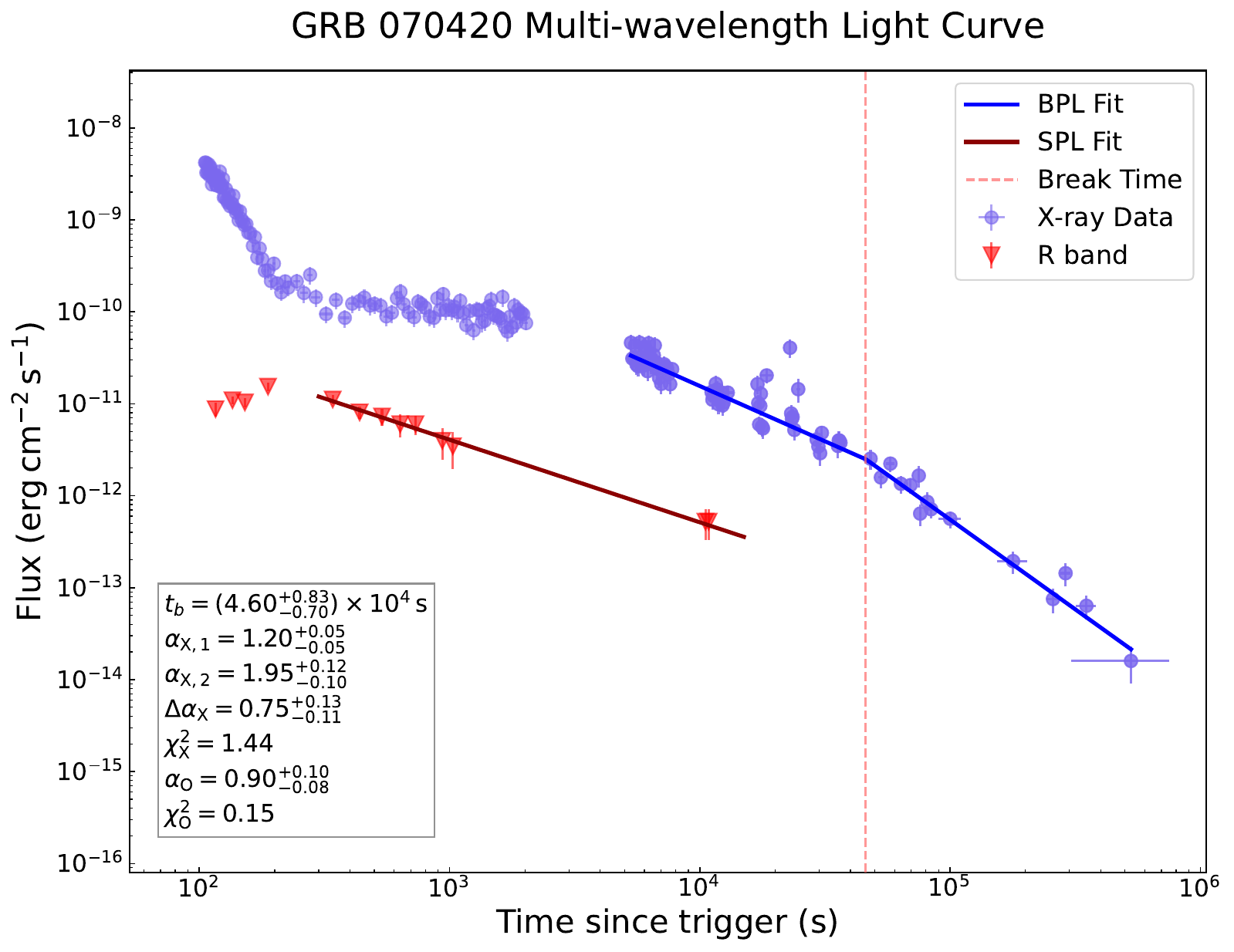}}\\
\resizebox{45mm}{!}{\includegraphics[]{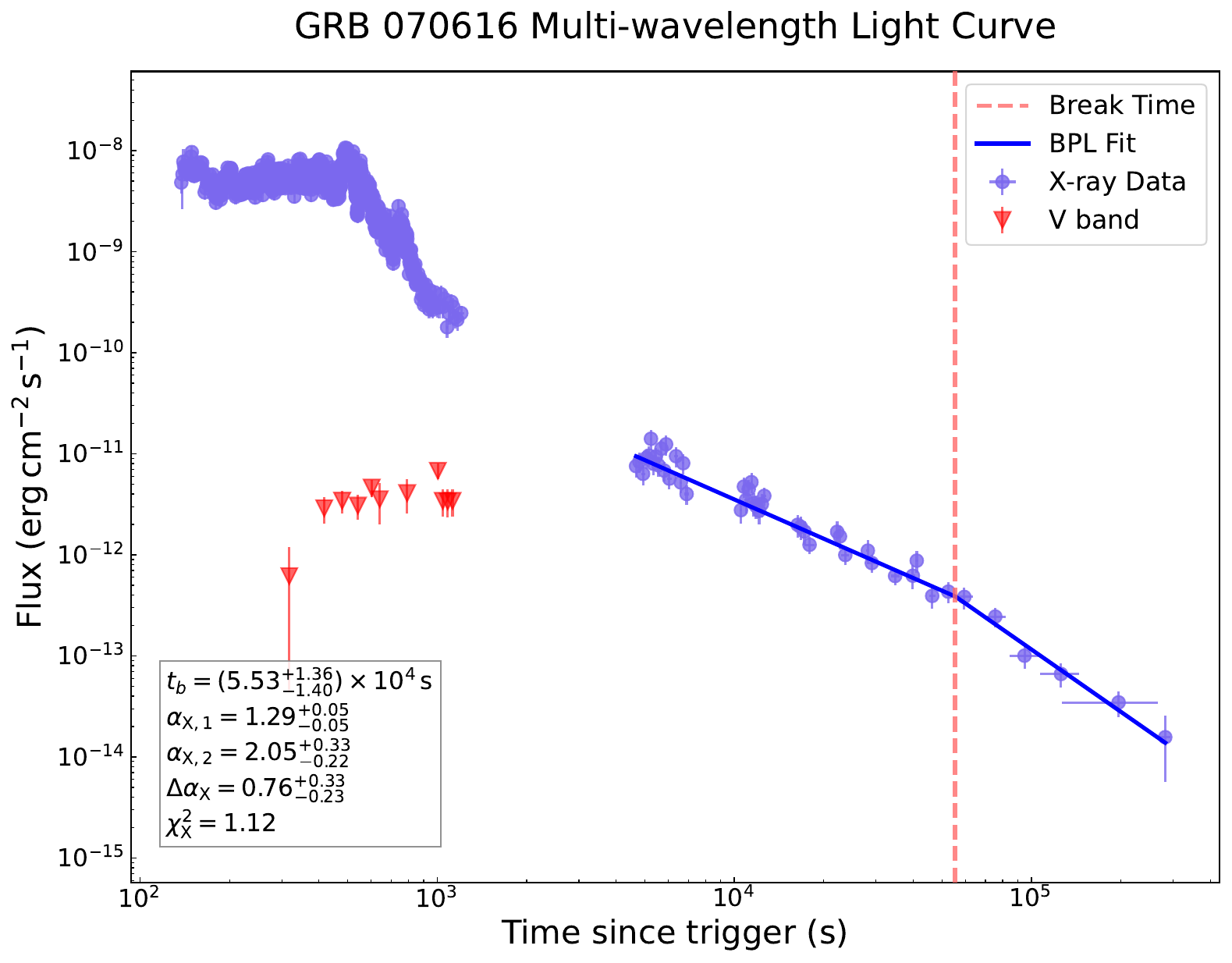}}%
\resizebox{45mm}{!}{\includegraphics[]{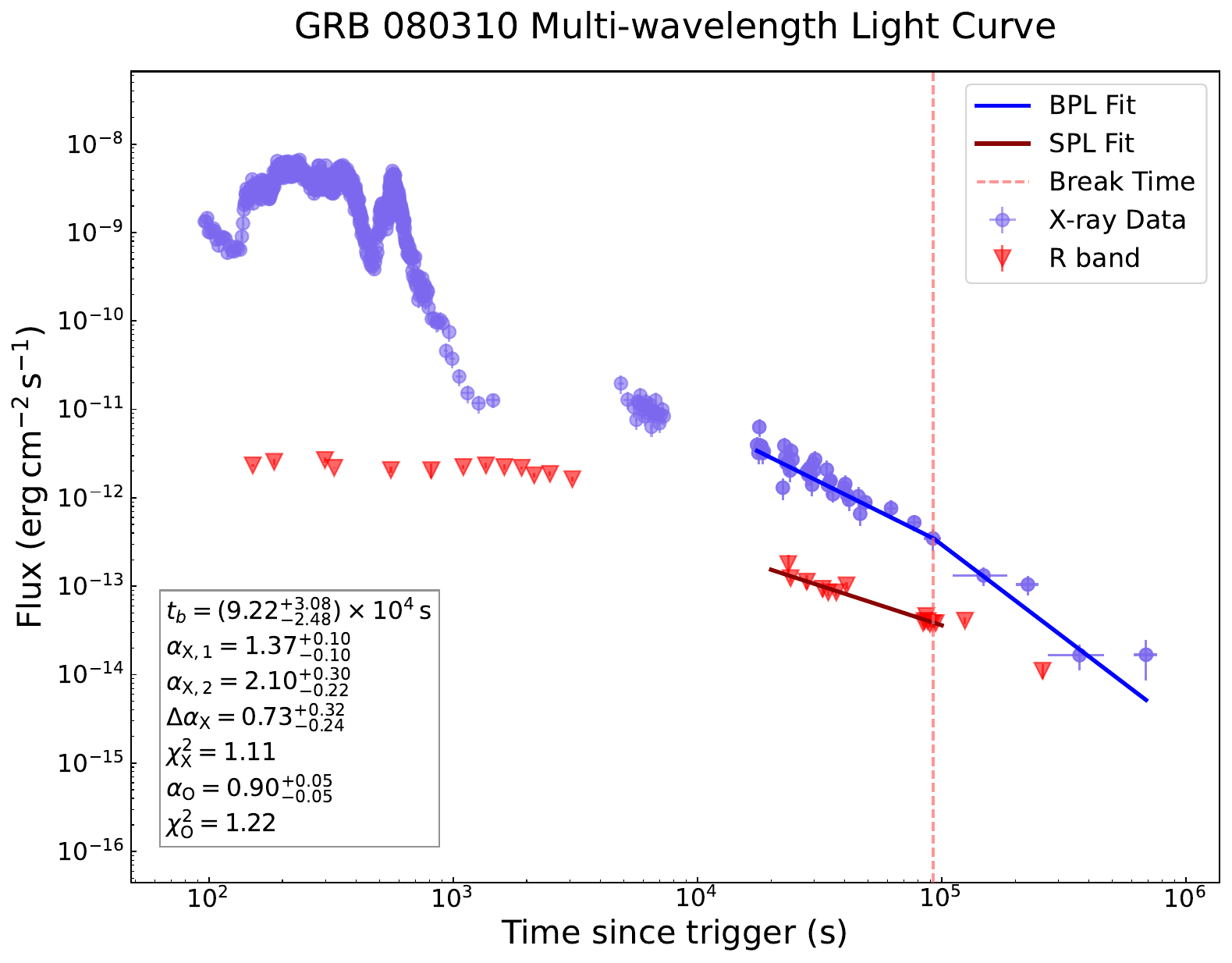}}%
\resizebox{45mm}{!}{\includegraphics[]{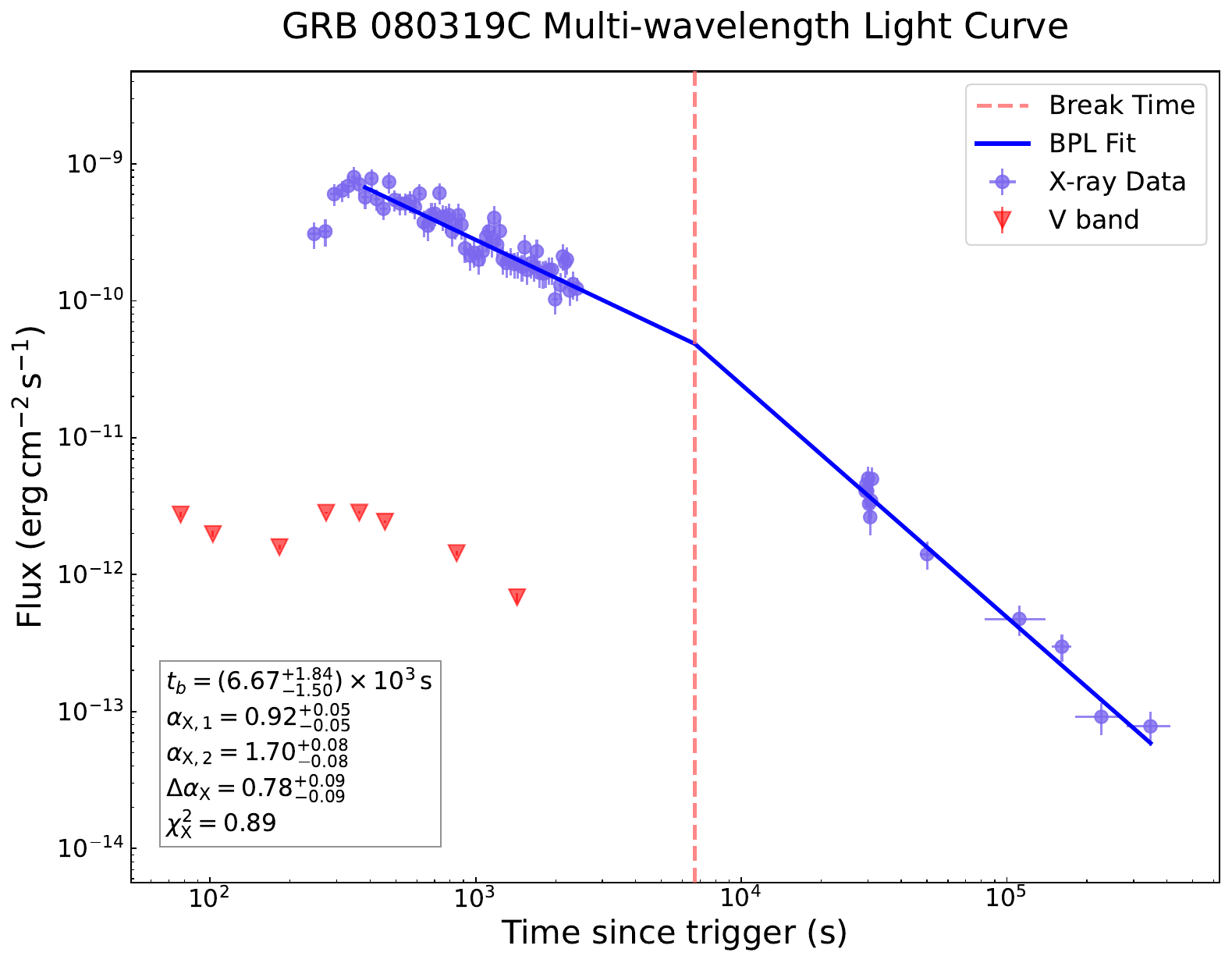}}%
\resizebox{45mm}{!}{\includegraphics[]{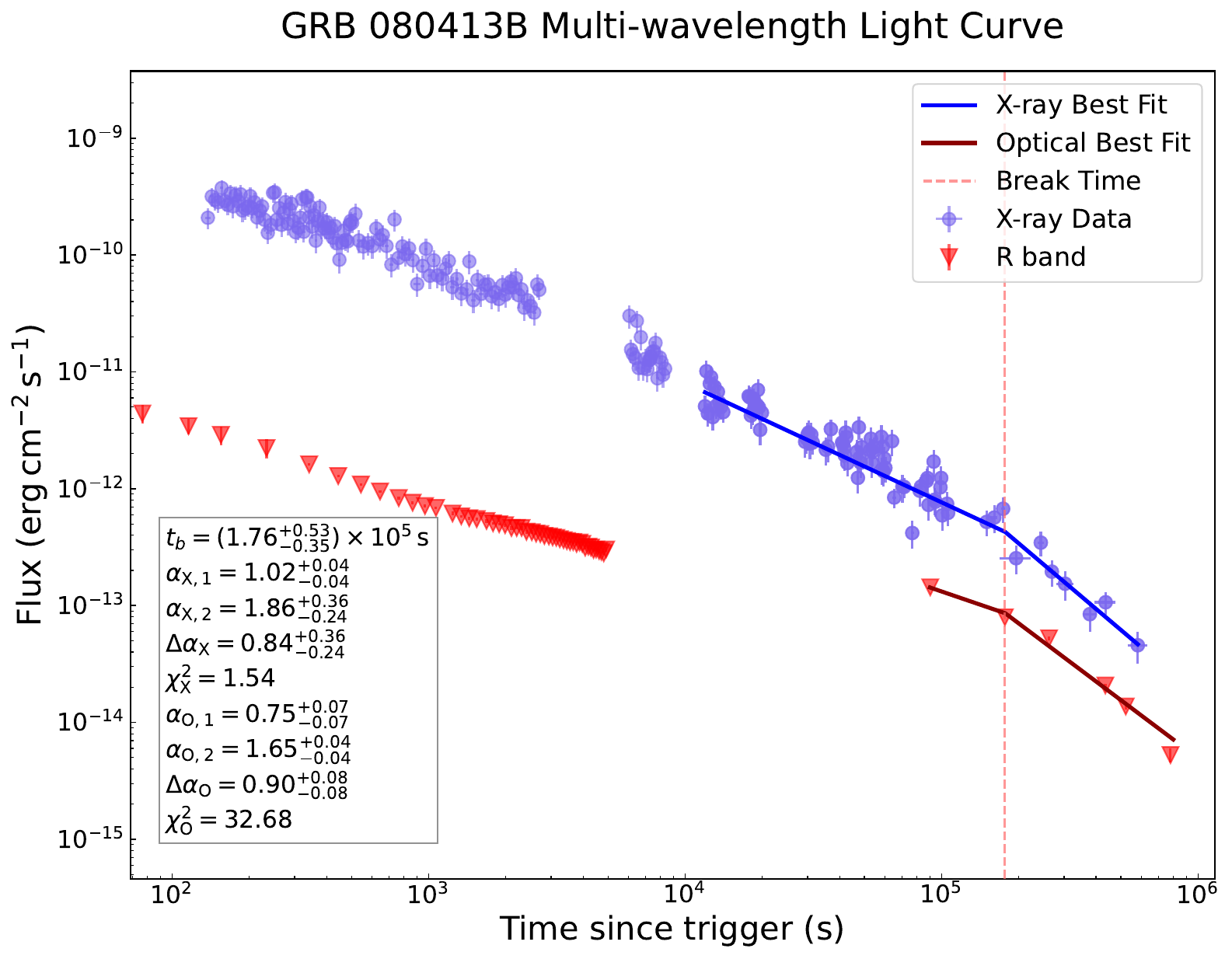}}\\
\resizebox{45mm}{!}{\includegraphics[]{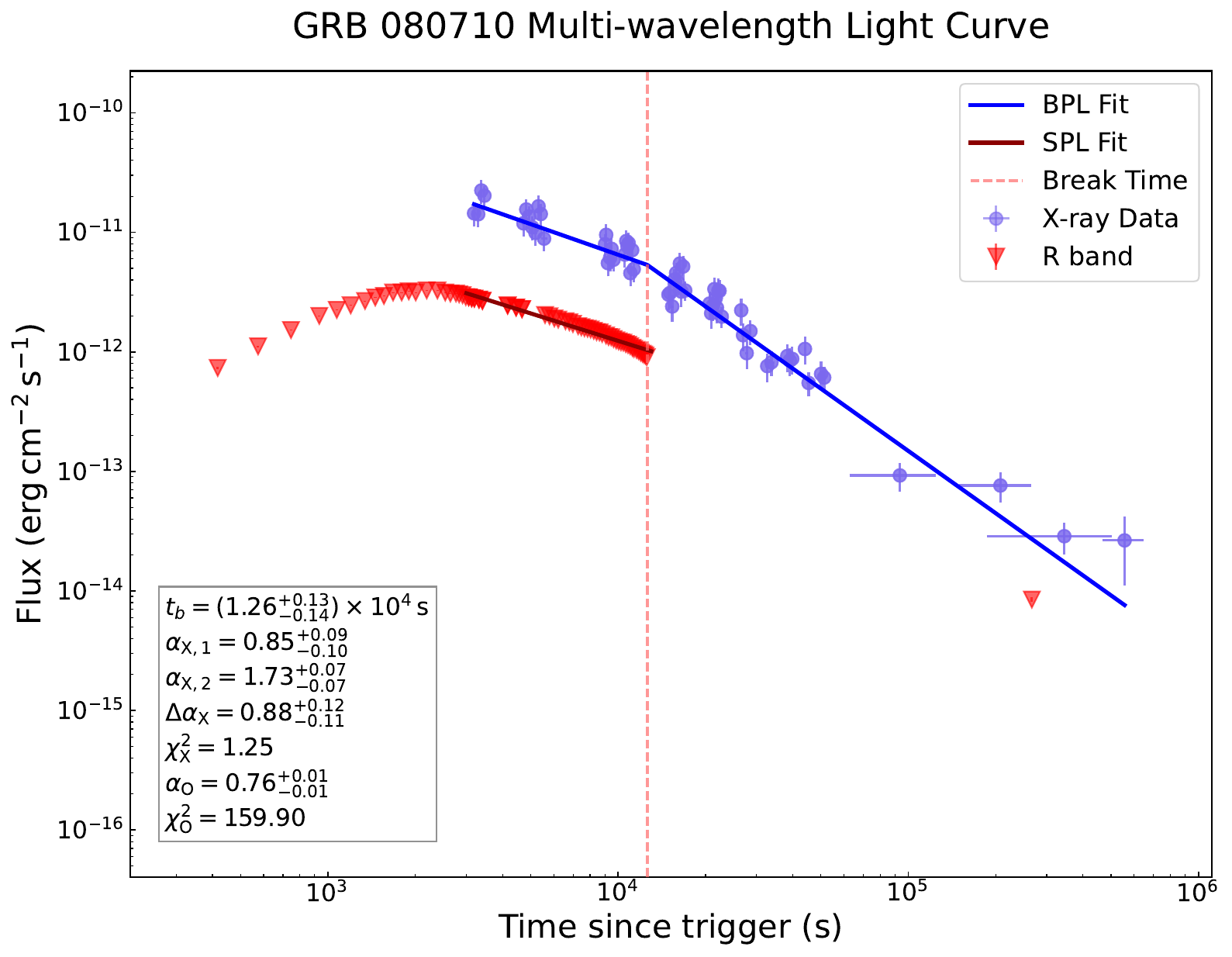}}%
\resizebox{45mm}{!}{\includegraphics[]{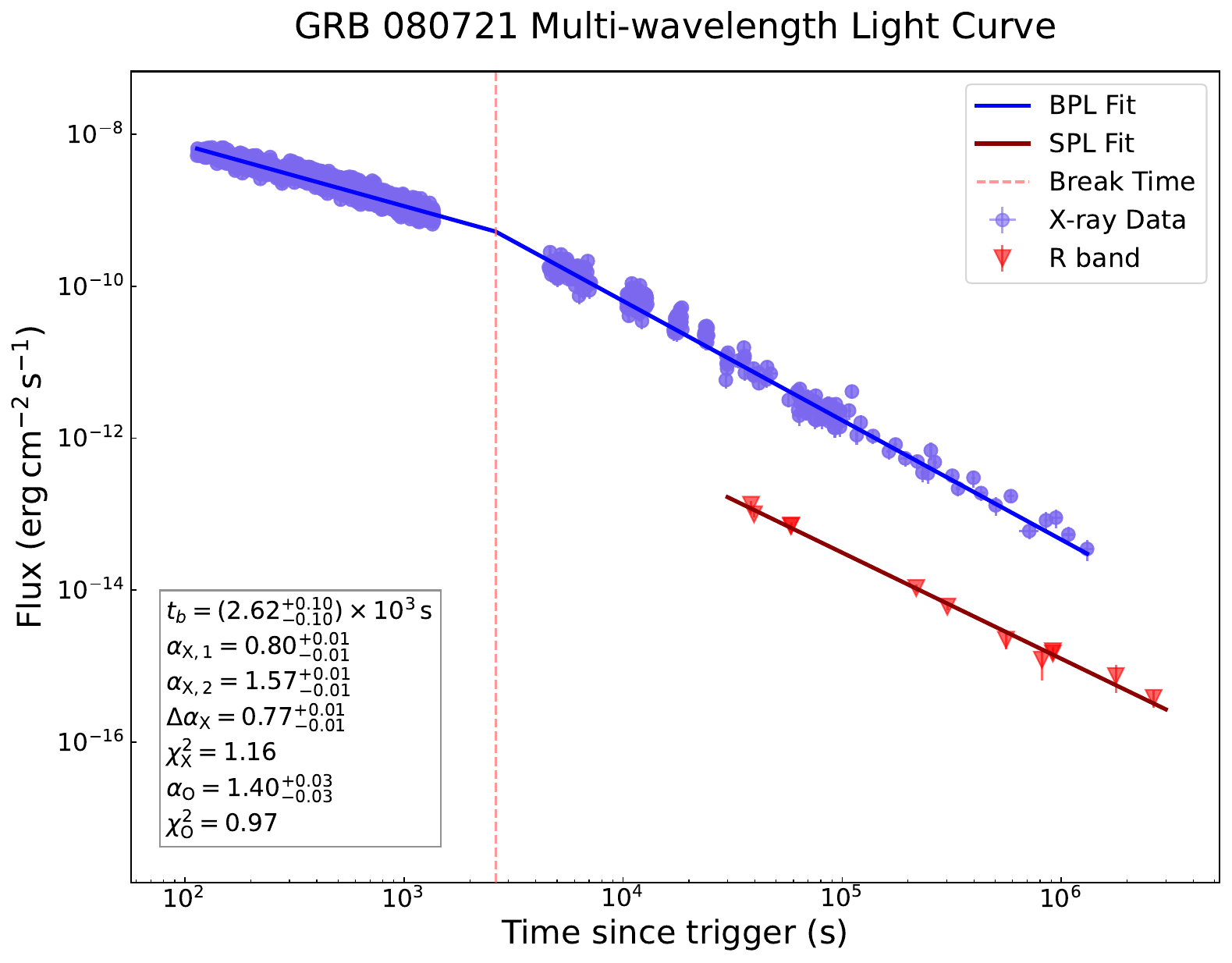}}%
\resizebox{45mm}{!}{\includegraphics[]{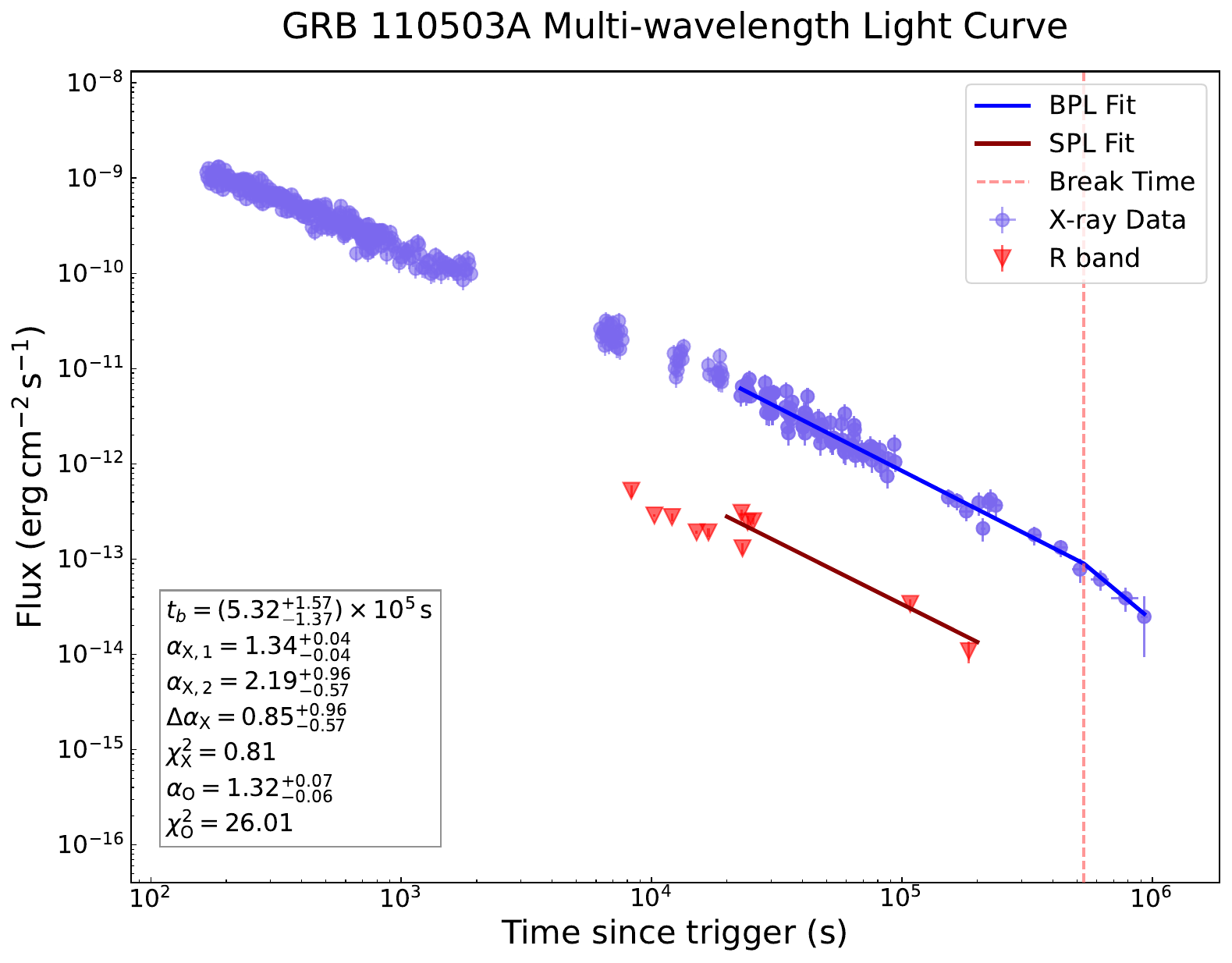}}%
\resizebox{45mm}{!}{\includegraphics[]{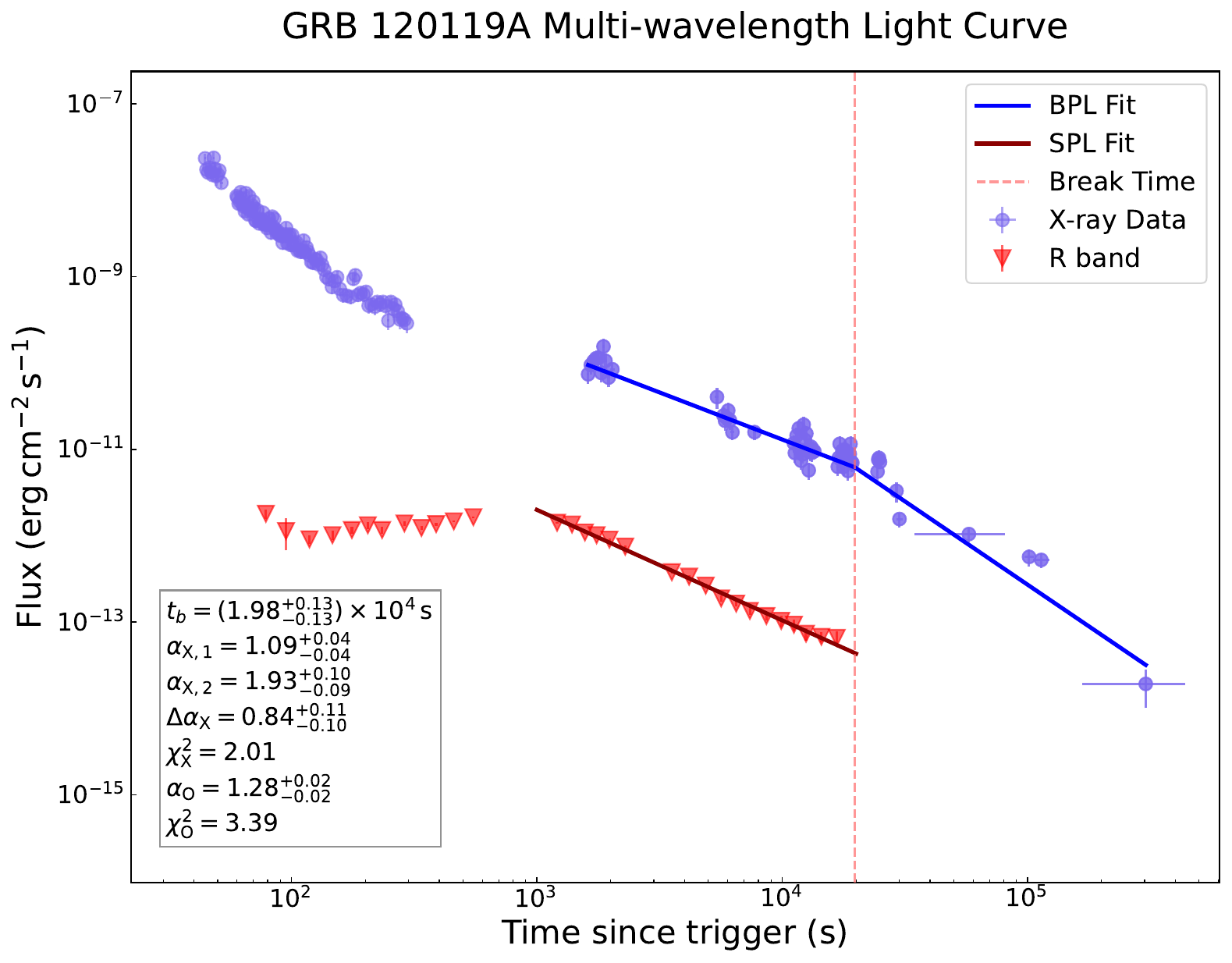}}\\
\resizebox{45mm}{!}{\includegraphics[]{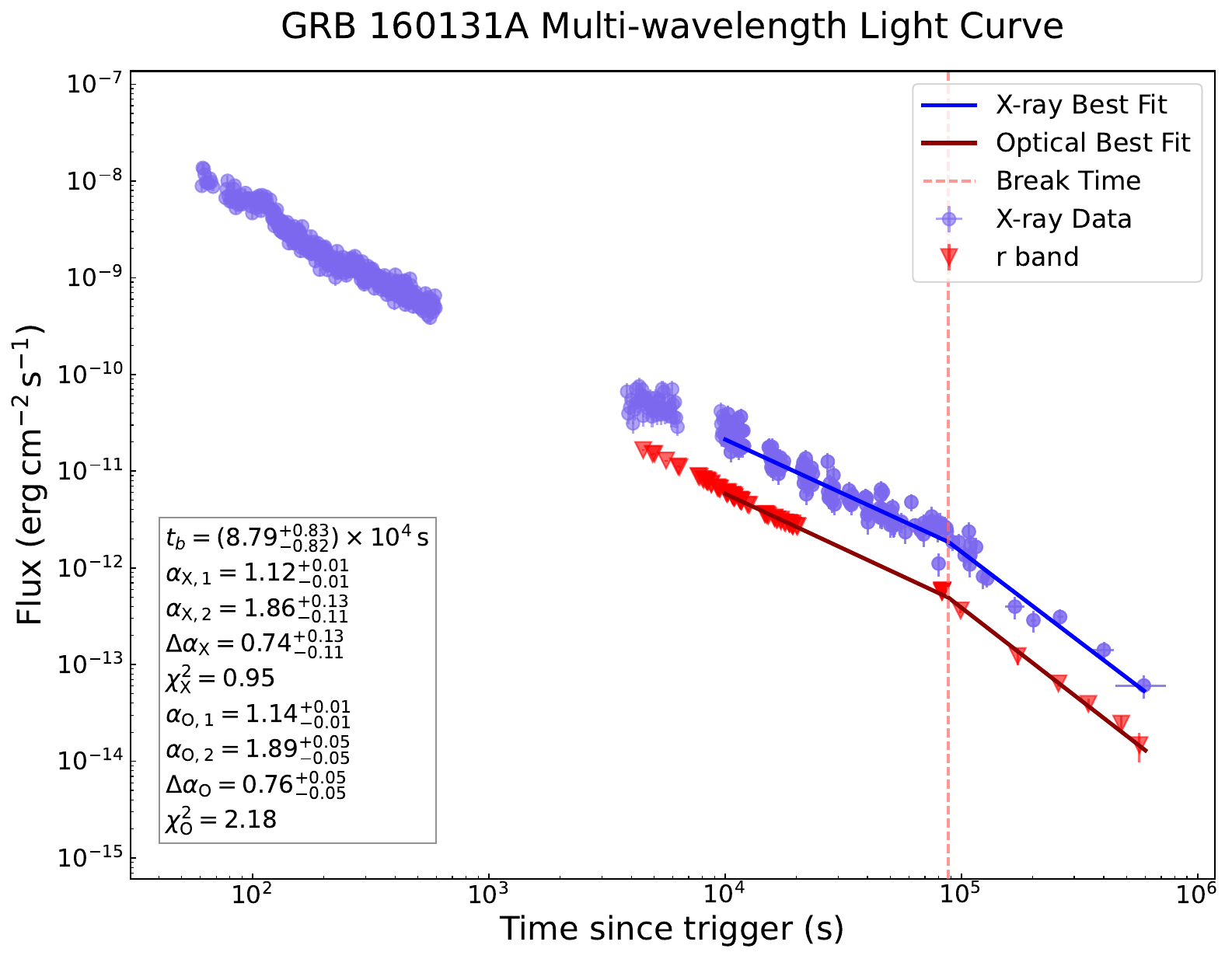}}%
\resizebox{45mm}{!}{\includegraphics[]{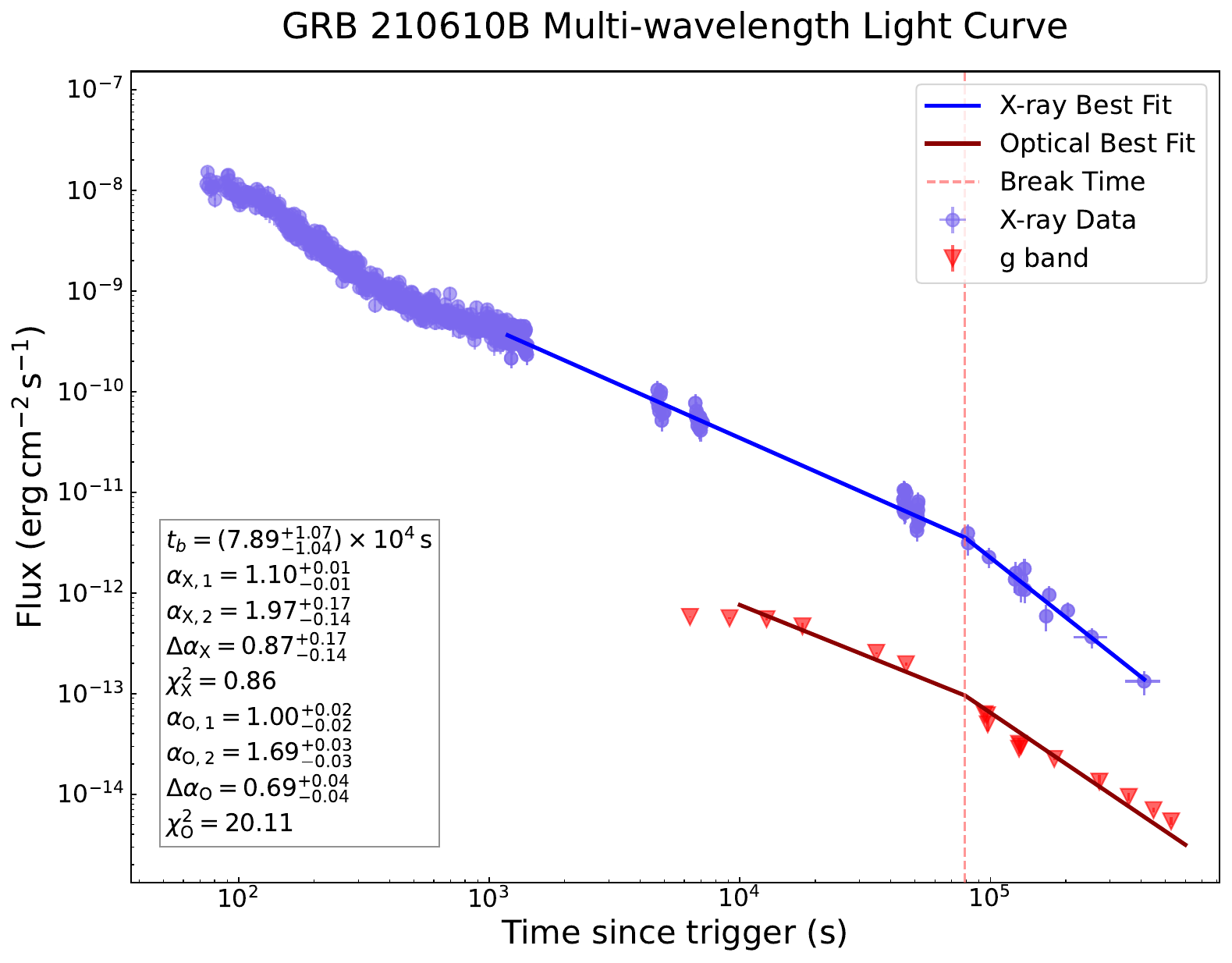}}
\caption{(Continued)}
\end{figure}

\clearpage
\begin{figure*}
\centering
\resizebox{45mm}{!}{\includegraphics[]{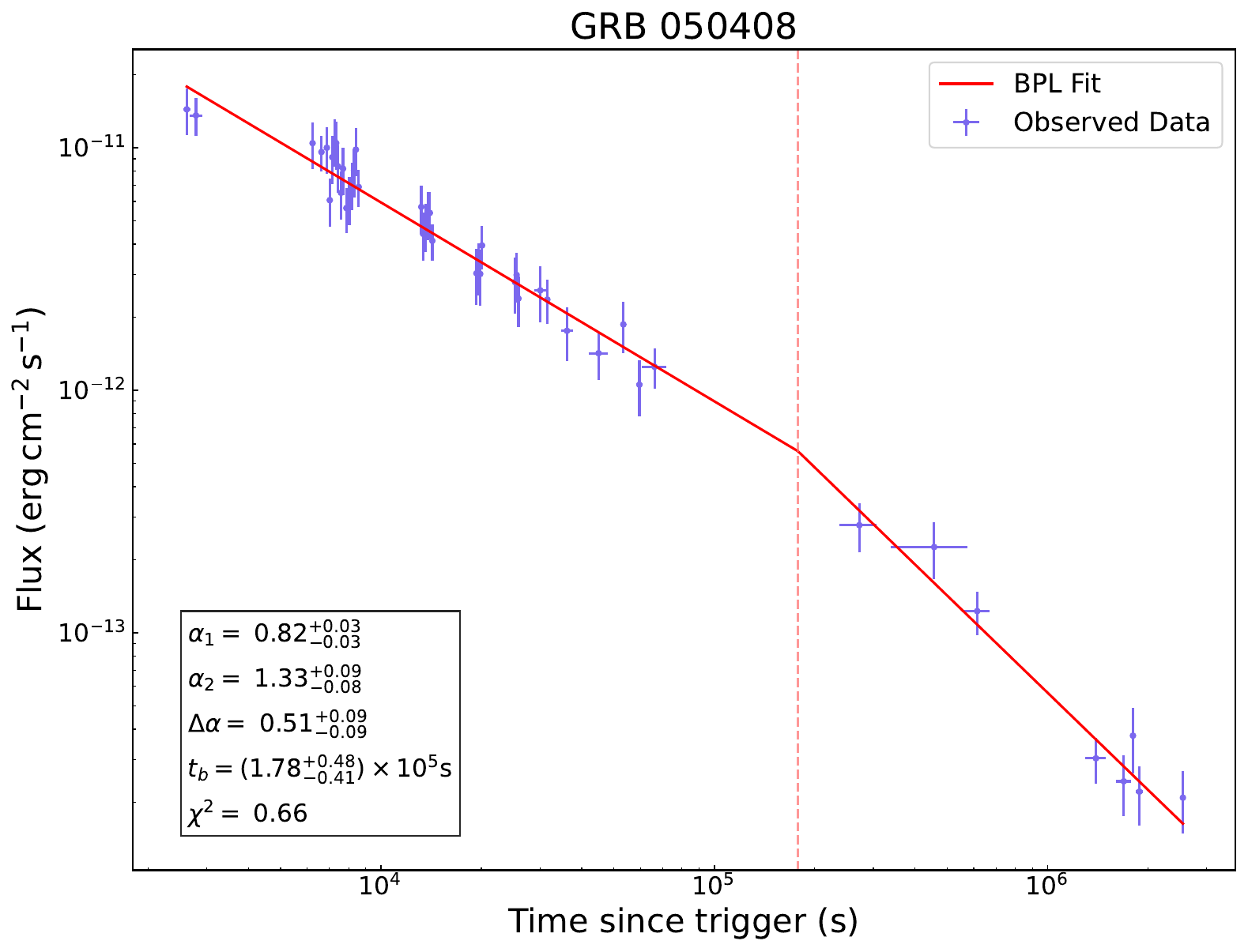}}%
\resizebox{45mm}{!}{\includegraphics[]{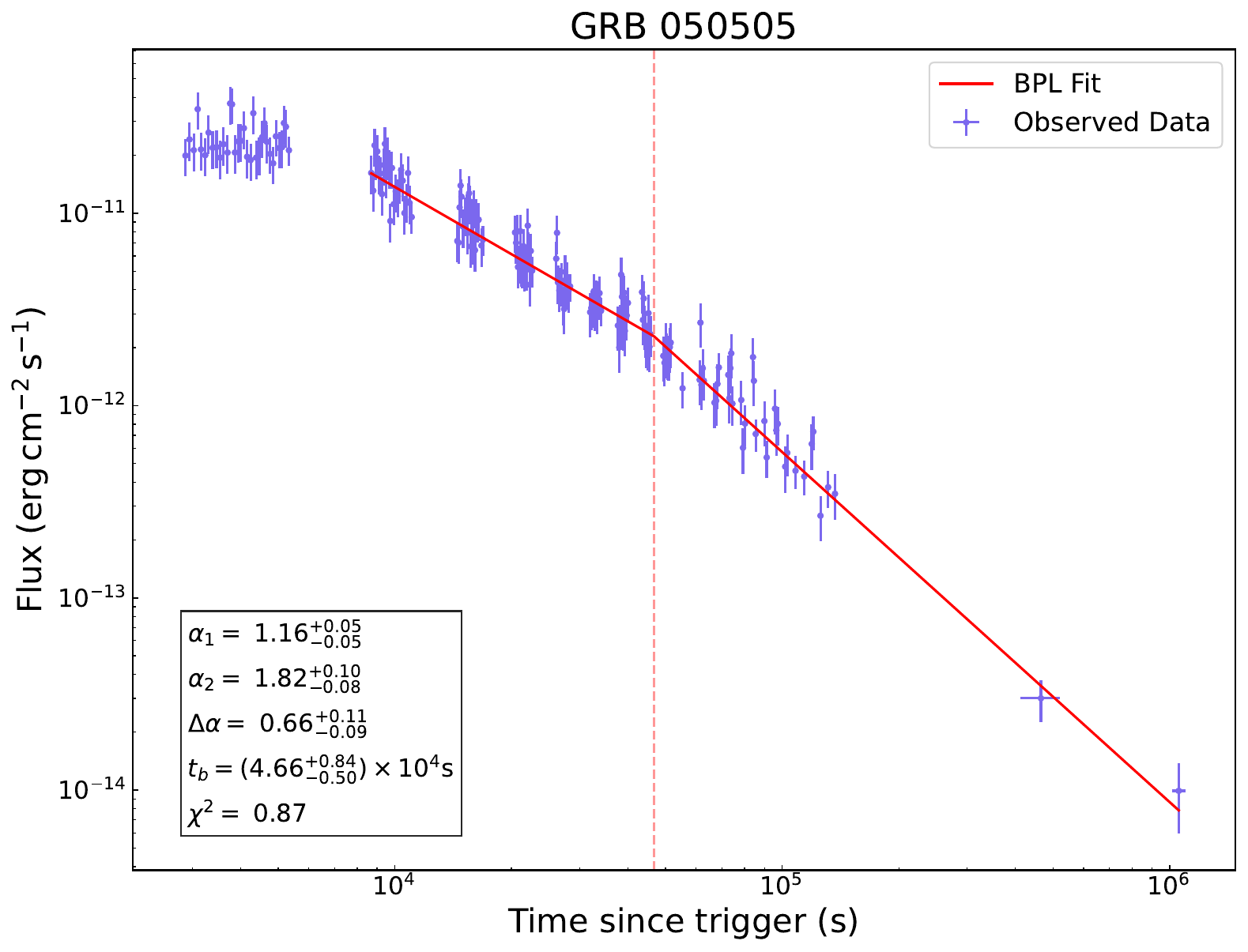}}%
\resizebox{45mm}{!}{\includegraphics[]{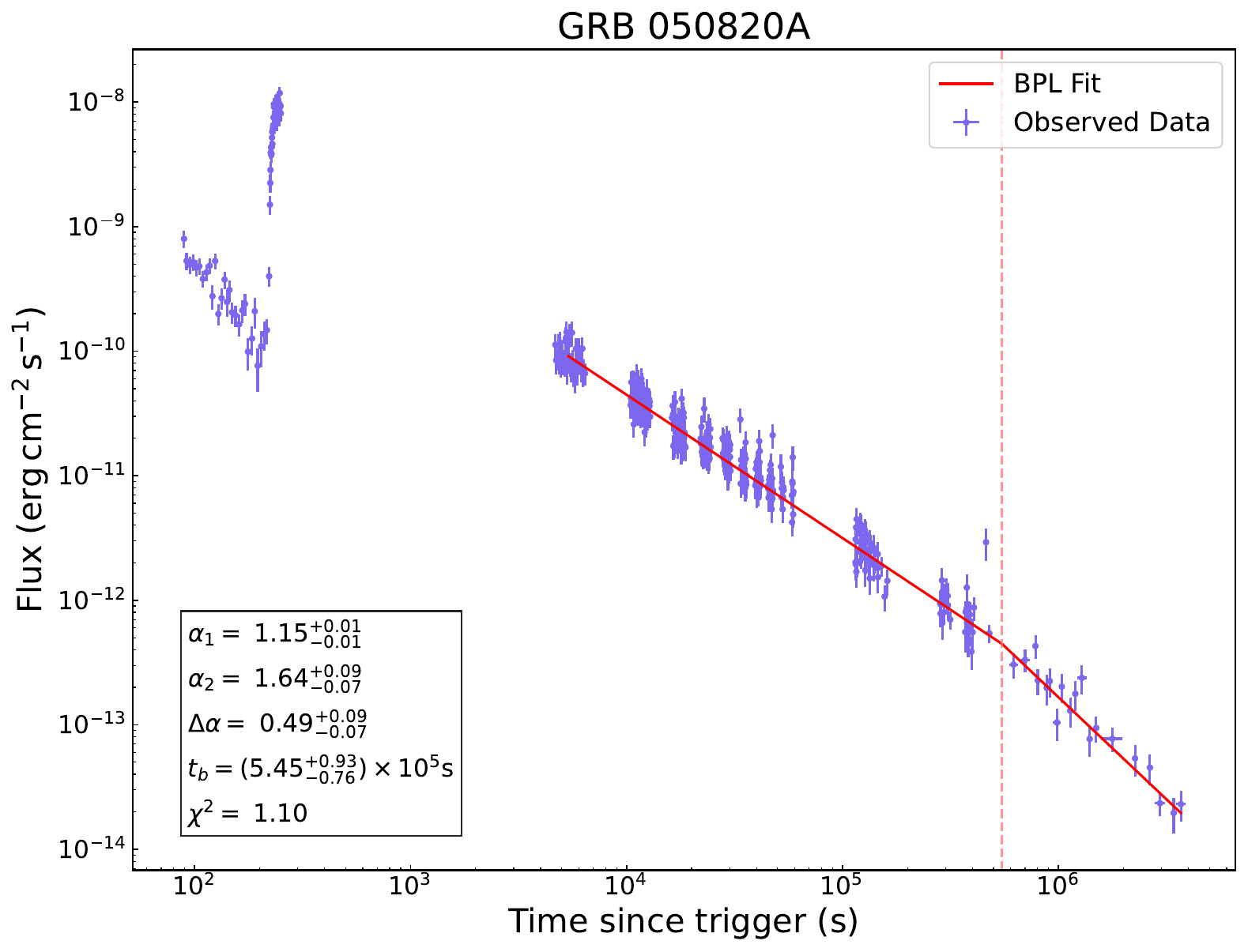}}%
\resizebox{45mm}{!}{\includegraphics[]{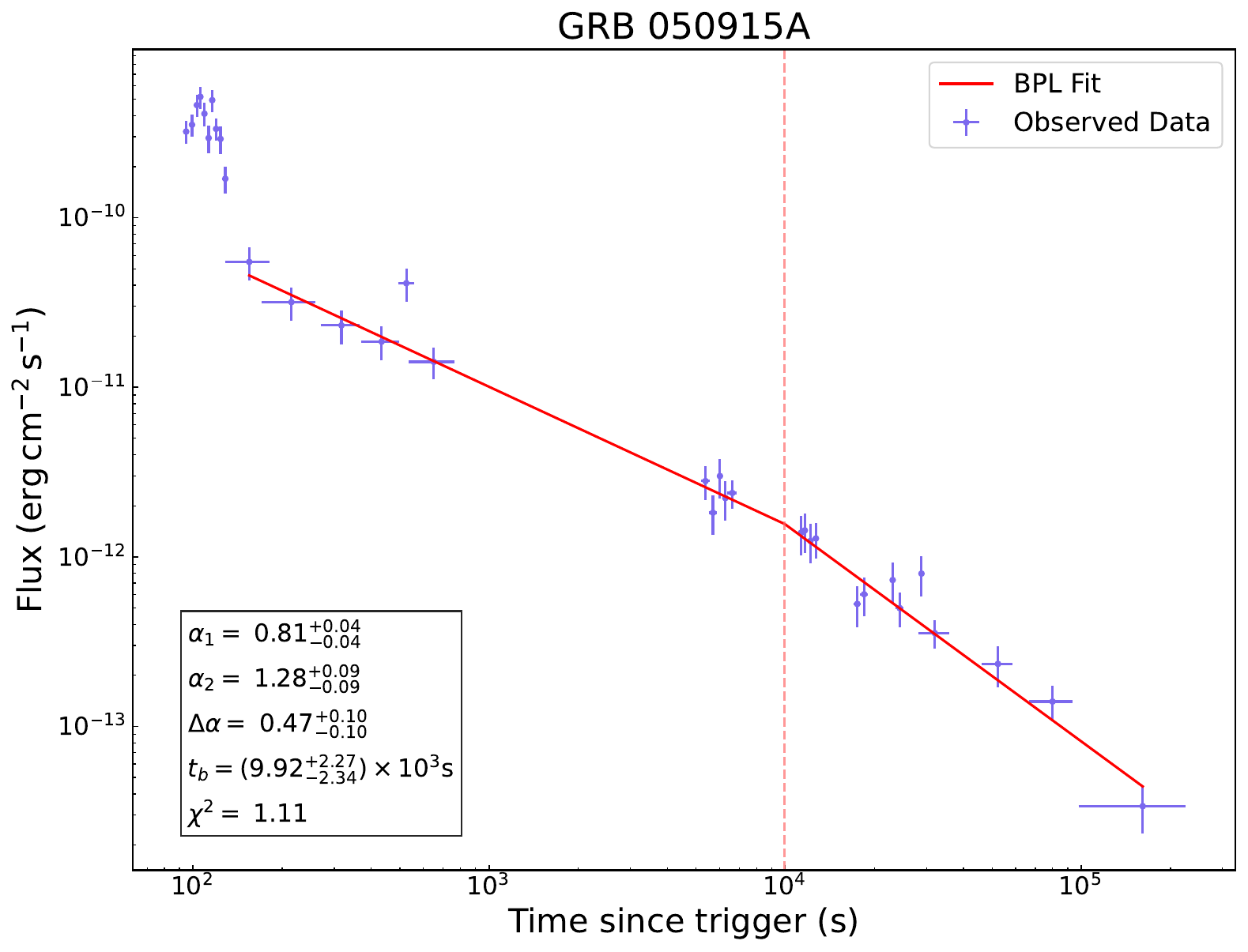}}\\
\resizebox{45mm}{!}{\includegraphics[]{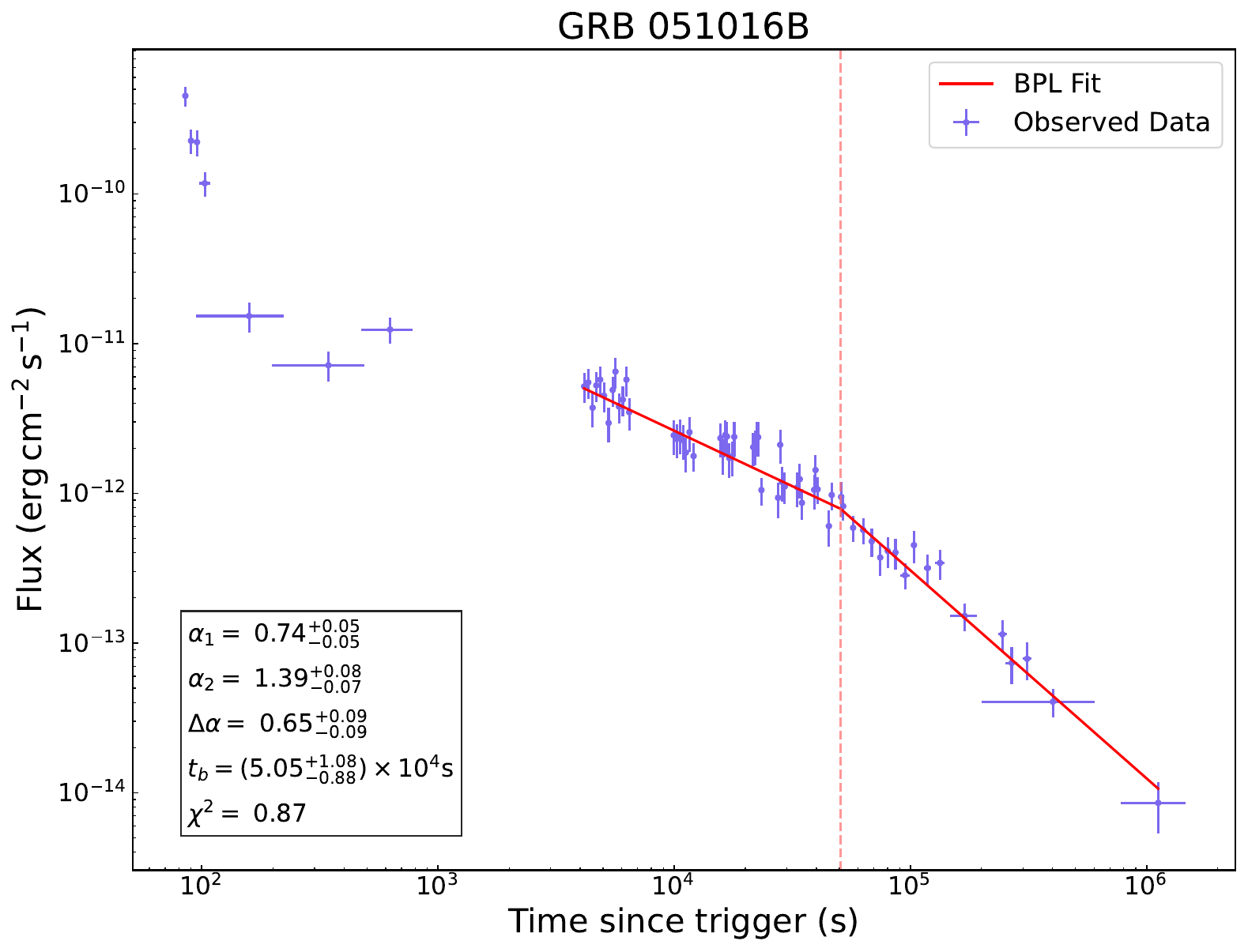}}%
\resizebox{45mm}{!}{\includegraphics[]{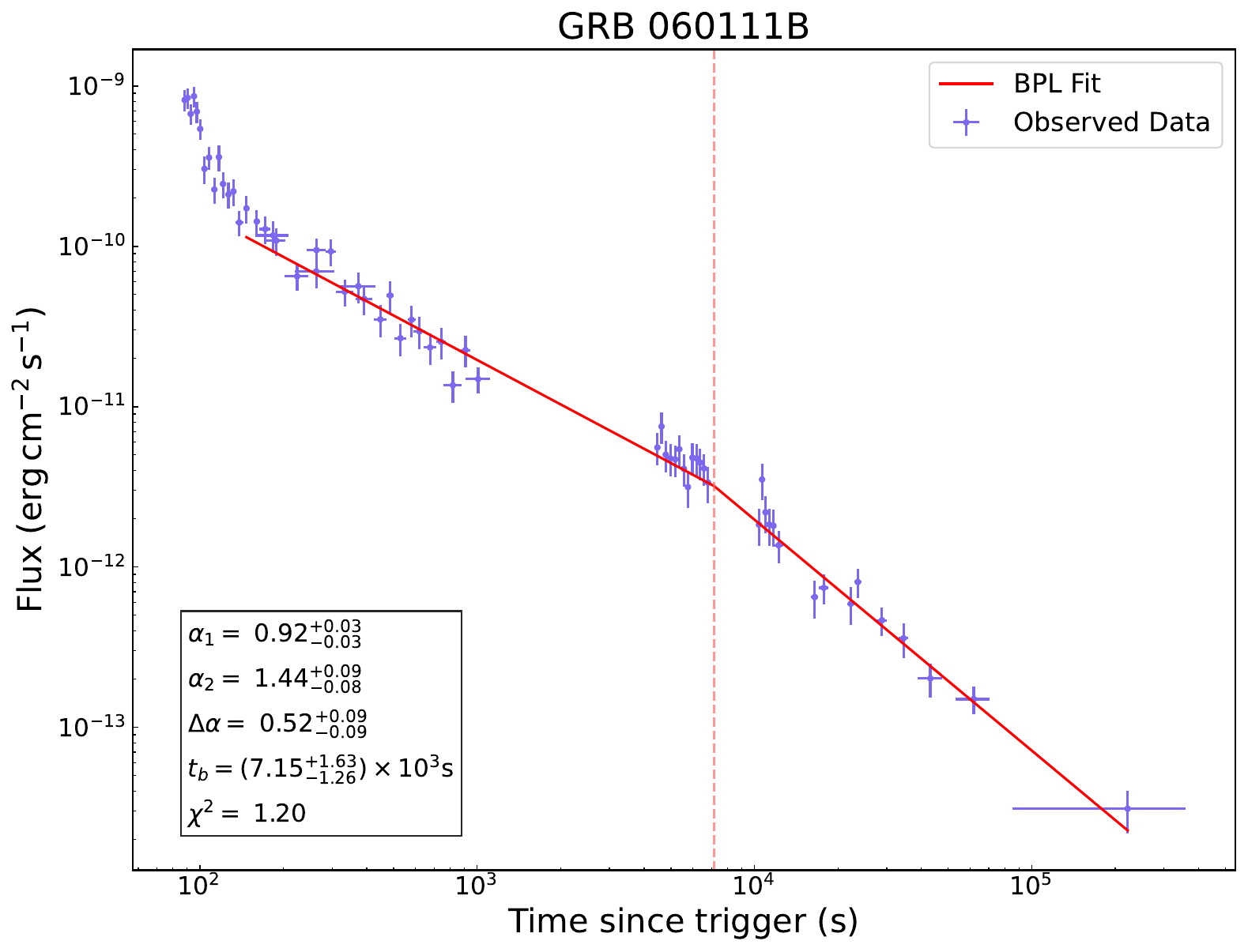}}%
\resizebox{45mm}{!}{\includegraphics[]{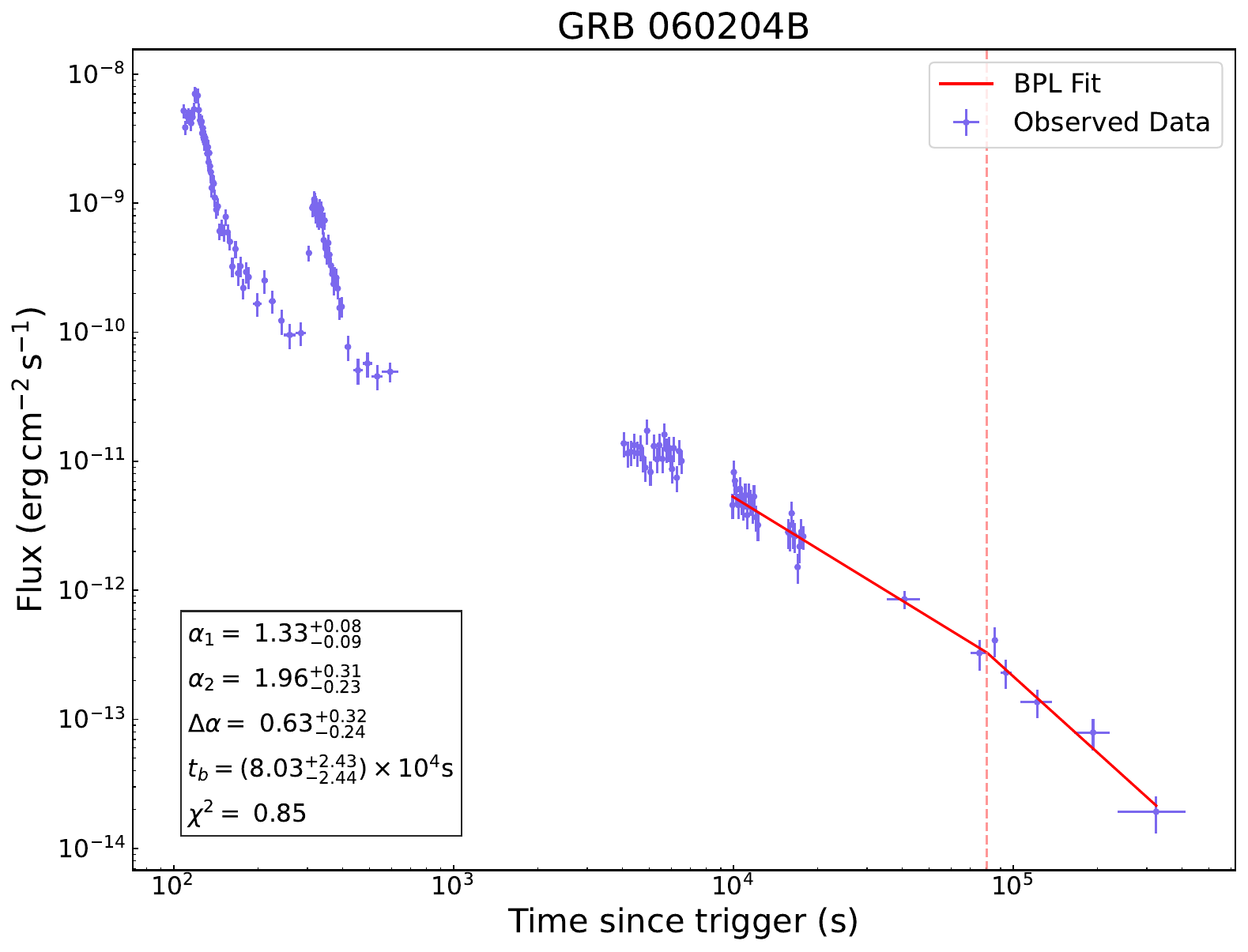}}%
\resizebox{45mm}{!}{\includegraphics[]{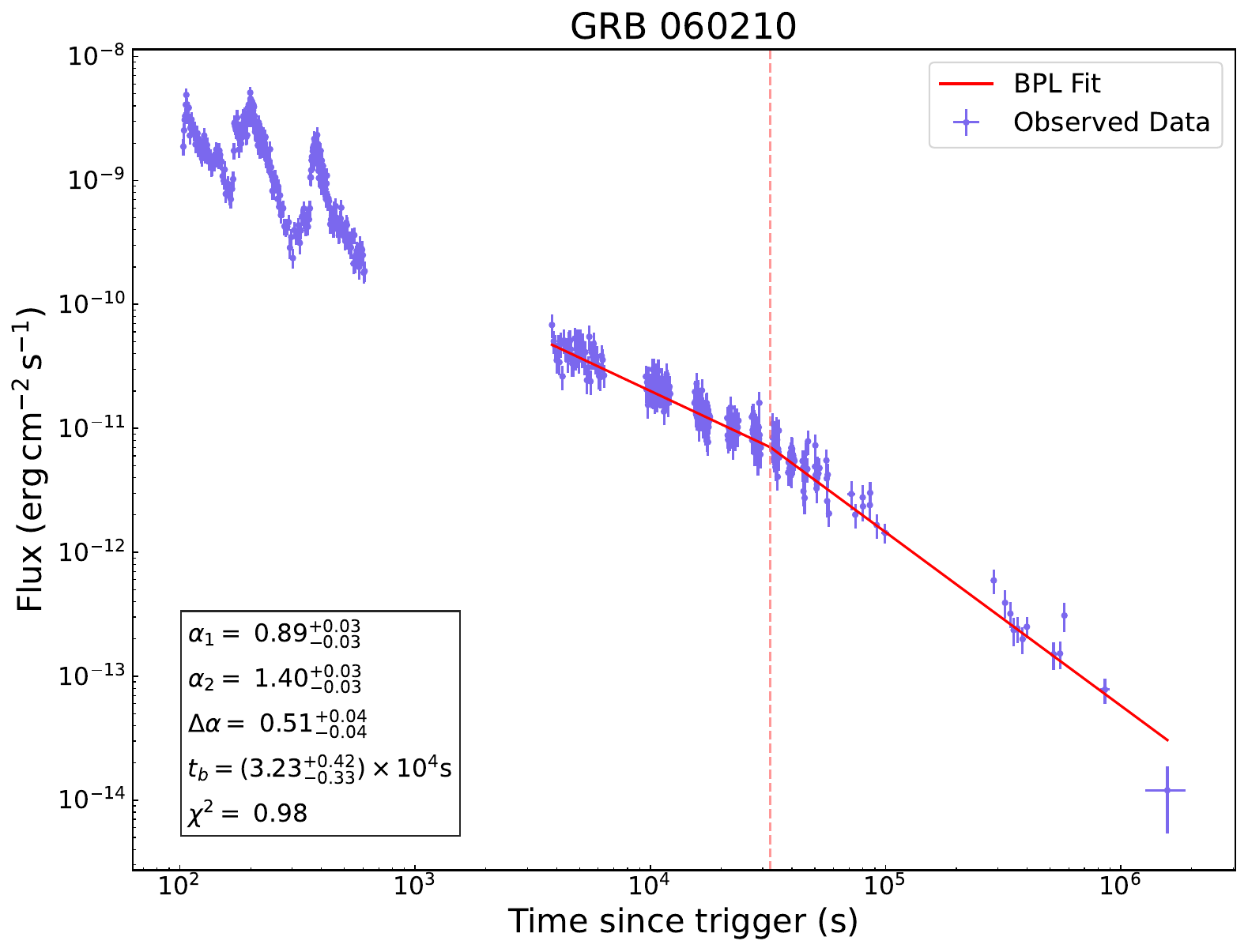}}\\
\resizebox{45mm}{!}{\includegraphics[]{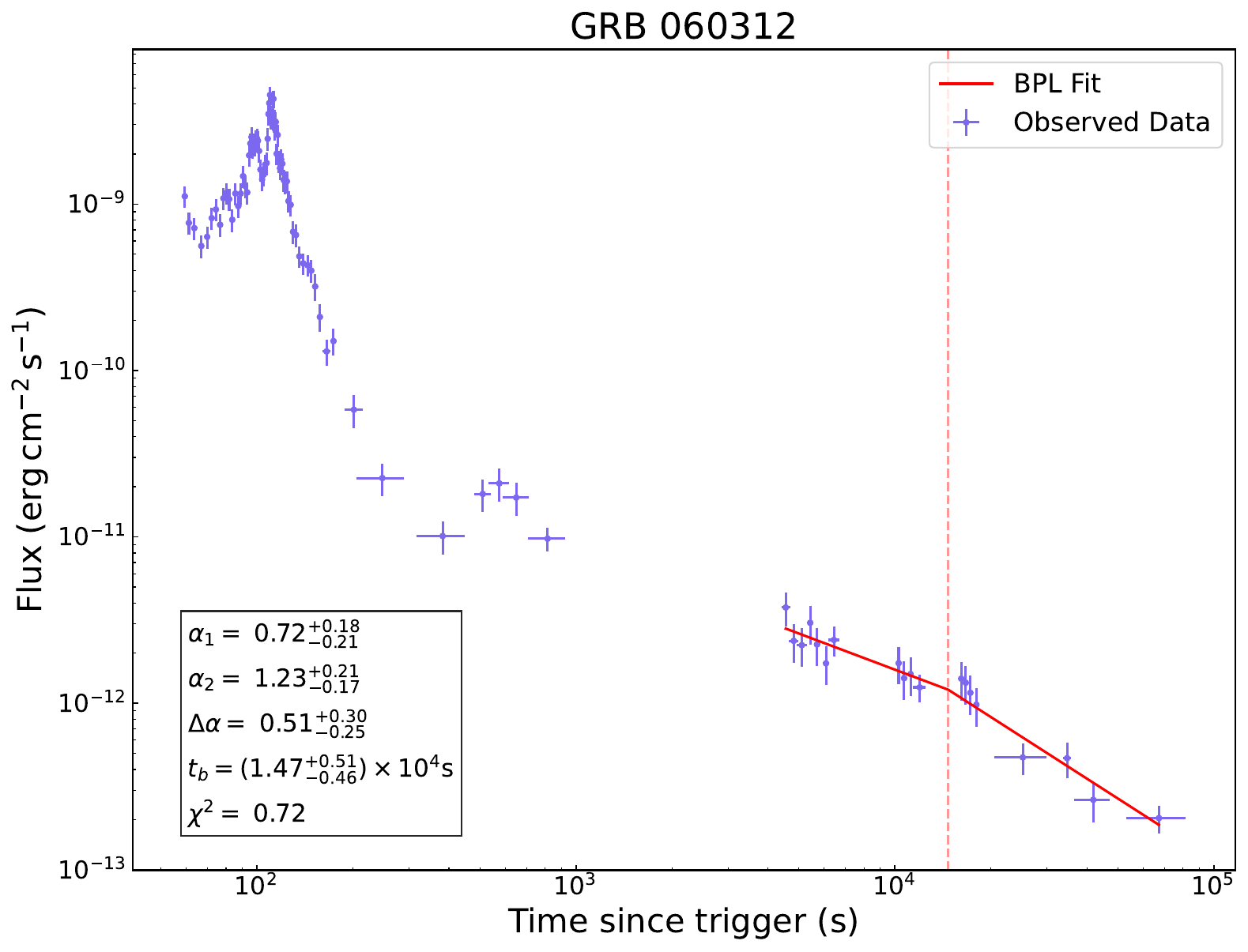}}%
\resizebox{45mm}{!}{\includegraphics[]{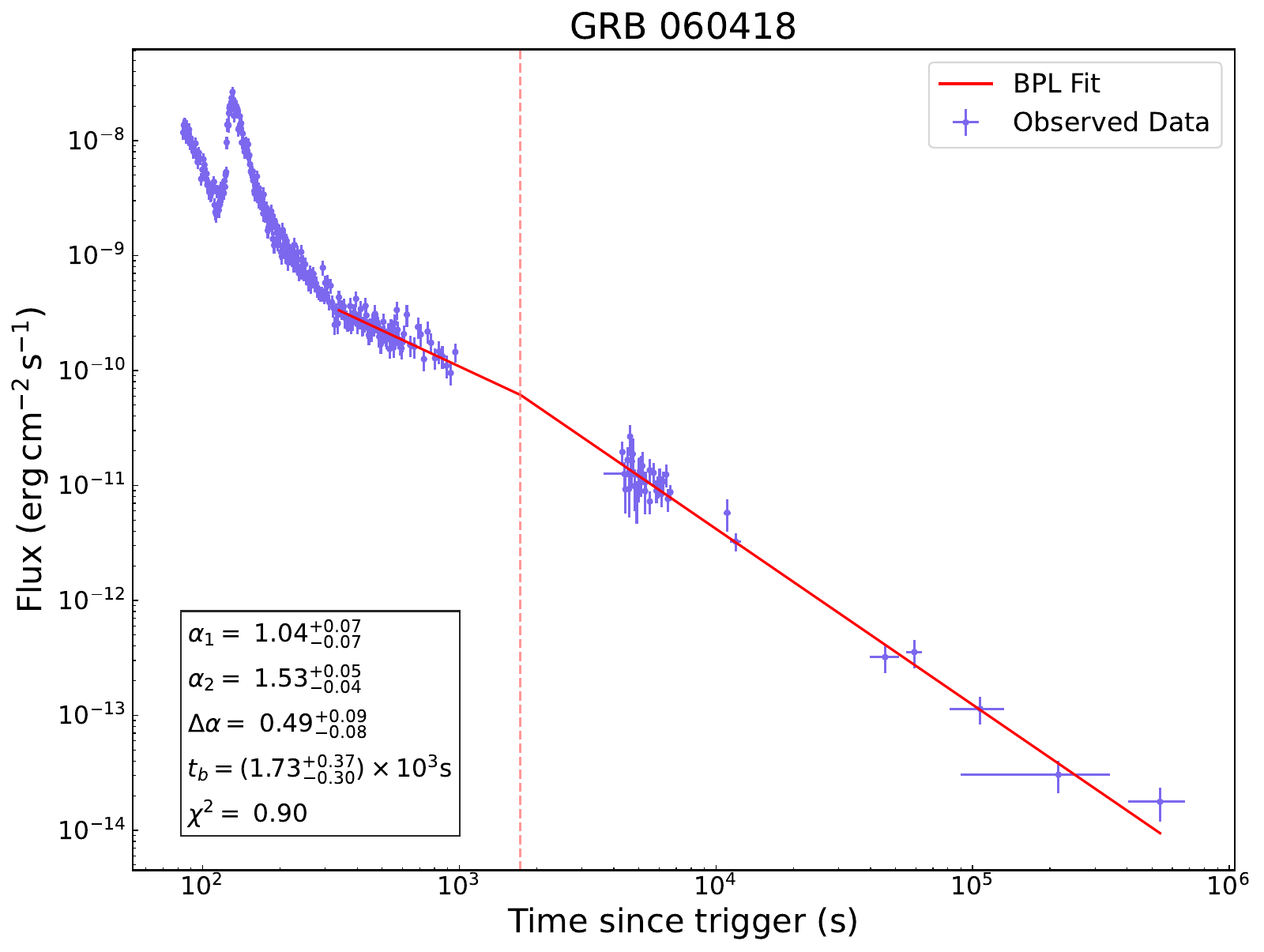}}%
\resizebox{45mm}{!}{\includegraphics[]{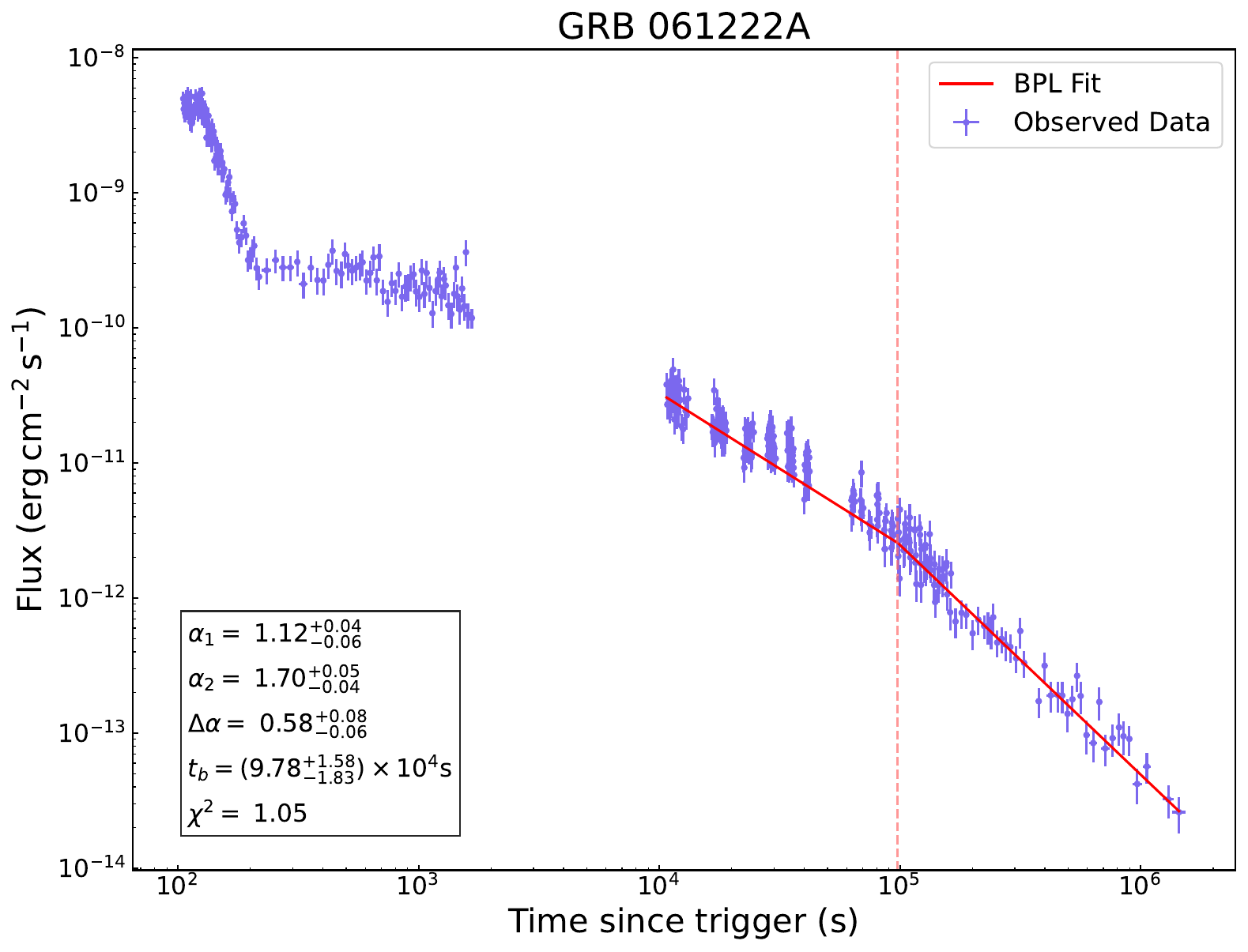}}%
\resizebox{45mm}{!}{\includegraphics[]{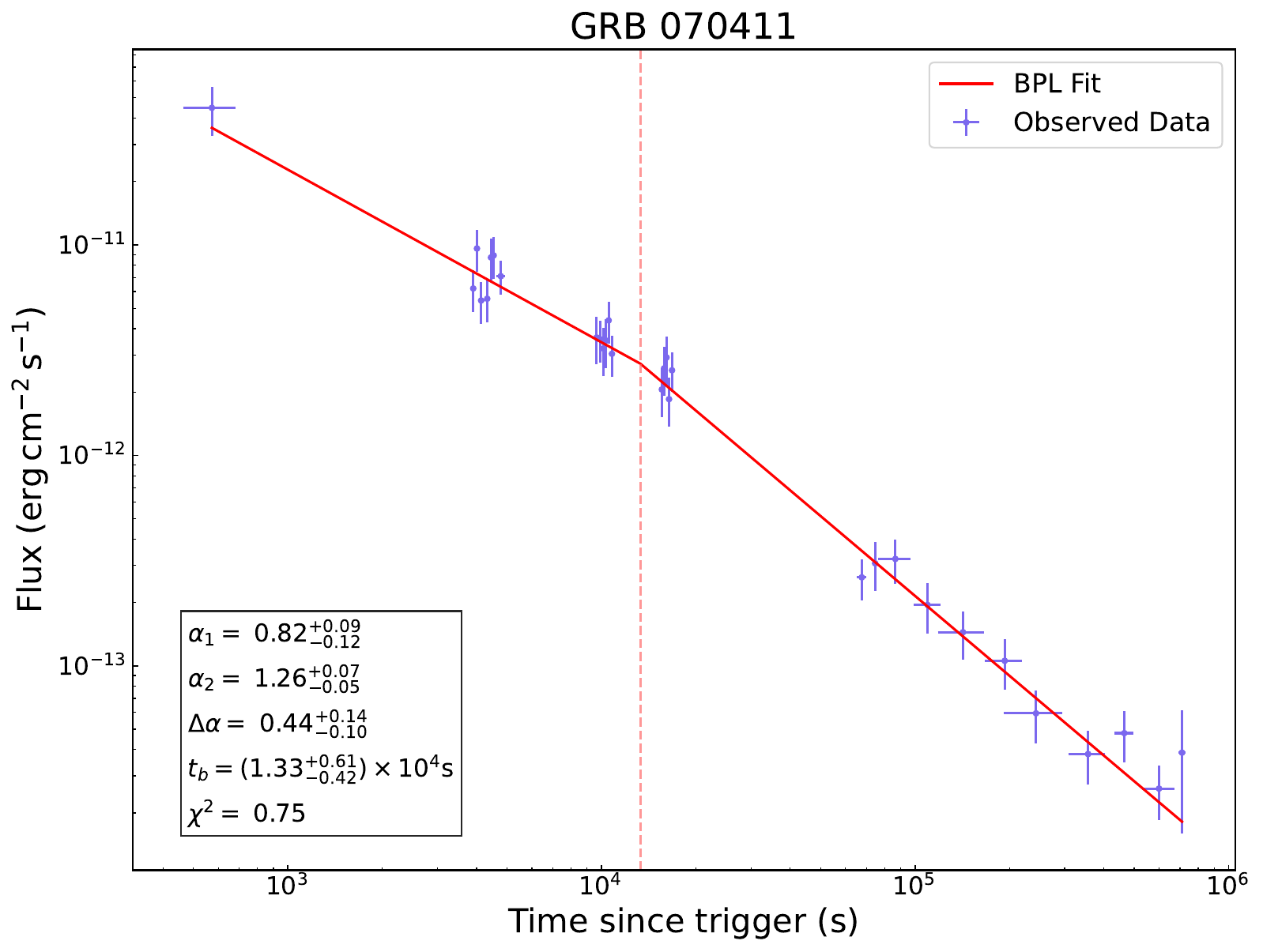}}\\
\resizebox{45mm}{!}{\includegraphics[]{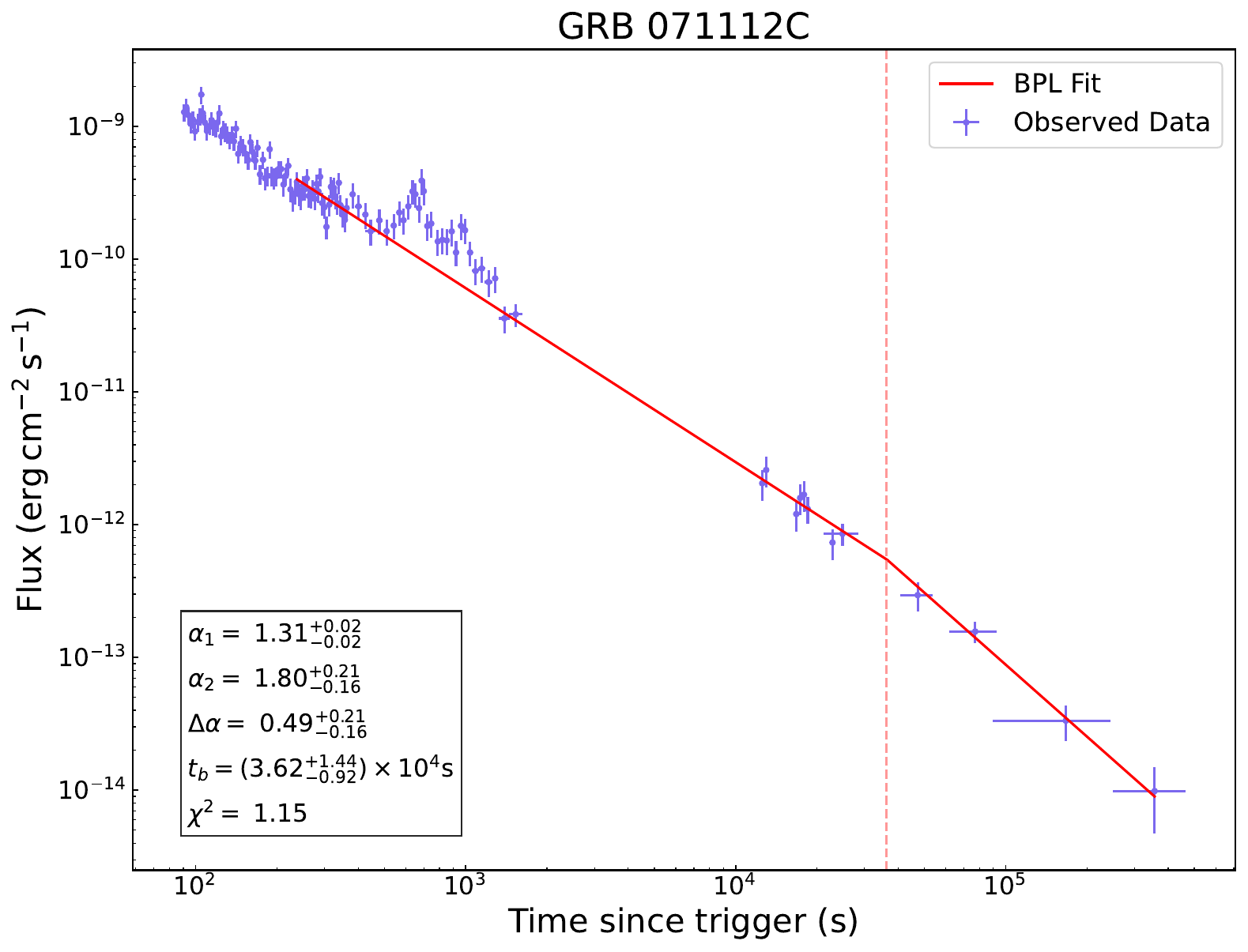}}%
\resizebox{45mm}{!}{\includegraphics[]{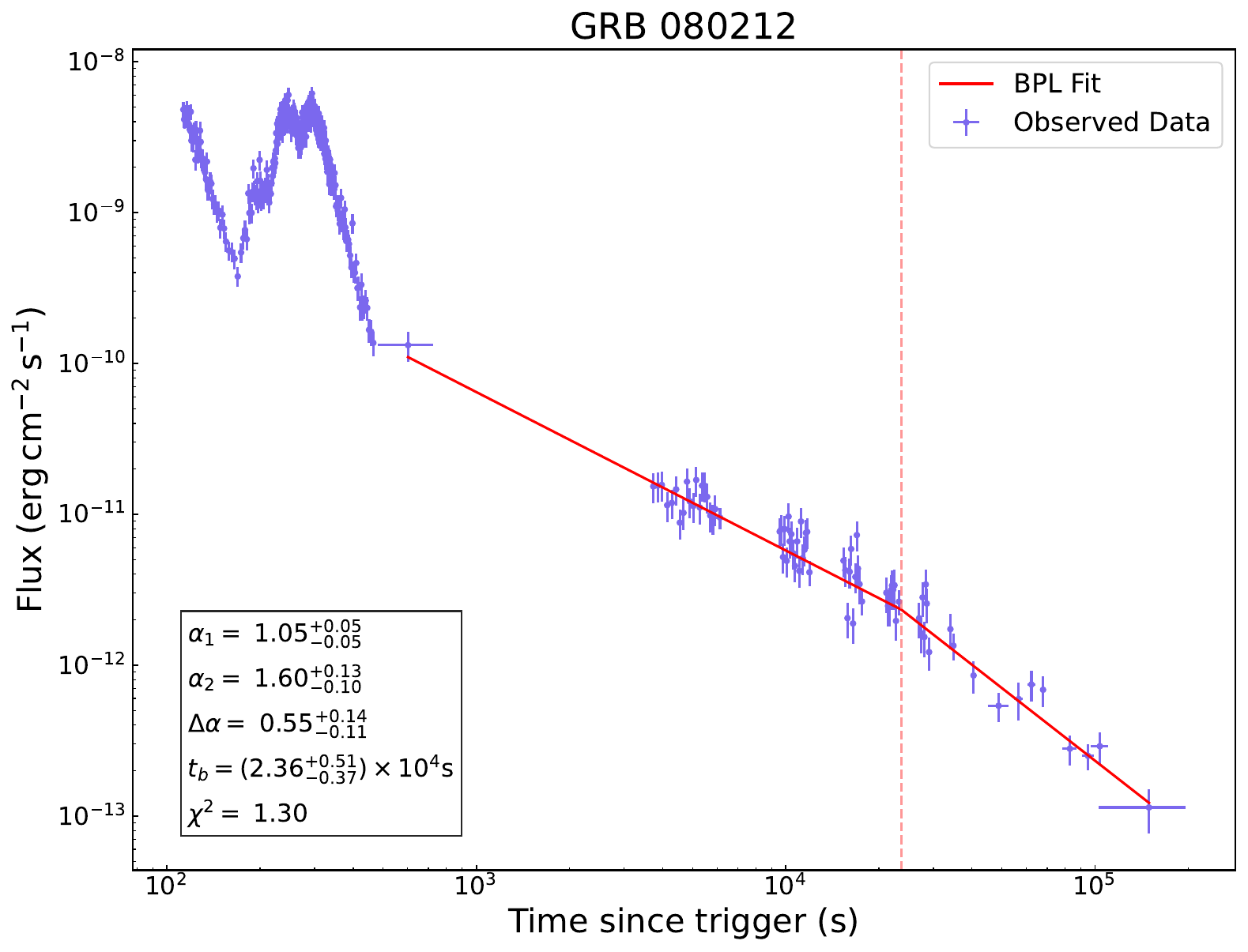}}%
\resizebox{45mm}{!}{\includegraphics[]{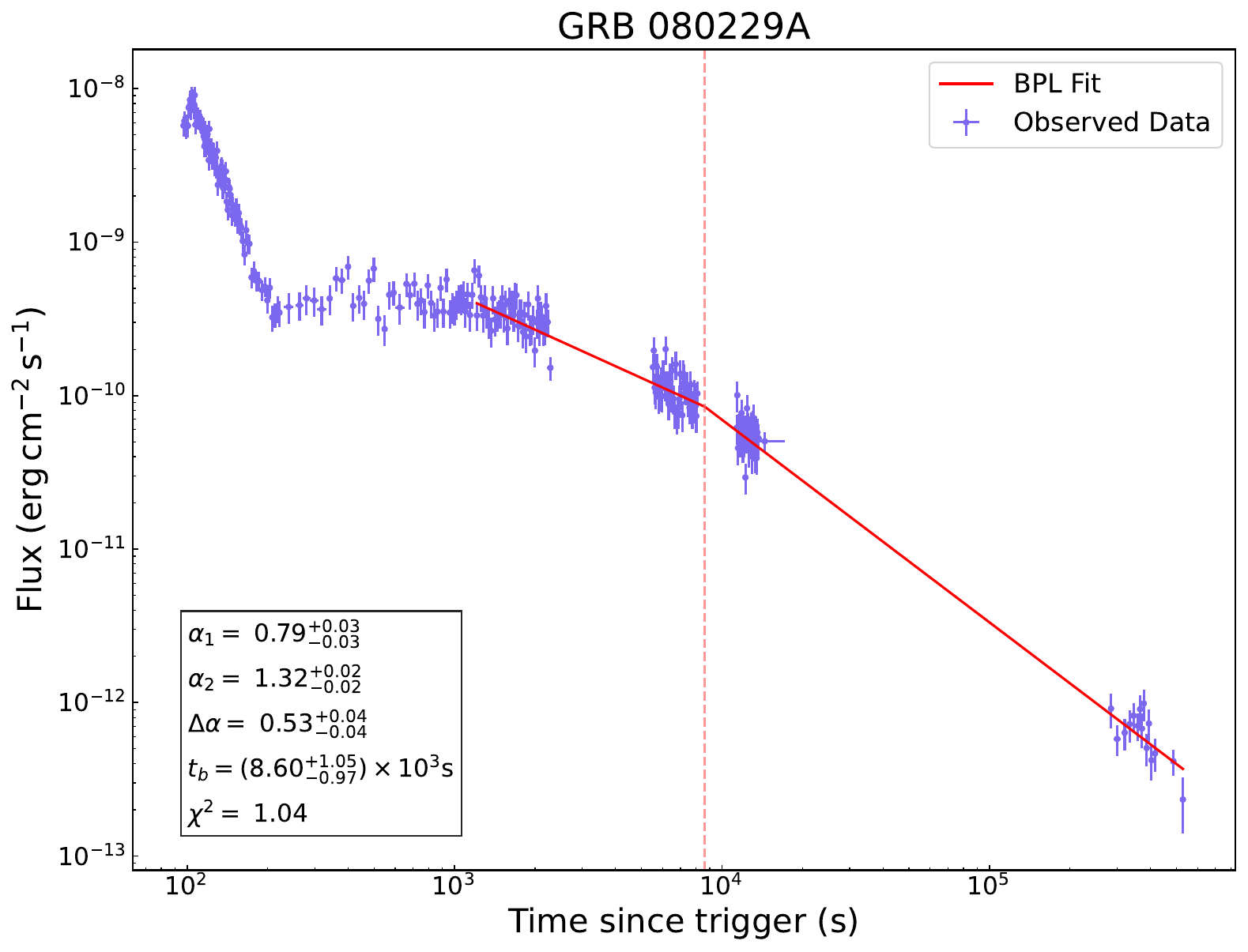}}%
\resizebox{45mm}{!}{\includegraphics[]{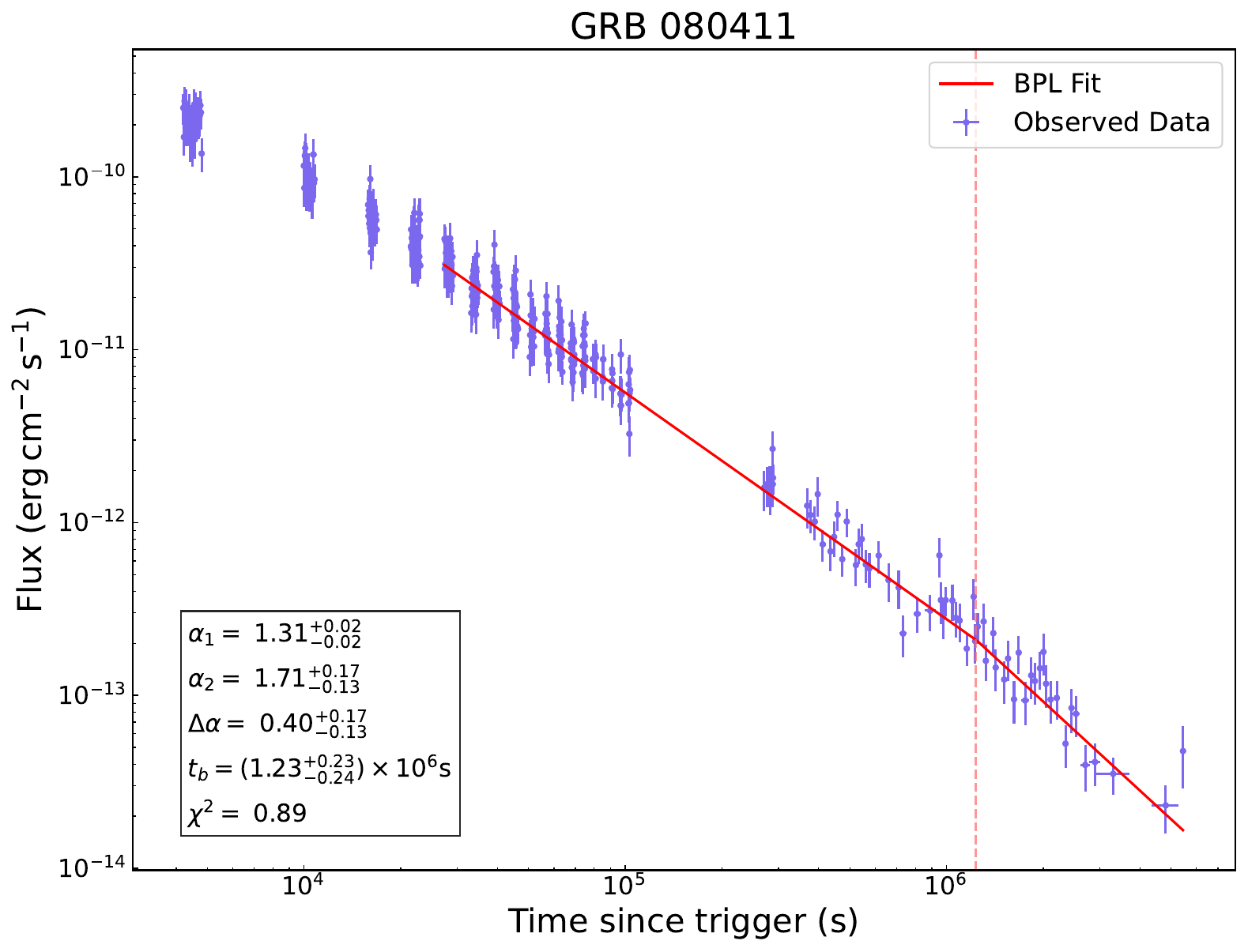}}\\
\resizebox{45mm}{!}{\includegraphics[]{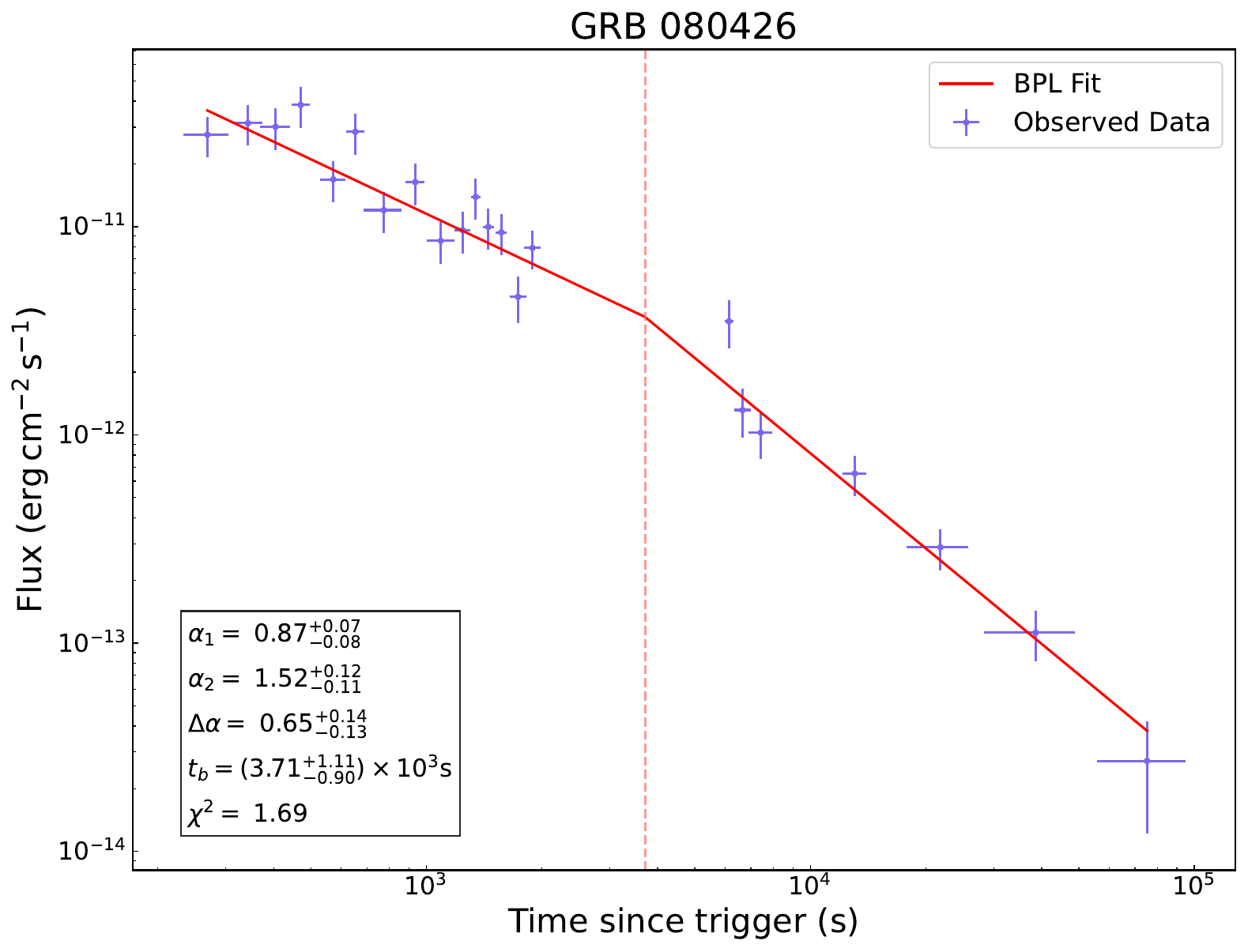}}%
\resizebox{45mm}{!}{\includegraphics[]{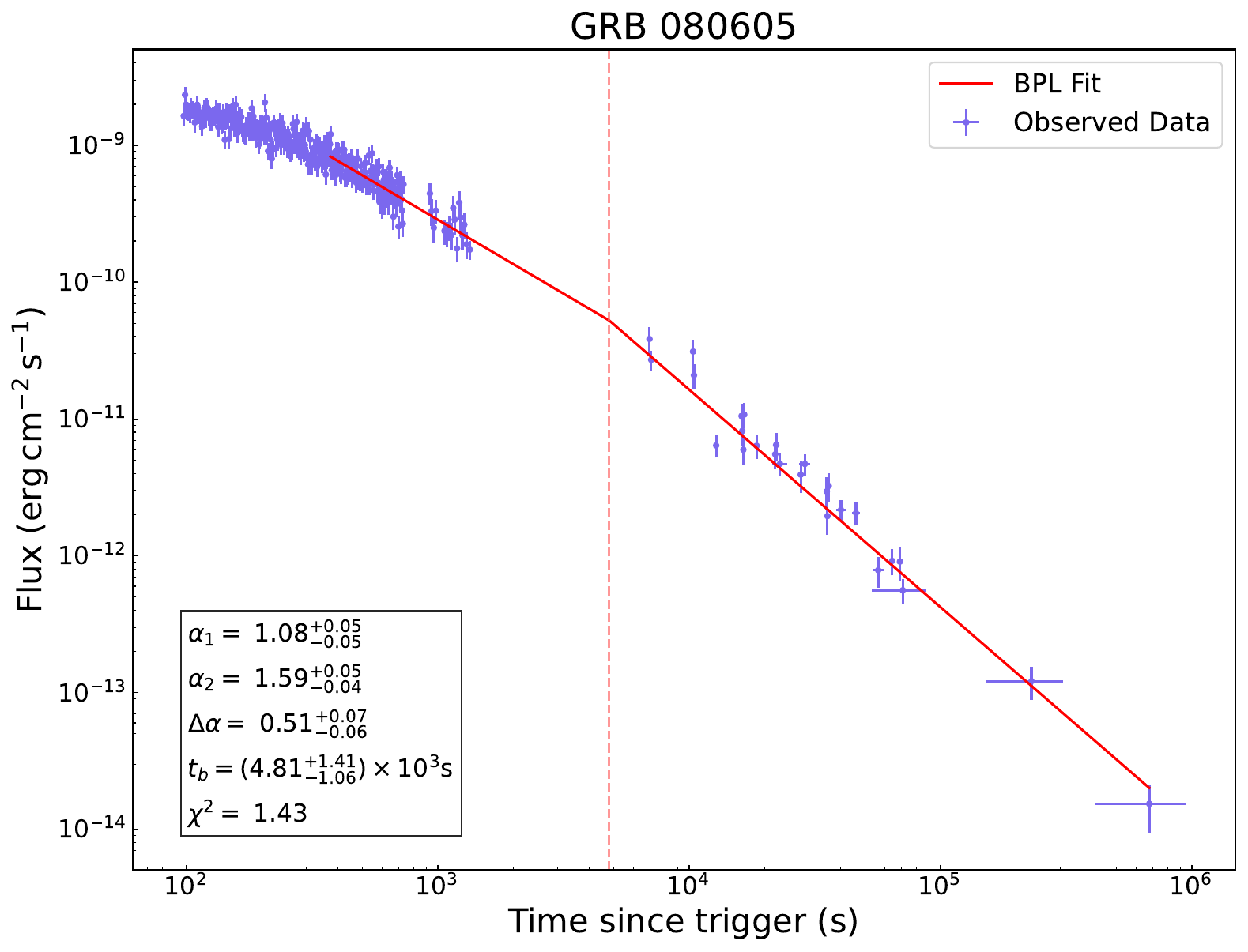}}%
\resizebox{45mm}{!}{\includegraphics[]{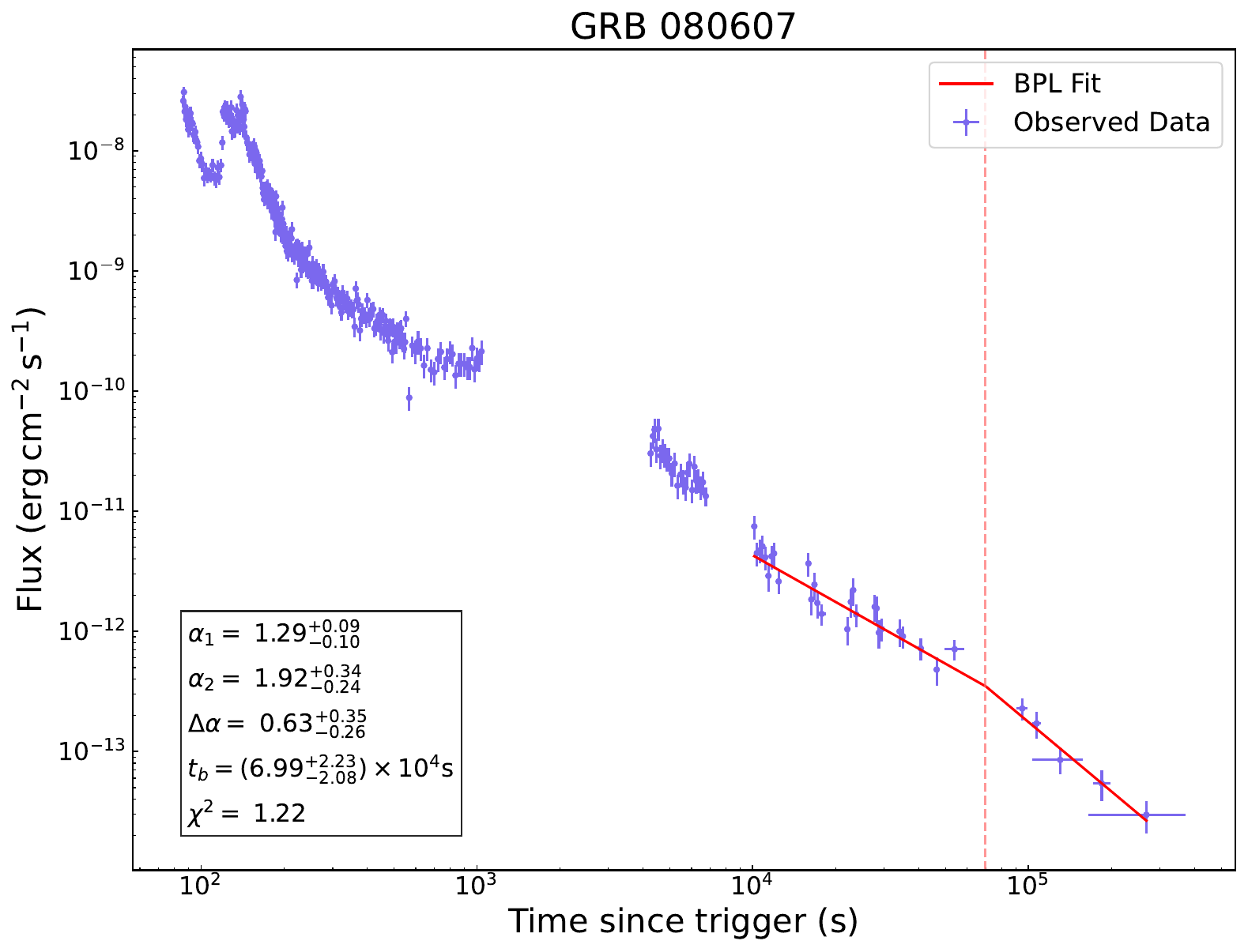}}%
\resizebox{45mm}{!}{\includegraphics[]{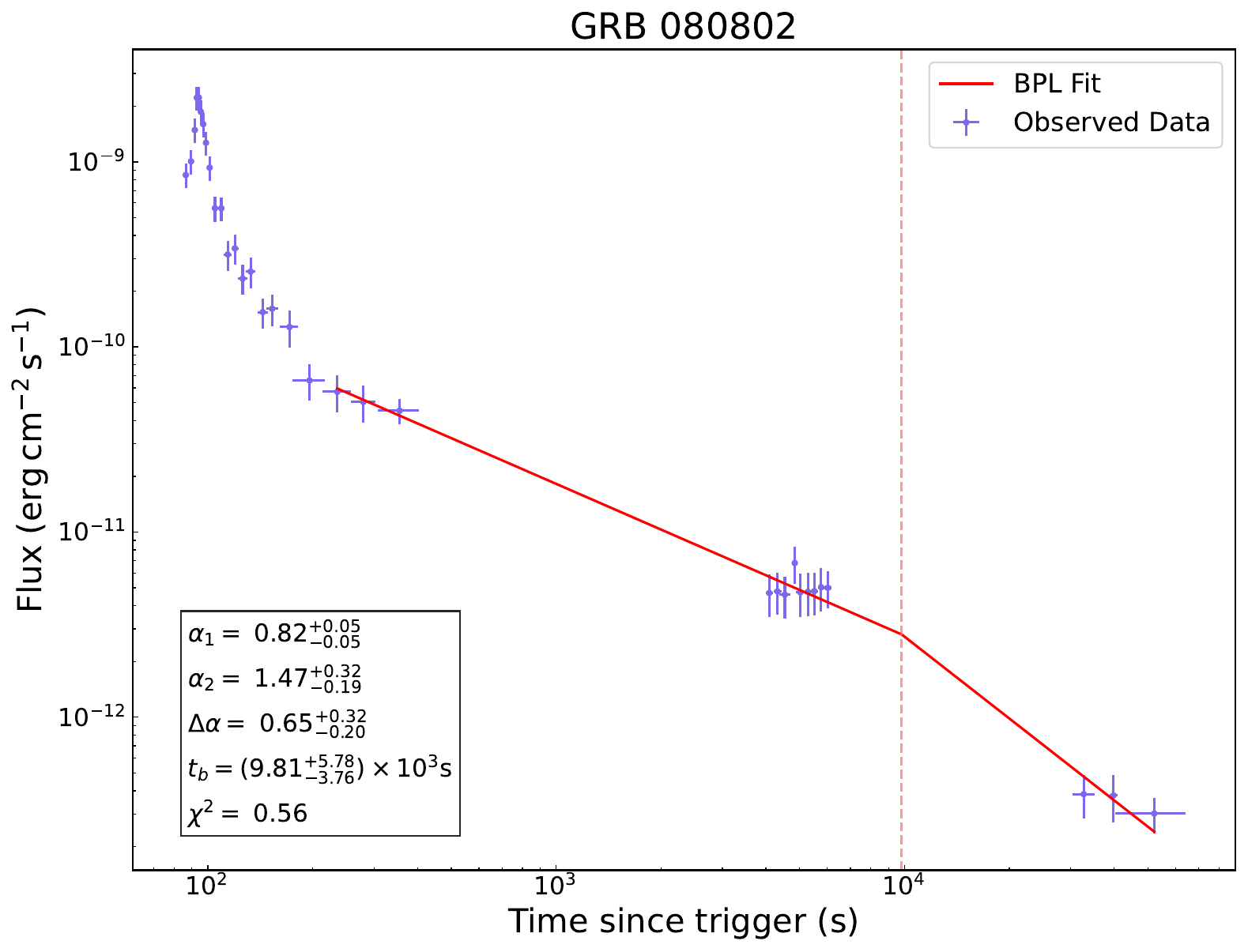}}\\
\resizebox{45mm}{!}{\includegraphics[]{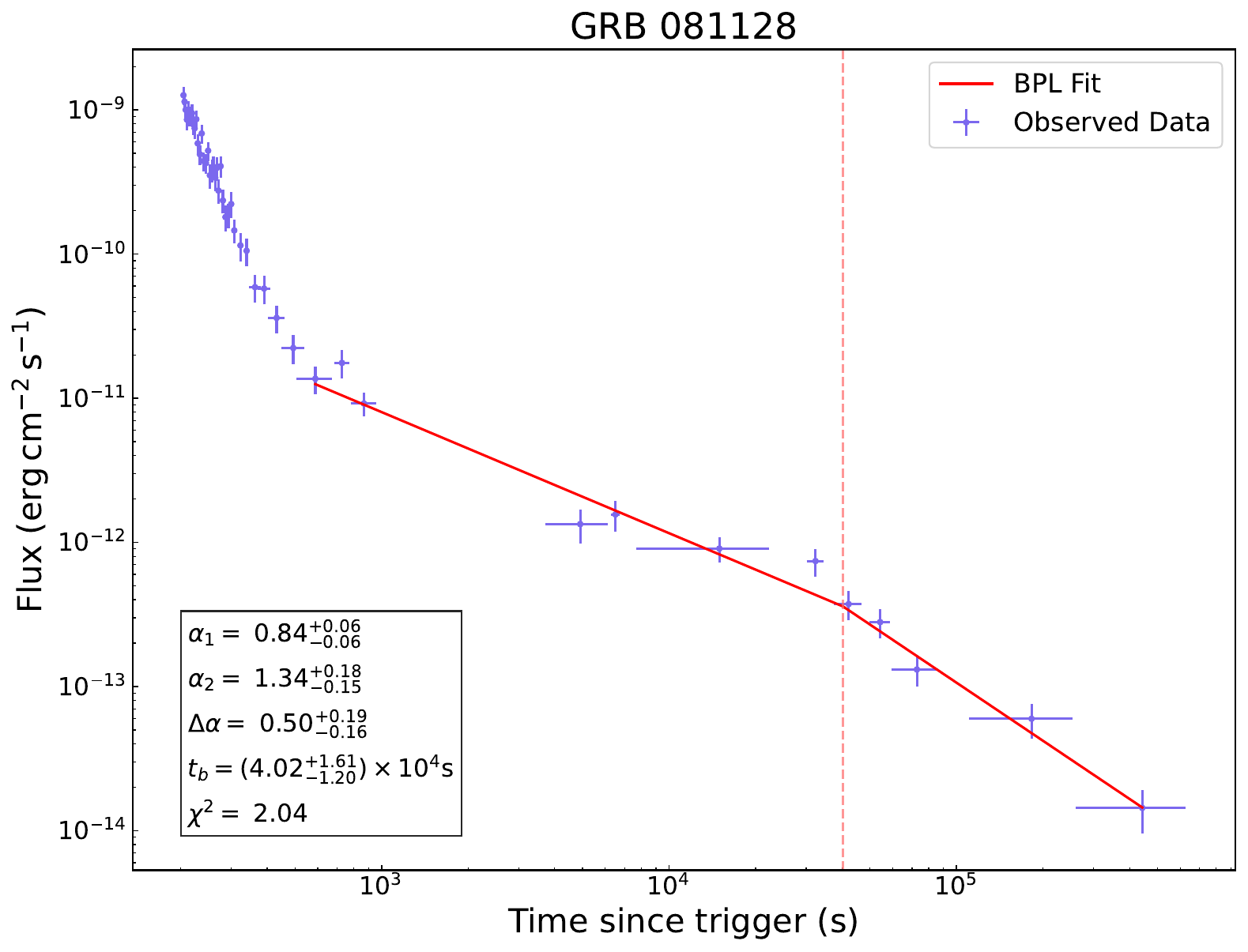}}%
\resizebox{45mm}{!}{\includegraphics[]{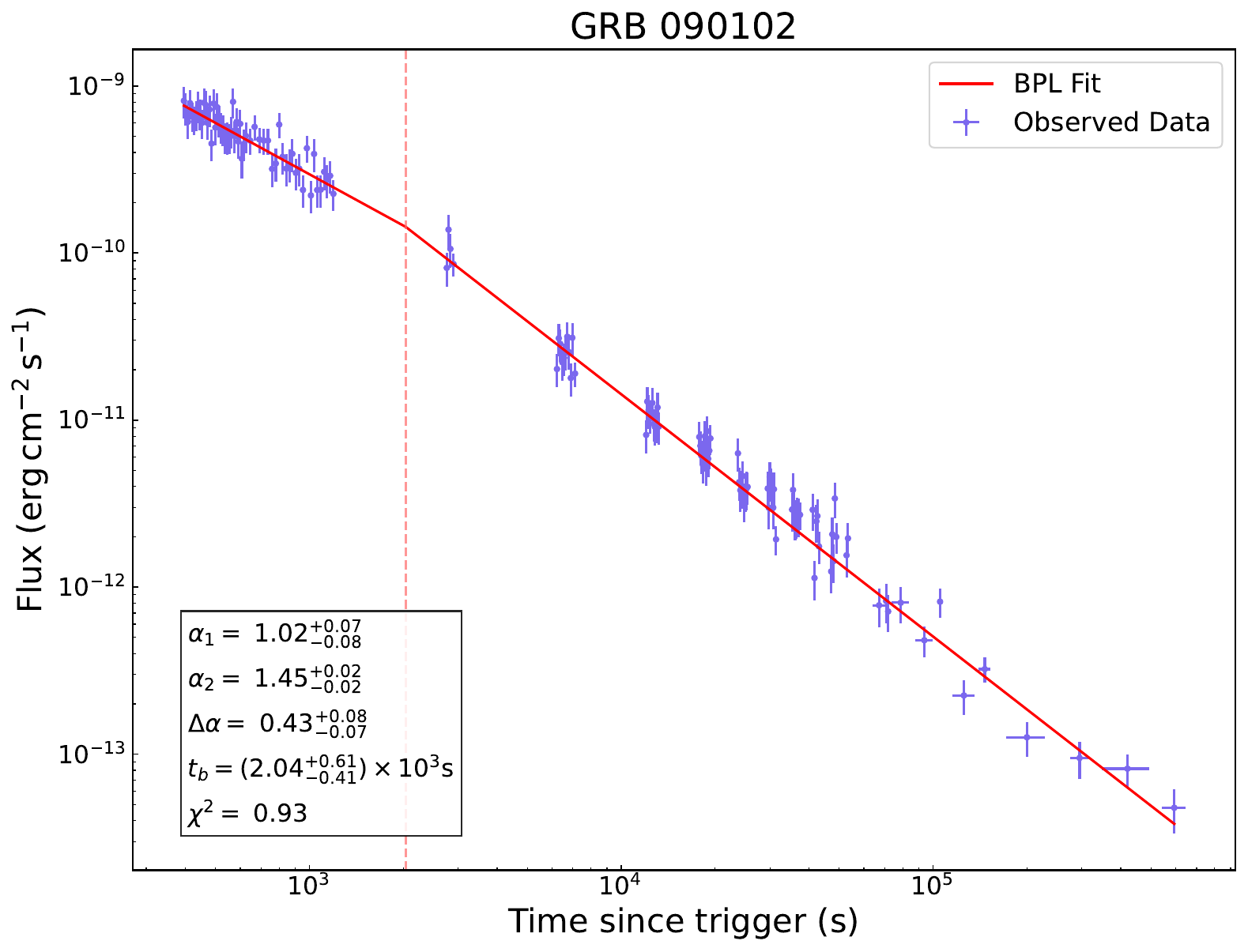}}%
\resizebox{45mm}{!}{\includegraphics[]{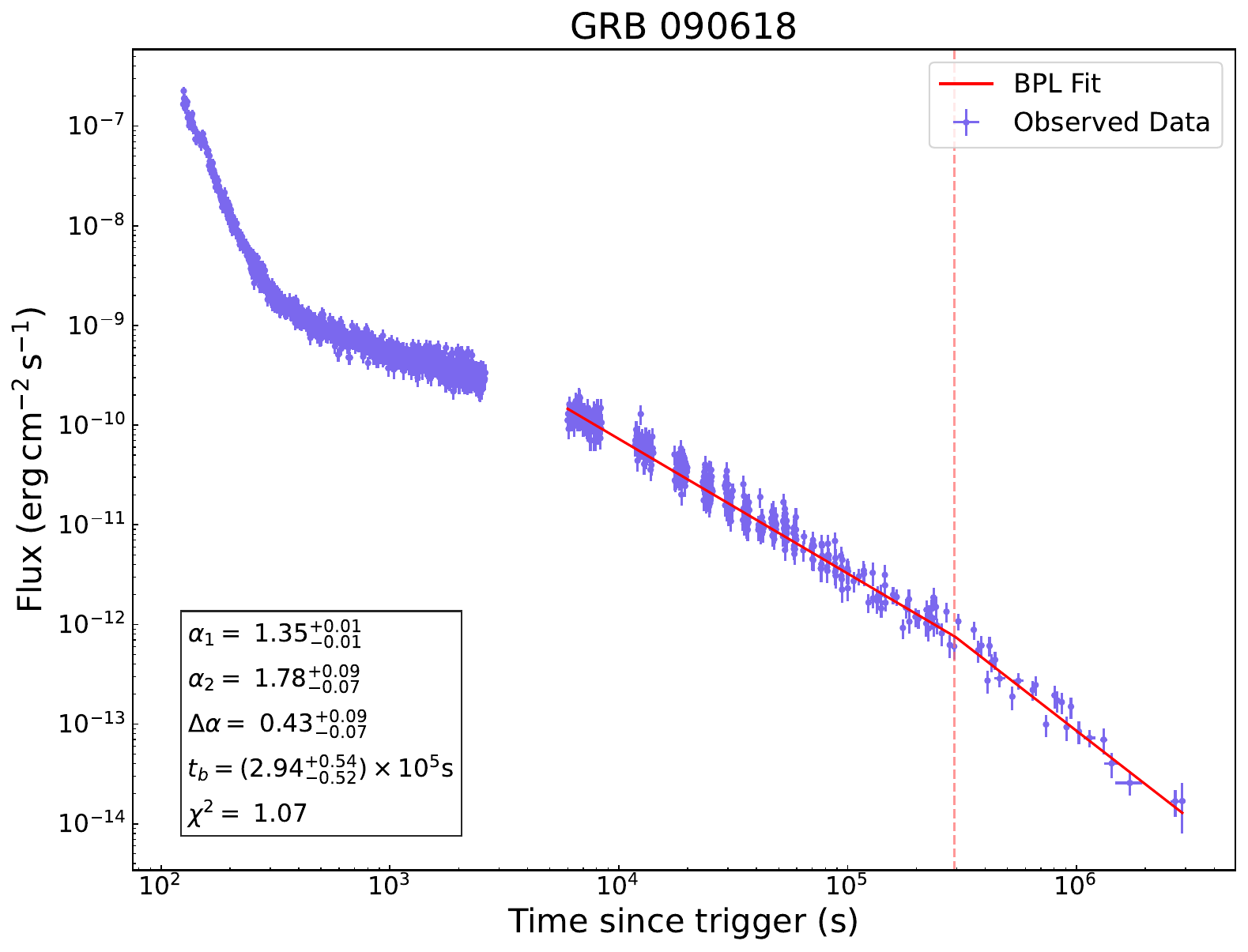}}%
\resizebox{45mm}{!}{\includegraphics[]{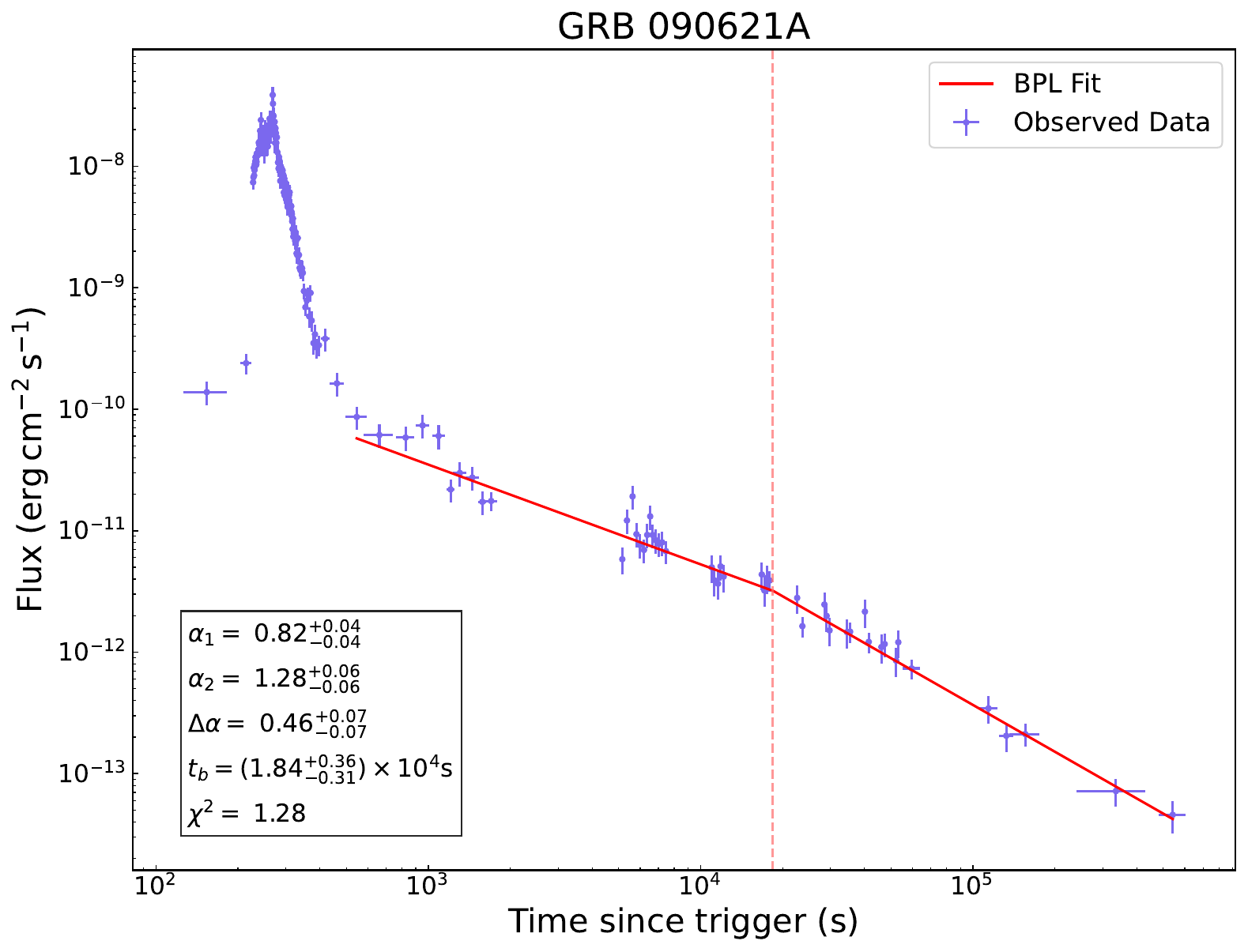}}\\
\resizebox{45mm}{!}{\includegraphics[]{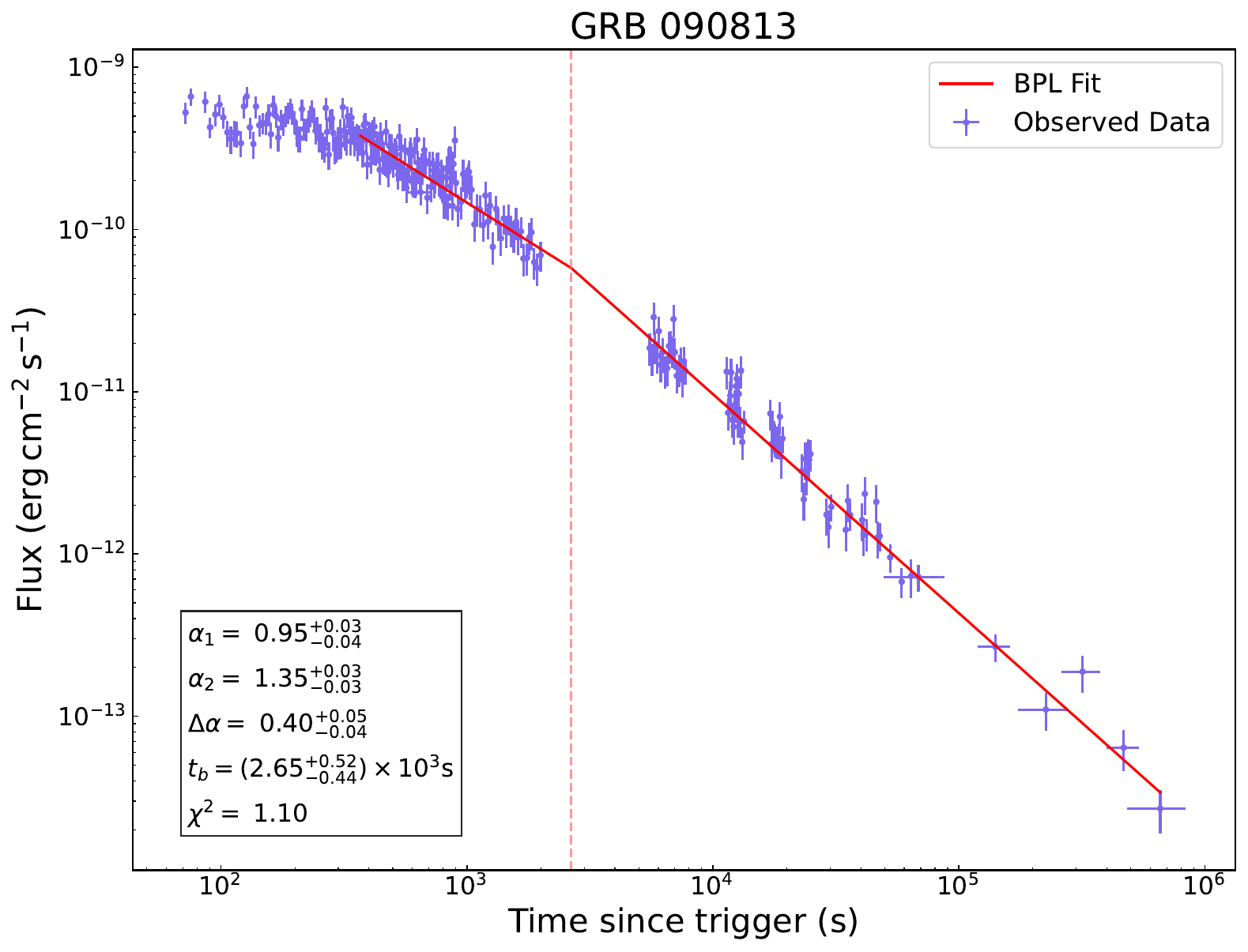}}%
\resizebox{45mm}{!}{\includegraphics[]{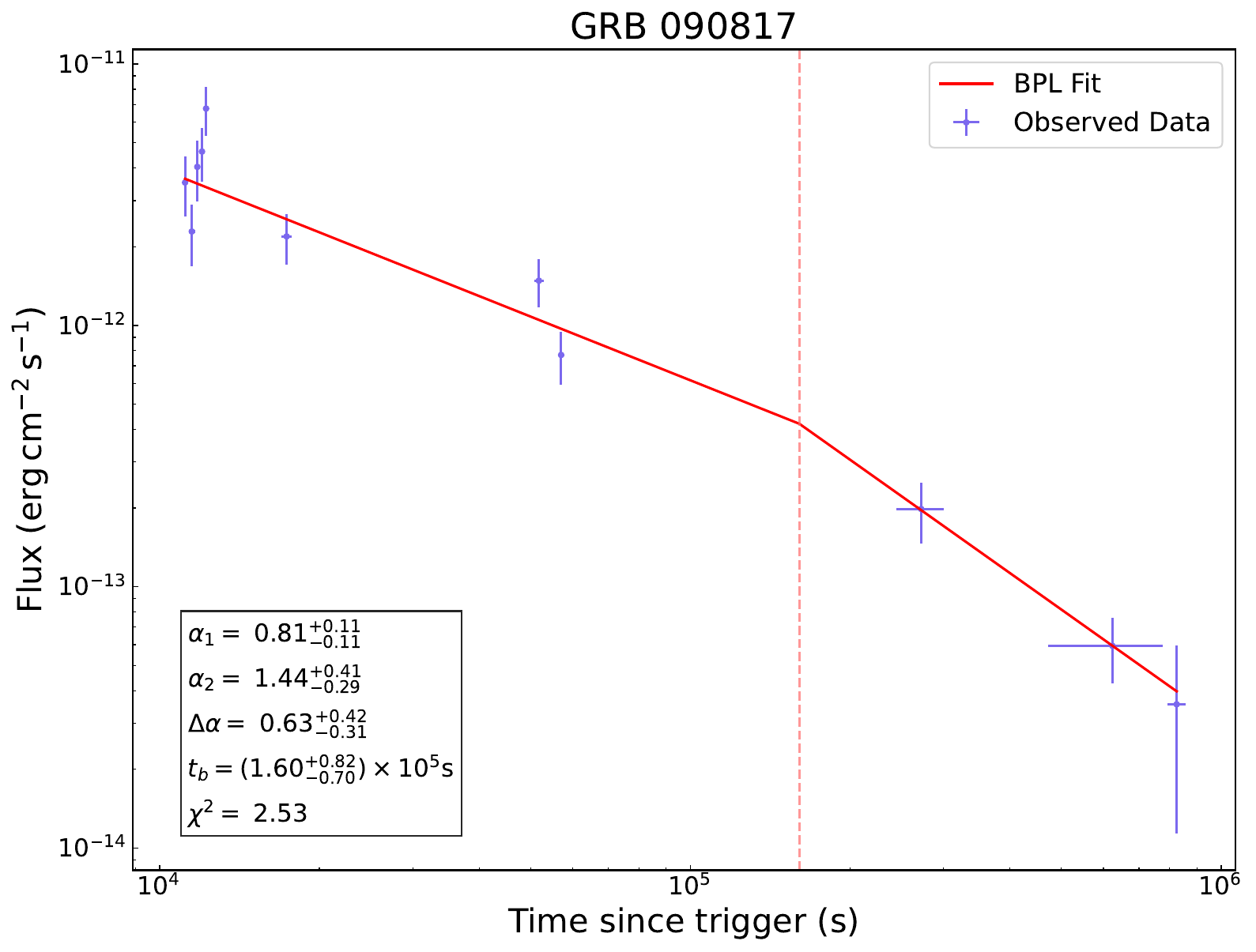}}%
\resizebox{45mm}{!}{\includegraphics[]{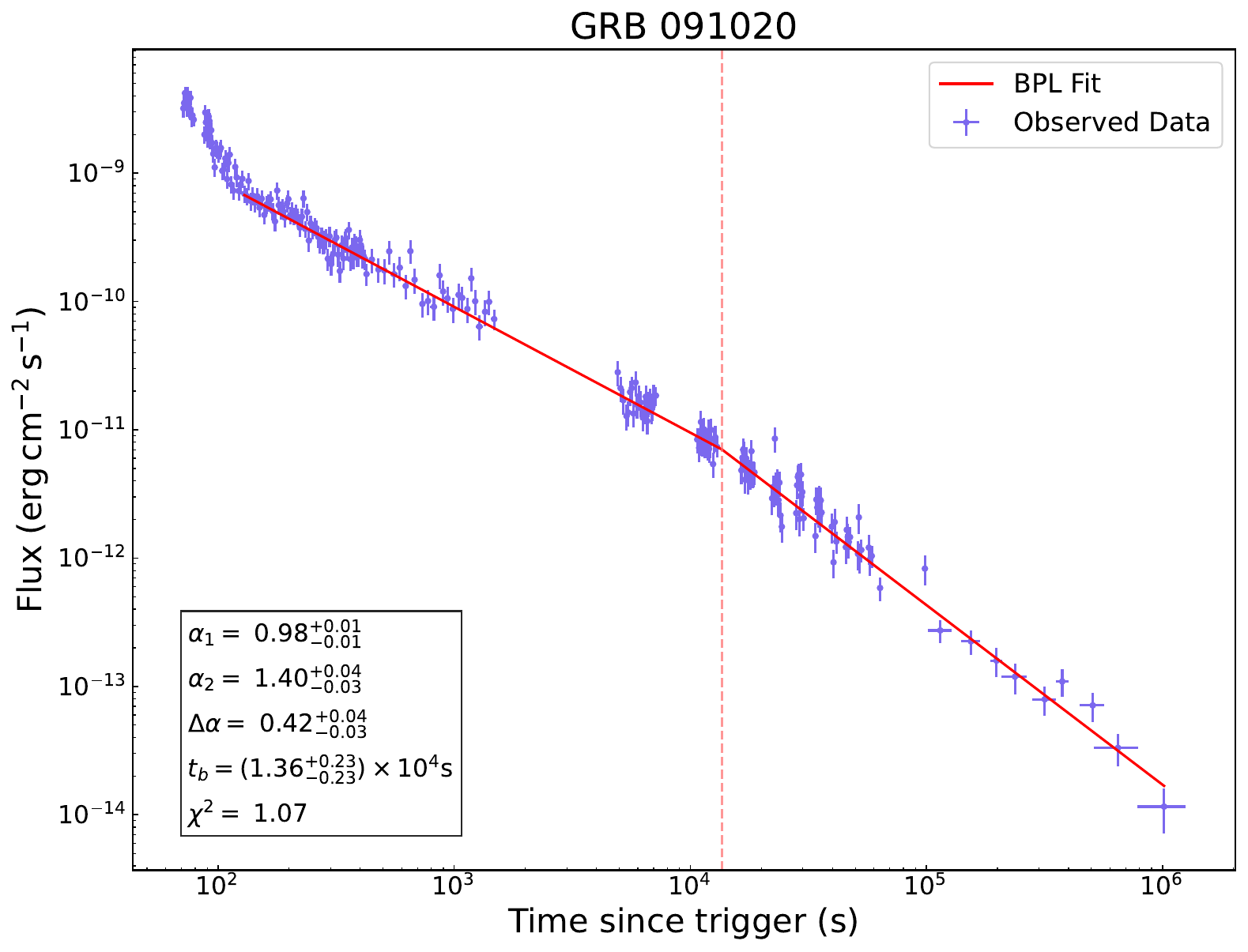}}%
\resizebox{45mm}{!}{\includegraphics[]{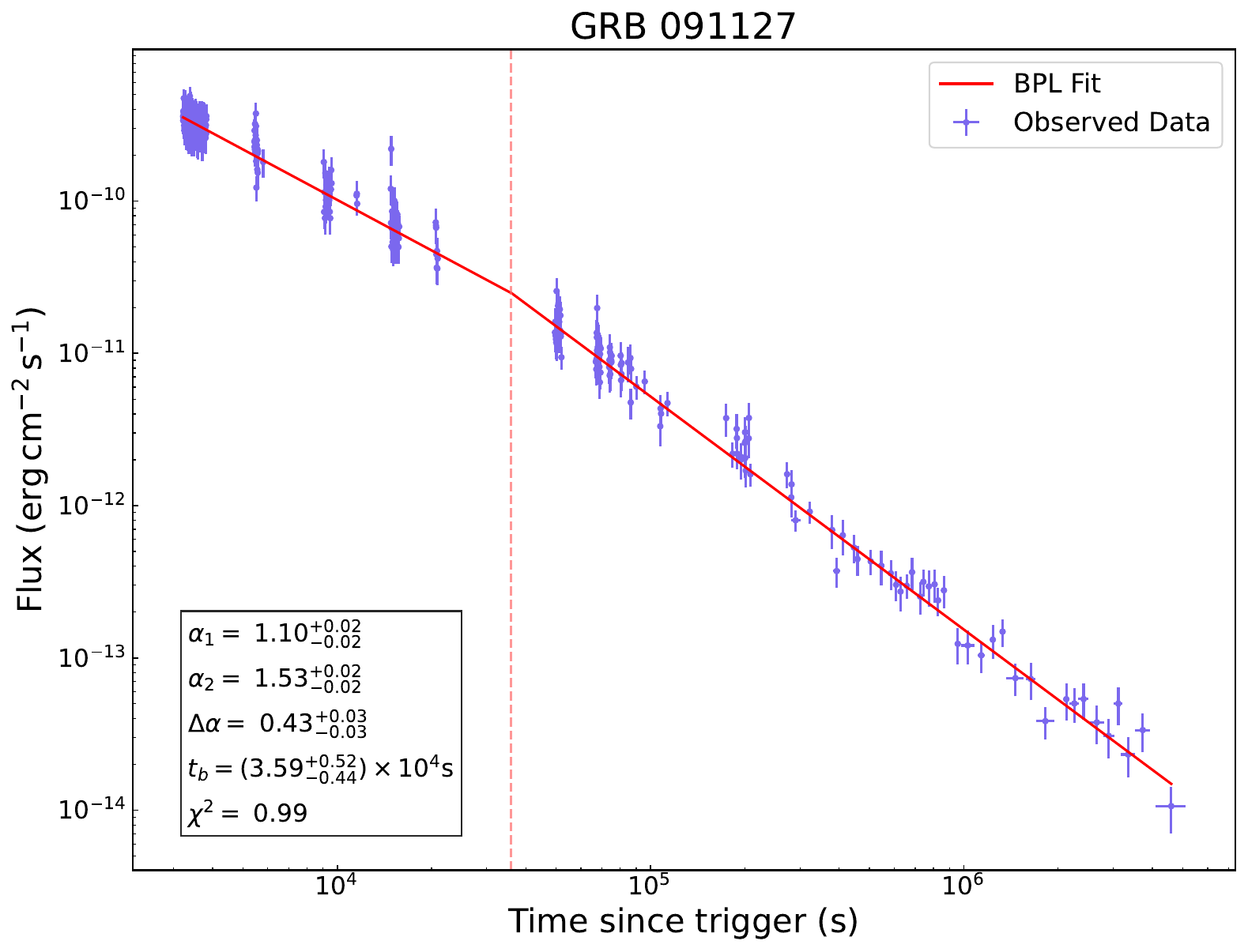}}\\
\caption{X-ray afterglow light curves for the 88 GRBs in our sample classified as having a wind circum-burst environment. Blue points show the Swift/XRT observational data, and the solid red curves represent the best-fit broken power-law models. Multi-band (X-ray and optical) light curves are appended at the end.}
\label{fig:figure2}
\end{figure*}

\clearpage
\addtocounter{figure}{-1}
\begin{figure*}
\centering
\resizebox{45mm}{!}{\includegraphics[]{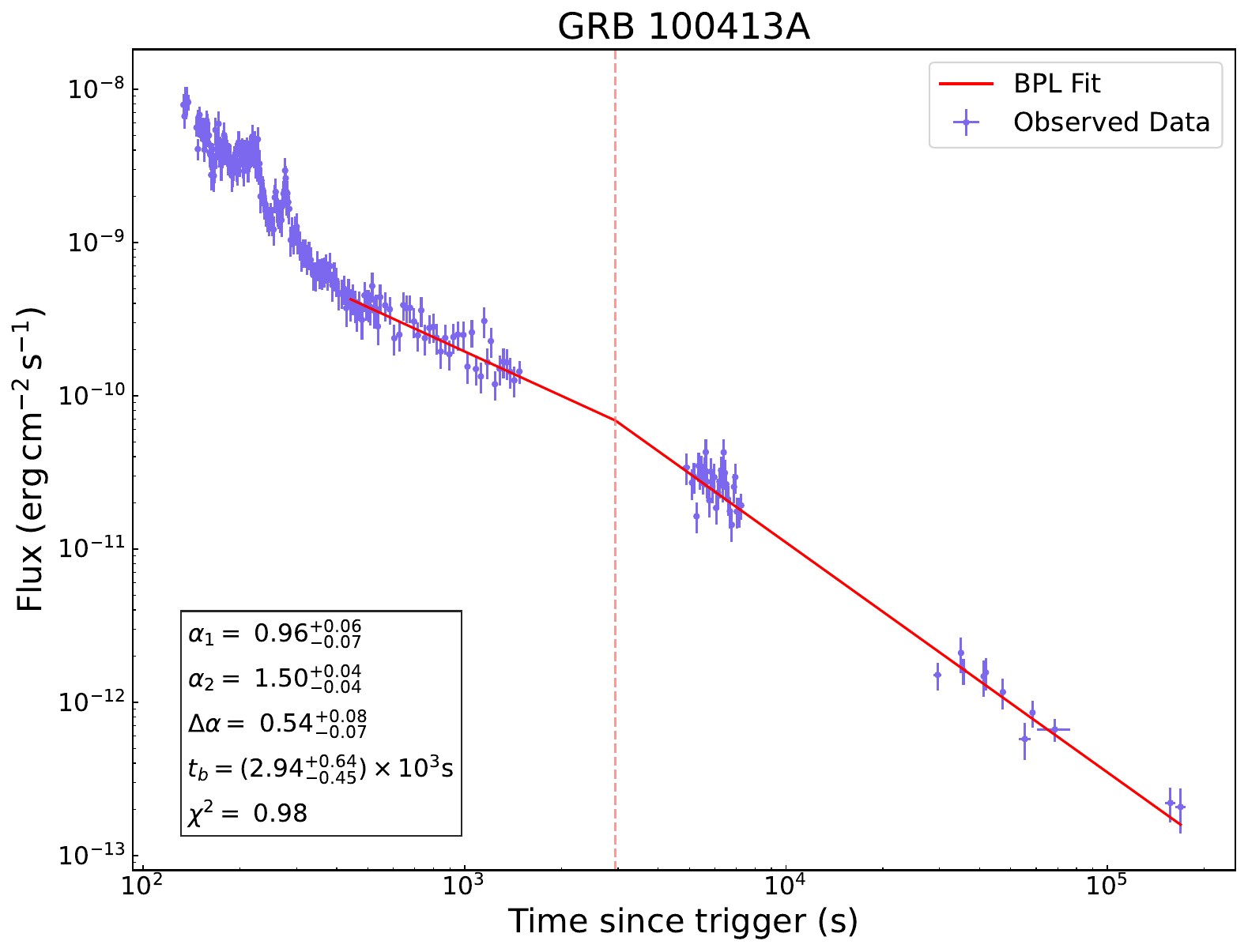}}%
\resizebox{45mm}{!}{\includegraphics[]{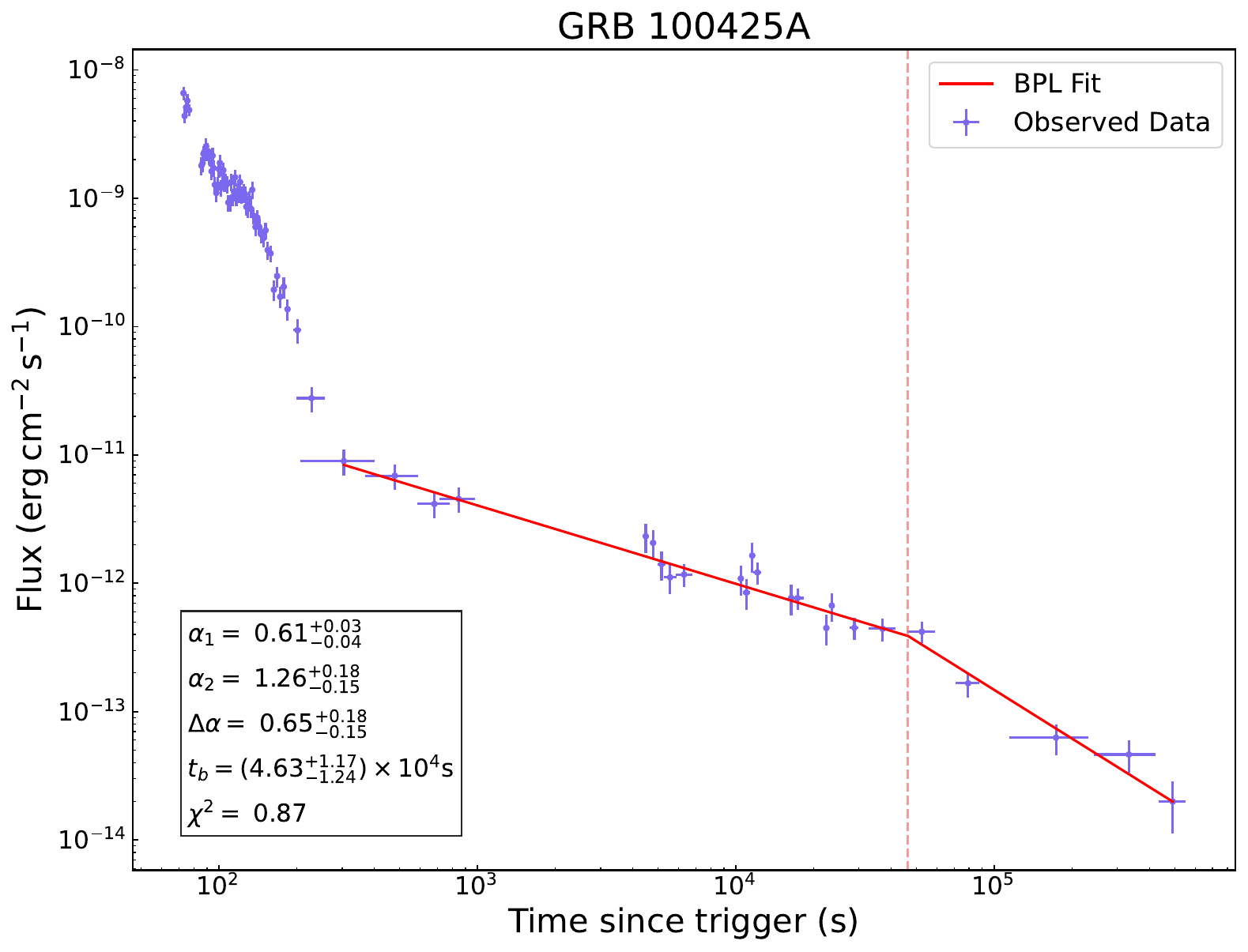}}%
\resizebox{45mm}{!}{\includegraphics[]{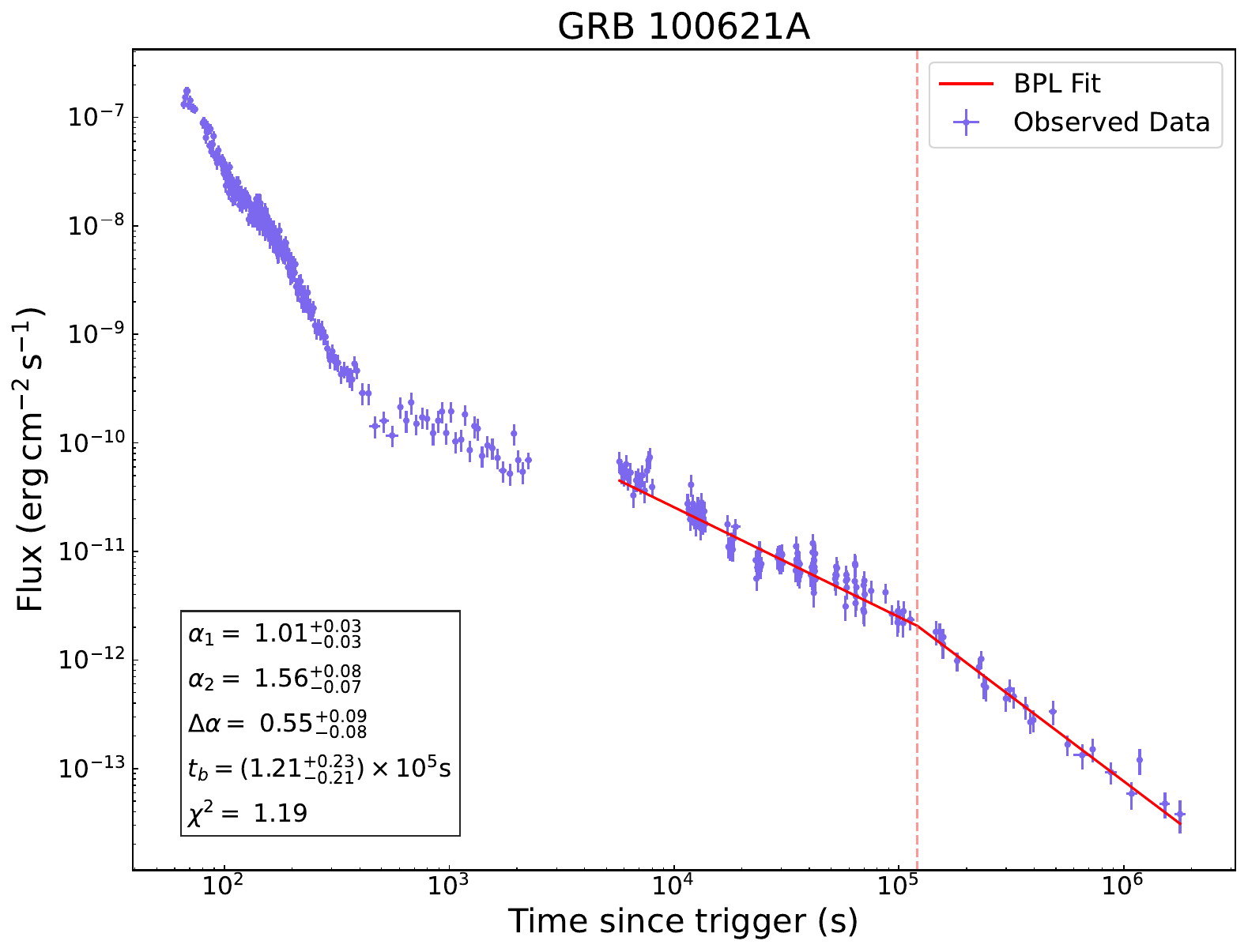}}%
\resizebox{45mm}{!}{\includegraphics[]{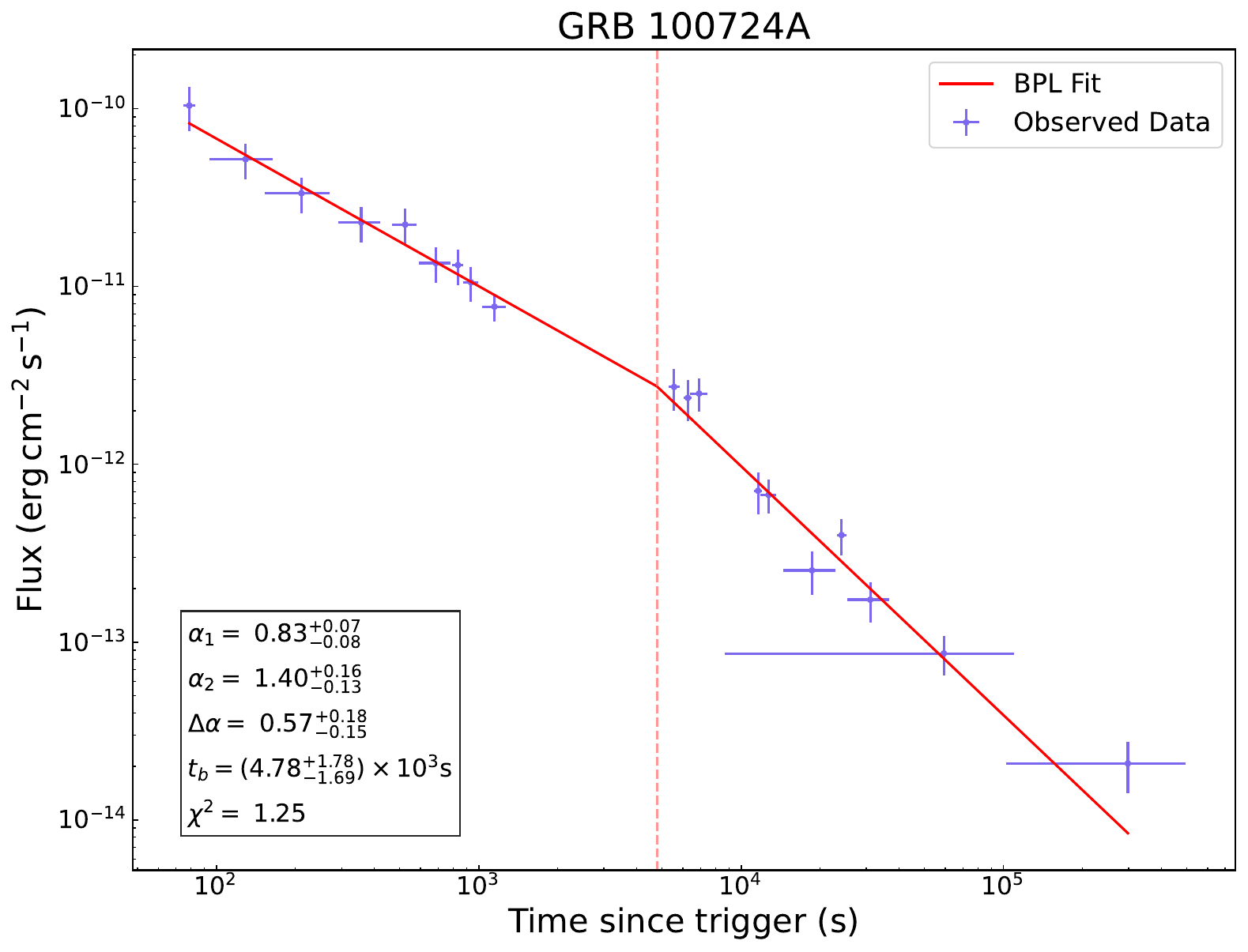}}\\
\resizebox{45mm}{!}{\includegraphics[]{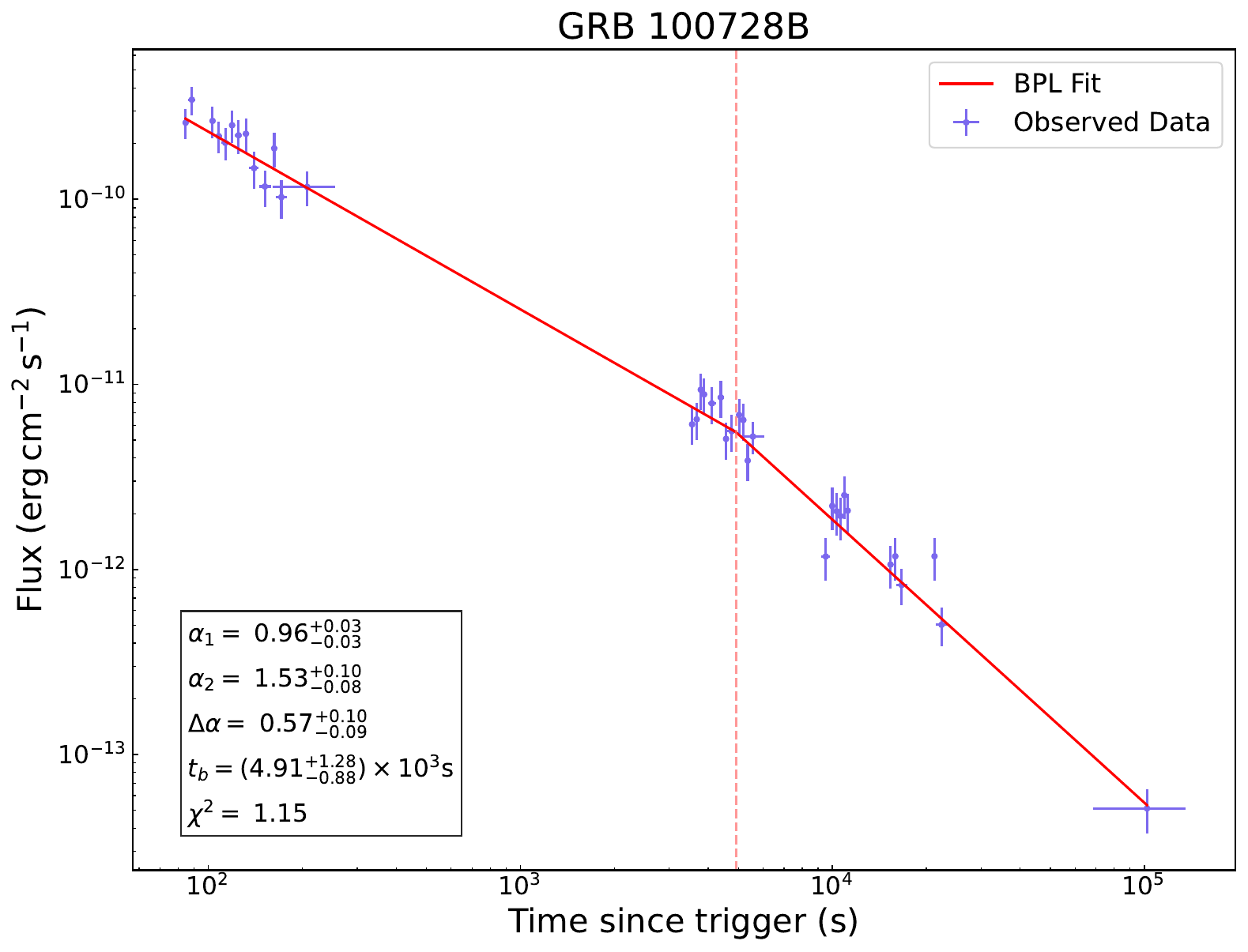}}%
\resizebox{45mm}{!}{\includegraphics[]{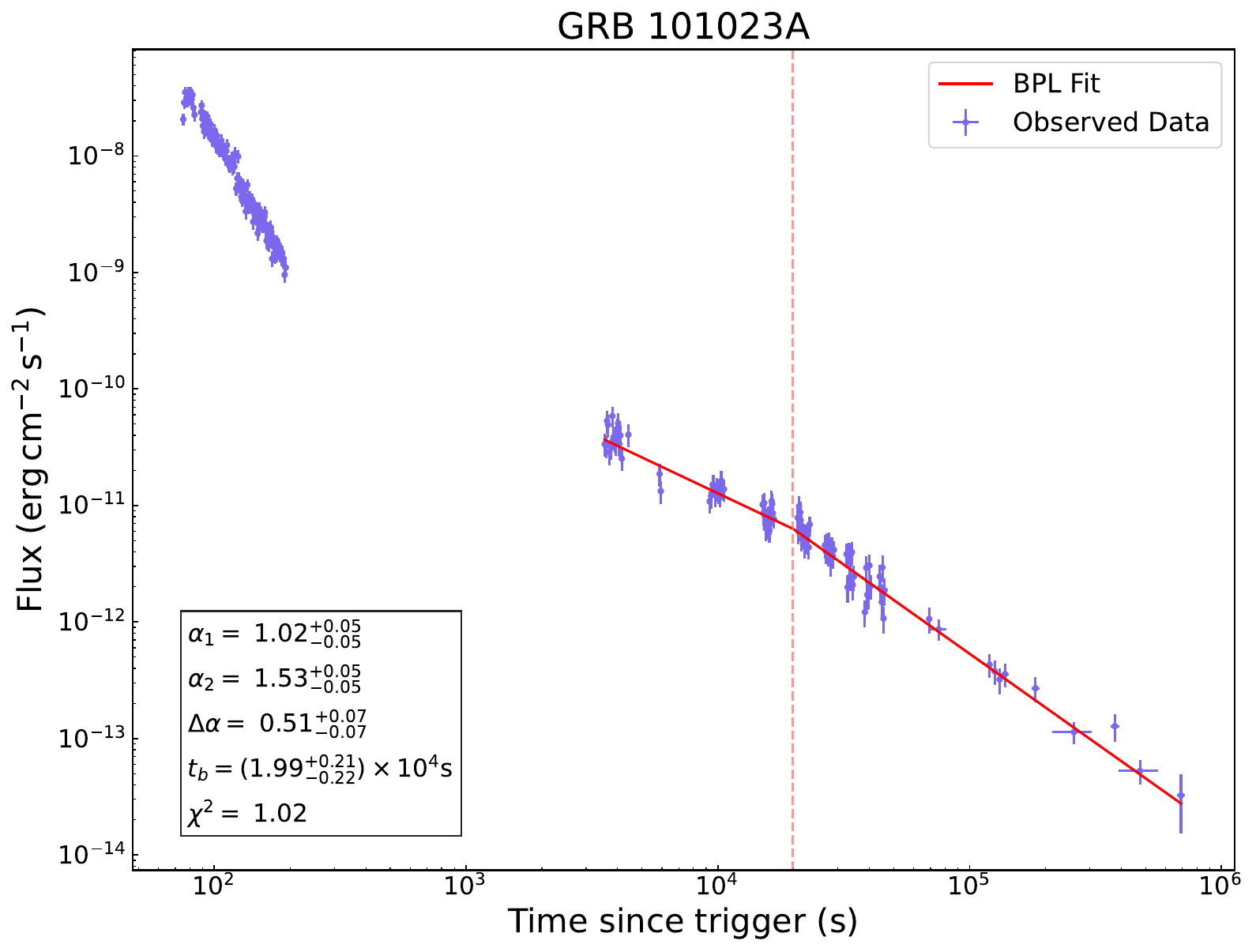}}%
\resizebox{45mm}{!}{\includegraphics[]{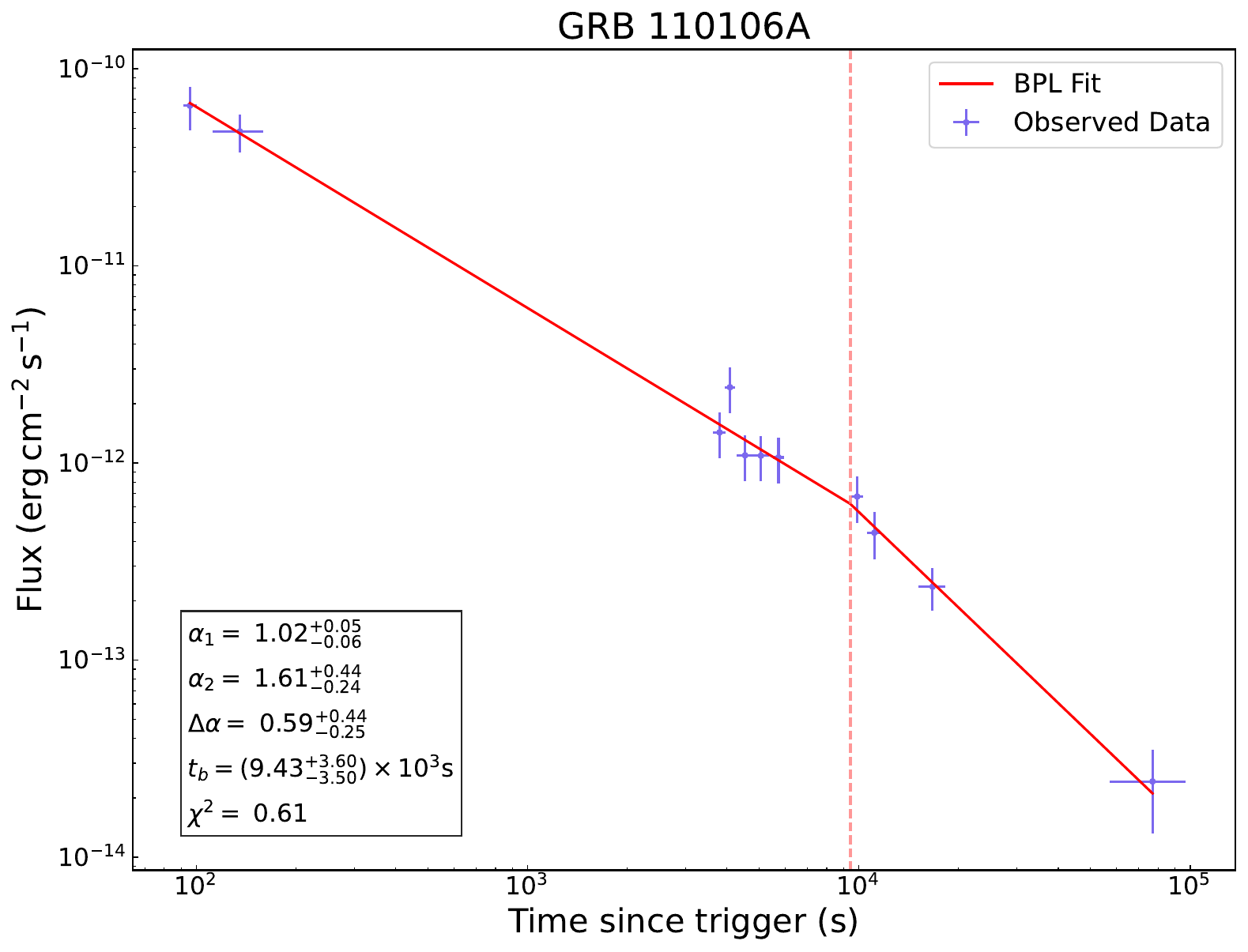}}%
\resizebox{45mm}{!}{\includegraphics[]{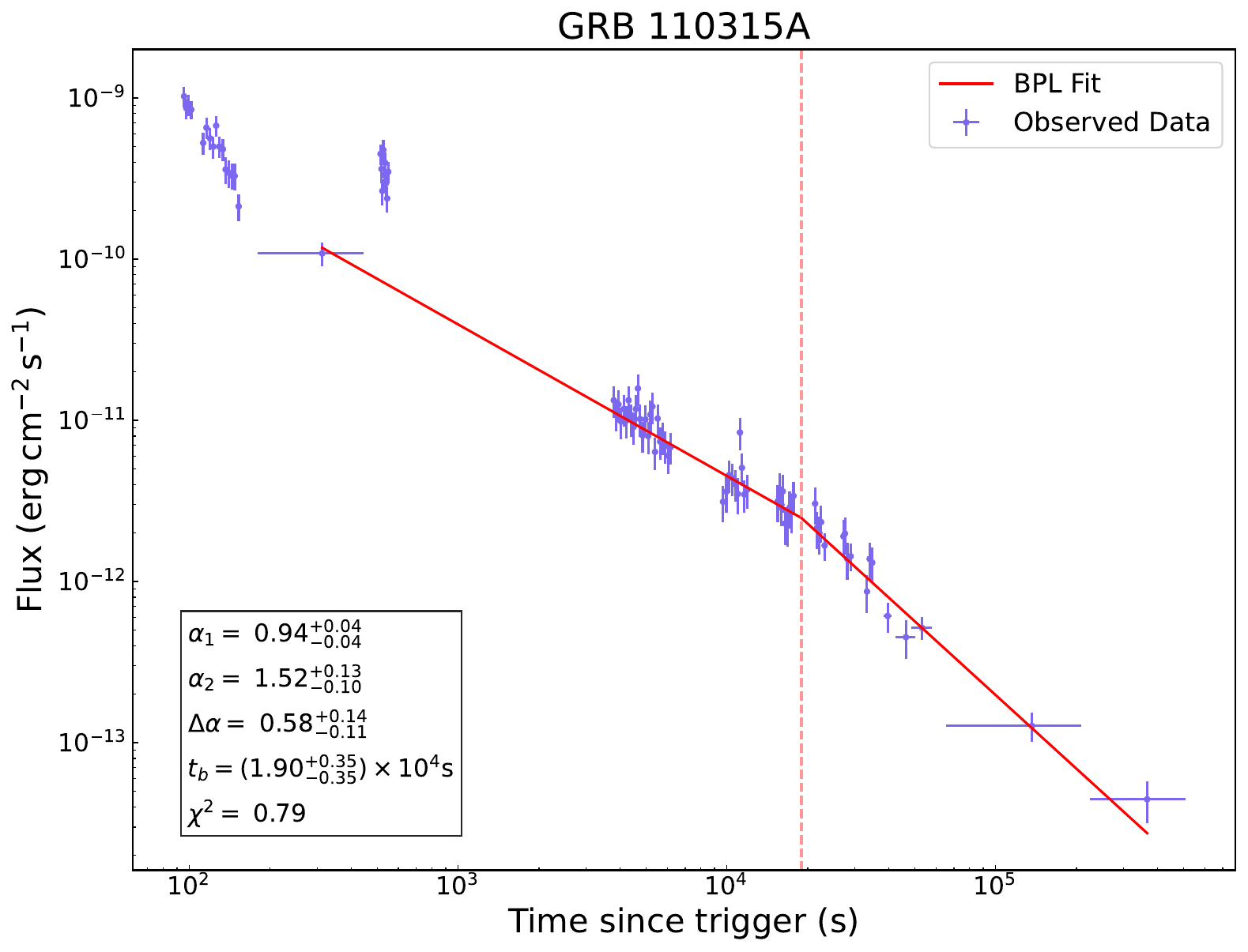}}\\
\resizebox{45mm}{!}{\includegraphics[]{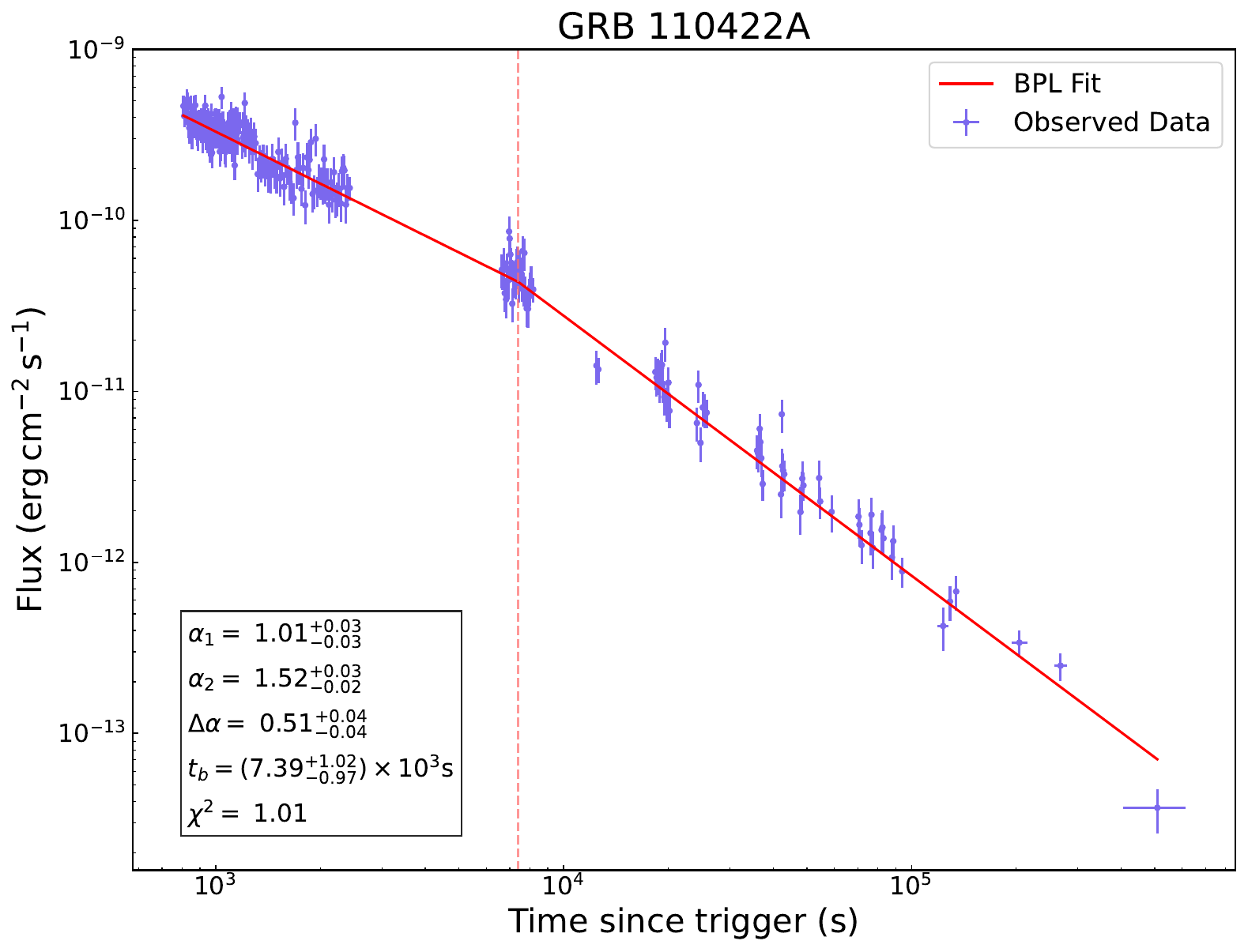}}%
\resizebox{45mm}{!}{\includegraphics[]{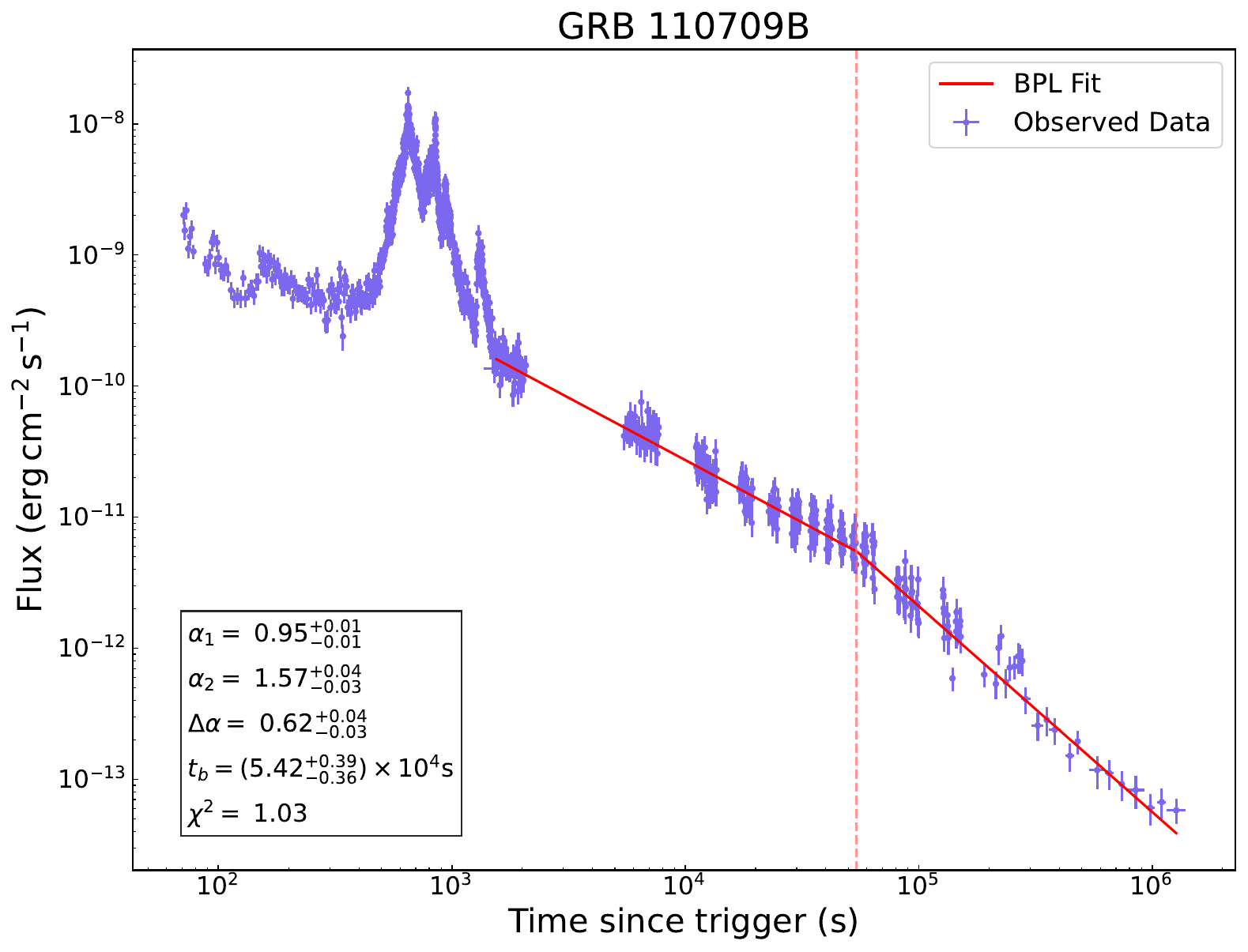}}%
\resizebox{45mm}{!}{\includegraphics[]{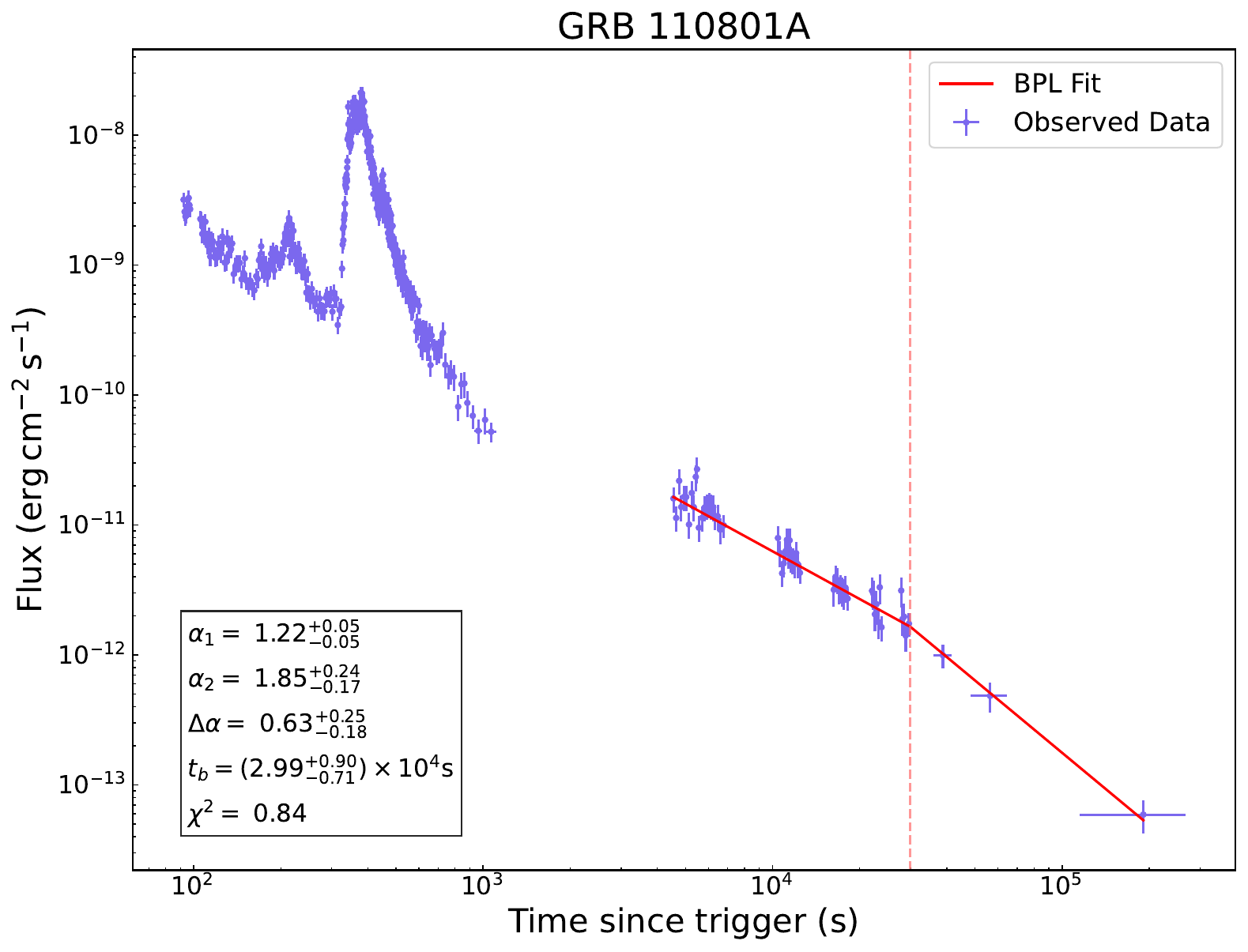}}%
\resizebox{45mm}{!}{\includegraphics[]{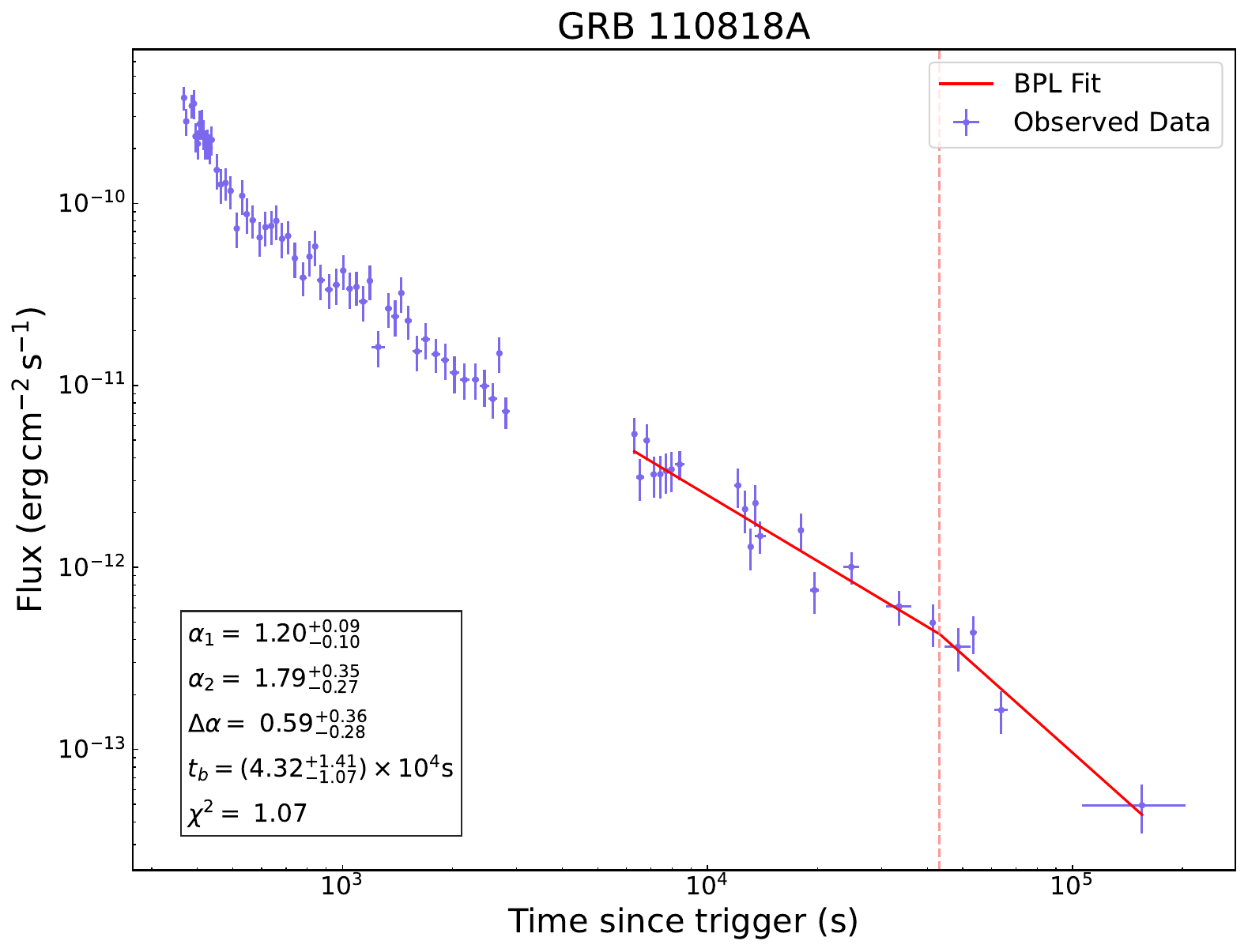}}\\
\resizebox{45mm}{!}{\includegraphics[]{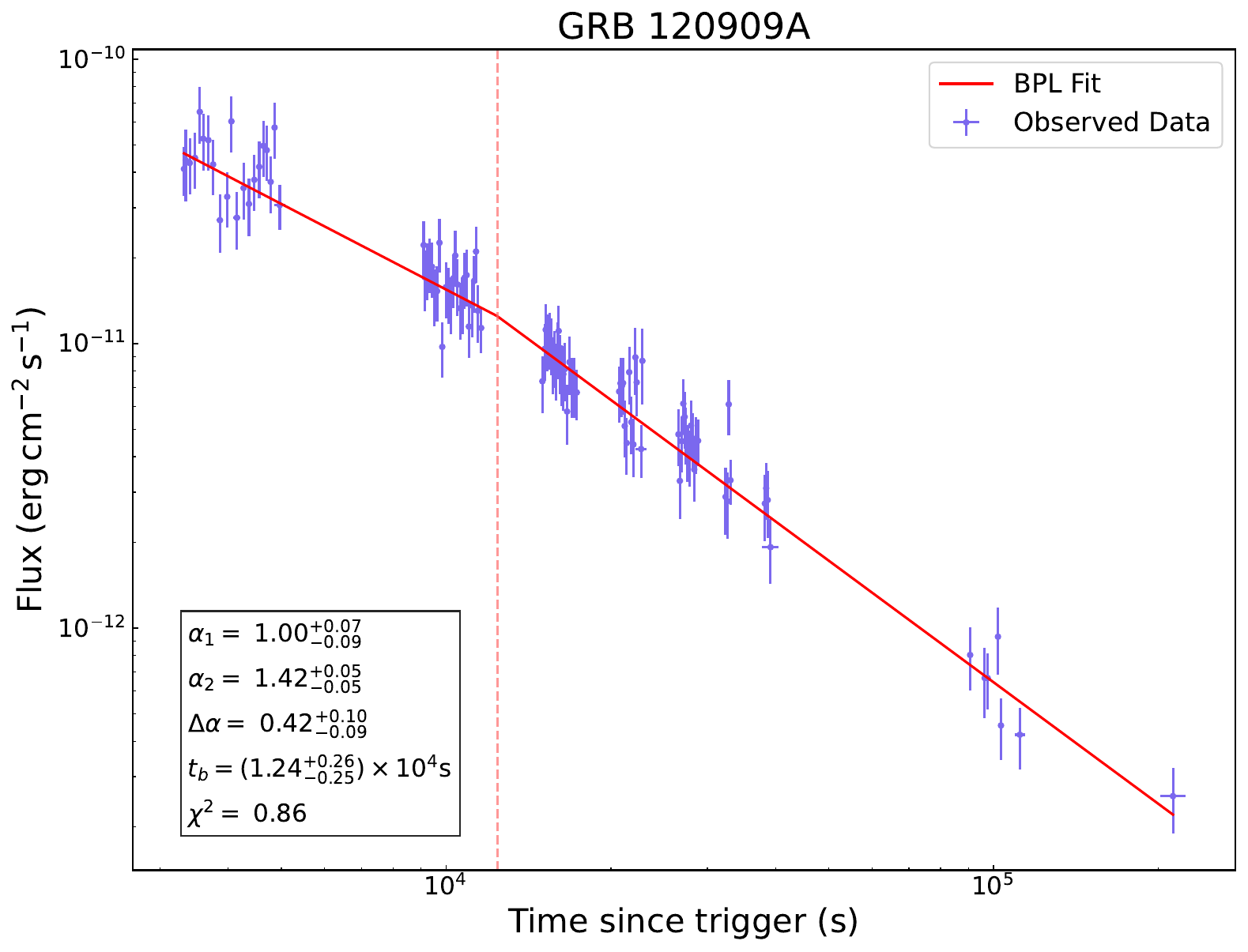}}%
\resizebox{45mm}{!}{\includegraphics[]{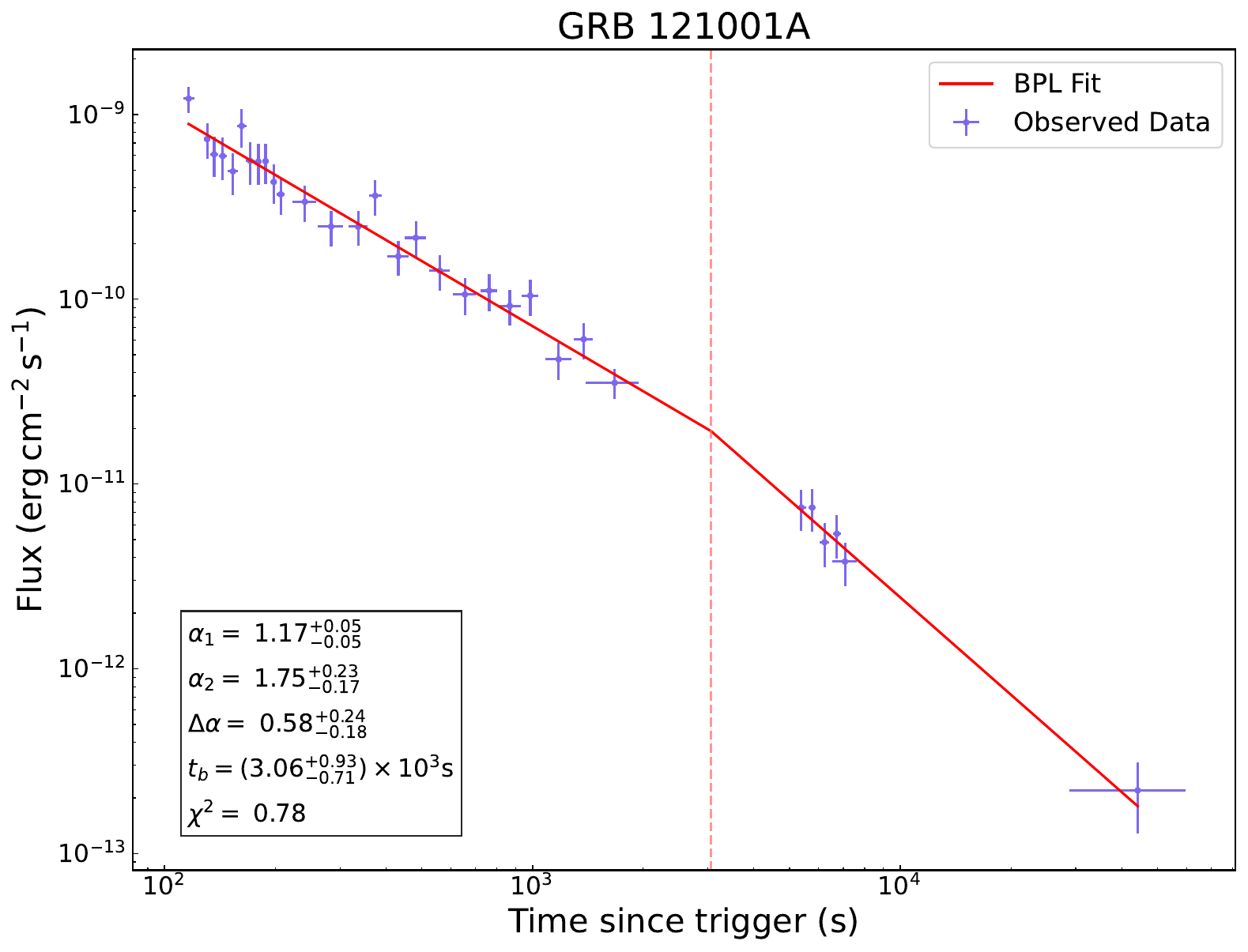}}%
\resizebox{45mm}{!}{\includegraphics[]{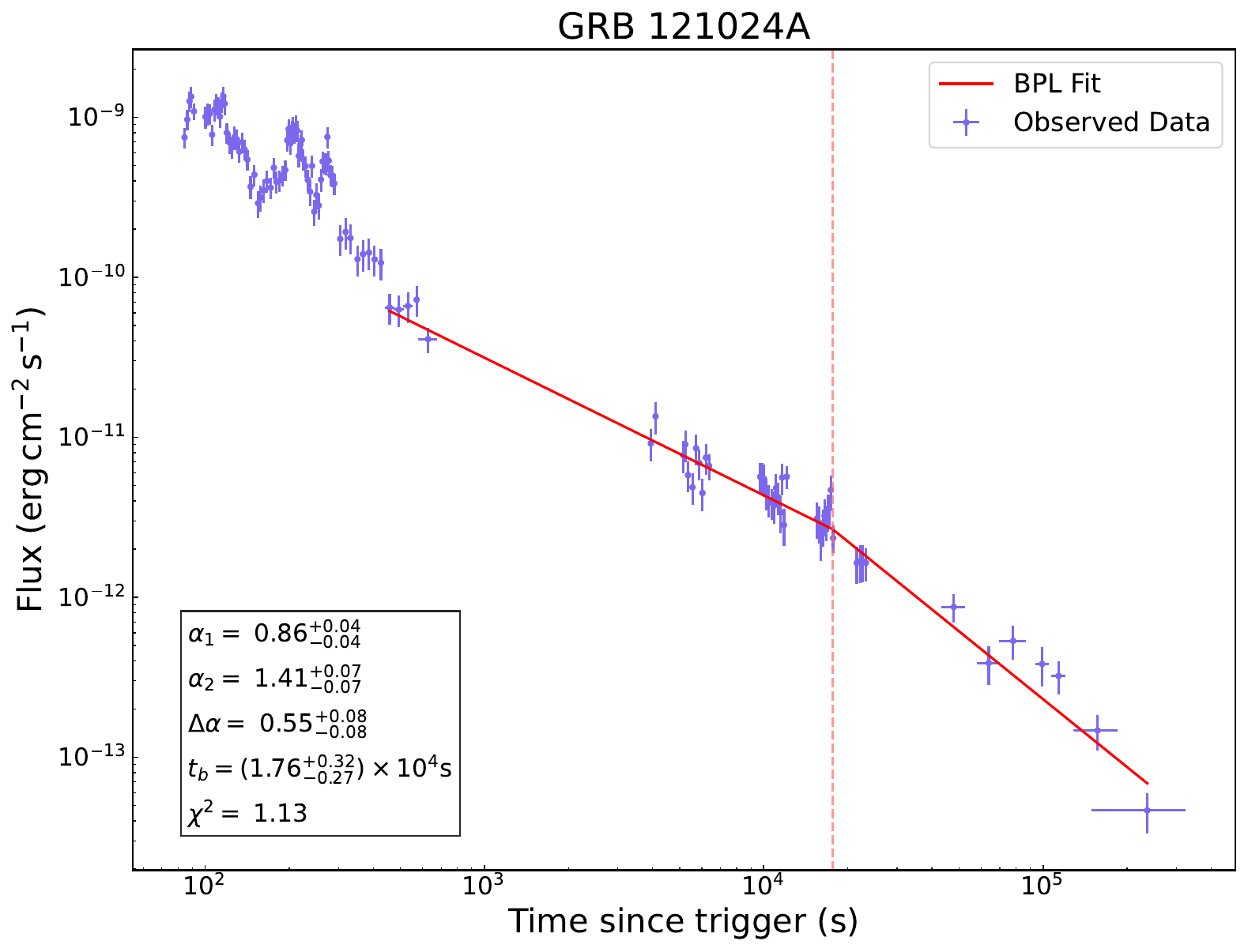}}%
\resizebox{45mm}{!}{\includegraphics[]{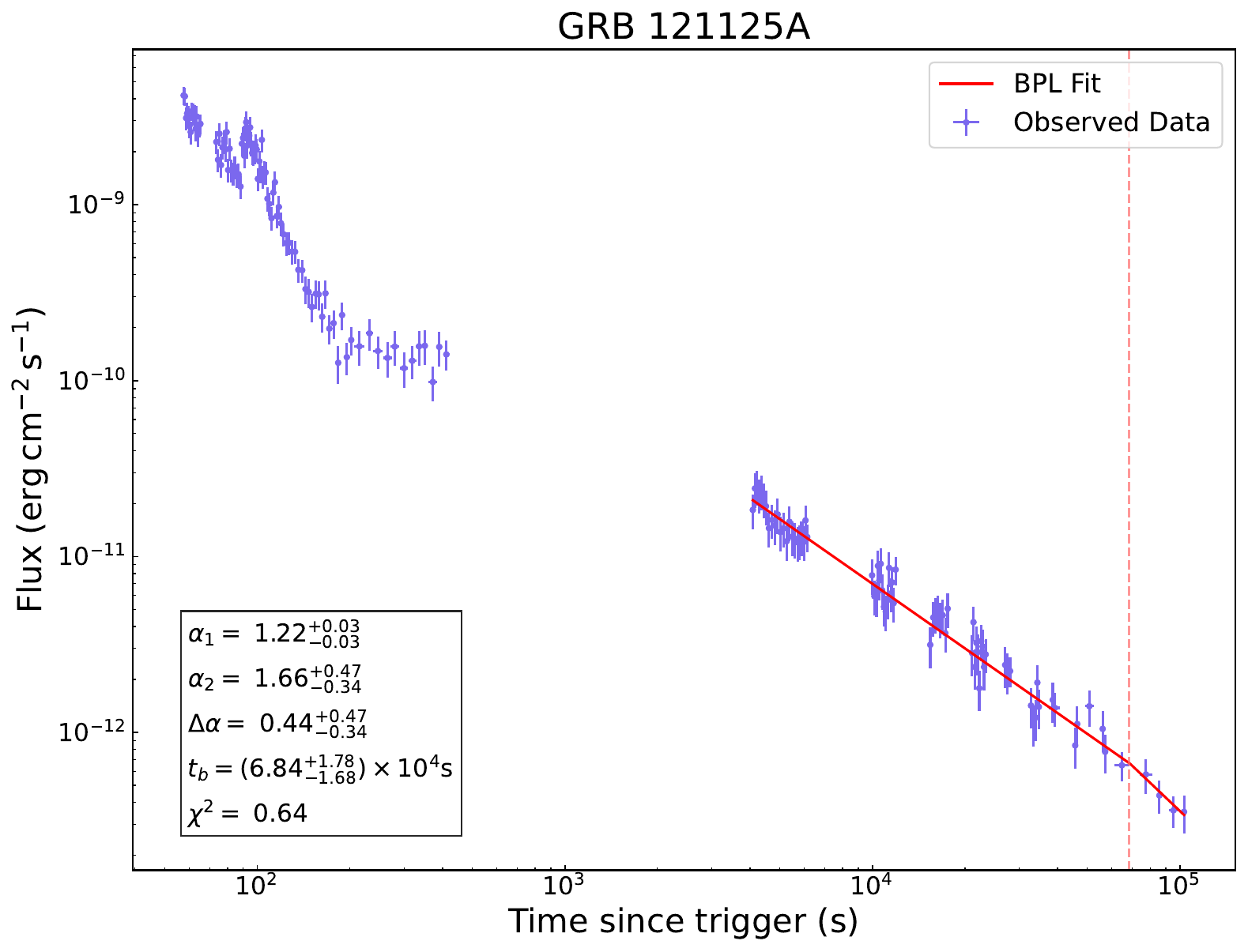}}\\
\resizebox{45mm}{!}{\includegraphics[]{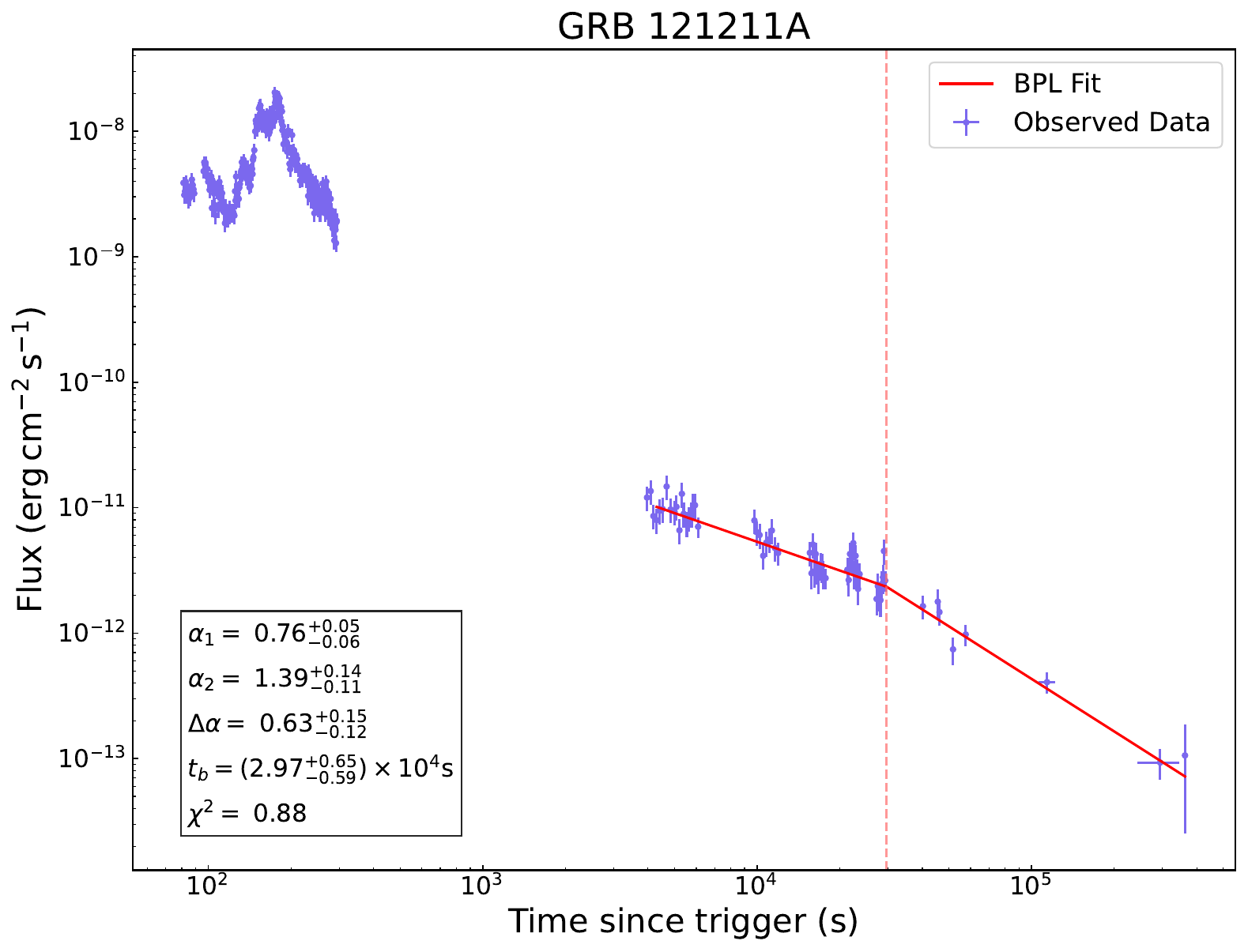}}%
\resizebox{45mm}{!}{\includegraphics[]{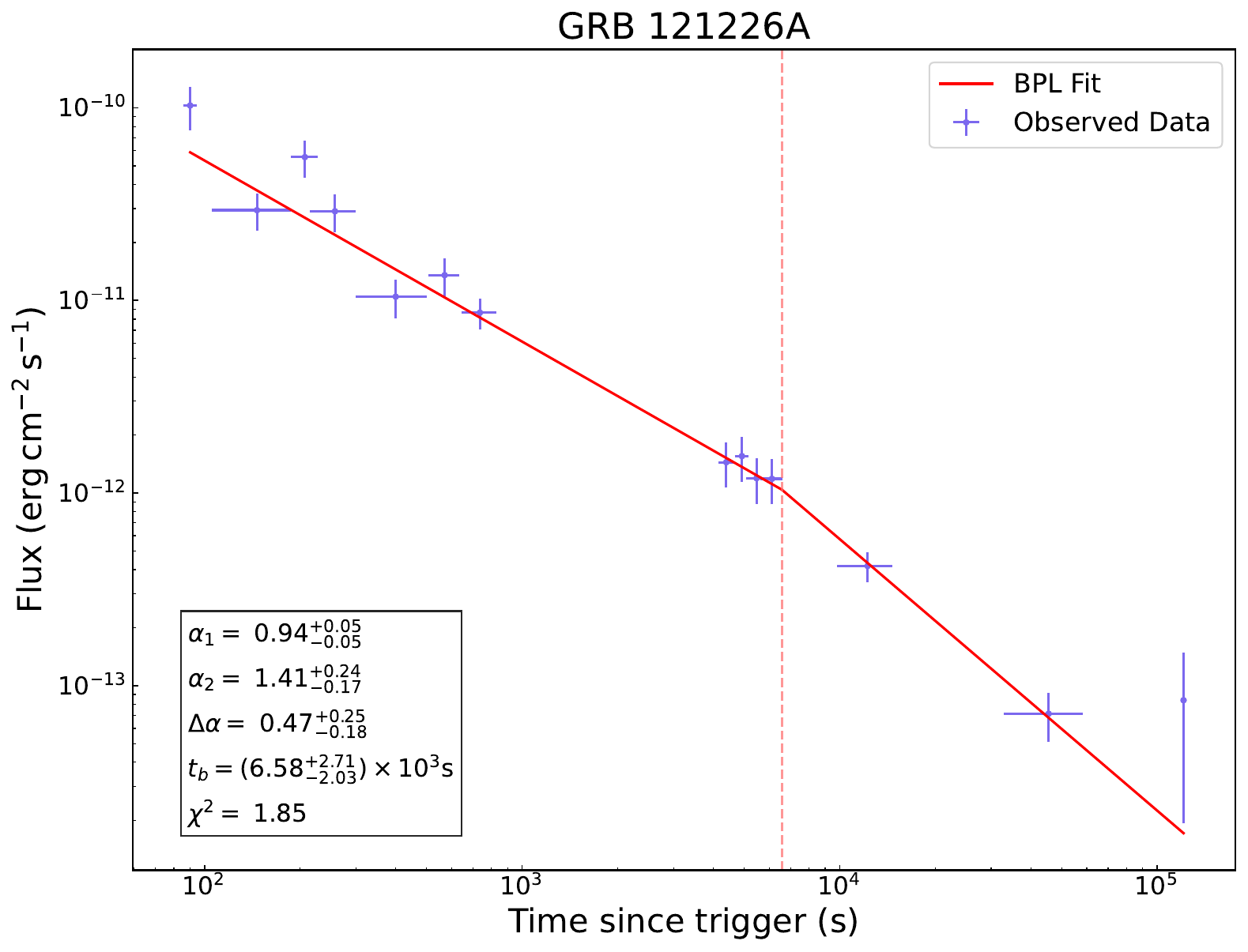}}%
\resizebox{45mm}{!}{\includegraphics[]{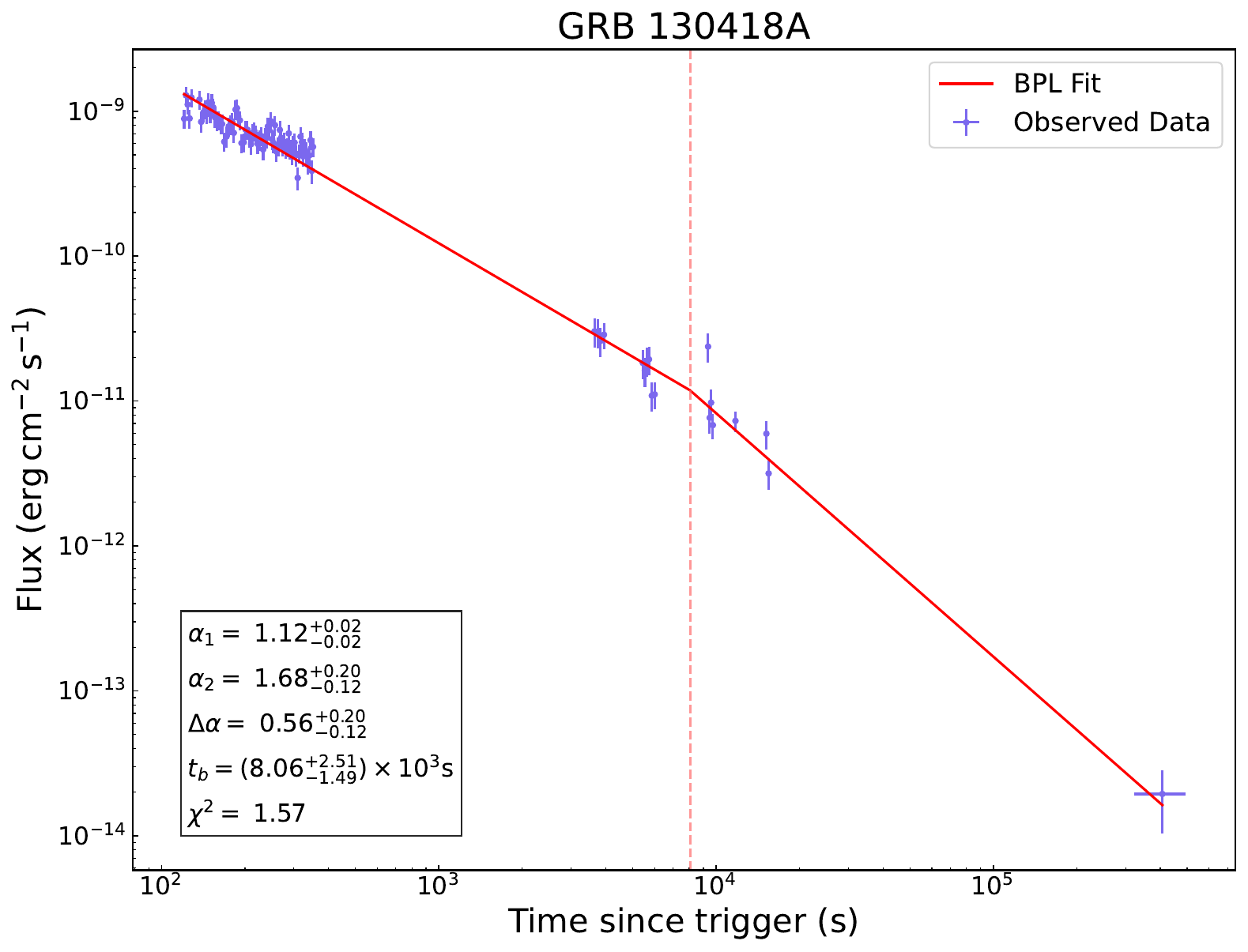}}%
\resizebox{45mm}{!}{\includegraphics[]{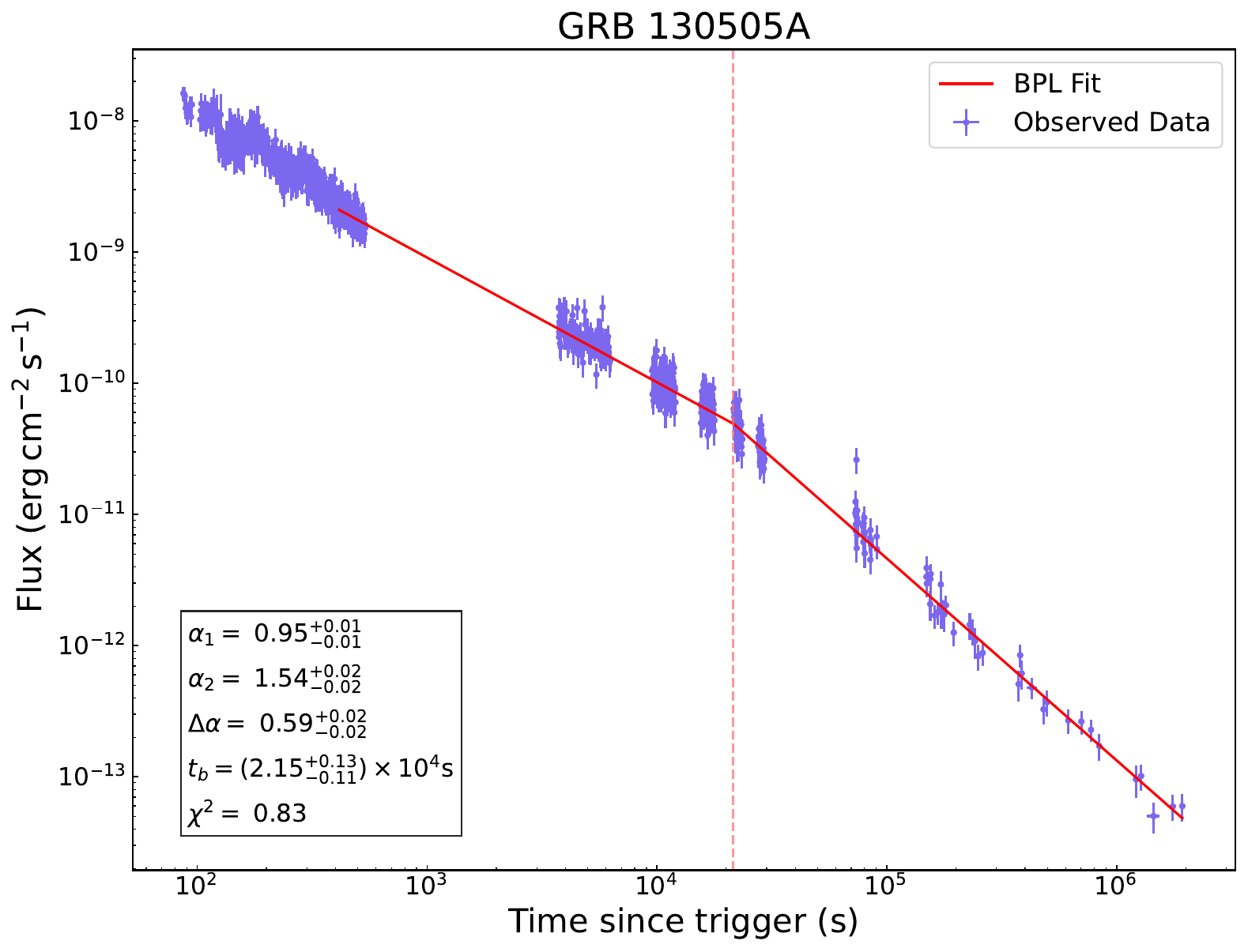}}\\
\resizebox{45mm}{!}{\includegraphics[]{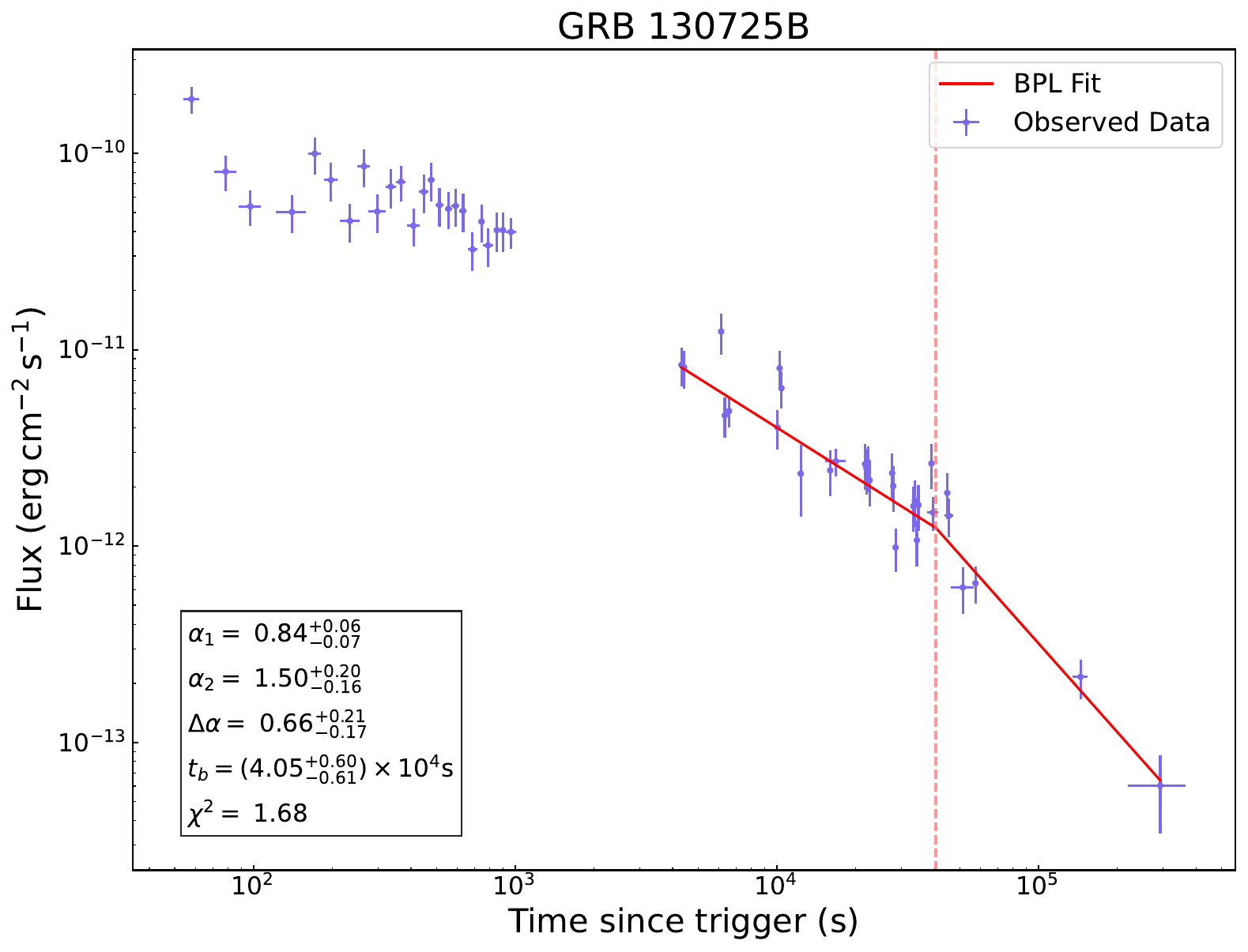}}%
\resizebox{45mm}{!}{\includegraphics[]{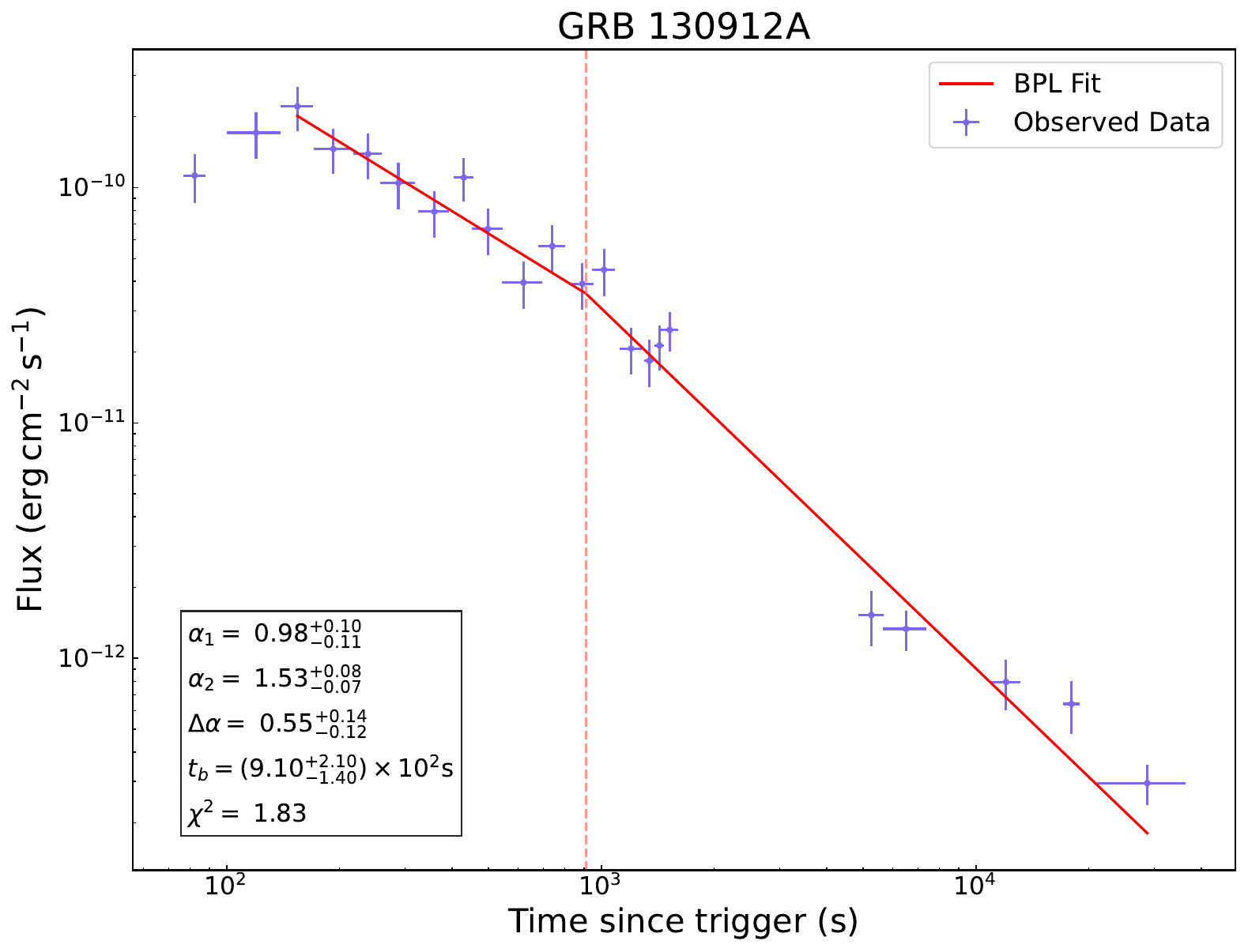}}%
\resizebox{45mm}{!}{\includegraphics[]{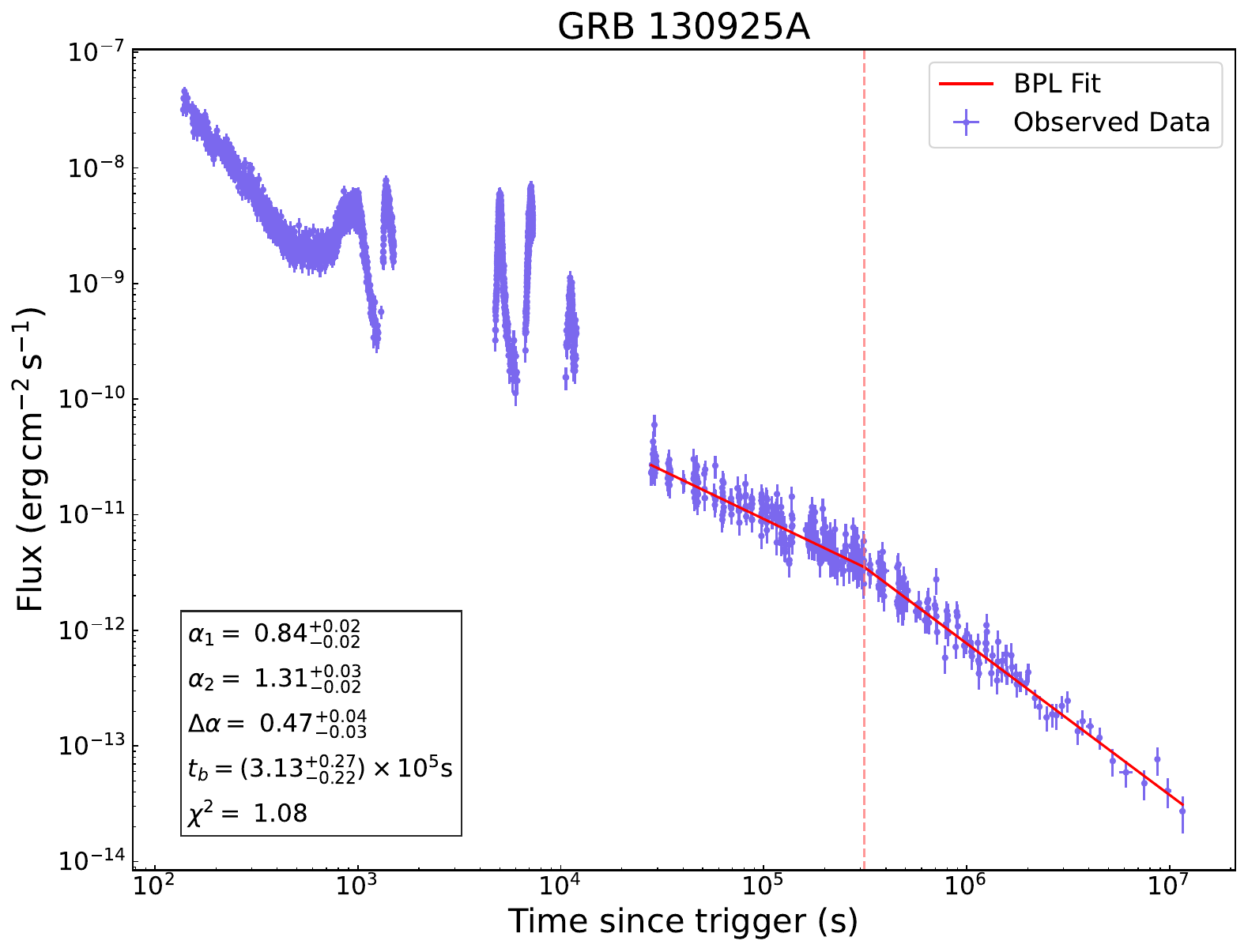}}%
\resizebox{45mm}{!}{\includegraphics[]{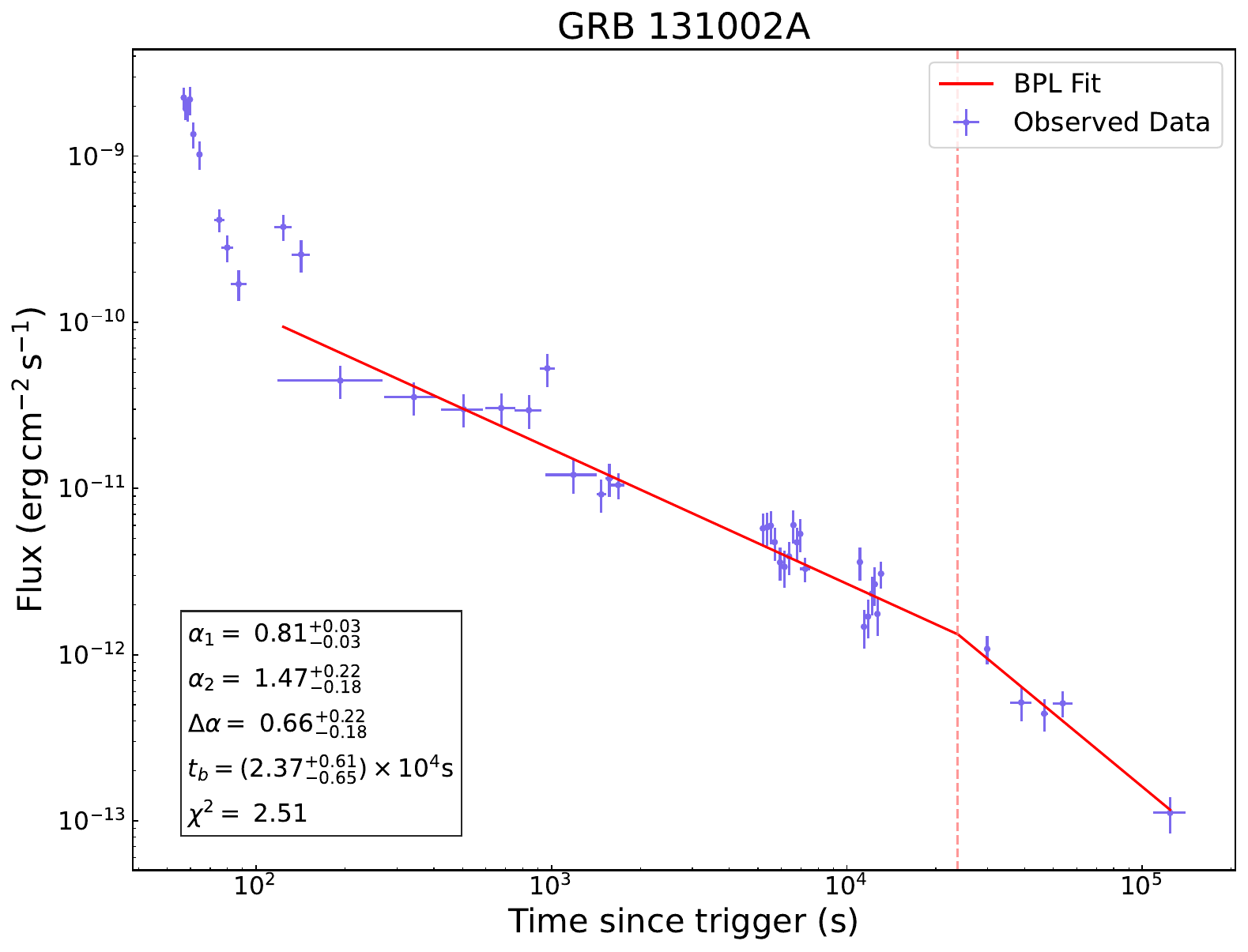}}\\
\resizebox{45mm}{!}{\includegraphics[]{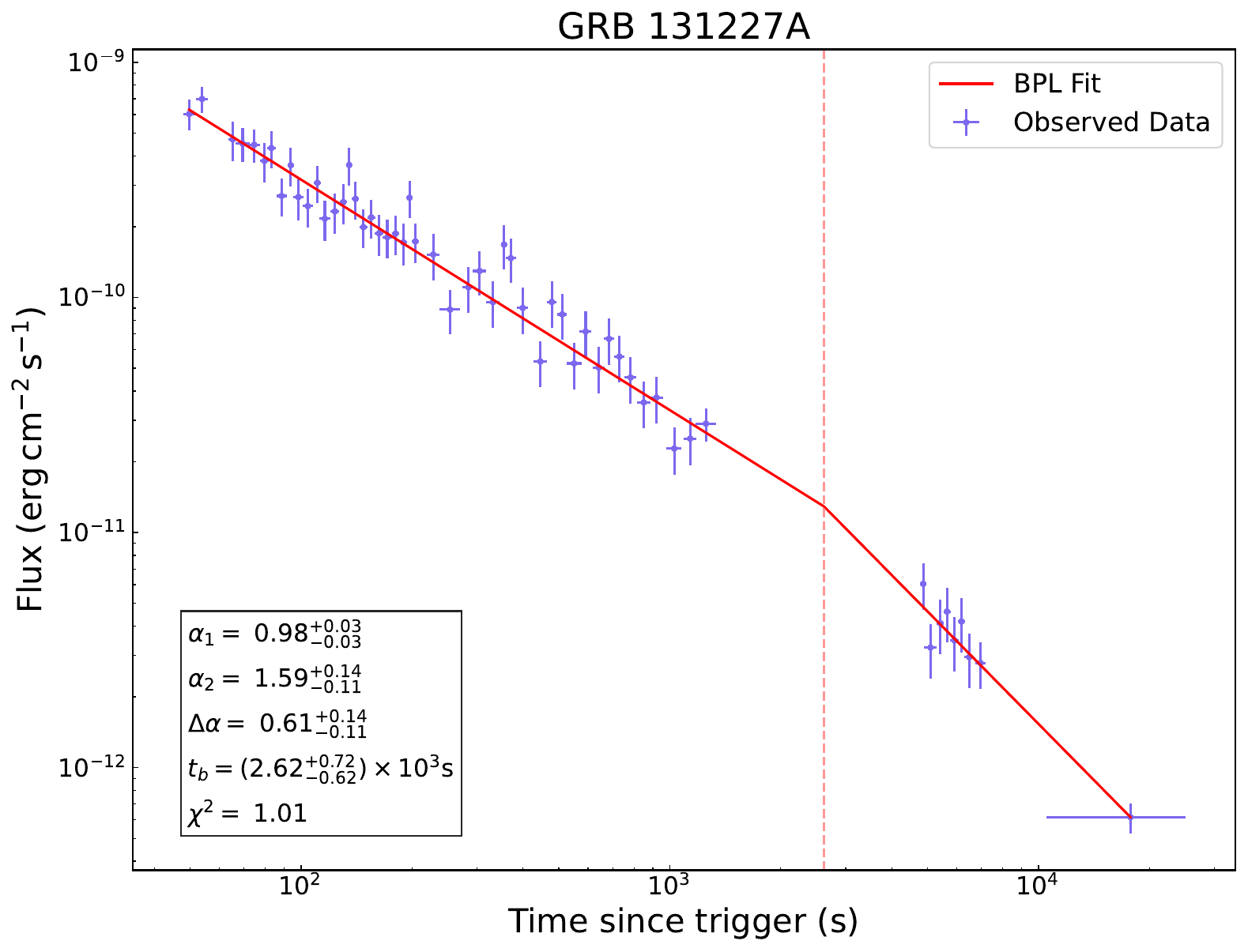}}%
\resizebox{45mm}{!}{\includegraphics[]{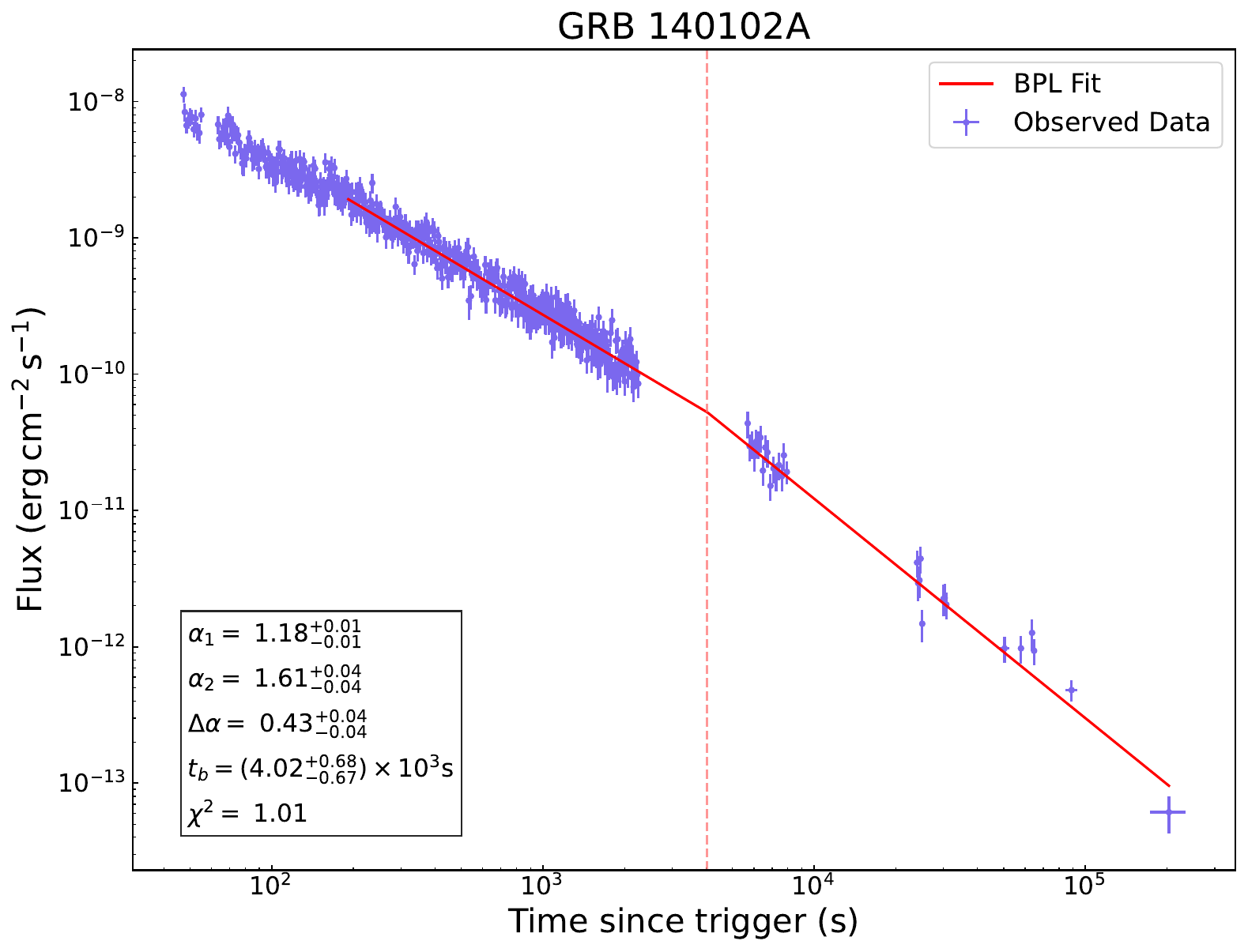}}%
\resizebox{45mm}{!}{\includegraphics[]{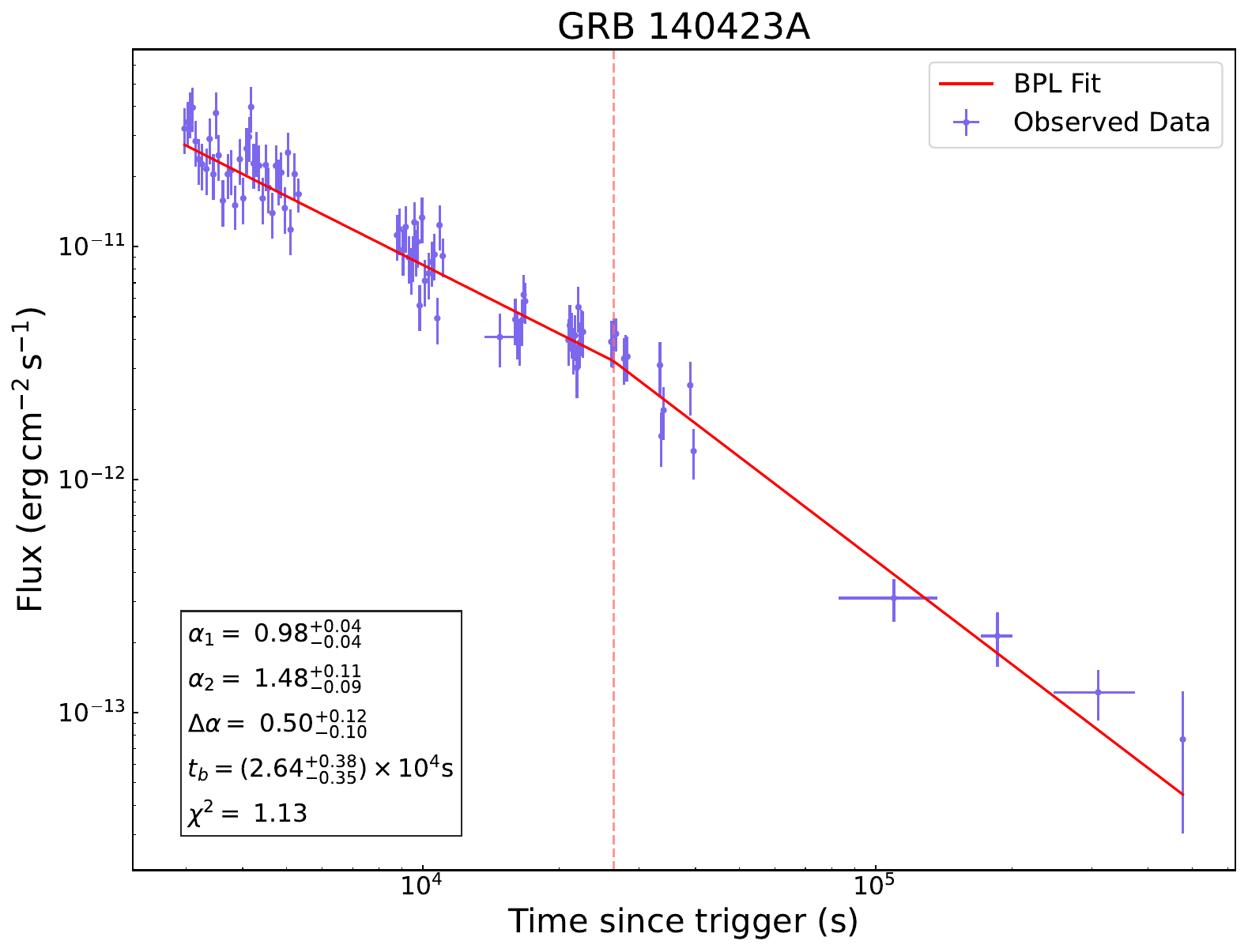}}%
\resizebox{45mm}{!}{\includegraphics[]{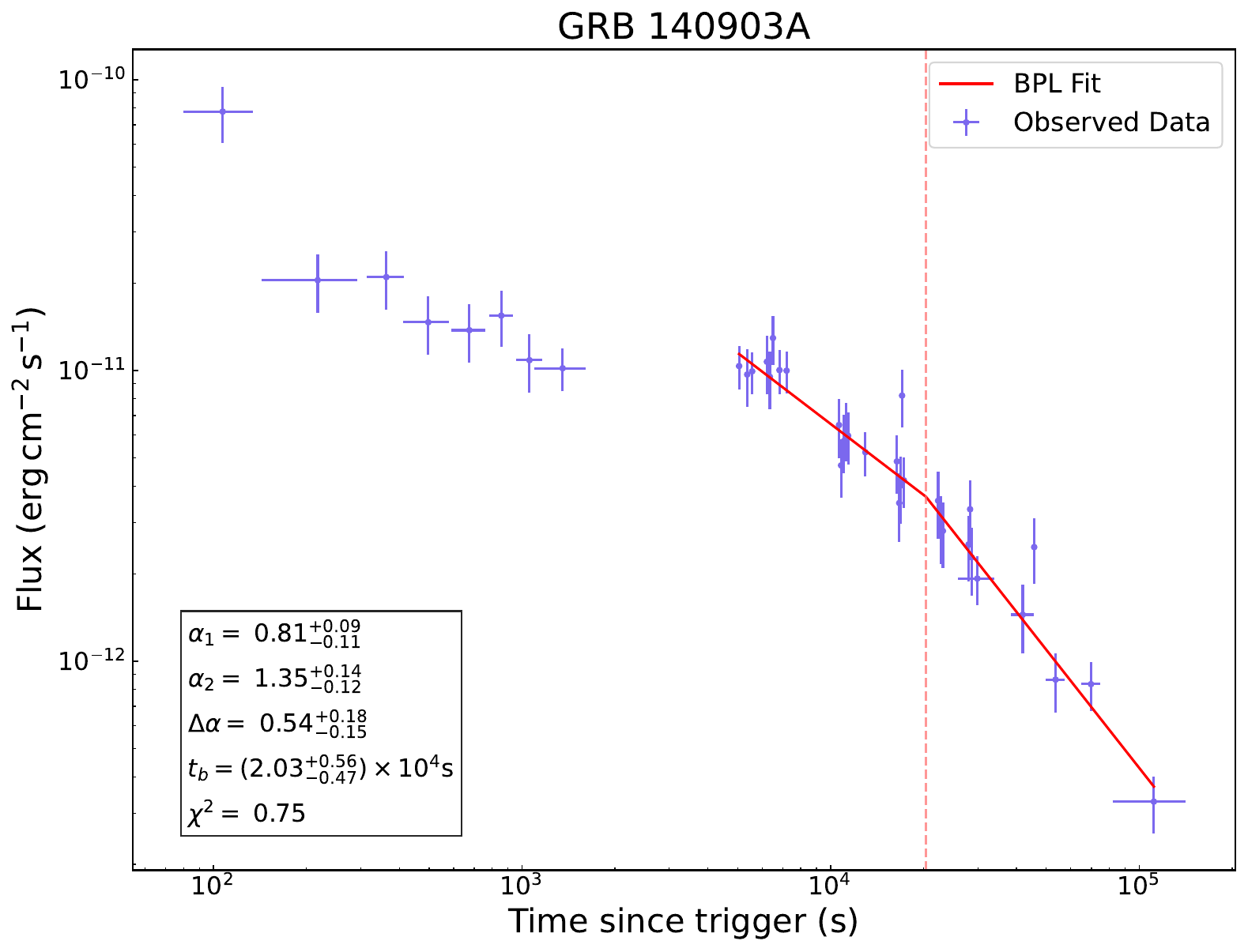}}\\
\caption{(Continued)}
\end{figure*}

\clearpage
\addtocounter{figure}{-1}
\begin{figure*}
\centering
\resizebox{45mm}{!}{\includegraphics[]{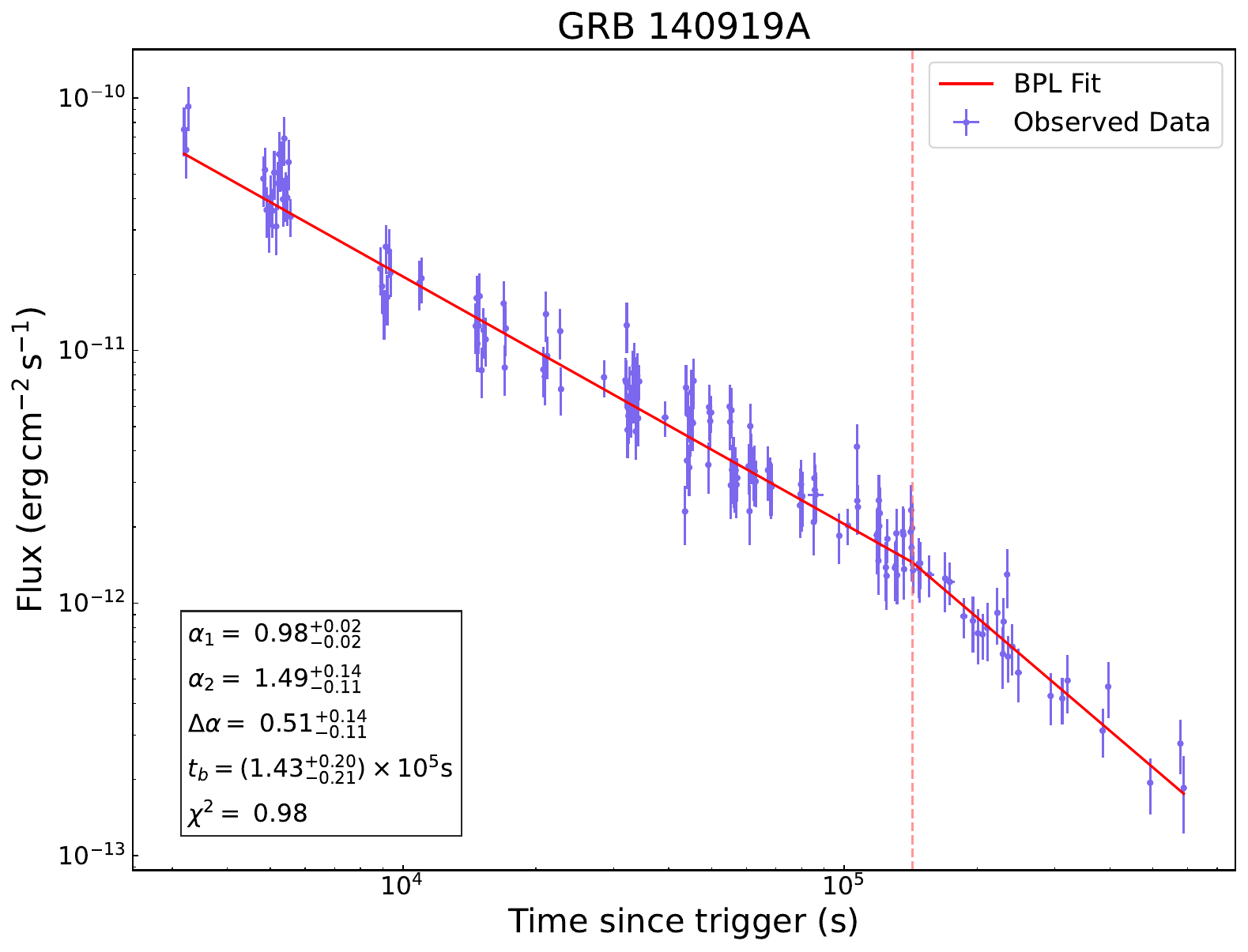}}%
\resizebox{45mm}{!}{\includegraphics[]{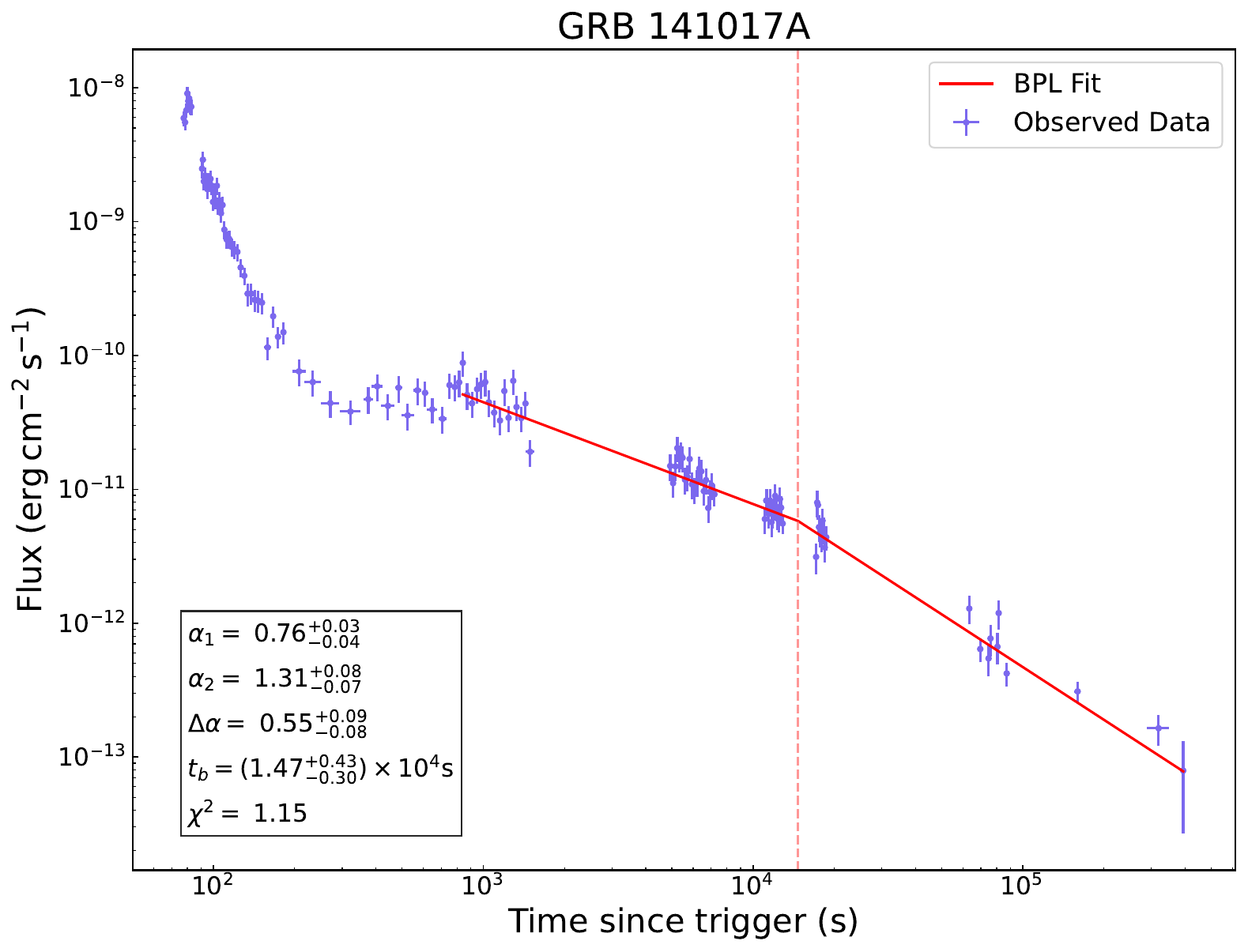}}%
\resizebox{45mm}{!}{\includegraphics[]{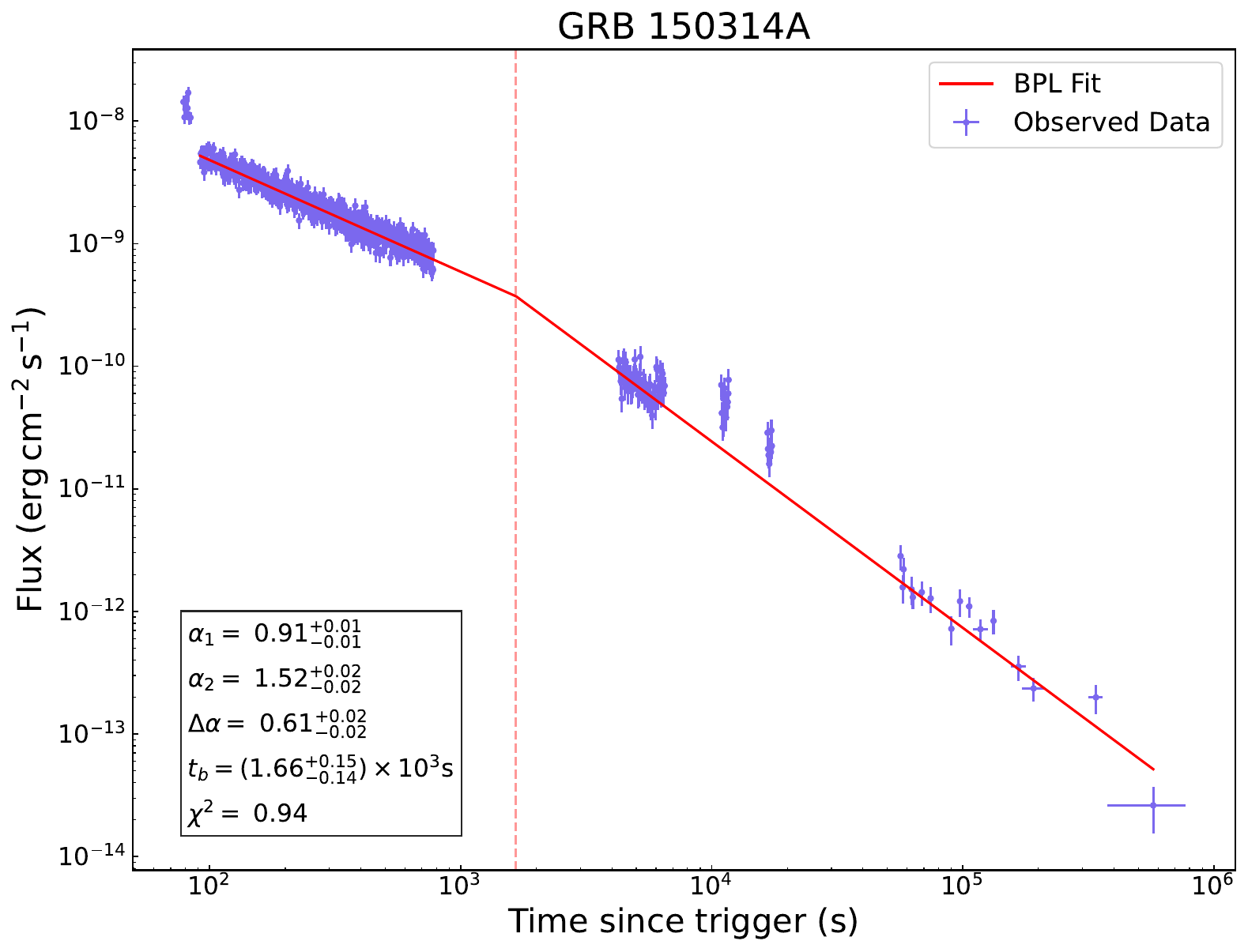}}%
\resizebox{45mm}{!}{\includegraphics[]{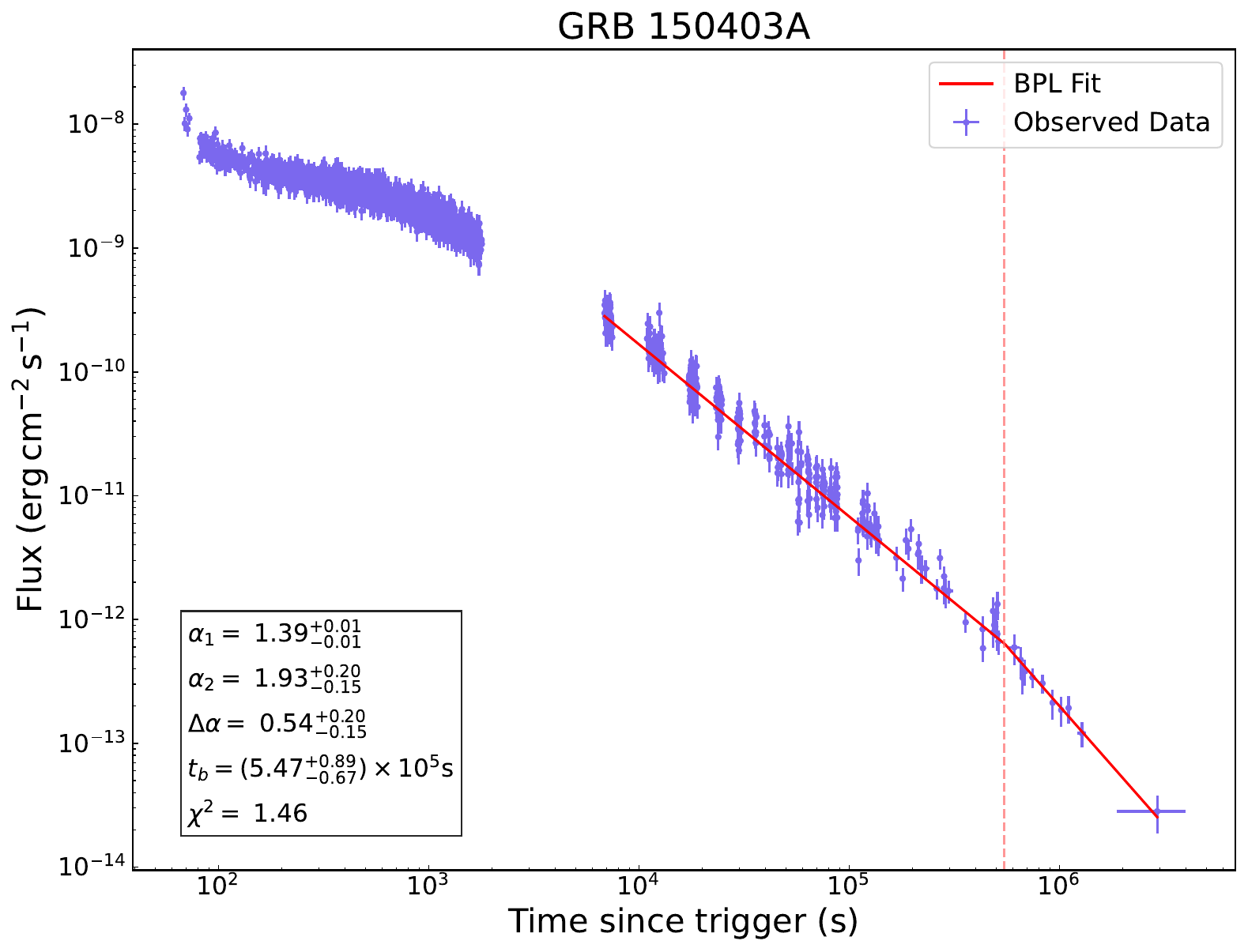}}\\
\resizebox{45mm}{!}{\includegraphics[]{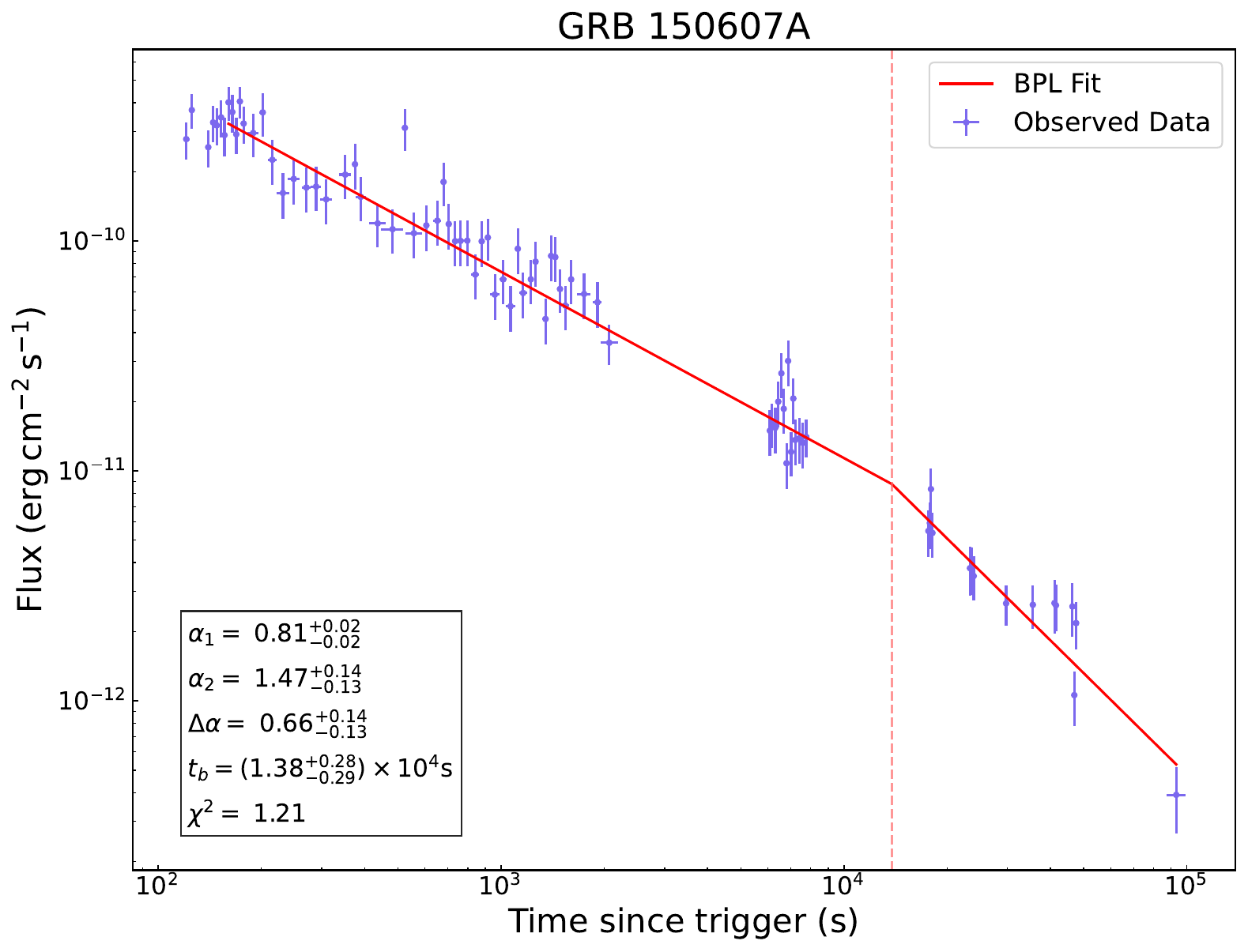}}%
\resizebox{45mm}{!}{\includegraphics[]{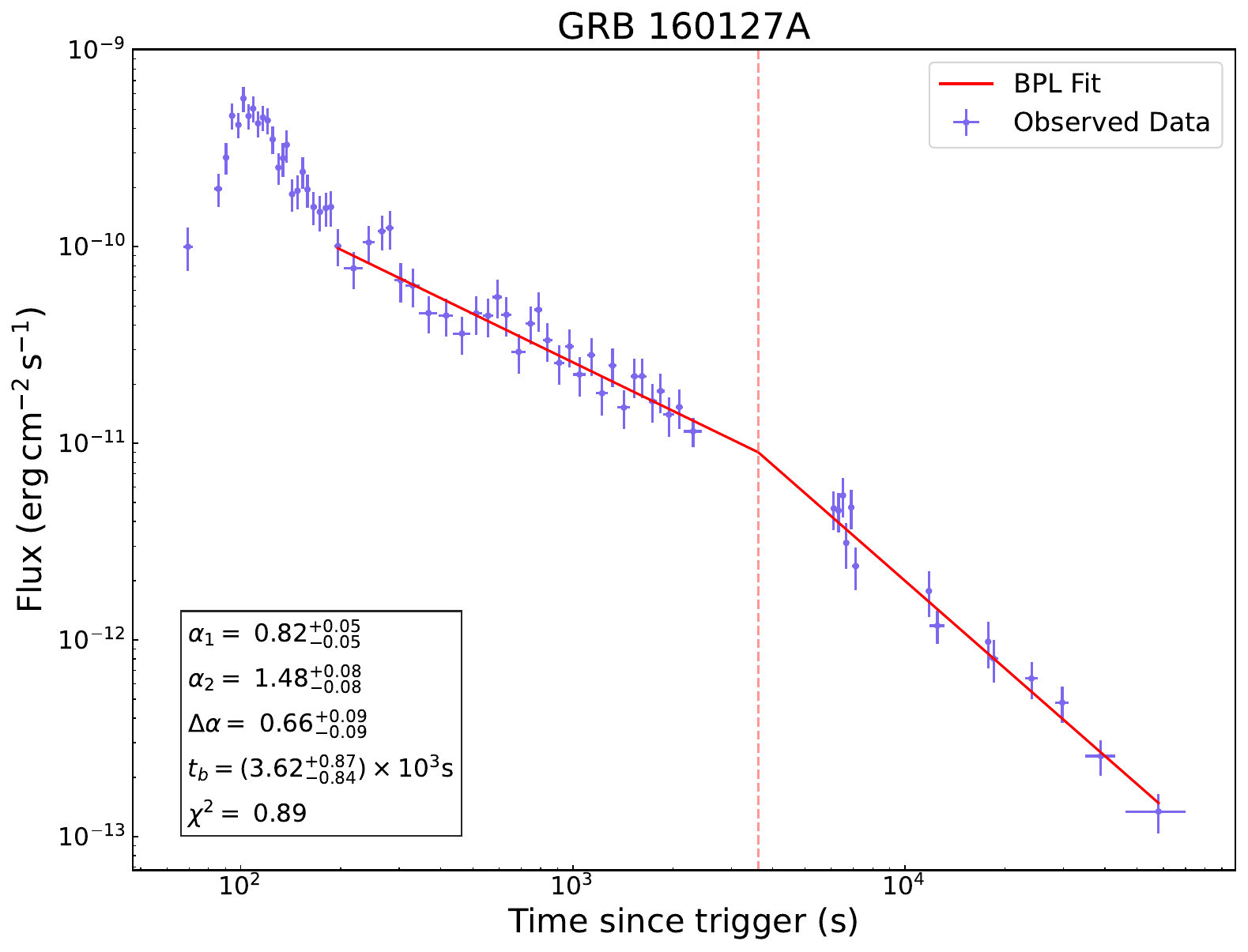}}%
\resizebox{45mm}{!}{\includegraphics[]{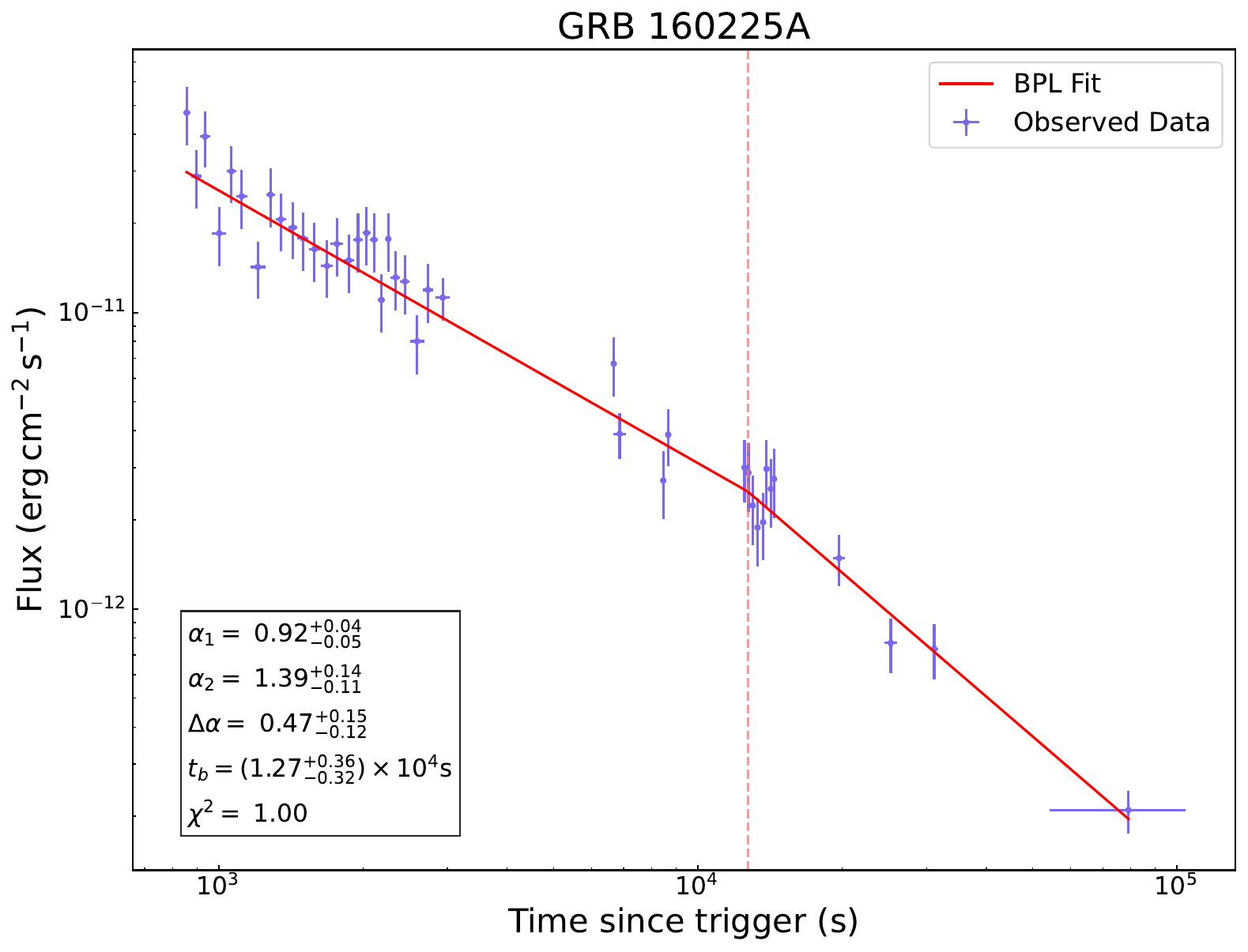}}%
\resizebox{45mm}{!}{\includegraphics[]{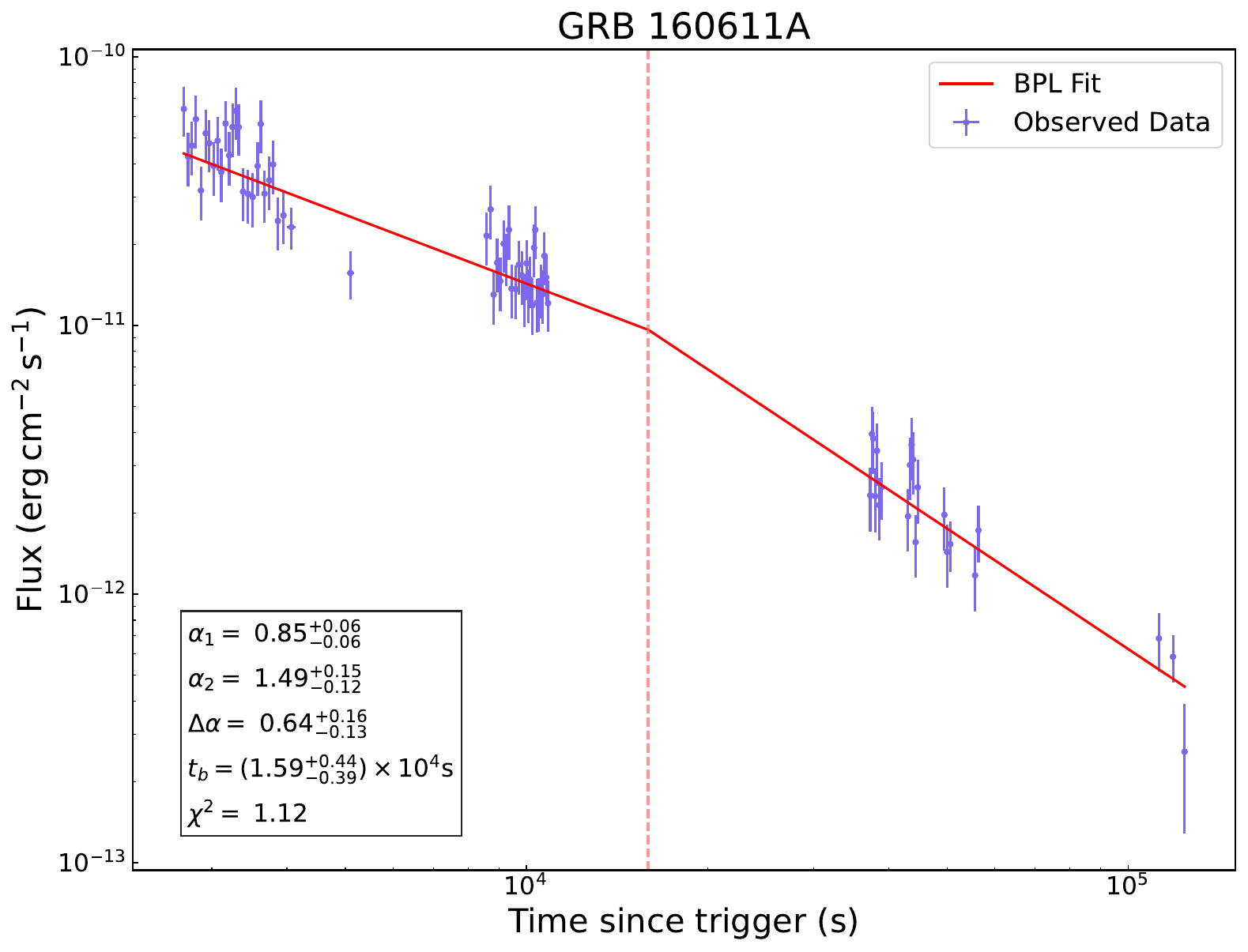}}\\
\resizebox{45mm}{!}{\includegraphics[]{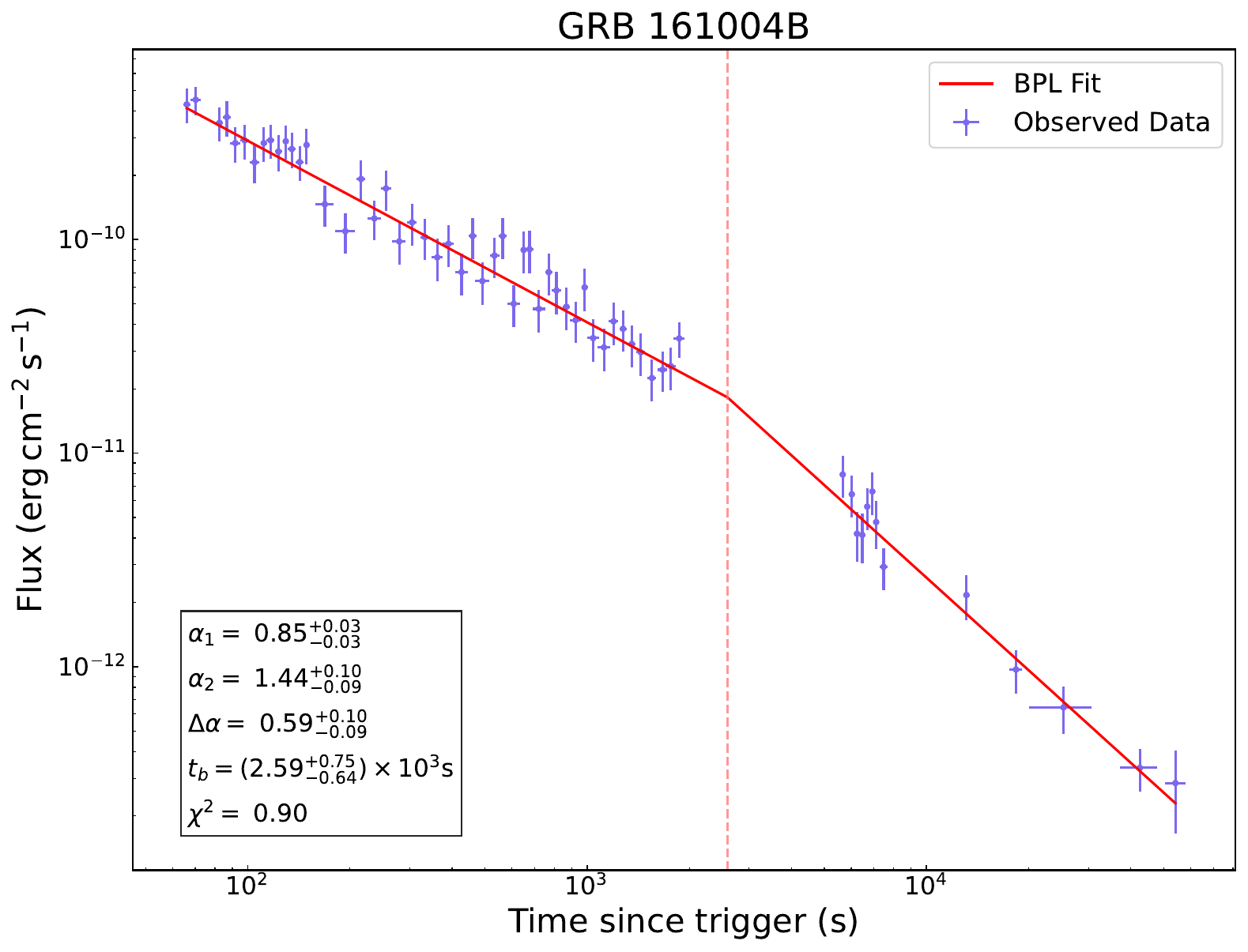}}%
\resizebox{45mm}{!}{\includegraphics[]{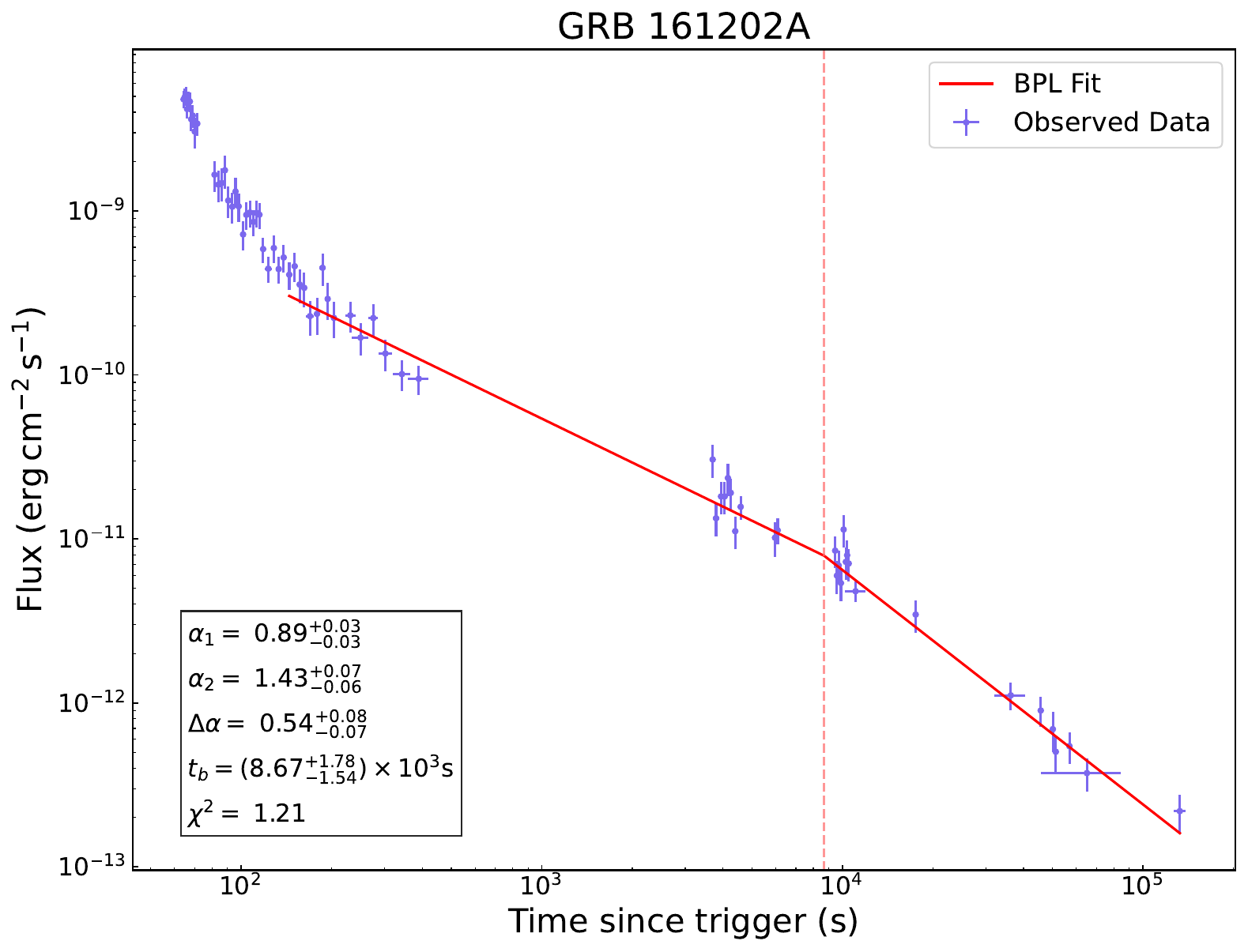}}%
\resizebox{45mm}{!}{\includegraphics[]{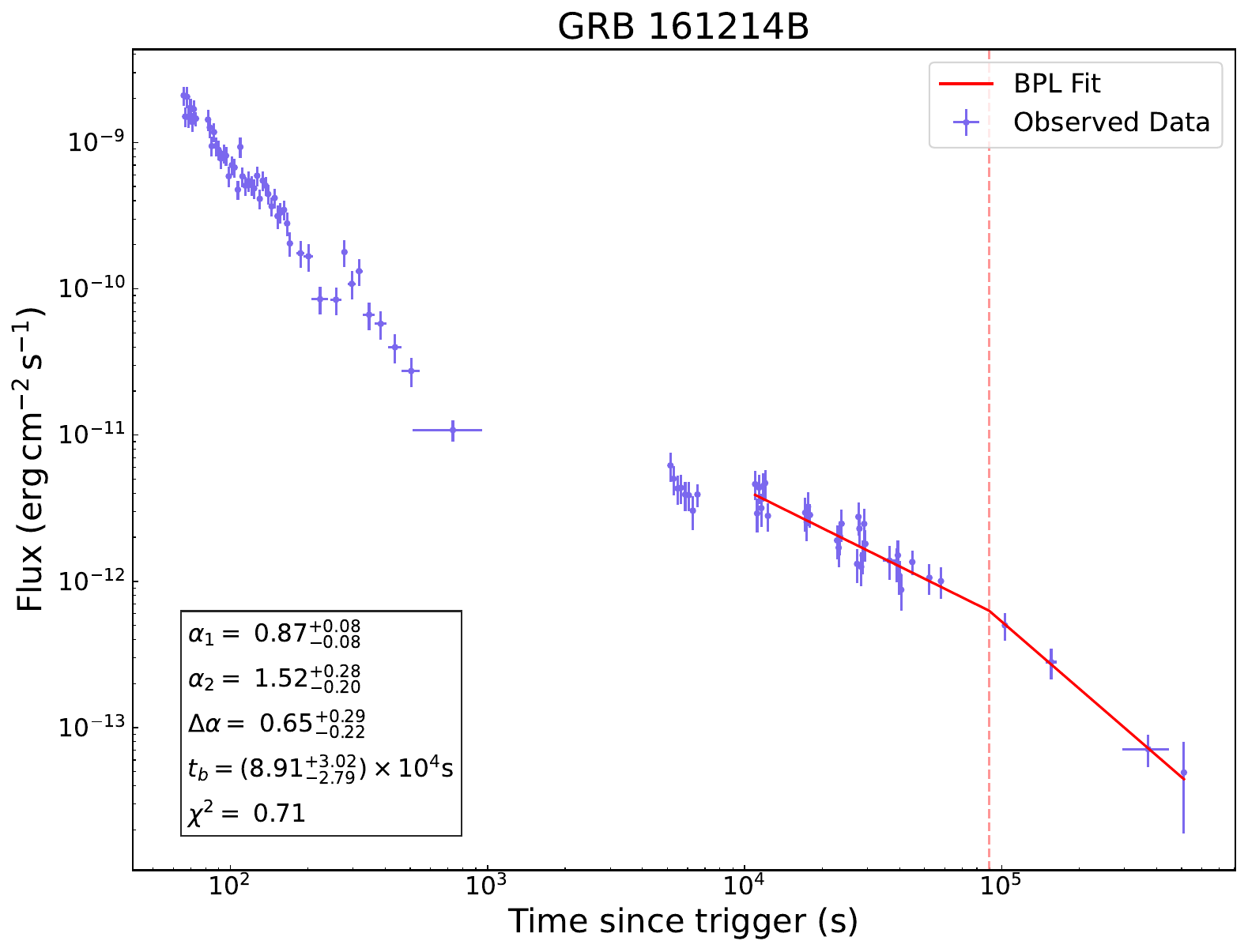}}%
\resizebox{45mm}{!}{\includegraphics[]{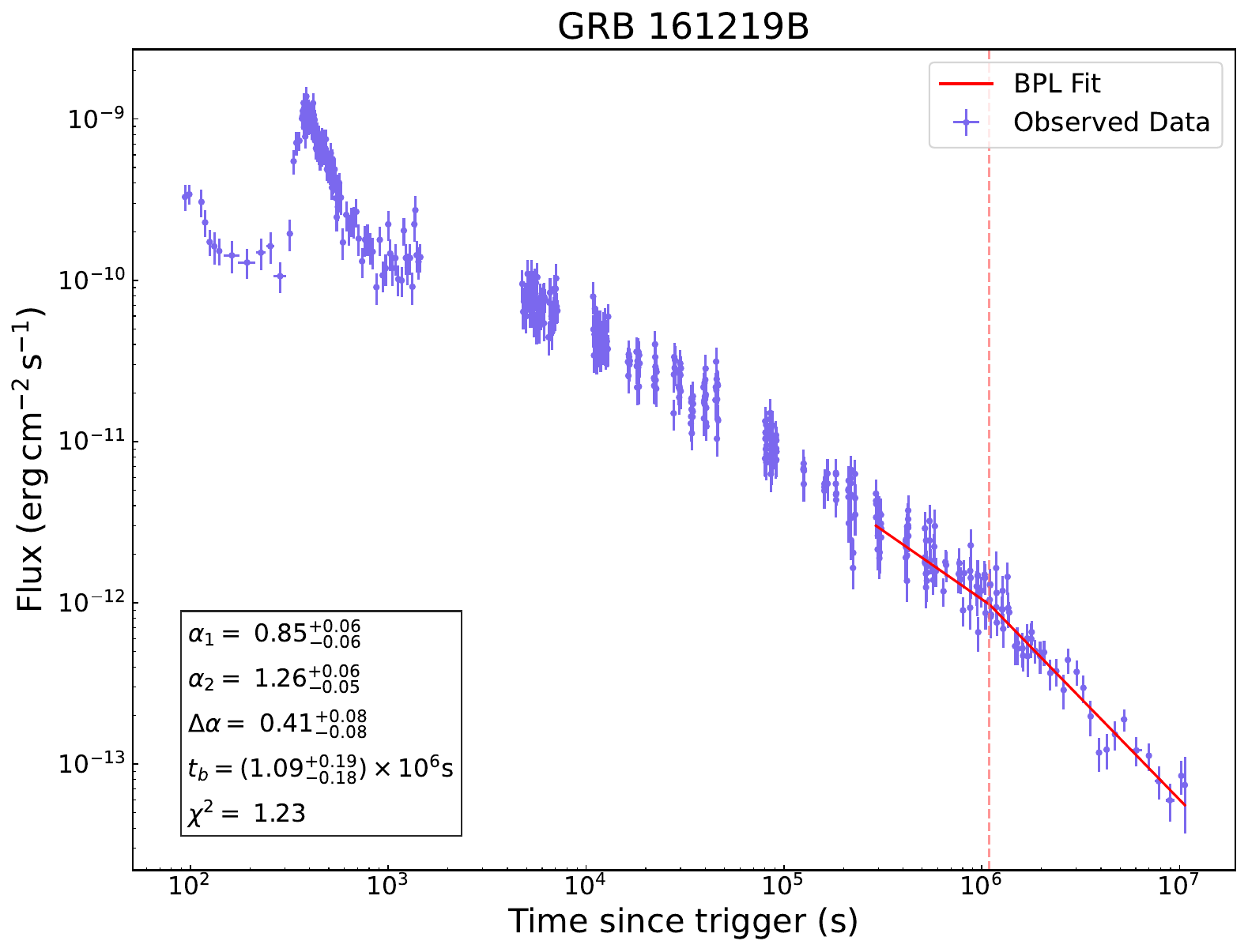}}\\
\resizebox{45mm}{!}{\includegraphics[]{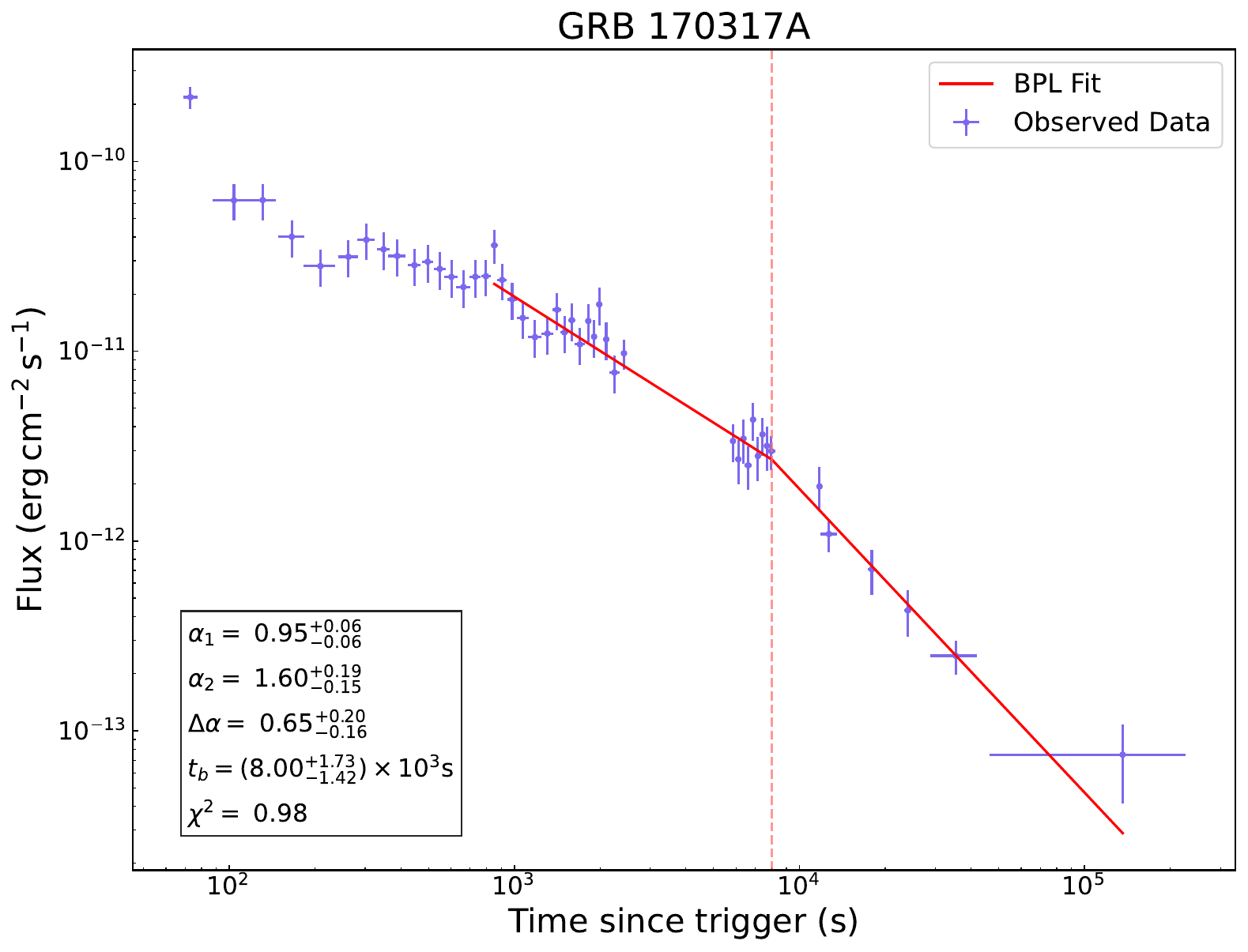}}%
\resizebox{45mm}{!}{\includegraphics[]{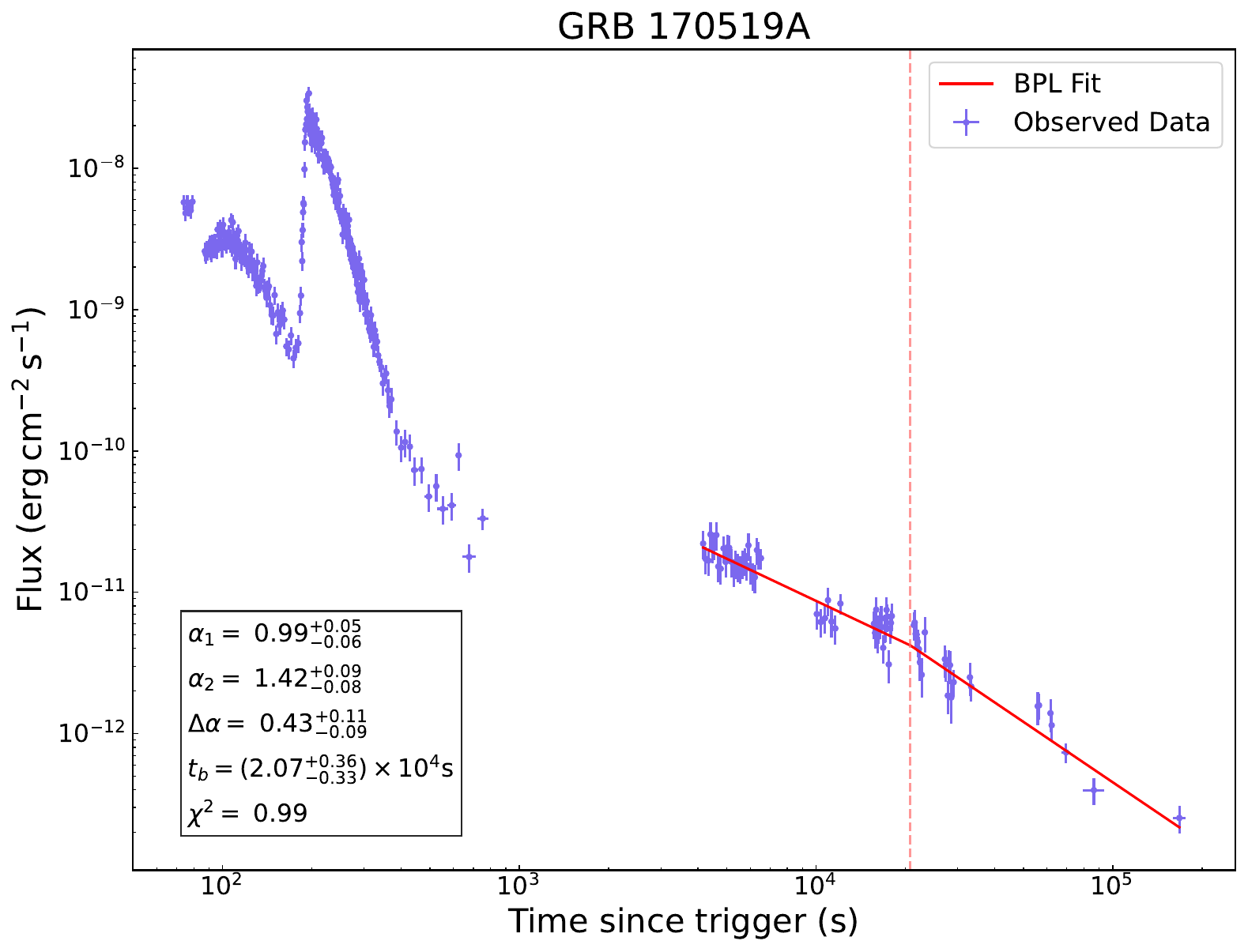}}%
\resizebox{45mm}{!}{\includegraphics[]{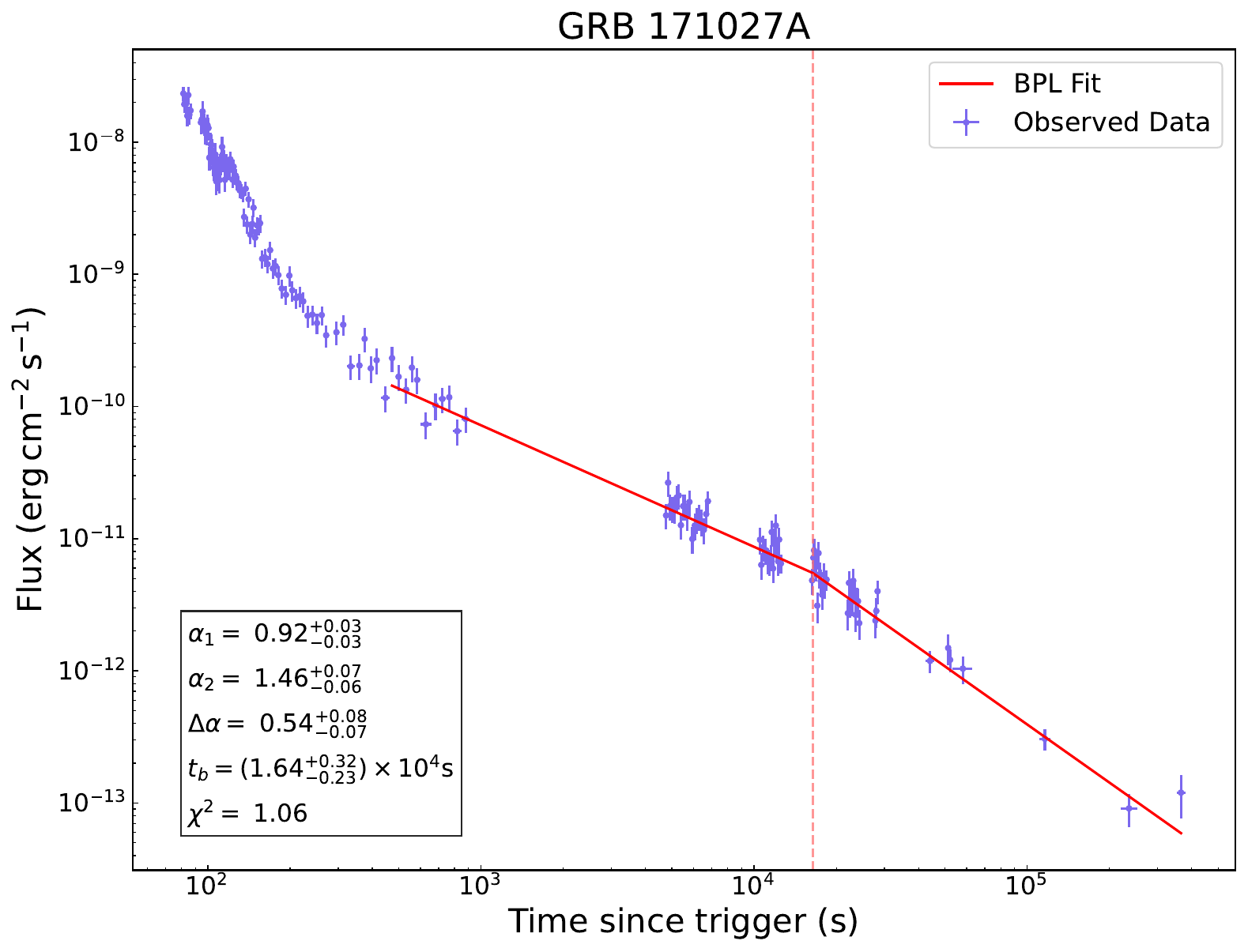}}%
\resizebox{45mm}{!}{\includegraphics[]{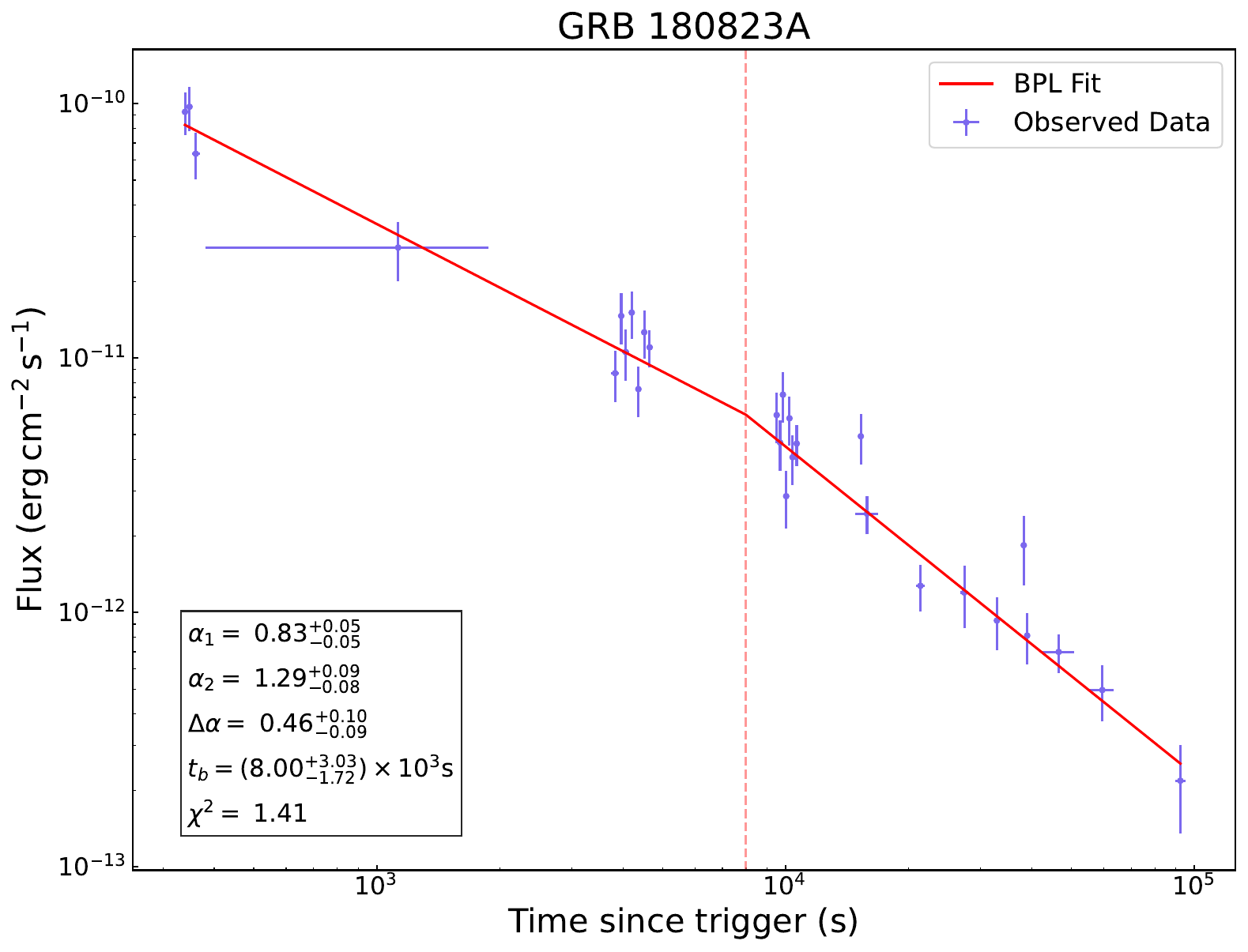}}\\
\resizebox{45mm}{!}{\includegraphics[]{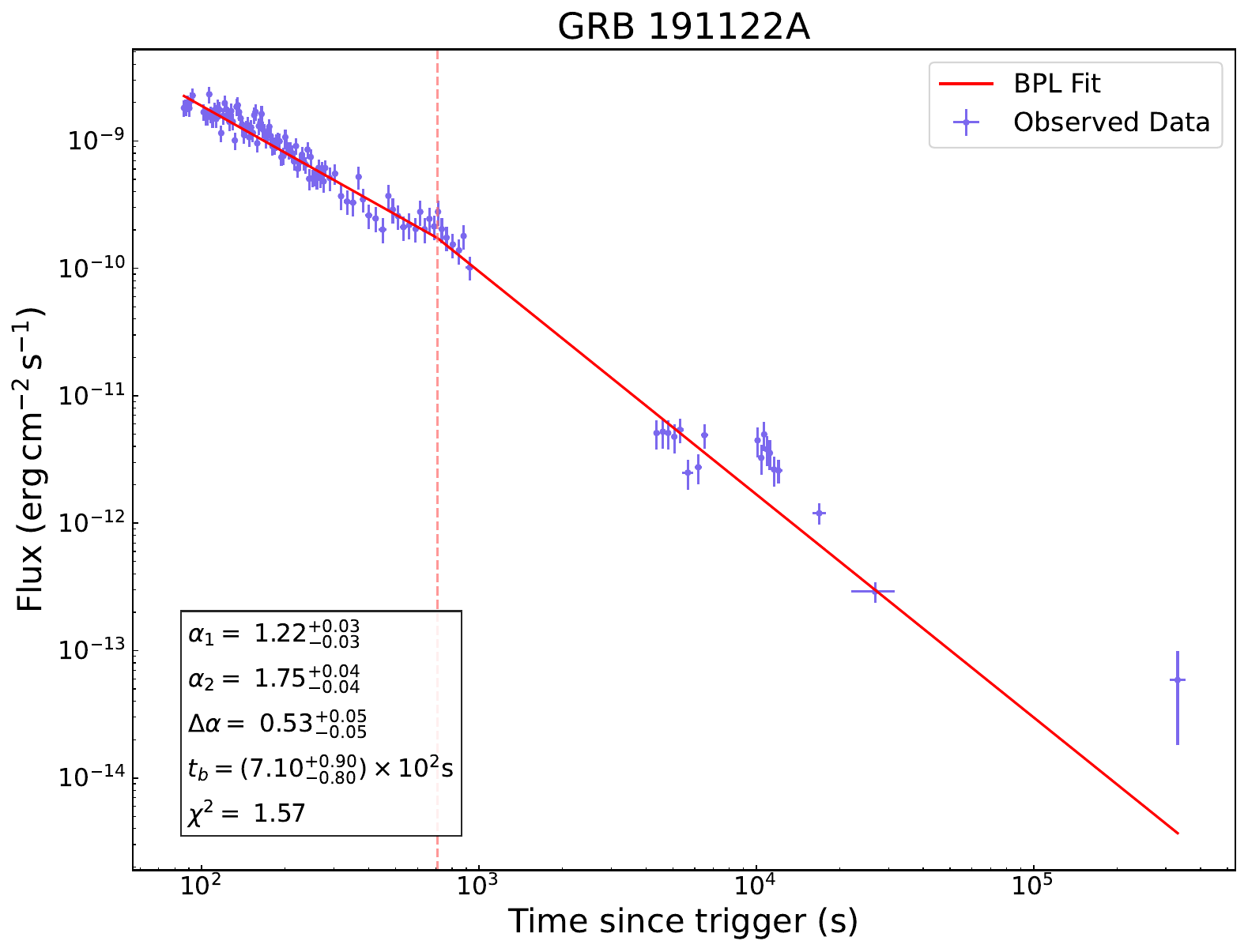}}%
\resizebox{45mm}{!}{\includegraphics[]{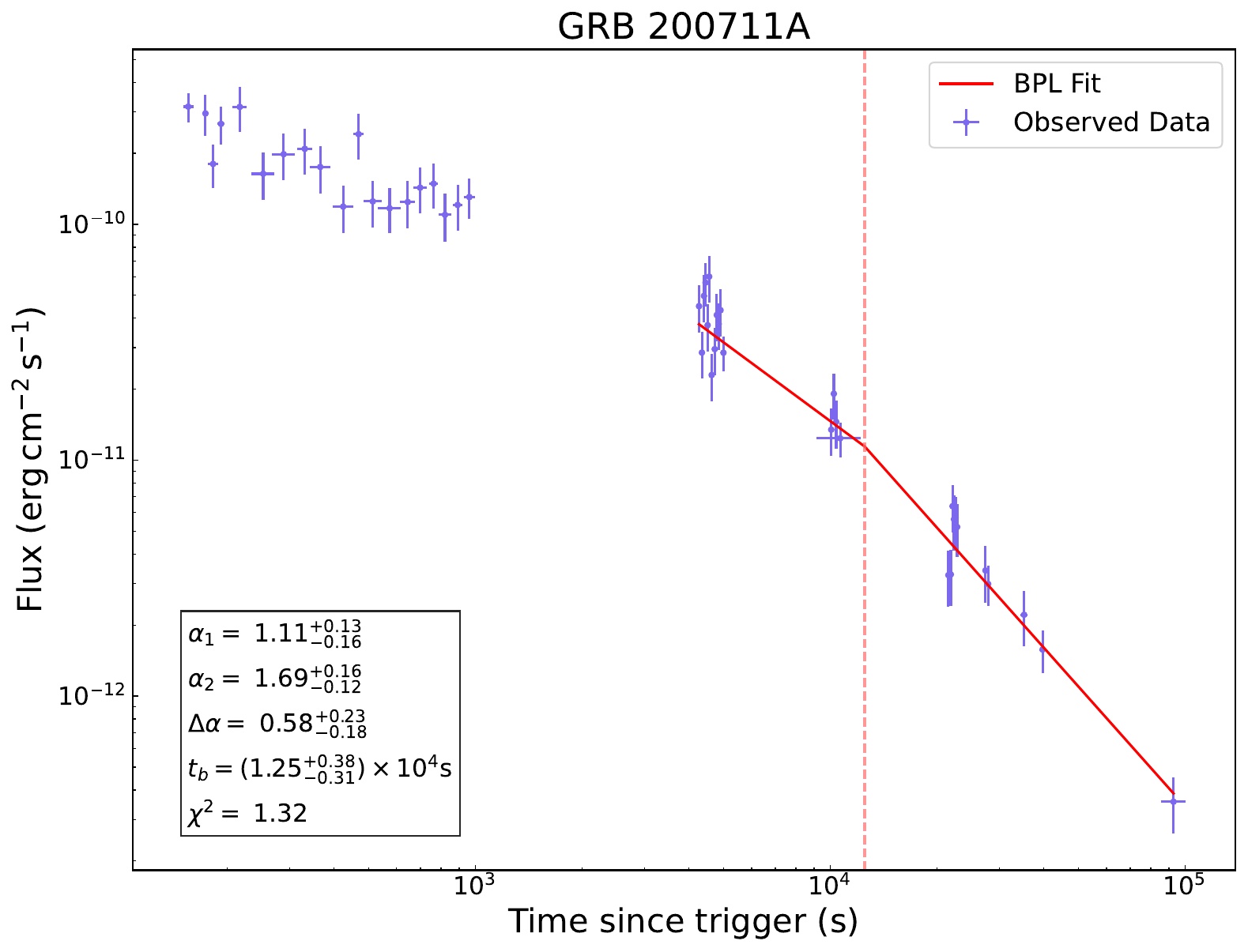}}%
\resizebox{45mm}{!}{\includegraphics[]{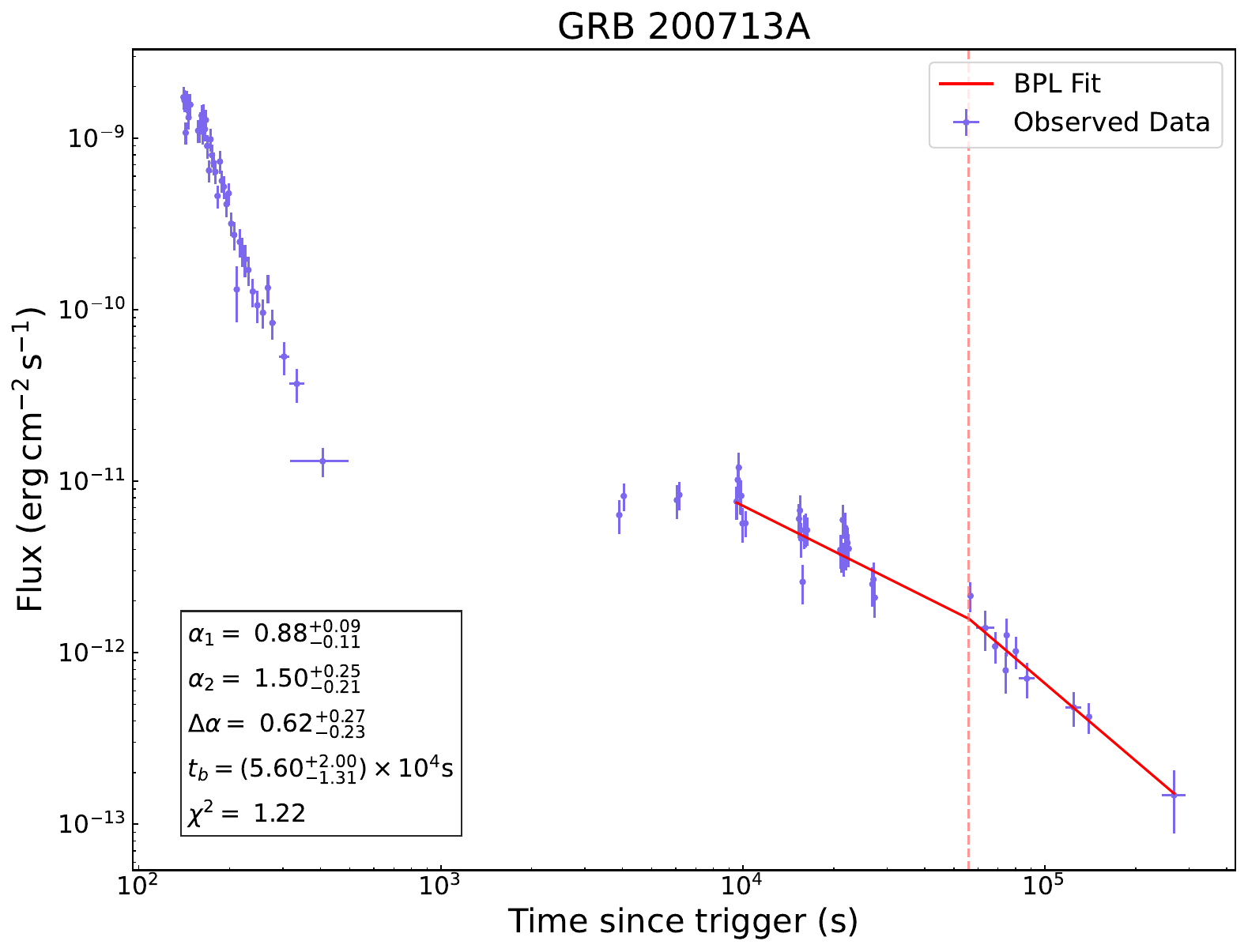}}%
\resizebox{45mm}{!}{\includegraphics[]{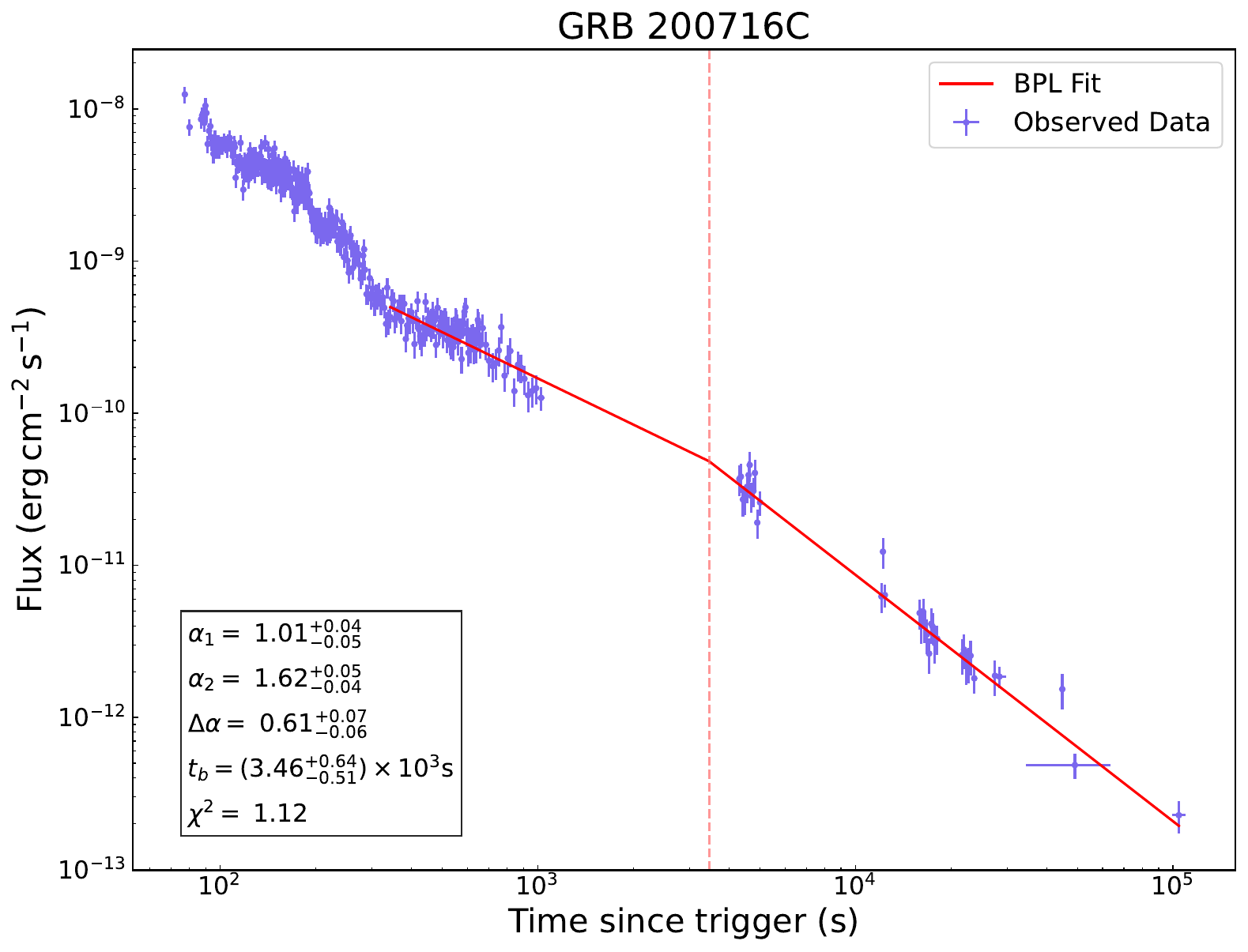}}\\
\resizebox{45mm}{!}{\includegraphics[]{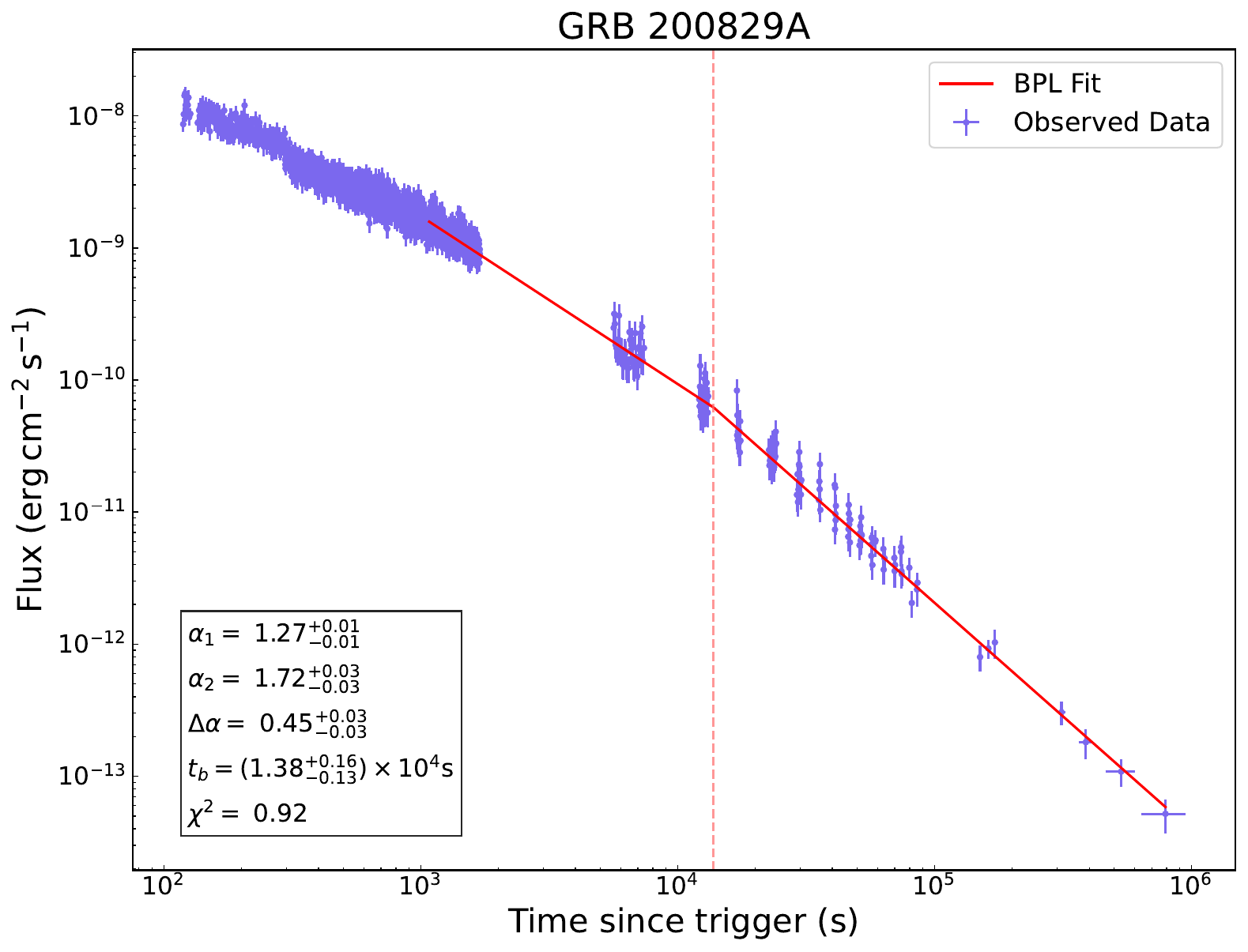}}%
\resizebox{45mm}{!}{\includegraphics[]{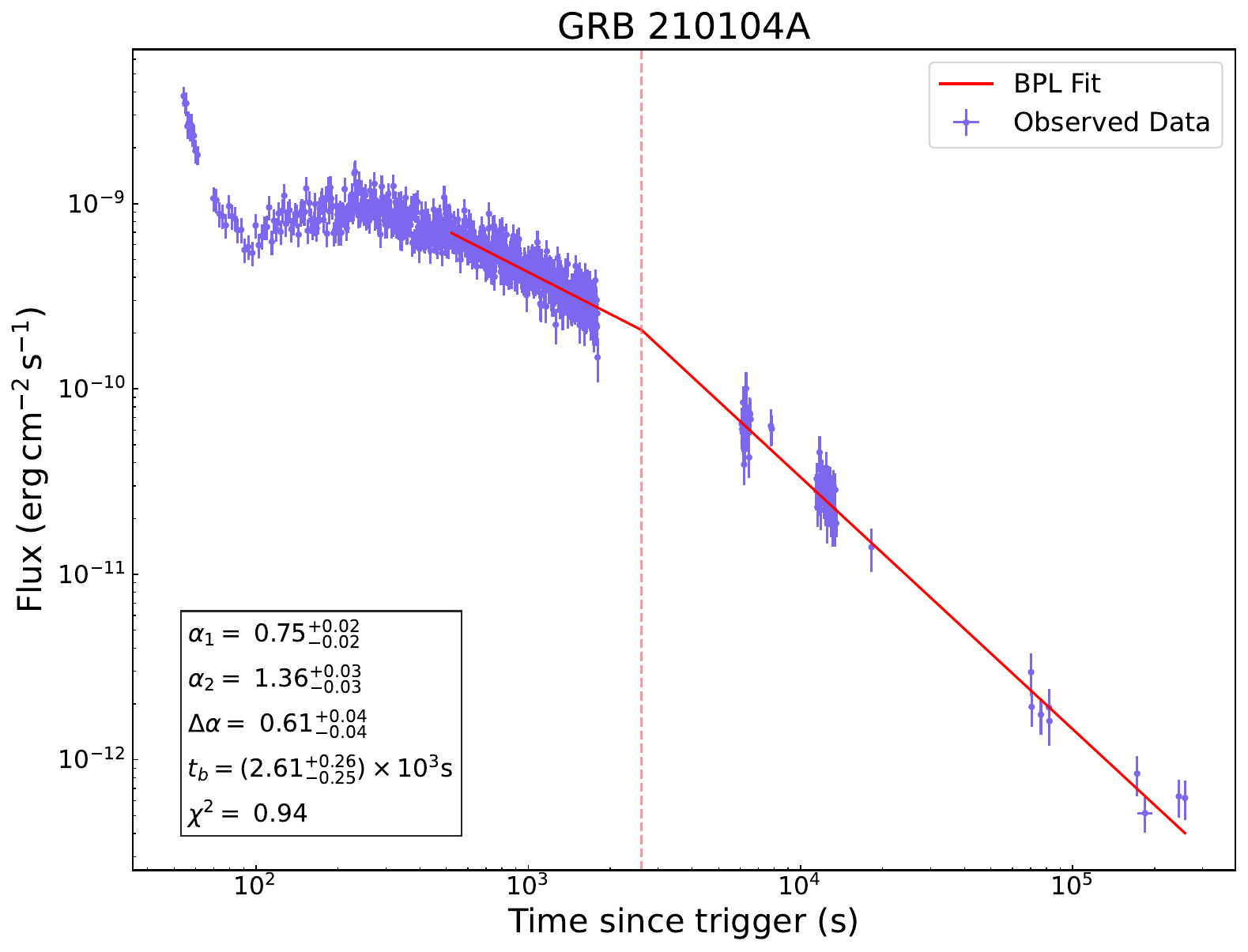}}%
\resizebox{45mm}{!}{\includegraphics[]{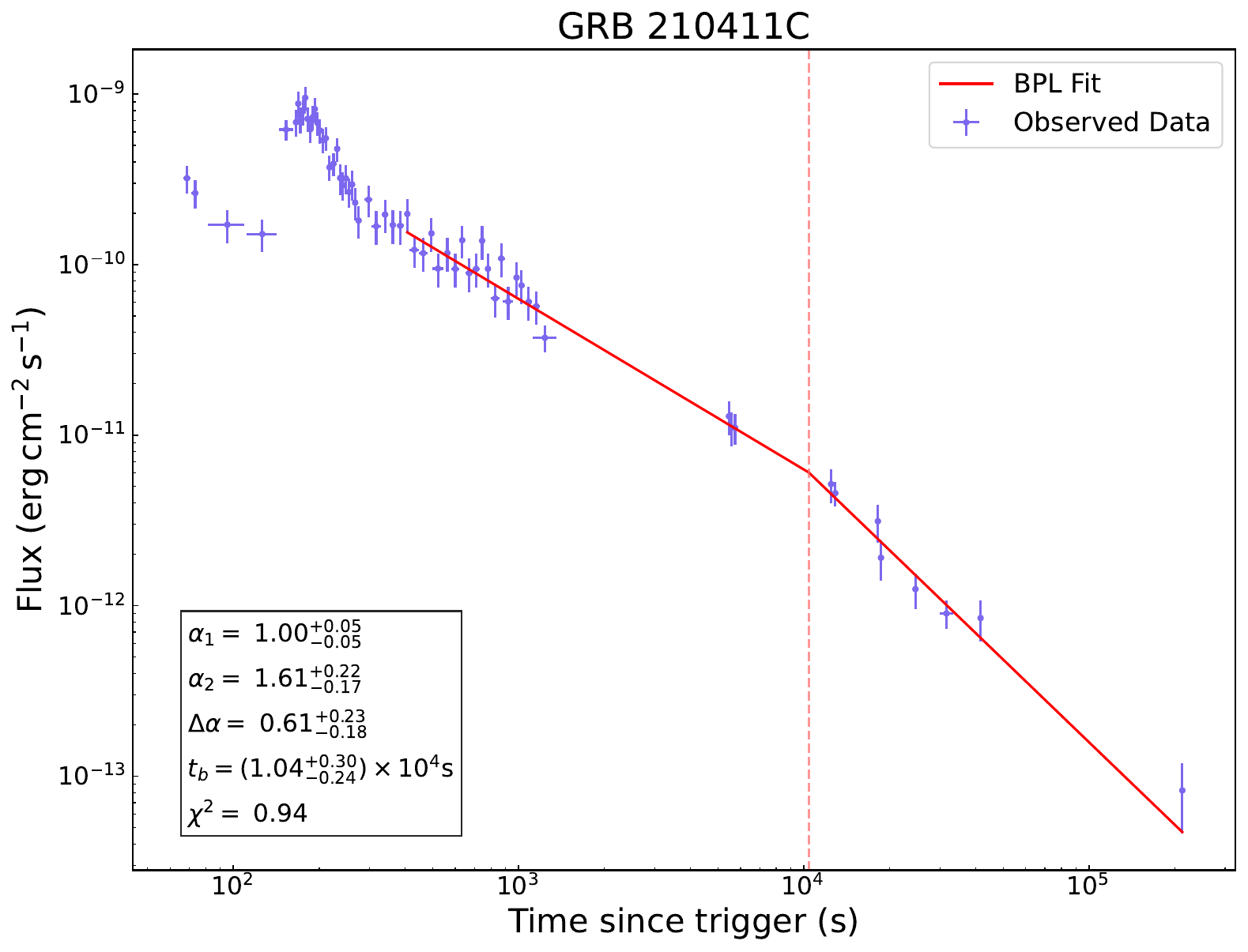}}%
\resizebox{45mm}{!}{\includegraphics[]{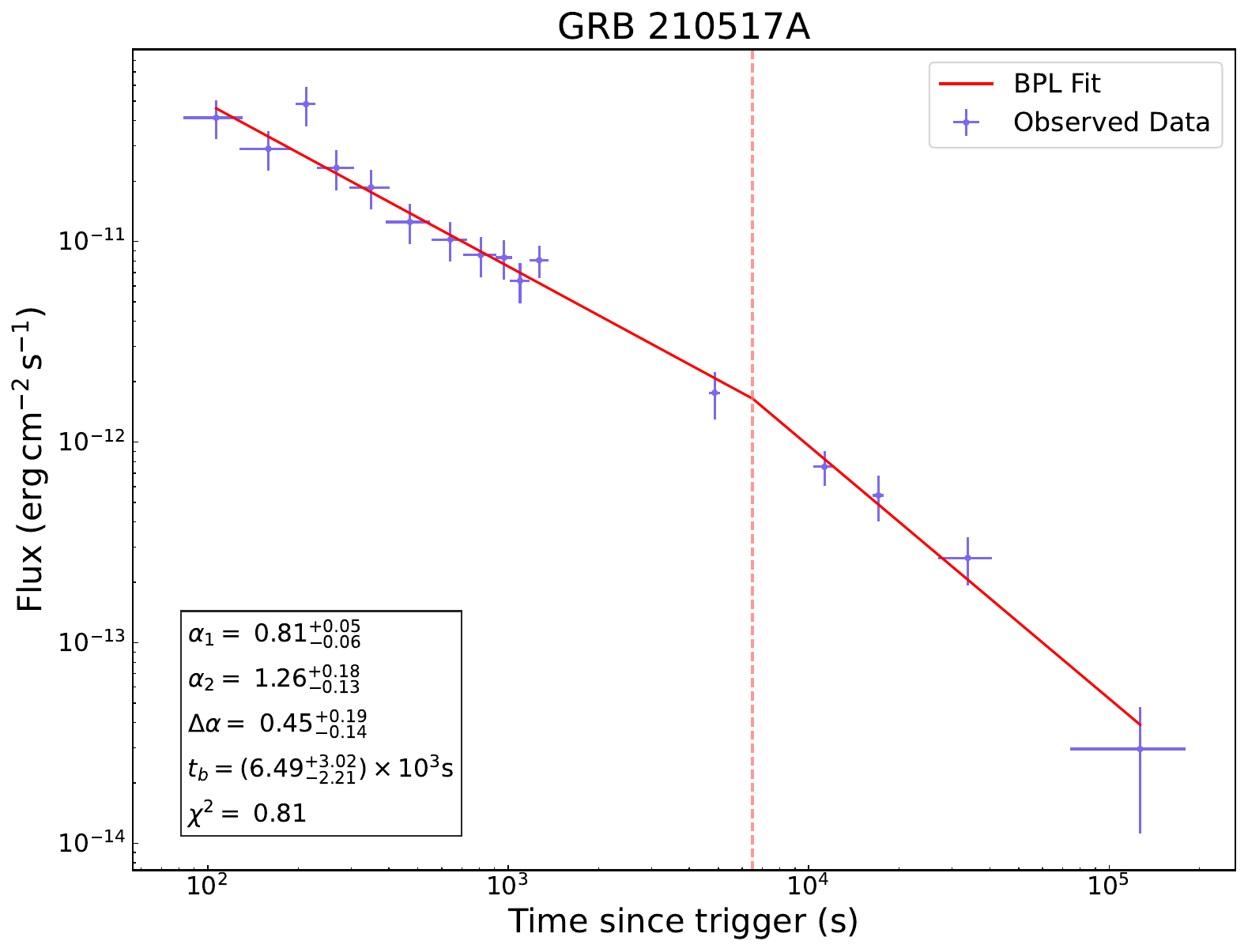}}\\
\resizebox{45mm}{!}{\includegraphics[]{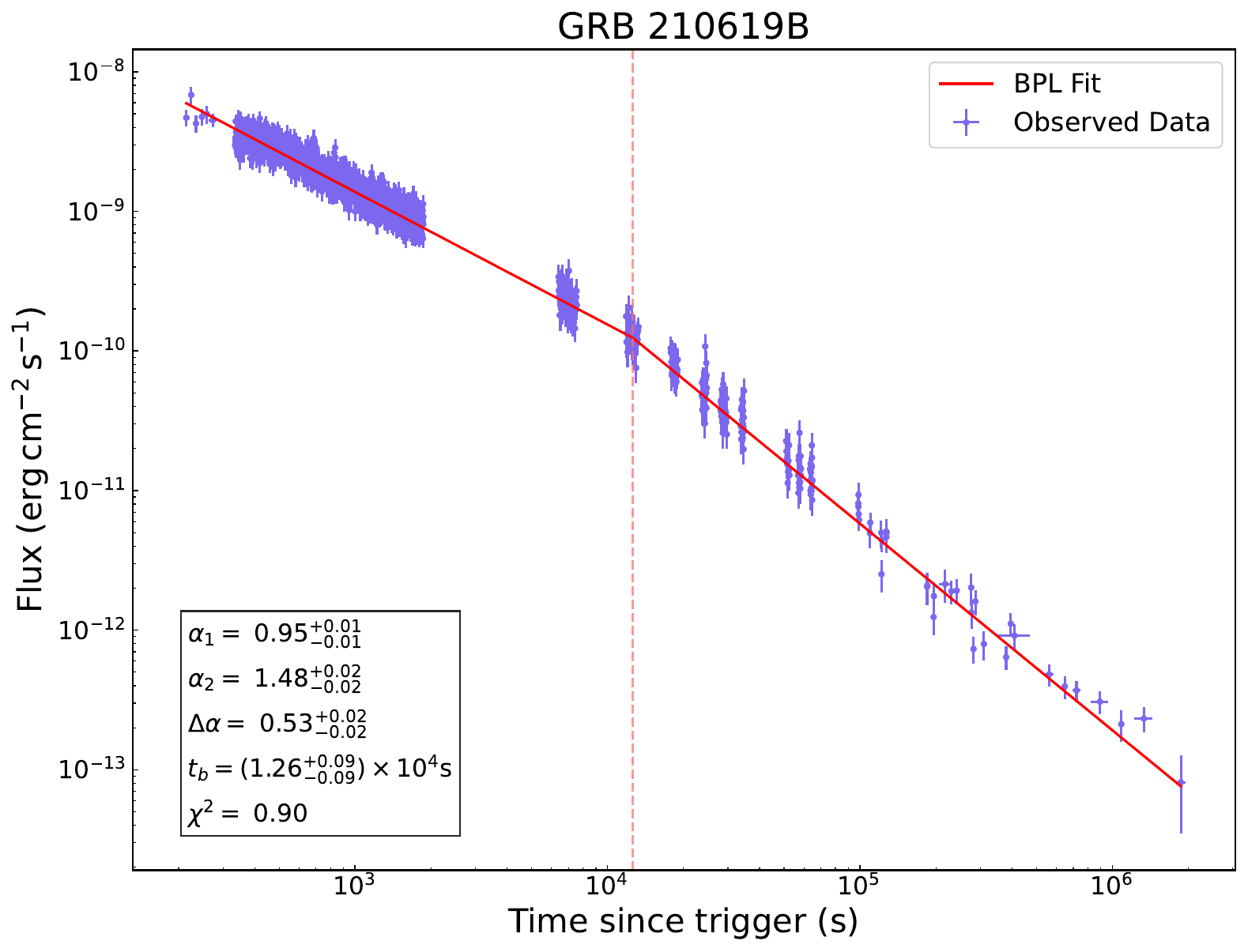}}%
\resizebox{45mm}{!}{\includegraphics[]{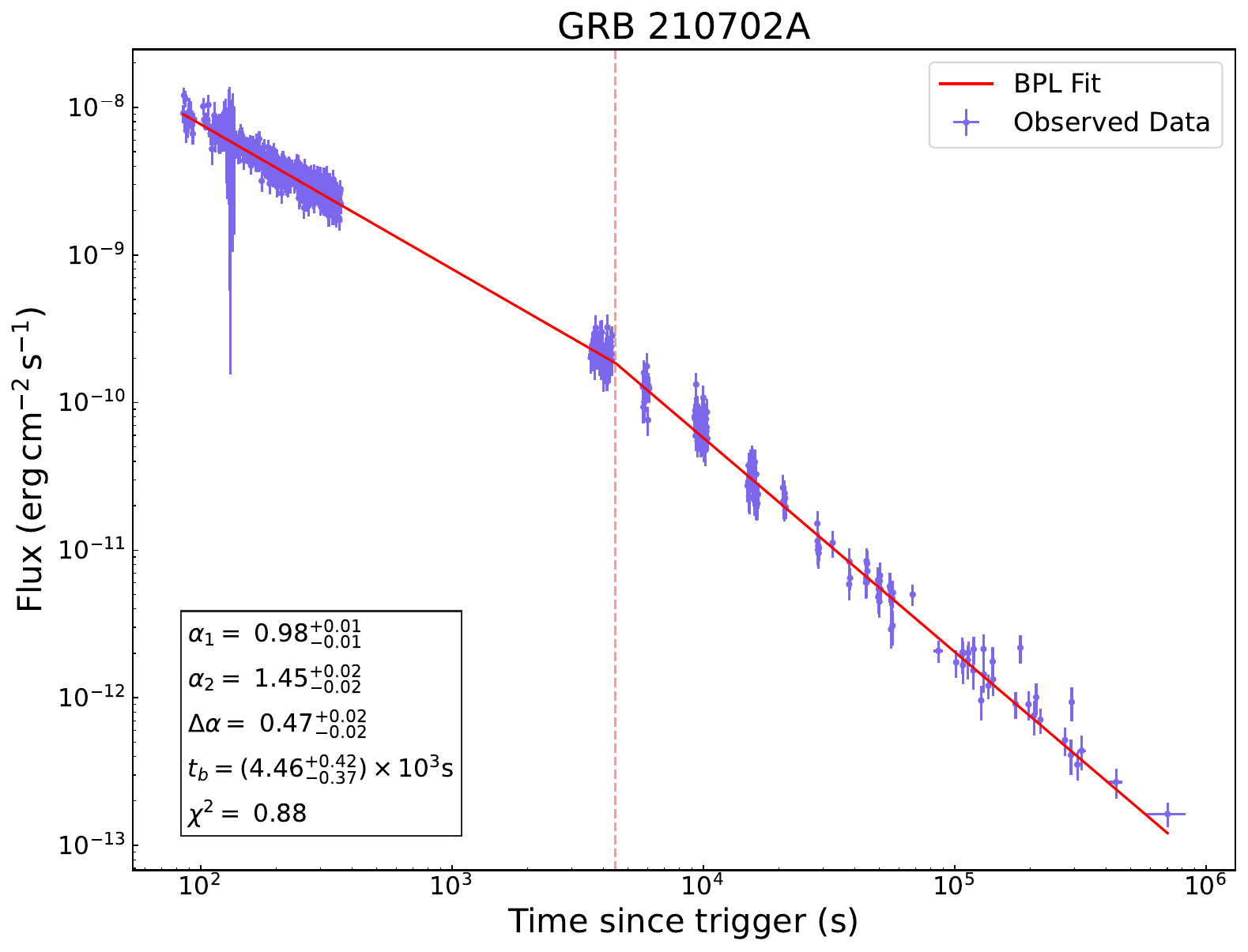}}%
\resizebox{45mm}{!}{\includegraphics[]{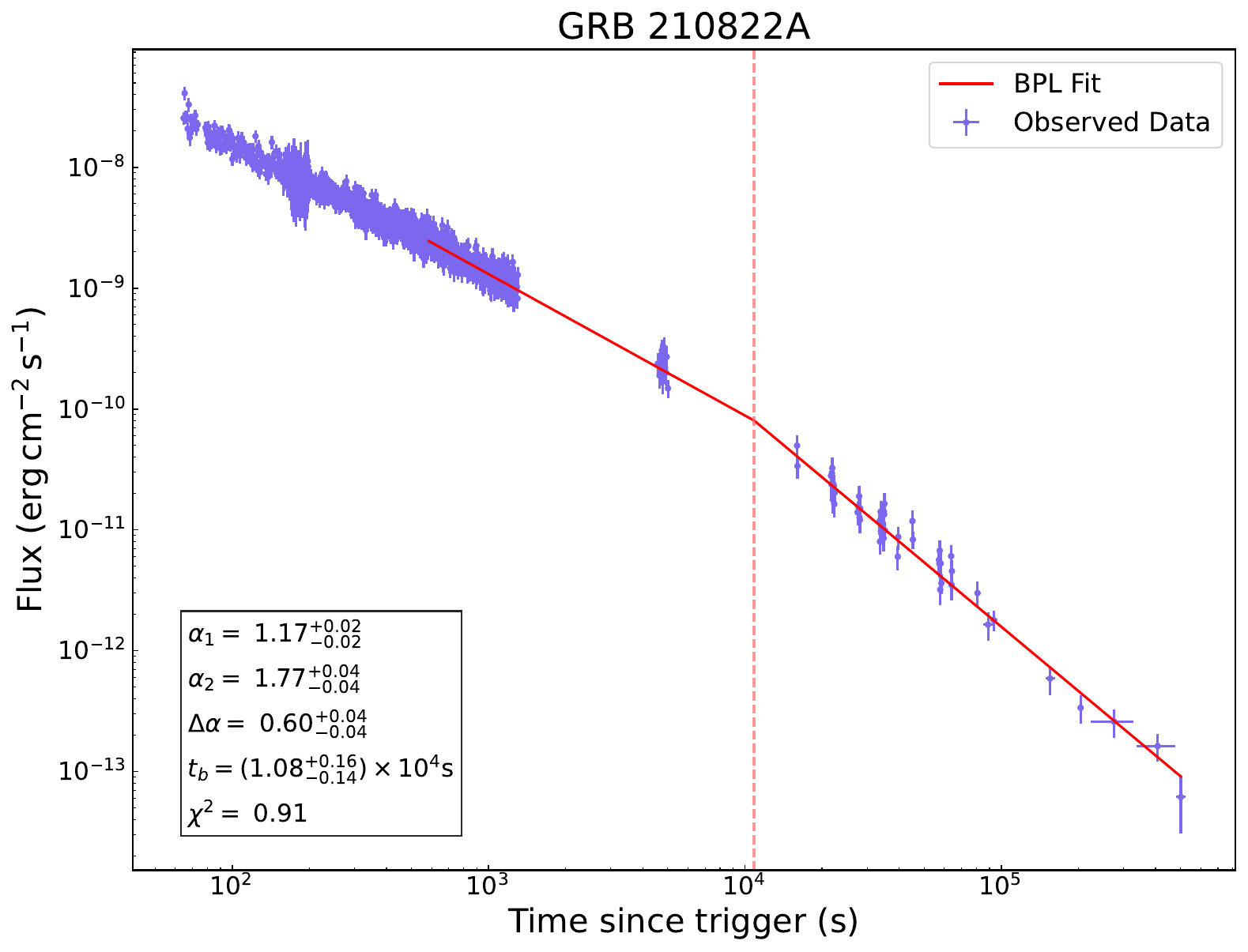}}%
\resizebox{45mm}{!}{\includegraphics[]{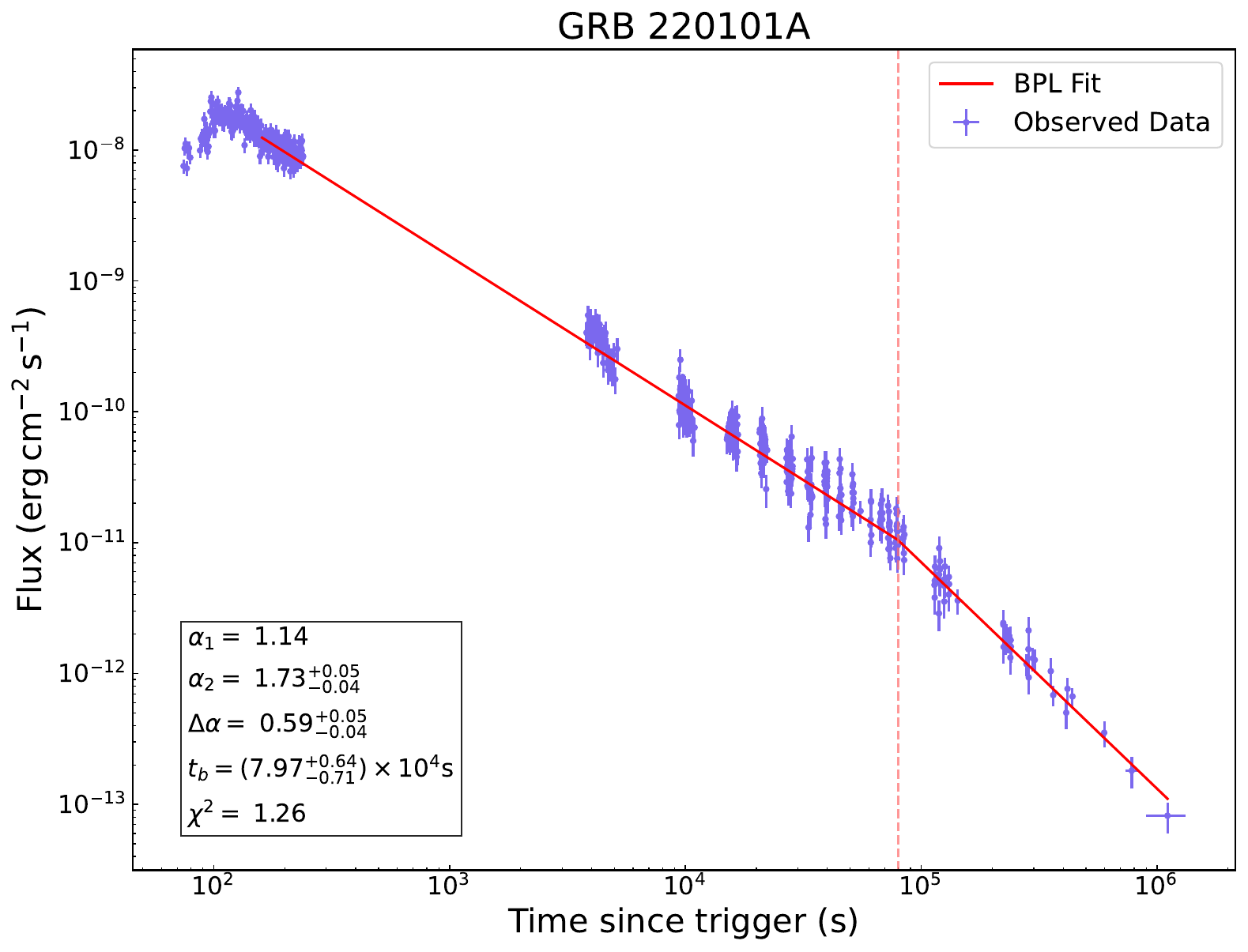}}\\
\caption{(Continued)}
\end{figure*}

\clearpage
\addtocounter{figure}{-1}
\begin{figure*}
\centering
\resizebox{45mm}{!}{\includegraphics[]{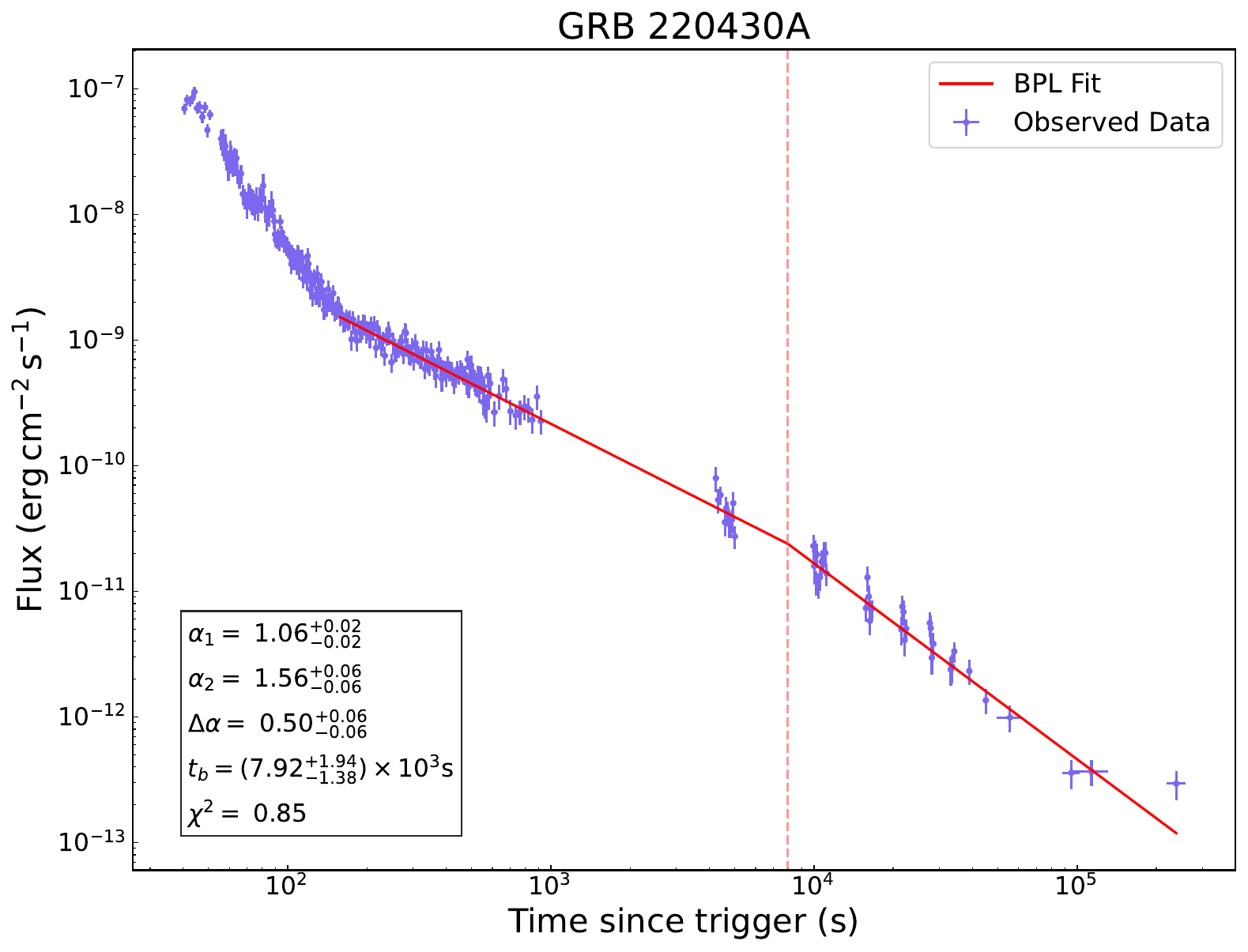}}%
\resizebox{45mm}{!}{\includegraphics[]{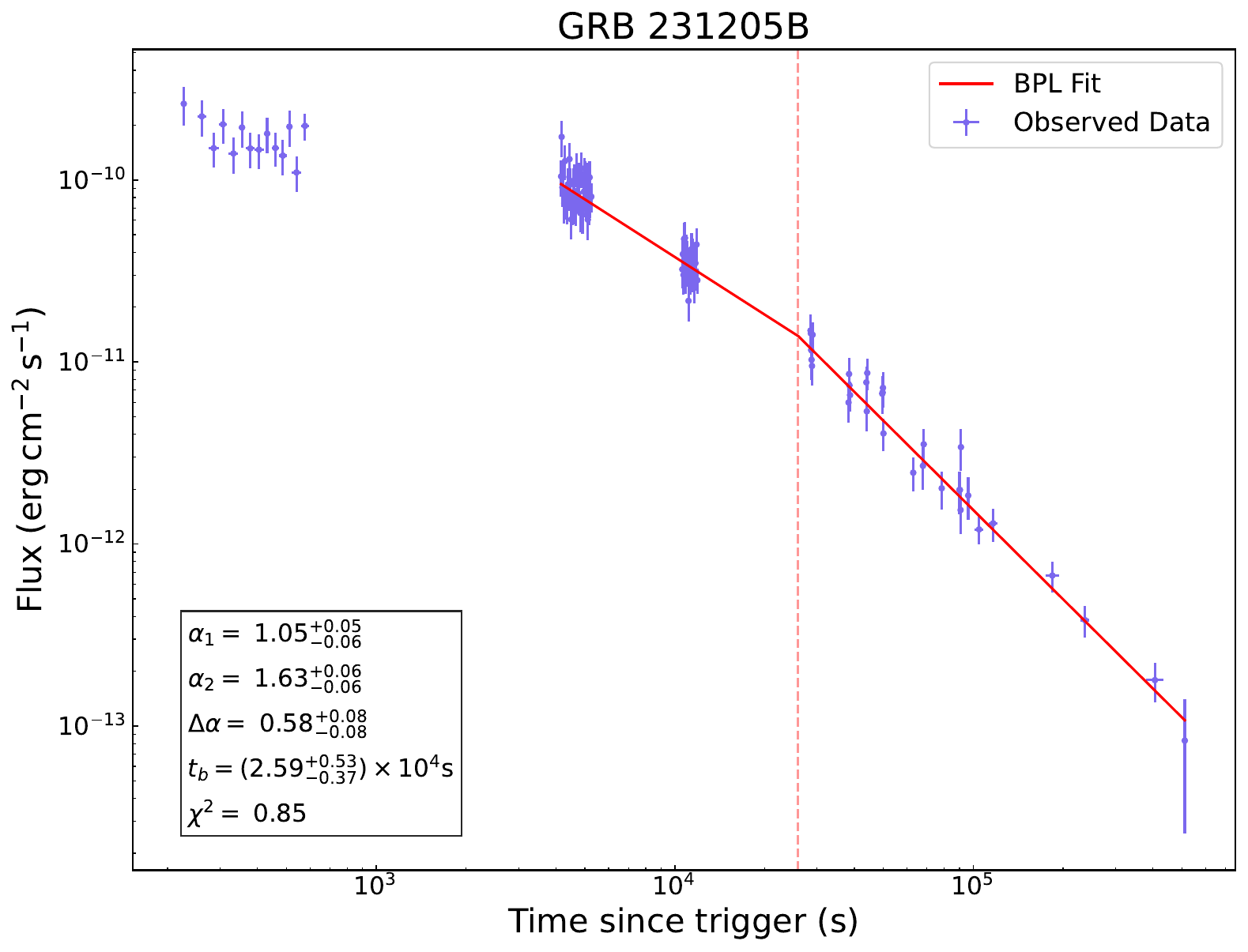}}%
\resizebox{45mm}{!}{\includegraphics[]{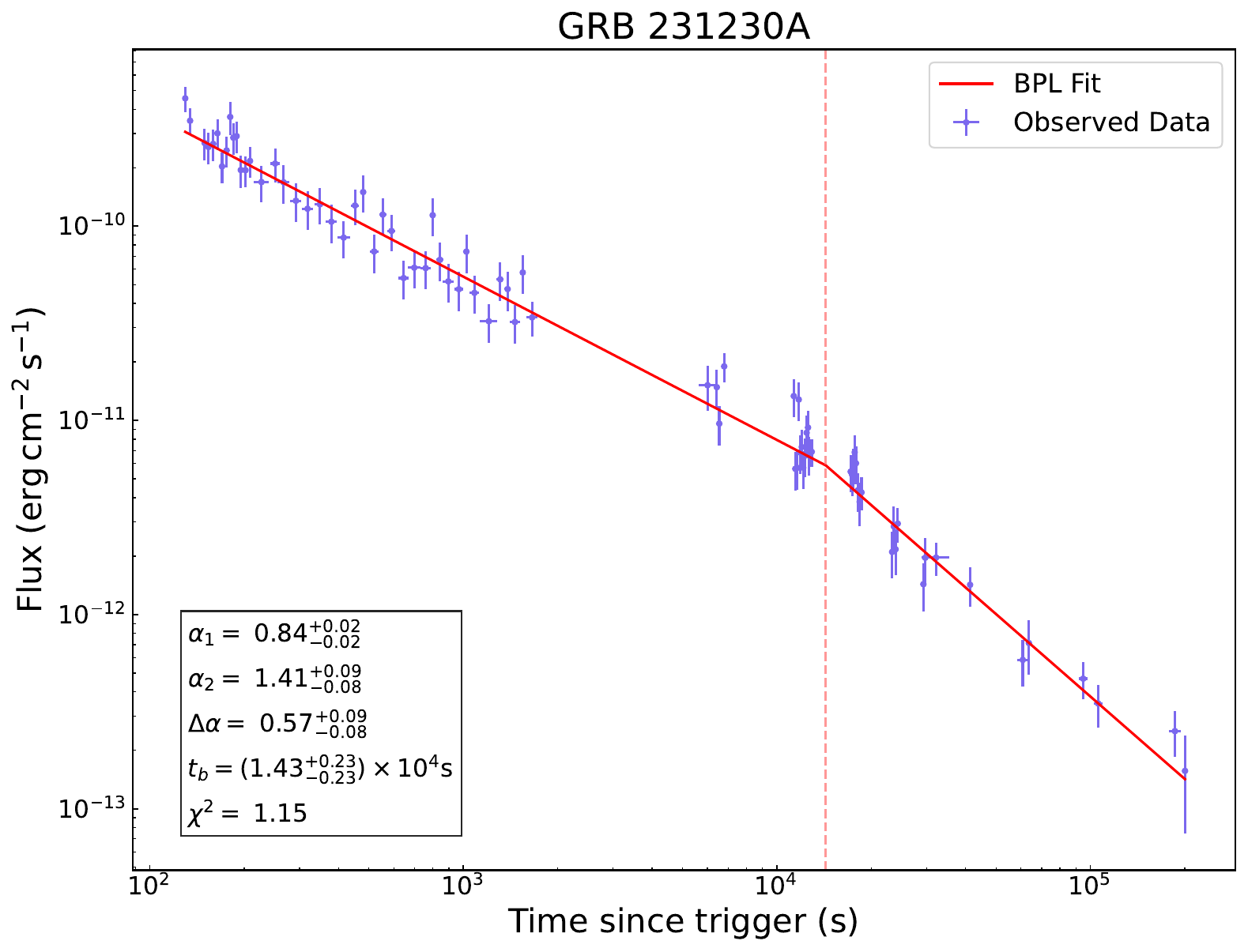}}%
\resizebox{45mm}{!}{\includegraphics[]{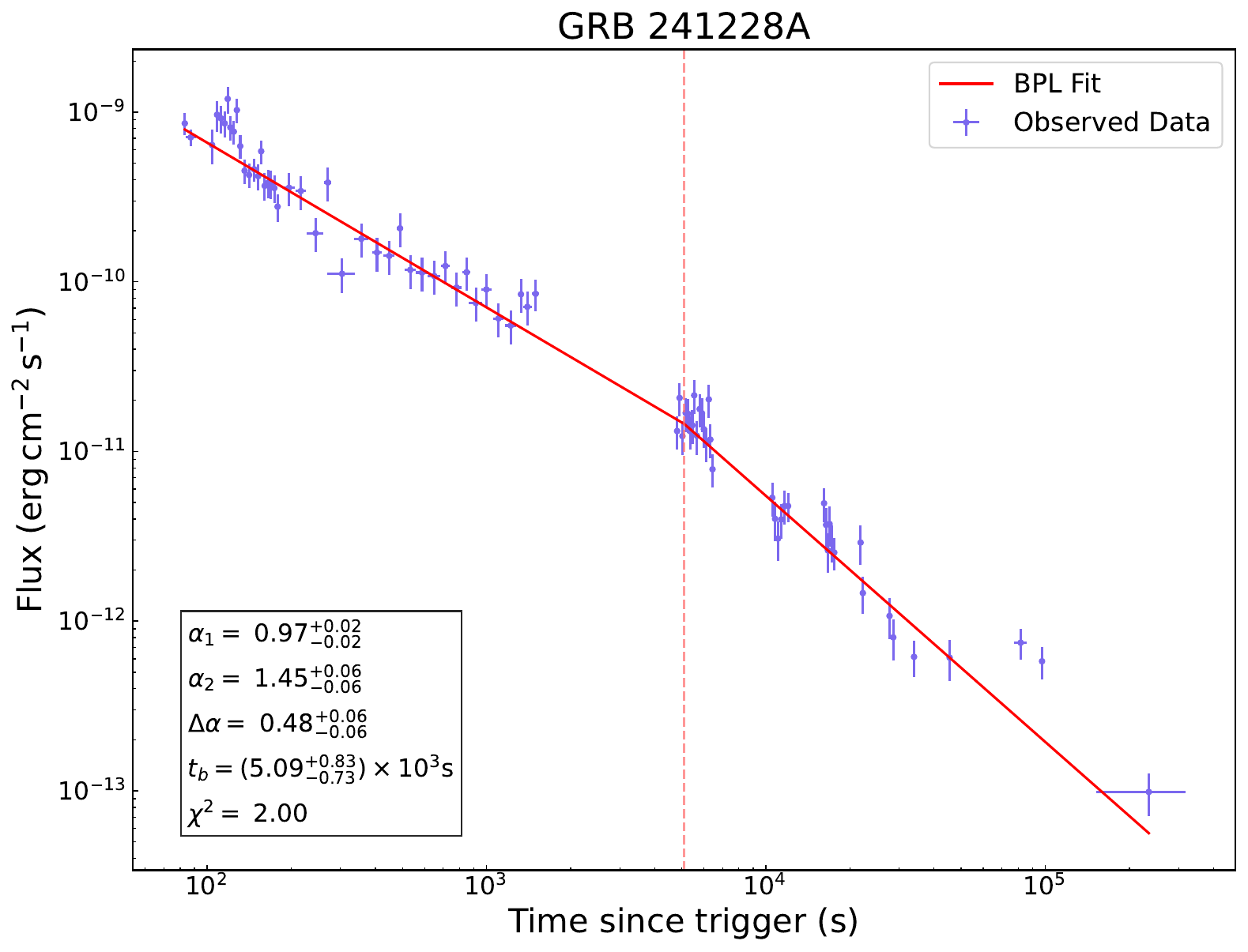}}\\
\resizebox{45mm}{!}{\includegraphics[]{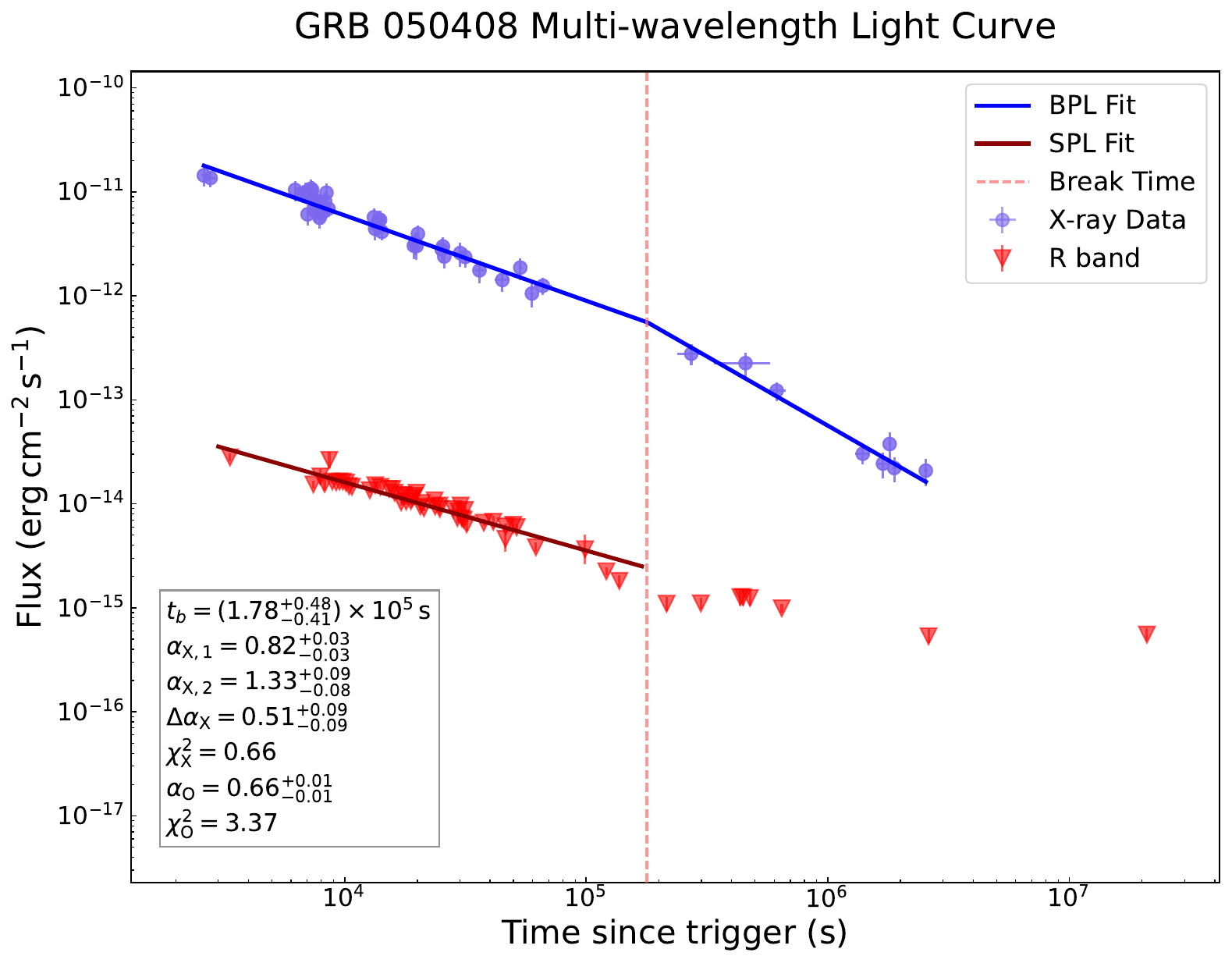}}%
\resizebox{45mm}{!}{\includegraphics[]{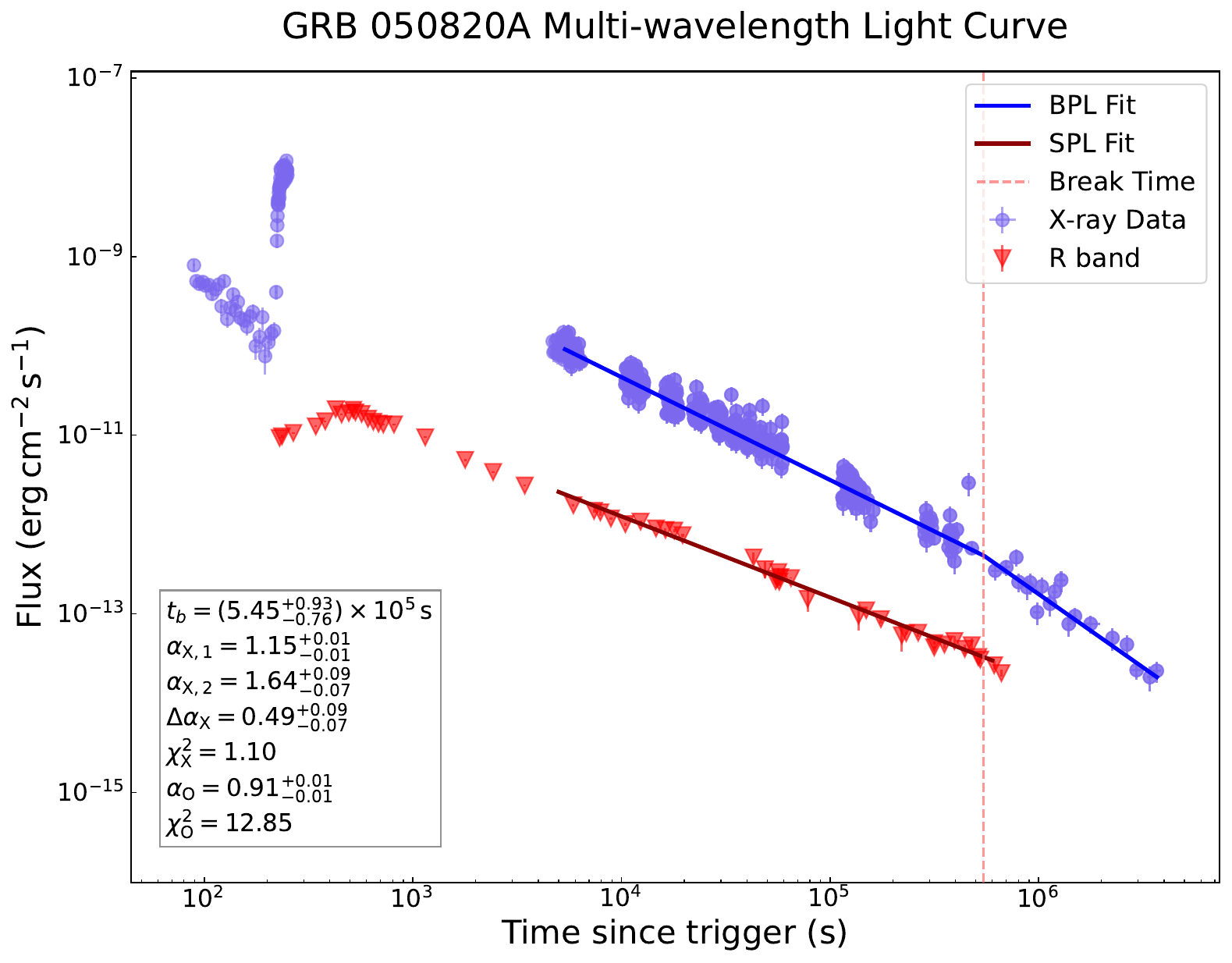}}%
\resizebox{45mm}{!}{\includegraphics[]{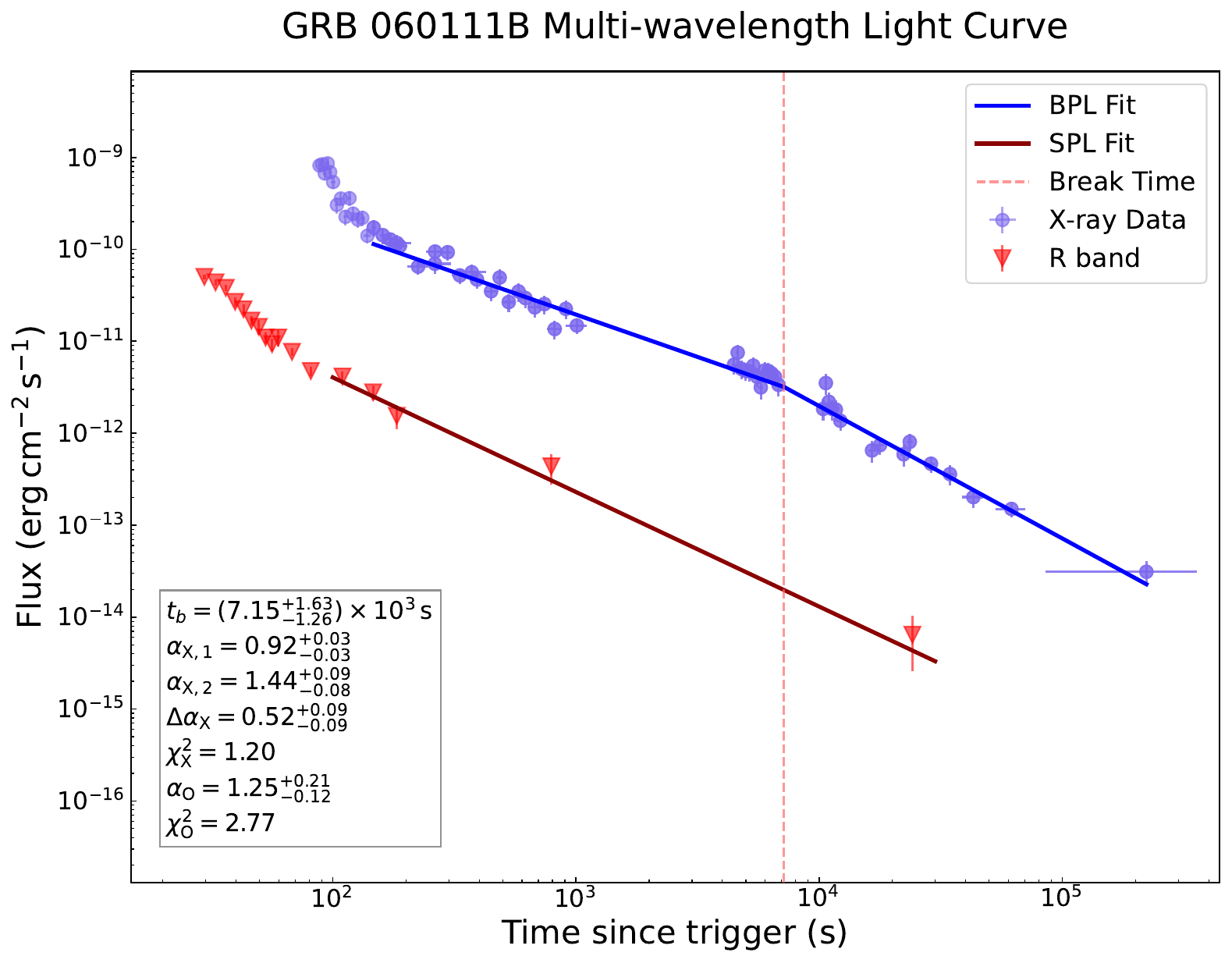}}%
\resizebox{45mm}{!}{\includegraphics[]{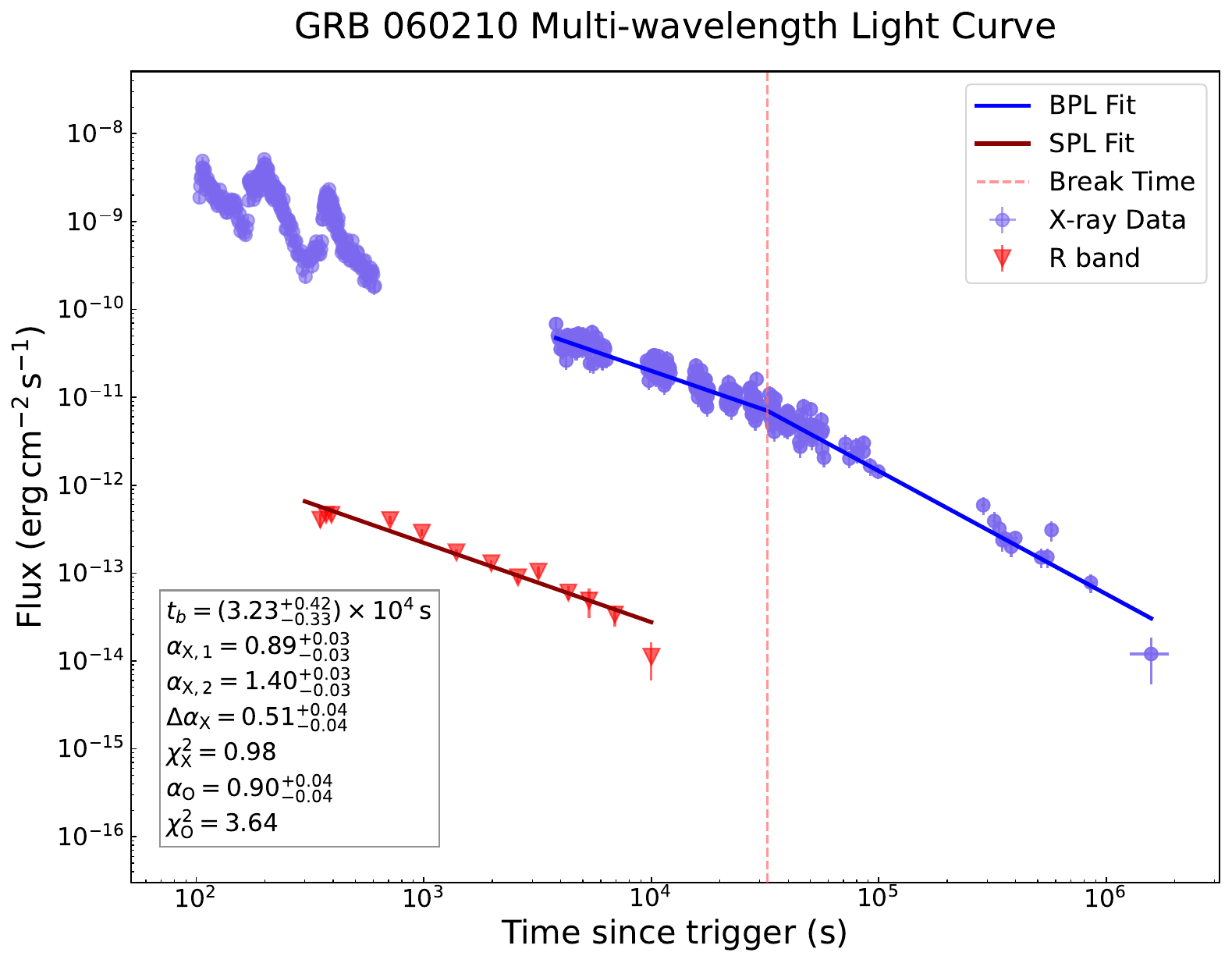}}\\
\resizebox{45mm}{!}{\includegraphics[]{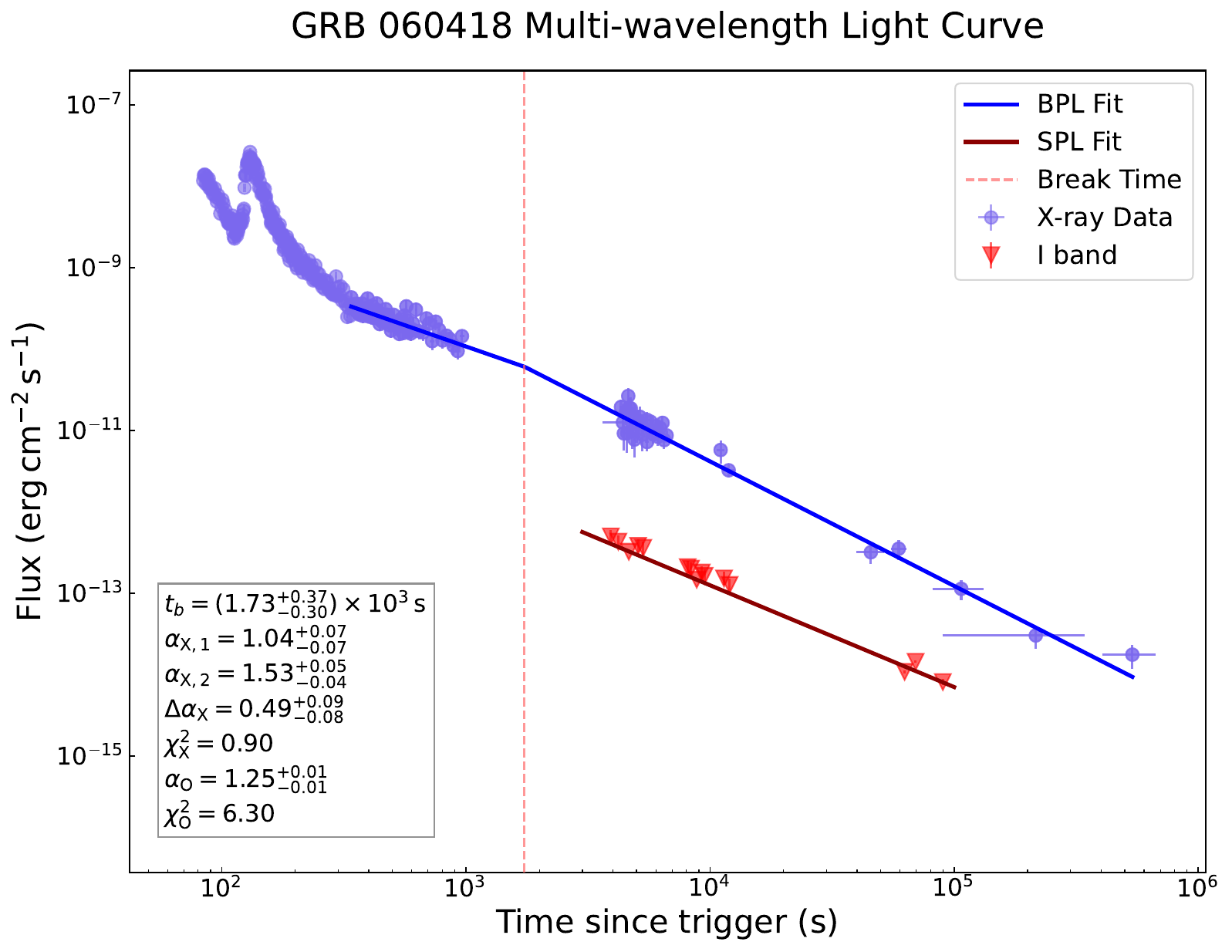}}%
\resizebox{45mm}{!}{\includegraphics[]{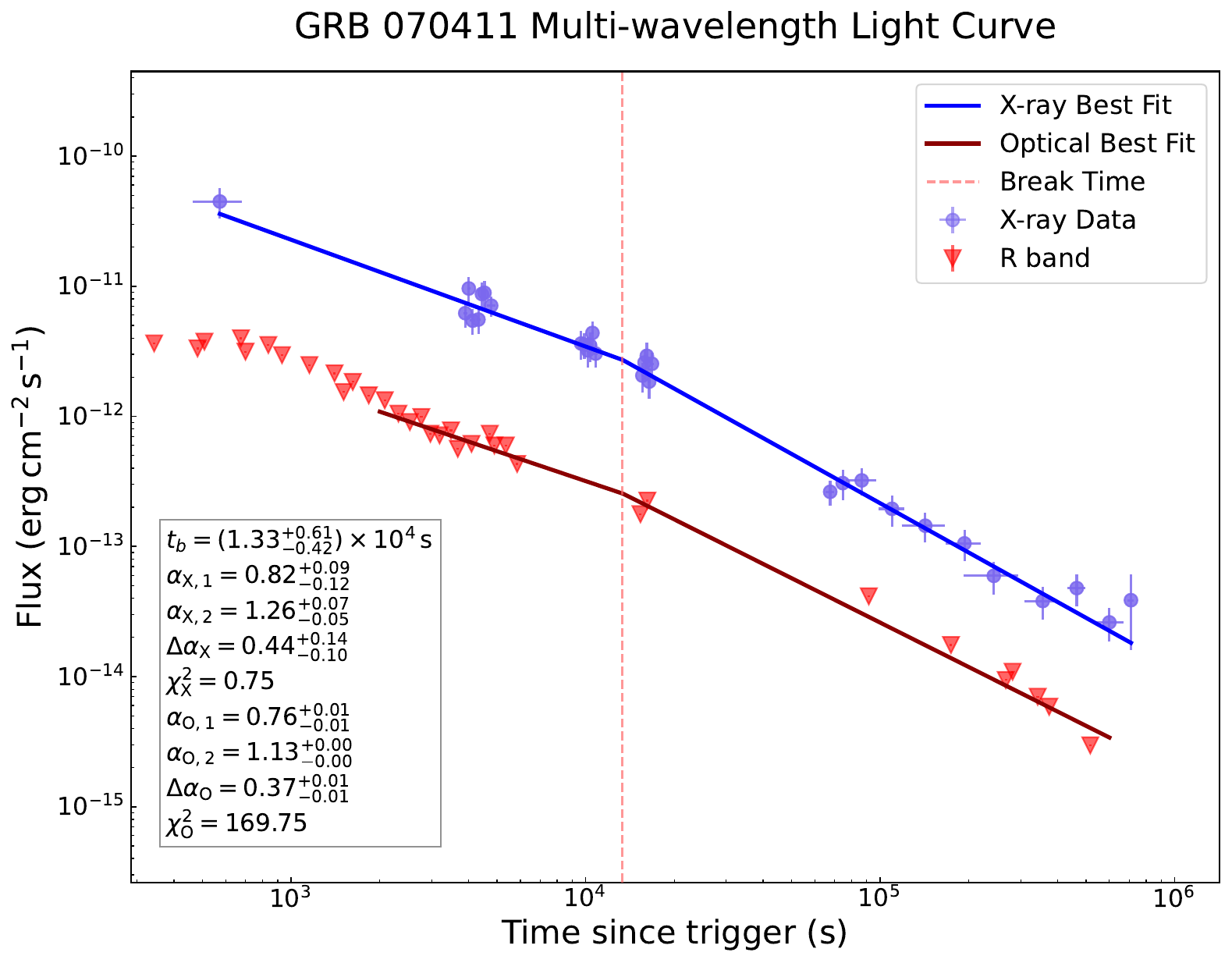}}%
\resizebox{45mm}{!}{\includegraphics[]{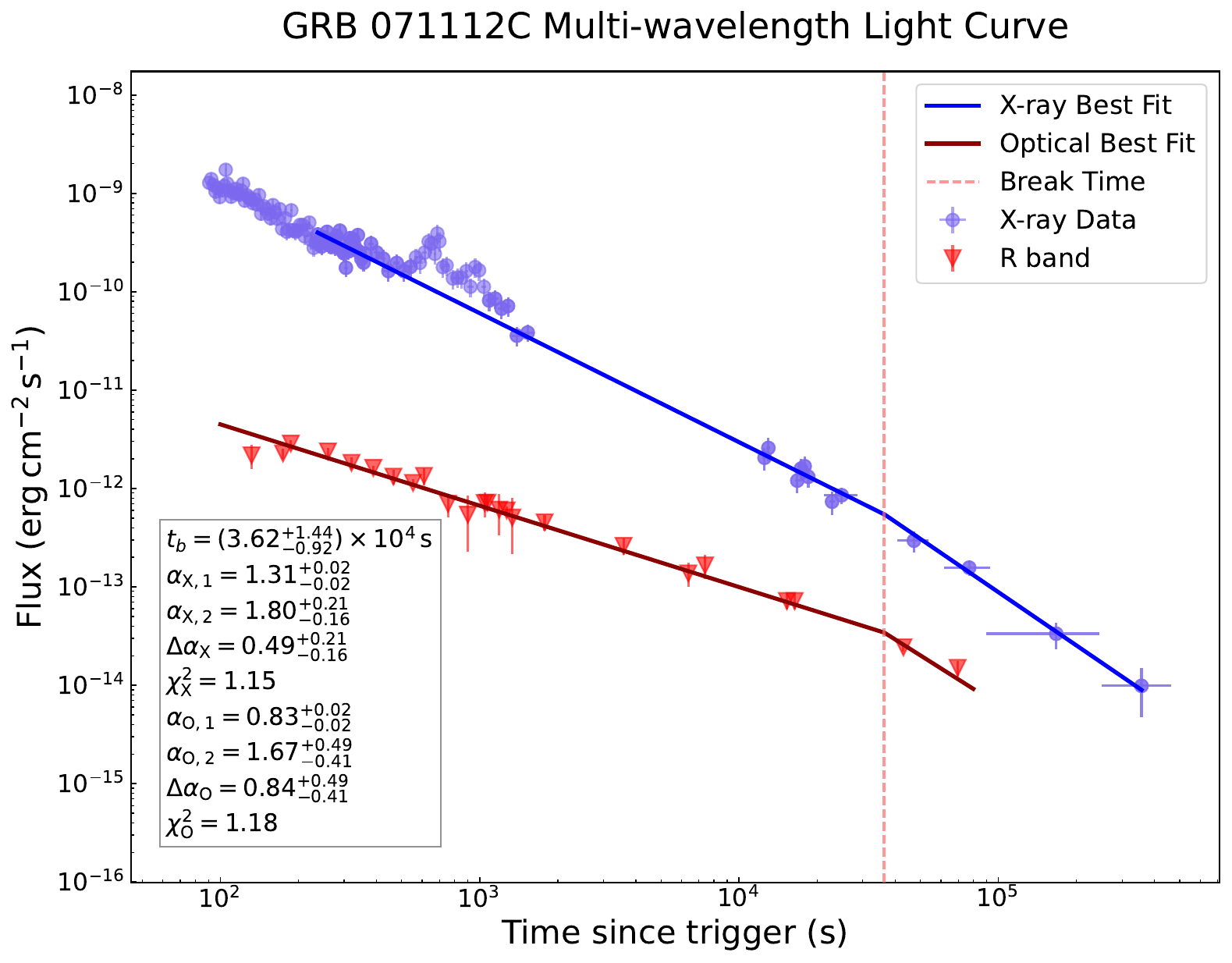}}%
\resizebox{45mm}{!}{\includegraphics[]{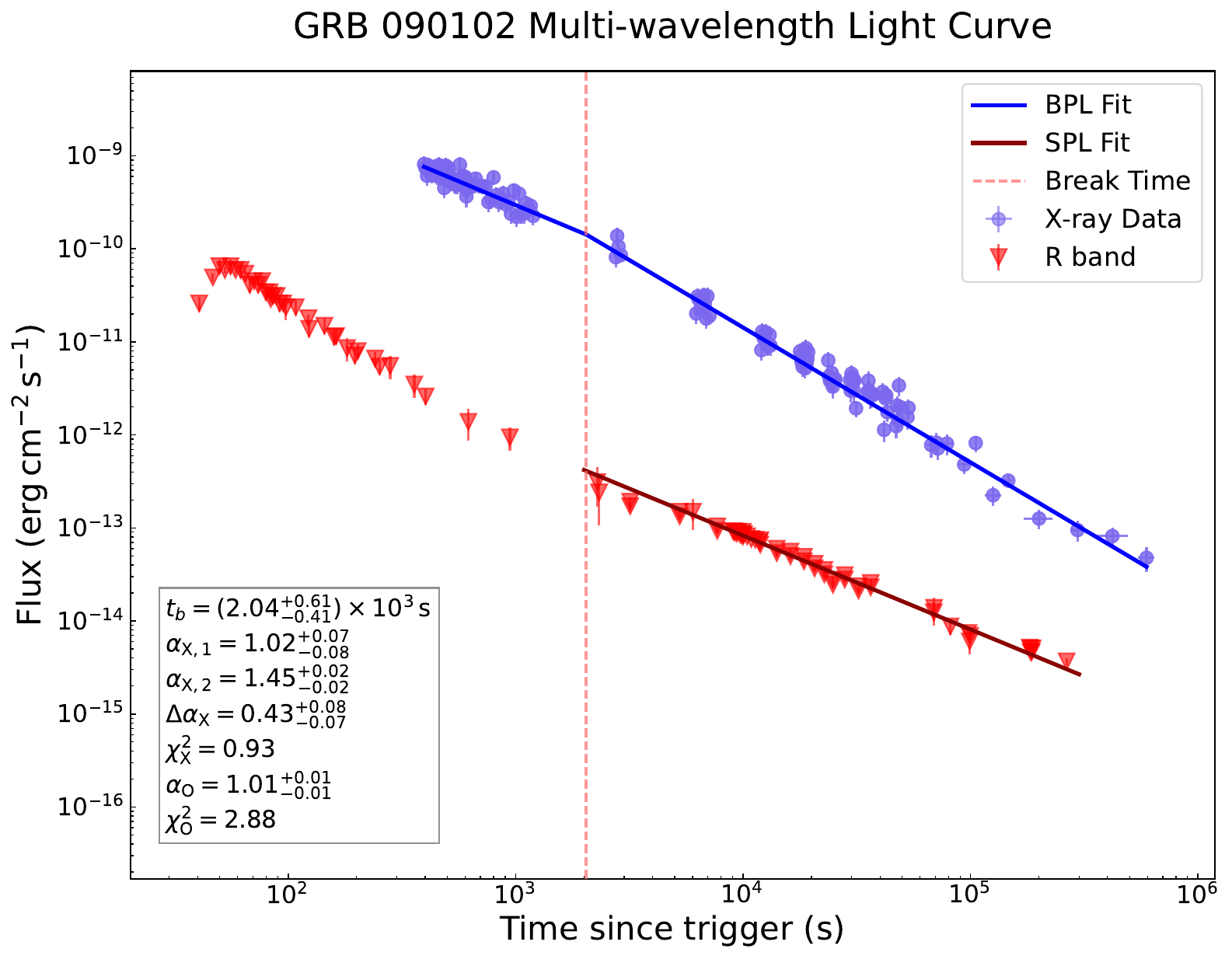}}\\
\resizebox{45mm}{!}{\includegraphics[]{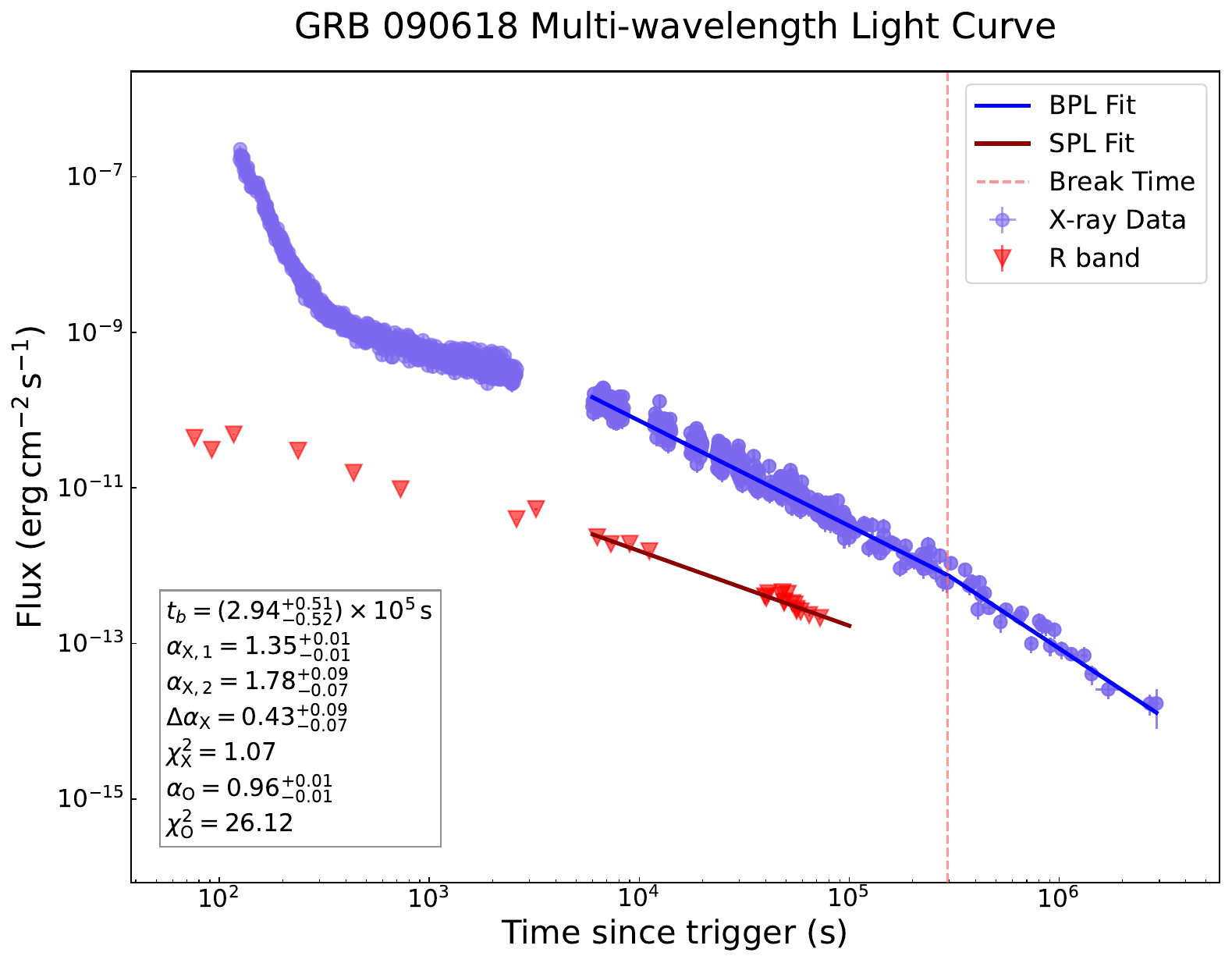}}%
\resizebox{45mm}{!}{\includegraphics[]{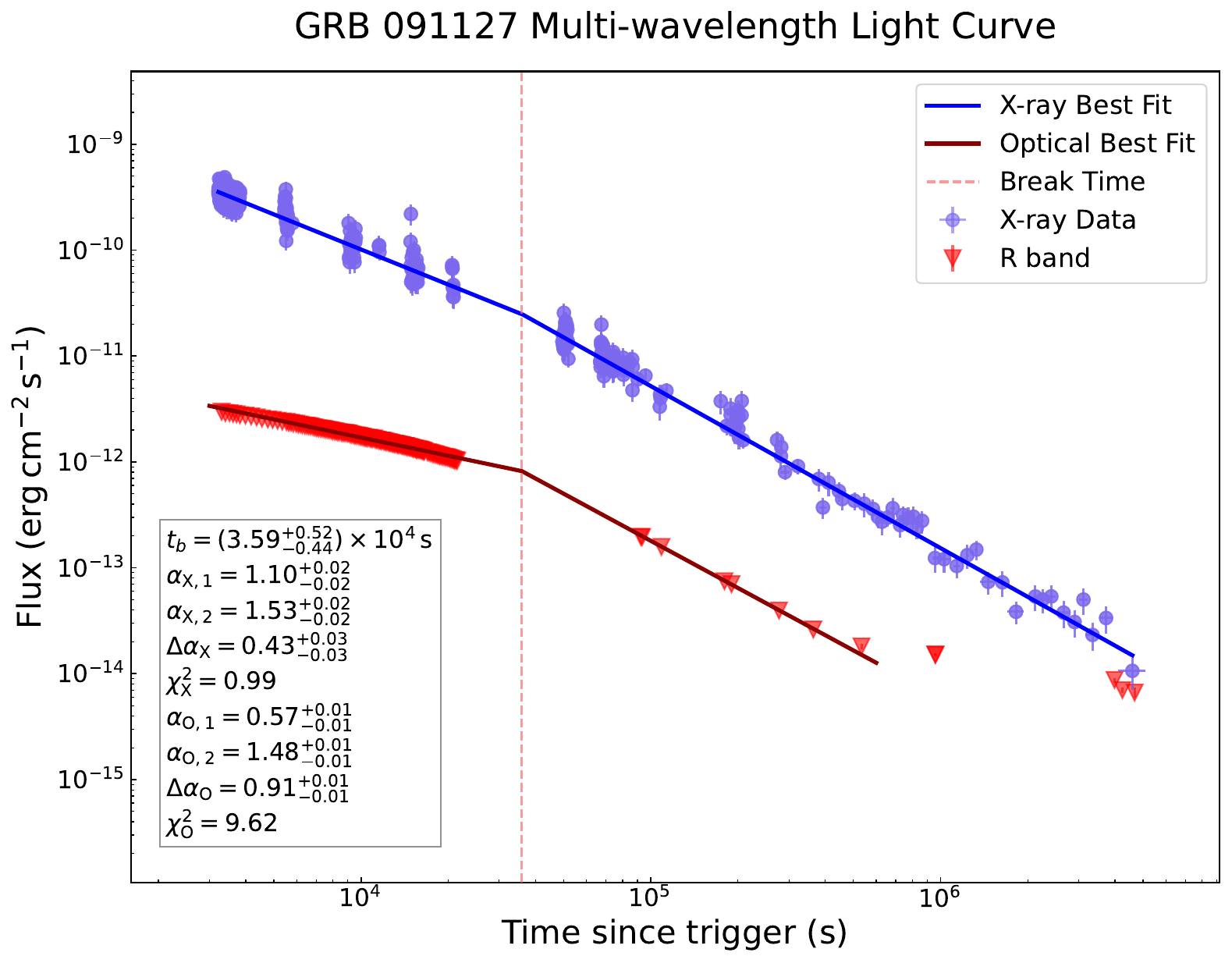}}%
\resizebox{45mm}{!}{\includegraphics[]{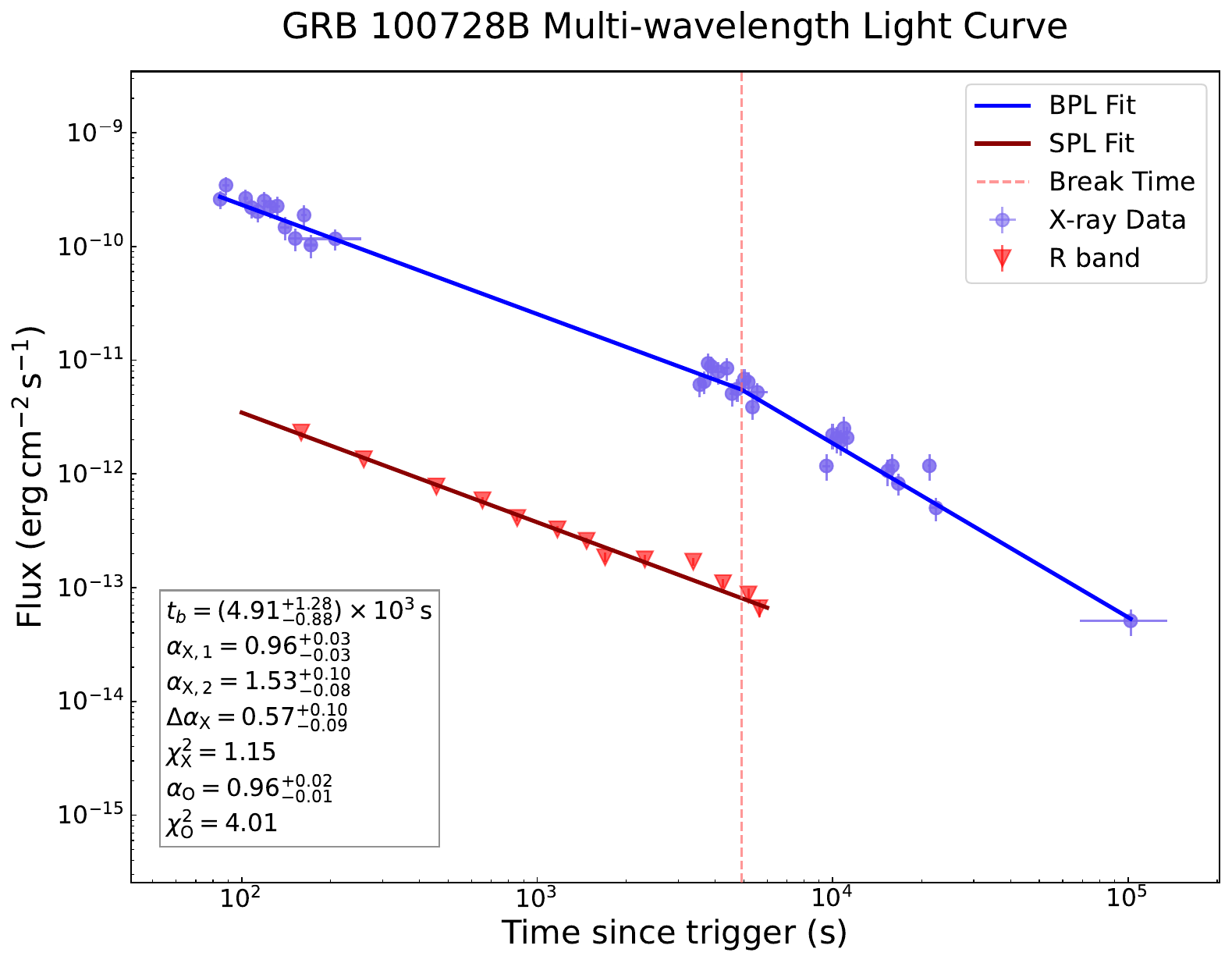}}%
\resizebox{45mm}{!}{\includegraphics[]{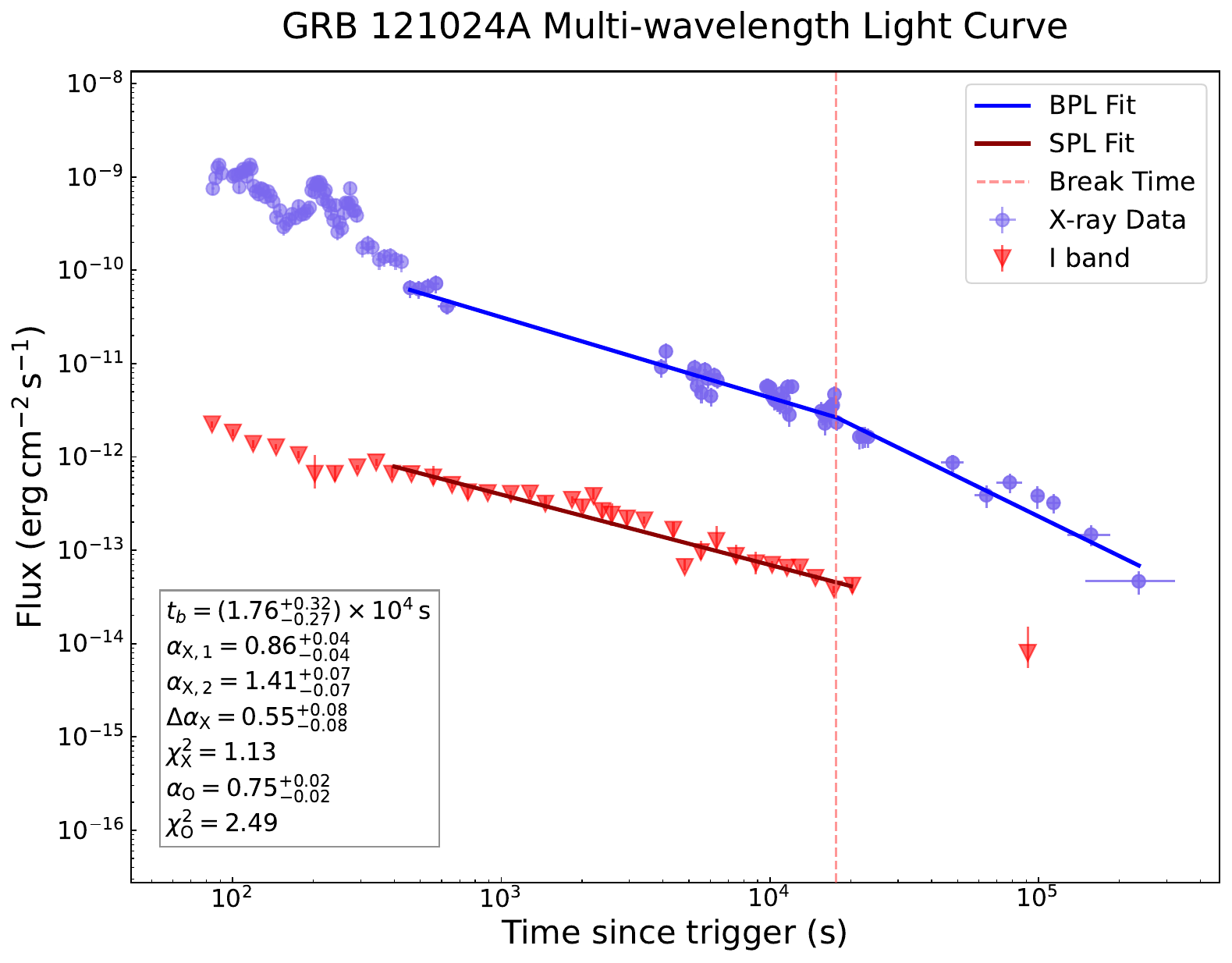}}\\
\resizebox{45mm}{!}{\includegraphics[]{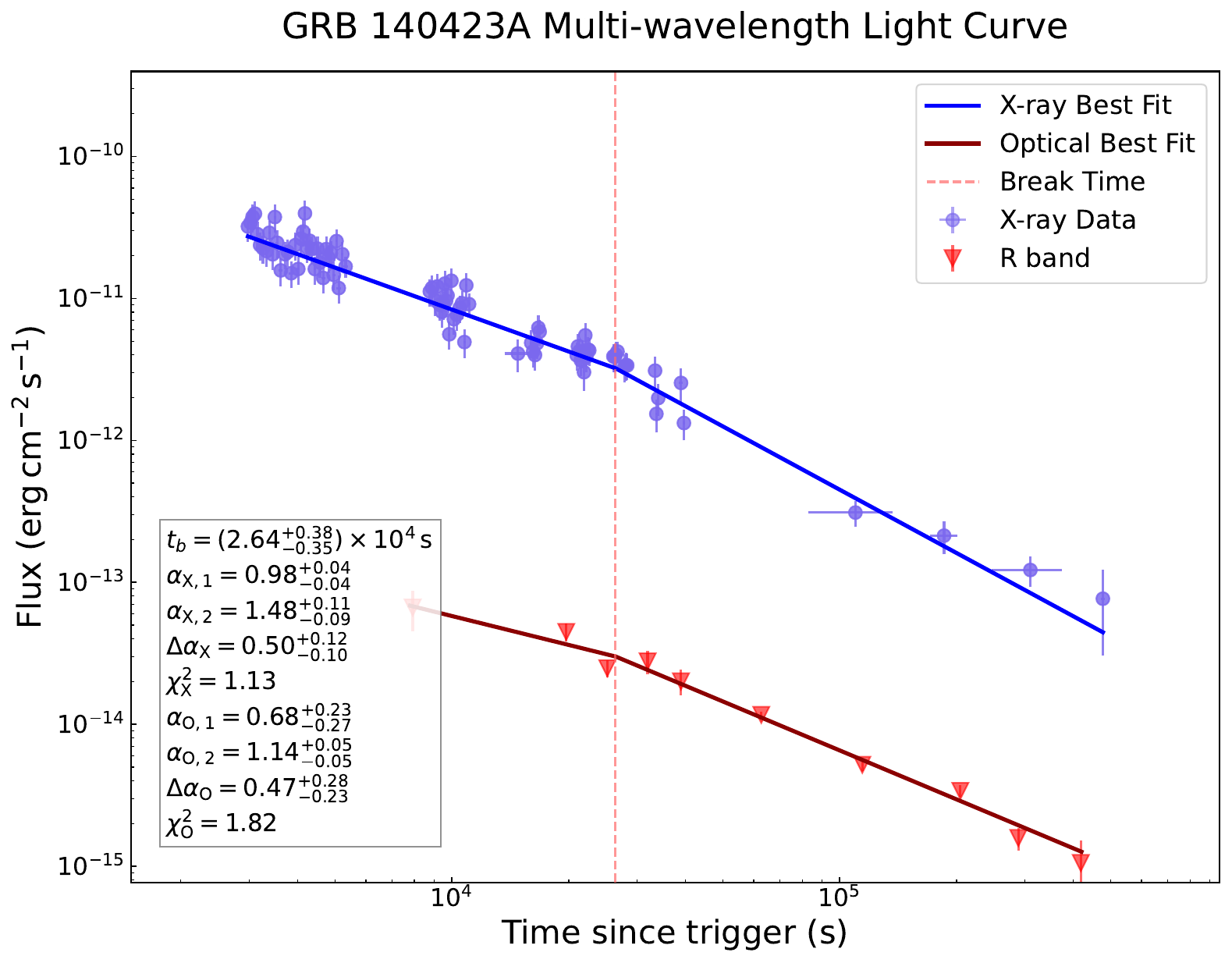}}%
\resizebox{45mm}{!}{\includegraphics[]{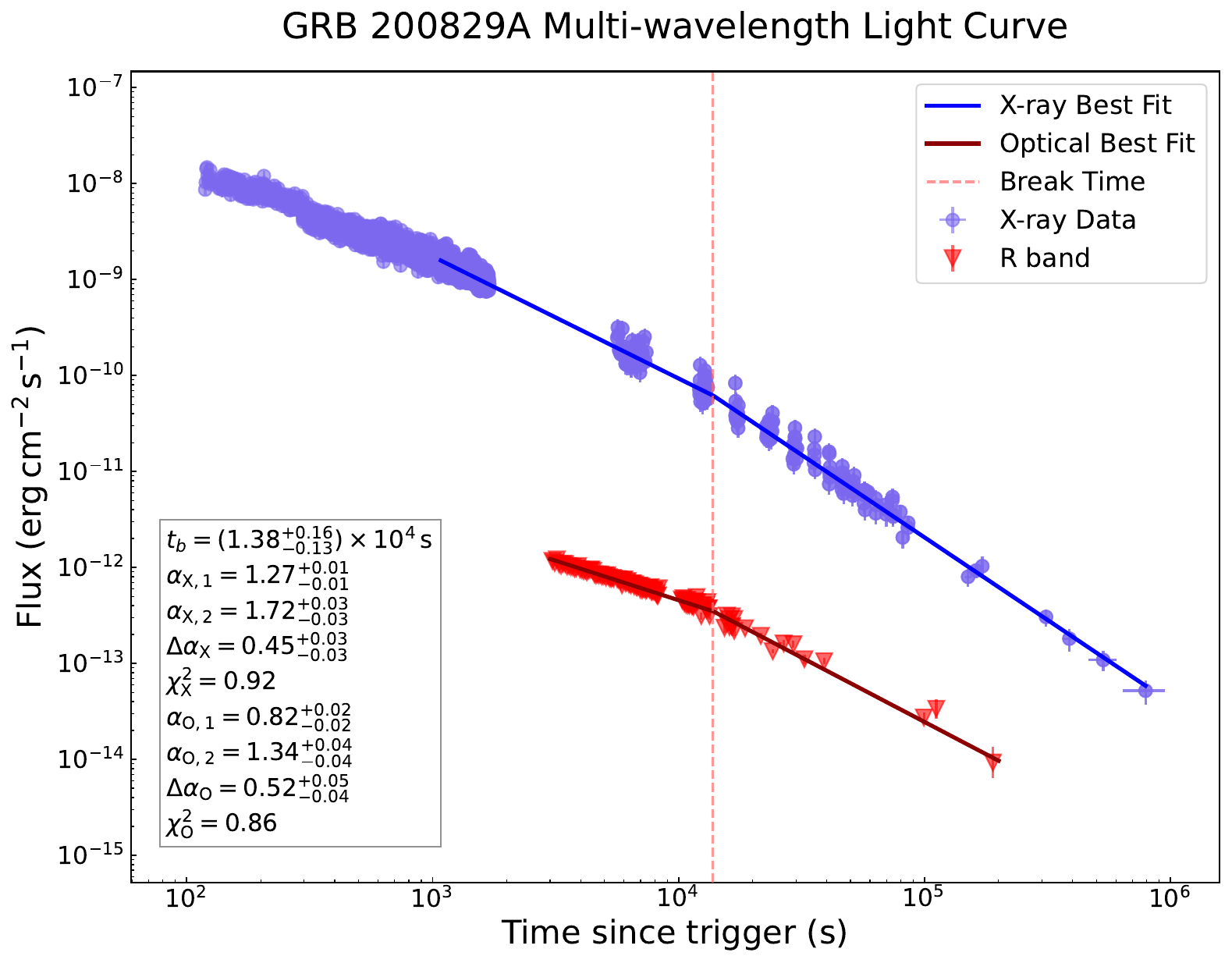}}%
\resizebox{45mm}{!}{\includegraphics[]{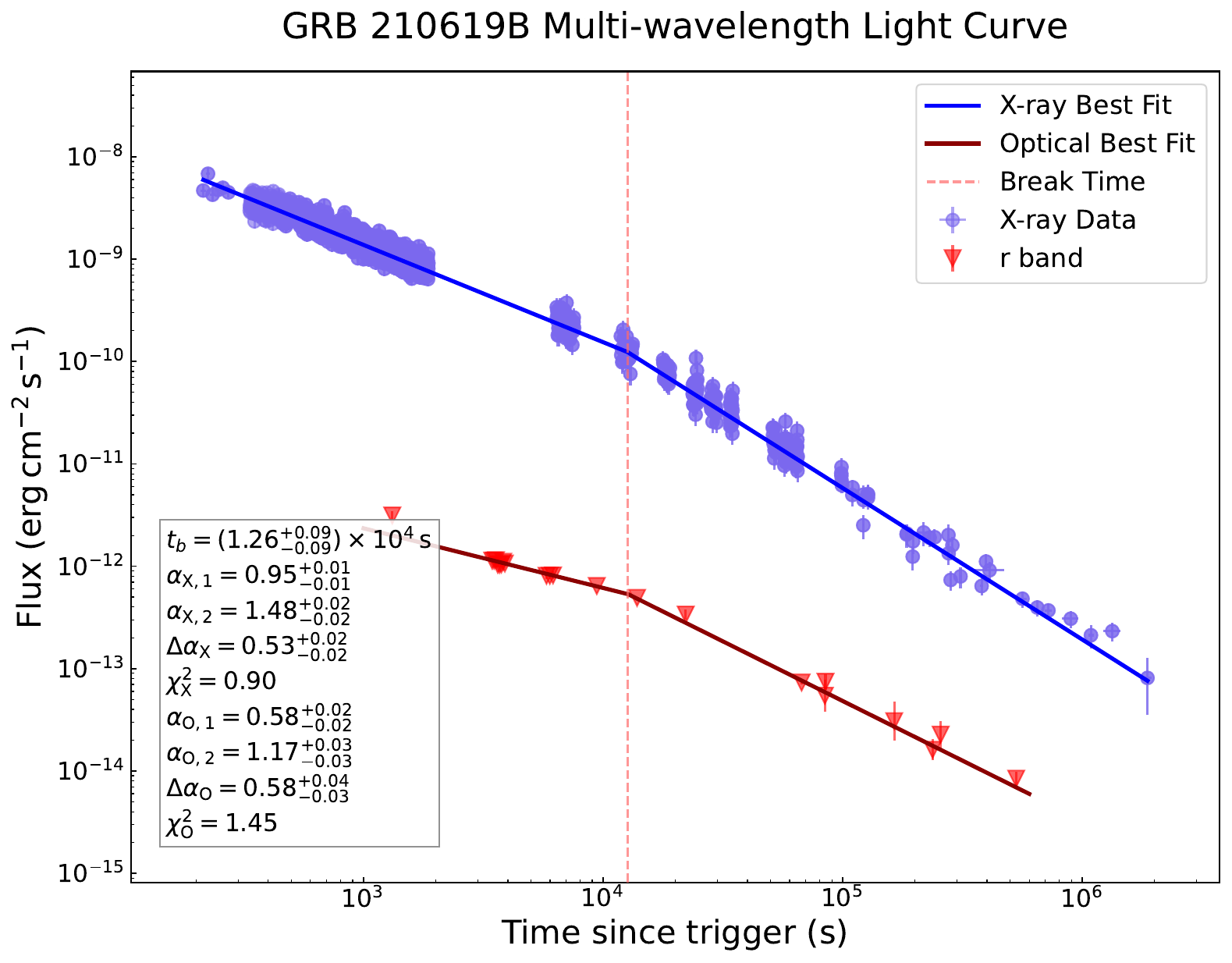}}
\caption{(Continued)}
\end{figure*}

\begin{figure*}
\centering
\resizebox{100mm}{!}{\includegraphics[]{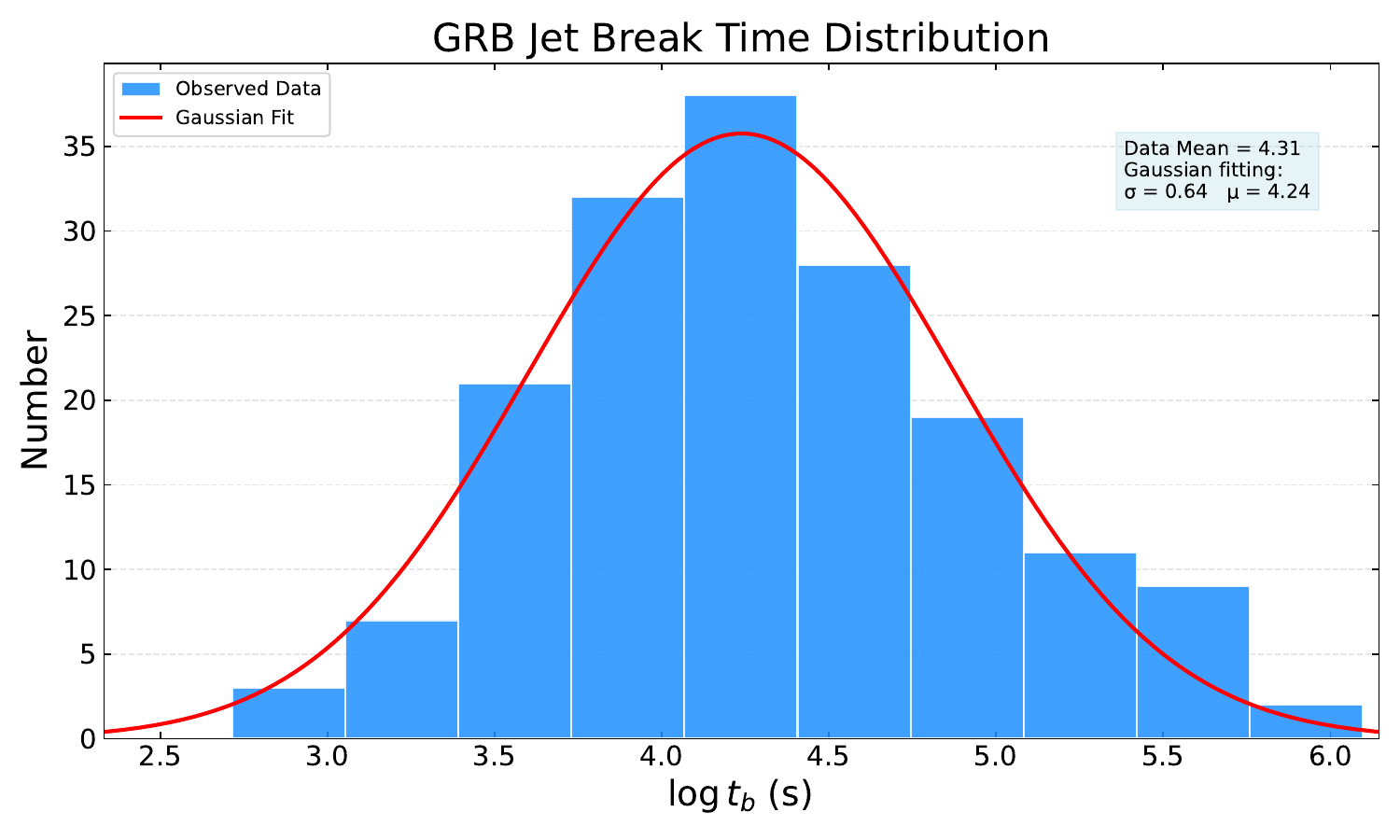}}%
\caption{Distribution of jet break times for the 170 GRBs in our sample. The histogram shows the observed distribution, and the red curve represents a Gaussian fit to the data.}
\label{fig:figure3}
\end{figure*}

\begin{figure*}
\centering
\resizebox{80mm}{!}{\includegraphics[]{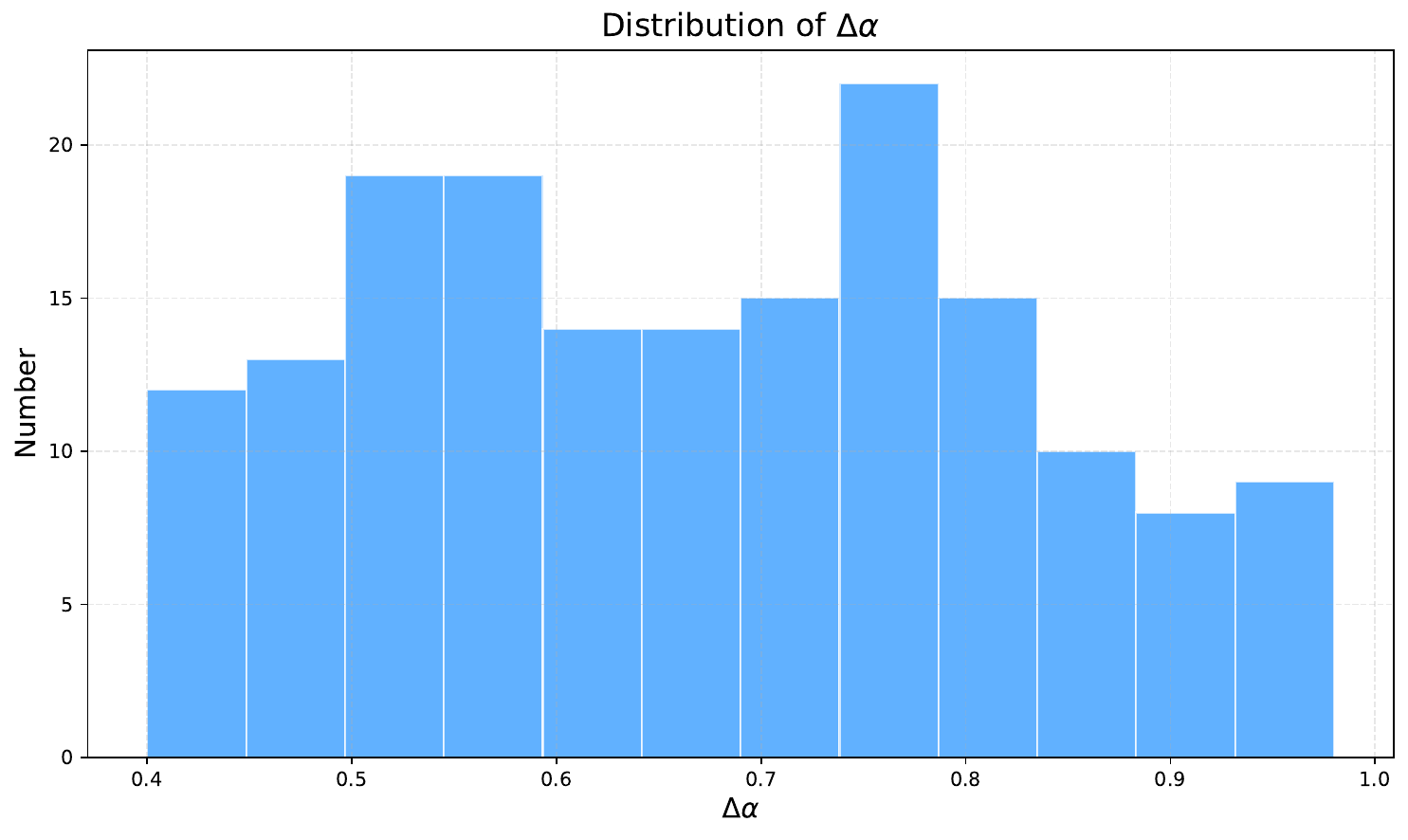}}
\resizebox{80mm}{!}{\includegraphics[]{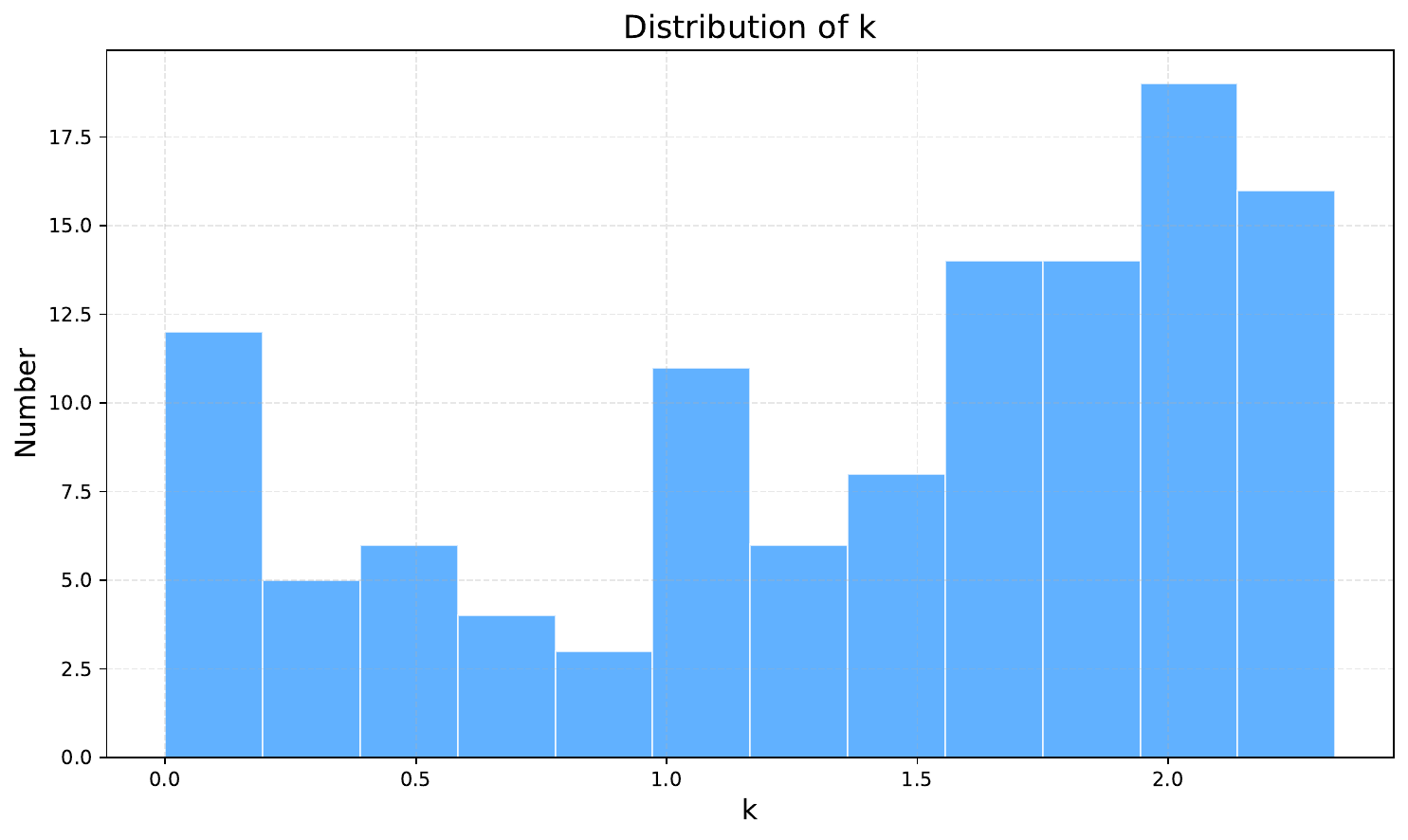}}\\
\caption{Distributions of the temporal decay index change $\Delta\alpha$ (left) and the density profile index $k$ (right) for our sample, where $k$ is derived from $\Delta\alpha$ using the relation $\Delta\alpha = (3 - k)/(4 - k)$.}
\label{fig:figure4}
\end{figure*}

\begin{figure*}
\centering
\resizebox{60mm}{!}{\includegraphics[]{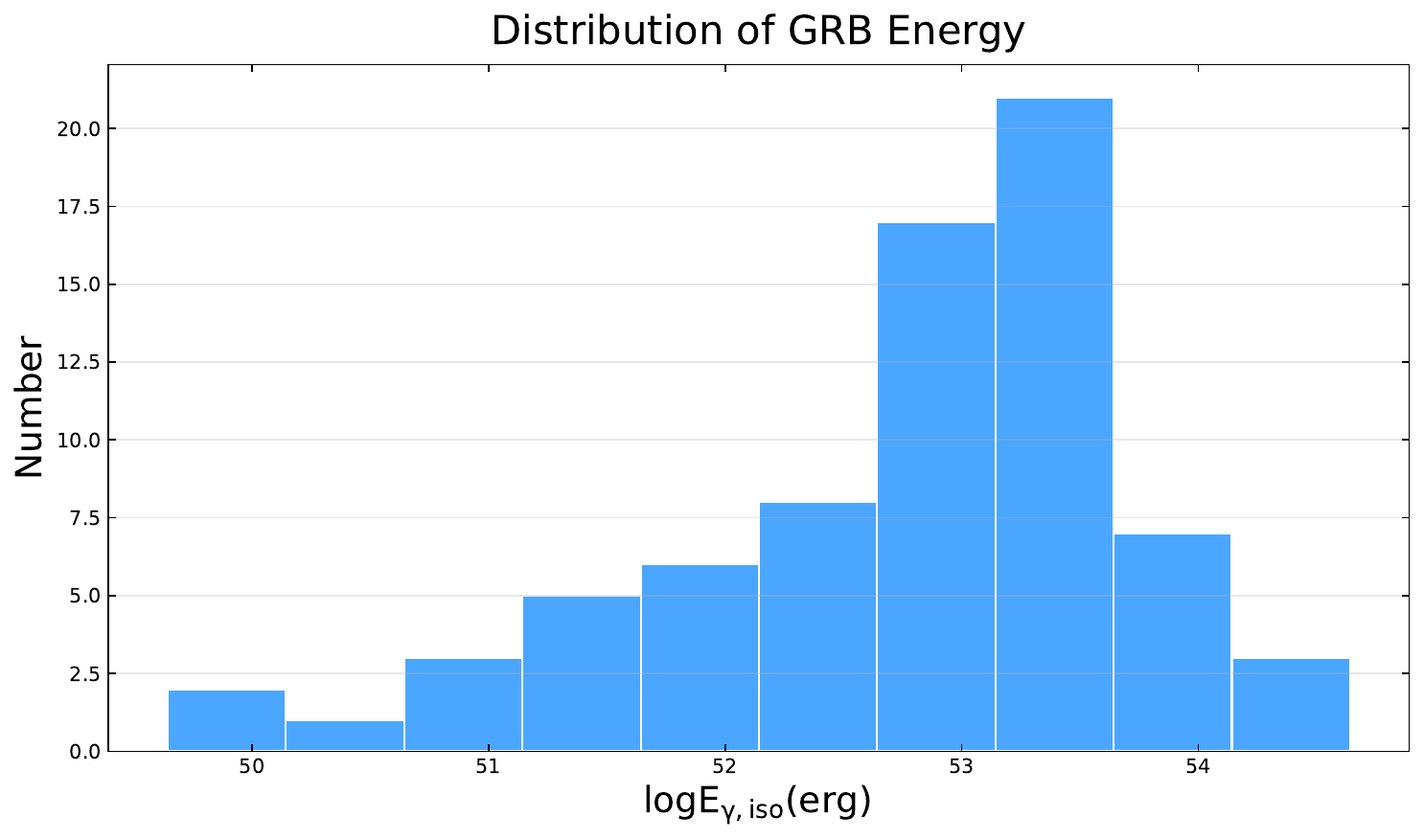}}%
\resizebox{60mm}{!}{\includegraphics[]{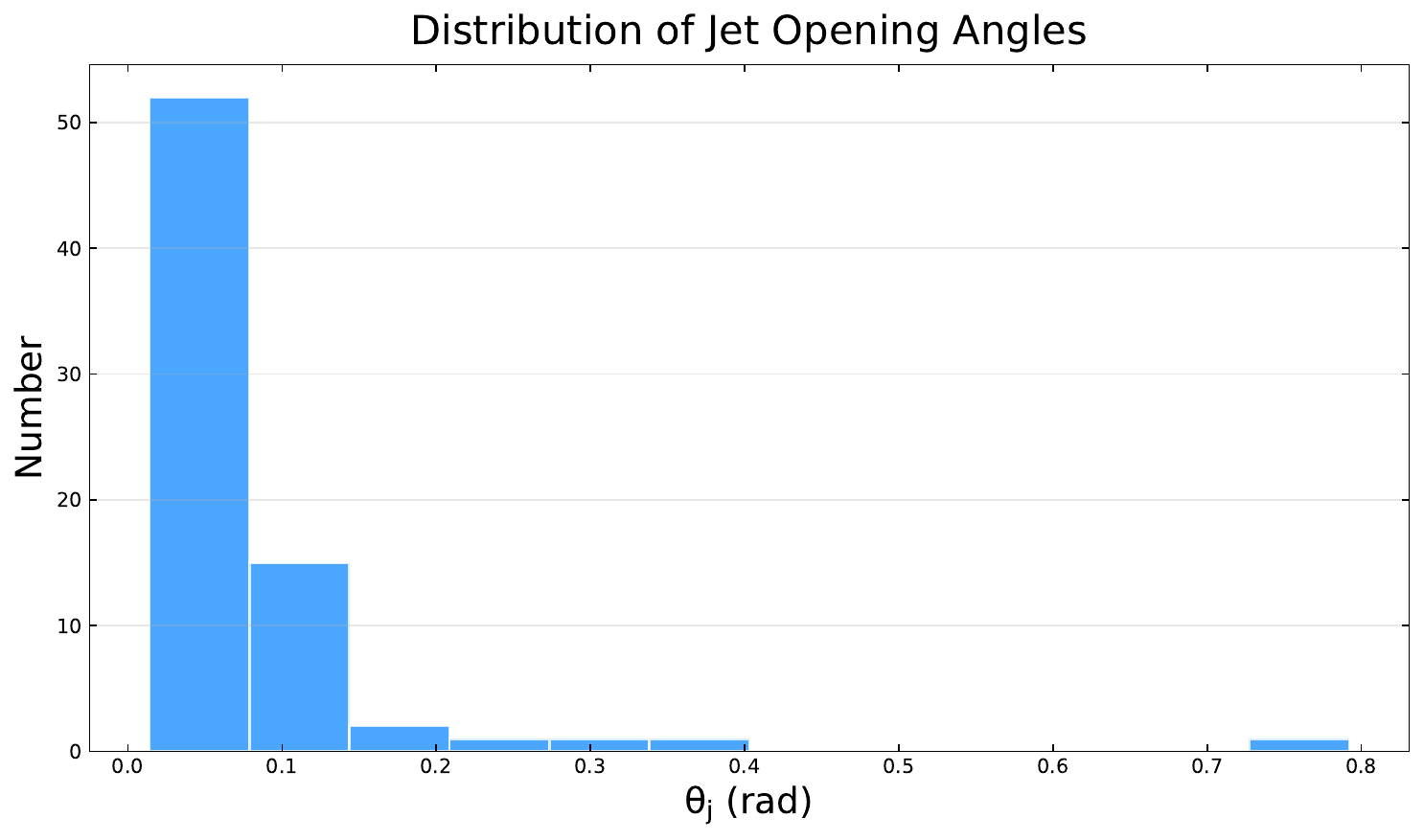}}%
\resizebox{60mm}{!}{\includegraphics[]{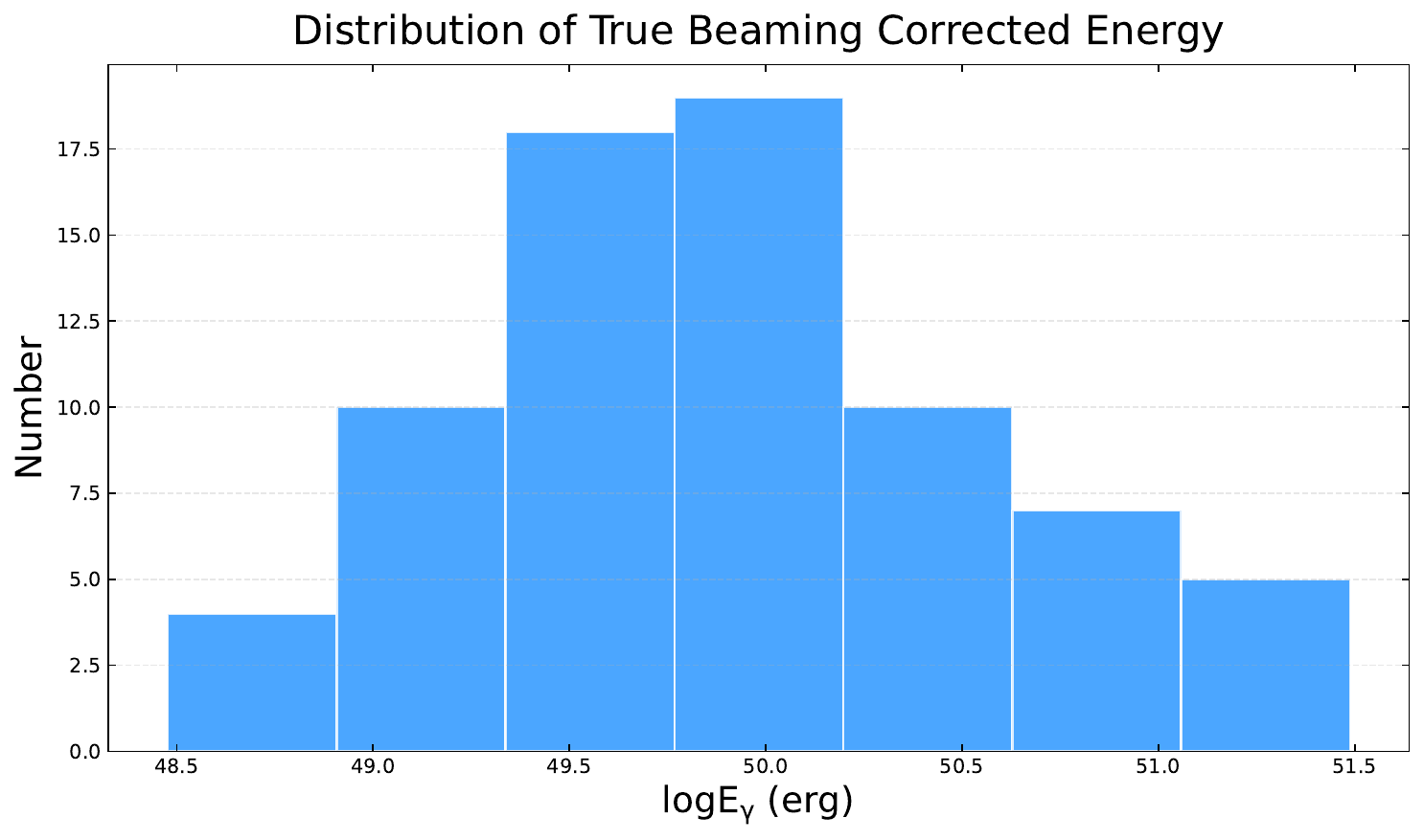}}%
\caption{Distributions of derived physical parameters for GRBs with redshifts: the isotropic gamma-ray energy ($E_{\gamma,\mathrm{iso}}$), the jet half-opening angle ($\theta_j$), and the true beaming-corrected gamma-ray energy ($E_{\gamma} = E_{\gamma,\mathrm{iso}}(1 - \cos \theta_j)$).}
\label{fig:figure5}
\end{figure*}

\begin{figure*}
\centering
\resizebox{80mm}{!}{\includegraphics[]{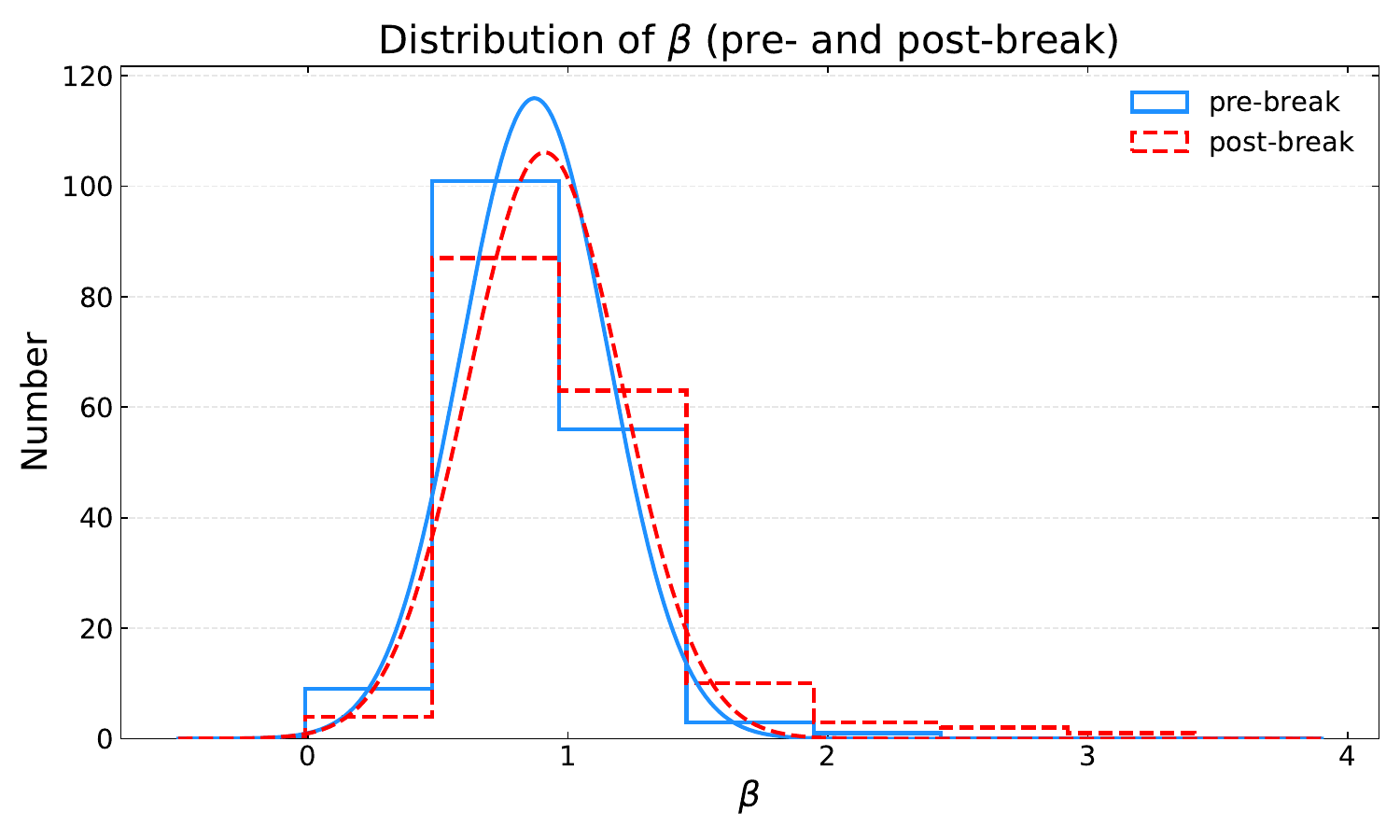}}\\
\caption{Distributions of the spectral index $\beta$ in the pre- and post-jet break phases for the full sample of 170 GRBs. The histograms show the $\beta$ values derived from X-ray spectra.}
\label{fig:figure6}
\end{figure*}

\begin{figure*}
\centering
\resizebox{100mm}{!}{\includegraphics[]{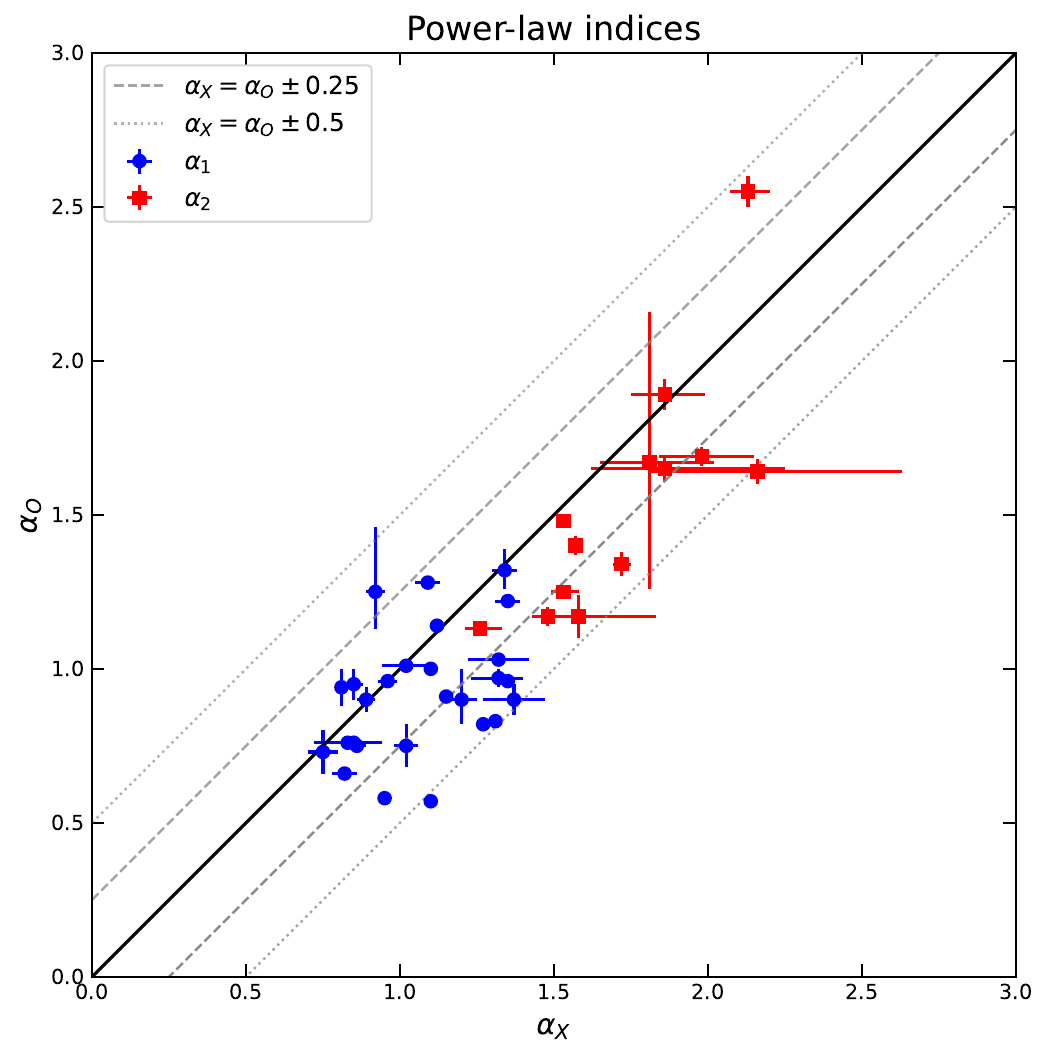}}\\
\caption{Power-law indices for the light curve decay. The black solid line indicates where the X-ray and optical temporal indices are equal, and the dashed lines indicate where $\alpha_x = \alpha_o \pm 0.25$, which is expected for a standard forward-shock model with an ISM and a wind environment. The dotted lines are $\alpha_x = \alpha_o \pm 0.5$, which is expected by the energy injection case for an ISM and a wind environment.}
\label{fig:figure7}
\end{figure*}

\begin{deluxetable*}{ccccccccccc}[htpb]
	\label{tab:Fitting Results}
    \tablewidth{0pt}
	\tablecaption{\textbf{Fitting Results of All Samples}}
	\tablehead{
		\colhead{GRB} & \colhead{${T}_{90}$} & \colhead{${F_{b}}^{a}$} & \colhead{${t_{b}}^{a}$} & \colhead{${\alpha_{1}}^{b}$} & \colhead{${\alpha_{2}}^{b}$} & \colhead{$\Delta\alpha$}
		& \colhead{k} & \colhead{${\rm CBM}^{c}$} & \colhead{$\chi^2_{r}$}\\
		\colhead{} & \colhead{(s)} & \colhead{$(\times 10^{-12}$erg/cm$^2$/s)} & \colhead{$(\times 10^{3}$s)} & \colhead{} & \colhead{} & \colhead{} & \colhead{} & \colhead{} & \colhead{}}
	\startdata
050128 & 19.2 & $5.07^{+0.89}_{-0.72}$ & $22.82^{+3.18}_{-3.00}$ & $1.00^{+0.02}_{-0.02}$ & $1.72^{+0.14}_{-0.12}$ & $0.72^{+0.14}_{-0.12}$ & $0.43^{+1.79}_{-1.53}$ & ISM & 0.82 \\
050315 & 95.6 & $1.29^{+0.12}_{-0.13}$ & $169.25^{+18.51}_{-13.44}$ & $0.68^{+0.02}_{-0.02}$ & $1.57^{+0.09}_{-0.08}$ & $0.89^{+0.09}_{-0.08}$ & uncertain & ISM & 0.94 \\
050318 & 32 & $8.55^{+2.38}_{-1.79}$ & $11.67^{+1.86}_{-1.75}$ & $1.04^{+0.10}_{-0.12}$ & $1.89^{+0.10}_{-0.08}$ & $0.85^{+0.16}_{-0.13}$ & uncertain & ISM & 0.88 \\
050408 & - & $0.56^{+0.15}_{-0.12}$ & $178.28^{+48.16}_{-41.11}$ & $0.82^{+0.03}_{-0.03}$ & $1.33^{+0.09}_{-0.08}$ & $0.51^{+0.09}_{-0.09}$ & $1.96^{+0.37}_{-0.37}$ & wind & 0.66 \\
050505 & 58.9 & $2.30^{+0.41}_{-0.49}$ & $46.61^{+8.42}_{-4.97}$ & $1.16^{+0.05}_{-0.05}$ & $1.82^{+0.10}_{-0.08}$ & $0.66^{+0.11}_{-0.09}$ & $1.06^{+0.95}_{-0.78}$ & wind-like & 0.87 \\
050713B & 54.2 & $0.72^{+0.17}_{-0.13}$ & $318.18^{+56.54}_{-59.23}$ & $0.98^{+0.03}_{-0.03}$ & $1.88^{+0.36}_{-0.25}$ & $0.90^{+0.36}_{-0.25}$ & uncertain & ISM & 0.95 \\
050721 & 98.4 & $0.68^{+0.19}_{-0.17}$ & $56.51^{+13.87}_{-13.42}$ & $0.75^{+0.18}_{-0.19}$ & $1.51^{+0.35}_{-0.26}$ & $0.76^{+0.40}_{-0.32}$ & uncertain & ISM & 0.65 \\
050726 & 49.9 & $8.72^{+1.21}_{-1.04}$ & $7.14^{+0.76}_{-0.75}$ & $0.92^{+0.02}_{-0.02}$ & $1.86^{+0.10}_{-0.09}$ & $0.94^{+0.10}_{-0.09}$ & uncertain & ISM & 1.05 \\
050820A & 26 & $0.45^{+0.09}_{-0.08}$ & $544.67^{+93.25}_{-76.19}$ & $1.15^{+0.01}_{-0.01}$ & $1.64^{+0.09}_{-0.07}$ & $0.49^{+0.09}_{-0.07}$ & $2.04^{+0.35}_{-0.27}$ & wind & 1.1 \\
050915A & 52 & $1.57^{+0.50}_{-0.36}$ & $9.92^{+2.27}_{-2.34}$ & $0.81^{+0.04}_{-0.04}$ & $1.28^{+0.09}_{-0.09}$ & $0.47^{+0.10}_{-0.10}$ & $2.11^{+0.36}_{-0.36}$ & wind & 1.11 \\
050922C & 4.5 & $0.18^{+0.10}_{-0.05}$ & $96.94^{+27.94}_{-26.82}$ & $1.35^{+0.04}_{-0.04}$ & $2.16^{+0.46}_{-0.34}$ & $0.81^{+0.46}_{-0.34}$ & uncertain & ISM & 1.54 \\
051016B & 4 & $0.79^{+0.17}_{-0.15}$ & $50.51^{+10.76}_{-8.84}$ & $0.74^{+0.05}_{-0.05}$ & $1.39^{+0.08}_{-0.07}$ & $0.65^{+0.09}_{-0.09}$ & $1.14^{+0.73}_{-0.73}$ & wind-like & 0.87 \\
051221A & 1.4 & $0.04^{+0.02}_{-0.01}$ & $405.69^{+100.63}_{-106.83}$ & $1.32^{+0.08}_{-0.08}$ & $2.00^{+0.59}_{-0.46}$ & $0.68^{+0.60}_{-0.47}$ & $0.88^{+5.86}_{-4.59}$ & ISM-like & 1.12 \\
051227 & 114.6 & $0.91^{+0.61}_{-0.36}$ & $7.78^{+3.58}_{-2.90}$ & $0.96^{+0.08}_{-0.09}$ & $1.69^{+0.28}_{-0.19}$ & $0.73^{+0.29}_{-0.21}$ & $0.30^{+3.98}_{-2.88}$ & ISM & 1.21 \\
060111B & 58.8 & $3.20^{+0.78}_{-0.67}$ & $7.15^{+1.63}_{-1.26}$ & $0.92^{+0.03}_{-0.03}$ & $1.44^{+0.09}_{-0.08}$ & $0.52^{+0.09}_{-0.09}$ & $1.92^{+0.39}_{-0.39}$ & wind & 1.2 \\
060116 & 105.9 & $0.20^{+0.12}_{-0.07}$ & $77.01^{+33.91}_{-31.13}$ & $0.86^{+0.10}_{-0.10}$ & $1.64^{+0.47}_{-0.30}$ & $0.78^{+0.48}_{-0.32}$ & uncertain & ISM & 0.93 \\
060204B & 139.4 & $0.33^{+0.24}_{-0.12}$ & $80.34^{+24.31}_{-24.36}$ & $1.33^{+0.08}_{-0.09}$ & $1.96^{+0.31}_{-0.23}$ & $0.63^{+0.32}_{-0.24}$ & $1.30^{+2.34}_{-1.75}$ & wind-like & 0.85 \\
060210 & 255 & $7.05^{+0.90}_{-0.91}$ & $32.33^{+4.23}_{-3.27}$ & $0.89^{+0.03}_{-0.03}$ & $1.40^{+0.03}_{-0.03}$ & $0.51^{+0.04}_{-0.04}$ & $1.96^{+0.17}_{-0.17}$ & wind & 0.98 \\
060312 & 50.6 & $1.21^{+0.45}_{-0.36}$ & $14.65^{+5.12}_{-4.60}$ & $0.72^{+0.18}_{-0.21}$ & $1.23^{+0.21}_{-0.17}$ & $0.51^{+0.30}_{-0.25}$ & $1.96^{+1.25}_{-1.04}$ & wind & 0.72 \\
060313 & 0.74 & $5.05^{+0.64}_{-0.59}$ & $7.60^{+0.82}_{-0.81}$ & $0.71^{+0.04}_{-0.04}$ & $1.65^{+0.07}_{-0.07}$ & $0.94^{+0.08}_{-0.08}$ & uncertain & ISM & 0.66 \\
060418 & 103.1 & $61.34^{+17.15}_{-15.05}$ & $1.73^{+0.37}_{-0.30}$ & $1.04^{+0.07}_{-0.07}$ & $1.53^{+0.05}_{-0.04}$ & $0.49^{+0.09}_{-0.08}$ & $2.04^{+0.35}_{-0.31}$ & wind & 0.9 \\
060510A & 20.4 & $20.82^{+2.12}_{-1.91}$ & $25.18^{+1.75}_{-1.63}$ & $0.79^{+0.02}_{-0.02}$ & $1.51^{+0.03}_{-0.03}$ & $0.72^{+0.04}_{-0.04}$ & $0.43^{+0.51}_{-0.51}$ & ISM & 1.0 \\
060614 & 108.7 & $1.52^{+0.19}_{-0.19}$ & $111.83^{+8.12}_{-7.11}$ & $1.32^{+0.10}_{-0.10}$ & $2.13^{+0.07}_{-0.06}$ & $0.81^{+0.12}_{-0.12}$ & uncertain & ISM & 0.72 \\
060807 & 54 & $5.41^{+0.88}_{-0.73}$ & $15.40^{+1.51}_{-1.52}$ & $0.98^{+0.08}_{-0.09}$ & $1.77^{+0.05}_{-0.05}$ & $0.79^{+0.10}_{-0.09}$ & uncertain & ISM & 1.21 \\
060927 & 22.5 & $3.69^{+1.26}_{-0.84}$ & $4.19^{+0.94}_{-0.82}$ & $0.75^{+0.05}_{-0.05}$ & $1.58^{+0.25}_{-0.15}$ & $0.83^{+0.25}_{-0.16}$ & uncertain & ISM & 0.51 \\
061004 & 6.2 & $0.83^{+0.44}_{-0.31}$ & $20.48^{+7.41}_{-5.90}$ & $0.96^{+0.16}_{-0.19}$ & $1.75^{+0.24}_{-0.18}$ & $0.79^{+0.31}_{-0.24}$ & uncertain & ISM & 1.25 \\
061121 & 81.3 & $18.48^{+1.37}_{-1.37}$ & $21.36^{+1.55}_{-1.41}$ & $0.81^{+0.02}_{-0.02}$ & $1.54^{+0.03}_{-0.02}$ & $0.73^{+0.04}_{-0.03}$ & $0.30^{+0.55}_{-0.41}$ & ISM & 0.88 \\
061222A & 71.4 & $2.57^{+1.04}_{-0.55}$ & $97.75^{+15.85}_{-18.34}$ & $1.12^{+0.04}_{-0.06}$ & $1.70^{+0.05}_{-0.04}$ & $0.58^{+0.08}_{-0.06}$ & $1.62^{+0.45}_{-0.34}$ & wind & 1.05 \\
070318 & 74.6 & $0.27^{+0.04}_{-0.04}$ & $269.13^{+37.40}_{-33.15}$ & $0.84^{+0.03}_{-0.03}$ & $1.79^{+0.26}_{-0.23}$ & $0.95^{+0.26}_{-0.23}$ & uncertain & ISM & 1.76 \\
070411 & 121.5 & $2.73^{+1.25}_{-0.88}$ & $13.30^{+6.13}_{-4.19}$ & $0.82^{+0.09}_{-0.12}$ & $1.26^{+0.07}_{-0.05}$ & $0.44^{+0.14}_{-0.10}$ & $2.21^{+0.45}_{-0.32}$ & wind & 0.75 \\
070412 & 33.8 & $0.69^{+0.24}_{-0.18}$ & $26.99^{+5.84}_{-5.54}$ & $1.01^{+0.08}_{-0.09}$ & $1.96^{+0.31}_{-0.20}$ & $0.95^{+0.32}_{-0.22}$ & uncertain & ISM & 0.61 \\
070420 & 76.5 & $2.50^{+0.68}_{-0.58}$ & $45.99^{+8.30}_{-7.05}$ & $1.20^{+0.05}_{-0.05}$ & $1.95^{+0.12}_{-0.10}$ & $0.75^{+0.13}_{-0.11}$ & $0.00^{+2.08}_{-1.76}$ & ISM & 1.44 \\
070521 & 37.9 & $5.57^{+0.73}_{-0.70}$ & $18.46^{+1.49}_{-1.35}$ & $1.24^{+0.07}_{-0.07}$ & $2.16^{+0.12}_{-0.11}$ & $0.92^{+0.14}_{-0.13}$ & uncertain & ISM & 1.11 \\
070616 & 402.4 & $0.39^{+0.19}_{-0.11}$ & $55.11^{+14.28}_{-13.71}$ & $1.29^{+0.05}_{-0.05}$ & $2.05^{+0.34}_{-0.23}$ & $0.76^{+0.34}_{-0.24}$ & uncertain & ISM & 1.12 \\
071112C & 15 & $0.55^{+0.27}_{-0.21}$ & $36.21^{+14.37}_{-9.19}$ & $1.31^{+0.02}_{-0.02}$ & $1.80^{+0.21}_{-0.16}$ & $0.49^{+0.21}_{-0.16}$ & $2.04^{+0.81}_{-0.62}$ & wind & 1.15 \\
080212 & 123 & $2.33^{+0.55}_{-0.51}$ & $23.63^{+5.10}_{-3.67}$ & $1.05^{+0.05}_{-0.05}$ & $1.60^{+0.13}_{-0.10}$ & $0.55^{+0.14}_{-0.11}$ & $1.78^{+0.69}_{-0.54}$ & wind & 1.3 \\
080218B & 6.2 & $0.19^{+0.07}_{-0.05}$ & $70.60^{+19.19}_{-17.75}$ & $0.93^{+0.04}_{-0.04}$ & $1.81^{+0.34}_{-0.28}$ & $0.88^{+0.34}_{-0.28}$ & uncertain & ISM & 0.78 \\
080229A & 64 & $85.00^{+11.14}_{-10.07}$ & $8.60^{+1.05}_{-0.97}$ & $0.79^{+0.03}_{-0.03}$ & $1.32^{+0.02}_{-0.02}$ & $0.53^{+0.04}_{-0.04}$ & $1.87^{+0.18}_{-0.18}$ & wind & 1.04 \\
080310 & 365 & $0.35^{+0.20}_{-0.13}$ & $95.23^{+30.83}_{-24.78}$ & $1.37^{+0.10}_{-0.10}$ & $2.10^{+0.30}_{-0.22}$ & $0.73^{+0.32}_{-0.24}$ & $0.30^{+4.39}_{-3.29}$ & ISM & 1.11 \\
080319C & 34 & $47.94^{+14.95}_{-12.13}$ & $6.77^{+1.78}_{-1.51}$ & $0.93^{+0.05}_{-0.05}$ & $1.70^{+0.08}_{-0.07}$ & $0.77^{+0.09}_{-0.09}$ & uncertain & ISM & 0.89 \\
080411 & 56 & $0.21^{+0.07}_{-0.05}$ & $1234.57^{+230.69}_{-237.47}$ & $1.31^{+0.02}_{-0.02}$ & $1.71^{+0.17}_{-0.13}$ & $0.40^{+0.17}_{-0.13}$ & $2.33^{+0.47}_{-0.36}$ & wind & 0.89 \\
080413B & 8 & $0.43^{+0.12}_{-0.11}$ & $176.27^{+52.75}_{-35.50}$ & $1.02^{+0.04}_{-0.04}$ & $1.86^{+0.36}_{-0.24}$ & $0.84^{+0.36}_{-0.24}$ & uncertain & ISM & 1.54 \\
080426 & 1.7 & $3.68^{+1.28}_{-1.04}$ & $3.71^{+1.11}_{-0.90}$ & $0.87^{+0.07}_{-0.08}$ & $1.52^{+0.12}_{-0.11}$ & $0.65^{+0.14}_{-0.13}$ & $1.14^{+1.14}_{-1.06}$ & wind-like & 1.69 \\
080605 & 20 & $52.44^{+20.78}_{-16.23}$ & $4.81^{+1.41}_{-1.06}$ & $1.08^{+0.05}_{-0.05}$ & $1.59^{+0.05}_{-0.04}$ & $0.51^{+0.07}_{-0.06}$ & $1.96^{+0.29}_{-0.25}$ & wind & 1.43 \\
080607 & 79 & $0.35^{+0.21}_{-0.13}$ & $69.85^{+22.32}_{-20.79}$ & $1.29^{+0.09}_{-0.10}$ & $1.92^{+0.34}_{-0.24}$ & $0.63^{+0.35}_{-0.26}$ & $1.30^{+2.56}_{-1.90}$ & wind-like & 1.22 \\
080710 & 120 & $5.33^{+0.86}_{-0.74}$ & $12.61^{+1.33}_{-1.39}$ & $0.85^{+0.09}_{-0.10}$ & $1.73^{+0.07}_{-0.07}$ & $0.88^{+0.12}_{-0.11}$ & uncertain & ISM & 1.25 \\
080721 & 16.2 & $523.20^{+19.94}_{-18.65}$ & $2.62^{+0.10}_{-0.10}$ & $0.80^{+0.01}_{-0.01}$ & $1.57^{+0.01}_{-0.01}$ & $0.77^{+0.01}_{-0.01}$ & uncertain & ISM & 1.16 \\
080802 & 176 & $2.80^{+1.40}_{-0.95}$ & $9.81^{+5.78}_{-3.76}$ & $0.82^{+0.05}_{-0.05}$ & $1.47^{+0.32}_{-0.19}$ & $0.65^{+0.32}_{-0.20}$ & $1.14^{+2.61}_{-1.63}$ & wind-like & 0.56 \\
080903 & 66 & $50.06^{+16.01}_{-12.18}$ & $1.48^{+0.31}_{-0.25}$ & $1.43^{+0.04}_{-0.04}$ & $2.38^{+0.15}_{-0.12}$ & $0.95^{+0.16}_{-0.13}$ & uncertain & ISM & 1.17 \\
081128 & 100 & $0.36^{+0.18}_{-0.12}$ & $40.22^{+16.13}_{-12.02}$ & $0.84^{+0.06}_{-0.06}$ & $1.34^{+0.18}_{-0.15}$ & $0.50^{+0.19}_{-0.16}$ & $2.00^{+0.76}_{-0.64}$ & wind & 2.04 \\
	\enddata
\end{deluxetable*}

\addtocounter{table}{-1}
\begin{deluxetable*}{ccccccccccc}[htpb]
    \tablewidth{0pt}
	\tablecaption{\textbf{Fitting Results of All Samples}}
	\tablehead{
		\colhead{GRB} & \colhead{${T}_{90}$} & \colhead{${F_{b}}^{a}$} & \colhead{${t_{b}}^{a}$} & \colhead{${\alpha_{1}}^{b}$} & \colhead{${\alpha_{2}}^{b}$} & \colhead{$\Delta\alpha$}
		& \colhead{k} & \colhead{${\rm CBM}^{c}$} & \colhead{$\chi^2_{r}$}\\
		\colhead{} & \colhead{(s)} & \colhead{$(\times 10^{-12}$erg/cm$^2$/s)} & \colhead{$(\times 10^{3}$s)} & \colhead{} & \colhead{} & \colhead{} & \colhead{} & \colhead{} & \colhead{}}
	\startdata
    081211 & 3.5 & $0.42^{+0.27}_{-0.18}$ & $19.21^{+9.35}_{-7.79}$ & $0.87^{+0.09}_{-0.09}$ & $1.62^{+0.45}_{-0.28}$ & $0.75^{+0.46}_{-0.29}$ & $0.00^{+7.36}_{-4.64}$ & ISM & 1.43 \\
081221 & 34 & $0.23^{+0.06}_{-0.05}$ & $229.39^{+47.02}_{-34.70}$ & $1.22^{+0.03}_{-0.03}$ & $2.13^{+0.35}_{-0.33}$ & $0.91^{+0.35}_{-0.33}$ & uncertain & ISM & 1.14 \\
081230 & 60.7 & $0.24^{+0.09}_{-0.06}$ & $93.98^{+29.46}_{-23.66}$ & $0.91^{+0.04}_{-0.04}$ & $1.71^{+0.48}_{-0.28}$ & $0.80^{+0.48}_{-0.28}$ & uncertain & ISM & 1.29 \\
090102 & 27 & $143.13^{+47.96}_{-43.08}$ & $2.04^{+0.61}_{-0.41}$ & $1.02^{+0.07}_{-0.08}$ & $1.45^{+0.02}_{-0.02}$ & $0.43^{+0.08}_{-0.07}$ & $2.25^{+0.25}_{-0.22}$ & wind & 0.93 \\
090201 & 83 & $0.52^{+0.20}_{-0.17}$ & $174.57^{+55.15}_{-39.09}$ & $0.98^{+0.12}_{-0.14}$ & $1.73^{+0.22}_{-0.16}$ & $0.75^{+0.26}_{-0.20}$ & $0.00^{+4.16}_{-3.20}$ & ISM & 0.63 \\
090516 & 181.008 & $6.94^{+1.13}_{-0.80}$ & $15.78^{+1.36}_{-1.55}$ & $0.77^{+0.07}_{-0.08}$ & $1.75^{+0.05}_{-0.05}$ & $0.98^{+0.09}_{-0.09}$ & uncertain & ISM & 1.08 \\
090618 & 113.2 & $0.76^{+0.24}_{-0.16}$ & $293.73^{+54.16}_{-52.24}$ & $1.35^{+0.01}_{-0.01}$ & $1.78^{+0.09}_{-0.07}$ & $0.43^{+0.09}_{-0.07}$ & $2.25^{+0.28}_{-0.22}$ & wind & 1.07 \\
090621A & 219.236 & $3.22^{+0.70}_{-0.57}$ & $18.36^{+3.58}_{-3.14}$ & $0.82^{+0.04}_{-0.04}$ & $1.28^{+0.06}_{-0.06}$ & $0.46^{+0.07}_{-0.07}$ & $2.15^{+0.24}_{-0.24}$ & wind & 1.28 \\
090813 & 7.1 & $58.17^{+13.11}_{-11.15}$ & $2.65^{+0.52}_{-0.44}$ & $0.95^{+0.03}_{-0.04}$ & $1.35^{+0.03}_{-0.03}$ & $0.40^{+0.05}_{-0.04}$ & $2.33^{+0.14}_{-0.11}$ & wind & 1.1 \\
090817 & 220(Fermi) & $0.42^{+0.28}_{-0.17}$ & $160.47^{+81.67}_{-70.39}$ & $0.81^{+0.11}_{-0.11}$ & $1.44^{+0.41}_{-0.29}$ & $0.63^{+0.42}_{-0.31}$ & $1.30^{+3.07}_{-2.26}$ & wind-like & 2.53 \\
091020 & 34.6 & $7.04^{+1.71}_{-1.22}$ & $13.60^{+2.32}_{-2.32}$ & $0.98^{+0.01}_{-0.01}$ & $1.40^{+0.04}_{-0.03}$ & $0.42^{+0.04}_{-0.03}$ & $2.28^{+0.12}_{-0.09}$ & wind & 1.07 \\
091127 & 7.1 & $24.94^{+4.98}_{-4.43}$ & $35.92^{+5.23}_{-4.38}$ & $1.10^{+0.02}_{-0.02}$ & $1.53^{+0.02}_{-0.02}$ & $0.43^{+0.03}_{-0.03}$ & $2.25^{+0.09}_{-0.09}$ & wind & 0.99 \\
100413A & 191 & $68.99^{+17.94}_{-17.28}$ & $2.94^{+0.64}_{-0.45}$ & $0.96^{+0.06}_{-0.07}$ & $1.50^{+0.04}_{-0.04}$ & $0.54^{+0.08}_{-0.07}$ & $1.83^{+0.38}_{-0.33}$ & wind & 0.98 \\
100425A & 37 & $0.39^{+0.09}_{-0.07}$ & $46.30^{+11.67}_{-12.37}$ & $0.61^{+0.03}_{-0.04}$ & $1.26^{+0.18}_{-0.15}$ & $0.65^{+0.18}_{-0.15}$ & $1.14^{+1.47}_{-1.22}$ & wind-like & 0.87 \\
100621A & 63.6 & $2.06^{+0.48}_{-0.39}$ & $120.97^{+22.89}_{-20.78}$ & $1.01^{+0.03}_{-0.03}$ & $1.56^{+0.08}_{-0.07}$ & $0.55^{+0.09}_{-0.08}$ & $1.78^{+0.44}_{-0.40}$ & wind & 1.19 \\
100724A & 1.4 & $2.74^{+1.52}_{-0.91}$ & $4.78^{+1.78}_{-1.69}$ & $0.83^{+0.07}_{-0.08}$ & $1.40^{+0.16}_{-0.13}$ & $0.57^{+0.18}_{-0.15}$ & $1.67^{+0.97}_{-0.81}$ & wind & 1.25 \\
100725B & 200 & $2.25^{+0.41}_{-0.38}$ & $32.17^{+5.26}_{-4.76}$ & $0.74^{+0.05}_{-0.06}$ & $1.49^{+0.10}_{-0.09}$ & $0.75^{+0.12}_{-0.10}$ & $0.00^{+1.92}_{-1.60}$ & ISM & 0.96 \\
100728B & 12.1 & $5.52^{+1.59}_{-1.35}$ & $4.91^{+1.28}_{-0.88}$ & $0.96^{+0.03}_{-0.03}$ & $1.53^{+0.10}_{-0.08}$ & $0.57^{+0.10}_{-0.09}$ & $1.67^{+0.54}_{-0.49}$ & wind & 1.15 \\
101023A & 80.8 & $6.32^{+1.09}_{-0.86}$ & $19.89^{+2.12}_{-2.24}$ & $1.02^{+0.05}_{-0.05}$ & $1.53^{+0.05}_{-0.05}$ & $0.51^{+0.07}_{-0.07}$ & $1.96^{+0.29}_{-0.29}$ & wind & 1.02 \\
110102A & 264 & $0.54^{+0.19}_{-0.10}$ & $224.23^{+36.47}_{-45.04}$ & $1.28^{+0.02}_{-0.02}$ & $2.08^{+0.23}_{-0.19}$ & $0.80^{+0.23}_{-0.19}$ & uncertain & ISM & 0.99 \\
110106A & 4.3 & $0.62^{+0.41}_{-0.21}$ & $9.43^{+3.60}_{-3.50}$ & $1.02^{+0.05}_{-0.06}$ & $1.61^{+0.44}_{-0.24}$ & $0.59^{+0.44}_{-0.25}$ & $1.56^{+2.62}_{-1.49}$ & wind & 0.61 \\
110315A & 77 & $2.48^{+0.61}_{-0.45}$ & $18.96^{+3.50}_{-3.48}$ & $0.94^{+0.04}_{-0.04}$ & $1.52^{+0.13}_{-0.10}$ & $0.58^{+0.14}_{-0.11}$ & $1.62^{+0.79}_{-0.62}$ & wind & 0.79 \\
110422A & 25.9 & $43.82^{+9.34}_{-6.62}$ & $7.39^{+1.02}_{-0.97}$ & $1.01^{+0.03}_{-0.03}$ & $1.52^{+0.03}_{-0.02}$ & $0.51^{+0.04}_{-0.04}$ & $1.96^{+0.17}_{-0.17}$ & wind & 1.01 \\
110503A & 10 & $0.09^{+0.05}_{-0.03}$ & $532.31^{+156.64}_{-136.56}$ & $1.34^{+0.04}_{-0.04}$ & $2.19^{+0.96}_{-0.57}$ & $0.85^{+0.96}_{-0.57}$ & uncertain & ISM & 0.81 \\
110709A & 44.7 & $40.19^{+4.83}_{-3.33}$ & $5.98^{+0.37}_{-0.51}$ & $0.77^{+0.02}_{-0.02}$ & $1.73^{+0.05}_{-0.05}$ & $0.96^{+0.05}_{-0.05}$ & uncertain & ISM & 1.07 \\
110709B & 55.6 & $5.48^{+0.46}_{-0.42}$ & $54.20^{+3.90}_{-3.61}$ & $0.95^{+0.01}_{-0.01}$ & $1.57^{+0.04}_{-0.03}$ & $0.62^{+0.04}_{-0.03}$ & $1.37^{+0.28}_{-0.21}$ & wind-like & 1.03 \\
110715A & 13 & $0.48^{+0.32}_{-0.17}$ & $467.48^{+180.59}_{-150.57}$ & $1.16^{+0.06}_{-0.06}$ & $1.98^{+1.07}_{-0.46}$ & $0.82^{+1.07}_{-0.46}$ & uncertain & ISM & 1.72 \\
110801A & 385 & $1.65^{+0.71}_{-0.50}$ & $29.89^{+8.98}_{-7.08}$ & $1.22^{+0.05}_{-0.05}$ & $1.85^{+0.24}_{-0.17}$ & $0.63^{+0.25}_{-0.18}$ & $1.30^{+1.83}_{-1.31}$ & wind-like & 0.84 \\
110818A & 103 & $0.43^{+0.20}_{-0.15}$ & $43.23^{+14.09}_{-10.69}$ & $1.20^{+0.09}_{-0.10}$ & $1.79^{+0.35}_{-0.27}$ & $0.59^{+0.36}_{-0.28}$ & $1.56^{+2.14}_{-1.67}$ & wind & 1.07 \\
120119A & 253.8 & $6.20^{+0.60}_{-0.46}$ & $19.75^{+1.29}_{-1.31}$ & $1.08^{+0.04}_{-0.04}$ & $1.93^{+0.10}_{-0.09}$ & $0.85^{+0.11}_{-0.10}$ & uncertain & ISM & 2.01 \\
120909A & 220.596 & $12.48^{+3.97}_{-2.79}$ & $12.40^{+2.64}_{-2.52}$ & $1.00^{+0.07}_{-0.09}$ & $1.42^{+0.05}_{-0.05}$ & $0.42^{+0.10}_{-0.09}$ & $2.28^{+0.30}_{-0.27}$ & wind & 0.86 \\
121001A & 147 & $19.31^{+7.43}_{-6.33}$ & $3.06^{+0.93}_{-0.71}$ & $1.17^{+0.05}_{-0.05}$ & $1.75^{+0.23}_{-0.17}$ & $0.58^{+0.24}_{-0.18}$ & $1.62^{+1.36}_{-1.02}$ & wind & 0.78 \\
121024A & 69 & $2.66^{+0.49}_{-0.43}$ & $17.65^{+3.23}_{-2.65}$ & $0.86^{+0.04}_{-0.04}$ & $1.41^{+0.07}_{-0.07}$ & $0.55^{+0.08}_{-0.08}$ & $1.78^{+0.40}_{-0.40}$ & wind & 1.13 \\
121031A & 226 & $6.97^{+0.68}_{-0.60}$ & $38.02^{+3.06}_{-3.44}$ & $0.80^{+0.05}_{-0.06}$ & $1.55^{+0.33}_{-0.26}$ & $0.75^{+0.34}_{-0.26}$ & $0.00^{+5.44}_{-4.16}$ & ISM & 0.92 \\
121125A & 52.2 & $0.67^{+0.29}_{-0.18}$ & $68.44^{+17.80}_{-16.80}$ & $1.22^{+0.03}_{-0.03}$ & $1.66^{+0.47}_{-0.34}$ & $0.44^{+0.47}_{-0.34}$ & $2.21^{+1.50}_{-1.08}$ & wind & 0.64 \\
121211A & 182 & $2.34^{+0.54}_{-0.40}$ & $29.66^{+6.50}_{-5.95}$ & $0.76^{+0.05}_{-0.06}$ & $1.39^{+0.14}_{-0.11}$ & $0.63^{+0.15}_{-0.12}$ & $1.30^{+1.10}_{-0.88}$ & wind-like & 0.88 \\
121226A & 1 & $1.04^{+0.50}_{-0.35}$ & $6.58^{+2.71}_{-2.03}$ & $0.94^{+0.05}_{-0.05}$ & $1.41^{+0.24}_{-0.17}$ & $0.47^{+0.25}_{-0.18}$ & $2.11^{+0.89}_{-0.64}$ & wind & 1.85 \\
130418A & 274.924 & $11.85^{+3.42}_{-3.56}$ & $8.06^{+2.51}_{-1.49}$ & $1.12^{+0.02}_{-0.02}$ & $1.68^{+0.20}_{-0.12}$ & $0.56^{+0.20}_{-0.12}$ & $1.73^{+1.03}_{-0.62}$ & wind & 1.57 \\
130505A & 88 & $49.35^{+3.26}_{-3.40}$ & $21.48^{+1.29}_{-1.12}$ & $0.95^{+0.01}_{-0.01}$ & $1.54^{+0.02}_{-0.02}$ & $0.59^{+0.02}_{-0.02}$ & $1.56^{+0.12}_{-0.12}$ & wind & 0.83 \\
130511A & 5.43 & $2.63^{+1.13}_{-0.82}$ & $4.86^{+1.62}_{-1.48}$ & $0.91^{+0.06}_{-0.07}$ & $1.66^{+0.23}_{-0.18}$ & $0.75^{+0.24}_{-0.19}$ & $0.00^{+3.84}_{-3.04}$ & ISM & 0.56 \\
130527A & 44 & $79.43^{+10.55}_{-10.09}$ & $1.34^{+0.18}_{-0.14}$ & $0.78^{+0.06}_{-0.06}$ & $1.70^{+0.08}_{-0.07}$ & $0.92^{+0.10}_{-0.09}$ & uncertain & ISM & 0.97 \\
130606A & 276.58 & $2.76^{+0.66}_{-0.55}$ & $20.26^{+3.46}_{-3.20}$ & $1.08^{+0.03}_{-0.03}$ & $1.79^{+0.14}_{-0.12}$ & $0.71^{+0.14}_{-0.12}$ & $0.55^{+1.66}_{-1.43}$ & ISM-like & 1.03 \\
130725B & 10 & $1.24^{+0.23}_{-0.20}$ & $40.52^{+6.05}_{-6.12}$ & $0.84^{+0.06}_{-0.07}$ & $1.50^{+0.20}_{-0.16}$ & $0.66^{+0.21}_{-0.17}$ & $1.06^{+1.82}_{-1.47}$ & wind-like & 1.68 \\
130912A & 0.28 & $35.47^{+8.91}_{-8.95}$ & $0.91^{+0.21}_{-0.14}$ & $0.98^{+0.10}_{-0.11}$ & $1.53^{+0.08}_{-0.07}$ & $0.55^{+0.14}_{-0.12}$ & $1.78^{+0.69}_{-0.59}$ & wind & 1.83 \\
130925A & 160.296 & $3.53^{+0.28}_{-0.29}$ & $312.98^{+26.94}_{-22.37}$ & $0.84^{+0.02}_{-0.02}$ & $1.31^{+0.03}_{-0.02}$ & $0.47^{+0.04}_{-0.03}$ & $2.11^{+0.14}_{-0.11}$ & wind & 1.08 \\
131002A & 55.59 & $1.33^{+0.43}_{-0.28}$ & $23.72^{+6.08}_{-6.50}$ & $0.81^{+0.03}_{-0.03}$ & $1.47^{+0.22}_{-0.18}$ & $0.66^{+0.22}_{-0.18}$ & $1.06^{+1.90}_{-1.56}$ & wind-like & 2.51 \\
131227A & 18 & $12.91^{+4.34}_{-3.20}$ & $2.62^{+0.72}_{-0.62}$ & $0.98^{+0.03}_{-0.03}$ & $1.59^{+0.14}_{-0.11}$ & $0.61^{+0.14}_{-0.11}$ & $1.44^{+0.92}_{-0.72}$ & wind-like & 1.01 \\
140102A & 65 & $52.93^{+13.28}_{-9.54}$ & $4.02^{+0.68}_{-0.67}$ & $1.18^{+0.01}_{-0.01}$ & $1.61^{+0.04}_{-0.04}$ & $0.43^{+0.04}_{-0.04}$ & $2.25^{+0.12}_{-0.12}$ & wind & 1.01 \\
140323A & 104.9 & $6.49^{+1.05}_{-1.13}$ & $29.28^{+3.34}_{-2.43}$ & $1.20^{+0.06}_{-0.06}$ & $2.11^{+0.17}_{-0.15}$ & $0.91^{+0.18}_{-0.16}$ & uncertain & ISM & 1.01 \\
140423A & 134 & $3.23^{+0.55}_{-0.48}$ & $26.37^{+3.77}_{-3.49}$ & $0.98^{+0.04}_{-0.04}$ & $1.48^{+0.11}_{-0.09}$ & $0.50^{+0.12}_{-0.10}$ & $2.00^{+0.48}_{-0.40}$ & wind & 1.13 \\
\enddata
\end{deluxetable*}

\addtocounter{table}{-1}
\begin{deluxetable*}{ccccccccccc}[htpb]
    \tablewidth{0pt}
	\tablecaption{\textbf{Fitting Results of All Samples}}
	\tablehead{
		\colhead{GRB} & \colhead{${T}_{90}$} & \colhead{${F_{b}}^{a}$} & \colhead{${t_{b}}^{a}$} & \colhead{${\alpha_{1}}^{b}$} & \colhead{${\alpha_{2}}^{b}$} & \colhead{$\Delta\alpha$}
		& \colhead{k} & \colhead{${\rm CBM}^{c}$} & \colhead{$\chi^2_{r}$}\\
		\colhead{} & \colhead{(s)} & \colhead{$(\times 10^{-12}$erg/cm$^2$/s)} & \colhead{$(\times 10^{3}$s)} & \colhead{} & \colhead{} & \colhead{} & \colhead{} & \colhead{} & \colhead{}}
	\startdata
    140614A & 720 & $2.95^{+0.64}_{-0.53}$ & $10.46^{+1.69}_{-1.66}$ & $0.81^{+0.08}_{-0.08}$ & $1.59^{+0.15}_{-0.12}$ & $0.78^{+0.17}_{-0.14}$ & uncertain & ISM & 1.07 \\
140626A & 16.4 & $0.51^{+0.24}_{-0.18}$ & $20.33^{+7.62}_{-6.96}$ & $0.79^{+0.21}_{-0.25}$ & $1.49^{+0.72}_{-0.40}$ & $0.70^{+0.76}_{-0.45}$ & $0.67^{+8.44}_{-5.00}$ & ISM-like & 0.97 \\
140903A & 0.3 & $3.69^{+1.14}_{-0.87}$ & $20.32^{+5.60}_{-4.71}$ & $0.81^{+0.09}_{-0.11}$ & $1.35^{+0.14}_{-0.12}$ & $0.54^{+0.18}_{-0.15}$ & $1.83^{+0.85}_{-0.71}$ & wind & 0.75 \\
140919A & 151.3 & $1.45^{+0.28}_{-0.20}$ & $142.64^{+19.93}_{-21.18}$ & $0.98^{+0.02}_{-0.02}$ & $1.49^{+0.14}_{-0.11}$ & $0.51^{+0.14}_{-0.11}$ & $1.96^{+0.58}_{-0.46}$ & wind & 0.98 \\
141017A & 55.7 & $5.82^{+1.34}_{-1.24}$ & $14.66^{+4.31}_{-3.00}$ & $0.76^{+0.03}_{-0.04}$ & $1.31^{+0.08}_{-0.07}$ & $0.55^{+0.09}_{-0.08}$ & $1.78^{+0.44}_{-0.40}$ & wind & 1.15 \\
141031A & 920 & $0.40^{+0.22}_{-0.16}$ & $220.91^{+98.23}_{-82.56}$ & $0.85^{+0.12}_{-0.14}$ & $1.73^{+0.62}_{-0.38}$ & $0.88^{+0.64}_{-0.40}$ & uncertain & ISM & 1.43 \\
150202A & 25.7 & $0.90^{+0.55}_{-0.34}$ & $15.07^{+5.51}_{-4.93}$ & $1.22^{+0.09}_{-0.09}$ & $2.06^{+0.55}_{-0.39}$ & $0.84^{+0.56}_{-0.40}$ & uncertain & ISM & 0.93 \\
150314A & 14.79 & $372.68^{+34.28}_{-31.01}$ & $1.66^{+0.15}_{-0.14}$ & $0.91^{+0.01}_{-0.01}$ & $1.52^{+0.02}_{-0.02}$ & $0.61^{+0.02}_{-0.02}$ & $1.44^{+0.13}_{-0.13}$ & wind-like & 0.94 \\
150403A & 40.9 & $0.64^{+0.13}_{-0.13}$ & $547.05^{+89.08}_{-66.73}$ & $1.39^{+0.01}_{-0.01}$ & $1.93^{+0.20}_{-0.15}$ & $0.54^{+0.20}_{-0.15}$ & $1.83^{+0.95}_{-0.71}$ & wind & 1.46 \\
150530A & 6.62 & $2.17^{+0.97}_{-0.69}$ & $6.30^{+2.42}_{-1.84}$ & $0.86^{+0.06}_{-0.07}$ & $1.55^{+0.27}_{-0.19}$ & $0.69^{+0.28}_{-0.20}$ & $0.77^{+2.91}_{-2.08}$ & ISM-like & 0.91 \\
150607A & 26.3 & $8.77^{+2.17}_{-1.62}$ & $13.81^{+2.77}_{-2.90}$ & $0.81^{+0.02}_{-0.02}$ & $1.47^{+0.14}_{-0.13}$ & $0.66^{+0.14}_{-0.13}$ & $1.06^{+1.21}_{-1.12}$ & wind-like & 1.21 \\
151029A & 8.95 & $0.29^{+0.18}_{-0.12}$ & $10.79^{+4.82}_{-3.47}$ & $1.07^{+0.06}_{-0.05}$ & $1.75^{+0.43}_{-0.35}$ & $0.68^{+0.43}_{-0.36}$ & $0.88^{+4.20}_{-3.52}$ & ISM-like & 1.55 \\
160127A & 6.16 & $8.99^{+2.71}_{-2.04}$ & $3.62^{+0.87}_{-0.84}$ & $0.82^{+0.05}_{-0.05}$ & $1.48^{+0.08}_{-0.08}$ & $0.66^{+0.09}_{-0.09}$ & $1.06^{+0.78}_{-0.78}$ & wind-like & 0.89 \\
160131A & 325 & $1.86^{+0.23}_{-0.19}$ & $87.90^{+8.31}_{-8.20}$ & $1.12^{+0.00}_{-0.00}$ & $1.86^{+0.13}_{-0.11}$ & $0.74^{+0.13}_{-0.11}$ & $0.15^{+1.92}_{-1.63}$ & ISM & 0.95 \\
160225A & 157 & $2.49^{+0.94}_{-0.61}$ & $12.72^{+3.65}_{-3.19}$ & $0.92^{+0.04}_{-0.05}$ & $1.39^{+0.14}_{-0.11}$ & $0.47^{+0.15}_{-0.12}$ & $2.11^{+0.53}_{-0.43}$ & wind & 1.0 \\
160227A & 316.5 & $1.93^{+0.64}_{-0.54}$ & $148.56^{+38.72}_{-32.13}$ & $0.83^{+0.09}_{-0.10}$ & $1.53^{+0.10}_{-0.09}$ & $0.70^{+0.14}_{-0.13}$ & $0.67^{+1.56}_{-1.44}$ & ISM-like & 1.6 \\
160504A & 53.9 & $2.07^{+0.47}_{-0.37}$ & $22.70^{+4.72}_{-4.77}$ & $0.66^{+0.08}_{-0.08}$ & $1.43^{+0.18}_{-0.15}$ & $0.77^{+0.20}_{-0.17}$ & uncertain & ISM & 0.67 \\
160611A & 34.1 & $9.64^{+2.70}_{-2.09}$ & $15.93^{+4.36}_{-3.89}$ & $0.85^{+0.06}_{-0.06}$ & $1.49^{+0.15}_{-0.12}$ & $0.64^{+0.16}_{-0.13}$ & $1.22^{+1.23}_{-1.00}$ & wind-like & 1.12 \\
160826A & 52.6 & $17.47^{+7.81}_{-5.47}$ & $1.36^{+0.42}_{-0.31}$ & $1.38^{+0.06}_{-0.06}$ & $2.09^{+0.20}_{-0.16}$ & $0.71^{+0.21}_{-0.17}$ & $0.55^{+2.50}_{-2.02}$ & ISM-like & 0.92 \\
160917A & 16 & $1.88^{+1.05}_{-0.61}$ & $10.36^{+3.51}_{-2.87}$ & $1.22^{+0.02}_{-0.02}$ & $2.03^{+0.55}_{-0.35}$ & $0.81^{+0.55}_{-0.35}$ & uncertain & ISM & 1.13 \\
161004B & 15.9 & $18.25^{+5.46}_{-4.21}$ & $2.59^{+0.75}_{-0.64}$ & $0.85^{+0.03}_{-0.03}$ & $1.44^{+0.10}_{-0.09}$ & $0.59^{+0.10}_{-0.09}$ & $1.56^{+0.59}_{-0.54}$ & wind & 0.9 \\
161017A & 216.3 & $3.10^{+0.47}_{-0.39}$ & $42.59^{+4.22}_{-4.55}$ & $1.02^{+0.04}_{-0.04}$ & $1.82^{+0.11}_{-0.10}$ & $0.80^{+0.12}_{-0.11}$ & uncertain & ISM & 1.02 \\
161202A & 118.496 & $7.93^{+2.03}_{-1.61}$ & $8.67^{+1.78}_{-1.54}$ & $0.89^{+0.03}_{-0.03}$ & $1.43^{+0.07}_{-0.06}$ & $0.54^{+0.08}_{-0.07}$ & $1.83^{+0.38}_{-0.33}$ & wind & 1.21 \\
161214B & 24.8 & $0.63^{+0.27}_{-0.17}$ & $89.14^{+30.25}_{-27.94}$ & $0.87^{+0.08}_{-0.08}$ & $1.52^{+0.28}_{-0.20}$ & $0.65^{+0.29}_{-0.22}$ & $1.14^{+2.37}_{-1.80}$ & wind-like & 0.71 \\
161219B & 6.94 & $0.98^{+0.20}_{-0.16}$ & $1089.59^{+185.74}_{-178.72}$ & $0.85^{+0.06}_{-0.06}$ & $1.26^{+0.06}_{-0.05}$ & $0.41^{+0.08}_{-0.08}$ & $2.31^{+0.23}_{-0.23}$ & wind & 1.23 \\
170317A & 11.94 & $2.69^{+0.66}_{-0.57}$ & $8.00^{+1.73}_{-1.42}$ & $0.95^{+0.06}_{-0.06}$ & $1.60^{+0.19}_{-0.15}$ & $0.65^{+0.20}_{-0.16}$ & $1.14^{+1.63}_{-1.31}$ & wind-like & 0.98 \\
170405A & 164.7 & $22.56^{+2.61}_{-2.44}$ & $5.50^{+0.48}_{-0.40}$ & $0.96^{+0.05}_{-0.06}$ & $1.79^{+0.08}_{-0.07}$ & $0.83^{+0.10}_{-0.09}$ & uncertain & ISM & 0.85 \\
170519A & 216.4 & $4.23^{+0.93}_{-0.76}$ & $20.71^{+3.63}_{-3.34}$ & $0.99^{+0.05}_{-0.06}$ & $1.42^{+0.09}_{-0.08}$ & $0.43^{+0.11}_{-0.09}$ & $2.25^{+0.34}_{-0.28}$ & wind & 0.99 \\
170711A & 31.3 & $1.04^{+0.33}_{-0.25}$ & $57.10^{+19.78}_{-19.62}$ & $0.62^{+0.06}_{-0.07}$ & $1.43^{+0.28}_{-0.23}$ & $0.81^{+0.29}_{-0.24}$ & uncertain & ISM & 1.21 \\
170810A & 152.4 & $2.86^{+0.62}_{-0.50}$ & $29.08^{+4.64}_{-4.23}$ & $1.00^{+0.03}_{-0.03}$ & $1.73^{+0.17}_{-0.13}$ & $0.73^{+0.17}_{-0.13}$ & $0.30^{+2.33}_{-1.78}$ & ISM & 1.56 \\
171027A & 96.6 & $5.50^{+1.06}_{-1.01}$ & $16.41^{+3.18}_{-2.27}$ & $0.92^{+0.03}_{-0.03}$ & $1.46^{+0.07}_{-0.06}$ & $0.54^{+0.08}_{-0.07}$ & $1.83^{+0.38}_{-0.33}$ & wind & 1.06 \\
171102B & 17.8 & $2.70^{+1.31}_{-0.96}$ & $18.50^{+6.38}_{-4.40}$ & $1.16^{+0.12}_{-0.14}$ & $2.06^{+0.51}_{-0.37}$ & $0.90^{+0.53}_{-0.39}$ & uncertain & ISM & 0.86 \\
180626A & 30.07 & $2.18^{+0.69}_{-0.45}$ & $41.58^{+10.18}_{-10.43}$ & $0.87^{+0.05}_{-0.05}$ & $1.58^{+0.19}_{-0.15}$ & $0.71^{+0.20}_{-0.16}$ & $0.55^{+2.38}_{-1.90}$ & ISM-like & 0.87 \\
180809B & 233.2 & $15.53^{+5.14}_{-4.34}$ & $58.41^{+13.19}_{-10.82}$ & $1.24^{+0.06}_{-0.05}$ & $2.20^{+0.17}_{-0.15}$ & $0.96^{+0.18}_{-0.16}$ & uncertain & ISM & 0.95 \\
180823A & 80.3 & $5.98^{+1.74}_{-1.80}$ & $8.00^{+3.03}_{-1.72}$ & $0.83^{+0.05}_{-0.05}$ & $1.29^{+0.09}_{-0.08}$ & $0.46^{+0.10}_{-0.09}$ & $2.15^{+0.34}_{-0.31}$ & wind & 1.41 \\
190109B & 5.92 & $1.38^{+0.55}_{-0.37}$ & $7.21^{+2.42}_{-2.01}$ & $0.85^{+0.05}_{-0.05}$ & $1.68^{+0.41}_{-0.27}$ & $0.83^{+0.41}_{-0.27}$ & uncertain & ISM & 1.0 \\
190203A & 96 & $236.49^{+28.12}_{-25.38}$ & $2.65^{+0.25}_{-0.24}$ & $0.93^{+0.02}_{-0.02}$ & $1.68^{+0.03}_{-0.03}$ & $0.75^{+0.04}_{-0.04}$ & $0.00^{+0.64}_{-0.64}$ & ISM & 0.96 \\
190828B & 66.6 & $2.92^{+0.78}_{-0.61}$ & $40.00^{+8.70}_{-8.21}$ & $0.95^{+0.03}_{-0.03}$ & $1.73^{+0.15}_{-0.13}$ & $0.78^{+0.15}_{-0.13}$ & uncertain & ISM & 0.94 \\
191122A & 153.3 & $172.24^{+31.12}_{-25.93}$ & $0.71^{+0.09}_{-0.08}$ & $1.22^{+0.03}_{-0.03}$ & $1.75^{+0.04}_{-0.04}$ & $0.53^{+0.05}_{-0.05}$ & $1.87^{+0.23}_{-0.23}$ & wind & 1.57 \\
200711A & 29.39 & $11.44^{+5.22}_{-3.80}$ & $12.50^{+3.82}_{-3.09}$ & $1.11^{+0.13}_{-0.16}$ & $1.69^{+0.16}_{-0.12}$ & $0.58^{+0.23}_{-0.18}$ & $1.62^{+1.30}_{-1.02}$ & wind & 1.32 \\
200713A & 48.98 & $1.58^{+0.56}_{-0.54}$ & $56.02^{+19.99}_{-13.07}$ & $0.88^{+0.09}_{-0.11}$ & $1.50^{+0.25}_{-0.21}$ & $0.62^{+0.27}_{-0.23}$ & $1.37^{+1.87}_{-1.59}$ & wind-like & 1.22 \\
200716C & 86 & $48.28^{+12.61}_{-11.04}$ & $3.46^{+0.64}_{-0.51}$ & $1.01^{+0.04}_{-0.05}$ & $1.62^{+0.05}_{-0.04}$ & $0.61^{+0.07}_{-0.06}$ & $1.44^{+0.46}_{-0.39}$ & wind-like & 1.12 \\
200829A & 13.04 & $62.24^{+9.77}_{-9.35}$ & $13.77^{+1.58}_{-1.30}$ & $1.27^{+0.01}_{-0.01}$ & $1.72^{+0.03}_{-0.03}$ & $0.45^{+0.03}_{-0.03}$ & $2.18^{+0.10}_{-0.10}$ & wind & 0.92 \\
201209A & 48 & $4.30^{+0.82}_{-0.63}$ & $94.74^{+13.01}_{-13.98}$ & $0.88^{+0.03}_{-0.03}$ & $1.62^{+0.11}_{-0.09}$ & $0.74^{+0.11}_{-0.09}$ & $0.15^{+1.63}_{-1.33}$ & ISM & 0.9 \\
210104A & 32.06 & $207.83^{+18.86}_{-17.37}$ & $2.61^{+0.26}_{-0.25}$ & $0.75^{+0.02}_{-0.02}$ & $1.36^{+0.03}_{-0.03}$ & $0.61^{+0.04}_{-0.04}$ & $1.44^{+0.26}_{-0.26}$ & wind-like & 0.94 \\
210112A & 107.6 & $44.69^{+5.47}_{-5.55}$ & $8.97^{+1.01}_{-0.86}$ & $0.72^{+0.03}_{-0.03}$ & $1.42^{+0.03}_{-0.03}$ & $0.70^{+0.04}_{-0.04}$ & $0.67^{+0.44}_{-0.44}$ & ISM-like & 1.22 \\
210207B & 95.69 & $27.73^{+2.68}_{-2.55}$ & $8.56^{+0.65}_{-0.63}$ & $0.74^{+0.02}_{-0.02}$ & $1.69^{+0.05}_{-0.05}$ & $0.95^{+0.05}_{-0.05}$ & uncertain & ISM & 0.8 \\
210308A & 5.3 & $27.01^{+4.36}_{-4.22}$ & $2.18^{+0.34}_{-0.28}$ & $0.86^{+0.03}_{-0.03}$ & $1.59^{+0.09}_{-0.08}$ & $0.73^{+0.09}_{-0.09}$ & $0.30^{+1.23}_{-1.23}$ & ISM & 0.95 \\
210410A & 52.88 & $3.82^{+1.65}_{-1.18}$ & $6.97^{+2.39}_{-1.95}$ & $1.06^{+0.04}_{-0.03}$ & $1.77^{+0.33}_{-0.24}$ & $0.71^{+0.33}_{-0.24}$ & $0.55^{+3.92}_{-2.85}$ & ISM-like & 0.87 \\
210411C & 12.8 & $6.05^{+2.27}_{-1.85}$ & $10.40^{+3.05}_{-2.40}$ & $1.00^{+0.05}_{-0.05}$ & $1.61^{+0.22}_{-0.17}$ & $0.61^{+0.23}_{-0.18}$ & $1.44^{+1.51}_{-1.18}$ & wind-like & 0.94 \\
\enddata
\end{deluxetable*}

\addtocounter{table}{-1}
\begin{deluxetable*}{ccccccccccc}[htpb]
    \tablewidth{0pt}
	\tablecaption{\textbf{Fitting Results of All Samples}}
	\tablehead{
		\colhead{GRB} & \colhead{${T}_{90}$} & \colhead{${F_{b}}^{a}$} & \colhead{${t_{b}}^{a}$} & \colhead{${\alpha_{1}}^{b}$} & \colhead{${\alpha_{2}}^{b}$} & \colhead{$\Delta\alpha$}
		& \colhead{k} & \colhead{${\rm CBM}^{c}$} & \colhead{$\chi^2_{r}$}\\
		\colhead{} & \colhead{(s)} & \colhead{$(\times 10^{-12}$erg/cm$^2$/s)} & \colhead{$(\times 10^{3}$s)} & \colhead{} & \colhead{} & \colhead{} & \colhead{} & \colhead{} & \colhead{}}
	\startdata
    210420B & 158.8 & $9.58^{+5.43}_{-4.10}$ & $6.05^{+2.69}_{-1.86}$ & $0.91^{+0.12}_{-0.13}$ & $1.67^{+0.18}_{-0.14}$ & $0.76^{+0.22}_{-0.18}$ & uncertain & ISM & 1.53 \\
210517A & 3.06 & $1.65^{+0.80}_{-0.57}$ & $6.49^{+3.02}_{-2.21}$ & $0.81^{+0.05}_{-0.06}$ & $1.26^{+0.18}_{-0.13}$ & $0.45^{+0.19}_{-0.14}$ & $2.18^{+0.63}_{-0.46}$ & wind & 0.81 \\
210610B & 69.38 & $3.57^{+0.68}_{-0.54}$ & $78.90^{+10.66}_{-10.44}$ & $1.10^{+0.01}_{-0.01}$ & $1.97^{+0.17}_{-0.14}$ & $0.87^{+0.17}_{-0.14}$ & uncertain & ISM & 0.86 \\
210619B & 60.9 & $124.07^{+10.66}_{-9.84}$ & $12.62^{+0.94}_{-0.85}$ & $0.95^{+0.01}_{-0.01}$ & $1.48^{+0.02}_{-0.02}$ & $0.53^{+0.02}_{-0.02}$ & $1.87^{+0.09}_{-0.09}$ & wind & 0.9 \\
210702A & 138.2 & $185.40^{+20.70}_{-19.29}$ & $4.46^{+0.42}_{-0.37}$ & $0.98^{+0.01}_{-0.01}$ & $1.45^{+0.02}_{-0.02}$ & $0.47^{+0.02}_{-0.02}$ & $2.11^{+0.07}_{-0.07}$ & wind & 0.88 \\
210722A & 50.2 & $12.21^{+2.34}_{-1.86}$ & $9.48^{+1.32}_{-1.40}$ & $0.82^{+0.04}_{-0.04}$ & $1.58^{+0.08}_{-0.07}$ & $0.76^{+0.09}_{-0.08}$ & uncertain & ISM & 0.97 \\
210818A & 73.56 & $322.33^{+63.93}_{-71.55}$ & $0.52^{+0.12}_{-0.07}$ & $0.70^{+0.12}_{-0.14}$ & $1.49^{+0.04}_{-0.04}$ & $0.79^{+0.15}_{-0.13}$ & uncertain & ISM & 1.21 \\
210822A & 180.8 & $80.46^{+18.76}_{-15.39}$ & $10.85^{+1.60}_{-1.42}$ & $1.17^{+0.02}_{-0.02}$ & $1.77^{+0.04}_{-0.04}$ & $0.60^{+0.04}_{-0.04}$ & $1.50^{+0.25}_{-0.25}$ & wind-like & 0.91 \\
220101A & 173.36 & $10.49^{+1.29}_{-0.95}$ & $79.69^{+6.35}_{-7.07}$ & $1.14^{+0.00}_{-0.00}$ & $1.73^{+0.05}_{-0.04}$ & $0.59^{+0.05}_{-0.04}$ & $1.56^{+0.30}_{-0.24}$ & wind & 1.26 \\
220430A & 43.1 & $24.08^{+6.60}_{-6.35}$ & $7.92^{+1.94}_{-1.38}$ & $1.06^{+0.02}_{-0.02}$ & $1.56^{+0.06}_{-0.06}$ & $0.50^{+0.06}_{-0.06}$ & $2.00^{+0.24}_{-0.24}$ & wind & 0.85 \\
221201A & 33.71 & $4.74^{+0.70}_{-0.67}$ & $36.26^{+5.25}_{-4.88}$ & $0.87^{+0.01}_{-0.01}$ & $1.79^{+0.29}_{-0.23}$ & $0.92^{+0.29}_{-0.23}$ & uncertain & ISM & 1.48 \\
230628E & 4.72 & $0.41^{+0.13}_{-0.10}$ & $47.22^{+12.51}_{-11.24}$ & $0.94^{+0.06}_{-0.06}$ & $1.61^{+0.36}_{-0.26}$ & $0.67^{+0.36}_{-0.27}$ & $0.97^{+3.31}_{-2.48}$ & ISM-like & 1.12 \\
231118A & 37.63 & $14.83^{+3.61}_{-3.08}$ & $8.56^{+1.70}_{-1.46}$ & $0.84^{+0.03}_{-0.03}$ & $1.63^{+0.11}_{-0.09}$ & $0.79^{+0.11}_{-0.09}$ & uncertain & ISM & 0.94 \\
231129A & 105.46 & $1.73^{+0.50}_{-0.34}$ & $30.59^{+6.24}_{-7.10}$ & $0.82^{+0.04}_{-0.04}$ & $1.66^{+0.29}_{-0.22}$ & $0.84^{+0.29}_{-0.22}$ & uncertain & ISM & 1.48 \\
231205B & 46.94 & $13.92^{+3.40}_{-3.25}$ & $25.88^{+5.29}_{-3.71}$ & $1.05^{+0.05}_{-0.06}$ & $1.63^{+0.06}_{-0.06}$ & $0.58^{+0.08}_{-0.08}$ & $1.62^{+0.45}_{-0.45}$ & wind & 0.85 \\
231214A & 27.8 & $6.68^{+3.29}_{-1.98}$ & $4.67^{+1.22}_{-0.98}$ & $1.30^{+0.08}_{-0.10}$ & $2.04^{+0.34}_{-0.24}$ & $0.74^{+0.35}_{-0.25}$ & $0.15^{+5.18}_{-3.70}$ & ISM & 1.61 \\
231230A & 15.2 & $5.89^{+1.21}_{-0.91}$ & $14.28^{+2.32}_{-2.31}$ & $0.84^{+0.02}_{-0.02}$ & $1.41^{+0.09}_{-0.08}$ & $0.57^{+0.09}_{-0.08}$ & $1.67^{+0.49}_{-0.43}$ & wind & 1.15 \\
240809A & 72.86 & $457.18^{+35.55}_{-34.23}$ & $2.87^{+0.20}_{-0.19}$ & $0.83^{+0.02}_{-0.02}$ & $1.57^{+0.02}_{-0.02}$ & $0.74^{+0.03}_{-0.03}$ & $0.15^{+0.44}_{-0.44}$ & ISM & 1.03 \\
241010A & 30.86 & $3.93^{+0.90}_{-0.73}$ & $51.71^{+9.15}_{-7.79}$ & $0.90^{+0.04}_{-0.04}$ & $1.64^{+0.10}_{-0.08}$ & $0.74^{+0.11}_{-0.09}$ & $0.15^{+1.63}_{-1.33}$ & ISM & 1.09 \\
241228A & 8.06 & $14.58^{+3.28}_{-2.56}$ & $5.09^{+0.83}_{-0.73}$ & $0.97^{+0.02}_{-0.02}$ & $1.45^{+0.06}_{-0.06}$ & $0.48^{+0.06}_{-0.06}$ & $2.08^{+0.22}_{-0.22}$ & wind & 2.0 \\
\enddata
\tablenotetext{a}{Parameters at the jet break: $F_b$ is the flux at the break time $t_b$.}
\tablenotetext{b}{Temporal decay indices: $\alpha_1$ (pre-break) and $\alpha_2$ (post-break).}
\tablenotetext{c}{Circum-burst medium (CBM) type as classified by the density profile index $k$. }
\end{deluxetable*}

\begin{table*}[htpb]
\centering
\caption{\textbf{Parameters of GRBs with Known Redshift}}
\label{tab:Parameters}
\begin{tabular*}{\textwidth}{@{\extracolsep{\fill}}ccccccc}
\toprule
GRB & redshift & $E_{\gamma,iso}$ & ${\theta_{j}}$ & ${E_{\gamma}}$ & CBM$^{a}$ & References$^{b}$\\
    &         & $(\times 10^{52}$erg) & (rad) & $(\times 10^{50}$erg) & &\\
\midrule
050315 & 1.949 & 3.3 $\pm$ 6.2 & 0.097 $\pm$ 0.023 & 1.55 $\pm$ 3.01 & ISM & 13 \\
050318 & 1.44 & 2.3 $\pm$ 0.16 & 0.04 $\pm$ 0.002 & 0.18 $\pm$ 0.03 & ISM & 1 \\
050408 & 1.236 & 1.3 & 0.131 $\pm$ 0.009 & 1.11 $\pm$ 0.15 & wind & 2 \\
050505 & 4.2748 & 17.6 $\pm$ 2.61 & 0.039 $\pm$ 0.002 & 1.37 $\pm$ 0.26 & wind-like & 3 \\
050820A & 2.612 & 103.36 $\pm$ 8.23 & 0.051 $\pm$ 0.002 & 13.67 $\pm$ 1.69 & wind & 1 \\
050915A & 2.5273 & 4.84 & 0.041 $\pm$ 0.002 & 0.4 & wind & 4 \\
050922C & 2.198 & 5.3 $\pm$ 1.7 & 0.059 $\pm$ 0.006 & 0.92 $\pm$ 0.35 & ISM & 3 \\
051016B & 0.94 & 0.1 & 0.188 $\pm$ 0.01 & 0.18 & wind-like & 5 \\
051221A & 0.5465 & 0.91 $\pm$ 0.13 & 0.044 $\pm$ 0.002 & 0.09 $\pm$ 0.01 & ISM-like & 3 \\
051227 & 0.8 & 0.029 & 0.327 $\pm$ 0.021 & 0.15 & ISM & 6 \\
060111B & 1.5 & 1.18 & 0.058 $\pm$ 0.003 & 0.2 & wind & 14 \\
060116 & 4.0 & 3.02 $\pm$ 0.278 & 0.028 $\pm$ 0.002 & 0.12 $\pm$ 0.02 & ISM & 15 \\
060204B & 2.3393 & 8.31 & 0.061 $\pm$ 0.005 & 1.55 & wind-like & 4 \\
060210 & 3.91 & 32.23 $\pm$ 1.84 & 0.031 $\pm$ 0.001 & 1.6 $\pm$ 0.15 & wind & 1 \\
060418 & 1.49 & 13.55 $\pm$ 2.71 & 0.022 $\pm$ 0.002 & 0.34 $\pm$ 0.08 & wind & 1 \\
060510A & 1.2 & 0.0741 $\pm$ 0.003 & 0.111 $\pm$ 0.009 & 0.05 $\pm$ 0.01 & ISM & 14 \\
060614 & 0.12 & 0.21 $\pm$ 0.09 & 0.273 $\pm$ 0.029 & 0.78 $\pm$ 0.37 & ISM & 7 \\
060927 & 5.47 & 12.02 $\pm$ 2.77 & 0.019 $\pm$ 0.002 & 0.21 $\pm$ 0.06 & ISM & 1 \\
061004 & 3.3 & 0.87 & 0.074 $\pm$ 0.012 & 0.24 & ISM & 6 \\
061121 & 1.314 & 23.5 $\pm$ 2.7 & 0.063 $\pm$ 0.007 & 4.65 $\pm$ 1.19 & ISM & 1 \\
061222A & 2.088 & 31.1 & 0.047 $\pm$ 0.002 & 3.44 & wind & 4 \\
070318 & 0.84 & 0.9 $\pm$ 0.2 & 0.073 $\pm$ 0.004 & 0.24 $\pm$ 0.06 & ISM & 7 \\
070411 & 2.95 & 10.0 $\pm$ 8.0 & 0.036 $\pm$ 0.008 & 0.64 $\pm$ 0.59 & wind & 7 \\
070521 & 2.0865 & 14.75 $\pm$ 0.46 & 0.014 $\pm$ 0.001 & 0.15 $\pm$ 0.02 & ISM & 3 \\
071112C & 0.823 & 4.14 & 0.069 $\pm$ 0.007 & 0.99 & wind & 4 \\
080310 & 2.43 & 5.89 $\pm$ 1.08 & 0.035 $\pm$ 0.002 & 0.36 $\pm$ 0.07 & ISM & 8 \\
080319C & 1.95 & 15.5 $\pm$ 1.6 & 0.02 $\pm$ 0.002 & 0.31 $\pm$ 0.06 & ISM & 3 \\
080411 & 1.03 & 23.5 & 0.106 $\pm$ 0.005 & 13.08 & wind & 4 \\
080413B & 1.1 & 1.61 $\pm$ 0.27 & 0.055 $\pm$ 0.008 & 0.24 $\pm$ 0.08 & ISM & 1 \\
080605 & 1.6398 & 24.0 $\pm$ 4.7 & 0.025 $\pm$ 0.002 & 0.72 $\pm$ 0.19 & wind & 4 \\
080607 & 3.036 & 217.0 $\pm$ 6.0 & 0.025 $\pm$ 0.002 & 6.71 $\pm$ 1.09 & wind-like & 3 \\
080710 & 0.85 & 1.68 $\pm$ 0.22 & 0.058 $\pm$ 0.002 & 0.28 $\pm$ 0.04 & ISM & 9 \\
080721 & 2.591 & 150.9 $\pm$ 6.3 & 0.046 $\pm$ 0.003 & 15.64 $\pm$ 2.02 & ISM & 3 \\
081221 & 2.26 & 39.7 $\pm$ 1.0 & 0.034 $\pm$ 0.003 & 2.35 $\pm$ 0.39 & ISM & 3 \\
081230 & 2.0 & 17.38 $\pm$ 1.6 & 0.048 $\pm$ 0.003 & 2.01 $\pm$ 0.33 & ISM & 14 \\
090102 & 1.547 & 23.5 & 0.02 $\pm$ 0.002 & 0.48 & wind & 4 \\
090201 & 2.1 & 95.5 $\pm$ 2.8 & 0.027 $\pm$ 0.001 & 3.55 $\pm$ 0.24 & ISM & 3 \\
090516 & 4.109 & 88.5 $\pm$ 19.2 & 0.034 $\pm$ 0.003 & 5.23 $\pm$ 1.55 & ISM & 3 \\
090618 & 0.54 & 20.0 & 0.082 $\pm$ 0.004 & 6.76 & wind & 10 \\
091020 & 1.71 & 6.85 & 0.043 $\pm$ 0.002 & 0.64 & wind & 4 \\
091127 & 0.49 & 1.63 $\pm$ 0.02 & 0.092 $\pm$ 0.003 & 0.69 $\pm$ 0.05 & wind & 3 \\
100413A & 3.9 & 10.99 & 0.023 $\pm$ 0.001 & 0.28 & wind & 6 \\
100425A & 1.755 & 0.0942 & 0.171 $\pm$ 0.011 & 0.14 & wind-like & 4 \\
100621A & 0.542 & 4.39 & 0.096 $\pm$ 0.005 & 2.03 & wind & 4 \\
100724A & 1.288 & 0.149 & 0.091 $\pm$ 0.008 & 0.06 & wind & 4 \\
100728B & 2.106 & 3.55 $\pm$ 0.36 & 0.038 $\pm$ 0.003 & 0.26 $\pm$ 0.04 & wind & 16 \\
110422A & 1.77 & 72.0 & 0.021 $\pm$ 0.001 & 1.52 & wind & 4 \\
110503A & 1.613 & 21.28 $\pm$ 0.95 & 0.058 $\pm$ 0.006 & 3.58 $\pm$ 0.75 & ISM & 3 \\
110709B & 2.09 & 26.9 & 0.042 $\pm$ 0.001 & 2.38 & wind-like & 4 \\
\bottomrule
\end{tabular*}

\vspace{6pt}
\begin{minipage}{\textwidth}
\small
\noindent\textsuperscript{a}Circum-burst medium (CBM) type as classified by the density profile index $k$.\\
\noindent\textsuperscript{b}Reference for $E_{\gamma,iso}$.
\end{minipage}
\end{table*}

\addtocounter{table}{-1}
\begin{table*}[htpb]
\centering
\caption{\textbf{Parameters of GRBs with Known Redshift (continued)}}
\begin{tabular*}{\textwidth}{@{\extracolsep{\fill}}ccccccc}
\toprule
GRB & redshift & $E_{\gamma,iso}$ & ${\theta_{j}}$ & ${E_{\gamma}}$ & CBM$^{a}$ & References$^{b}$\\
    &         & $(\times 10^{52}$erg) & (rad) & $(\times 10^{50}$erg) & &\\
\midrule
110715A & 0.82 & 4.97 $\pm$ 0.33 & 0.089 $\pm$ 0.011 & 1.97 $\pm$ 0.5 & ISM & 3 \\
110801A & 1.858 & 9.65 & 0.048 $\pm$ 0.004 & 1.1 & wind-like & 4 \\
110818A & 3.36 & 17.1 & 0.041 $\pm$ 0.003 & 1.43 & wind & 4 \\
120119A & 1.728 & 27.2 $\pm$ 3.63 & 0.023 $\pm$ 0.002 & 0.72 $\pm$ 0.17 & ISM & 1 \\
120909A & 3.93 & 37.8 & 0.024 $\pm$ 0.001 & 1.07 & wind & 4 \\
121024A & 2.298 & 5.0 & 0.048 $\pm$ 0.002 & 0.57 & wind & 4 \\
121211A & 1.023 & 0.63 $\pm$ 0.54 & 0.103 $\pm$ 0.023 & 0.33 $\pm$ 0.32 & wind-like & 11 \\
130418A & 1.218 & 9.34 & 0.037 $\pm$ 0.003 & 0.64 & wind & 4 \\
130505A & 2.27 & 438.0 $\pm$ 10.0 & 0.016 & 5.88 $\pm$ 0.23 & wind & 3 \\
130511A & 1.3033 & 0.418 & 0.14 $\pm$ 0.016 & 0.41 & ISM & 4 \\
130606A & 5.91 & 28.3 $\pm$ 5.2 & 0.014 $\pm$ 0.002 & 0.28 $\pm$ 0.08 & ISM-like & 11 \\
130925A & 0.347 & 15.0 $\pm$ 0.3 & 0.093 $\pm$ 0.002 & 6.46 $\pm$ 0.31 & wind & 3 \\
131227A & 5.3 & 14.2 & 0.019 $\pm$ 0.001 & 0.27 & wind-like & 4 \\
140423A & 3.26 & 43.8 $\pm$ 0.3 & 0.029 $\pm$ 0.001 & 1.8 $\pm$ 0.13 & wind & 3 \\
140903A & 0.351 & 0.0044 $\pm$ 0.0003 & 0.358 $\pm$ 0.025 & 0.03 & wind & 3 \\
150314A & 1.758 & 76.8 $\pm$ 2.1 & 0.014 & 0.74 $\pm$ 0.04 & wind-like & 3 \\
150403A & 2.06 & 116.4 $\pm$ 6.2 & 0.052 $\pm$ 0.002 & 15.8 $\pm$ 1.59 & wind & 3 \\
151029A & 1.423 & 0.288 $\pm$ 0.037 & 0.063 $\pm$ 0.011 & 0.06 $\pm$ 0.02 & ISM-like & 3 \\
160131A & 0.97 & 87.0 $\pm$ 6.6 & 0.084 $\pm$ 0.007 & 30.76 $\pm$ 5.3 & ISM & 3 \\
160227A & 2.38 & 5.56 $\pm$ 0.36 & 0.069 $\pm$ 0.008 & 1.34 $\pm$ 0.33 & ISM-like & 3 \\
161017A & 2.013 & 8.3 $\pm$ 1.1 & 0.106 $\pm$ 0.008 & 4.62 $\pm$ 0.9 & ISM & 3 \\
161219B & 0.1475 & 0.0116 $\pm$ 0.0008 & 0.792 $\pm$ 0.036 & 0.34 $\pm$ 0.04 & wind & 3 \\
170405A & 3.51 & 9.01 & 0.032 $\pm$ 0.002 & 0.45 & ISM & 12 \\
170519A & 0.818 & 0.2 $\pm$ 0.04 & 0.129 $\pm$ 0.009 & 0.17 $\pm$ 0.04 & wind & 16 \\
\bottomrule
\end{tabular*}

\vspace{6pt}
\begin{minipage}{\textwidth}
\small
\noindent\textsuperscript{a}Circum-burst medium (CBM) type as classified by the density profile index $k$.\\
\noindent\textsuperscript{b}Reference for $E_{\gamma,iso}$.
\end{minipage}

\vspace{6pt}
\begin{minipage}{\textwidth}
\small
\noindent\textbf{References.} 1. \cite{2017A&A...598A.112D}, 2. \cite{2018A&A...620A.190D}, 3. \cite{2020MNRAS.492.1919M}, 4. \cite{2016ApJS..227....7L}, 5. \cite{2012ApJ...745..168L}, 6. This paper, 7. \cite{2018MNRAS.477.2173S}, 8. \cite{2010ApJ...720.1513K}, 9. \cite{2016ApJ...832..136R}, 10. \cite{2011MNRAS.418.1511C}, 11. \cite{2017RAA....17...53X}, 12. \cite{2020ApJ...900..112Z}, 13. \cite{2006ApJ...638..920V}, 14. \cite{2023ApJ...943..126D}, 15. \cite{2019ApJS..245....1T}, 16. \cite{2018ApJ...863...50S}
\end{minipage}
\end{table*}


\end{document}